\documentclass[a4paper,12pt]{book}
\usepackage{psfrag}
\usepackage{amsmath}
\usepackage{amssymb}
\usepackage{slashed}
\usepackage{hyperref}
\usepackage{subcaption}
\usepackage{tikz-feynman}
\usepackage{axodraw4j}
\usepackage{braket}
\usepackage{chngcntr}
\usepackage{bbold}
\usepackage[backend=biber,style=numeric,sorting=none,maxbibnames=99]{biblatex}
\usepackage{graphicx}
\usepackage{epstopdf}
\usepackage{caption}
\usepackage{subcaption}
\usepackage{pdfpages}
\usepackage[titletoc]{appendix}
\graphicspath{{./images/}}
\counterwithout{footnote}{chapter}
\hypersetup{
pdfsubject={PhD Degree Thesis},
colorlinks=true,
citecolor=black,
filecolor=black,
linkcolor=black,
urlcolor=blue
}
\usepackage[left=3cm, right=2cm, top=3.5 cm, bottom=3.5 cm]{geometry}
\usepackage{axodraw4j}
\usepackage{pstricks}
\usepackage{color}
\usepackage{float}

\newcommand\numberthis[1][]{%
    \refstepcounter{equation}%
    \ifx#1\empty\else\label{eq:#1}\fi%
    \tag{\theequation}%
}

\providecommand{\U}[1]{\protect\rule{.1in}{.1in}}

\def\slashchar#1{\setbox0=\hbox{$#1$}
   \dimen0=\wd0
   \setbox1=\hbox{/} \dimen1=\wd1
   \ifdim\dimen0>\dimen1
      \rlap{\hbox to \dimen0{\hfil/\hfil}}
      #1
   \else
      \rlap{\hbox to \dimen1{\hfil$#1$\hfil}}
      /
   \fi}

\def\bei{\begin{itemize}}
\def\ei{\end{itemize}}

\def\beeq{\begin{eqnarray}} 
\def\beqa{\begin{eqnarray}}
\def\bea{\begin{eqnarray}}

\def\eea{\end{eqnarray}}
\def\eqa{\end{eqnarray}}
\def\eeeq{\end{eqnarray}}

\def\eqar{\end{array}}
\def\beqar{\begin{array}}

\def\beas{\begin{eqnarray*}}
\def\beqas{\begin{eqnarray*}}

\def\eqas{\end{eqnarray*}}
\def\eeas{\end{eqnarray*}}

\def\beq{\begin{equation}} 
\def\be{\begin{equation}}

\def\ee{\end{equation}}
\def\eq{\end{equation}}
\def\eeq{\end{equation}}

\def\beqd{\begin{displaymath}}
\def\eeqd{\end{displaymath}}
\def\eqd{\end{displaymath}}

\def\beeq{\begin{eqnarray}} \def\eeeq{\end{eqnarray}}

\newcommand{\fin}{\end{document}}

\allowdisplaybreaks

\newcommand{\MSb}{\overline{\rm MS}}

\newcommand{\DY}{\Delta Y}
\newcommand{\JPsi}{J/\psi}
\newcommand{\Yps}{\Upsilon}

\newcommand{\cM}{{\cal M}}

\newcommand{\Arrow}[1]{%
\parbox{#1}{\tikz{\draw[->](0,0)--(#1,0);}}
}

\begin{document}

\noindent

\includepdf[pages=-]{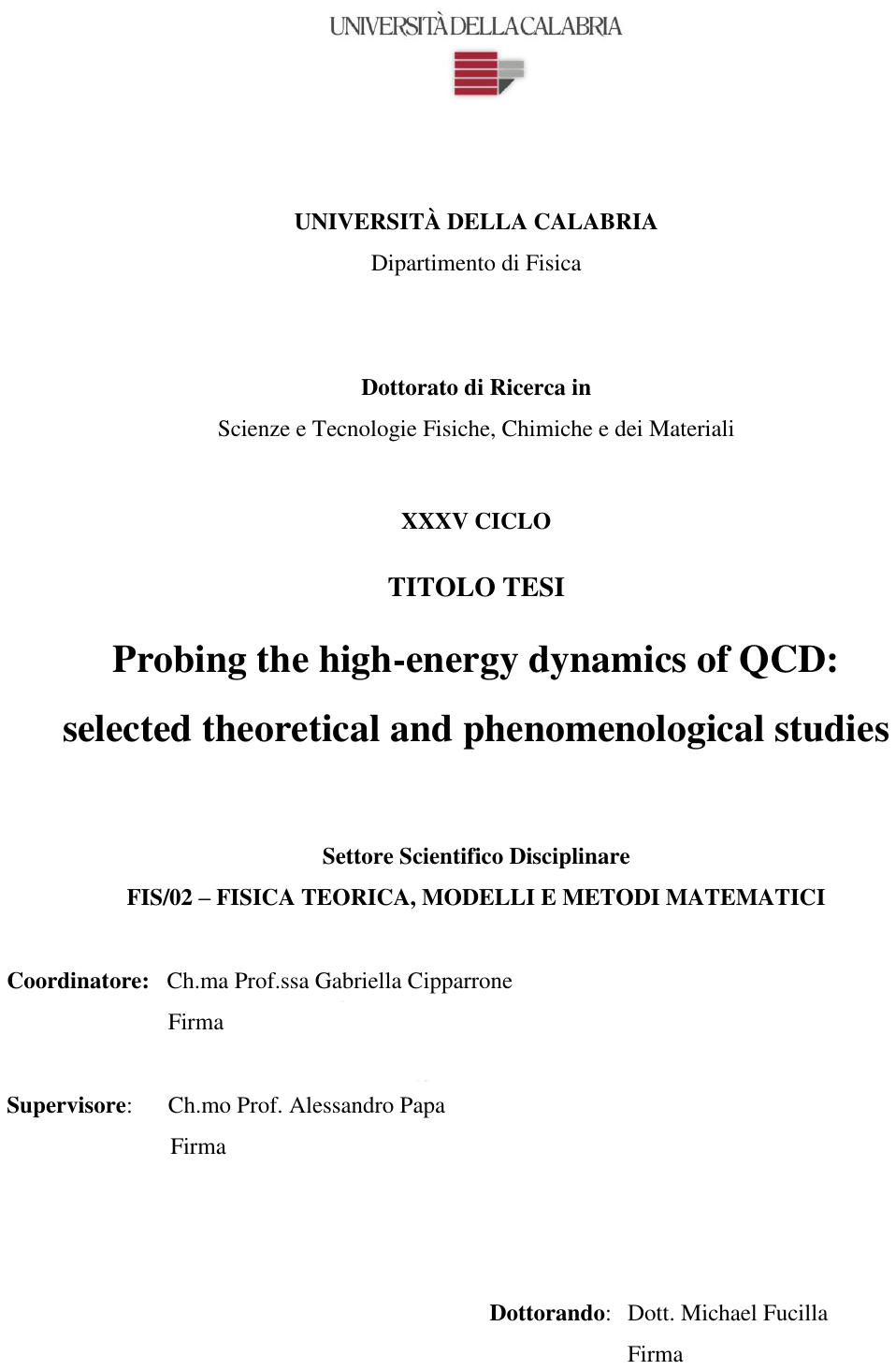}

\newpage%
\null\vspace{\stretch{1}}
\begin{flushright}
\emph{\textit{
I wish to report you, that I am still doing physics, but, I must admit, that only \\
nine hours per day. I still enjoy doing physics, I think, that we made a correct \\ decision, choosing physics as the way
of life. At least I have not felt bored even \\ a minute in my life. Every time when I put my formulae on the sheet of paper I feel \\ myself as a young guy from Gribov’s department,
who is hurrying up to meet you and \\ discuss, what I have understood. All our hopes, all our ideas, all our dreams \\ are
still with me and I am trying to share them with young physicists.\\
Eugene M. Levin}}{\textit{}} \cite{Bartels:2021zvx,Levin:2019pyo}
\thispagestyle{empty}
\end{flushright}
\vspace{\stretch{10}}\null

\chapter*{Acknowledgements}
First of all, I want to thank my supervisor Alessandro immensely for his availability, his attention and for all the knowledge he shared with me. I am sure that the passion for this field that he has passed on to me will never fade. I could not have written this manuscript without Alessandro's guidance and his inspiring way of approaching physics: strongly, formally and deeply. \\

A big thank you also goes to Samuel and Lech. During my visiting periods, I had the opportunity to work with them and with their young collaborators. I could not have had a more beautiful experience, from the professional and the personal point of view. Benefiting from their vast knowledge, I had the opportunity to expand my research interests and learn about multiple aspects of QCD, ranging from saturation physics to exclusive processes. I am proud to have the opportunity to continue my path in physics collaborating with them. I am grateful to the IJCLAB and NCBJ Warsaw for their hospitality and support. \\

I would also like to thank some colleagues I have worked and interacted with over the years. First of all, Francesco for having stimulated my interest in numerical analysis and phenomenology, sharing with me his experience and his tools. Subsequently, I would like to thank Emilie and Saad with whom I have had innumerable discussions, which have helped me grow enormously as a physicist. With Emilie we already have a nice paper together, with Saad I'm sure we will soon.
It is a pleasure to thank Victor S. Fadin, Dmitry Yu. Ivanov, Andrey Grabovsky, Mohammed M.A. Mohammed and Andrèe Dafne Bolognino for their collaboration in the devolpment of some of the papers on which this thesis is based. For interesting and inspiring discussions, I want also to thank: Renaud Boussarie, Maxim Nefedov, Valerio Bertone, Benoît Blossier, Cédric Mezrag, Jean-Philippe Lansberg, Michael Winn, Charlotte Van Hulse, Ronan McNulty, Marco Rossi, Jamal Jalilian-Marian and Michel Fontannaz. \\

 I would like to thank Leszek Motyka and Tuomas Lappi for accepting to be the referee of this work and for inspiring comments. \\

I want to thank all my family and friends who have supported me over the years. A special thanks to my friend Mario, who has been close to me, like a brother, in many difficult moments of these years. \\

I want to thank my mum, dad and Mattia from deep in my heart. Without their support, this journey and my life in general would have been much more difficult. \\

Last but not least, I dedicate this thesis and my efforts in writing it to Laura, love of my life and my \textit{place to be}.
\chapter*{Abstract}
The center-of-mass energies available at modern accelerators, such as the Large Hadron Collider (LHC), and at future generation accelerators, such as the Electron-Ion Collider (EIC) and Future Circular Collider (FCC), offer us a unique opportunity to investigate hadronic matter under the most extreme conditions ever reached. In particular, we can access the \textit{Regge-Gribov} (or \textit{semi-hard}) limit of QCD, characterized by the scale hierarchy $s \gg \{Q^2 \} \gg \Lambda_{{\rm{QCD}}}^2$, where $\sqrt{s}$ is the center-of-mass energy, $\{ Q \}$ a set of hard scales characterizing the process and $\Lambda_{{\rm{QCD}}}$ is the QCD mass scale. In this limit, large logarithmic corrections can affect both parton densities and hard scattering cross sections. The Balitsky-Fadin-Kuraev-Lipatov (BFKL) approach represents the established tool to resum to all orders, both in the leading (LLA) and the next-to-leading (NLA) approximation, these large-energy logarithmic contributions. However, it is well known that at very low values of the Bjorken-$x$, the density of partons, per unit transverse area, in hadronic wavefunctions becomes very large leading to the so-called \textit{saturation effects}. The evolution of densities is then described by non-linear generalizations of the BFKL equation. Among these, the most general is represented by the Balitsky-JIMWLK hierarchy of equations, which is needed to describe the scattering of a dilute projectile on a dense target, or also the scattering of two dense systems. The dense system condition can be achieved by a very small-$x$ proton, but is more easily achieved for large nuclei. \\

It is clear that a detailed comparison with experimental data requires precision predictions that can only be achieved in the next-to-leading logarithmic approximation or beyond. We face this task from two different perspectives. On the one hand developing analytical calculations that allow to increase the theoretical accuracy that can be reached in predictions, and on the other, by proposing phenomenological analyzes that can be directly tested experimentally. In particular, within the BFKL approach we calculate the full NLO impact factor for the Higgs production. This is the necessary ingredient to study the inclusive forward emissions of a Higgs boson in association with a backward identified jet. We claim that this result should necessarily supplement pure fixed-order calculations entering in the collinear factorization framework, which cannot be able to describe the entire kinematic spectrum in the Higgs-plus-jet channel. The result can be as well used to describe the inclusive hadroproduction of a forward Higgs in the limit of small Bjorken $x$. Moreover, using the knowledge of already known impact factors we propose a series of new semi-hard reactions that can be used to investigate BFKL dynamics at the LHC. We investigate all observables used so far to study BFKL, including: total cross sections, azimuthal coefficients, azimuthal distributions and $p_T$-differential distributions. In the context of linear evolution, we consider also the problem of extending BFKL beyond the NLLA. To this aim, we compute the Lipatov vertex in QCD with higher $\epsilon$-accuracy, where $\epsilon = (D-4)/2$. This ingredient enters the BFKL kernel at next-to-NLA (NNLLA) accuracy. In fact, the NNLLA formulation of BFKL requires not only two and three-loop calculations, but also higher $\epsilon$-accuracy of the one-loop results, for instance, in the part of the kernel containing the product of two one-loop Lipatov vertices. Finally, in the saturation framework, and more specifically in the Shockwave approach, we calculate the diffractive double hadron photo- or electroproduction cross sections with full NLL accuracy. These results are usable to detect saturation effects, at both the future EIC or already at LHC, using Ultra Peripheral Collisions.

\chapter*{Sintesi in lingua italiana}
Le energie nel centro di massa disponibili ai moderni acceleratori, come il Large Hadron Collider (LHC), e agli acceleratori di futura generazione, come l'Electron-Ion Collider (EIC) e il Future Circular Collider (FCC), ci offrono un'opportunità unica di indagare la materia adronica nelle condizioni più estreme mai raggiunte. In particolare, possiamo accedere al limite di \textit{Regge-Gribov} (o \textit{semi-duro}) della QCD, caratterizzato dalla gerarchia di scale $s \gg \{Q^2 \} \gg \Lambda_{{ \rm{QCD}}}^2$, dove $\sqrt{s}$ è l'energia nel centro di massa, $\{ Q \}$ un insieme di scale dure che caratterizzano il processo e $\Lambda_{{\rm{QCD}}}$ è la scala di massa della QCD. In questo limite, grandi correzioni logaritmiche entrano in gioco sia nelle funzioni di distribuzione partoniche che nelle sezioni d'urto dure. L'approccio BFKL è lo strumento appropriato per risommare a tutti gli ordini, sia nell'approssimazione dei logaritmi dominanti (LLA) che nell'approssimazione dei logaritmi sottodominanti (NLA), questi contributi logaritmici. Tuttavia, è ben noto che, a valori molto bassi della variabile di Bjorken $x$, la densità dei partoni, per unità di area trasversale, nelle funzioni d'onda adroniche diventa molto grande conducendo ai cosiddetti \textit{effetti di saturazione}. L'evoluzione delle densità deve essere allora descritta da generalizzazioni non lineari dell'equazione BFKL. Tra queste, la più generale è rappresentata dalla gerarchia di equazioni di Balitsky-JIMWLK, necessaria per descrivere la diffusione di un proiettile ``diluito" su un bersaglio denso o anche la diffusione di due sistemi densi. La condizione di sistema denso può essere raggiunta da un protone ad $x$ molte piccole, ma è più facilmente raggiungibile per grandi nuclei.  \\

Risulta chiaro che, un confronto dettagliato con i dati sperimentali, richiede delle predizioni di precisione che possono essere raggiunte solo nell'approssimazione dei logaritmi sottodominanti o oltre. Affrontiamo questo compito da due diverse prospettive, più precisamente, da un lato sviluppando calcoli analitici che consentano di aumentare l'accuratezza teorica raggiungibile nelle previsioni dall'altro proponendo analisi fenomenologiche direttamente verificabili sperimentalmente. In particolare, nell'ambito dell'approccio BFKL calcoliamo il fattore di impatto per la produzione di un bosone di Higgs, nell'approssimazione sottodominante. Questo è l'ingrediente necessario per studiare le emissioni inclusive di un bosone di Higgs in associazione ad un jet identificato. Crediamo che, questo risultato, dovrebbe necessariamente supportare i puri calcoli ad ordine fissato, i quali non possono descrivere l'intero spettro cinematico nel canale di Higgs più jet. Il risultato può essere utilizzato anche per descrivere l'adroproduzione inclusiva di un Higgs nel limite di basso $x$. Inoltre, utilizzando la conoscenza dei fattori di impatto già noti, proponiamo una serie di nuove reazioni semi-dure che possono essere utilizzate per studiare la dinamica BFKL ad LHC. Indaghiamo tutte le osservabili utilizzate finora per studiare BFKL, tra cui: sezioni d'urto totali, coefficienti azimutali, distribuzioni azimutali e distribuzioni differenziali nel $p_T$. Nel contesto dell'evoluzione lineare, consideriamo anche il problema dell'estensione di BFKL oltre l'approssimazione dei logaritmi sottodominanti. A tale scopo, calcoliamo il vertice di Lipatov in QCD con maggiore accuratezza in $\epsilon$. Questo ingrediente entra nel kernel BFKL con precisione successiva a quella sottodominante. Infatti, a questo grado di accuratezza, la formulazione di BFKL richiede non solo calcoli a due e tre loop, ma anche una maggiore accuratezza in $\epsilon$ dei risultati a un loop, per esempio, nella parte del kernel che contiene il prodotto di due vertici di Lipatov ad un loop. Infine, nel quadro della saturazione, e più specificamente nell'approccio Shockwave, calcoliamo le sezioni d'urto diffrattive della foto- o elettroproduzione di due adroni con precisione sottodominante. Questi risultati sono utilizzabili per rilevare effetti di saturazione, sia al futuro EIC che già al LHC, utilizzando collisioni ultra-periferiche.
\chapter*{List of Publications}
\section*{Regular articles}
\begin{itemize}

\item V.~S.~Fadin, M.~Fucilla and A.~Papa,
\textit{One-loop Lipatov vertex in QCD with higher $\epsilon$-accuracy},
JHEP \textbf{04} (2023), 137\\ 
\href{https://arxiv.org/abs/2302.09868}{arXiv:2302.09868},
\href{https://link.springer.com/article/10.1007/JHEP04(2023)137}{DOI: 10.1007/JHEP04(2023)137}. 

\item M.~Fucilla, A.~V.~Grabovsky, E.~Li, L.~Szymanowski and S.~Wallon,
\textit{NLO computation of diffractive di-hadron production in a saturation framework},
JHEP \textbf{03} (2023), 159 \\
\href{https://arxiv.org/abs/2211.05774}{arXiv:2211.05774},
\href{https://link.springer.com/article/10.1007/jhep03(2023)159}{DOI: 10.1007/JHEP03(2023)159}

\item F.~G.~Celiberto, M.~Fucilla, D.~Y.~Ivanov, M.~M.~A.~Mohammed and A.~Papa,
\textit{The next-to-leading order Higgs impact factor in the infinite top-mass limit},
JHEP \textbf{08} (2022), 092 \\
\href{https://arxiv.org/abs/2205.02681}{arXiv:2205.02681}, \href{https://link.springer.com/article/10.1007/JHEP08(2022)092}{DOI:10.1007/JHEP08(2022)092}. 

\item F.~G.~Celiberto, M.~Fucilla, M.~M.~A.~Mohammed and A.~Papa, \textit{Ultraforward production of a charmed hadron plus a Higgs boson in unpolarized proton collisions},
Phys. Rev. \textbf{D 105} (2022) no.11, 114056 \\
\href{https://arxiv.org/abs/2205.13429}{arXiv:2205.13429}, \href{https://journals.aps.org/prd/abstract/10.1103/PhysRevD.105.114056}{DOI:10.1103/PhysRevD.105.114056}. 

\item F.~G.~Celiberto and M.~Fucilla,
\textit{Diffractive semi-hard production of a $J/\psi $ or a $\Upsilon $ from single-parton fragmentation plus a jet in hybrid factorization}, Eur. Phys. J. C \textbf{82} (2022) no.10, 929 \\
\href{https://arxiv.org/abs/2202.12227}{arXiv:2202.12227},
\href{https://link.springer.com/article/10.1140/epjc/s10052-022-10818-8}{DOI:10.1140/epjc/s10052-022-10818-8}.

\item F.~G.~Celiberto, M.~Fucilla, D.~Y.~Ivanov, M.~M.~A.~Mohammed and A.~Papa,
\textit{Bottom-flavored inclusive emissions in the variable-flavor number scheme: A high-energy analysis},
Phys. Rev. \textbf{D 104} (2021) no.11, 114007 \\
\href{https://arxiv.org/abs/2109.11875}{arXiv:2109.11875},
\href{https://journals.aps.org/prd/abstract/10.1103/PhysRevD.104.114007}{DOI:10.1103/PhysRevD.104.114007}.

\item F.~G.~Celiberto, M.~Fucilla, D.~Y.~Ivanov and A.~Papa,
\textit{High-energy resummation in $\Lambda_c$ baryon production},
Eur. Phys. J. C \textbf{81} (2021) no.8, 780 \\
\href{https://arxiv.org/abs/2105.06432}{arXiv:2105.06432},
\href{https://link.springer.com/article/10.1140/epjc/s10052-021-09448-3}{DOI:10.1140/epjc/s10052-021-09448-3}.

\item A.~D.~Bolognino, F.~G.~Celiberto, M.~Fucilla, D.~Y.~Ivanov and A.~Papa,
\textit{``Inclusive production of a heavy-light dijet system in hybrid high-energy and collinear factorization,''}
Phys. Rev. \textbf{D 103} (2021) no.9, 094004 \\
\href{https://arxiv.org/abs/2103.07396}{arXiv:2103.07396},
\href{https://journals.aps.org/prd/abstract/10.1103/PhysRevD.103.094004}{DOI:10.1103/PhysRevD.103.094004}. 

\item A.~D.~Bolognino, F.~G.~Celiberto, M.~Fucilla, D.~Y.~Ivanov and A.~Papa,
\textit{``High-energy resummation in heavy-quark pair hadroproduction,''}
Eur. Phys. J. C \textbf{79} (2019) no.11, 939 \\
\href{https://arxiv.org/abs/1909.03068}{arXiv:1909.03068},
\href{https://link.springer.com/article/10.1140/epjc/s10052-019-7392-1}{DOI:10.1140/epjc/s10052-019-7392-1}. 

\end{itemize}

\section*{Review, community and white papers}

\begin{itemize}

\item M.~Begel, S.~Hoeche, M.~Schmitt, H.~W.~Lin, P.~M.~Nadolsky, C.~Royon, Y.~J.~Lee, S.~Mukherjee, C.~Baldenegro and J.~Campbell, \textit{et al.}
\textit{``Precision QCD, Hadronic Structure \& Forward QCD, Heavy Ions: Report of Energy Frontier Topical Groups 5, 6, 7 submitted to Snowmass 2021,''} \\
\href{https://arxiv.org/abs/2209.14872#}{arXiv:2209.14872}, .  \vspace{0.3 cm}

\item M.~Hentschinski, C.~Royon, M.~A.~Peredo, C.~Baldenegro, A.~Bellora, R.~Boussarie, F.~G.~Celiberto, S.~Cerci, G.~Chachamis and J.~G.~Contreras, \textit{et al.}
\textit{``White Paper on Forward Physics, BFKL, Saturation Physics and Diffraction,''} Acta Phys.Polon.B 54 (2023) \\
\href{https://arxiv.org/abs/2203.08129}{arXiv:2203.08129}. 
\href{https://www.actaphys.uj.edu.pl/index_n.php?I=R&V=54&N=3#A2}{DOI:10.5506/APhysPolB.54.3-A2} \vspace{0.3 cm}

\item J.~L.~Feng, F.~Kling, M.~H.~Reno, J.~Rojo, D.~Soldin, L.~A.~Anchordoqui, J.~Boyd, A.~Ismail, L.~Harland-Lang and K.~J.~Kelly, \textit{et al.}
\textit{``The Forward Physics Facility at the High-Luminosity LHC,''}
J. Phys. \textbf{G 50} (2023) no.3, 030501 \\
\href{https://arxiv.org/abs/2203.05090}{arXiv:2203.05090},
\href{https://www.sciencedirect.com/science/article/pii/S0370157322001235?via%3Dihub}{DOI: 10.1016/j.physrep.2022.04.004}. \vspace{0.3 cm}

\item L.~A.~Anchordoqui, A.~Ariga, T.~Ariga, W.~Bai, K.~Balazs, B.~Batell, J.~Boyd, J.~Bramante, M.~Campanelli and A.~Carmona, \textit{et al.}
\textit{``The Forward Physics Facility: Sites, experiments, and physics potential,''}
Phys. Rept. \textbf{968} (2022), 1-50
\href{https://arxiv.org/abs/2109.10905}{arXiv:2109.10905},
\href{https://iopscience.iop.org/article/10.1088/1361-6471/ac865e}{DOI:10.1088/1361-6471/ac865e}. 
\end{itemize}

\section*{Conference papers}
\begin{itemize}
\item M.~Fucilla,
\textit{``The Higgs impact factor at next-to-leading order,''} Difflowx2022. \\
\href{https://arxiv.org/abs/2212.01794}{arXiv:2212.01794}. \vspace{0.3 cm}

\item A.~D.~Bolognino, F.~G.~Celiberto, M.~Fucilla, D.~Y.~Ivanov, M.~M.~A.~Mohammed and A.~Papa,
\textit{``High-energy signals from heavy-flavor physics,''} Difflowx2022. \\
\href{https://arxiv.org/abs/2211.16818}{arXiv:2211.16818}. \vspace{0.3 cm}

\item M.~Fucilla, A.~V.~Grabovsky, L.~Szymanowski, E.~Li and S.~Wallon,
\textit{``Diffractive di-hadron production at NLO within the shockwave formalism,''} Difflowx2022. \\
\href{https://arxiv.org/abs/2211.04390}{arXiv:2211.04390}. \vspace{0.3 cm}

\item F.~G.~Celiberto, M.~Fucilla and A.~Papa,
\textit{``The high-energy limit of perturbative QCD: Theory and phenomenology,''} EPJ Web Conf. \textbf{270} (2022) 00001 \\
\href{https://arxiv.org/abs/2209.01372}{arXiv:2209.01372},
\href{https://doi.org/10.1051/epjconf/202227000001}{DOI:https://doi.org/10.1051/epjconf/202227000001}. \vspace{0.3 cm}

\item F.~G.~Celiberto and M.~Fucilla,
\textit{``Inclusive $J/\psi$ and $\Upsilon$ emissions from single-parton fragmentation in hybrid high-energy and collinear factorization,''} DIS2022. \\
\href{https://arxiv.org/abs/2208.07206}{arXiv:2208.07206},
\href{https://ui.adsabs.harvard.edu/abs/2022arXiv220807206C/abstract}{DOI:10.48550/arXiv.2208.07206}. \vspace{0.3 cm}

\item A.~D.~Bolognino, F.~G.~Celiberto, M.~Fucilla, D.~Y.~Ivanov, A.~Papa, W.~Sch\"afer and A.~Szczurek,
\textit{``Hadron structure at small-x via unintegrated gluon densities,''}
Rev. Mex. Fis. Suppl. \textbf{3} (2022) no.3, 0308109 \\
\href{https://arxiv.org/abs/2202.02513}{arXiv:2202.02513},
\href{https://rmf.smf.mx/ojs/index.php/rmf-s/article/view/6292}{DOI:10.31349/SuplRevMexFis.3.0308109}. \vspace{0.3 cm}

\item F.~G.~Celiberto, M.~Fucilla, D.~Y.~Ivanov, M.~M.~A.~Mohammed and A.~Papa,
``Higgs boson production in the high-energy limit of pQCD,''
PoS \textbf{PANIC2021} (2022), 352 \\
\href{https://arxiv.org/abs/2111.13090}{arXiv:2111.13090},
\href{https://pos.sissa.it/380/352}{DOI:10.22323/1.380.0352}. \vspace{0.3 cm}

\item A.~D.~Bolognino, F.~G.~Celiberto, M.~Fucilla, D.~Y.~Ivanov and A.~Papa,
\textit{``Heavy flavored emissions in hybrid collinear/high energy factorization,''}
PoS \textbf{EPS-HEP2021} (2022), 389 \\
\href{https://arxiv.org/abs/2110.12772}{arXiv:2110.12772},
\href{https://ui.adsabs.harvard.edu/abs/2021arXiv211012772D/abstract}{DOI:10.48550/arXiv.2110.12772}. \vspace{0.3 cm}

\item F.~G.~Celiberto, M.~Fucilla, A.~Papa, D.~Y.~Ivanov and M.~M.~A.~Mohammed,
\textit{``Higgs-plus-jet inclusive production as stabilizer of the high-energy resummation,''} 
PoS \textbf{EPS-HEP2021} (2022), 589 \\
\href{https://arxiv.org/abs/2110.09358}{arXiv:2110.09358},
\href{https://pos.sissa.it/398/589}{DOI:10.22323/1.398.0589}. \vspace{0.3 cm}

\item F.~G.~Celiberto, M.~Fucilla, D.~Y.~Ivanov, M.~M.~A.~Mohammed and A.~Papa,
\textit{``BFKL phenomenology: resummation of high-energy logs in inclusive processes,''}
SciPost Phys. Proc. \textbf{10} (2022), 002 \\
\href{https://arxiv.org/abs/2110.12649}{arXiv:2110.12649}, 
\href{https://scipost.org/10.21468/SciPostPhysProc.10.002}{DOI:10.21468/SciPostPhysProc.10.002}. \vspace{0.3 cm}

\item A.~D.~Bolognino, F.~G.~Celiberto, M.~Fucilla, D.~Y.~Ivanov and A.~Papa,
\textit{``Hybrid high-energy/collinear factorization in a heavy-light dijets system reaction,''}
SciPost Phys. Proc. \textbf{8} (2022), 068 \\ 
\href{https://arxiv.org/abs/2107.12120}{arXiv:2107.12120},
\href{https://scipost.org/10.21468/SciPostPhysProc.8.068}{DOI:10.21468/SciPostPhysProc.8.068}. \vspace{0.3 cm}

\item A.~D.~Bolognino, F.~G.~Celiberto, M.~Fucilla, D.~Y.~Ivanov, B.~Murdaca and A.~Papa,
\textit{``Inclusive production of two rapidity-separated heavy quarks as a probe of BFKL dynamics,''}
PoS \textbf{DIS2019} (2019), 067 \\
\href{https://arxiv.org/abs/1906.05940}{arXiv:1906.05940},
\href{https://pos.sissa.it/352/067}{DOI:10.22323/1.352.0067}. \vspace{0.3 cm}
   
\end{itemize}
\tableofcontents
\addcontentsline{toc}{chapter}{Introduction}
\chapter*{Introduction}
\begin{flushright}
\emph{QCD \textit{is the most perfect and non-trivial of the \\ established microscopic theories of~physics.\\ John C. Collins~\cite{Collins:2011zzd}}}
\end{flushright}

\textit{Quantum chromodynamics} (QCD) is a quantum field theory describing the \textit{Strong interaction}, one of the four fundamental forces of Nature. It is based on the non-Abelian gauge group $SU(N_c)$, where $N_c = 3$ is the number of quark colors~\cite{Fritzsch:1973pi}. Although QCD is a theory that has developed over a long period through a combination of theoretical and experimental efforts, its birth year is usually considered to be 1973, when David Gross and Frank Wilczek, and independently David Politzer, discovered the \textit{Asymptotic freedom}, \textit{i.e.} the remarkable property for which interactions between color sources become asymptotically weaker as the energy scale increases and the corresponding length scale decreases. From that point on, the continued predictive successes of the theory definitely established QCD as the theory of strong interactions. However, to date, a full understanding of the theory has not been achieved. \\

One of the main aspects that makes the theory tremendously complicated is the \textit{Color confinement}: the phenomenon that color-charged particles (such as quarks and gluons) cannot be isolated, and therefore cannot be directly observed. There is not yet an analytic proof of color confinement in any non-Abelian gauge theory, but it is well established from lattice QCD calculations and decades of experiments. The phenomenon can be understood qualitatively by noting that the force-carrying gluons of QCD have color charge, unlike the photons of \textit{Quantum electrodynamics} (QED). Whereas the electric field between electrically charged particles decreases rapidly as those particles are separated, the gluon field between a pair of color charges forms a narrow flux tube (or string) between them. Because of this behavior of the gluon field, the strong force between the particles is constant regardless of their separation. Therefore, as two color charges are separated, at some point it becomes energetically favorable for a new quark–antiquark pair to appear, rather than extending the tube further. This tells us that the theory, which is asymptotically free at high-energy (short distance), is strongly coupled at low-energy (large distance). At high-energy, since the coupling of the theory is small, the interaction processes can be described through the perturbation theory, we speak in this case of \textit{pertutbative} QCD (pQCD). On the other hand, at low energies a perturbative approach is not possible; in this case, the most common approach is the \textit{Lattice} QCD (LQCD), based on numerical calculations on a discretized space-time lattice. \\

It is clear that, when collision processes involving hadronic particles are experimentally investigated, both dynamics (long and short distance) are involved. Among the greatest achievements of pQCD, there are the so-called \textit{factorization theorems}, which allow us to separate the total
process into a \textit{hard part}, a cross-section for the scattering of partons computable in perturbation theory, and some non-perturbative parts, encoding the long-distance dynamics. In the factorization approach, the proton beams are treated as collections of so-called partons: quasi-free quarks and gluons, whose (non-perturbative) distributions, the parton densities contain the long distance dynamics and have to be either fit from experimental data or evaluated using LQCD. Factorization implies that these quantities must be universal, so that, extracted in a process, they can then be used for the description of other reactions. Concerning the hard part, in any scattering process, at least two scales are involved: the center-of-mass of the whole process, $\sqrt{s}$, and the so-called hard scale, $Q$. The hard scale must satisfy the relation 
\begin{equation*}
    Q^2 > \Lambda_{{\rm{QCD}}}^2 \; ,
\end{equation*}
where $\Lambda_{{\rm{QCD}}}$ is the QCD mass scale, to ensure a perturbative treatment in $\alpha_s (Q^2)$. Although the perturbative approach is a very powerful tool, it also presents some pitfalls. Among these, one of the most important is related to the appearance of logarithmic corrections, depending on the kinematic scales involved and entering the perturbative series with a power increasing along with the order. It is clear that, if in certain kinematics these logarithms are large, they can compensate the smallness of the strong coupling, $\alpha_s$. In these cases, a resummation to all orders of the perturbative series becomes mandatory. \\

An important example of this phenomenon is related to the cancellation of infrared (IR) divergences in QCD for IR-safe observables. These are typically dealt with through the \textit{dimensional regularization} procedure, which leads us to divergent quantities of the type
\begin{equation}
    \frac{1}{\epsilon} \left( Q^2 \right)^{\epsilon} = \frac{1}{\epsilon} + \ln Q^2 + \mathcal{O}(\epsilon) \; ,
\end{equation}
which appear when considering collinear gluon dynamics in the massless quark limit. Here, $\epsilon=(D-4)/2 \epsilon$. In an IR-safe observable, the pole cancels but the logarithm remain. At any order in perturbation theory we thus have corrections of the type $\alpha_s^n \ln^p Q^2$. In the the so-called \textit{Bjorken} limit, $Q^2 \rightarrow \infty$ and moderately small Bjorken $x$, given by $x = (Q^2 / s)$, the logarithms can compensate the smallness of the coupling constant. The resummation of such logarithms leads to the Dokshitzer-Gribov-Lipatov-Altarelli-Parisi~\cite{Gribov:1972ri,Dokshitzer:1977sg,Altarelli:1977zs} (DGLAP) evolution equations. \\

Another situation that can lead to the appearance of large logarithmic corrections is the existence of a hierarchy among the scales involved. A very famous case is that of the so-called \textit{Regge-Gribov} or \textit{semi-hard} region, where
\begin{equation*}
    s \gg Q^2 \gg \Lambda_{{\rm{QCD}}}^2 \; .
\end{equation*}
In this case, the logarithmic corrections are of the type $ \ln (s / Q^2)$. Suppose we consider a proton-proton collision, where the partonic center-of-mass energy is $\sqrt{\hat{s}} = \sqrt{x_1 x_2 s}$, with $x_{1,2}$ the Bjorken fraction associated to the two colliding protons, we then have
\begin{equation*}
    \ln \left( \frac{s}{Q^2} \right) = \ln \left( \frac{1}{x_1} \right) + \ln \left( \frac{\hat{s}}{Q^2} \right) + \ln \left( \frac{1}{x_2} \right) \; .
\end{equation*}
All logarithms on the right-hand side can be potentially large. Those of small-$x$ appear in the evolution of parton densities, while those of the ratio $\hat{s}/Q^2$ in the partonic cross sections. As we will see, from a physical point of view, they are always related to the occurrence of large rapidity intervals. The Balitsky-Fadin-Kuraev-Lipatov (BFKL)~\cite{Fadin:1975cb,Kuraev:1976ge,Kuraev:1977fs,Balitsky:1978ic} approach represents the established tool to resum to all orders, both in the leading (LLA) and the next-to-leading (NLLA) approximation, these large-energy logarithmic contributions. In the BFKL framework, the cross section of hadronic processes can be expressed as the convolution of two impact factors, related to the transition from each colliding particle to the respective final-state object, and a process-independent Green’s function. The BFKL approach has proven to be very robust as it is able to predict rapid growth of the $\gamma^{*} p$ cross section at increasing energy and its consistent with pre-QCD results from Regge theory. It is also very powerful, being able to be applied in many different contexts. \\

\begin{figure}
\begin{picture}(400,215)
\put(137,10){\includegraphics[width=0.4\textwidth]{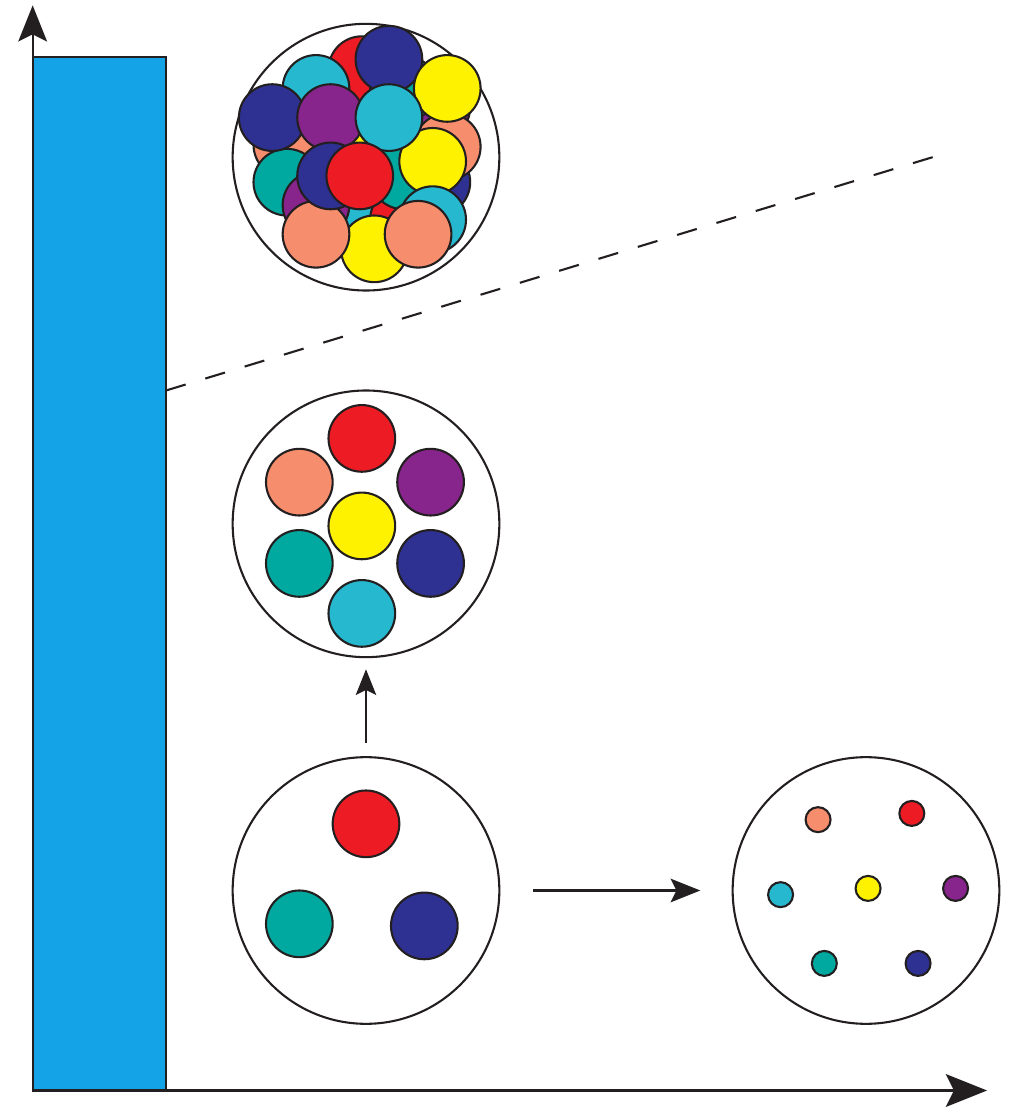}}
\put(235,185){Saturation}
\put(150,68){\rotatebox{90}{Non-perturbative}}
\put(174,71){\rotatebox{90}{\scalebox{0.8}{BFKL}}}
\put(230,60){\scalebox{0.8}{DGLAP}}
\put(265,155){$Q_s$}
\put(285,0){$\ln Q^2$}
\put(75,190){$Y = \ln (1/x)$}
\end{picture}
\caption{The parton distribution in the transverse plane as a function of $\ln (1/x)$ and $\ln Q^2$.}
\label{Int:Int:DGLAPvsBFKL}
\end{figure}
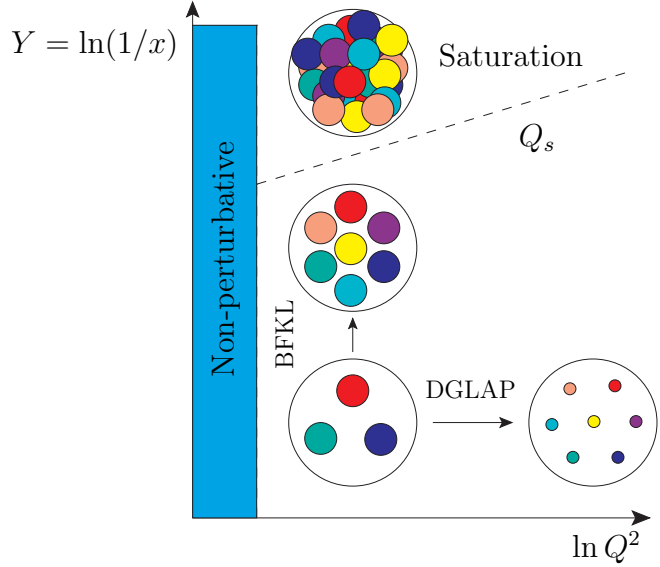

One of the major theoretical problems of the BFKL formalism is its inconsistency with the Froissart bound, which states that the total cross sections cannot grow in $s$ faster than $ k \times \ln^2 s$, with $k$ a constant. This limit is explicitly violated by the power-like behavior in $s$ of BFKL-resummed total cross sections. When we refer to the small-$x$ logarithms, the violation of Froissart bound is physically interpretable as an infinite growth of the gluon density at small-$x$. The hadronic systems appears to become a denser and denser gluon medium, until at some point it becomes infinitely dense. This scenario seems to suggest that at a certain point some \textit{saturation effects} must intervene to slow this growth. As we shall see, these effects are theoretically described by non-linear generalizations of the BFKL equation. Among these, the most general is represented by the \textit{Balitsky-JIMWLK hierarchy of equations}, derived, in the so-called Shockwave approach, by Balitsky~\cite{Balitsky:2001re,Balitsky:1995ub,Balitsky:1998kc,Balitsky:1998ya} and, in the so-called Color Glass Condensate (CGC) approach, by Jalilian-Marian, Iancu, McLerran, Weigert, Leonidov and Kovner (JIMWLK)~\cite{Jalilian-Marian:1997qno,Jalilian-Marian:1997jhx,Jalilian-Marian:1997ubg,Jalilian-Marian:1998tzv,Kovner:2000pt,Weigert:2000gi,Iancu:2000hn,Iancu:2001ad,Ferreiro:2001qy}. These approaches are needed to describe the scattering of a dilute projectile on a dense target, or also the scattering of two dense systems. The dense system condition can be achieved by a very small-$x$ proton, but is more easily achieved for nuclei. A schematic representation of the proton/nucleus structure in terms of its constituents is shown in Fig.~\ref{Int:Int:DGLAPvsBFKL}. \\

The center-of-mass energies available at modern accelerators, such as the Large Hadron Collider (LHC), and at future generation accelerators, such as the Electron-Ion Collider (EIC) and Future Circular Collider (FCC), offer us a unique opportunity to test the theoretical framework illustrated above, investigating hadronic matter under the most extreme conditions ever reached. This requires making theoretical predictions that are as accurate as possible and this is exactly the aim of this work. The thesis is divided into three parts. In the first and in the second part we consider the linear regime, well described by the BFKL approach. In the first part, we focus on aspects more related to phenomenology, such as calculations of next-to-leading order impact factors and numerical studies. In the second part, we focus on more formal aspects related to the extension of BFKL beyond the NLLA. The third and final part is entirely dedicated to saturation in which we investigate a new process with full NLL accuracy. More specifically, the thesis is organized as follows:
\begin{itemize}
    \item In the first chapter, we introduce the problem of scattering in the Regge limit of QCD. We briefly describe the pre-QCD approach and then move on to the formulation of the BFKL approach. We first develop the approach pedagogically in the leading logarithmic approximation (LLA), and then, the chapter closes with a discussion of the approach in the next-to-leading logarithm approximation (NLLA). 
    \item In the second chapter, we present an example of full NLO computation of an impact factor. In particular, we consider the forward Higgs boson impact factor, obtained in the infinite top-mass limit. This is the necessary ingredient to describe the inclusive hadroproduction of a forward Higgs in the limit of small Bjorken $x$, as well to study the inclusive forward emissions of a Higgs boson in association with a backward identified jet. We claim that this result should necessarily supplement pure fixed-order calculations entering in the collinear factorization framework, which cannot be able to describe the entire kinematic spectrum in the Higgs-plus-jet channel. We corroborate this claim in the third chapter, showing a comparison between fixed-order results and resummed predictions, in the case of Higgs $p_T$-distribution.
    \item The third chapter is devoted to the BFKL phenomenology and, in particular, to the study, in the NLLA, of processes featuring a forward-plus-backward two particle final state configuration. With particular reference to processes involving the production of the Higgs boson or of bound states of heavy quarks, we will show which observables are useful for testing the dynamics of high-energy QCD and what are the phenomenological challenges. We will present numerical predictions for all these observables that can compared with data coming from the LHC.
    \item The fourth chapter focuses on more formal aspects. In particular, we briefly discuss the NNLLA formulation of the BFKL approach and then we compute one of the ingredient needed in this construction, \textit{i.e.} the one-loop Lipatov vertex with higher accuracy in the dimensional regularization parameter $\epsilon = (D-4)/2$. 
    \item In the fifth chapter, we briefly introduce the saturation picture and discuss the non-linear extension of the BFKL approach in the so-called Shockwave formalism. We derive, in this context, the B-JIMWLK evolution equation for the dipole operator. 
    \item In the sixth and last chapter, we apply the Shockwave approach to the study, in the full next-to-leading logarithmic approximation, of the cross-sections of diffractive double hadron photo- or electroproduction, on a nucleon or a nucleus. The results are usable, to detect saturation effects, at both the future Electron-Ion-Collider (EIC) or already at LHC, using Ultra Peripheral Collisions (UPC).
\end{itemize}

\newpage
\thispagestyle{empty}
\addcontentsline{toc}{chapter}{BFKL in the NLLA: theory and phenomenology}
\vspace*{\fill}
    \begin{center}
      { \Huge \textbf{Part I}} \vspace{0.3 cm} \\
      { \Huge \textbf{BFKL in the NLLA: theory and phenomenology}}
    \end{center}
\vspace*{\fill}
\chapter{High-energy scattering in the Regge limit}
\begin{flushright}
\emph{\textit{The effort to understand the universe is one of the very few things that lifts human \\ life a little above the level of farce, and gives it some of the grace of tragedy. \\ Steven Weinberg~\cite{Weinberg:1977ji}}}
\end{flushright}

In the first section of this chapter the Regge theory will be introduced, following the standard presentations in books~\cite{Forshaw:1997qp,Collins:1977rt}; it represents a very general framework in which to discuss the scattering of particles at high center-of-mass energies. The successive section is focused on the discussion of the BFKL approach~\cite{Fadin:1975cb,Kuraev:1976ge,Kuraev:1977fs,Balitsky:1978ic}, which gives the description of pQCD-scattering amplitudes in the region of large $s$ and fixed momentum transfer $t$, $s \gg |t|$ (Regge region), with various colour states in the $t$-channel. For this part, we will follow Refs.~\cite{Forshaw:1997qp,Fadin:1998sh}.
\section{Regge theory and the Pomeron}
Before the advent of the field-theoretical approach to strong interactions, physicists sought to extract as much information as possible about scattering amplitudes of strongly interacting particles by studying the consequences of a number of postulates on
the $S$-matrix,  whose $ab_{th}$ element is the overlap between the $in$-state (free particles state as $t\longrightarrow -\infty$), $\ket{a}$, and the $out$-state (free particles state as $t\longrightarrow + \infty$), $\ket{b}$,
\begin{equation}
S_{ab}=\braket{b_{out}|a_{in}}.
\end{equation}
This type of approach is known as $S$-\textit{matrix bootstrap}.
\subsection{The $S$-matrix bootstrap}
The most general postulates, which can be imposed on the $S$-matrix, are
\begin{enumerate}
\item \textbf{The $S$-matrix is Lorentz invariant}. \\ This implies that can be expressed as a function of the (Lorentz invariant) scalar products of the incoming and outgoing momenta. For a two-particle scattering the most natural choice is to use the Mandelstam variables $s,t,u$. Since only two of them are independent, a two-particle scattering amplitude can be expressed as $\mathcal{A}\left(s,t\right)$, \textit{i.e.} a function of only $s,t$ (obviously, also other choices are possible).  
\item \textbf{The $S$-matrix is unitary}. \\
This means that
\begin{equation}
SS^{\dagger} = S^{\dagger}S = \mathbb{1}.
\end{equation}
As known, this is a statement of conservation of probability. The scattering amplitude, $A_{ab}$, for scattering from an $in$-state $\ket{a}$ to an $out$-state $\ket{b}$ is related to the $S$-matrix element by
\begin{equation*}
    S_{ab} = \delta_{ab} + i \; 2 \pi \delta^4 \left( \sum_a p_a - \sum_b p_b \right) A_{ab} \; .
\end{equation*}
An immediate consequence of the unitary of the scattering matrix are the Cutkosky rules, which read
\begin{equation}
\label{Int:Eq:Cutkosky}
2 \Im m \mathcal{A}_{ab}=(2\pi)^{4} \delta^{4} \left( \sum_{a} p_a -\sum_{b} p_b \right) \sum_{c} \mathcal{A}_{ac} \mathcal{A}^{\dagger}_{cb} \; ,
\end{equation}  
where $p_a,p_b$ are the 4-momenta of the particles in the states $\ket{a},\ket{b}$, respectively.
The Cutkosky rules allow to determine the imaginary part of an amplitude by considering the scattering amplitudes of the incoming and outgoing states into all possible ``intermediate'' states. 
When we consider the special case of forward elastic amplitude, $\mathcal{A}_{aa}$, 
we get the famous \textit{optical theorem}. It states that the total cross-section for the scattering of two particles in the state $\ket{a}$ is related to the imaginary of the elastic scattering amplitude by the relation 
\begin{equation}
\label{Int:Eq:OpticalThe}
2 \Im m \mathcal{A}_{aa} \left(s,0 \right)=(2\pi)^{4} \sum_{n} \delta^{4} \left( \sum_{f} p_f -\sum_{b} p_b \right) |\mathcal{A}_{a\longrightarrow n}|^{2}= F \sigma_{tot},
\end{equation} 
where $F$ is the flux factor.
\item  \textbf{Analyticity} \\
The $S$-matrix is an analytic function of Lorentz invariants (regarded as complex variables), with only those singularities required by unitarity \cite{Forshaw:1997qp}. \\
It can be shown that this property is a consequence of causality\footnote{This connection has a very deep meaning and is discussed, for instance, in the section 6.6 of \cite{Duncan:2021cf}.}. Analiticity has a number of important and useful consequences. One is that, combined with unitarity, it allows to establish the existence of an $s$-plane singularity structure of the amplitude $\mathcal{A} \left(s,t\right)$ (there are $s$-plane cuts with branch points corresponding to the physical thresholds). Moreover, it enables to reconstruct the real part of an amplitude from its imaginary part using dispersion relations.
\end{enumerate}
From these postulates, coupled with the spectrum of elementary particles, one can develop a set of conditions for amplitudes. In fact, unitarity relates the imaginary parts of amplitudes to sums of the products of other amplitudes, and dispersion relations then allow to determine the corresponding real parts. To underline the beauty of this approach, we want to observe that no assumptions about the underlying quantum field theory are made. All this information descend from fundamental principles of Nature.\\
As mentioned earlier, this approach uses dispersion relations to reconstruct the amplitudes, which involve integrations over the complex $s$ variable, extending to infinity. This means that we need to know the asymptotic behavior in $s$ of the amplitudes; this is the goal of \textit{Regge theory}.
\subsection{Regge theory}
The Italian physicist T. Regge showed that it is useful to regard the angular momenta, $l$, as a complex variable, when discussing solutions of the Schr\"odinger equation for non-relativistic potential scattering \cite{Regge:1959it}. He has demonstrated that, for a wide class of potentials, the only singularities of the scattering amplitude in the complex $l$ plane are poles, called ``Regge poles'' after him. When these poles occur for positive integer values of $l$, they correspond to bound states or resonances, and they are also important for determining dispersion properties of the amplitudes. \\

His methods is also applicable in high-energy elementary particle physics. Regge theory predicts that, for a great variety of processes, the high-energy behaviour of a scattering amplitude $\mathcal{A}\left(s,t\right)$ will be
\begin{equation}
\label{Int:Eq:AsynInS}
\mathcal{A}\left(s,t\right)\sim s^{\alpha(t)}.
\end{equation}  
\begin{figure}
\begin{picture}(417,175)
\put(157,0){\includegraphics[width=0.33\textwidth]{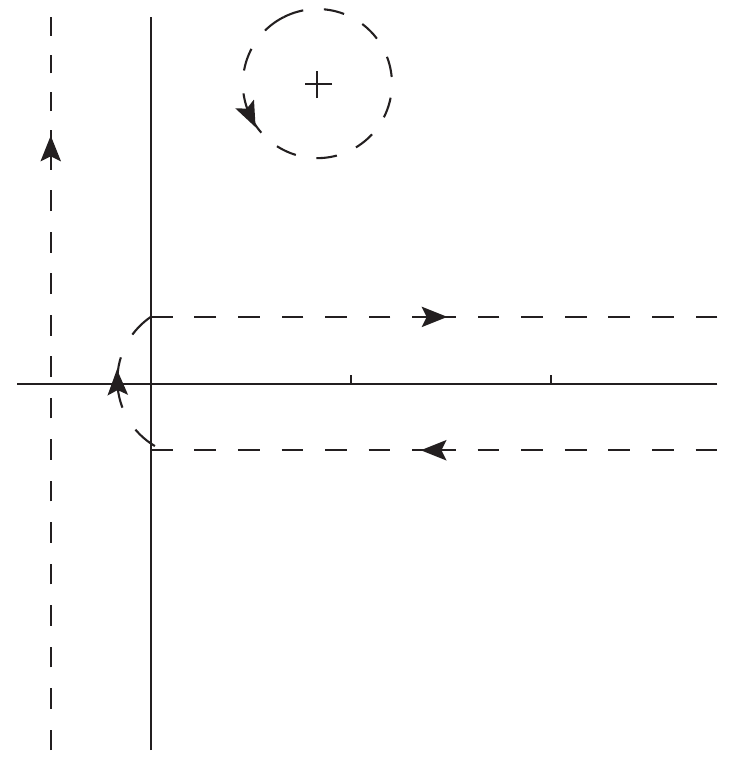}}
\put(240,100){$C$}
\put(150,100){$C'$}
\put(150,65){$- \frac{1}{2}$}
\put(226,67){$1$}
\put(267,67){$2$}
\put(214,130){$a_{n_{\eta}}$}
\put(265,145){$l$-plane}
\end{picture}
\caption{Sommerfeld-Watson transformation.}
\label{Int:Fig:SommWats}
\end{figure}
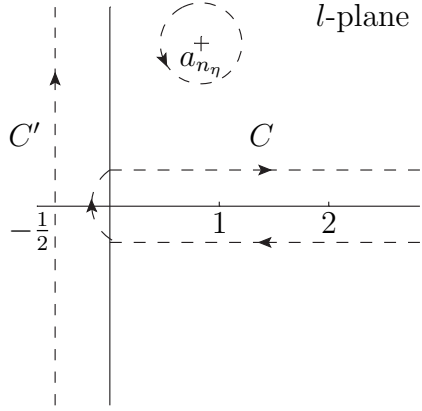
To find this high-energy behaviour, one must consider the partial-wave expansion of the amplitude. Through this expansion, the amplitude can be expressed as a series of Legendre polynomials $P_l \left(1+\frac{2s}{t} \right)$,
\begin{equation}
\mathcal{A} \left(s,t\right)= \sum_{l=0}^{\infty} \left(2l+1\right)a_l\left(t\right)P_l\left(1+\frac{2s}{t}\right),
\end{equation}
where $a_l\left(t\right)$ are called partial-wave amplitudes. 
Sommerfeld~\cite{Sommerfeld:1961pa} rewrote this partial-wave expansion in terms of a contour integral in the complex angular momentum $(l)$ plane as
\begin{equation}
\mathcal{A}\left(s,t\right)= \frac{1}{2i} \oint_{C} dl \frac{\left(2l+1\right)}{\sin \pi l} \sum_{\eta=\pm1} \frac{\eta+e^{-i\pi l}}{2} a^{(\eta)} \hspace{-0.05 cm} \left(l,t\right)P_l \left(l,1+\frac{2s}{t}\right),
\end{equation}
where $a(l,t)$ and $P\left(l,1+\frac{2s}{t}\right)$ are the analytic continuations in $l$ of the partial waves $a_l(t)$, and of the Legendre polynomials $P_l\left(1+\frac{2s}{t}\right)$, respectively. The contour $C$ surrounds the positive real axis as shown in Fig.~\ref{Int:Fig:SommWats}. \\
It is important to note the presence of the index $\eta$, called \textit{signature}, which assumes the two possible values $1$ and $-1$. The signature means parity with respect to the substitution $ \cos \theta_{t} \leftrightarrow -\cos \theta_{t} $, where $\theta_t$ is the $t$-channel-scattering angle. As we will see, the separation into partial waves with a definite signature represents an important operation for the development of the BFKL approach. \\
Now, we need to carry out the so-called \textit{Sommerfeld-Watson transformation}, \textit{i.e.} to deform the path $C$ into the path $C'$, which runs parallel to the imaginary axis with $\Re \; l = - 1/2$, encircling any poles\footnote{In principles also cuts can be presented.} that the function $a^{( \eta )} \left(l,t \right)$ may have at $l= a_{n_{\eta}} (t)$ and picking up $2 \pi i \; \times $ the residue of that pole (see Fig.~\ref{Int:Fig:SommWats}). After this transformation, for the particular case of simple poles (in the $l$-plane) and in the Regge kinematical region $|s|\gg t$, one finds that
\begin{equation}
\label{Int:Eq:AsynAmplitude}
\mathcal{A} \left(s,t\right) \xrightarrow{s \rightarrow \infty} \frac{\eta+e^{-i\pi \alpha(t)}}{2} \beta(t) s^{\alpha(t)},
\end{equation}
where $\alpha(t)$ represents the position of the leading Regge pole in the $l$-plane; it is not a fixed value, but rather a function of the transferred momentum $t$.
More in general, in the case of non-simple poles and cuts, the expression of the amplitude in Eq.~(\ref{Int:Eq:AsynAmplitude}) will be different. \\ 
\begin{figure}
\begin{picture}(417,145)
\put(157,20){\includegraphics[width=0.3\textwidth]{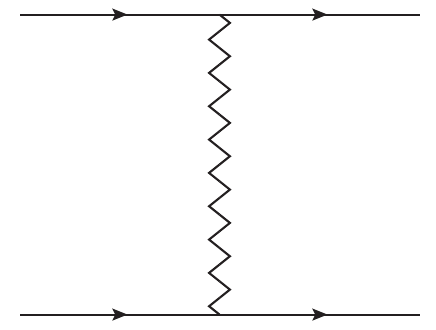}}
\put(240,70){$\longleftarrow$ Reggeon}
\put(142,117){$ a $}
\put(142,23){$ b $}
\put(305,118){$ c $}
\put(305,23){$ d $}
\put(214,135){$ \gamma_{ac} (t)$}
\put(214,5){$ \gamma_{bd} (t)$}
\end{picture}
\caption{A Regge exchange diagram.}
\label{Int:Fig:Reggefigure}
\end{figure}
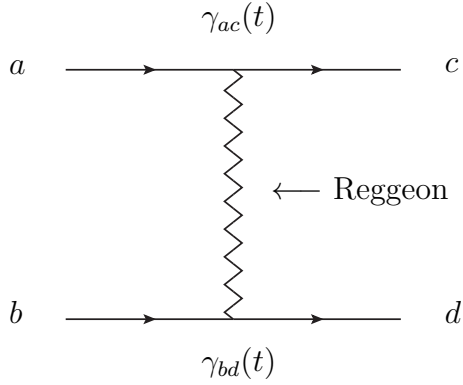
\hspace{-0.2 cm} The amplitude given by Eq.~(\ref{Int:Eq:AsynAmplitude}) can be seen as the exchange in the $t$-channel of an object with an effective ``angular momentum'' equal to $\alpha(t)$. It is called a Reggeon, and $\alpha(t)$ is called ``Regge trajectory''. A Reggeon-exchange amplitude can be thought as the superposition of amplitudes for the exchange of all possible particles in the $t$-channel, with quantum number determined by those of colliding particles. The amplitude can be factorized as shown in Fig.~\ref{Int:Fig:Reggefigure} into a coupling $\gamma_{ac}(t)$ of the Reggeon between the particles $a$ and $c$, a similar coupling $\gamma_{bd}(t)$ between particles $b$ and $d$, and a universal contribution from the Reggeon exchange. The couplings $\gamma$ are functions of $t$ only, and hence, the Reggeon exchange determines completely the behaviour in $s$ of the whole amplitude.
\subsection{Reggeization and Pomeron}
Now, it is possible to give a formal definition of Reggeization. A particle of mass $M$ and spin $J$ is said to ``Reggeize'' if the amplitude, $\mathcal{A}$, for a process involving the exchange in the $t$-channel of the quantum numbers of that particle, behaves asymptotically in $s$ as 
\begin{equation*}
\mathcal{A} \propto s^{\alpha(t)}, 
\end{equation*}
where $\alpha(t)$ is the trajectory and $\alpha(M^2)=J$, so that the particles itself lies on the trajectory. \\ 

In the large-$s$ limit, a hadronic process is governed by the exchange of one or more Reggeons in the $t$-channel.  The exchange of Reggeons instead of particles gives rise to scattering amplitudes of the type of Eq.~(\ref{Int:Eq:AsynInS}). Using the optical theorem together with the Eq.~(\ref{Int:Eq:AsynInS}) we can obtain the asymptotic behaviour of the total cross section of that process, which reads
\begin{equation}
\sigma_{\rm{tot}} \propto s^{(\alpha(0)-1)}.
\end{equation}  
In 1965 Pomeranchuk proved from general assumptions that in any scattering process in which there is charge exchange, the cross section vanishes asymptotically (the Pomeranchuk's theorem~\cite{Pomeranchuk:1956po}). Foldy and Peierls proved the converse, that is, if for a particular scattering process the cross section does not fall as $s$ increases, then that process must be dominated by the exchange of vacuum quantum numbers~\cite{Foldy:1963ip}.
It is observed experimentally that the total cross sections do not vanish asymptotically. In fact they rise slowly as $s$ increases. If we are to attribute this rise to the exchange of a single Regge pole, then it follows that the exchange is that of a Reggeon which carries the quantum numbers of the vacuum, called Pomeron ($\alpha_P(0)>1$) in honour of its inventor Pomeranchuk. The physical particles which would provide the resonances for integer values of the Pomeron trajectory have not been identified, but in QCD, natural candidates are hypothetical bound states of gluons (glueballs). \\

We want to remark once again that the notions of Reggeon and Pomeron that we have introduced are not linked to the perturbative approach. From now on, we will be interested in studying the so-called semi-hard processes in pQCD, \textit{i.e.} those processes characterized by the scale hierarchy\footnote{Here $t$ plays the role of a generic hard scale of the process.}
\begin{equation}
    s \gg t \gg \Lambda_{QCD}^2 \; .
    \label{Int:Eq:Semihard}
\end{equation}
In this context, we will talk about gluon Reggeization in pQCD and we will construct the so-called hard Pomeron (or BFKL Pomeron). These objects, which allow us to find the asymptotic behavior of the scattering amplitudes in pQCD, should not be confused with the corresponding soft objects which are of non-perturbative origin. The connection between the two objects concerns only the quantum numbers exchanged in the $t$-channel and the first condition in~(\ref{Int:Eq:Semihard}).  
\section{BFKL approach}
The BFKL equation is an integral equation that determines the behaviour at high energy $\sqrt{s}$ of the perturbative QCD amplitudes in which vacuum quantum numbers are exchanged in the $t$-channel. It was derived in the LLA, which means collection of all terms of the type $ \alpha_s^n \ln^n s $. This approximation leads to an increase of cross sections. In fact, calculated in LLA, the total cross $\sigma_{tot}^{LLA}$ grows at large center-of-mass energies as
\begin{equation}
\label{Int:Eq:AsyPomerSing}
\sigma_{tot}^{LLA} \sim \frac{s^{\omega_0}}{\sqrt{\ln s}},
\end{equation}
where $\omega_0 = (g^2 N \ln 2)/\pi^2$, with $N$ the number of colors in QCD, is the LLA position of the rightmost singularity in the complex momentum plane of the $t$-channel partial wave with vacuum quantum numbers (Pomeron singularity).
The BFKL equation works also in the NLLA (next-to-leading logarithmic approximation), in which one has to resum also all the terms of the type $\alpha_s^{n+1} \ln^{n}s$ in the perturbative series. To derive the BFKL equation, the Reggeization of the gluon in QCD is fundamental.
In this section, first, the BFKL equation will be treated in the LLA, then the NLLA case will be discussed.          
\subsection{Reggeization of the gluon in pQCD}
\label{Int:subsec:Reggeization of the gluon in pQCD}
Reggeization of the gluon in QCD has a very deep meaning. Obviously, it means that there exists a Reggeon with gluon quantum numbers, negative signature and trajectory
\begin{equation}
\label{Int:Eq:trajectory}
\alpha(t)=1+\omega(t) \; ,
\end{equation}
passing through one at $t=0$, but it also establishes that only this Reggeon gives the leading contribution, in each order of the perturbation theory, to amplitudes with the gluon quantum numbers exchanged in the $t$-channel. \\
Considering an elastic scattering process $ A+B \longrightarrow A'+B'$ with $s \gg |t|$, in which
\begin{equation}
s=(p_A+p_B)^2 \; ,  \hspace{2cm}  t=q^2 \; ,  \hspace{2cm} q= p_A-p_{A'} \; ,
\end{equation} 
the Reggeization implies that the amplitude with the gluon quantum numbers in $t$-channel has the factorized form
\begin{equation}
\label{Int:Eq:ReggeizedAmp}
\mathcal{A}_{AB}^{A'B'}= \Gamma^i_{A'A} \left( \frac{s}{-t} \right)^{ \alpha (t) } \left[ -1 + e^{-i \pi \alpha (t)} \right] \Gamma^i_{B'B} = \Gamma^i_{A'A} \frac{s}{t} \left[ \left(\frac{s}{-t}\right)^{\omega(t)} + \left( \frac{-s}{-t} \right)^{\omega(t)} \right] \Gamma^i_{B'B} \; ,
\end{equation} 
where $\Gamma_{P'P}^i$ (with $P=A,B$) represent the particle-particle-Reggeon (PPR) vertices (they do not depend on $s$) and $i$ is the colour index. 
It is not difficult to observe that the amplitude in Eq.~(\ref{Int:Eq:ReggeizedAmp}) is exactly of the form shown in Figure~\ref{Int:Fig:Reggefigure}, in fact,  there are two couplings of particles with the Reggeon and a universal contribution which behaves as $s^{\alpha(t)}$. The second term between square brackets comes from the contribution of the $u$-channel\footnote{Since $s+u+t \simeq s + u = \sum_i m_{i}^2 \simeq 0$, we have $u \simeq -s $.}. It can also be seen that this amplitude is antisymmetric under the transformation $s \rightarrow u \simeq -s$, which, in the Regge limit, is equivalent to negative signature. \\
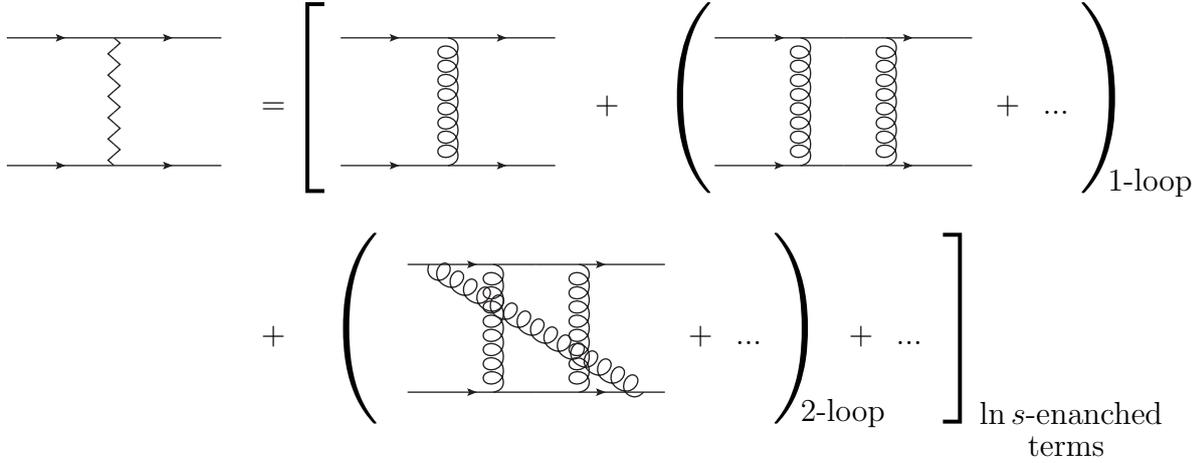
\begin{figure}
\begin{picture}(417,195)
\put(0,102){\includegraphics[scale=0.5]{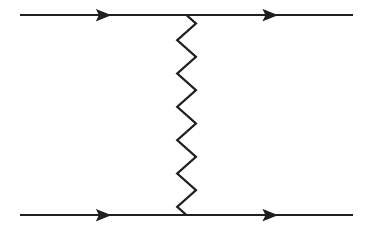}}
\put(100,125){$=$}
\put(110,125){\scalebox{2}{\Bigg [ }}
\put(125,102){\includegraphics[scale=0.5]{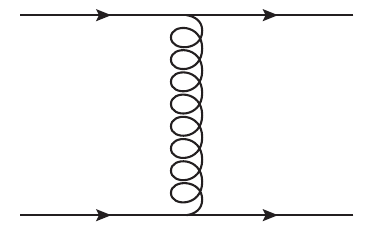}}
\put(225,125){$+$}
\put(250,125){\scalebox{2}{\Bigg ( }}
\put(265,102){\includegraphics[scale=0.5]{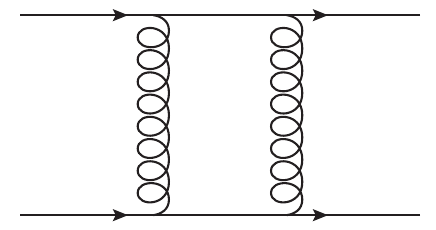}}
\put(375,125){$+ \hspace{0.3 cm}...$ \scalebox{2}{\Bigg )}}
\put(417,96){1-loop}
\put(100,38){$+$}
\put(125,38){\scalebox{2}{\Bigg ( }}
\put(150,15){\includegraphics[scale=0.5]{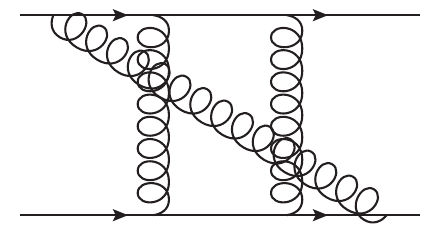}}
\put(260,38){$+ \hspace{0.3 cm}...$ \scalebox{2}{\Bigg )}}
\put(302,9){2-loop}
\put(320,38){$+ \hspace{0.3 cm}...$}
\put(355,38){\scalebox{2}{\Bigg ] }}
\put(365,8){ $\ln s$-enanched }
\put(387,-4){terms}
\end{picture}
  \caption{Schematic representation of the Reggeized gluon in pQCD. At Born level it is given by the exchange of a single gluon in the $t$-channel, while, at higher orders, it is represented by the exchange of a certain number of gluons which combine in such a way as to give the exchange of an octet Reggeized gluon in the $t$-channel.}
  \label{Int:Fig:PertExpReggeon}
\end{figure}
The analytic form in Eq.~(\ref{Int:Eq:ReggeizedAmp}) is the result of an infinite resummation of diagrams, this means that the pQCD Reggeized gluon possesses a perturbative expansion in terms of normal gluons (see Fig.~\ref{Int:Fig:PertExpReggeon}).  Each of the terms in the expansion in Fig.~\ref{Int:Fig:PertExpReggeon} represents a term in the $\omega(t)$-expansion (or equivalently $\alpha_s$-expansion) of the expression in Eq.~(\ref{Int:Eq:ReggeizedAmp}). \\
To give a clearer idea of the Reggeization, we will now extract the Regge trajectory at one-loop accuracy. To do this, we need to consider a ``reference" amplitude, compute it at least up to 1-loop accuracy in the high-energy limit and then compare it with the following expanded version of Eq.~(\ref{Int:Eq:ReggeizedAmp}):
\begin{equation}
    \left( \mathcal{A} \right)_{12}^{1'2'} =  \Gamma_{1'1}^a \frac{s}{t} \left[ 2 + \omega (t) \ln \left( \frac{-s}{-t} \right) + \omega (t) \ln \left( \frac{s}{-t} \right) \right] \Gamma_{2'2}^a + ... \; .
    \label{Int:Eq:ReggeizedAmpExp}
\end{equation}
Let us consider the quark-quark scattering amplitude with gluon quantum number exchange in the $t$-channel (see Fig.~\ref{Int:Fig:LOqqqq}). We choose the momenta of the two incoming quarks as
\begin{equation}
    p_1^{\mu} = \frac{\sqrt{s}}{2} \left( 1, \vec{0}, 1 \right) \; , \hspace{0.5 cm} p_2^{\mu} = \frac{\sqrt{s}}{2} \left( 1, \vec{0}, -1 \right) \; , 
\end{equation}
and use $q$ to denote the exchanged momentum carried by the gluon in the $t$-channel, so that
\begin{equation}
    q = p_1 - p_{1'} = p_{2'} - p_2 \hspace{0.3 cm} \;  {\rm{and}} \; \; \; t = q^2 \; .
    \label{Int:Eq:ReggeRegion}
\end{equation}
We are interested in studying the process in the Regge kinematical region, where 
\begin{equation}
   2 p_1 \cdot p_2 = s \gg |t| = \vec{q}^{\; 2} \; .   
\end{equation}
We choose to work in the covariant Feynman gauge and the form of the leading order contribution to the amplitude is
\begin{equation}
    \mathcal{A}_{12}^{1'2'} = (-i) i g t_{ji}^a \bar{u} (p_1-q) \gamma^{\mu} u (p_1) \left( \frac{-i g_{\mu \nu} \delta^{ab}}{q^2} \right)i g t_{kn}^b \bar{u} (p_2+q) \gamma^{\nu} u (p_2) \; .
    \label{Int:Eq:GeneralQuarkAmp}
\end{equation}
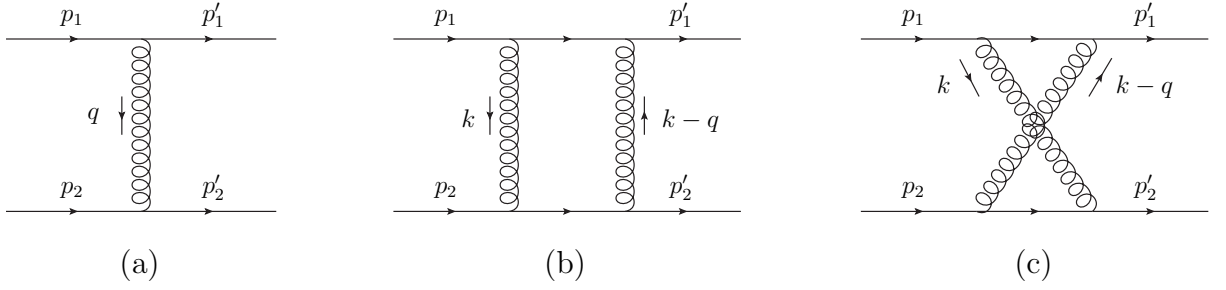
\begin{figure}
\begin{picture}(420,125)
\put(0,30){\includegraphics[scale=0.45]{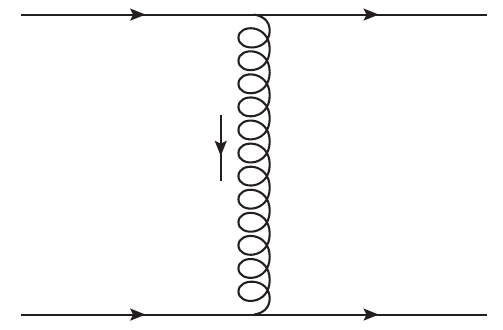}}
\put(47,10){(a)}
\put(35,68){\scalebox{0.8}{$q$}}
\put(25,40){\scalebox{0.8}{$p_2$}}
\put(25,105){\scalebox{0.8}{$p_1$}}
\put(78,40){\scalebox{0.8}{$p_2'$}}
\put(78,105){\scalebox{0.8}{$p_1'$}}
\put(145,30){\includegraphics[scale=0.45]{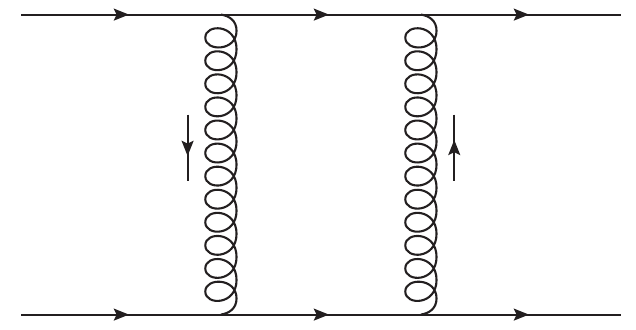}}
\put(206,10){(b)}
\put(175,65){\scalebox{0.8}{$k$}}
\put(250,65){\scalebox{0.8}{$k-q$}}
\put(165,40){\scalebox{0.8}{$p_2$}}
\put(165,105){\scalebox{0.8}{$p_1$}}
\put(253,40){\scalebox{0.8}{$p_2'$}}
\put(253,105){\scalebox{0.8}{$p_1'$}}
\put(320,30){\includegraphics[scale=0.45]{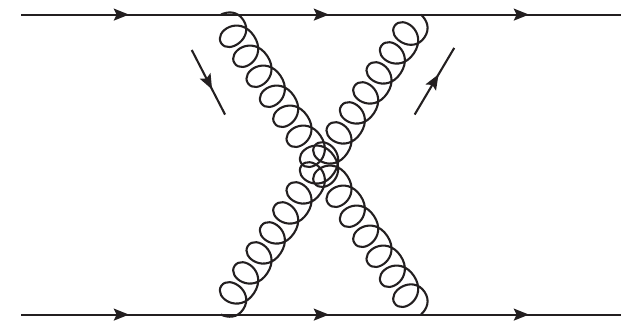}}
\put(382,10){(c)}
\put(353,78){\scalebox{0.8}{$k$}}
\put(420,78){\scalebox{0.8}{$k-q$}}
\put(340,40){\scalebox{0.8}{$p_2$}}
\put(340,105){\scalebox{0.8}{$p_1$}}
\put(427,40){\scalebox{0.8}{$p_2'$}}
\put(427,105){\scalebox{0.8}{$p_1'$}}
\end{picture}
  \caption{In (a) the leading order diagram contributing to the quark-quark scattering; (b) and (c) are the two next-leading order diagrams contributing to the Regge trajectory in the LLA.}
  \label{Int:Fig:LOqqqq}
\end{figure}
\hspace{-0.2 cm}Now, it is important to stress that in the Regge kinematical region, any components of $q$ is small compared to $\sqrt{s}$, and we can make the replacement\footnote{The normalization of spinors is $u^{\dagger} (p_1) u (p_1) = 2 E_{p_1} \delta_{\lambda_{1'}, \lambda_1}$.}
\begin{equation}
    \bar{u} (p_1-q) \gamma^{\mu} u (p_1) \simeq \bar{u} (p_1) \gamma^{\mu} u (p_1) = 2 \delta_{\lambda_{1'} , \lambda_1} p_1^{\mu} \; .
    \label{Int:Eq:EikonalApprox}
\end{equation}
This is called \textit{eikonal approximation} and it is correct when the gauge exchanged particle is relatively soft. The final result for the amplitude can be written as
\begin{equation}
    \mathcal{A}_{12}^{1'2'} = \Gamma_{1' 1}^{a} \left( \frac{2s}{t} \right) \Gamma_{2' 2}^{a} \; ,
\end{equation}
where
\begin{equation}
    \Gamma_{1' 1}^{a} = g t_{ji}^a \delta_{\lambda_{1'} \lambda_1} \; ,  \hspace{1 cm} \Gamma_{2' 2}^{a} = g t_{kn}^a \delta_{\lambda_{2'} \lambda_2}
\end{equation}
are the effective quark-quark-Reggeon vertices. \\
At this point it is interesting to note that this result could also be obtained starting from Eq.~(\ref{Int:Eq:GeneralQuarkAmp}) and performing the following replacement in the gluon propagator
\begin{equation}
  \frac{g^{\mu \nu}}{t} = \frac{1}{t} \left( g_{\perp \perp}^{\mu \nu} + \frac{2 p_1^{\mu} p_2^{\nu} + 2 p_1^{\nu} p_2^{\mu}}{s} \right) \rightarrow \frac{1}{t} \frac{2 p_1^{\nu} p_2^{\mu}}{s} = \frac{2 s}{t} \left(-\frac{p_2^{\mu}}{s} \right) \left(-\frac{p_1^{\nu}}{s} \right) \; .
\end{equation}
This replacement is known as \textit{Gribov trick} and represents a general method for extracting the high-energy behavior of QCD amplitudes in which relatively soft gluons are exchanged in the $t$-channel. The application of Gribov's trick produces an overall factor $2s/t$ and it causes the upper (lower) vertex to contract with a term $-p_2^{\mu}/s$ ($-p_1^{\nu}/s$). These contractions generate the effective vertices mentioned above, which are energy-independent. If we look only at the upper part of the diagram (a) in Fig.~\ref{Int:Fig:LOqqqq} it is as if the effective vertex can be obtained by writing the normal quark-quark-gluon vertex and assigning an ``effective polarization" (usually called \textit{nonsense polarization}), $-p_2^{\mu}/s$, to the $t$-channel gluon. \\
In order to compute the Regge trajectory, we need to compute one-loop corrections the $\mathcal{A}_{12}^{1'2'}$. In the LLA and in a covariant gauge there are no contributions from one-loop graphs which contain corrections to propagators or vertices, but only from the two diagrams denoted by (b) and (c) in Fig.~\ref{Int:Fig:LOqqqq}. We will denote the first as $\mathcal{A}_{12,{\rm{box}}}^{1'2'}$ and the second as $\mathcal{A}_{12,{\rm{cross}}}^{1'2'}$. 
The ``cross" diagram will be obtained form the ``box" one by using the crossing symmetry, \textit{i.e.} 
\begin{equation}
    \mathcal{A}_{12,{\rm{cross}}}^{1'2'} = - \mathcal{A}_{12,{\rm{box}}}^{1'2'} (s \rightarrow u \simeq -s) \; .
\label{Int:CrossSymRel}
\end{equation}
Eq.~(\ref{Int:CrossSymRel}) requires a number of remarks. As it is easy to observe, when we calculate the ``box" and the ``cross" diagram, in general, they have a different color structure. If the ``box" is proportional to
\begin{equation}
    (t^b t^a)_{ji} (t^b t^a)_{kn} \; , 
\end{equation}
the ``cross" is proportional to 
\begin{equation}
    (t^a t^b)_{ji} (t^b t^a)_{kn} \; .
\end{equation}
With these structures, in general the two amplitudes have no definite signature and Eq.~(\ref{Int:CrossSymRel}) is not valid. What is convenient to do is to separate each of the two amplitudes into two parts with a definite signature, using the identities
\begin{equation}
    (t^b t^a)_{ji} = \frac{1}{2} [t^b, t^a]_{ji} + \frac{1}{2} \{ t^b, t^a \}_{ji} \;, 
\end{equation}
\begin{equation}
    (t^a t^b)_{ji} = \frac{1}{2} [t^a, t^b]_{ji} + \frac{1}{2} \{ t^a, t^b \}_{ji} = - \frac{1}{2} [t^b, t^a]_{ji} + \frac{1}{2} \{ t^b, t^a \}_{ji} \; .
\end{equation}
Now, the parts of the amplitudes depending on the commutator, as far as the color structure is concerned, differ by one sign, while those depending on the anticommutator are equal. If we strip off the color factors, the amplitude of the ``cross" diagram is obtained from the ``box" through the substitution $s \rightarrow u \simeq - s$ which produces a global minus sign of difference in the real part of the two amplitudes. This last minus sign is compensated with that which comes from the color structure only for the part of amplitude proportional to the commutator. In the part of amplitude proportional to the anticommutator, it generates a ``fatal" cancellation of the dominant ($\ln (s)$-enhanced) contribution. 
The part of amplitude containing the commutator, which we will henceforth refer to as \textit{negative signature} contributions, dominate in the high-energy approximation and satisfies the relation in Eq.~(\ref{Int:CrossSymRel}). To all orders it produces the behavior in Eq.~(\ref{Int:Eq:ReggeizedAmp}). From now on, speaking of gluon quantum numbers, we will always refer to the part of the octet with negative signature and will therefore make use of definite parity with respect to $s \rightarrow u $ transformation of this contribution. \\
The ``box" diagram can be easily calculated using dispersive techniques. Cutting the diagram in the $s$-channel and using Cutkosky rules we have\footnote{We will always use $\sum_{\{ f \}}$ to denote the summation over all relevant quantum numbers of the intermediate state.}
\begin{equation}
   \Im \mathcal{A}_{12,{\rm{box}}}^{1'2'} =\frac{1}{2} \sum_{ \{ f \} } \int d \Phi_{2} \mathcal{A}_{12}^{1'2'} (k) \mathcal{A}_{12}^{1'2' \dagger} (q-k) \; ,
\end{equation}
where\footnote{We work in dimension $D=4+2\epsilon$. A non-zero $\epsilon$ is introduced to regularize IR-divergences.}
\begin{equation}
d \Phi_2 = \frac{d^D l_1 }{(2 \pi)^{D-1}} \frac{d^D l_2}{(2 \pi)^{D-1}} \delta (l_1^2) \delta (l_2^2) (2 \pi)^D \delta^D (p_1 + p_2 - l_1 - l_2) 
\label{Int:Eq:PhaseSpace2}
\end{equation}
is the two-body phase space. In Eq.~(\ref{Int:Eq:PhaseSpace2}), $l_1$ and $l_2$ represent the momenta of the two ``cut" particles. Because of the $\delta^D (p_1 + p_2 - l_1 - l_2)$, one integration can be done immediately and we can change variables from the remaining one to the momenta $k$ exchanged in the $t$-channel; we obtain
\begin{equation}
    d \Phi_2 = \frac{1}{(2 \pi)^{D-2}} d^D k \delta ((p_1-k)^2) \delta ((p_2+k)^2) \; . 
\end{equation}
Now, it is convenient to introduce the \textit{Sudakov decomposition} for the momentum $k$:
\begin{equation}
    k^{\mu} = \beta p_1^{\mu} + \alpha p_2^{\mu} + k_{\perp}^{\mu} \; ,
\end{equation}
where $k_{\perp}=(0, \vec{k}, 0)$ is a four-vector, completely transverse with respect to the plane identified by the two light-cone vectors $p_1$ and $p_2$. In terms of these variables, the phase space becomes 
\begin{equation*}
 d \Phi_{2} = \frac{s}{2 (2 \pi)^{D-2}} d \alpha d \beta d^{D-2} k_{\perp} \delta (-s(1-\beta) \alpha + k_{\perp}^{2} )  \delta ( s (1+\alpha) \beta + k_{\perp}^{2} ) 
\end{equation*}
\begin{equation}
 \simeq \frac{s}{2 (2 \pi)^{D-2}} d \alpha d \beta d^{D-2} k_{\perp} \delta (-s \alpha + k_{\perp}^{2} )  \delta ( s \beta + k_{\perp}^{2}) \; ,
 \label{Int:Fig:PhaseSpace2Part}
\end{equation}
where, in the last step, we have used that in the kinematics we are interested in
\begin{equation}
    \alpha \ll 1 \; , \hspace{0.5 cm} \beta \ll 1 \; , \hspace{0.5 cm} \; k^2 = k_{\perp}^{2} = -\vec{k}^2 , \hspace{0.5 cm} (k+q)^2 = (k+q)_{\perp}^{2} = -(\vec{k}+\vec{q} \; )^2 \; .
\end{equation}
Now, it is very simple to see that $\Im \mathcal{A}_{12,{\rm{box}}}^{1'2'}$ takes the form
\begin{equation}
    \Im \mathcal{A}_{12,{\rm{box}}}^{1'2'} = g \delta_{\lambda_1 \lambda_{1'}} g \delta_{\lambda_2 \lambda_{2'}} (t_{li}^a t_{mn}^a) (t_{jl}^b t_{km}^b) \left( \frac{s}{t} \right) \frac{g^2}{ (2 \pi)^{D-2}} \int d^{D-2} k_{\perp} \frac{t}{k_{\perp}^{2} (k-q )_{\perp}^{2}}  \; .
\end{equation}
Since we are interested in gluon quantum numbers we have to perform the projection on the negative signature and the octet colour
state in the $t$-channel. This is easily achieved through the replacement 
\begin{equation}
    (t^b t^a)_{ji} \rightarrow \frac{1}{2} (t^b t^a - t^a t^b)_{ji} = \frac{1}{2} (T^c)_{ab} t_{ji}^c \; .
    \label{Int:Eq:ProjectionNegSign}
\end{equation}
After this, the color structure becomes
\begin{equation}
    \frac{1}{2} (T^c)_{ab} t_{ji}^c (t^b t^a)_{kn} = - \frac{C_A}{4} t_{ji}^a t_{kn}^a \; ,
\end{equation}
and hence we obtain
\begin{equation}
     \Im \mathcal{A}_{12,{\rm{box}}}^{1'2'} = \Gamma_{1'1}^a \frac{s}{t} \Gamma_{2'2}^a (-\pi) \left[ \frac{g^2t}{(2\pi)^{(D-1)}} \frac{N}{2} \int \frac{d^{D-2}k_{\perp}}{k_{\perp}^2 \left(q-k\right)_{\perp}^2} \right] \; . 
\end{equation}
The mere existence of this imaginary part tells us that the full amplitude must be proportional to 
\begin{equation}
    \ln \left( \frac{-s}{-t} \right) = \ln \left( \frac{s}{-t} \right) - i \pi \; ,
\end{equation}
and hence, we have
\begin{equation}
     \mathcal{A}_{12,{\rm{box}}}^{1'2'} = \Gamma_{1'1}^a \frac{s}{t} \Gamma_{2'2}^a \ln \left( \frac{s}{t} \right) \left[ \frac{g^2t}{(2\pi)^{(D-1)}} \frac{N}{2} \int \frac{d^{D-2}k_{\perp}}{k_{\perp}^2 \left(q-k\right)_{\perp}^2} \right] \; . 
\end{equation}
Using the Eq.~(\ref{Int:CrossSymRel}) we can immediately write down the full (LO+NLO) amplitude, which reads
\begin{equation*}
    \mathcal{A}_{12}^{1'2'} + \mathcal{A}_{12,\rm{box}}^{1'2'} + \mathcal{A}_{12,\rm{cross}}^{1'2'} = \mathcal{A}_{12}^{1'2'} + \mathcal{A}_{12,\rm{box}}^{1'2'} - \mathcal{A}_{12,\rm{box}}^{1'2'} (s \rightarrow u \simeq -s)
\end{equation*}
\begin{equation}
    =  \Gamma_{1'1}^a \frac{s}{t} \left[ 2 + \omega^{(1)} (t) \ln \left( \frac{-s}{-t} \right) + \omega^{(1)} (t) \ln \left( \frac{s}{-t} \right) \right] \Gamma_{2'2}^a \; ,
\label{Int:Eq:qqScaHighEnergy}
\end{equation}
where
\begin{equation}
\omega^{(1)}(t) \equiv \frac{g^2t}{(2\pi)^{(D-1)}} \frac{N}{2} \int \frac{d^{D-2}k_{\perp}}{k_{\perp}^2 \left(q-k\right)_{\perp}^2} \; .
\end{equation}
The last definition is not casual. By comparing (\ref{Int:Eq:qqScaHighEnergy}) and (\ref{Int:Eq:ReggeizedAmpExp}), we understand immediately that, the deviation from one of the gluon trajectory is
\begin{equation}
\label{Int:Eq:ReggeTraj1loop}
\omega (t) \simeq \omega^{(1)}(t) = \frac{g^2t}{(2\pi)^{(D-1)}} \frac{N}{2} \int \frac{d^{D-2}k_{\perp}}{k_{\perp}^2 \left(q-k\right)_{\perp}^2} \; .
\end{equation}
A curious reader might reasonably wonder why the relation $ \omega(t) \simeq \omega^{(1)} (t)$. It indicates that we have extracted only the one-loop contribution, $\omega^{(1)}(t)$, to the complete Regge trajectory, $\omega(t)$. What are the sub-dominant contributions to Regge trajectory, it will become clear in the next sections when we will discuss BFKL in the NLLA. For the moment, is important to remind that, although we have made a resummation to all orders, we have collected, order by order, only those terms in which the number of powers of $\alpha_s$ is compensated by an equal number of logarithms of the ratio $s/t$. \\
The integral in Eq.~(\ref{Int:Eq:ReggeTraj1loop}) can be evaluated by using standard techniques for Feynman integrals, we get
\begin{equation}
\label{Int:Eq:ReggeTraj1loopInt}
\omega^{(1)} (t) = \frac{g^2t}{(2\pi)^{(D-1)}} \frac{N}{2} \int \frac{d^{D-2}k_{\perp}}{k_{\perp}^2 \left(q-k\right)_{\perp}^2} = -\frac{g^2N \Gamma(1-\epsilon)}{(4\pi)^{2+\epsilon}} \frac{[\Gamma(\epsilon)]^2}{\Gamma(2\epsilon)} (\vec{q}^{\; 2})^{\epsilon} \; .
\end{equation}
From Eq.~(\ref{Int:Eq:ReggeTraj1loopInt}), it is easy to observe that, for the gluon, $\omega(0)=0$, hence $\alpha(0)=1$; this shows that the gluon lies on the Reggeon trajectory with its quantum numbers. 
\subsection{The Lipatov vertex}
\label{Int:Sec:LipatovVertex}
The second key ingredient to build the BFKL approach in the leading-logarithmic approximation is the so-called \textit{Lipatov vertex}. It is an effective non-local vertex. To understand its nature, Let us consider the quark-quark scattering again. Let us imagine we want to demonstrate explicitly that the form (\ref{Int:Eq:ReggeizedAmp}) is correct at two-loop level. If we want to use again dispersive techniques, cutting diagrams in the $s$-channel, we should compute a series of diagrams that we can split into two categories\footnote{In analogy with before, we do not get contribution of the form $\alpha_s^2 \ln^2 \left( s/t \right)$ from graph which consist of vertex or self-energy corrections on the one-loop diagrams.}:
\begin{figure}
\begin{subfigure}{.5\textwidth}
\centering
\includegraphics[width=0.90\textwidth]{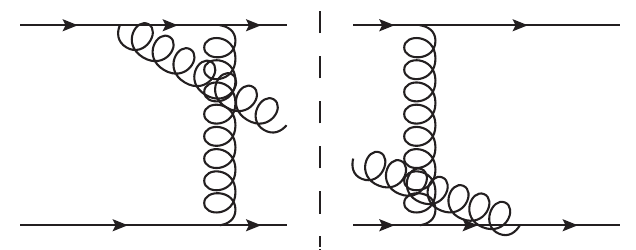}
\caption{}
\end{subfigure}
\begin{subfigure}{.5\textwidth}
\centering
\includegraphics[width=0.90\textwidth]{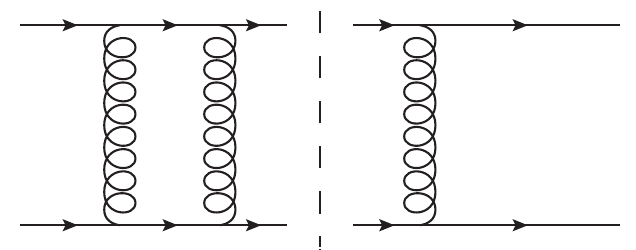}
\caption{}
\end{subfigure}
\caption{The two types of diagram which contribute to the imaginary part of the elastic amplitude, $\Im A_{12}^{1'2'}$, at this perturbative order.}
\label{Int:Fig:CutvsUnCut}
\end{figure}
\begin{itemize}
    \item Diagrams in which the cut goes through one gluon line and two quark lines (see diagram (a) in Fig.~\ref{Int:Fig:CutvsUnCut}); we have 25 diagrams of this kind.
    \item Diagrams in which the cut only goes through the two quark lines (see diagram (b) in Fig.~\ref{Int:Fig:CutvsUnCut}); we have 4 diagrams of this kind.
\end{itemize}
Let us concentrate on the first class for the moment. It is clear that, for this class of diagram, the contribution to the imaginary part of the elastic amplitude will read\footnote{I denoted this contribution with the label ``ladder" because (within this order), as we will see, this contributions are organizable as an effective ladder of pure gluons.}
\begin{equation}
   \Im \mathcal{A}_{12,{\rm{ladder}}}^{1'2'} = \frac{1}{2} \sum_{\{ f \}} \int d \Phi_{3} \mathcal{A}_{12}^{1'2'3, \sigma} (q_1 , q_2) (\mathcal{A}_{12}^{1'2'3, \gamma} (q_1-q,q_2-q))^{\dagger} \; ,
   \label{Int:Eq:Amp12Ladder}
\end{equation}
where $3$ is labelling the additional particle (gluon in this case). Since the gluon can be emitted by one of the four quark lines or by the $t$-channel gluon, the inelastic amplitude $\mathcal{A}_{12}^{1'2'3, \sigma}$ has five contributions. These contributions are shown in Fig.~\ref{Int:Fig:RRgIn5cont}.
\begin{figure}
\begin{subfigure}{.325\textwidth}
\centering
\includegraphics[width=0.90\textwidth]{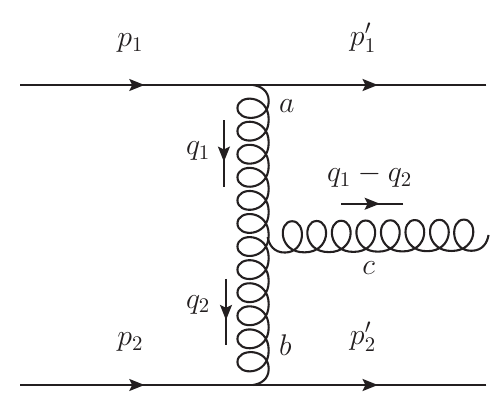}
\caption{}
\end{subfigure}
\begin{subfigure}{.335\textwidth}
\centering
\includegraphics[width=0.90\textwidth]{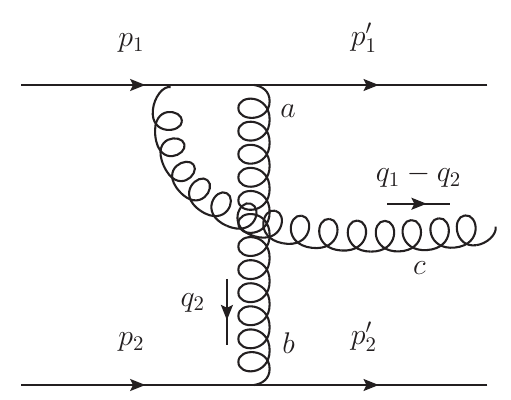}
\caption{}
\end{subfigure}
\begin{subfigure}{.325\textwidth}
\centering
\includegraphics[width=0.96\textwidth]{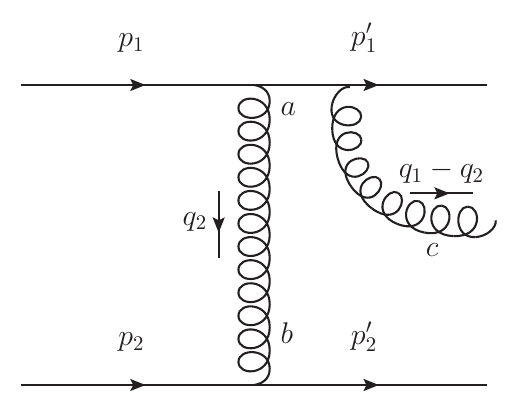}
\caption{}
\end{subfigure}
\begin{center}
\begin{subfigure}{.325\textwidth}
\includegraphics[width=0.96\textwidth]{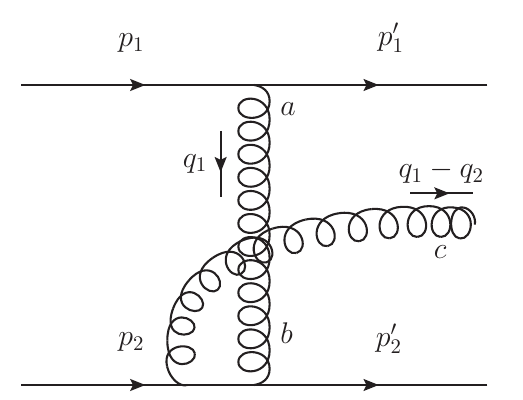}
\caption{}
\end{subfigure}
\begin{subfigure}{.325\textwidth}
\includegraphics[width=0.96\textwidth]{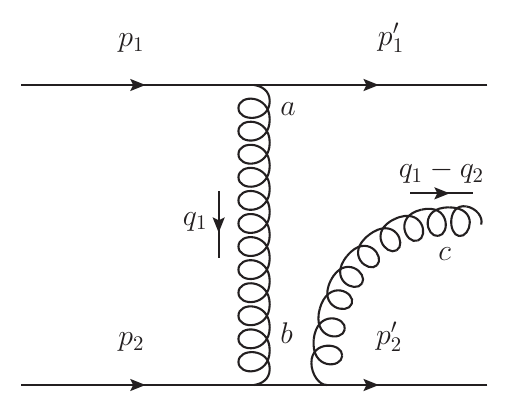}
\caption{}
\end{subfigure}
\end{center}
\caption{The five diagrams contributing to the inelastic process $q+q \rightarrow q+g+q$.}
\label{Int:Fig:RRgIn5cont}
\end{figure}
Let us start by computing the diagram (a). Using the Gribov trick for the $t$-channel gluons, it can be immediately written as
\begin{equation}
    \mathcal{A}_{12,(a)}^{1'2'3} = -2 s \Gamma_{1'1}^{a} \frac{1}{t_1} g (T^c)_{ab} \left[ 2 \frac{p_{1,\rho} p_{2,\delta}}{s} A^{\rho \delta \sigma} (q_2, q_1-q_2) \varepsilon_{\sigma}^* \right] \frac{1}{t_2} \Gamma_{2'2}^{b} \; ,
\end{equation}
where 
\begin{equation*}
  A^{\rho \delta \sigma} (q_2, q_1-q_2) = g^{\rho \sigma} (q_1 - 2 q_2)^{\rho} + g^{\rho \delta} (q_1 + q_2)^{\sigma} - g^{\rho \sigma} (2 q_1 - q_2)^{\delta}  
\end{equation*}
is the tensor structure appearing in the triple-gluon vertex. Now, we introduce the usual Sudakov decomposition for gluons momenta:
\begin{equation}
    q_1 = \beta_{q_1} p_1 + \alpha_{q_1} p_2 + q_{1, \perp} \; , \hspace{1 cm}  q_2 = \beta_{q_2} p_1 + \alpha_{q_2} p_2 + q_{2, \perp} \; .
\end{equation}
The dominant logarithms come from the region where
\begin{equation}
    1 \gg \beta_{q_1} \gg \beta_{q_2} \; , \hspace{0.5 cm}  1 \gg |\alpha_{q_2}| \gg |\alpha_{q_1}| \; , \hspace{0.5 cm} t_1=q_1^2 = -\vec{q}_1^{\; 2} \; , \hspace{0.5 cm} t_2=q_2^2 = -\vec{q}_2^{\; 2} \; .
\end{equation}
It is easy to understand that the first three conditions automatically imply also that
\begin{equation}
    1 \gg \beta_{q_1} - \beta_{q_2} > 0 , \hspace{0.5 cm} 1 \gg \alpha_{q_1} - \alpha_{q_2} > 0 \; , 
\end{equation}
and hence the three particles through the cut are strongly ordered in rapidity. This kinematical constraint is very important and it is known as \textit{multi-Regge kinematics}. \\ From the on-shell condition of the emitted gluon, we also have
\begin{equation}
    (q_1-q_2)^2 = -\beta_{q_1} \alpha_{q_2} s - (\vec{q}_1-\vec{q}_2)^{2} = 0  \implies  \beta_{q_1} \alpha_{q_2} s = - (\vec{q}_1-\vec{q}_2)^{2}  \; .
\end{equation}
That said, we understand that
\begin{equation}
    2 \frac{p_{1,\rho} p_{2,\delta}}{s} A^{\rho \delta \sigma} (q_2, q_1-q_2) \simeq - \alpha_{q_2} p_2^{\sigma} - \beta_{q_1} p_1^{\sigma} + (q_1+q_2)_{\perp}^{\sigma} 
\end{equation}
and hence
\begin{equation}
    \mathcal{A}_{12,(a)}^{1'2'3} = 2 s \Gamma_{1'1}^{a} \frac{1}{t_1} g (T^c)_{ab} \varepsilon_{\sigma}^* \left[ \alpha_{q_2} p_2^{\sigma} + \beta_{q_1} p_1^{\sigma} - (q_1+q_2)_{\perp}^{\sigma} \right] \frac{1}{t_2} \Gamma_{2'2}^{b} \; .
\end{equation}
Let us now consider diagram (b), we first observe that the quark propagator appearing in this diagram, in the multi-Regge kinematics, can be approximated as  
\begin{equation}
    \frac{i (\hat{p}_1-\hat{q}_1+\hat{q}_2)}{(p_1-q_1+q_2)^2} \simeq \frac{i \hat{p}_1 }{\alpha_{q_2} s} \; ,
\label{Int:Eq:ApproxQuarkProp}
\end{equation}
where $\hat{p} = \gamma^{\mu} p_{\mu} $. The approximation in (\ref{Int:Eq:ApproxQuarkProp}) is justified by the fact that any components of $q_1-q_2$ is negligible compared to the longitudinal component of $p_1$.
After this observation is clear that, by using again the Gribov trick for the $t$-channel gluon propagator, we get
\begin{equation}
    \mathcal{A}_{12,(b)}^{1'2'3} = (t^a t^c)_{ji} t^a_{kn} 2 s g \bar{u} (p_1-q_1) \frac{\hat{p}_2}{s} u(p_1) \frac{1}{t_1} g \varepsilon_{\sigma}^* \left[ \frac{2 p_1^{\sigma} t_1}{\alpha_{q_2} s} \right] \frac{1}{t_2} \bar{u} (p_2+q_2) \frac{\hat{p}_2}{s} u(p_2) \; .
\end{equation}
The diagram (c) is computed in the same way, but with respect to the previous one it has a minus sign and the color structure in front is $(t^c t^a)_{ji} t^a_{kn}$. The sum of the two gives
\begin{equation*}
   \mathcal{A}_{12,(b)}^{1'2'3} + \mathcal{A}_{12,(c)}^{1'2'3} = (t^a t^c - t^c t^a)_{ji} t^a_{kn} 2 s g \bar{u} (p_1-q_1) \frac{\hat{p}_2}{s} u(p_1) \frac{1}{t_1} g \varepsilon_{\sigma}^* \left[ \frac{2 p_1^{\sigma} t_1}{\alpha_{q_2} s} \right] \frac{1}{t_2} \bar{u} (p_2+q_2) \frac{\hat{p}_2}{s} u(p_2) 
\end{equation*}
\begin{equation}
    = 2 s \Gamma^{a}_{1'1} \frac{1}{t_1} g \left( T^c \right)_{ab} \varepsilon_{\sigma}^* \left[ - \frac{2 p_1^{\sigma} t_1}{\alpha_{q_2} s} \right] \frac{1}{t_2} \Gamma^{b}_{2'2} \; .
    \label{Int:Eq:Lip(b)+(c)}
\end{equation}
The sum of the two remaining diagrams, $\mathcal{A}_{12,(d)}^{1'2'3} + \mathcal{A}_{12,(e)}^{1'2'3}$, can be obtained from (\ref{Int:Eq:Lip(b)+(c)}) by adding a minus sign and performing the substitution: $p_1 \leftrightarrow p_2, \; t_1 \leftrightarrow t_2, \alpha_2 \leftrightarrow - \beta_1$. The contribution is
\begin{equation}
    \mathcal{A}_{12,(d)}^{1'2'3} + \mathcal{A}_{12,(e)}^{1'2'3} = 2 s \Gamma^{a}_{1'1} \frac{1}{t_1} g \left( T^c \right)_{ab} \varepsilon_{\sigma}^* \left[ - \frac{2 p_2^{\sigma} t_2}{\beta_{q_1} s} \right] \frac{1}{t_2} \Gamma^{b}_{2'2} \; .
    \label{Int:Eq:Lip(d)+(e)}
\end{equation}
Hence, the complete inelastic amplitude $\mathcal{A}_{12}^{1'2'3}$ is
\begin{equation*}
    \mathcal{A}_{12}^{1'2'3} = 2 s \Gamma_{1'1}^{a} \frac{1}{t_1} g (T^c)_{ab} \varepsilon_{\sigma}^* \left[ p_2^{\sigma} \left( \alpha_{q_2} - \frac{2 t_2}{\beta_{q_1} s} \right) + p_1^{\sigma} \left( \beta_{q_1} - \frac{2 t_1}{\alpha_{q_2} s} \right) - (q_1+q_2)_{\perp}^{\sigma} \right] \frac{1}{t_2} \Gamma_{2'2}^{b}
\end{equation*}
\begin{equation}
    \equiv 2 s \Gamma_{1'1}^{a} \frac{1}{t_1} g (T^c)_{ab} \varepsilon_{\sigma}^* C^{\sigma} (q_2, q_1) \frac{1}{t_2} \Gamma_{2'2}^{b} \; .
    \label{Int:Eq:Inelastic2->3}
\end{equation}
The structure in the middle of the two propagators, 
\begin{equation}
    \gamma_{ab}^c (q_1, q_2) = g (T^c)_{ab} \varepsilon_{\sigma}^* C^{\sigma} (q_2, q_1) \; ,
\end{equation}
is the previously mentioned Lipatov effective vertex and its structure is schematically represented in Fig.~\ref{Int:Fig:RRgInTermLip}. Defining $k=q_1-q_2$ and re-expressing the Lorentz structure inside the Lipatov vertex in terms of scalar products, we obtain
\begin{equation*}
     C^{\sigma} (q_2, q_1) =  - q_{1 \perp}^{\sigma} - q_{2 \perp}^{\sigma} - \frac{p_1^{\sigma}}{2 p_1 \cdot k} \left( k_{\perp}^2 - 2 q_{1 \perp}^2 \right) + \frac{p_2^{\sigma}}{2 p_2 \cdot k} \left( k_{\perp}^2 - 2 q_{2 \perp}^2 \right)  
\end{equation*}
\begin{equation}
     = - q_{1}^{\sigma} - q_{2}^{\sigma} + p_1^{\sigma} \left( \frac{q_1^2}{p_1 \cdot k} + \frac{2 p_2 \cdot k}{p_1 \cdot p_2} \right) - p_2^{\sigma} \left( \frac{q_2^2}{p_2 \cdot k} + \frac{2 p_1 \cdot k}{p_1 \cdot p_2} \right)  \; .
     \label{Int:Eq:TensLipatVer}
\end{equation}
Before proceeding further, let us take a little digression to discuss some properties of the vertex. It is simple to observe that
\begin{equation}
    C^{\sigma} (q_2, q_1) k_{\sigma} = 0 \; ,    \label{Int:Eq:GaugeInvarianceLipatovVertex}
\end{equation}
which means that the vertex is gauge invariant. \\ In physical light-cone gauges, the vertex simplifies considerably. Let us choose the following gauge:
\begin{equation}
     \varepsilon (k) \cdot p_2 = \varepsilon (k) \cdot k = 0 \; .
\end{equation}
From the first condition $\varepsilon (k)$ has no $p_1$ component, the second automatically fixes the one along $p_2$ and we have
\begin{equation}
    \varepsilon_{\sigma} (k) = \varepsilon_{\perp \sigma} (k) - p_{2 \sigma} \frac{\varepsilon_{\perp} \cdot k_{\perp}}{p_2 \cdot k}  \; .
    \label{Int:Eq:LightConeGauge1}
\end{equation}
Using (\ref{Int:Eq:LightConeGauge1}), we find the form of the Lipatov vertex in this gauge,
\begin{equation}
    \gamma_{ab}^c (q_1, q_2) = g (T^c)_{ab} \varepsilon_{\sigma}^* \left[ q_{1 \perp}^{\sigma} - k_{\perp}^{\mu} \frac{2 q_{1 \perp}^2}{k_{\perp}^2} \right] \; .
\end{equation}
In the light-cone gauge defined by
\begin{equation}
    \varepsilon (k) \cdot p_1 = \varepsilon (k) \cdot k = 0 \; ,
\end{equation}
the vertex becomes 
\begin{equation}
   \gamma_{ab}^c (q_1, q_2) = g (T^c)_{ab} \varepsilon_{\sigma}^* \left[ q_{2 \perp}^{\sigma} + k_{\perp}^{\mu} \frac{2 q_{2 \perp}^2}{k_{\perp}^2} \right] \; . 
\end{equation}
\begin{figure}
\begin{picture}(417,205)
\put(17,102){\includegraphics[scale=0.5]{images/RRGa.pdf}}
\put(150,140){$+$}
\put(167,102){\includegraphics[scale=0.5]{images/RRGb.pdf}}
\put(300,140){$+$}
\put(317,102){\includegraphics[scale=0.5]{images/RRGc.pdf}}
\put(0,38){$+$}
\put(17,0){\includegraphics[scale=0.5]{images/RRGd.pdf}}
\put(150,38){$+$}
\put(167,0){\includegraphics[scale=0.5]{images/RRGe.pdf}}
\put(300,38){$ \equiv $}
\put(317,0){\includegraphics[scale=0.5]{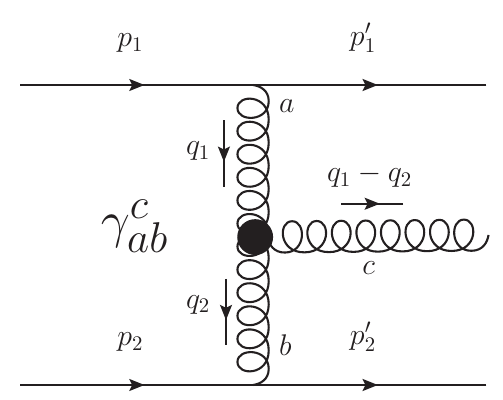}}
\end{picture}
\caption{Schematic representation of the Lipatov effective vertex.}
\label{Int:Fig:RRgInTermLip}
\end{figure}
Let us now return to the calculation of the imaginary part of the elastic amplitude (Eq.~(\ref{Int:Eq:Amp12Ladder})). First of all, we have to compute the contraction between the two Lorentz structures appearing in the Lipatov vertex\footnote{Recall that Hermitian conjugation requires the reversal of the direction of momentum in the right hand effective vertex.},
\begin{equation}
    C^{\sigma} (q_2, q_1) C_{\sigma} (q-q_2,q-q_1) = 2 q_{\perp}^2 - 2 \frac{(q_2-q)_{\perp}^2 q_{1 \perp}^2}{(q_1-q_2)_{\perp}^2} - 2 \frac{(q_1-q)_{\perp}^2 q_{2 \perp}^2}{(q_1-q_2)_{\perp}^2} \; .
\end{equation}
In the last equation we have neglected the longitudinal component $\beta_q$($\alpha_q$) of the net transverse momentum exchanged in the $q$-$q$ scattering, with respect to $\beta_{q_1}$($\alpha_{q_2}$). As for the color structure and other factors, we have
\begin{equation}
     \frac{2 s^2 g^6 \delta_{\lambda_1 \lambda_{1'}} \delta_{\lambda_2 \lambda_{2'}}}{q_{1 \perp}^2 q_{2 \perp}^2 (q_1-q)_{\perp}^2 (q_2-q)_{\perp}^2} f_{abc} f_{dec} (t^d t^a)_{ji} (t^e t^b)_{kn}  \; .
\end{equation}
Performing again the projection on negative signature and $t$-channel color octet through the substitution:
\begin{equation}
    (t^e t^b)_{kn} \rightarrow \frac{1}{2} (t^e t^b - t^b t^e)_{kn} \; ,
\end{equation}
we get
\begin{equation}
 -  \frac{s^2 g^6 \delta_{\lambda_1 \lambda_{1'}} \delta_{\lambda_2 \lambda_{2'}} C_A^2}{4 q_{1 \perp}^2 q_{2 \perp}^2 (q_1-q)_{\perp}^2 (q_2-q)_{\perp}^2} (t^a)_{ji} (t^a)_{kn}  \; .
\end{equation}
For the imaginary part of the elastic amplitude, we get
\begin{equation*}
    \Im \mathcal{A}_{12,{\rm{ladder}}}^{1'2'} = - \frac{s^2 g^6 \delta_{\lambda_1 \lambda_{1'}} \delta_{\lambda_2 \lambda_{2'}}}{2} C_A^2 (t^a)_{ji} (t^a)_{kn} 
\end{equation*}
\begin{equation}
  \times  \int d \Phi_3 \frac{1}{q_{1 \perp}^2 q_{2 \perp}^2 (q_1-q)_{\perp}^2 (q_2-q)_{\perp}^2} \left[  q_{\perp}^2 -  \frac{(q_2-q)_{\perp}^2 q_{1 \perp}^2}{(q_1-q_2)_{\perp}^2} - \frac{(q_1-q)_{\perp}^2 q_{2 \perp}^2}{(q_1-q_2)_{\perp}^2} \right] \; .
\end{equation}
Now, we can write down the three-particles phase space in terms of Sudakov variables of $q_1$ and $q_2$,
\begin{equation*}
     d \Phi_3 = \frac{1}{(2 \pi)^{2 D-3}} \left( \frac{s}{2} \right)^2 d \beta_{q_1} d \beta_{q_2} d \alpha_{q_1} d \alpha_{q_2} d^{D-2} q_{1 \perp} d^{D-2} q_{2 \perp} 
\end{equation*}
\begin{equation}
   \times \delta (-\alpha_{q_1} s + q_{1 \perp}^2) \delta (\beta_{q_2} s + q_{2 \perp}^2) \delta (-\beta_{q_1} \alpha_{q_2} s + (q_1-q_2)_{\perp}^2) \; ,
\end{equation}
and obtain\footnote{The motivation for the bounds in the integration over $\beta_1$ is that, in the kinematic region we are interested in, $ 1 \gg \beta_1 \gg \beta_2 > 0 $.}
\begin{equation*}
    \Im \mathcal{A}_{12,{\rm{ladder}}}^{1'2'} = - \frac{s g^6 \delta_{\lambda_1 \lambda_{1'}} \delta_{\lambda_2 \lambda_{2'}} C_A^2 (t^a)_{ji} (t^a)_{kn} }{8 (2 \pi)^{2D-3}} \int_{-t/s}^1 \frac{d \beta_{q_1}}{\beta_{q_1}} d^{D-2} q_{1 \perp} d^{D-2} q_{2 \perp} 
\end{equation*}
\begin{equation}
  \times \left[ \frac{ q_{\perp}^2 }{q_{1 \perp}^2 q_{2 \perp}^2 (q_1-q)_{\perp}^2 (q_2-q)_{\perp}^2} - \frac{1}{q_{1 \perp}^2 (q_2-q)_{\perp}^2 (q_1-q_2)_{\perp}^2} - \frac{1}{q_{2 \perp}^2 (q_1-q)_{\perp}^2 (q_1-q_2)_{\perp}^2} \right] \; .
  \label{Int:Eq:LadderPart}
\end{equation}
The first term in the previous expression looks promising, in fact, since the two transverse momentum integration are decoupled, we can easily obtain a factor $\omega^2(t)$. The other two terms, on the other hand, are far from promising, but we still have to include the contributions that come from the second family of diagrams (see (b) of Fig.~\ref{Int:Fig:CutvsUnCut}). Let us consider the diagram (b) of Fig.~\ref{Int:Fig:CutvsUnCut}, we have
\begin{equation*}
    \Im A_{12, (b)}^{1'2'} = \frac{1}{2} \sum_{\{ f \}} \int d \Phi_2 \mathcal{A}_{12,{\rm{box}}}^{1'2'} (q_2) \mathcal{A}_{12}^{1'2'} (q-q_2)
\end{equation*}
\begin{equation}
   = - \frac{s g^6 \delta_{\lambda_1 \lambda_{1'}} \delta_{\lambda_2 \lambda_{2'}} C_A^2 t_{ji}^a t_{kn}^a}{16 (2 \pi)^{2D-3}} \ln \left( \frac{s}{-t} \right) \int d^{D-2} q_{1 \perp} d^{D-2} q_{2 \perp} \left[ \frac{1}{q_{1 \perp}^2 (q_2-q)_{\perp}^2 (q_1-q_2)_{\perp}^2} \right] \; .
\label{Int:Eq:(b)UnCutDiag}
\end{equation}
The last equality in Eq.~(\ref{Int:Eq:(b)UnCutDiag}) is obtained by exploiting the expression for $d \Phi_2$, as in Eq.~(\ref{Int:Fig:PhaseSpace2Part}), but in terms of the Sudakov variables of $q_2$ and expressing the Regge trajectory $\omega (q_2^2)$, contained in $\mathcal{A}_{12, {\rm{box}}}^{1'2'} (q_2)$, as an integral over the variable $q_1$\footnote{As usual, we also project on negative signature and $t$-channel color octet.}. Another point that needs to be clarified concerns $\ln (s/(-t))$. We have only considered the real part of the $\mathcal{A}_{12, {\rm{box}}}^{1'2'}$ amplitude, which is dominant in the high-energy limit. 
However, the reader may be confused by the fact that, if we keep full amplitude, it seems that $\Im \mathcal{A}_{12}^{1'2'}$ acquires an imaginary part, which is absurd since it is real by definition. The illusory presence of this imaginary part is due to the fact that we are taking all contributions through the cut separately. If we put all the diagrams together, a contribution will also appear where the box is to the right of the cut and its imaginary part will remove exactly the one previously mentioned. \\
The diagram in which the ``box" is replaced by the ``cross" gives exactly the same contribution, while, it is simple to observe that the contribution of the two diagrams in which the single gluon is at the right of the cut are obtained by the exchange $q_1 \leftrightarrow q_2$. In this way, we obtain that the total contribution due to the four diagrams in which the cut crosses only the quark lines is
\begin{equation*}
    \Im A_{12,\text{unladder}}^{1'2'} =  - \frac{s g^6 \delta_{\lambda_1 \lambda_{1'}} \delta_{\lambda_2 \lambda_{2'}} C_A^2 t_{ji}^a t_{kn}^a}{8 (2 \pi)^{2D-3}} \ln \left( \frac{s}{-t} \right)
\end{equation*}
\begin{equation}
    \times \int d^{D-2} q_{1 \perp} d^{D-2} q_{2 \perp} \left[ \frac{1}{q_{1 \perp}^2 (q_2-q)_{\perp}^2 (q_1-q_2)_{\perp}^2} + \frac{1}{q_{2 \perp}^2 (q_1-q)_{\perp}^2 (q_1-q_2)_{\perp}^2} \right] \; .
\end{equation}
This contribution exactly cancels the ``unwanted" part in Eq.~(\ref{Int:Eq:LadderPart}). Finally, we obtain that the imaginary part of the two-loop correction to the amplitude $\mathcal{A}_{12}^{1'2'}$ is
\begin{equation*}
    \Im A_{12}^{1'2'(2)} =  - \frac{s g^6 \delta_{\lambda_1 \lambda_{1'}} \delta_{\lambda_2 \lambda_{2'}} C_A^2 (t^a)_{ji} (t^a)_{kn} }{8 (2 \pi)^{2D-3}} 
\end{equation*}
\begin{equation}
  \times  \int_{-t/s}^1 \frac{d \beta_{q_1}}{\beta_{q_1}} d^{D-2} q_{1 \perp} d^{D-2} q_{2 \perp} 
    \left[ \frac{ q_{\perp}^2 }{q_{1 \perp}^2 q_{2 \perp}^2 (q_1-q)_{\perp}^2 (q_2-q)_{\perp}^2} \right] \; .
\end{equation}
With few efforts, we can rewrite this as
\begin{equation}
    \Im A_{12}^{1'2'(2)} = \Gamma_{1'1}^a \frac{s}{t} \Gamma_{2'2}^a \frac{\omega(t)^2}{2} \left[ - 2 \pi \ln \left( \frac{s}{-t} \right) \right] \; .
\end{equation}
We can use dispersion relation to reconstruct the full amplitude. First, we observe that the term in the square bracket is the imaginary part of $\ln^2 ((-s)/(-t))$ and then we add the contribution coming from the crossed diagrams in which $s$ is replaced by $u$\footnote{As usual, the minus we get from the factor $s$ t
hat goes in $u=-s$ is compensated by a color factor and hence the net effect is only in the sign inside the logarithm.}:
\begin{equation}
    A_{12}^{1'2'(2)} = \Gamma_{1'1}^a \frac{s}{t} \Gamma_{2'2}^a \frac{\omega(t)^2}{2} \left[ \ln^2 \left( \frac{-s}{-t} \right) + \ln^2 \left( \frac{s}{-t} \right) \right] \; .
\end{equation}
This is exactly the third term in the $\omega(t)$ (or equivalently $\alpha_s$)-expansion of Eq.~(\ref{Int:Eq:ReggeizedAmp}) and it therefore confirms the Reggeization ansatz up to 2-loop. It is important to note that it was necessary to include the contribution of diagrams in which the gluon does not pass through the cut.
In the context of this dispersive technique, the contributions obtained when the gluon is cut, are generated by the interference of two inelastic amplitudes at tree level, while the contributions in which the gluon is not cut can be interpreted as coming from radiative corrections to the inelastic amplitude at the previous perturbative order. The systematic inclusion of these terms, to all orders in perturbation theory, will be discussed in the next section. 
\subsection{Multi-Regge form of the inelastic $2 \rightarrow 2 + n $ amplitude}
In this section, we want to establish the last ingredient that allows to construct the BFKL approach, \textit{i.e.} the general form of the inelastic $2 \rightarrow 2 + n$ amplitude in the Regge limit. This form is a generalization of the one appearing in Eq.~(\ref{Int:Eq:Inelastic2->3}).
\subsubsection{A useful gauge trick}
Let us once again consider the five diagrams that contribute to the Lipatov vertex, Fig.~\ref{Int:Fig:RRgIn5cont}. Let us imagine that we remove the lower quark line, we are left with the three diagrams corresponding to the emission of a gluon either from the upper quark lines or from the central gluon. In the sum of these three diagrams, all but the bottom gluon are on-shell and this means that we have the Ward identity
\begin{equation}
        q_2^{\nu} \mathcal{M}_{\nu}^{\sigma} (q_1, q_2) = 0  \implies \mathcal{A}_{12,(a)+(b)+(c)}^{1'2'3 (\text{w. l. l.})} \sim p_2^{\nu} \mathcal{M}_{\nu}^{\sigma} (q_1, q_2) = - \frac{1}{\alpha_{q_2}} q_{2 \perp}^{\nu} \mathcal{M}_{\nu}^{\sigma} (q_1, q_2) \; .
\end{equation}
The second implication is correct as the $p_2$-component of $\mathcal{M}$ is negligible. Now, since in the eikonal approximation $\mathcal{M}$ has no transverse component from diagrams (b) and (c) in Fig.~\ref{Int:Fig:RRgIn5cont}, the only diagram contributing is (a). Now, we can repeat the discussion removing the upper part of the diagram and obtain
\begin{equation}
        q_1^{\mu} \mathcal{N}_{\mu}^{\sigma} (q_1, q_2) = 0 \implies \mathcal{A}_{12,(a)+(e)+(d)}^{1'2'3 (\text{w. u. l.})} \sim p_1^{\mu} \mathcal{N}_{\mu}^{\sigma} (q_1,q_2) = - \frac{1}{\beta_{q_1}} q_{1 \perp}^{\mu} \mathcal{N}_{\mu}^{\sigma} (q_1,q_2) \;,
\end{equation}
where $\mathcal{N}$ is the analog of $\mathcal{M}$ for the lower line. Now, since in the eikonal approximation $\mathcal{N}$ has no transverse component from diagrams of the type (b) and (c) in Fig.~\ref{Int:Fig:RRgIn5cont}, again, the only diagram contributing is of the type (a). This observation is very important, because it means that we can organize the structure of the amplitude in (\ref{Int:Eq:Inelastic2->3}), in such a way to make only the genuine ladder-type diagram contributing. Let us be more precise, the amplitude is  
\begin{equation}
    \mathcal{A}_{12}^{1'2'3} = 2 s \Gamma_{1'1}^{a} \frac{1}{t_1} g (T^c)_{ab} \varepsilon_{\sigma}^* C^{\sigma} (q_2, q_1) \frac{1}{t_2} \Gamma_{2'2}^{b} = 2 s \left( \frac{2 p_{1}^{\mu} p_{2}^{\nu}}{s} \right) \Gamma_{1'1}^{a} \frac{1}{t_1} g (T^c)_{ab} \varepsilon_{\sigma}^* \Gamma_{\mu \nu}^{\sigma} (q_1, q_2) \frac{1}{t_2} \Gamma_{2'2}^{b} \; ,
    \label{Int:Eq:A2to3Rev}
\end{equation}
where we have defined the quantity
\begin{equation}
    \Gamma_{\mu \nu}^{\sigma} (q_1, q_2) \equiv \frac{2 p_{2, \mu} p_{1,\nu}}{s} C^{\sigma} (q_2, q_1) \; .
    \label{Eq:Int:CinTermofGamma}
\end{equation}
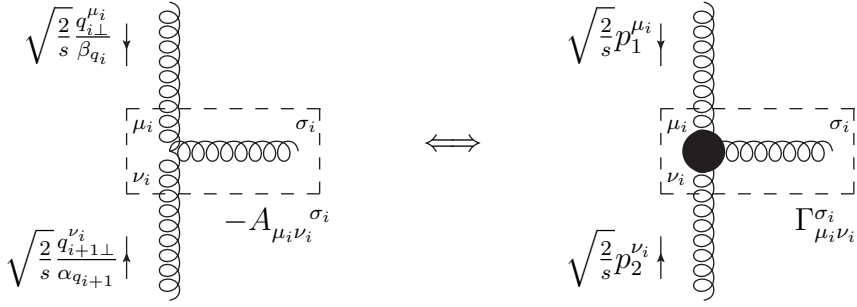
\begin{figure}
\begin{picture}(420,125)
\put(100,0){\includegraphics[scale=0.5]{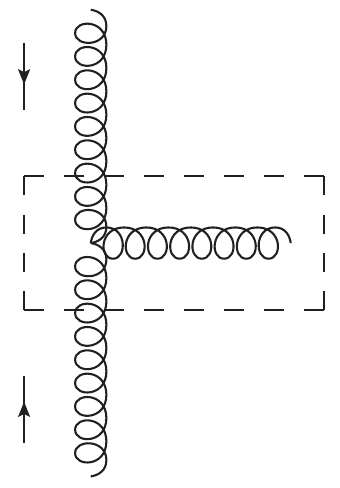}}
\put(67,99){$ \sqrt{\frac{2}{s}} \frac{q_{i \perp}^{\mu_i}}{\beta_{q_i}}$}
\put(108,69){$ _{\mu_i} $}
\put(108,49){$ _{\nu_i} $}
\put(169,69){$ _{\sigma_i} $}
\put(142,29){$ -A_{\mu_i \nu_i}^{\; \; \; \; \; \; \sigma_i} $}
\put(60,15){$ \sqrt{\frac{2}{s}} \frac{q_{i+1 \perp}^{\nu_i}}{\alpha_{q_{i+1}}}$}
\put(217,59){$ \Longleftrightarrow $}
\put(300,0){\includegraphics[scale=0.5]{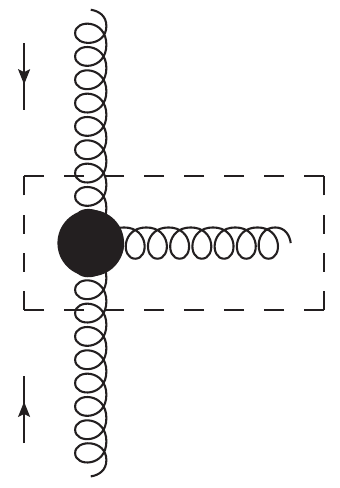}}
\put(270,99){$\sqrt{\frac{2}{s}} {\tiny p_{1}^{\mu_i}}$}
\put(308,69){$ _{\mu_i} $}
\put(308,49){$ _{\nu_i} $}
\put(369,69){$ _{\sigma_i} $}
\put(356,29){$ \Gamma_{\mu_i \nu_i}^{\sigma_i} $}
\put(270,15){$\sqrt{\frac{2}{s}} {\tiny p_{2}^{\nu_i}}$}
\end{picture}
\caption{Schematic representation of the gauge trick.}
\label{Int:Fig:GaugeTrick}
\end{figure}
In the last equality in Eq.~(\ref{Int:Eq:A2to3Rev}), we have isolated the factors $p_{1, \mu}$ and $p_{2,\nu}$ appearing in the previous Ward identities. From the previous discussion, it is clear that, the same amplitude must be obtained considering only the diagram involving the triple gluon vertex (\textit{i.e.} $-A_{\mu \nu}^{ \; \; \; \; \sigma} (q_2,q_1-q_2)$ instead of $\Gamma_{\mu \nu}^{\sigma} (q_1, q_2)$ in Eq.~(\ref{Int:Eq:A2to3Rev})), but performing at the same time the replacement
\begin{equation}
    \left( \frac{2 p_{1}^{\mu} p_{2}^{\nu} }{s} \right) \longrightarrow \left( \frac{2 q_{1 \perp}^{\mu} q_{2 \perp}^{\nu}}{\beta_{q_1} \alpha_{q_2} s} \right) \; .
\end{equation}
Hence, we must have
\begin{equation*}
    \mathcal{A}_{12}^{1'2'3} = - 2 s \left( \frac{2 q_{1 \perp}^{\mu} q_{2 \perp}^{\nu}}{\beta_{q_1} \alpha_{q_2} s} \right) \Gamma_{1'1}^{a} \frac{1}{t_1} g (T^c)_{ab} \varepsilon_{\sigma}^* A_{\mu \nu}^{ \; \; \; \; \sigma} (q_2,q_1-q_2) \frac{1}{t_2} \Gamma_{2'2}^{b} \; .
\end{equation*}
By substituting the explicit expression of $A_{\mu \nu}^{ \; \; \; \; \sigma}$ and using the Sudakov decomposition for momenta $q_1$ and $q_2$, we can verify the relation explicitly. After a bit of algebra, we find
\begin{equation*}
   \mathcal{A}_{12}^{1'2'3} = 2 s \Gamma_{1'1}^{a} \frac{1}{t_1} g (T^c)_{ab} 
\end{equation*}
\begin{equation*}
    \times \varepsilon_{\sigma}^* \left[ p_2^{\sigma} \left( \alpha_{q_2} - \frac{2 t_2}{\beta_{q_1} s} \right) + p_1^{\sigma} \left( \beta_{q_1} - \frac{2 t_1}{\alpha_{q_2} s} \right) - (q_1+q_2)_{\perp}^{\sigma} + \frac{(q_{1 \perp}^2 - q_{2 \perp}^2)}{\beta_{q_1} \alpha_{q_2} s} (q_1-q_2)^{\sigma} \right] \frac{1}{t_2} \Gamma_{2'2}^{b} \; .
\end{equation*}
We recover exactly the same expression, apart for a gauge term that evidently vanishes because
\begin{equation}
    \varepsilon_{\sigma}^* (q_1-q_2)^{\sigma} = \varepsilon_{\sigma}^* (q_1-q_2) (q_1-q_2)^{\sigma} = 0 \; .
    \label{Int:Eq:GaugeInvarianceLipatov}
\end{equation}
We have therefore established that, apart from a gauge term, the following equality
\begin{equation}
     \frac{2 p_{1}^{\mu} p_{2}^{\nu}}{s} \Gamma_{\mu \nu}^{\sigma} (q_1, q_2) =  - \frac{2 q_{1 \perp}^{\mu} q_{2 \perp}^{\nu}}{\beta_{q_1} \alpha_{q_2} s} A_{\mu \nu}^{ \; \; \; \; \sigma} (q_2,q_1-q_2) \; 
\end{equation}
holds. In the next section, we will need this relation applied to the $i$-th vertex of the scale, \textit{i.e.}
\begin{equation}
    \frac{2 p_{1}^{\mu_i} p_{2}^{\nu_i}}{s} \Gamma_{\mu_i \nu_i}^{\sigma_i} (q_1, q_2) =  - \frac{2 q_{i \perp}^{\mu_i} q_{i+1 \perp}^{\nu_i}}{\beta_{q_i} \alpha_{q_{i+1}} s} A_{\mu_i \nu_i}^{ \; \; \; \; \; \sigma_i} (q_{i+1},q_i-q_{i+1}) \; . 
   \label{Int:Eq:LadderToUnladder}
\end{equation}
A schematic representation of this trick is shown in Fig.~\ref{Int:Fig:GaugeTrick}.
\subsubsection{Proof of multiperipheral form of inelastic amplitude à la Gribov-Levin-Ryskin}
\begin{figure}
\begin{picture}(400,200)
\put(150,0){\includegraphics[scale=0.5]{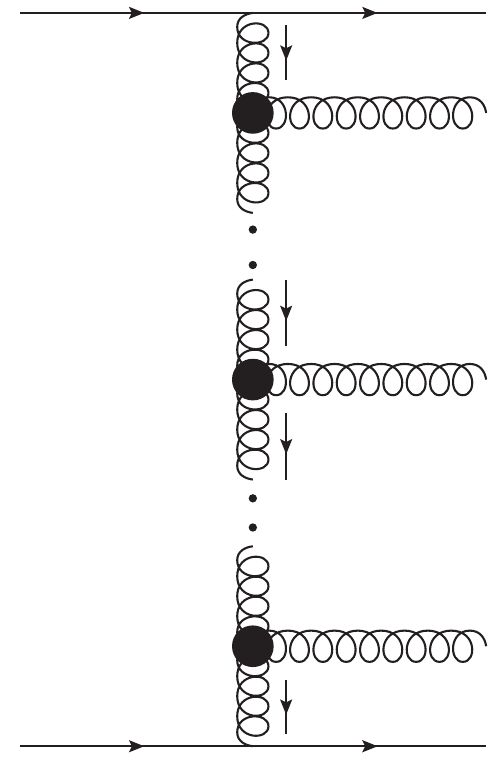}}
\put(175,190){$ p_{1} $}
\put(223,170){$ q_{1} $}
\put(223,107){$ q_{i} $}
\put(155,87){$ \gamma^{P_i}_{c_i c_{i+1}} \rightarrow $}
\put(223,74){$ q_{i+1} $}
\put(223,13){$ q_{n} $}
\put(175,16){$ p_{2} $}
\end{picture}
\caption{Schematic representation of $\mathcal{A}_{12}^{1'2'...n+2}$.}
\label{Int:Fig:Inelastic2-2n}
\end{figure}
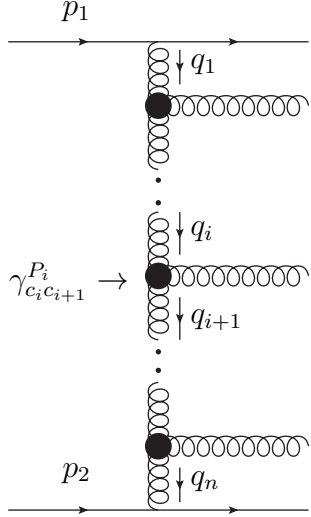
In this section, we want to prove that the inelastic $2 \rightarrow 2 + n$ amplitude, generalizing Eq.~(\ref{Int:Eq:Inelastic2->3}), assumes the following form:
\begin{equation}
    \mathcal{A}_{12}^{1'2'+ n} = 2 s \Gamma_{1'1}^{c_1} \left( \prod_i^n \gamma^{P_i}_{c_i c_{i+1}} (q_i, q_{i+1}) \frac{1}{t_i} \right) \frac{1}{t_{n+1}} \Gamma_{2'2}^{c_{n+1}} \; ,
    \label{Int:Eq:InelGLR}
\end{equation}
which is schematically represented in Fig.~\ref{Int:Fig:Inelastic2-2n}. 
As a first observation, we note that the eikonal approximation we have illustrated in the section \ref{Int:subsec:Reggeization of the gluon in pQCD} holds for particles of any spin and therefore also for gluons. In this case, it is still valid at every central vertex due to the strong ordering in rapidity required by the multi-Regge kinematics, \textit{i.e.}
\begin{equation}
1 \gg \beta_{q_i} \gg \beta_{q_{i+1}} \gg \frac{-t}{s}  \; , \hspace{0.5 cm} 1 \gg | \alpha_{q_{i+1}} | \gg | \alpha_{q_{i}}| \gg \frac{-t}{s}  \; . 
\label{Int:Eq:MultiReggeKinematicsN}
\end{equation}
To prove the form (\ref{Int:Eq:InelGLR}), we will follow the elegant derivation in \cite{Gribov:1984tu}. Consider the amplitude for two quarks to scatter into two quarks plus $n$ gluons. As described before, if we cut the $i_{th}$ vertical propagator the amplitude is split into an upper part $\mathcal{M}_{\mu} (p_1, q_1,...,q_i)$ and a lower part $\mathcal{N}_{\nu} (p_2, q_i, ..., q_n)$ (see Fig.~\ref{Int:Fig:CutAmpliGLR}). As seen in the previous sections, the following two Ward identities 
\begin{equation}
    q_{i \perp}^{\mu} \mathcal{M}_{\mu} = - \beta_{q_i} p_2^{\mu} \mathcal{M}_{\mu} \; , \hspace{1 cm}   q_{i \perp}^{\nu} \mathcal{N}_{\nu} = - \alpha_{q_i} p_1^{\nu} \mathcal{N}_{\mu} 
\end{equation}
hold. This means that we can replace the numerator of the cut gluon propagator by
\begin{equation}
    \frac{2 q_{i \perp}^{\mu} q_{i \perp}^{\nu} }{\beta_{q_i} \alpha_{q_i} s} \;.
    \label{Int:Eq:GribovRepl}
\end{equation}
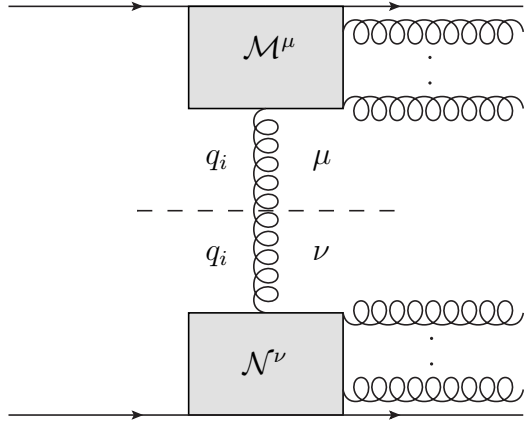
\begin{figure}
\begin{picture}(400,175)
\put(120,0){\includegraphics[scale=0.6]{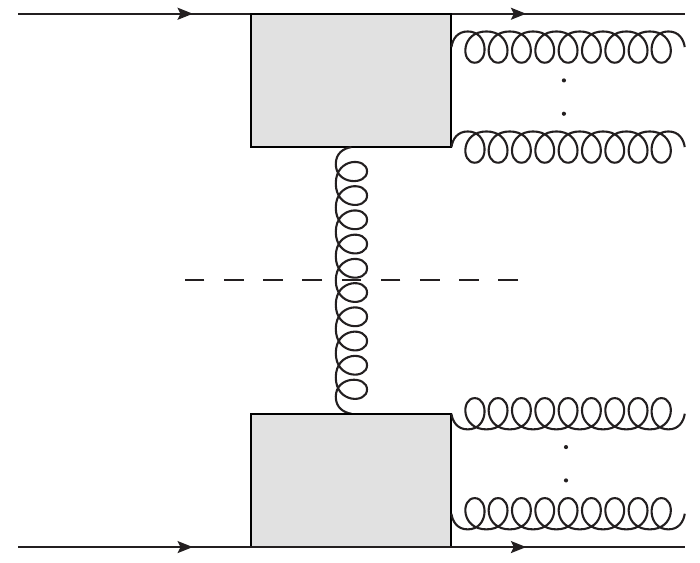}}
\put(199,99){$ q_{i} $}
\put(199,63){$ q_{i} $}
\put(239,99){$ \mu $}
\put(239,63){$ \nu $}
\put(213,139){$ \mathcal{M}^{\mu} $}
\put(213,20){$ \mathcal{N}^{\nu} $}
\end{picture}
\caption{$t$-channel cut of the amplitude $\mathcal{A}_{12}^{1'2'+n}$.}
\label{Int:Fig:CutAmpliGLR}
\end{figure}
It is evident that this procedure can be repeated for all vertical gluon lines and we can perform the same manipulations. Therefore, we end up with an amplitude that is a genuine uncrossed ladder in which the metric tensor in the propagator (approximated à la Gribov) can be written as (\ref{Int:Eq:GribovRepl}). \\
We can associate a factor $\sqrt{\frac{2}{s}} q_{i \perp}^{\mu_i} / \alpha_{q_i}$ to the top of the $i_{th}$ vertical gluon line and a factor $\sqrt{\frac{2}{s}} q_{i \perp}^{\nu_i} / \beta_{q_i}$ to the bottom of the same line. In this way, the amplitude immediately becomes
\begin{equation}
    \mathcal{A}_{12}^{1'2'+n} = 2 s \Gamma_{1'1}^{c_1} \left( \prod_i^n (-g) \varepsilon_{\sigma_i}^{*} T^{P_i}_{c_i c_{i+1}} \frac{2 q_{i \perp}^{\mu_i} q_{i+1 \perp}^{\nu_i} }{\alpha_{q_{i+1}} \beta_{q_i} s} A_{\mu_i \nu_i}^{\; \; \; \; \; \sigma_i} (q_{i+1}, q_i-q_{i+1}) \frac{1}{t_i} \right) \frac{1}{t_{n+1}} \Gamma_{2'2}^{c_{n+1}} \; .
    \label{Int:Eq:InelGLRGaugeTrick}
\end{equation}
Using Eqs.~(\ref{Eq:Int:CinTermofGamma}, \ref{Int:Eq:LadderToUnladder}), we obtain exactly the result (\ref{Int:Eq:InelGLR}). \vspace{0.2 cm} \\
\textit{Genuine gluon ladders only} \vspace{0.2 cm} \\
\begin{figure}
\begin{picture}(400,250)
\put(40,50){\includegraphics[scale=0.5]{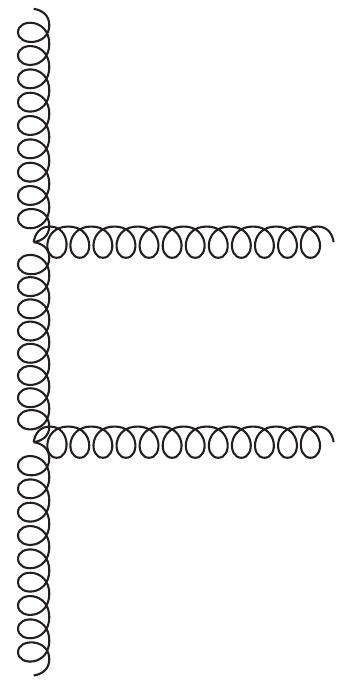}}
\put(40,30){(a)}
\put(22,192){$ q_{i-1} $}
\put(62,192){$ \mu $}
\put(126,156){$ \sigma_{i-1} $}
\put(22,132){$ q_{i} $}
\put(62,132){$ \tau $}
\put(126,106){$ \sigma_{i} $}
\put(22,82){$ q_{i+1} $}
\put(62,82){$ \nu $}
\put(180,50){\includegraphics[scale=0.5]{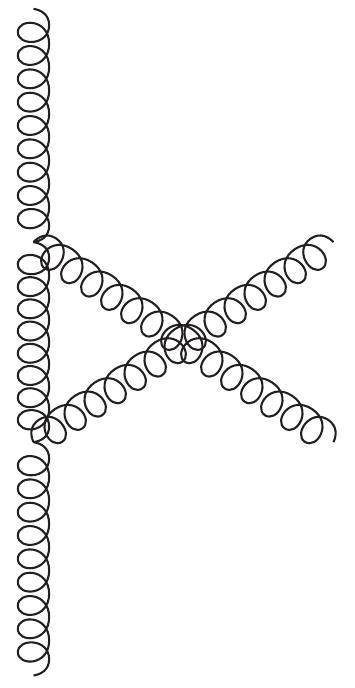}}
\put(180,30){(b)}
\put(162,192){$ q_{i-1} $}
\put(202,192){$ \mu $}
\put(264,156){$ \sigma_{i-1} $}
\put(162,132){$ \tilde{q}_{i} $}
\put(196,132){$ \tau $}
\put(264,106){$ \sigma_{i} $}
\put(162,82){$ q_{i+1} $}
\put(202,82){$ \nu $}
\put(320,150){\includegraphics[scale=0.5]{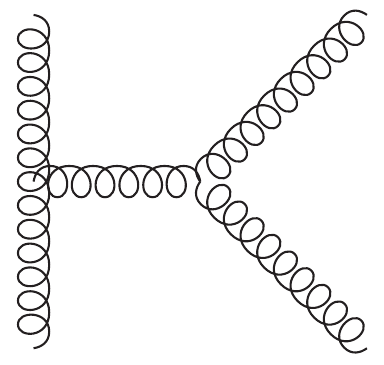}}
\put(320,135){(c)}
\put(302,212){$ q_{i-1} $}
\put(342,212){$ \mu $}
\put(352,232){$ q_{i-1} - q_i $}
\put(412,232){$ \sigma_{i-1} $}
\put(302,172){$ q_{i+1} $}
\put(342,172){$ \nu $}
\put(352,152){$ q_{i} - q_{i+1} $}
\put(412,152){$ \sigma_{i} $}
\put(320,30){\includegraphics[scale=0.5]{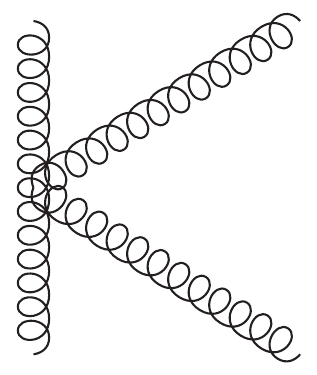}}
\put(320,10){(d)}
\put(302,102){$ q_{i-1} $}
\put(342,102){$ \mu $}
\put(394,102){$ \sigma_{i-1} $}
\put(302,42){$ q_{i+1} $}
\put(342,42){$ \nu $}
\put(394,30){$ \sigma_{i} $}
\end{picture}
\caption{The diagram (a) represents the genuine ladder contribution, while, all other contributions are the non-genuine ones.}
\label{Int:Fig:VariousLadder}
\end{figure}
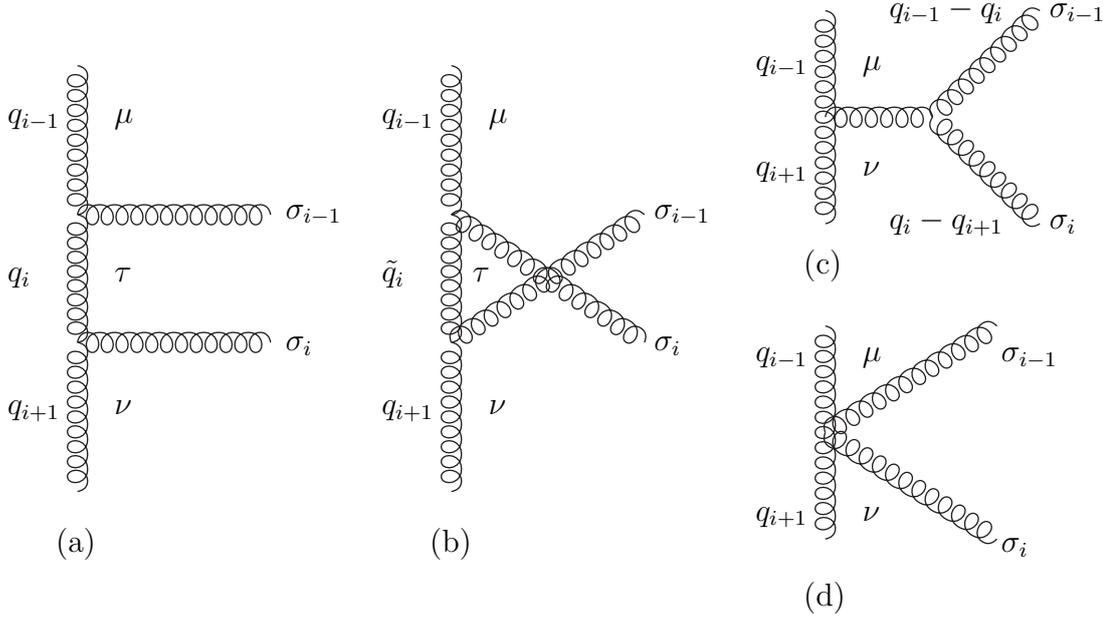
Everything we have shown works if, when the gauge trick is applied to the $t$-channel gluons, the only dominant diagrams are the genuine gluon ladder contributions (see Fig.~(\ref{Int:Fig:VariousLadder})).
Each of the diagrams contains multiple terms, we focus on a few of these to demonstrate the suppression mechanism of non-genuine contributions. 
Since they are useless for the purposes of the discussion below, we neglect color factors. \\
Let us start by considering the diagram (a), we are going to focus on terms proportional to $p_1^{\sigma_{i-1}} p_2^{\sigma_i}$ and $q_{i-1 \perp}^{\sigma_{i-1}} q_{i+1 \perp}^{\sigma_{i}}$. The upper vertex of the diagram has a contribution proportional to
\begin{equation*}
    - \frac{2 q_{i-1 \perp}^{\mu} q_{i \perp}^{\tau}}{\beta_{q_{i-1}} \alpha_{q_i} s} g_{\mu \tau} (q_{i-1}+q_{i})^{\sigma_{i-1}} \sim \frac{t}{\beta_{q_{i-1}} \alpha_{q_i} s} (\beta_{q_{i-1}} p_1^{\sigma_{i-1}}) = \beta_{q_{i-1}} p_1^{\sigma_{i-1}} \; ,
\end{equation*}
while the lower has a term of the form
\begin{equation*}
    - \frac{2 q_{i \perp}^{\tau} q_{i+1 \perp}^{\nu}}{\beta_{q_i} \alpha_{q_{i+1}} s} g_{\tau \nu} (q_{i}+q_{i+1})^{\sigma_i} \sim \frac{t}{\beta_{q_i} \alpha_{q_{i+1}} s} (\alpha_{q_{i+1}} p_2^{\sigma_{i}}) = \alpha_{q_{i+1}} p_2^{\sigma_{i}} \; .
\end{equation*}
In the last two equalities of previous expressions we have used that $\beta_{q_{i-1}} \sim \frac{t}{\alpha_{q_i} s}$\footnote{Please note that these relations come from the on-shell conditions of outgoing particles.}. The total contribution goes like
\begin{equation}
    \beta_{q_{i-1}} \alpha_{q_{i+1}} p_1^{\sigma_{i-1}} p_2^{\sigma_{i}} \; .
\label{Int:Eq:p1p2Cro}
\end{equation}
The transverse contribution can be immediately extracted and goes like
\begin{equation}
  \left( \frac{t}{\beta_{q_{i-1}} \alpha_{q_i} s} \right) \left( \frac{t}{\beta_{q_i} \alpha_{q_{i+1}} s} \right)  q_{i-1 \perp}^{\sigma_{i-1}} q_{i+1 \perp}^{\sigma_{i}} \sim q_{i-1 \perp}^{\sigma_{i-1}} q_{i+1 \perp}^{\sigma_{i}} \; .
\end{equation}
We must also note that we have a propagator of the order
\begin{equation}
    \frac{1}{q_i^2} \sim \frac{1}{t} \sim \frac{1}{\beta_{q_{i-1}} \alpha_{q_{i}} s} \; .
\end{equation}
Let us move to diagram (b). We observe that, when we work on a genuine ladder, we are always dealing with terms of type
\begin{equation*}
    - \frac{2 q_{i \perp}^{\mu_i} q_{i+1 \perp}^{\nu_i} }{ \beta_{q_i} \alpha_{q_{i+1}} s} A_{\mu_i \nu_i}^{\; \; \; \; \; \sigma_i} (q_{i+1}, q_i-q_{i+1}) 
\end{equation*}
\begin{equation}
    = \frac{2 q_{i \perp}^{\mu_i} q_{i+1 \perp}^{\nu_i} }{ \beta_{q_i} \alpha_{q_{i+1}} s} \left[ - g_{\mu_i \nu_i} (q_i + q_{i+1})^{\sigma_i} + g_{\mu_i}^{\sigma_i} (2 q_i - q_{i+1})_{\nu_i} + g_{\nu_i}^{\sigma_i} (2 q_{i+1} - q_i)_{\mu_i} \right] \; .
\label{Int:Eq:ClosestNeiLadder}
\end{equation}
Unlike this, when we work with crossed ladders, the momenta entering the upper and lower vertex are not simply the ``closest neighbors" $q_i$ and $q_{i+1}$. The generalization of formula (\ref{Int:Eq:ClosestNeiLadder}) in this latter case is
\begin{equation*}
    - \frac{2 q_{i \perp}^{\mu} q_{j \perp}^{\nu} }{ \beta_{q_i} \alpha_{q_{j}} s} A_{\mu \nu}^{\; \; \; \sigma} ( q_j, q_i-q_j )
\end{equation*}
\begin{equation}
    = \frac{2 q_{i \perp}^{\mu} q_{j \perp}^{\nu} }{ \beta_{q_i} \alpha_{q_j} s} \left[ - g_{\mu \nu} (q_i + q_j)^{\sigma} + g_{\mu}^{\sigma} (2 q_i - q_j)_{\nu} + g_{\nu}^{\sigma} (2 q_{j} - q_i)_{\mu} \right] \; .
\label{Int:Eq:NonClosestNeiLadder}
\end{equation}
By using this formula, we are able to see that the diagram (b) has a contribution proportional to
\begin{equation}
    \left( \frac{\beta_{q_i}}{\beta_{q_{i-1}}} \right)^2 \beta_{q_{i-1}} \alpha_{q_{i+1}} p_1^{\sigma_{i-1}} p_2^{\sigma_{i}}  \; .
\label{Int:Eq:p1p2UnCro}
\end{equation}
We immediately realize that Eq.~(\ref{Int:Eq:p1p2UnCro}) has a suppressing\footnote{Remind the strong order in the Sudakov variables in Eq.~(\ref{Int:Eq:MultiReggeKinematicsN}).} factor $ (\beta_{q_i} / \beta_{q_{i-1}} )^2 $ with respect to the corresponding term in Eq.~(\ref{Int:Eq:p1p2Cro}). Furthermore, also the propagator is strongly suppressed, in fact
\begin{equation}
    \frac{1}{\tilde{q}_i^{\; 2}} = \frac{1}{(q_{i-1} + q_{i+1} - q_{i})^2} \sim \frac{1}{\beta_{q_{i-1}} \alpha_{q_{i+1}} s} \ll \frac{1}{\beta_{q_{i-1}} \alpha_{q_{i}} s} \; .
\end{equation}
We now pass to diagram (c), it is obtained by the contraction of two triple gluon vertices and the usual ``effective" polarization of $t$-channel gluons. We can easily find that it has a contribution proportional to
\begin{equation}
    \frac{\beta_{q_i}}{\beta_{q_{i-1}}} q_{i-1 \perp}^{\sigma_{i-1}} q_{i+1 \perp}^{\sigma_{i}} \; ,
\end{equation}
which again has a suppression factor and, moreover, the gluon propagator in this case gives
\begin{equation}
    \frac{1}{(q_{i-1}-q_{i+1})^2} = \frac{1}{\alpha_{q_{i+1}} \beta_{q_{i-1}} s} \ll \frac{1}{ \beta_{q_i} \alpha_{q_{i+1}} s} \; .
\end{equation}
The last diagram is (d); since in this case we do not have a propagator, we can multiply and divide by a factor $t$ in such a way as to make the factor related to the denominator equal to that of (a). In this way, we obtain a term of the form
\begin{equation}
    \frac{t}{ \beta_{q_{i-1}} \alpha_{q_{i+1}} s } q_{i-1 \perp}^{\mu} q_{i-1 \perp}^{\nu} \sim  \frac{ \beta_{q_i} }{ \beta_{q_{i-1}} } q_{i-1 \perp}^{\mu} q_{i-1 \perp}^{\nu}  \; ,
\end{equation}
which has the usual suppression factor. \vspace{0.1 cm} \\ 
\textit{Absence of fermions} \\
Another important point is the absence of diagrams containing fermions (see Fig.~\ref{Int:Fig:VariousLadderFermion}). 
The reason for this absence is to be found in the fact that a particle of spin 1/2 is being exchanged in the $t$-channel. A particle of this spin generates a less strong $s$-dependence than that of a spin one boson such as the gluon. \\ 
We can understand this suppression as before. In the case where fermions cross, (b) in Fig.~\ref{Int:Fig:VariousLadderFermion}, or are produced by the splitting of a single outgoing gluon, (c) in Fig.~\ref{Int:Fig:VariousLadderFermion}, we certainly have a suppression factor coming from a propagator. For the diagram (a) in Fig.~\ref{Int:Fig:VariousLadderFermion} a simple computation shows that we have again a suppression factor of $\beta_{q_i} / \beta_{q_{i-1}}$. 
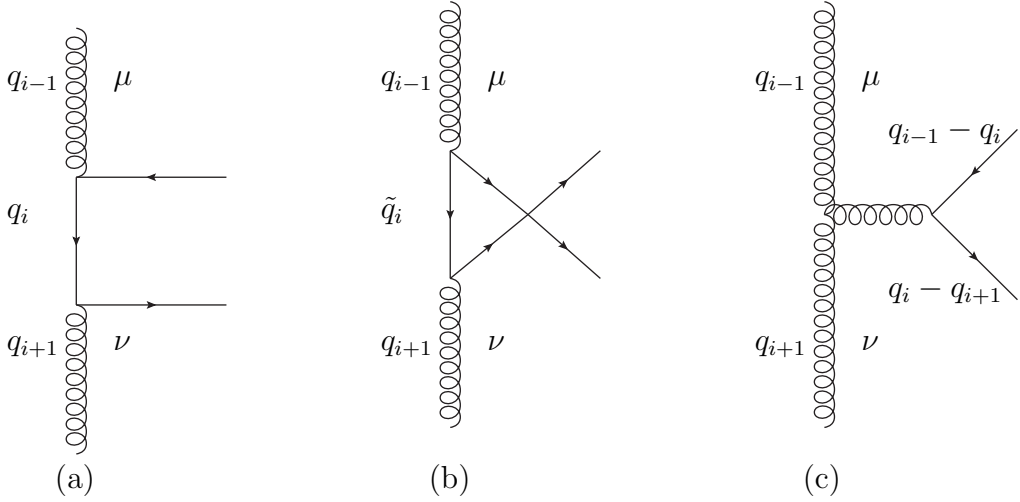
\begin{figure}
\begin{picture}(400,220)
\put(40,30){\includegraphics[scale=0.5]{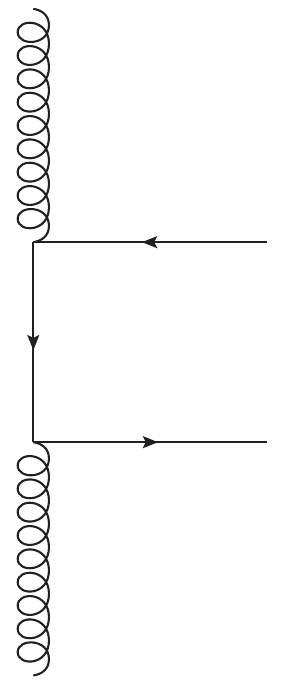}}
\put(40,20){(a)}
\put(22,172){$ q_{i-1} $}
\put(62,172){$ \mu $}
\put(22,122){$ q_{i} $}
\put(22,72){$ q_{i+1} $}
\put(62,72){$ \nu $}
\put(180,40){\includegraphics[scale=0.5]{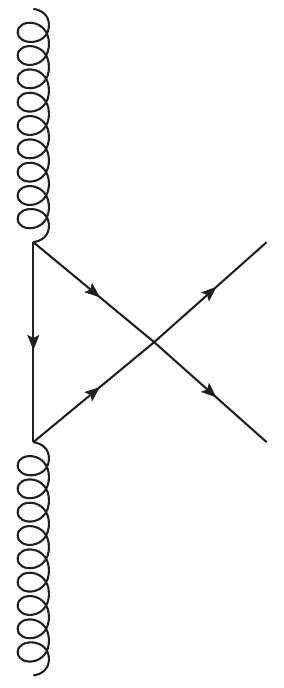}}
\put(180,20){(b)}
\put(162,172){$ q_{i-1} $}
\put(202,172){$ \mu $}
\put(162,122){$ \tilde{q}_{i} $}
\put(162,72){$ q_{i+1} $}
\put(202,72){$ \nu $}
\put(320,40){\includegraphics[scale=0.5]{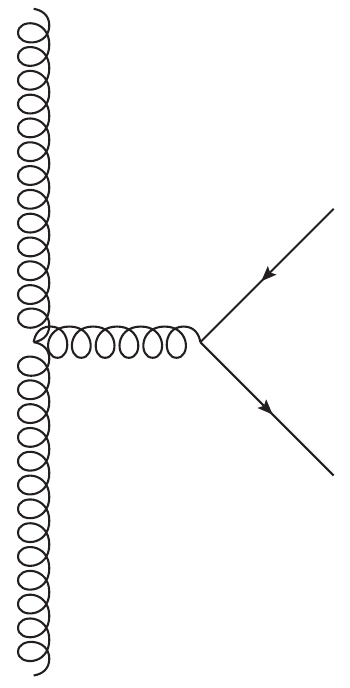}}
\put(320,20){(c)}
\put(302,172){$ q_{i-1} $}
\put(342,172){$ \mu $}
\put(352,152){$ q_{i-1} - q_i $}
\put(302,72){$ q_{i+1} $}
\put(342,72){$ \nu $}
\put(352,92){$ q_{i} - q_{i+1} $}
\end{picture}
\caption{Diagrams involving fermions.}
\label{Int:Fig:VariousLadderFermion}
\end{figure}
\subsubsection{Reggeization in all sub-channels}
In the previous subsection, we constructed the tree-level amplitude for two quarks to scatter to two quarks and $n$ gluons in the multi-Regge kinematics. This amplitude, multiplied by its complex conjugate and integrated over the phase space of $n+2$ particles will contribute to the imaginary part of the amplitude of the Reggeized gluon. Obviously, as seen explicitly in the section \ref{Int:Sec:LipatovVertex}, this is not sufficient to reconstruct all the dominant part of the amplitude in the high-energy limit. In fact, in the two-loop calculation it is necessary to insert some corrections in which the cut only goes through the two quark lines. It is important to find a systematic method of including these loop corrections in our generalized form for the amplitude in Eq.~(\ref{Int:Eq:InelGLR}). \\
\begin{figure}
\begin{picture}(400,210)
\put(105,110){\includegraphics[scale=0.5]{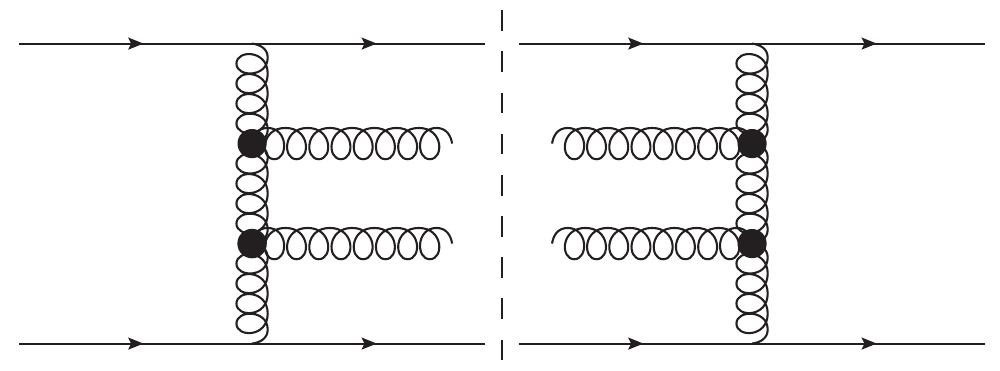}}
\put(105,0){\includegraphics[scale=0.5]{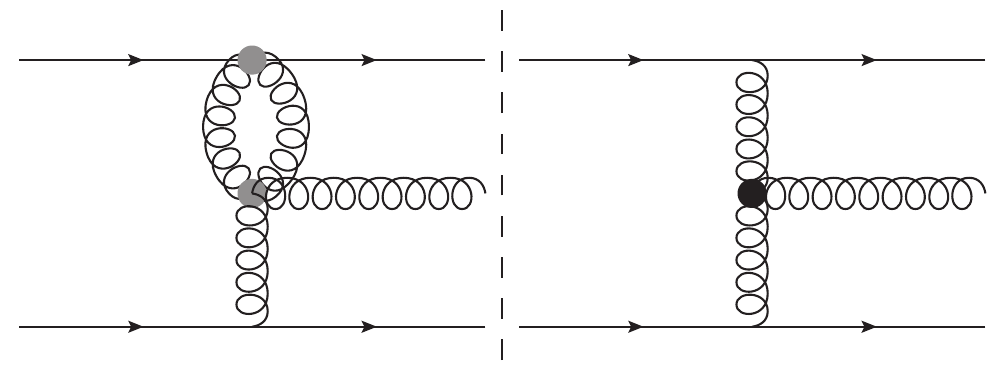}} 
\put(360,40){$+$}
\put(375,40){...}
\end{picture}
\caption{In the upper diagram the contribution to the order $g^8$ which is obtained from the square of the amplitude $\mathcal{A}_{12}^{1'2'+2}$. In the bottom diagram, the contribution to the order $g^8$ in which one gluon is not cut. This contribution can be included by reggeizing the corresponding propagator in the amplitude $\mathcal{A}_{12}^{1'2'3}$.}
\label{Int:Fig:ReshuffleContribution}
\end{figure}
Let us start with a very simple observation. If we focus only on the real part of the amplitude $\mathcal{A}_{A B}^{A'B'}$ (which is dominant in the high-energy approximation), the Reggeization ansatz in Eq.~(\ref{Int:Eq:ReggeizedAmp}) becomes
\begin{equation}
    \Re \mathcal{A}_{AB}^{A'B'} = \Gamma_{A'A}^{i} \frac{2 s}{t} \left( \frac{s}{-t} \right)^{\omega(t)} \Gamma_{B'B}^{i} \; .
\end{equation}
Basically, it can be obtained by the following replacement in the $t$-channel gluon propagator
\begin{equation}
    \frac{-i g_{\mu \nu} \delta^{ab}}{t} \rightarrow \frac{-i g_{\mu \nu} \delta^{ab}}{t} \left( \frac{s}{-t} \right)^{\omega(t)} \;.
\end{equation}
Now, the crucial point of the whole Reggeization is that, including these loop corrections is equivalent to Reggeize all the $t$-channel gluons in the inelastic amplitude $\mathcal{A}_{12}^{1'2'+n}$, \textit{i.e.} performing the replacement
\begin{equation}
    \frac{1}{t_i} \rightarrow \frac{1}{t_i} \left( \frac{s_i}{s_R} \right)^{\omega (t_i)} 
\end{equation}
in Eq.~(\ref{Int:Eq:InelGLR}), where $s_i$ is the center-of-mass energy coming into the $i_{{\rm{th}}}$ section of the gluon ladder and we have introduced a scale $s_R$ which is of the order of $-t$. The scale $s_R$ is irrelevant in LLA as its variation produces sub-leading effects. \\ 
Therefore, the final \textbf{multi}-\textbf{Regge exchange} amplitude for $2 \rightarrow 2+n$ is\footnote{It is important to observe that this simple factorized form is strictly valid only for the real part of the amplitudes. However, as we will see, only the real part is important within LLA and NLLA accuracy. The real part will be implied in subsequent formulas in which we remove the symbol $\Re$.} 
\begin{equation}
    \Re \mathcal{A}_{12}^{1'2'+n} = 2s \Gamma_{1' 1}^{c_1} \left( \prod_{i=1}^{n} \gamma_{c_i c_{i+1}}^{P_i}(q_i,q_{i+1}) \left( \frac{s_i}{s_R} \right)^{\omega(t_i)} \frac{1}{t_i} \right) \frac{1}{t_{n+1}} \left( \frac{s_{n+1}}{s_R} \right)^{\omega(t_{n+1})} \Gamma_{2' 2}^{c_{n+1}} \; .
    \label{Int:Eq:FinMulInelastic}
\end{equation}
A proof can be achieved, in the LLA, using Regge theory techniques, exploiting unitarity in all possible final state sub-channels.
We will not delve into the intricacies of this proof, but we want just to give some intuitive arguments. Previously, we proved the Regge form of the scattering amplitude for two quarks to two quarks up to the order $g^6$ (2-loop). Suppose we want to consider the order $g^8$ (3-loop), for sure we will have to consider the inelastic amplitude
\begin{equation}
     \mathcal{A}_{12}^{1' 2' 3 4} = 2s \Gamma_{1' 1}^{c_1} \frac{1}{t_1} \left( \frac{s_1}{s_R} \right)^{\omega(t_1)} \gamma_{c_1 c_2}^{P_1}(q_1,q_2) \frac{1}{t_2} \left( \frac{s_2}{s_R} \right)^{\omega(t_2)} \gamma_{c_2 c_3}^{P_2}(q_2,q_3) \frac{1}{t_{3}} \left( \frac{s_{3}}{s_R} \right)^{\omega(t_{3})} \Gamma_{2' 2}^{c_{3}} \; .
     \label{Int:Eq:Sub-Regge2->4}
\end{equation}
An immediate observation is that, if we perturbatively expand the Regge factors up to the non-trivial first order, we find a term of order $g^4$ plus three terms of order $g^6$. When we square and integrate over the phase space this $2 \rightarrow 4$ amplitude to reconstruct the contribution to the imaginary part of the $2 \rightarrow 2$ amplitude, only the term of order $g^4$ in Eq.~(\ref{Int:Eq:Sub-Regge2->4}) will produce a term contributing within $g^8$ (see top diagram in Fig.~\ref{Int:Fig:ReshuffleContribution}). Hence, the Reggeization of this contribution will only impact the next perturbative order. Now, let us take a step back and look at the form of the inelastic $2 \rightarrow 3$ amplitude, we have
\begin{equation}
    \mathcal{A}_{12}^{1' 2' 3} = 2s \Gamma_{1' 1}^{c_1} \frac{1}{t_1} \left( \frac{s_1}{s_R} \right)^{\omega(t_1)} \gamma_{c_1 c_2}^{P_1}(q_1,q_2) \frac{1}{t_2} \left( \frac{s_2}{s_R} \right)^{\omega(t_2)} \Gamma_{2' 2}^{c_{2}} \; .
    \label{Int:Eq:Sub-Regge2->3}
\end{equation}
\begin{figure}
\begin{picture}(400,120)
\put(160,10){\includegraphics[scale=0.5]{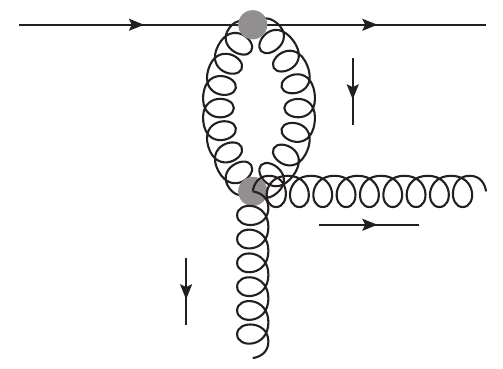}}
\put(290,70){\bigg \}}
\put(300,70){$s_1 = (p_1 - q_2)^2$}
\put(190,105){$p_1$}
\put(235,105){$p_1-q_1$}
\put(236,33){$q_1-q_2$}
\put(250,75){$q_1$}
\put(190,27){$q_2$}
\end{picture}
\caption{Upper section of the diagram in which one gluon is not cut.}
\label{Int:Fig:KinematicsInLadderSection}
\end{figure}
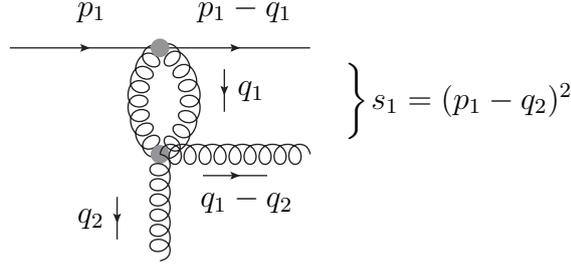
Obviously, as before, this form contains a factor $g^3$ as well as two factors $g^5$; this means that, when we square we produce four interference terms which are of the order $g^8$ and will impact in our 3-loop computation. A schematic representation of some of these terms is presented in the bottom diagram of Fig.~\ref{Int:Fig:ReshuffleContribution}. These terms reproduce exactly the corrections of order $g^8$ which, using this dispersive technique, would have been obtained from diagrams in which a gluon is not cut. There is therefore a reshuffle, which allows us to automatically generate the contributions in which there are loops that are not cut when the $2 \rightarrow 2$ amplitude is dispersively reconstructed using Cutkosky's rules. \\
Let us look more closely at the contribution in the bottom of Fig.~\ref{Int:Fig:ReshuffleContribution} and study the section of the diagram shown in Fig.~\ref{Int:Fig:KinematicsInLadderSection}. In this section of the diagram, the momentum exchanged is $q_1$ and the energy available in the center of mass is
\begin{equation}
    s_1 = (p_1-q_2)^{2} \simeq - \alpha_{q_2} s \; .
\end{equation}
This energy is much lower than $s$, but is much greater than the typical transverse scale $s_R$ that we introduced, in fact
\begin{equation}
    s_1 \simeq - \alpha_{q_2} s = \frac{1}{\beta_{q_1}} \left( -\beta_{q_1} \alpha_{q_2} s \right) \sim \frac{1}{\beta_{q_1}} (-t) \sim \frac{1}{\beta_{q_1}} s_R \; .
\end{equation}
We therefore have a scale hierarchy which, in the spirit of Reggeization, will produce logarithmic corrections of the ratio $s_1/s_R$. To all orders in perturbation theory, the effect will be to Reggeize the propagator with momenta $q_1$, \textit{i.e.}
\begin{equation}
    \frac{1}{t_1} \rightarrow \frac{1}{t_1} \left( \frac{s}{s_R} \right)^{\omega (t_1)} \; .
\end{equation}
Obviously, the same reasoning extended to all subsections of the ladder, leads us precisely to the form in Eq.~(\ref{Int:Eq:FinMulInelastic}). The final form of the inelastic amplitude, schematized in Fig.~(\ref{Int:Fig:Inelastic2-2nRegge}), it is therefore a ``half ladder'', whose vertical gluons are themselves Reggeized gluons (and therefore they are themselves an infinite superposition of diagrams).
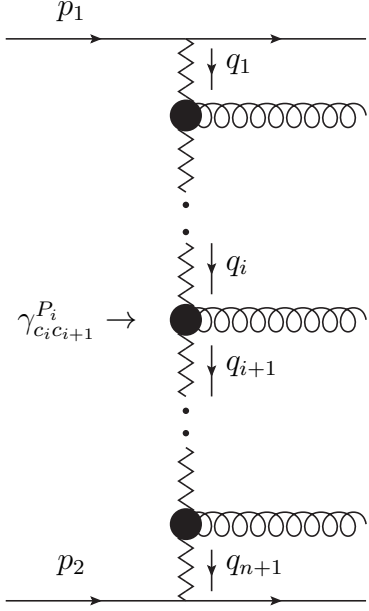
\begin{figure}
\begin{picture}(400,230)
\put(150,0){\includegraphics[scale=0.6]{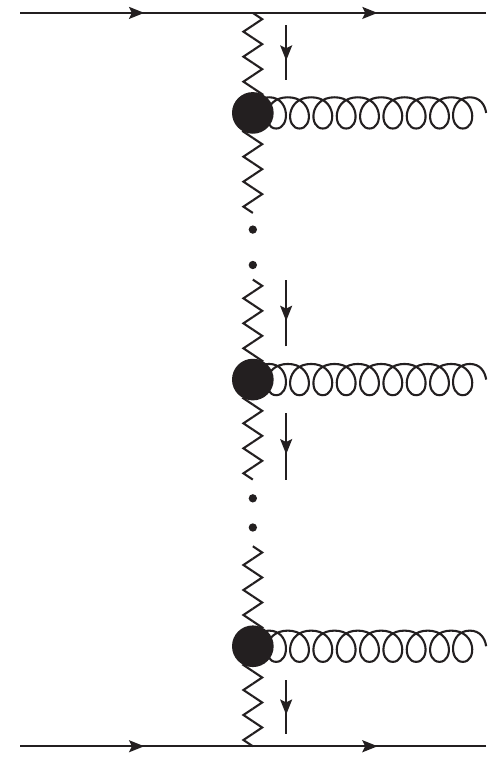}}
\put(175,225){$ p_{1} $}
\put(238,205){$ q_{1} $}
\put(238,129){$ q_{i} $}
\put(160,107){$ \gamma^{P_i}_{c_i c_{i+1}} \rightarrow $}
\put(238,91){$ q_{i+1} $}
\put(238,16){$ q_{n+1} $}
\put(175,16){$ p_{2} $}
\end{picture}
\caption{Schematic representation of $\mathcal{A}_{12}^{1'2'+n}$ after we include the Reggeization in all sub-channels.}
\label{Int:Fig:Inelastic2-2nRegge}
\end{figure}
\subsection{BFKL equation}
\begin{figure}
\begin{picture}(400,240)
\put(140,0){\includegraphics[width=0.35\textwidth]{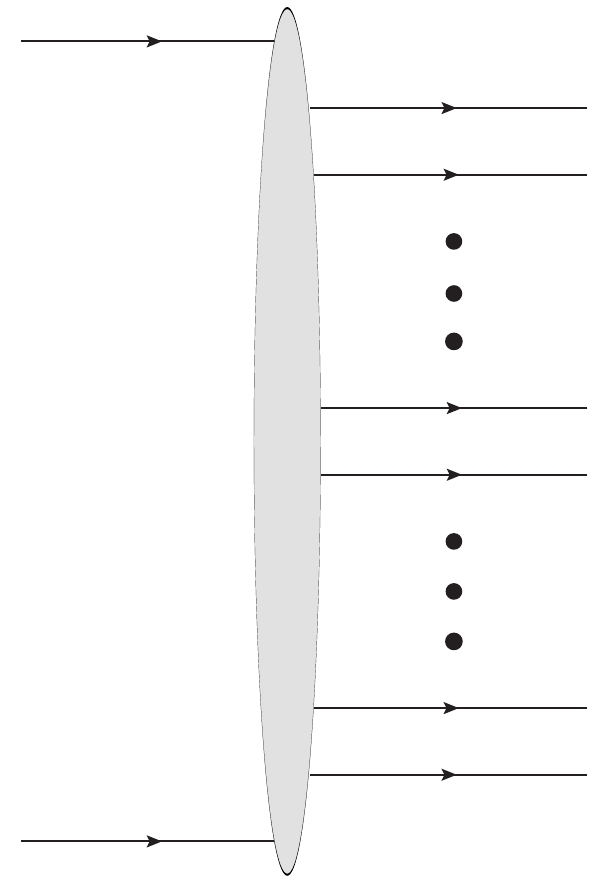}}
\put(115,10){$p_2$}
\put(115,220){$p_1$}
\put(312,205){$k_0$}
\put(312,185){$k_1$}
\put(312,125){$k_{i-1}$}
\put(312,105){$k_i$}
\put(342,115){$\bigg \} \; \; s_i = (k_{i-1} + k_i)^2$}
\put(312,45){$k_{n}$}
\put(312,25){$k_{n+1}$}
\end{picture}
\caption{Production of $n+2$ particles in the multi-Regge kinematics}
\label{Int:Fig:MRKn+2}
\end{figure}
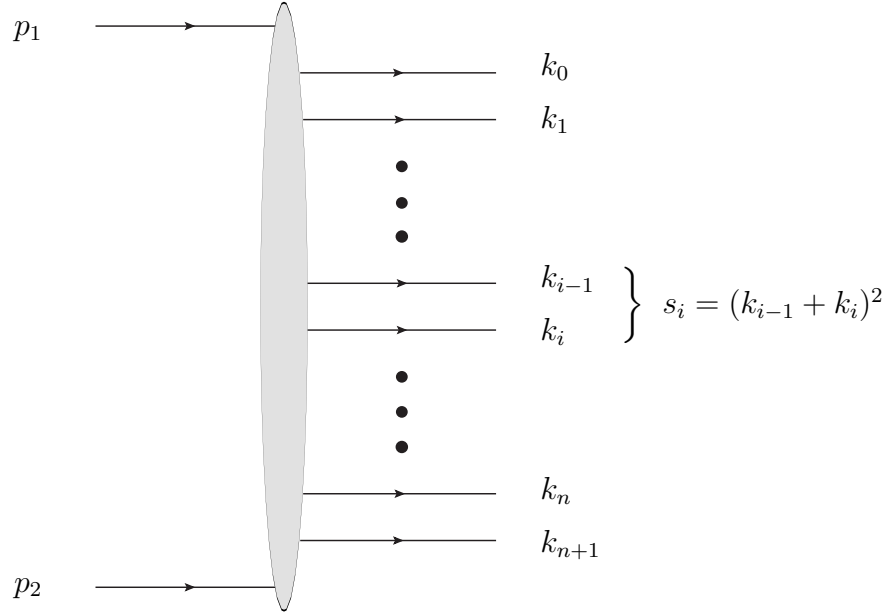

Let us now consider the BFKL equation in the LLA. Instead of directly considering the Pomeron channel, \textit{i.e.} imaginary part of the forward-scattering amplitude ($t = 0$ and vacuum quantum numbers in the $t$-channel), we will keep more general. Indeed, it should be noted that the BFKL approach gives the description of QCD-scattering amplitudes in the region of large $s$, for any fixed
momentum transfer $t$ (\textit{i.e.} not growing with $s$) and for various color state exchanges in the $t$-channel. \\
The imaginary part of the scattering amplitude $\mathcal{A}_{12}^{1'2'}$ (of the process $1+2 \longrightarrow 1'+2'$) in the $s$-channel can be written, using the Cutkosky rules (\ref{Int:Eq:Cutkosky}), as
\begin{equation}
\label{Unitary}
\Im_s \mathcal{A}_{12}^{1'2'} = \frac{1}{2} \sum_{n=0}^{\infty} \sum_{\{f\}} \int \mathcal{A}_{12}^{\tilde{1} \tilde{2} + n} \left(\mathcal{A}_{1'2'}^{\tilde{1} \tilde{2}+n}\right)^* d\Phi_{n+2} \; ,
\end{equation}
where $\mathcal{A}_{AB}^{\tilde{A} \tilde{B}+n}$ is the amplitude introduced in Eq.~(\ref{Int:Eq:FinMulInelastic}) for the production of $n+2$ particles having momenta $k_i$, with $i=0,1,...,n,n+1$ (we set $p_{\tilde{1}}=k_0$ and $p_{\tilde{2}}=k_{n+1}$), $d\Phi_{n+2}$ is the element of the intermediate particle phase of space, $\sum_{{f}}$ means sum over all the discrete quantum numbers of the intermediate particles (see Figure~\ref{Int:Fig:MRKn+2}).
\subsubsection{Kinematics} 
As before, we introduce the light-cone vectors $p_1$ and $p_2$ ($p_1^2=p_2^2=0$), so that the center-of-mass energy is $s=2 p_1 \cdot p_2$. Using the Sudakov decomposition, the momenta $k_i$ can be parametrized as
\begin{equation}
k_i= \beta_i p_1+\alpha_i p_2+ k_{i\perp}, \hspace{2cm} s\alpha_i \beta_i = k_i^2-k_{i\perp}^2 = k_i^2 + \vec{k}_i^{\; 2}.
\end{equation}
To extract the dominant contribution ($\sim s$) in Eq.~(\ref{Unitary}) we should rely on the MRK. By definition, in this kinematics, transverse momenta of produced particles are limited and their Sudakov variables $\alpha_i$ and $\beta_i$ are strongly ordered in the rapidity space, having so\footnote{Differently from the previous section, we give the definition of multi-Regge kinematics in terms of the longitudinal Sudakov variables of the outgoing particles. The two sets of conditions are fully equivalent.}
\begin{equation}
\label{Int:Eq:StrongOrderingN+2}
\alpha_{n+1} \gg \alpha_{n} \gg \alpha_{n-1} ... \gg \alpha_0  \; , \hspace{1cm}  \beta_0 \gg \beta_1 \gg \beta_2 ... \gg \beta_{n+1} \; .
\end{equation}
In this case, we have that
\begin{equation}
    \alpha_{n+1} \simeq 1 , \;  \beta_0 \simeq 1 \; , \hspace{1 cm} {\rm{and}}  \hspace{1 cm} \alpha_{0} \simeq \frac{\vec{k}_0^{\; 2}}{s}, \; \beta_{n+1} \simeq \frac{\vec{k}_{n+1}^{\; 2}}{s} \; .
\end{equation}
In the MRK the squared invariant masses $s_{i,j}=(k_i+k_j)^2$ of any pair of produced particles $i$ and $j$ are large (see Fig.~\ref{Int:Fig:MRKn+2}). In particular we have
\begin{equation}
    s_{i-1, i} \equiv s_i = (k_{i-1} + k_{i})^2 \simeq s \beta_{i-1} \alpha_i = \frac{\beta_{i-1}}{\beta_{i}} (k_i^2 + \vec{k}_i^{\; 2}) \; .
    \label{Int:Eq:SquaredInvariantMasses}
\end{equation}
The momenta $q_i$ of the $t$-channel Reggeons can be expressed as follows:
\begin{equation}
\begin{split}
q_i & = p_1 - \sum_{j=0}^{i-1} k_j = \left( 1- \sum_{j=0}^{i-1} \beta_j \right) p_1 - \left( \sum_{j=0}^{i-1} \alpha_j \right) p_2 - \sum_{j=0}^{i-1} k_{j \perp} \\ &
    = \left( \sum_{j=i}^{n+1} \beta_j \right) p_1 - \left( \sum_{j=0}^{i-1} \alpha_j \right) p_2 - \sum_{j=0}^{i-1} k_{j \perp} \simeq \beta_i p_1 - \alpha_{i-1} p_2 - \sum_{j=0}^{i-1} k_{j \perp} \; ,
\end{split}
\end{equation}
where we first used the conservation of momenta along $p_1$-direction,
\begin{equation}
    1 = \sum_{j=0}^{n+1} \beta_j
\end{equation}
and then conditions of MRK in Eq.~(\ref{Int:Eq:StrongOrderingN+2}) to simplify the expression. The $i_{{\rm{th}}}$ momentum squared is $t_i=q_i^2 \simeq q_{i \perp}^2 = - \vec{q}_i^{\; 2}$ and is predominantly transverse. \\
We now recall the form of the inelastic amplitude $2 \rightarrow 2+n$, introduced in the previous section (see Fig.~\ref{Int:Fig:Inelastic2-2nRegge}):
\begin{equation}
\label{Ampl.Ver}
\mathcal{A}_{1 2}^{\tilde{1} \tilde{2} +n} = 2s \Gamma_{\tilde{1} 1}^{c_1} \left( \prod_{i=1}^{n} \gamma_{c_i c_{i+1}}^{P_i}(q_i,q_{i+1}) \left( \frac{s_i}{s_R} \right)^{\omega(t_i)} \frac{1}{t_i} \right) \frac{1}{t_{n+1}} \left( \frac{s_{n+1}}{s_R} \right)^{\omega(t_{n+1})} \Gamma_{\tilde{2} 2}^{c_{n+1}} ,
\end{equation}
where
\begin{equation}
\gamma_{c_i c_{i+1}}^{G_i} \left(q_i,q_{i+1}\right) = g T_{c_i c_{i+1}}^{d_i} \varepsilon_{\mu}^*(k_i)C^{\mu}(q_{i+1},q_i), 
\end{equation}
and the Lorentz structure ($C^{\mu}$) is
\begin{equation}
\label{Cmu}
C^{\mu}(q_{i+1},q_{i})=-q_i^{\mu}-q_{i+1}^{\mu}+p_1^{\mu} \left( \frac{q_i^2}{k_i \cdot p_1}+2 \frac{k_i \cdot p_2}{p_1 \cdot p_2} \right) -p_2^{\mu} \left( \frac{q_{i+1}^2}{k_i \cdot p_2}+2\frac{k_i \cdot p_1}{p_1 \cdot p_2} \right).
\end{equation} 
The current conservation property $(k_i)_{\mu} C^{\mu}$ (see Eq.~(\ref{Int:Eq:GaugeInvarianceLipatovVertex})) permits us to choose an arbitrary gauge for each of the produced gluons. It is easy to observe that the other amplitude appearing in the Eq.~(\ref{Unitary}) ($\mathcal{A}_{1'2'}^{\tilde{1} \tilde{2}+n}$) can be obtained from Eq.~(\ref{Ampl.Ver}) by the substitutions
\begin{equation*}
 1 \longrightarrow 1' \; , \hspace{0.5 cm} 2 \longrightarrow 2' \; , \hspace{0.5 cm} q_i \longrightarrow q'_i \equiv q_i-q , 
\end{equation*}
with
\begin{equation}
     q = p_1-p_{1'} \simeq q_{\perp}.
\end{equation}
\subsubsection{Color decomposition} 
Of all the possible color states exchanged in the $t$-channel, we would like to extract the two cases of greatest physical relevance, that of singlet (Pomeron channel) and that of anti-symmetrical octet (Reggeized gluon channel). For this purpose, we have to decompose the product of two adjoint $SU(3)$ representations (the two gluons in $t$-channel) into sum of irreducible representations, \textit{i.e.}
\begin{equation}
    8 \otimes 8 = 1^{+} \oplus 8_A^{-} \oplus 8_S^{+} \oplus (10 \oplus \overline{10})^{-} \oplus 27^{+} \; ,
    \label{Int:Eq:DecOntoIrre}
\end{equation}
where the superscript denotes the signature of the representation and the symmetric and anti-symmetric parts of the color octet have been decoupled. In practice, this corresponds to decomposing the product of two matrices in the adjoint representation of $SU(3)$ in the following way:    
\begin{equation}
\label{Int:Eq:ColorDecAdj}
T_{c_i c_{i+1}}^{d_i} \left(T_{c'_i c'_{i+1}}^{d_i}\right)^* = \sum_R c_R \braket{c_i c'_i| \hat{\mathcal{P}}_R|c_{i+1} c'_{i+1}},
\end{equation}
where $\hat{\mathcal{P}}_R$ are the projection operators of two-gluon colour states in the $t$-channel in the unitary condition (\ref{Unitary}) on the irreducible representation $R$ of the colour group. For the singlet representation, we have
\begin{equation}
    c_0 = N \; , \hspace{1.5 cm} \braket{c_i c_i'|\hat{\mathcal{P}}_{0}|c_{i+1} c'_{i+1}} = \frac{\delta^{c_i c_i'} \delta^{c_{i+1} c_{i+1}'}}{N^2-1} \; ,
\end{equation}
while for the color octet
\begin{equation}
    c_8 = \frac{N}{2} \; , \hspace{1.5 cm} \braket{c_i c_i'|\hat{\mathcal{P}}_{8}|c_{i+1} c'_{i+1}} = \frac{f_{a c_i c_i'} f_{a c_{i+1} c_{i+1}'}}{N} \; .
    \label{Int:Eq:DecColoctet}
\end{equation}
It is easy to check that these are the correct projectors; for example, in the case of singlet we have
\begin{equation}
   c_0 \braket{c_i c_i'|\hat{\mathcal{P}}_{0}|c_{i+1} c'_{i+1}} = N \frac{\delta^{c_i c_i'} \delta^{c_{i+1} c_{i+1}'}}{N^2-1} =  \begin{cases}
   2 C_A^{\;2} C_F & \hspace{1 cm} {\rm{if}} \hspace{0.1 cm} c_i = c_i' \hspace{0.1 cm} {\rm{and}} \hspace{0.1 cm}  c_{i+1} = c'_{i+1} \\
   0 & \hspace{1 cm} {\rm{otherwise}} 
   \end{cases} \; ,
\end{equation}
where in the upper term the summation is taken over $c_i$ and $c_{i+1}$.
On the other hand, if we are exchanging a singlet ($c_i = c_i'$ and $c_{i+1} = c'_{i+1}$), on the right side of the equation (\ref{Int:Eq:ColorDecAdj}) we have 
\begin{equation}
    T_{c_i c_{i+1}}^{d_i} \left(T_{c_i c_{i+1}}^{d_i}\right)^* = f_{d_i c_{i} c_{i+1}} f_{d_i c_{i} c_{i+1}} =  2 C_A^{\;2} C_F \; .
\end{equation}
The octet case can be verified in a similar way by remembering that in Eq.~(\ref{Int:Eq:DecColoctet}) we are referring to the anti-symmetric part and that therefore this component must be extracted. To do this, we can apply the substitution (\ref{Int:Eq:ProjectionNegSign}) to the $SU(3)$ matrices in the adjoint representation. 
The introduced decomposition brings us to
\begin{equation}
\sum_{G_i} \gamma_{c_i c_{i+1}}^{G_i} \left(q_i,q_{i+1}\right) \left(\gamma_{c'_i c'_{i+1}}^{G_i} \left(q_i,q_{i+1}\right) \right)^* = \sum_R \braket{c_i c'_i| \hat{\mathcal{P}}_R |c_{i+1} c'_{i+1}} 2(2\pi)^{D-1} \mathcal{K}_r^{(R)} \left( \vec{q}_i, \vec{q}_{i+1}; \vec{q} \right) \; ,
\end{equation}
where the sum is taken over color and polarization states of the produced gluon and
\begin{equation}
\label{Kernel1}
\begin{split}
\mathcal{K}_{r}^{(R)} \left(\vec{q}_i,\vec{q}_{i+1}; \vec{q} \right) = & -\frac{g^2 c_R}{2(2\pi)^{D-1}}C^{\mu}(q_{i+1},q_i)C_{\mu}(q_{i+1}-q,q_i-q) \\ & = \frac{g^2c_R}{(2\pi)^{D-1}} \left( \frac{\vec{q}_i^{\; 2}(\vec{q}_{i+1}-\vec{q})^2 + \vec{q}_{i+1}^{\; 2}(\vec{q}_i-\vec{q})^{\; 2}}{\left( \vec{q}_i-\vec{q}_{i+1} \right)^2}-\vec{q}^{\; 2} \right).
\end{split}
\end{equation}
Therefore, from Eq.~(\ref{Int:Eq:ColorDecAdj}) it is possible to decompose the scattering amplitude $\mathcal{A}_{12}^{1'2'}$ as
\begin{equation}
\label{Inv1}
\mathcal{A}_{12}^{1'2'} = \sum_R (\mathcal{A}_R)_{12}^{1'2'} ,
\end{equation} 
where $(\mathcal{A}_R)_{12}^{1'2'}$ is the part of the scattering amplitude corresponding to the definite irreducible representation $R$ of the colour group in the $t$-channel. 
\subsubsection{Partial wave expansion}
Now, it is necessary to introduce the Mellin transformation $f_R(\omega, \vec{q})_{AB}^{A'B'}$ (partial wave) of the imaginary part of the amplitude, that following \cite{Fadin:1998sh}, we define by
\begin{equation}
\label{Int:Eq:MellinTransf}
f_R(\omega,\vec{q})_{12}^{1'2'} = \int_{s_0}^{\infty}  \frac{d s}{s} \left( \frac{s}{s_0} \right)^{-\omega} \left[ \frac{ \Im_s (\mathcal{A}_R)^{1'2'}_{1 2}}{s} \right] \; . 
\end{equation}     
Its inverse is given by
\begin{equation}
\label{Int:Eq:PartialWaveExp}
\Im_s (\mathcal{A}_R)^{1'2'}_{12} = \frac{s}{2 \pi i} \int_{\delta-i\infty}^{\delta+i\infty} d\omega \left( \frac{s}{s_0}\right)^{\omega} f_R(\omega,\vec{q})_{12}^{1'2'} = \frac{s}{2 \pi i} \oint_C d\omega \left(\frac{s}{s_0}\right)^{\omega}f_R(\omega,\vec{q})_{1 2}^{1'2'} \; ,
\end{equation} 
where $\delta + i \Im ({\omega})$ is an axis, on the real axis of the complex plane of the variable $\omega$, which lies to the right of all singularities of $f_R (\omega, \vec{q})$ and the closed surfaces $C$ is obtained by adding to this axis an infinite semicircle which extends in the left region of the complex plane (which is in the direction in which the real part of $\omega$ becomes negative). The second equality holds due to the fact that the contribution on the semicircle is zero since the integrand vanishes exponentially when the real part of $\omega$ tends to minus infinity. \\
Using dispersion relations it is possible to reconstruct the total amplitude. Hence, the expression for the total amplitude in the form of a partial wave expansion is
\begin{equation}
\label{Int:Eq:PartialWaveExpFull}
(\mathcal{A}_R)_{12}^{1'2'} = \frac{s}{2 \pi} \int_{\delta-i\infty}^{\delta+i\infty} \frac{d\omega}{\sin(\pi \omega)} \left[\left(\frac{-s}{s_0}\right)^{\omega}- \eta \left(\frac{s}{s_0}\right)^{\omega}\right] f_R(\omega, \vec{q})_{AB}^{A'B'} ,
\end{equation} 
where $\eta$ is the signature and coincides with the symmetry of the representation $R$. \\ 
It is important to observe that in the octet representation we have to add the Born contribution to the dispersion relation, \textit{i.e.} 
\begin{equation}
\label{Int:Eq:PartialWaveExpFullPlusBorn}
(\mathcal{A}_R)_{12}^{1'2'} = (\mathcal{A}_R^{B})_{12}^{1'2'} + \frac{s}{2 \pi} \int_{\delta-i\infty}^{\delta+i\infty} \frac{d\omega}{\sin(\pi \omega)} \left[\left(\frac{-s}{s_0}\right)^{\omega}- \eta \left(\frac{s}{s_0}\right)^{\omega}\right] f_R(\omega, \vec{q})_{AB}^{A'B'} \; .
\end{equation} 
\subsubsection{Phase space}
From Eq.~(\ref{Int:Eq:SquaredInvariantMasses}), imposing the on-shellness of outgoing gluons, we have
\begin{equation}
    \left( \prod_{i=1}^{n} \vec{k}_i^{\; 2} \right)^{-1} = \left( \prod_{i=1}^{n} \frac{\beta_i}{\beta_{i-1}} s_i \right)^{-1} \implies \left( \prod_{i=1}^{n+1} s_i \right) \left( \prod_{i=1}^{n} \vec{k}_i^{\; 2} \right)^{-1} = s_{n+1} \frac{\beta_0}{\beta_n} \simeq s \; .
\end{equation}
We can manipulate this further to get
\begin{equation}
\label{Int:Eq:SInSi}
    s = \left( \prod_{i=1}^{n+1} s_i \right) \left( \prod_{i=1}^{n} \vec{k}_i^{\; 2} \right)^{-1} = \left( \prod_{i=1}^{n+1} \frac{s_i}{\sqrt{\vec{k}_{i-1}^{\; 2} \vec{k}_{i}^{\; 2}}} \right) \sqrt{\vec{q}_{1}^{\; 2} \vec{q}_{n+1}^{\; 2}} \; .
\end{equation}
The phase of space element of $n+2$ particles reads
\begin{equation*}
d \Phi_{n+2} = (2 \pi)^D \delta^D \left( p_1 + p_2 - \sum_{i=0}^{n+1} k_i \right) \left( \prod_{i=0}^{n+1} \delta (k_i^2) \frac{d^D k_i}{(2 \pi)^{D-1}} \right)
\end{equation*} 
\begin{equation*}
= \frac{2}{s} (2 \pi)^D \delta \left( \hspace{-0.1 cm} 1 - \sum_{i=0}^{n+1} \alpha_i \hspace{-0.1 cm} \right) \delta \left( \hspace{-0.1 cm} 1-\sum_{i=0}^{n+1} \beta_i \hspace{-0.1 cm} \right) \delta^{(D-2)} \left( \sum_{i=0}^{n+1} k_{i\perp} \hspace{-0.1 cm} \right) \hspace{-0.1 cm} \frac{s}{2} \delta (\alpha_0 \beta_0 s + k_{0 \perp}^2) d \alpha_0 d \beta_0 \frac{d^{D-2} k_{0 \perp}}{(2 \pi)^{D-1}} 
\end{equation*}
\begin{equation*}
    \times \left( \prod_{i=1}^{n} \frac{s}{2} \delta (\alpha_{i} \beta_{i} s + k_{i \perp}^2) d \alpha_{i} d \beta_{i} \frac{d^{D-2} k_{i \perp}}{(2 \pi)^{D-1}} \right) \frac{s}{2} \delta (\alpha_{n+1} \beta_{n+1} s + k_{n+1 \perp}^2) d \alpha_{n+1} d \beta_{n+1} \frac{d^{D-2} k_{n+1 \perp}}{(2 \pi)^{D-1}} 
\end{equation*}
\begin{equation}
\label{Int:Eq:phasespaceN+2}
\begin{split}
= \frac{2}{s} (2 \pi)^D \delta \left( \hspace{-0.1 cm} 1 - \sum_{i=0}^{n+1} \alpha_i \hspace{-0.1 cm} \right) \delta \left( \hspace{-0.1 cm} 1-\sum_{i=0}^{n+1} \beta_i \hspace{-0.1 cm} \right) \delta^{(D-2)} \left( \sum_{i=0}^{n+1} k_{i\perp} \hspace{-0.1 cm} \right) \frac{d\beta_{0}}{2 \beta_{0}} \frac{d \alpha_{n+1}}{2 \alpha_{n+1}} \prod_{i=1}^{n} \frac{d \beta_i}{2 \beta_i} \prod_{i=0}^{n+1} \frac{d^{D-2}k_{i\perp}}{(2\pi)^{D-1}} \; .
\end{split}
\end{equation} 
In performing the integration over phase space, exploiting the MRK, the limits of integration in the variables $\beta_i$'s must be restricted according to the condition $\beta_i < \beta_{i-1}$. In the following, we need to consider
\begin{equation}
    \frac{d s}{s} d \Phi_{n+2} = \frac{d s}{2 s^2} (2 \pi)^D \prod_{i=1}^{n} \frac{d \beta_i}{2 \beta_i} \prod_{i=0}^{n} \frac{d^{D-2}k_{i\perp}}{(2\pi)^{D-1}} \;.
\end{equation}
We perform the following changes of variables
\begin{equation}
    \{ \beta_1, ..., \beta_n, s \} \rightarrow \{ s_1, ..., s_n, s_{n+1} \} \; , \hspace{1 cm}  \{ k_{0 \perp}, ..., k_{n \perp} \} \rightarrow \{ q_1, ..., q_{n+1} \}
\end{equation}
by making use of relations
\begin{equation}
 \left \{  s_i = \frac{\beta_{i-1}}{\beta_{i}} s_R \; , i =1,...,n \right \} \; , \hspace{0.2 cm} s_{n+1} = \beta_n s \; , \hspace{0.2 cm} k_{0 \perp} = - q_{1 \perp} \; , \hspace{0.2 cm} \bigg \{  k_{i \perp} =  q_{i \perp} -  q_{i+1 \perp} \; , i = 1,..., n \bigg \}  
\end{equation}
and get
\begin{equation}
\label{Int:Eq:PhaseSpaceInSqi}
    \frac{ds}{s} d \Phi_{n+2} = \frac{2 \pi}{s} \left( \prod_{i=1}^{n+1} \frac{d s_i}{s_i} \right) \left( \prod_{i=1}^{n+1} \frac{d^{D-2} q_i }{(2 \pi)^{D-1}} \right) \; ,
\end{equation}
where the integration over variables $s_i$ is extended between $s_R$ and infinity.
\subsubsection{Generalized BFKL equation}
We can consider the $n_{{\rm{th}}}$ contribution to the partial wave amplitude, implicitly defined by the relation
\begin{equation}
\label{Int:Eq:NContPartialWave}
\begin{split}
f_R(\omega,\vec{q})_{1 2}^{1' 2'} & = \sum_{n=0}^{\infty} f_R^{(n)}(\omega, \vec{q})_{1 2}^{1' 2'} \; ,
\end{split}
\end{equation}
which reads
\begin{equation*}
    f_R^{(n)}(\omega, \vec{q})_{1 2}^{1' 2'} = \frac{1}{2} \int_{s_0}^{\infty} \frac{ds}{s} d \Phi_{n+2} \left( \frac{s}{s_0} \right)^{-\omega} \frac{1}{s} \sum_{f} \mathcal{A}_{12}^{\tilde{1} \tilde{2} + n} \left(\mathcal{A}_{1'2'}^{\tilde{1} \tilde{2}+n}\right)^*
\end{equation*}
\begin{equation*}
    = 2 \int_{s_0}^{\infty} \frac{ds}{s} d \Phi_{n+2} \left( \frac{s}{s_0} \right)^{-\omega} s \left( \prod_{i=1}^{n} \braket{c_i c_i'|\hat{\mathcal{P}}_{R}|c_{i+1} c'_{i+1}} 2 (2 \pi)^{D-1} \mathcal{K}_r^{(R)} (\vec{q}_i, \vec{q}_{i+1}; \vec{q}) \right) 
\end{equation*}
\begin{equation}
\label{Int:Eq:NContPartialWaveExpli}
    \times \left( \prod_{i=1}^{n+1} \left( \frac{s_i}{s_R} \right)^{\omega(t_i) + \omega(t_i') } \frac{1}{\vec{q}_i^{\; 2} (\vec{q}_i-\vec{q})^2} \right) \left( \sum_{\tilde{1}} \Gamma_{\tilde{1} 1'}^{*} \Gamma_{\tilde{1} 1} \right) \left( \sum_{\tilde{2}} \Gamma_{\tilde{2} 2'}^{*} \Gamma_{\tilde{2} 2} \right) \; .
\end{equation}
Now, we observe that
\begin{equation}
    \braket{c_i c_i'|\hat{\mathcal{P}}_{R}|c_{i+1} c_{i+1}'} = \braket{c_i c_i'|\hat{\mathcal{P}}_{R} \hat{\mathcal{P}}_{R}|c_{i+1} c_{i+1}'} =\sum_{\nu} \braket{c_i c_i'|\hat{\mathcal{P}}_{R}|\nu} \braket{\nu|\hat{\mathcal{P}}_{R}|c_{i+1} c_{i+1}'} \; ,
\end{equation}
where the sum over the index $\nu$ is extended to all states of the irreducible representation $R$. Using this, we can write
\begin{equation}
\label{Int:Eq:ProjectTreatment}
\begin{split}
    & \sum_{\nu , \nu'} \braket{c_1 c_1'|\hat{\mathcal{P}}_{R}|\nu} \left[ \braket{ \nu |\hat{\mathcal{P}}_{R}| c_2 c_2' } \braket{ c_2 c_2' |\hat{\mathcal{P}}_{R}| c_3 c_3' } ... \braket{c_n c_n'|\hat{\mathcal{P}}_{R}|\nu'} \right] \braket{ \nu' |\hat{\mathcal{P}}_{R}| c_{n+1} c_{n+1}' } \\ &
    = \sum_{\nu , \nu'} \braket{c_1 c_1'|\hat{\mathcal{P}}_{R}|\nu} \braket{ \nu |\nu'} \braket{ \nu' |\hat{\mathcal{P}}_{R}| c_{n+1} c_{n+1}' } = \sum_{\nu} \braket{c_1 c_1'|\hat{\mathcal{P}}_{R}|\nu}  \braket{ \nu |\hat{\mathcal{P}}_{R}| c_{n+1} c_{n+1}' } \; .
\end{split}
\end{equation}
Starting from (\ref{Int:Eq:NContPartialWaveExpli}), using Eqs. (\ref{Int:Eq:SInSi}, \ref{Int:Eq:ProjectTreatment}) and performing the change of variables (\ref{Int:Eq:PhaseSpaceInSqi}), we obtain
\begin{equation*}
    f_R^{(n)} (\omega, \vec{q})_{1 2}^{1' 2'} = \frac{1}{(2 \pi)^{D-2}} \int \left( \prod_{i}^{n+1} \frac{d^{D-2} q_{i \perp}}{\vec{q}_i^{\; 2} (\vec{q}_i-\vec{q})^2}  d \left( \frac{s_i}{s_R} \right) \left( \frac{s_i}{s_R} \right)^{\omega(t_i) + \omega (t_i') - \omega - 1} \right)  \left( \frac{s_0}{ \sqrt{\vec{q}_1^{\; 2} \vec{q}_{n+1}^{\; 2}}  } \right)^{\omega}
\end{equation*}
\begin{equation}
  \times  \sum_{\nu} I_{1'1}^{(R, \nu)} \left( \prod_{i=1}^{n} \mathcal{K}_r^{(R)} (\vec{q}_i, \vec{q}_{i+1} ; \vec{q}) \right) I_{2'2}^{(R, \nu)} \; ,
  \label{Int:Eq:PartialWaveN}
\end{equation}
where
\begin{equation}
\label{Int:Eq:ImpactFactorsGen}
    I_{1'1}^{(R, \nu)} = \sum_{\tilde{1}} \braket{c_1 c_1'|\hat{\mathcal{P}}_{R}|\nu} \Gamma_{\tilde{1} 1}^{c_1} \left( \Gamma_{\tilde{1} 1}^{c_1'*} \right)  \; , \hspace{1.5 cm} I_{2'2}^{(R, \nu)} = \sum_{\tilde{2}} \Gamma_{\tilde{2} 2}^{c_{n+1}} \Gamma_{\tilde{2} 2}^{c_{n+1}'*}  \braket{ \nu |\hat{\mathcal{P}}_{R}| c_{n+1} c_{n+1}' }  \; .
\end{equation}
The quantity $I_{i'i}^{(R, \nu)}$ in (\ref{Int:Eq:ImpactFactorsGen}) are usually called \textit{impact factors}. It is important to keep in mind that for the singlet representation the index $\nu$ takes only one value, hence it can be omitted and
\begin{equation}
\label{singletrepres}
\braket{cc'| \hat{\mathcal{P}}_0|0} = \frac{\delta_{cc'}}{\sqrt{N^2-1}} ,
\end{equation}  
whereas for the antisymmetrical octet (gluon) representation $\nu$ coincides with the gluon colour index and
\begin{equation}
\braket{cc'| \hat{\mathcal{P}}_8|a} = \frac{f_{acc'}}{\sqrt{N}} .
\end{equation}   
We can now perform the integration over the variables $(s_i/s_R)$'s to find
\begin{equation*}
    f_R^{(n)} (\omega, \vec{q})_{1 2}^{1' 2'} = \frac{1}{(2 \pi)^{D-2}} \int \left( \prod_{i}^{n+1} \frac{d^{D-2} q_{i \perp}}{\vec{q}_i^{\; 2} (\vec{q}_i-\vec{q})^2} \frac{1}{\omega - \omega(t_i) + \omega (t_i')} \right)
\end{equation*}
\begin{equation}
    \times \sum_{\nu} I_{1'1}^{(R, \nu)} \left( \prod_{i=1}^{n} \mathcal{K}_r^{(R)} (\vec{q}_i, \vec{q}_{i+1} ; \vec{q}) \right) I_{2'2}^{(R, \nu)} \; ,
\end{equation}
where we neglected terms of order $\omega \ln s_R$, since the essential integration region in (\ref{Int:Eq:PartialWaveExp}) is the one where $\omega \sim (\ln s)^{-1}$. In particular, the exact lower bound of integration in the invariant masses has been neglected as it produces sub-dominant effects. Finally, from the last expression, we can reconstruct the complete partial wave and cast it in the form
\begin{gather}
\label{Int:Eq:PartialwaveexpCom}
f_R(\omega,\vec{q})_{1 2}^{1' 2'} = \sum_{n=0}^{\infty} f_R^{(n)}(\omega, \vec{q})_{1 2}^{1' 2'} \nonumber \\ = \frac{1}{(2 \pi)^{D-2}} \int \frac{d^{D-2}q_{A \perp}}{\vec{q}_A^{\; 2}(\vec{q}_A-\vec{q})^2} \frac{d^{D-2}q_{B \perp}}{\vec{q}_B^{\; 2}(\vec{q}_B-\vec{q})^2}  \sum_{\nu} I_{1' 1}^{(R,\nu)} G_{\omega}^{(R)}(\vec{q}_A, \vec{q}_B; \vec{q}) I_{2' 2}^{(R,\nu)} \; .
\end{gather}
The function $G_{\omega}^{(R)}$, called Green function for the scattering of two Reggeized gluons, is defined as
\begin{equation}
\begin{split}
& G_{\omega}^{(R)}(\vec{q}_A, \vec{q}_B; \vec{q}) = \sum_{n=0}^{\infty} \int \left( \prod_{i=1}^{n+1} \frac{d^{D-2}q_{i \perp}}{\vec{q}_i^{\; 2}(\vec{q}_i-\vec{q})^{2} \left( \omega-\omega(t_i)-\omega(t'_i) \right)} \right) \\ & \times \left( \prod_{i=1}^{n} \mathcal{K}_r^{(R)}(\vec{q}_i,\vec{q}_{i+1};\vec{q}) \right) \vec{q}_A^{\;2}(\vec{q}_A-\vec{q})^2  \vec{q}_B^{\;2}(\vec{q}_B-\vec{q})^2 \delta^{D-2}(q_{1\perp}-q_{A\perp}) \delta^{D-2}(q_{n+1\perp}-q_{B\perp}).
\end{split}
\end{equation}
The function $G_{\omega}^{(R)}$ is the only quantity depending on $\omega$, and hence it determines the $s$-behaviour of the scattering amplitude. It is the perturbative solution of the integral equation (see Fig.~\ref{I.equation}):    
\begin{figure}
\begin{picture}(400,150)
\put(10,0){\includegraphics[scale=0.4]{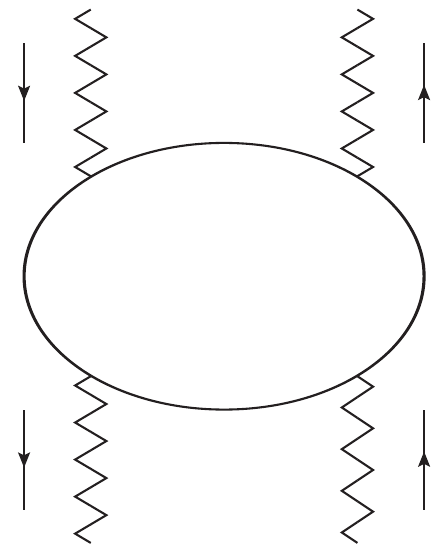}}
\put(45,50){$G_{\omega}$}
\put(0,17){$q_2$}
\put(0,87){$q_1$}
\put(100,17){$q_2-q$}
\put(100,87){$q_1-q$}
\put(135,50){\scalebox{1.5}{=}}
\put(160,50){$q_1$}
\put(170,35){\includegraphics[scale=0.4]{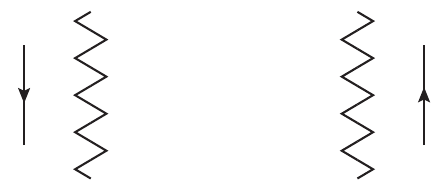}}
\put(260,50){$q_1-q$}
\put(305,50){\scalebox{1.5}{+}} 
\put(330,00){\includegraphics[scale=0.4]{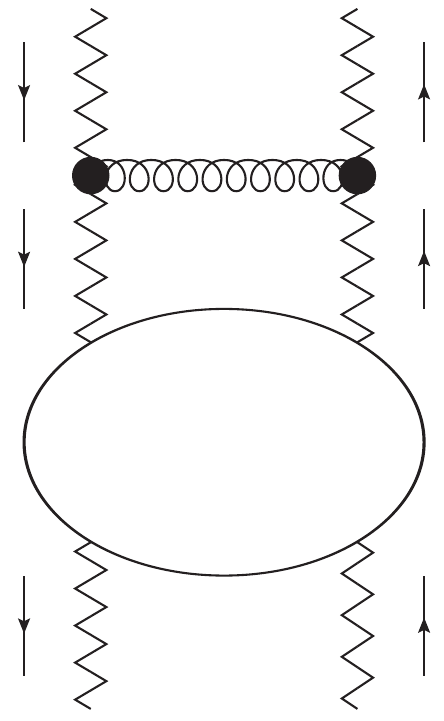}}
\put(320,17){$q_2$}
\put(320,87){$q_1'$}
\put(320,120){$q_1$}
\put(420,17){$q_2-q$}
\put(420,87){$q_1'-q$}
\put(420,120){$q_1-q$}
\put(365,50){$G_{\omega}$}
\end{picture}
\caption{Schematic representation of the BFKL equation.}
\label{I.equation}
\end{figure}
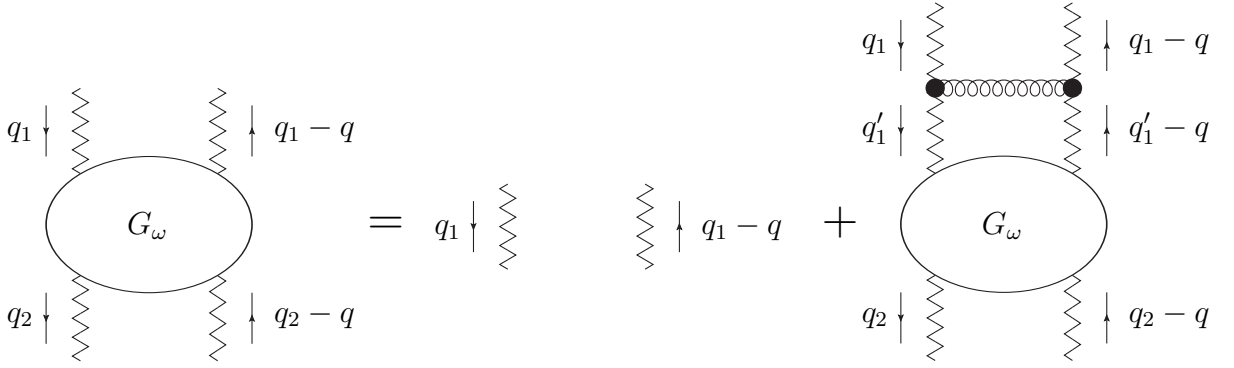
\begin{equation}
\label{BFKLequation}
\omega G_{\omega}^{(R)}(\vec{q}_1,\vec{q}_2;\vec{q})= \vec{q}_1^{\; 2}(\vec{q}_1-\vec{q})^{2} \delta^{(D-2)}(\vec{q}_1-\vec{q}_2) + \int \frac{d^{D-2}q'_{1 \perp}}{\vec{q}_1^{\; ' 2}(\vec{q}_1^{\; '}-\vec{q})^2} \mathcal{K}^{(R)}(\vec{q}_1, \vec{q}_1^{\; '} ; \vec{q}) G_{\omega}^{(R)}(\vec{q}_1^{\; '} , \vec{q}_2; \vec{q}),
\end{equation}
where the function $\mathcal{K}^{(R)}(\vec{q}_1, \vec{q}_1^{\; '}; \vec{q})$, called kernel of the integral function, has the following expression:
\begin{equation}
\label{kernel}
\begin{split}
\mathcal{K}^{(R)}(\vec{q}_1, \vec{q}_2; \vec{q}) = & \left[ \omega(q_{1\perp}^2) + \omega((q_1-q)_{\perp}^2) \right] \vec{q}_1^{\;2} \left( \vec{q}_1-\vec{q} \right)^{2} \delta^{(D-2)} (\vec{q}_1-\vec{q}_2) + \mathcal{K}_r^{(R)}(\vec{q}_1, \vec{q}_2; \vec{q}) \; .
\end{split}
\end{equation}
Two terms can be distinguished in Eq.~(\ref{kernel}); the first of them is called ``virtual'' and it is expressed in terms of the Regge trajectory of the gluon, while the second, which is related to the production of real particles, is given by the Eq.~(\ref{Kernel1}) (see the Fig.~\ref{Int:Fig:BFKLKernel}). \\
The Eq.~(\ref{BFKLequation}), in case of $R=0$ (singlet) and $t=0$, is called \textit{BFKL equation}. In the most general case, it is called \textit{generalized BFKL equation}. It is an iterative equation. 
The kernel $\mathcal{K}^{(R)}$ and the Green function $G_{\omega}^{(R)}$ can be seen as operators which act on the space of the transverse momentum defined by
\begin{equation*}
\hat{\vec{q}} \ket{\vec{q}_i} = \vec{q}_i \ket{\vec{q}_i}, 
\end{equation*}
\begin{equation}
\label{Int:MomRepr}
\braket{\vec{q}_1 \; |\vec{q}_2}= \delta^{(2)} \left( \vec{q}_1-\vec{q}_2 \right), \quad \quad \braket{A|B}=\braket{A|\vec{k} \; }\braket{ \; \vec{k} \; |B} = \int d^2 \vec{k} \; A(\vec{k} \;) B(\vec{k} \;).
\end{equation}
In this representation, the equation (\ref{BFKLequation}) can be written in the operator form:
\begin{equation}
\omega \hat{G}_{\omega}^{(R)} = 1 + \hat{ \mathcal{K} }^{(R)} \hat{G}_{\omega}^{(R)} \; ,
\label{Int:Eq:BFKLOPEform}
\end{equation} 
which implies
\begin{equation}
\hat{G}_{\omega}^{(R)} = \frac{1}{\omega-\hat{\mathcal{K}}^{(R)}} \; .
\end{equation}
\begin{figure}
\begin{picture}(400,110)
\put(30,0){\includegraphics[scale=0.6]{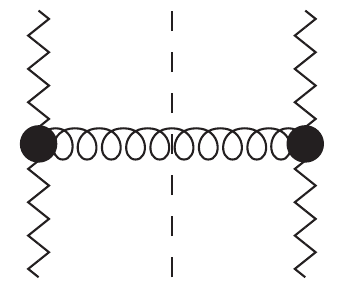}}
\put(150,37){+}
\put(180,0){\includegraphics[scale=0.6]{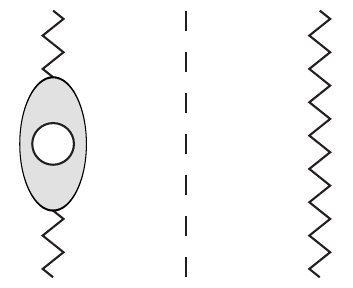}}
\put(300,37){+} 
\put(330,0){\includegraphics[scale=0.6]{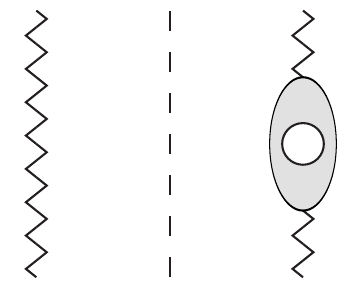}}
\end{picture}
\caption{Schematic representation of the LO BFKL kernel.}
\label{Int:Fig:BFKLKernel}
\end{figure}
\subsection{Introduction to BFKL cross section and impact factors}
We now focus on the case of the exchange of vacuum quantum numbers in the $t$-channel, considering an elastic process, $A+B \rightarrow A+B$, with the additional simplification of forward scattering ($q=0$). We can relate the imaginary part of this amplitude to the total cross section of the process $A+B \rightarrow X$ by using the optical theorem (\ref{Int:Eq:OpticalThe})\footnote{Note that the flux factor, for $\sqrt{s}$ much larger than the masses of the incoming particles, becomes $F \approx 2s$.},
\begin{equation}
\label{Int:Eq:HighEneOpThe}
\sigma_{AB}(s) = \frac{\Im_s \mathcal{A}_{AB}^{AB}}{s} \; .
\end{equation} 
From Eqs. (\ref{Int:Eq:PartialWaveExp}) and (\ref{Int:Eq:PartialwaveexpCom}),
\begin{equation}
\label{Int:Eq:TotCross}
\sigma_{AB}(s) = \int_{\delta-i\infty}^{\delta+i\infty} \frac{d\omega}{2 \pi i} \frac{1}{(2 \pi)^{D-2}} \int d^{D-2} q_{A\perp} d^{D-2} q_{B\perp} \left( \frac{s}{s_0} \right)^{\omega} \frac{\Phi_A(\vec{q}_A)}{(\vec{q}_A^{\;2})^2}G_{\omega}(\vec{q}_A, \vec{q}_B) \frac{\Phi_B(-\vec{q}_B)}{(\vec{q}_B^{\;2})^2} ,
\end{equation}  
where $\Phi_A(\vec{q}_A) \equiv I_{AA}^{(0)}$ , $\Phi_B(\vec{q}_B) \equiv I_{BB}^{(0)}$ are the impact factors projected on the color singlet representation. Their expressions are obtained from (\ref{Int:Eq:ImpactFactorsGen}), remembering that in the singlet representation and in the case of an elastic process, the color projector are given by the Eq.~(\ref{singletrepres}) with $c=c'$; hence:
\begin{equation}
\label{Int:Eq:ImpactFactorsSin1Par}
\Phi_{A}(\vec{q}_A) = \frac{1}{\sqrt{N^2-1}} \sum_{\tilde{A},c} |\Gamma_{\tilde{A}A}^{c}|^{2} ; \quad \vec{q}_A = - \vec{p}_{\tilde{A}} \; ,
\end{equation} 
\begin{equation}
\Phi_{B}(\vec{q}_B) = \frac{1}{\sqrt{N^2-1}} \sum_{\tilde{B},c} |\Gamma_{\tilde{B}B}^{c}|^{2} ; \quad \vec{q}_B = - \vec{p}_{\tilde{B}} \; .
\end{equation}
\begin{figure}
\begin{picture}(400,200)
\put(130,0){\includegraphics[width=0.4\textwidth]{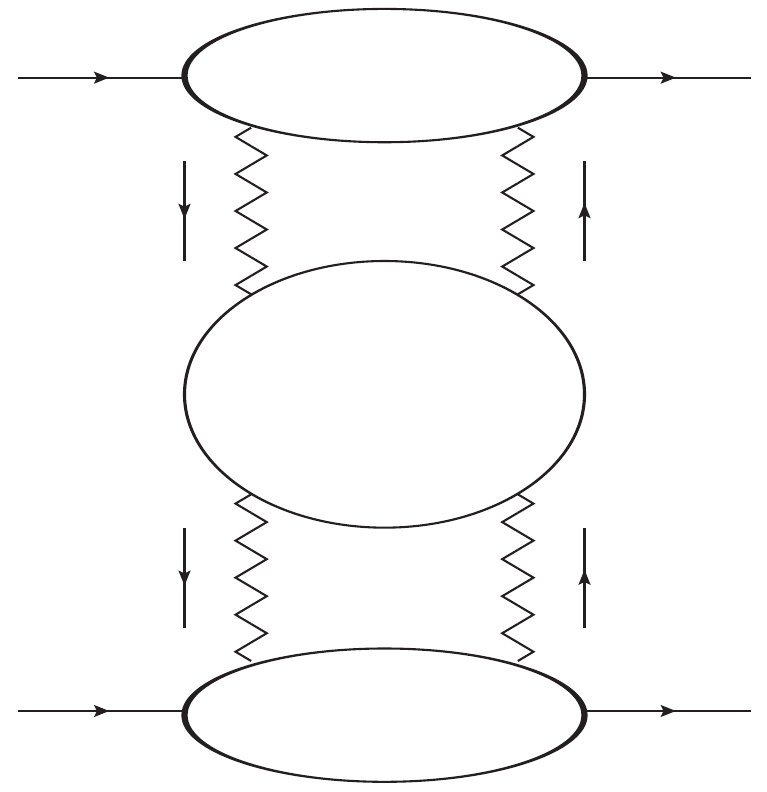}}
\put(210,165){\scalebox{1.2}{$\Phi_{AA}$}}
\put(210,90){\scalebox{1.2}{$G_{\omega}$}}
\put(210,15){\scalebox{1.2}{$\Phi_{BB}$}}
\put(112,17){$p_B$}
\put(112,167){$p_A$}
\put(316,17){$p_B$}
\put(316,167){$p_A$}
\put(157,50){$q_B$}
\put(157,136){$q_A$}
\put(276,50){$q_B$}
\put(276,136){$q_A$}
\end{picture}
\caption{Schematic representation of the factorized amplitude.}
\label{convolution}
\end{figure}
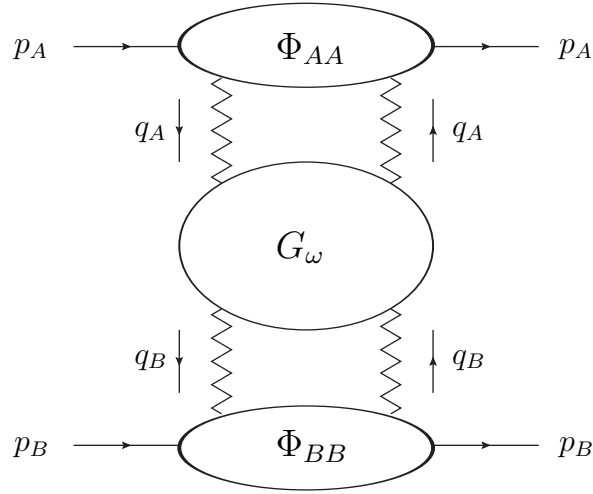
\hspace{-0.3 cm} At this point, it is very interesting to understand that in the BFKL approach the amplitude for a diffusion process at high energy can be written as the convolution of the Green function $G$ for two Reggeized gluon with the impact factors $\Phi_{A'A}$ and $\Phi_{B'B}$ of the colliding particles (see Fig.~\ref{convolution}).
The BFKL Green function is universal and completely governs the $s$-behavior of the amplitude, instead the impact factors depend on the particular process under analysis and they are $s$-independent. \\
It is important to make a clarification on the definition of impact factor. The one given above is not the most general possible, in fact, only one particle of the intermediate state across the $s$-channel cut has been included in its definition. In general the intermediate state can be a system of particles, and in this case, within the LLA, the definition generalizes to~\cite{Fadin:1998fv}
\begin{equation}
\label{Int:Eq:ImpactFactSingGen}
    \Phi_{AA} (\vec{q}_A) = \braket{cc|\hat{\mathcal{P}}_0|0} \sum_{\{ f \}} \int \frac{d s_{PR}}{2 \pi} d \rho_f \Gamma_{\{ f \} A}^c \left( \Gamma_{\{ f \} A}^{c'}  \right)^* \; .
\end{equation}
In the case of a single particle in the intermediate state the integration becomes completely trivial and the expression (\ref{Int:Eq:ImpactFactSingGen}) reduces to the one in Eq.~(\ref{Int:Eq:ImpactFactorsSin1Par}).
\subsection{Solution of BFKL equation at $t=0$}
The $s$-behaviour of the total cross section (\ref{Int:Eq:AsyPomerSing}) is determined by solving the BFKL equation (\ref{BFKLequation}) and finding the leading singularity $\omega_0$. Then, performing the anti-Mellin transformation, it is possible to get the asymptotic behaviour of cross sections.
First of all, the Eqs.~(\ref{BFKLequation},~\ref{kernel}) must be taken in the case of $R=0$ (singlet) and $t=0$. Redefining
\begin{equation}
\frac{G_{\omega}^{(0)} (\vec{q}_1,\vec{q}_2;\vec{0})}{\vec{q}_1^{\;2} \vec{q}_2^{\; 2}} \equiv G_{\omega}\left(\vec{q}_1,\vec{q}_2\right),
\end{equation}    
\begin{equation}
\frac{\mathcal{K}^{(0)}(\vec{q}_1,\vec{q}_2,\vec{0})}{\vec{q}_1^{\; 2} \vec{q}_2^{\; 2}} \equiv \mathcal{K} (\vec{q}_1,\vec{q}_2),
\end{equation}
and hence
\begin{equation}
\frac{\mathcal{K}_{r}^{(0)}(\vec{q}_1,\vec{q}_2,\vec{0})}{\vec{q}_1^{\; 2} \vec{q}_2^{\; 2}} \equiv \mathcal{K}_{r} (\vec{q}_1,\vec{q}_2) = \frac{g^2N}{(2\pi)^{D-1}} \frac{2}{(\vec{q}_1-\vec{q}_2)^2}  \; ,
\label{Int:Eq:RealBornKert0}
\end{equation}
it straightforward to find that the BFKL equation takes the following form:
\begin{equation}
\omega G_{\omega}\left(\vec{q}_1,\vec{q}_2\right)=\delta^{(2)}\left(\vec{q}_1-\vec{q}_2\right)+ \int d^{D-2} k \;  \mathcal{K}(\vec{q}_1, \vec{k}) G_{\omega} (\vec{k}, \vec{q}_2)  ,
\end{equation}
that can be rewritten as
\begin{equation}
\omega G_{\omega}\left(\vec{q}_1,\vec{q}_2\right)=\delta^{(2)}\left(\vec{q}_1-\vec{q}_2\right)+\mathcal{K} \bullet G_{\omega} \left( \vec{q}_1, \vec{q}_2 \right) \; ,
\end{equation}
where we have defined the convolution operator as
\begin{equation}
    A \bullet B \; (q_1, q_2) = \int d^{D-2} k \; A (q_1,k) B(k,q_2) \; .
\end{equation}
The expression of the kernel function reads
\begin{equation}
    \mathcal{K} (\vec{q}_1, \vec{q}_2) = \frac{\mathcal{K}^{(0)} (\vec{q}_1, \vec{q}_2)}{\vec{q}_1^{\; 2} \vec{q}_2^{\; 2}} = 2 \omega (- \vec{q}_1^{\; 2}) \delta^{D-2} (\vec{q}_1 - \vec{q}_2) +  \frac{g^2N}{(2\pi)^{D-1}} \frac{2}{(\vec{q}_1-\vec{q}_2)^2} \; .
\end{equation}
Finding the Green function $G_{\omega}$ means solving the eigenvalue problem for the kernel $\mathcal{K}$. In fact, if the equation
\begin{equation}
\label{eigenvalueproblem}
\mathcal{K} \bullet \phi_i(\vec{q}) =  \lambda_i \phi_i(\vec{q}),
\end{equation}
is solved, then, the Green function can be reconstructed through the spectral representation
\begin{equation}
G_{\omega} \left( \vec{q}_1 , \vec{q}_2 \right)= \sum_i \frac{\phi_i(\vec{q}_1)\phi_i^*(\vec{q}_2)}{\omega-\lambda_i} \; .
\end{equation}
It is important to stress that the BFKL kernel can be seen as an operator acting on a space of complex functions, defined on a vectorial space, $\mathcal{V}$. Now, it is easy to see that the kernel is rotationally invariant\footnote{Please note that the Jacobian of a rotation is one.}, \textit{i.e.}
\begin{equation}
    \mathcal{K} (R \vec{q}_1, R \vec{q}_2) = \mathcal{K} (\vec{q}_1, \vec{q}_2) \; , \hspace{0.5 cm} \vec{q}_1, \vec{q}_2 \in \mathcal{V} \; ,
\end{equation}
with $R$ a generic rotation operator in $\mathcal{V}$.
The consequence of rotational invariance is that the kernel admits a set of eigenfunctions of the type 
\begin{equation}
    \phi_{\gamma}^{n} (\vec{q}) = f_{\gamma} (|\vec{q} \;|) Y_n (\phi) \; ,
\end{equation}
where $\phi$ on the right hand side denotes the set of angular variables and $Y_n$ is a spherical harmonic which, in dimension two, is nothing more than an exponential function $e^{i n \phi}$. \\
The kernel is also a scale invariant operator, in fact, given a $\lambda \in {\rm I\!R}$,
\begin{equation}
     \mathcal{K} ( \lambda \vec{q}_1, \lambda \vec{q}_2) = \lambda^{-2} \mathcal{K} (\vec{q}_1, \vec{q}_2) \; , \hspace{0.5 cm} \vec{q}_1, \vec{q}_2 \in \mathcal{V} \; .
\end{equation}
The scale invariance forces the radial function $f_{\gamma} (|\vec{q} \;|)$ to be a power of $|\vec{q} \;|$, we will take them as
\begin{equation}
    f_{\gamma} (|\vec{q} \;|) = (\vec{q}^{\; 2})^{\gamma} \; .
\end{equation}
The last step is to fix the value of $\gamma$ and the overall normalization by requiring that these functions form an orthonormal complete sets, \textit{i.e.}
\begin{equation}
\int d^2 \vec{q} \; \phi_{\nu}^n (\vec{q}) \phi_{\nu'}^{n'} (\vec{q}) = \delta (\nu-\nu') \delta (n-n').
\end{equation}
By imposing this last condition, it is simple to see that $\gamma$ must be equal to $ i \nu - 1/2 $ with $\nu \in {\rm I\!R}$ and hence, the set of eigenfunctions that solves the problem is 
\begin{equation}
\label{Int:Eq:LOBFKLEigenfuntions}
\left \{ \phi_{\nu}^n (\vec{q} \;)= \frac{1}{\pi \sqrt{2}} (\vec{q}^{\;2} )^{-\frac{1}{2}+i\nu} e^{in \theta} \; : \; \nu \in {\rm I\!R} \; , \; n \in \mathbb{Z} \right \}.
\end{equation}
To find the corresponding eigenvalues we have to make the kernel act on a test eigenfunction. We consider separately virtual corrections and real emissions in dimensional regularization with $d=2+2 \epsilon$. To this purpose, we introduce the ``continuation” of the LO BFKL eigenfunctions to non-integer dimensions, 
\begin{equation}
\label{Int:Eq:LOBFKLeigenCont}
    \left( \vec{q}^{\; 2} \right)^{\gamma} e^{in \theta} \rightarrow \left( \vec{q}^{\; 2} \right)^{\gamma- \frac{n}{2}} \left( \vec{q} \cdot \vec{l} \; \right)^n \; ,
\end{equation}
where $\vec{l}$ lies only in the first two of the $2+2 \epsilon$ transverse space dimensions, \textit{i.e.} $\vec{l} = \vec{e}_1 + i \vec{e}_2 $ with $\vec{e}_{1,2}^{\; 2} = 1$, $\vec{e}_1 \cdot \vec{e}_2 = 0$. In the limit $\epsilon \rightarrow 0$ the r.h.s. of Eq.~(\ref{Int:Eq:LOBFKLeigenCont}) reduces to the LO BFKL eigenfunction, since we have
\begin{equation}
e^{i n \theta} = (\cos \theta + i \sin \theta)^n = \left( \frac{q_x+i q_y}{|q|} \right)^n = \left( \frac{(\vec{q} \cdot \vec{l})^n}{(\vec{q}^{\; 2})^{n/2}} \right) \; .
\end{equation}
For the virtual part we have
\begin{equation*}
    \mathcal{K}_v \bullet \phi_{\nu}^n (\vec{q}) =  \int \frac{d^{D-2} k}{\pi \sqrt{2}} (\vec{k}^{\; 2})^{-1/2 + i \nu} \left( \frac{(\vec{k} \cdot \vec{l})^n}{(\vec{k}^{\; 2})^{n/2}} \right) \; 2 \omega (- \vec{q}^{\; 2}) \delta^{D-2} (\vec{q}-\vec{k}) 
\end{equation*}
\begin{equation}
    = - \frac{2 g^2 N (\vec{q}^{\; 2})^{\epsilon}}{(4 \pi)^{2+\epsilon}} \frac{\Gamma (1-\epsilon) \Gamma^2 (\epsilon)}{\Gamma (2 \epsilon)} \phi_{\nu}^{n} (\vec{q}) \; .
\end{equation}
We see that the virtual part contains a pole in $\epsilon$; this will be cancelled by the real part, ensuring the infrared finiteness in the singlet case. For the real part, we have
\begin{equation*}
    \mathcal{K}_r \bullet \phi_{\nu}^n (\vec{q}) =  \int \frac{d^{D-2} k}{\pi \sqrt{2}} (\vec{k}^{\; 2})^{-1/2 + i \nu} \left( \frac{(\vec{k} \cdot \vec{l})^n}{(\vec{k}^{\; 2})^{n/2}} \right) \; \frac{g^2 N}{(2 \pi)^D} \frac{2}{(\vec{q}-\vec{k})^{2}} \; .
\end{equation*}
To solve this integral, we perform a Feynman parametrization to obtain
\begin{equation}
   \mathcal{K}_r \bullet \phi_{\nu}^n (\vec{q}) = \frac{2 g^2 N}{(2 \pi)^{3+2 \epsilon}} \frac{\Gamma (\frac{3}{2} + \frac{n}{2} - i \nu)}{\Gamma (\frac{1}{2} + \frac{n}{2} - i \nu)} \int_0^1 dx \int \frac{d^{D-2} k}{\pi \sqrt{2}} \frac{(1-x)^{-1/2+n/2-i \nu}  (\vec{k} \cdot \vec{l})^n}{\left[ \vec{k}^{\; 2} - 2 x \vec{q} \cdot \vec{k} + x \vec{q}^{\; 2} \right]^{3/2 + n/2 - i \nu}} \; .
\end{equation}
If we perform the shift $\vec{k} \rightarrow \vec{k}+x\vec{q} $, in the numerator we obtain a factor
\begin{equation}
\label{binomialExpSolKer}
    (\vec{k} \cdot \vec{l} + x \vec{q} \cdot \vec{l})^n = \sum_j \binom{n}{j} (\vec{k} \cdot \vec{l})^j x^{n-j} (\vec{q} \cdot \vec{l})^{n-j} \; .
\end{equation}
Only the $j=0$ term in Eq.~(\ref{binomialExpSolKer}) gives a non-vanishing contribution to the integral and therefore,
\begin{equation*}
    \mathcal{K}_r \bullet \phi_{\nu}^n (\vec{q}) = \frac{2 g^2 N}{(2 \pi)^{3+2 \epsilon}} \frac{\Gamma (\frac{3}{2} + \frac{n}{2} - i \nu)}{\Gamma (\frac{1}{2} + \frac{n}{2} - i \nu)} (\vec{q}^{\; 2})^{n/2} e^{in \phi} \int_0^1 dx \; x^n (1-x)^{-1/2+n/2-i \nu} 
\end{equation*}
\begin{equation}
    \times \int \frac{d^{D-2} k}{\pi \sqrt{2}} \frac{1}{\left[ \vec{k}^{\; 2} + x(1-x) \vec{q}^{\; 2}  \right]^{3/2+n/2-i \nu}} \; .
\end{equation}
Both integrations are now simple and we get
    \begin{equation}
    \mathcal{K}_r \bullet \phi_{\nu}^n  (\vec{q}) = \frac{2 g^2 N (\vec{q}^{\; 2})^{\epsilon}}{(4 \pi)^{2+\epsilon}} \frac{2 \Gamma (\epsilon) \Gamma(\frac{1}{2} + \frac{n}{2} - i \nu-\epsilon) \Gamma(\frac{1}{2} + \frac{n}{2} + i \nu+\epsilon)}{\Gamma(\frac{1}{2} + \frac{n}{2} - i \nu) \Gamma(\frac{1}{2} + \frac{n}{2} + i \nu + 2\epsilon)} \phi_{\nu}^{n} (\vec{q}) \; .
\end{equation}
Combining real and virtual parts, we get the full action of the kernel on its eigenfunctions:
\begin{equation*}
    \mathcal{K} \bullet \phi_{\nu}^n  (\vec{q}) = \frac{2 g^2 N (\vec{q}^{\; 2})^{\epsilon}}{(4 \pi)^{2+\epsilon}} \left( \frac{2 \Gamma (\epsilon) \Gamma(\frac{1}{2} + \frac{n}{2} - i \nu-\epsilon) \Gamma(\frac{1}{2} + \frac{n}{2} + i \nu+\epsilon)}{\Gamma(\frac{1}{2} + \frac{n}{2} - i \nu) \Gamma(\frac{1}{2} + \frac{n}{2} + i \nu + 2\epsilon)}- \frac{\Gamma (1-\epsilon) \Gamma^2 (\epsilon)}{\Gamma(2 \epsilon)} \right) \phi_{\nu}^{n} (\vec{q}) \; .
\end{equation*}
Expanding in $\epsilon$ and up to constant order, we get 
\begin{equation}
    \mathcal{K} \bullet \phi_{\nu}^n = \bar{\alpha}_s \chi(n, \nu) \equiv \omega_n (\nu) \; ,
\end{equation}
where
\begin{equation}
\label{Int:Eq:AlphaStrongBar}
    \bar{\alpha}_s = \frac{\alpha_s N}{\pi} 
\end{equation}
and
\begin{equation*}
    \chi(n, \nu) =  2 \psi (1) -\psi ( 1/2 + n/2 + i \nu)-\psi ( 1/2 + n/2 - i \nu)
\end{equation*}
\begin{equation}
\label{Int:Eq:CharacteristicFunction}
 = 2 (-\gamma_E - \Re [\psi((n+1)/2 + i\nu)]).  
\end{equation}
The function $\chi(n, \nu)$ in Eq.~(\ref{Int:Eq:CharacteristicFunction}) characterizes the behavior predicted by the BFKL approach in the leading-logarithmic approximation, it is a remarkable achievement of L. Lipatov and it is therefore known as \textit{Lipatov characteristic function}. \\
As anticipated, from the knowledge of the eigenvalues and the eigenfunctions, we can immediately write a spectral representation for the Green, $G_{\omega} \left( \vec{q}_1,\vec{q}_2 \right)$, which reads 
\begin{equation}
\label{sol}
G_{\omega} \left( \vec{q}_1,\vec{q}_2 \right) = \sum_n^{\infty} \int_{-\infty}^{\infty} d\nu \left( \frac{q_1^2}{q_2^2} \right)^{i \nu} \frac{e^{in(\theta_1-\theta_2)}}{2 \pi^2 q_1 q_2} \frac{1}{\omega-\bar{\alpha}_s \chi(n,\nu)} .
\end{equation}
Since $\nu$ is a continuous variable, we do not obtain an isolated pole which we can associate with the intercept of the Pomeron. One is interested in the leading $\ln s$ behaviour which means the singularity with the largest real part in the $\omega$-plane. This allows to make a number of simplifications. Since the function $\chi ( n, \nu)$ decreases with increasing $n$, it is possible to restrict the sum over $n$ in Eq.~(\ref{sol}) to the case where $n=0$. Furthermore, $\chi(0, \nu)$ decreases with increasing $|\nu|$, so one can expand $\chi(0, \nu)$ as a power series in $\nu$ and keep only the first two terms. One obtains
\begin{equation}
\chi(0, \nu)= 4\ln2 - 14 \zeta (3) \nu^2 +... \; .
\end{equation}
In this approximation 
\begin{equation}
\label{approxsol}
G_{\omega} \left( \vec{q}_1, \vec{q}_2 \right) \approx \frac{1}{\pi q_1 q_2} \int_{-\infty}^{\infty} \frac{d\nu}{2 \pi} \left( \frac{q_1^2}{q_2^2} \right)^{i \nu} \frac{1}{(\omega - \omega_0 + a^2 \nu^2)} ,
\end{equation}
with
\begin{equation}
\omega_0 = 4\bar{\alpha}_s \ln2 
\end{equation}
being the position of the leading singularity (the branch point of the cut) and 
\begin{equation}
a^2 = 14 \bar{\alpha}_s \zeta(3).
\end{equation} 
Eq.~(\ref{approxsol}) can be inverted, performing the anti-Mellin transform,
\begin{equation}
\tilde{G}_{s}(\vec{q}_1, \vec{q}_2) = \frac{1}{2 \pi i} \oint_{C} d\omega \left(\frac{s}{s_0}\right)^{\omega} \frac{1}{\pi q_1 q_2} \int_{-\infty}^{+\infty} \frac{d\nu}{2 \pi} \left( \frac{q_1^2}{q_2^2} \right)^{i \nu} \frac{1}{\omega-\omega_0+a^2 \nu^2},
\end{equation}
where the contour $C$ is to the right of the $\omega$-plane singularity of the integrand function. Using the residue theorem, one finds
\begin{equation}
\begin{split}
\tilde{G}_{s}(\vec{q}_1, \vec{q}_2)& = \frac{1}{2 \pi^2 q_1 q_2} \int_{-\infty}^{+\infty} d\nu \left( \frac{s}{s_0} \right)^{\omega_0 - a^2\nu^2} \left( \frac{q_1^2}{q_2^2} \right)^{i \nu} \\ & = \frac{1}{2 \pi^2 q_1 q_2} \int_{-\infty}^{+\infty} d\nu e^{\left( \omega_0 - a^2 \nu^2 \right) \ln\left( \frac{s}{s_0} \right)+i \nu \ln \left( \frac{q_1^2}{q_2^2} \right)} \; .
\end{split}
\end{equation}
An exponential factor, independent from $\nu$, can be taken out the integral, while, in the remaining term one can complete the square to obtain:
\begin{equation}
\begin{split}
\tilde{G}_{s}(\vec{q}_1, \vec{q}_2) = & \frac{1}{2 \pi^2 q_1 q_2} \left( \frac{s}{s_0} \right)^{\omega_0} \int_{-\infty}^{+\infty} d\nu \\ & \times \exp \left(-a^2 \ln\left( \frac{s}{s_0} \right) \nu^2 + i \ln \left(\frac{q_1}{q_2}\right)\nu + \frac{\ln^2\left(\frac{q_1^2}{q_2^2}\right)}{4a^2\ln\left(\frac{s}{\vec{q}_2^{\; 2}}\right)} - \frac{\ln^2\left(\frac{q_1^2}{q_2^2}\right)}{4a^2\ln\left(\frac{s}{\vec{q}_2^{\; 2}}\right)} \right) \\ & = \frac{1}{2 \pi^2 q_1 q_2} \left( \frac{s}{s_0} \right)^{\omega_0} \int_{-\infty}^{+\infty} d\nu \\ & \times \exp \left( - \frac{\ln^2\left(\frac{q_1^2}{q_2^2}\right)}{4a^2\ln\left(\frac{s}{\vec{q}_2^{\; 2}}\right)} \right) \exp \left( - \left(a \sqrt{\ln\left( \frac{s}{s_0} \right)} \nu - \frac{i \ln \left(\frac{q_1^2}{q_2^2} \right)}{2a \sqrt{\ln \left(\frac{s}{s_0}\right)}} \right)^2 \right) \; .
\end{split}
\end{equation}
Applying the substitution
\begin{equation}
z = \nu - i \frac{\ln \left( \frac{q_1^2}{q_2^2} \right)}{2a^2 \ln \left( \frac{s}{s_0} \right)} \; ,
\end{equation} 
and remembering the solution of the Gaussian integral, 
\begin{equation}
\int_{-\infty}^{\infty} dz e^{-Az^2} = \sqrt{\frac{\pi}{A}} \; ,
\end{equation}
the result is
\begin{equation}
\tilde{G}_{s}(\vec{q}_1, \vec{q}_2) \approx \frac{1}{\sqrt{\vec{q}_1^{\; 2} \vec{q}_2^{\; 2}}} \left( \frac{s}{s_0}\right)^{\omega_0} \frac{1}{\sqrt{\pi \ln(s/s_0)}} \frac{1}{2 \pi a} \exp \left(-\frac{\ln^2(\vec{q}_1^{\; 2}/\vec{q}_2^{\; 2})}{4a^2 \ln(s/s_0)} \right).
\end{equation}
The predicted $s$-behaviour (\ref{Int:Eq:AsyPomerSing}) of the cross sections appears from this expression\footnote{Note that the $s$-dependences appear only in this ``piece'' of the cross section.}, in fact, using the power series expansion of the exponential, one has:
\begin{equation}
\begin{split}
\tilde{G}_{s}(\vec{q}_1, \vec{q}_2) \approx & \frac{1}{\sqrt{\vec{q}_1^{\; 2} \vec{q}_2^{\; 2}}} \left( \frac{s}{s_0}\right)^{\omega_0} \frac{1}{\sqrt{\pi \ln(s/s_0)}} \\ & \frac{1}{2 \pi a} \left[1 + \left(-\frac{\ln^2(\vec{q}_1^{\; 2}/\vec{q}_2^{\; 2})}{4a^2 \ln(s/s_0)} \right) + \frac{1}{2} \left(-\frac{\ln^2(\vec{q}_1^{\; 2}/\vec{q}_2^{\; 2})}{4a^2 \ln(s/s_0)} \right)^2 + ... \right].
\end{split}
\end{equation}
$\alpha_s$ is fixed, increasing $s$, all terms of the expansion are suppressed with respect to one, and the leading contribution is
\begin{equation}
\tilde{G}_{s}(\vec{q}_1, \vec{q}_2) \approx \frac{1}{\sqrt{\vec{q}_1^{\; 2} \vec{q}_2^{\; 2}}} \left( \frac{s}{s_0}\right)^{\omega_0} \frac{1}{\sqrt{\pi \ln(s/s_0)}} \frac{1}{2 \pi a},
\end{equation}
that shows exactly the behaviour in Eq.~(\ref{Int:Eq:AsyPomerSing}).  
\subsection{BFKL equation in the NLLA}
In the previous sections, we built the BFKL approach in the LLA; the goal of this section, is instead to describe the approach in the NLLA.
In this approximation we can use an approach which coincides in the main features with that used in the LLA \cite{Fadin:1989kf}. In general, the program of the calculations is analogous to that in LLA. The final goal is the elastic scattering amplitude, which has to be restored from its $s$- and $u$-channel imaginary parts. The $s$-channel imaginary part is given by the unitarity relation (\ref{Unitary}). It can be shown that, Eqs.~(\ref{Inv1})-(\ref{Int:Eq:PartialWaveExpFull}), expressing the elastic scattering amplitudes in terms of their $s$-channel imaginary parts, remain unchanged. \\
Although conceptually the approach does not differ much from what has been described previously, the technical effort to complete the calculation program is enormous. For this reason, in the following, we will limit ourselves to a simple introductory description, referring to \cite{Fadin:1998sh} (and reference therein) for all the details.
\subsubsection{Gluon Reggeization in the NLLA}
\begin{figure}
\begin{picture}(400,280)
\put(30,227){ $\bullet$ \; $\omega^{(1)} (t) \; \longrightarrow \; \omega^{(2)} (t)$}
\put(230,200){\includegraphics[width=0.05\textwidth]{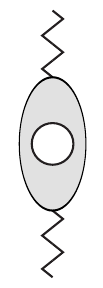}}
\put(255,247){1-loop}
\put(375,247){2-loop}
\put(30,137){ $\bullet$ $ \Gamma_{P'P}^{c (0)} \; \longrightarrow \; \Gamma_{P'P}^{c (1)} $}
\put(300,227){$\longrightarrow$}
\put(350,200){\includegraphics[width=0.05\textwidth]{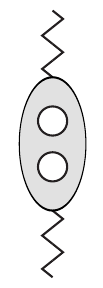}}
\put(210,120){\includegraphics[width=0.15\textwidth]{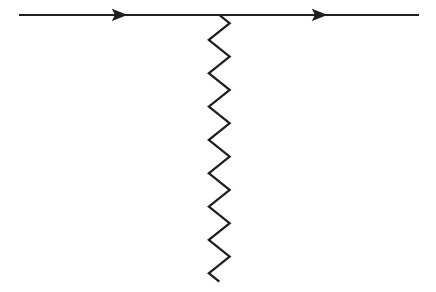}}
\put(260,148){Born}
\put(300,143){$\longrightarrow$}
\put(330,120){\includegraphics[width=0.15\textwidth]{images/PPRVertex.pdf}}
\put(380,148){1-loop}
\put(30,57){ $\bullet$ $ \gamma_{c_i c_{i+1}}^{G_i (0)} \; \longrightarrow \; \gamma_{c_i c_{i+1}}^{G_i (1)}  $}
\put(235,30){\includegraphics[width=0.1\textwidth]{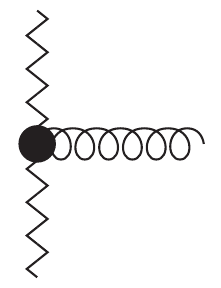}}
\put(255,77){Born}
\put(300,57){$\longrightarrow$}
\put(355,30){\includegraphics[width=0.1\textwidth]{images/LipatovVertex.pdf}}
\put(375,77){1-loop}
\end{picture}
\caption{Next-to-leading corrections to Regge trajectory, PPR vertices, and central gluon production vertex in the multi-Regge kinematics.}
\label{MRK}
\end{figure}
In the NLLA, Eq.~(\ref{Int:Eq:ReggeizedAmp}) has been proved in the first three orders of perturbation theory and assumed to be valid to all orders. Only later, it has been proved to all orders of perturbation theory. Demonstrating Reggeization in the NLLA is again a very challenging topic, we refer to \cite{Ioffe:2010zz} (and reference therein) for full details on this topic. \\
Instead, we want to focus on how the Regge trajectory can be extracted in this approximation. To this scope, Let us consider again the quark-quark scattering and let us expand the Eq.~(\ref{Int:Eq:ReggeizedAmp}) up to $\alpha_s^3$:
\begin{equation*}
    \Gamma^a_{1'1} \frac{s}{t} \left[ \left(\frac{s}{-t}\right)^{\omega(t)} + \left( \frac{-s}{-t} \right)^{\omega(t)} \right] \Gamma^a_{2'2} \simeq  \bigg \{ \Gamma^{a(0)}_{1'1} \frac{2s}{t} \Gamma^{a(0)}_{2'2} \bigg \}_{\color{red} _{LLA}} 
\end{equation*}
\begin{equation*}
  + \bigg \{ \Gamma^{a(0)}_{1'1} \frac{s}{t} \left[ \omega^{(1)}(t) \ln \left(\frac{s}{-t}\right) + \omega^{(1)}(t) \ln \left( \frac{-s}{-t} \right) \right] \Gamma^{a(0)}_{2'2} \bigg \}_{\color{red} _{LLA}} \hspace{-0.4 cm} + \hspace{0.1 cm} \bigg \{ \Gamma^{a(1)}_{1'1} \frac{2s}{t} \Gamma^{a(0)}_{2'2} + \Gamma^{a(0)}_{1'1} \frac{2s}{t} \Gamma^{a(1)}_{2'2} \bigg \}_{\color{blue} _{NLLA}}  
\end{equation*}
\begin{equation*}
    + \bigg \{ \Gamma^{a(0)}_{1'1} \frac{s}{t} \left[ \frac{(\omega^{(1)}(t))^2}{2} \ln^2 \left(\frac{s}{-t}\right) + \frac{(\omega^{(1)}(t))^2}{2} \ln^2 \left( \frac{-s}{-t} \right) \right] \Gamma^{a(0)}_{2'2} \bigg \}_{\color{red} _{LLA}}
\end{equation*}
\begin{equation*}
   +\bigg \{ \Gamma^{a(1)}_{1'1} \frac{s}{t} \omega^{(1)}(t) \left[ \ln \left(\frac{s}{-t}\right) + \ln \left( \frac{-s}{-t} \right) \right] \Gamma^{a(0)}_{2'2} + \Gamma^{a(0)}_{1'1} \frac{s}{t} \omega^{(1)}(t) \left[  \ln \left(\frac{s}{-t}\right) +  \ln \left( \frac{-s}{-t} \right) \right] \Gamma^{a(1)}_{2'2}  
\end{equation*}
\begin{equation}
   + \Gamma^{a(0)}_{1'1} \frac{s}{t} \left[ \omega^{(2)}(t) \ln \left(\frac{s}{-t}\right) + \omega^{(2)}(t) \ln \left( \frac{-s}{-t} \right) \right] \Gamma^{a(0)}_{2'2} \bigg \}_{\color{blue} _{NLLA}} \hspace{-0.4 cm} + \hspace{0.1 cm} \bigg \{ \Gamma^{a(2)}_{1'1} \frac{2s}{t} \Gamma^{a(0)}_{2'2} + ... \bigg \}_{\color{purple} _{NNLLA}} \; . 
\label{Int:Eq:ReggeExpAlpha3}
\end{equation}
In Eq.~(\ref{Int:Eq:ReggeExpAlpha3}) we have restored the superscript (i) to denote the $i$-loops correction to a certain quantity and we have introduced a subscript to clarify within which approximation a given term contributes. Analyzing the Eq.~(\ref{Int:Eq:ReggeExpAlpha3}), we have that
\begin{itemize}
    \item In the first line we have the only contribution of order $\alpha_s$, which is proportional to the Born effective vertices and obviously contributes within the LLA.
    \item In the second line we have three terms of order $\alpha_s^2$. The first is proportional to the $1$-loop Regge trajectory and contains a logarithm of the energy which compensates the additional $\alpha_s$ factor, hence it contributes within the LLA. The second and the third are both proportional to the $1$-loop correction to one of the two effective vertices; compared to the Born contribution they have an extra $\alpha_s$ power and no large energy logarithm to compensate; they contribute within the NLLA. 
    \item In the remaining lines we have a collection of terms of order $\alpha_s^3$. The first is proportional to the squared of the $1$-loop Regge trajectory and contains two logarithms of the energy which compensate the additional $\alpha_s^2$ factor, hence it contributes within the LLA. The second and the third contain the $1$-loop Regge trajectory, the $1$-loop correction of one of the two effective vertices and a logarithm of the energy which compensates one $\alpha_s$ factor; they contribute within the NLLA. The fourth term is proportional to the so called $2$-loop correction to the Regge trajectory and contains a large logarithm of the energy; it contributes within NLLA. Finally, we have terms proportional to the product of $1$-loop corrections to the effective vertices or terms proportional to the $2$-loops correction to one of the two effective vertices. They contribute within the NNLLA.
\end{itemize}
At this point, it is clear how to proceed. First, we have to go back to our $1$-loop computation of the quark-quark scattering and we have to include also those corrections that we have discarded in LLA since they do not generate large logarithms of energy. Among these we have contributions from the box and cross diagrams in Fig.~\ref{Int:Fig:LOqqqq} which come from regions of the phase space which we have ignored previously, but also vertex corrections which, in Feynman gauge, do not produce large logarithms of the energy. \\
By comparing this result with the correction of order $\alpha_s$ in Eq.~(\ref{Int:Eq:ReggeExpAlpha3}), we can extract the effective vertices up to $1$-loop accuracy. In general, they assume the following form\footnote{As always, we are thinking in terms of the quark-quark scattering amplitude, but nothing prevents us from considering the case of quark-gluon or gluon-gluon. The agreement of results extracted from different amplitudes is a further confirmation of the Reggeization.}
\begin{equation}
\label{PPRNLLA}
\Gamma_{P'P} = \delta_{\lambda_P, \lambda_{P'}} \Gamma_{P'P}^{(+)} + \delta_{\lambda_P, -\lambda_{P'}} \Gamma_{P'P}^{(-)} \; ,
\end{equation}  
where in the first term of Eq.~(\ref{PPRNLLA}), as before, the helicity is conserved, but there is a second term in which the helicity is not conserved~\cite{Fadin:1992zt,Fadin:1993rc,Fadin:1993qb,Fadin:1994fj,Fadin:1995km}. \\
Once these vertex corrections are found, the only ingredient to extract is $\omega^{(2)} (t)$. To do this, we need to compute the $2$-loop correction to the quark-quark scattering amplitude, in the high-energy limit, and select those contributions that give rise to energy logarithms and compare it with Eq.~(\ref{Int:Eq:ReggeExpAlpha3}). The result can be found in \cite{Fadin:1998sh}, it was derived in \cite{Fadin:1995km,Fadin:1994uz,Kotsky:1996xm,Fadin:1995xg,Fadin:1996tb}. 
\subsubsection*{Multi-Regge kinematics}
In the MRK and NLLA, as well as in LLA, only the amplitudes with the gluon quantum numbers in the channels with momentum transfers $q_i$ do contribute. The appearance in the right hand side of Eq.~(\ref{Unitary}) of amplitudes with quantum numbers in $t_i$-channels different from the gluon ones leads to loss of at least two large logarithms and therefore can be ignored in NLLA. As before, the key point in the calculation of the amplitudes contributing in the unitarity relation (\ref{Unitary}) is the gluon Reggeization. In MRK the real parts of the contributing amplitudes (only these parts are relevant in NLLA because the LLA amplitudes are real) are presented in the same form (\ref{Int:Eq:PartialWaveExpFull}) as in LLA. \\
In this kinematics, we need the two-loop contribution $\omega^{(2)}(t)$ to the gluon Regge trajectory $\omega(t)$ and the corrections to the real parts of the PPR-vertices that we discussed previously. Moreover, we need also to compute the one-loop corrections to the Lipatov vertex~\cite{Fadin:1993rc,Fadin:2000yp}. The calculation program for these corrections is schematized in Fig.~\ref{MRK}. 
\subsubsection*{Quasi-multi-Regge kinematics}
Probably, the most interesting feature of the NLLA is the appearance of a new kinematics, in fact, in the NLLA, MRK is not the only kinematics that contributes to the unitarity relation (\ref{Unitary}). Since we have the possibility to ``lose" one large logarithm (in comparison with LLA), the limitation of the strong ordering (\ref{Int:Eq:StrongOrderingN+2}) in the rapidity space cannot be implied more. Any (but only one) pair of the produced particles can have a fixed (not increasing with $s$) invariant mass, \textit{i.e.} components of this pair can have rapidities of the same order. This kinematics was called quasi-multi-Regge kinematics (QMRK).
\begin{figure}
\begin{picture}(400,260)
\put(30,217){ $\bullet$ $ \Gamma_{P'P}^{c (0)} \; \longrightarrow \; \Gamma_{ \{f \} P}^{c (0)} $}
\put(210,200){\includegraphics[width=0.15\textwidth]{images/PPRVertex.pdf}}
\put(30,137){ $\bullet$ $ \gamma_{c_i c_{i+1}}^{G_i (0)} \; \longrightarrow \; \gamma_{c_i c_{i+1}}^{G G (0)}  $}
\put(300,227){$\longrightarrow$}
\put(330,200){\includegraphics[width=0.15\textwidth]{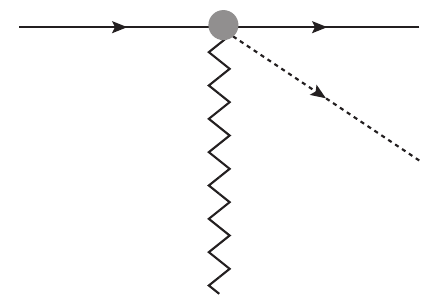}}
\put(235,120){\includegraphics[width=0.1\textwidth]{images/LipatovVertex.pdf}}
\put(300,143){$\longrightarrow$}
\put(355,120){\includegraphics[width=0.1\textwidth]{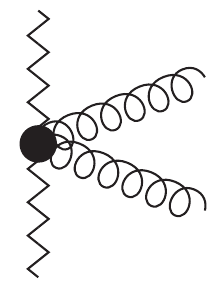}}
\put(30,57){ $\bullet$ $ \gamma_{c_i c_{i+1}}^{G_i (0)} \; \longrightarrow \; \gamma_{c_i c_{i+1}}^{Q Q (0)}  $}
\put(235,30){\includegraphics[width=0.1\textwidth]{images/LipatovVertex.pdf}}
\put(300,57){$\longrightarrow$}
\put(355,30){\includegraphics[width=0.1\textwidth]{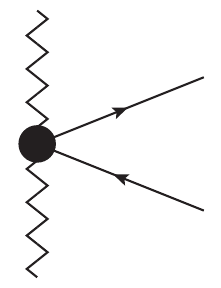}}
\end{picture}
\caption{Next-to-leading corrections to the vertices PPR and RRG in the quasi-multi-Regge kinematics.}
\label{QMRK}
\end{figure}
We can deal with this kinematics by including, together with the production of a gluon, the production of more complicated states in Reggeon-Reggeon (RR) and Reggeon-particle (RP) collisions, which were neglected in the LLA. In particular, we must
\begin{itemize}
    \item Include the possibility of producing gluon-gluon (GG)~\cite{Fadin:1989kf,Fadin:1996nw,Fadin:1996zv} and quark-antiquark ($\bar{Q}Q$)~\cite{Catani:1990xk,Catani:1990eg,Camici:1996st,Camici:1997ta,Fadin:1997hr} states in Reggeon-Reggeon collisions.  
    \item Include the possibility of producing states with larger number of particles in the Reggeon-particle collisions in the fragmentation region of one of the initial particles.
\end{itemize}
The calculation program for these corrections is schematized in Fig.~\ref{QMRK}. 
\subsubsection{BFKL kernel and impact factors in the NLLA}
As already mentioned above, the partial wave (\ref{Int:Eq:MellinTransf}) can be presented in the same form (\ref{Int:Eq:PartialwaveexpCom}), but with modified impact factors and Green function. This fact is very important because it confirms that also in NLLA the amplitudes can be factorized as in Fig.~\ref{convolution}. \\
\begin{figure}
\begin{picture}(400,110)
\put(30,0){\includegraphics[scale=0.5]{images/ReakKern.pdf}}
\put(55,85){$ \mathcal{K}_{RRG}^{(1)} $}
\put(115,0){\includegraphics[scale=0.5]{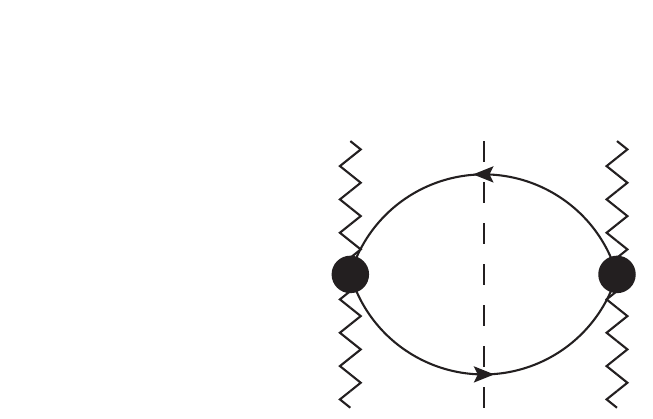}}
\put(212,85){$ \mathcal{K}_{RRQQ}^{(0)} $}
\put(270,0){\includegraphics[scale=0.5]{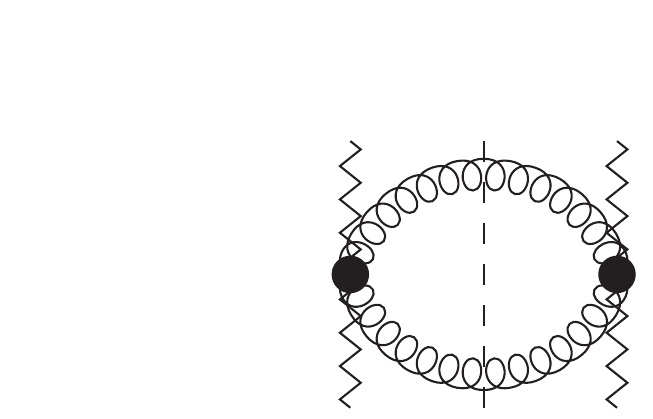}}
\put(367,85){$ \mathcal{K}_{RRGG}^{(0)} $}
\end{picture}
\caption{Schematic representation of the real part of the NLO BFKL kernel. In the first contribution, one Lipatov vertex is to be taken at one-loop.}
\label{Int:Fig:RealKernelcorrectionsNLO}
\end{figure}
In the definition of the impact factors (\ref{Int:Eq:ImpactFactorsGen}) we have to include radiative corrections to the PPR vertices and the contribution of the ``excited" states in the fragmentation region. Eq. (\ref{BFKLequation}) for the Green function remains unchanged, as well as the representation (\ref{kernel}) of the kernel, but the gluon trajectory has to be taken in the two-loop approximation: 
\begin{equation}
\omega(t)= \omega^{(1)}(t)+ \omega^{(2)}(t),
\end{equation}
and the part, related with the real particle production, must contain, together with the contribution from the one-gluon production in the RR collisions (calculated to one-loop level), contributions from the two-gluon and quark-antiquark productions. The one-gluon contribution must be calculated with the one-loop accuracy, whereas the two-gluon and quark-antiquark contributions have to be taken in the Born approximation.
For the forward scattering case the real production takes the form (see Fig.~\ref{Int:Fig:RealKernelcorrectionsNLO})
\begin{equation}
\mathcal{K}_r \left( \vec{q}_1, \vec{q}_2 \right) = \mathcal{K}_{RRG}^{(1)} \left( \vec{q}_1, \vec{q}_2 \right) + \mathcal{K}_{RRGG}^{(0)} \left( \vec{q}_1, \vec{q}_2 \right) + \mathcal{K}_{RRQ\bar{Q}}^{(0)} \left( \vec{q}_1, \vec{q}_2 \right).
\end{equation}
There is an important point to stress, the appearance of a new kinematics introduces technical problems. Calculating the two-gluon production contribution to the kernel and the contribution to the impact factor from the gluon production in the fragmentation region, we meet divergencies of the integrals over invariant masses of the produced particles at upper limits\footnote{In the next chapter we will refer to these as \textit{rapidity divergences}.}. The reason for the divergencies is the absence of a natural bound between MRK and QMRK. In order to give a precise meaning to the corresponding contributions and to treat them carefully, an artificial bound, which we denote as $s_{\Lambda}$, is introduced~\cite{Fadin:1998fv}. Obviously, the dependence on this artificial parameter disappears in final results. \\
As a consequence of what has just been explained, the definition of impact factor at the next-to-leading order, in the singlet case, reads
\begin{equation*}
\Phi_{AA}(\vec q_1; s_0) = \left( \frac{s_0}
{\vec q_1^{\:2}} \right)^{\omega( - \vec q_1^{\:2})}
\sum_{\{f\}}\int\theta(s_{\Lambda} -
s_{AR})\frac{ds_{AR}}{2\pi}\ d\rho_f \ \Gamma_{\{f\}A}^c
\left( \Gamma_{\{f\}A}^{c^{\prime}} \right)^* 
\langle cc^{\prime} | \hat{\cal P}_0 | 0 \rangle
\end{equation*}
\begin{equation}
-\frac{1}{2}\int d^{D-2}q_2\ \frac{\vec q_1^{\:2}}{\vec q_2^{\:2}}
\: \Phi_{AA}^{(0)}(\vec q_2)
\: {\cal K}^{(0)}_r (\vec q_2, \vec q_1)\:\ln\left(\frac{s_{\Lambda}^2}
{s_0(\vec q_2 - \vec q_1)^2} \right)~.
\label{Int:Eq:ImpactproNext}
\end{equation}
The first term in Eq.~(\ref{Int:Eq:ImpactproNext}) is very similiar to that in Eq.~(\ref{Int:Eq:ImpactFactSingGen}); the important difference with respect to the leading-order definition is the Heaviside theta that creates the separation between MRK and QMRK and automatically regularizes the aforementioned rapidity divergence ($s_{AR}$ goes to infinity). In this scheme, a contribution in which there is the emission of an additional gluon is included in the ``QMRK part" of the impact factor only if $s_{AR} < s_{\Lambda}$, \textit{i.e.} only if the particle (or the system of particle) that we had at the leading order and the additional emitted gluon are not strongly separated in rapidity. \\
The second term comes instead from the MRK contribution to the partial wave amplitude Eq.~(\ref{Int:Eq:PartialWaveN}) in the NLLA. Consistently with how we define the QMRK kinematics using the cut-off $s_{\Lambda}$, we must now impose that the integration over invariant masses $s_{i}$ in MRK kinematics is performed by imposing $s_i > s_{\Lambda}$. 
When we compute in the LLA, the lower bound of integration over invariant masses produces negligible contributions within this approximation. Within the NLLA, terms from the lower bound of integration must be included. In the specific case of the invariant mass $s_{AR}$, the produced term is exactly the one in the second line of Eq.~(\ref{Int:Eq:ImpactproNext}). \\
The $s_{\Lambda}$ parameter introduced must be understood as tending to infinity. Both expressions in the first and second line of Eq.~(\ref{Int:Eq:ImpactproNext}) are divergent, but in their combination any dependence from $s_{\Lambda}$ disappears and the result is finite. As it is easy to understand, also the contribution to the kernel from the emission of two gluons has a term similar to the second in Eq.~(\ref{Int:Eq:ImpactproNext}). For the complete calculation of the separation between MRK and QMRK see \cite{Fadin:1998sh}.

\chapter{Next-leading order forward Higgs impact factor}
\begin{flushright}
\emph{We have made the discovery of a new particle, a completely new particle, \\ which is most probably very different from all the other particles. \\ Rolf-Dieter Heuer}
\end{flushright}
\label{Chap:HiggsImp}
Precision physics in the Higgs sector has been one of the main challenges in recent years. The pure fixed-order calculations entering the collinear factorization framework, which have been pushed up to N$^3$LO, are not able to describe the entire kinematic spectrum. In particular conditions, they must be necessarily supplemented by all-order resummations. In the Regge kinematical region, large energy-type logarithms spoil the perturbative behavior of the series and must be resummed to all orders. The BFKL approach, that we have presented in the previous chapter, allows us to achive this resummation, in the NLLA, but, as already said, it requires the knowledge of impact factors. In this chapter, we calculate the next-to-leading order correction to the impact factor (vertex) for the production of a forward Higgs boson, in the infinite top-mass limit. We obtain the result both in the momentum representation and as superposition of the eigenfunctions of the leading-order BFKL kernel. As already mentioned earlier, this impact factor allows to describe the inclusive hadroproduction of a forward Higgs in the limit of small Bjorken $x$, as well as for the more interesting study of the inclusive forward emissions of a Higgs boson in association with a backward identified object. \\

The chapter contains five sections. In the first section, we introduce the necessary ingredients and the program of computations. In the second and third section, we calculate real and virtual corrections, respectively. In the fourth, we perform the projection onto the eigenfunctions of the LO BFKL kernel and we show the cancellation of divergences. Lastly, in the fifth section we summarize and discuss future perspectives. The material in this chapter is based on Ref.~\cite{Celiberto:2022fgx,Fucilla:2022whr}.

\section{LO and NLO in a nutshell}

\subsection{The leading-order impact factor for forward Higgs production}
\label{ssec:LO_IF}

\begin{figure}
\begin{center}
\includegraphics[scale=0.50]{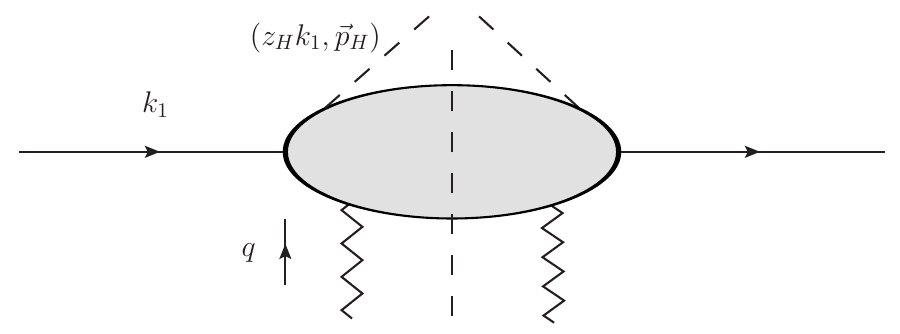}
\end{center}
\caption{Schematic description of the Higgs impact factor: $z_H$ is the fraction of longitudinal momentum of the initial parton carried by the outgoing Higgs and $\vec{p}_H$ is its transverse momentum.}
\label{HiggsImpactFactorRepres}
\end{figure}
Since we are interested in the description of the inclusive production of
a Higgs in proton-proton collisions, say in the forward region, the setup sketched 
in the previous chapter must be suitably modified. First of all, we need to consider 
processes in which the parton $A$ can be a gluon or any of the active quarks and 
must then take the convolution of the partonic impact factor with the corresponding PDF. Secondly, the integration over the intermediate states $\{f\}$ in~(\ref{Int:Eq:ImpactproNext}) must exclude the state of the Higgs, so that the resulting impact factor will be differential in the Higgs kinematics (see Fig.~\ref{HiggsImpactFactorRepres}).

Let us illustrate the procedure in the simplest case of the leading-order (LO) contribution.
In the infinite top-mass approximation, the Higgs field couples to QCD {\it via} the 
effective Lagrangian~\cite{Ellis:1975ap,Shifman:1979eb},
\begin{equation}
\mathcal{L}_{ggH} = - \frac{g_H}{4} F_{\mu \nu}^{a} F^{\mu \nu,a} H \; ,
\label{EffLagrangia}
\end{equation}
where $H$ is the Higgs field, $F_{\mu \nu}^a = \partial_{\mu} A_{\nu}^a - \partial_{\nu} A_{\mu}^a  + g f^{abc} A_{\mu}^b A_{\nu}^c$ is the field strength tensor, 
\begin{equation}
g_H = \frac{\alpha_s}{3 \pi v} \left( 1 + \frac{11}{4} \frac{\alpha_s}{\pi} \right) + {\cal O} (\alpha_s^3)
\label{gH}
\end{equation}
is the effective coupling~\cite{Dawson:1990zj,Ravindran:2002} and $v^2 = 1/(G_F \sqrt{2})$ with $G_F$ the Fermi constant. Feynman rules deriving from this theory can be found in Appendix~\ref{AppendixA1}.
We introduce the Sudakov decomposition for the Higgs and Reggeon momenta:
\begin{equation}
  p_H = z_H k_1 + \frac{m_H^2+ \vec{p}_H^{\; 2}}{z_H s} k_2 + p_{H \perp}\; , \hspace{0.5 cm} p_H^2 = m_H^2 \; ,
 \end{equation}
 \begin{equation}
  q = - \alpha_q k_2 + q_{\perp} \; , \hspace{0.5 cm} q^2 = -\vec{q}^{\; 2} \; .
  \end{equation} 
At LO, the impact factor takes contribution only from the case when the initial-state parton (particle $A$) is a gluon, for which $k_A=k_1$. For the polarization vector of all gluons in the external lines involved in the calculation we will use the light-cone gauge,
\begin{equation}
\varepsilon (k) \cdot k_2 = 0 \;,
\label{ExternalGluonGauge}
\end{equation}
which leads to the following Sudakov decomposition:
\begin{equation}
\varepsilon(k)=-\frac{(k_\perp \cdot \varepsilon_\perp(k))}{(k\cdot k_2)}k_2+\varepsilon_\perp(k)\;,
\end{equation}
so that for the initial-state gluon we have $\varepsilon (k_1) = \varepsilon_{\perp}(k_1)$.
The transverse polarization vectors have the properties
$$
\left( \varepsilon_{\perp}^*(k_1, \lambda_1) \cdot \varepsilon_{\perp}(k_1, \lambda_2) \right) =
\left( \varepsilon^*(k_1, \lambda_1) \cdot \varepsilon(k_1, \lambda_2) \right) = - \delta_{\lambda_1,\,
\lambda_2},
$$
\begin{equation}\label{217}
\sum_{\lambda} \varepsilon_{\perp}^{*\mu}(k, \lambda) \varepsilon_{\perp}^{\nu}(k, \lambda) =
- g_{\perp\perp}^{\mu\nu}~,
\end{equation}
where the index $\lambda$ enumerates the independent polarizations of gluon (sometimes it will be omitted for simplicity, but it should be always understood),
$g^{\mu\nu}$ is the metric tensor in the full space and 
$g_{\perp\perp}^{\mu\nu}$ the one in the transverse subspace, 
\begin{equation}
g_{\perp\perp}^{\mu\nu} = g^{\mu\nu} - 
\frac{k_1^{\mu} k_2^{\nu} + k_2^{\mu}k_1^{\nu}}
{(k_1 \cdot k_2)}~.
\label{metricTensorDec}
\end{equation}

\begin{figure}
\begin{center}
\includegraphics[scale=0.50]{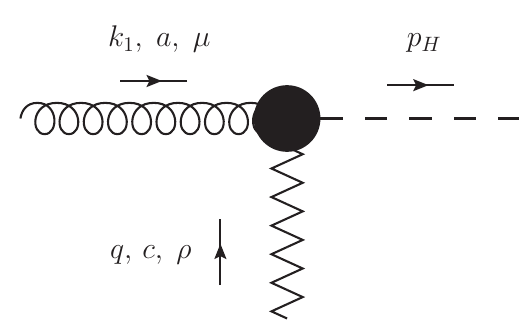} \hspace{1.5 cm} \includegraphics[scale=0.50]{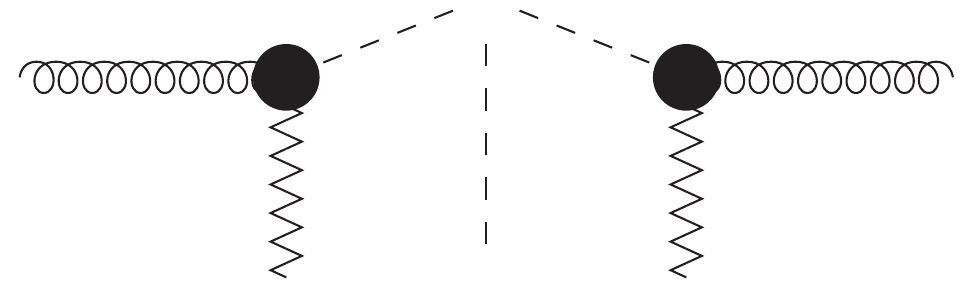}
\end{center}
\caption{Born gluon-Reggeon-Higgs effective vertex (left) and schematic description of the single contribution to the impact factor LO (right). We draw the Higgs boson above the cut to emphasize that we do not integrate over its kinematic variables.}
\label{BornVertex}
\end{figure}

The LO impact factor before differentiation in the kinematic variables of the Higgs
is given by 
\begin{equation}
    \Phi_{gg}^{\{ H \}(0)} (\vec{q} \; ) = \frac{\langle
cc^{\prime} | \hat{\cal P}_{0} | 0 \rangle}{2(N^2-1)} \sum_{a, \lambda } \int \frac{d s_{gR}}{2 \pi} d \rho_H  \Gamma_{\{ H \} g}^{ac(0)} (q) \left( \Gamma_{\{ H \} g}^{ac'(0)} (q) \right)^{*} \; ,
\label{Higgsssec:LO_IF1}
\end{equation}
where the only state contributing to the intermediate state $\{f\}$ is the Higgs boson
and the particle $A$ is identified with an on-shell gluon. Here, 
\begin{equation}
    \Gamma_{\{ H \} g}^{ac(0)} (q) = \frac{g_H}{2} \delta^{ac} \ \bigl(q_{ \perp}\cdot  \varepsilon_{\perp} (k_1)\bigr) \; ,
    \label{gRHBorn}
\end{equation}
is the high-energy gluon-Reggeon-Higgs (gRH) Born vertex (see Fig.~\ref{BornVertex}). The
overall factor $1/2(N^2-1)$ comes from the average over the polarization and color states of the incoming gluon. The vertex in Eq.~(\ref{gRHBorn}) is obtained from the Higgs effective theory Feynman rules, taking for the Reggeon in the $t$-channel the ``nonsense'' polarization $(-k_2^{\rho}/s)$. We stress again that this effective polarization arises from the fact that, for gluons in the $t$-channel that connect strongly separated regions in rapidity, the Gribov trick,
\begin{equation}
g^{\rho \nu} = g_{\perp\perp}^{\rho \nu} +
2 \frac{k_1^{\rho} k_2^{\nu} + k_2^{\rho} k_1^{\nu}}{s} \longrightarrow 2 \frac{k_2^{\rho} k_1^{\nu}}{s} = 2 s \left(\frac{-k_2^{\rho}}{s} \right) \left( \frac{-k_1^{\nu}}{s} \right) ~ \; ,
\label{Gribov}
\end{equation}
can be used in the kinematic region relevant for the ``upper'' impact factor. Since the integration over the invariant mass $s_{PR}$ and over the phase space is completely trivial, we simply obtain
\begin{equation}
    \Phi_{gg}^{\{ H \}(0)} (\vec{q} \; ) = \frac{g_H^2}{8 \sqrt{N^2-1}} \vec{q}^{\; 2}\;.
\end{equation}
Hence, it is straightforward to construct the differential (in the kinematic variables of the Higgs) LO order impact factor. This is the only chapter of the thesis in which, to adapt our notation to Ref.~\cite{Celiberto:2022fgx}, we work in $D=4-2 \epsilon$ dimensions. The impact factor in this dimension reads\footnote{Not to burden our formulas, we write $\delta^{(2)}$ instead of $\delta^{(D-2)}$ in the R.H.S. and 
$d^2\vec p_H$ instead of $d^{D-2}\vec p_H$ in the L.H.S.; the latter convention will apply
also in the following. Moreover, if $g_H$ is meant to have dimension of an inverse mass, then in $D$-dimension it should be replaced by $g_H \mu^{-\epsilon}$, where $\mu$ is an arbitrary mass scale.} 
    \begin{equation}
    \frac{d\Phi_{gg}^{\{ H \}(0)} (\vec{q} \; )}{d z_H d^2 \vec{p}_H} = \frac{g_H^2}{8 (1-\epsilon) \sqrt{N^2-1}} \vec{q}^{\; 2} \delta (1-z_H) \delta^{(2)} (\vec{q}-\vec{p}_H) \; .
    \label{LOHiggsImp2}
\end{equation}
After convolution with the gluon PDF, we obtain the proton-initiated LO impact factor,
  \begin{equation}
    \frac{d \Phi_{PP}^{ \{ H \}(0)} (x_H, \vec{p}_H, \vec{q})}{d x_H d^2 \vec{p}_H} = \int_{x_H}^1 \frac{d z_H}{z_H} f_g \left( \frac{x_H}{z_H} \right) \frac{d \Phi_{gg}^{ \{H \}(0)} (z_H, \vec{p}_H, \vec{q})}{d z_H d^2 \vec{p}_H} = \frac{g_H^2 \vec{q}^{\; 2} f_g (x_H) \delta^{(2)} ( \vec{q} - \vec{p}_H) }{8 (1-\epsilon) \sqrt{N^2-1}} \; .
    \label{Factorization}
\end{equation}

\subsection{NLO computation in a nutshell}
\label{ssec:NLO_nutshell}

\begin{figure}
  \begin{center}
  \includegraphics[scale=0.40]{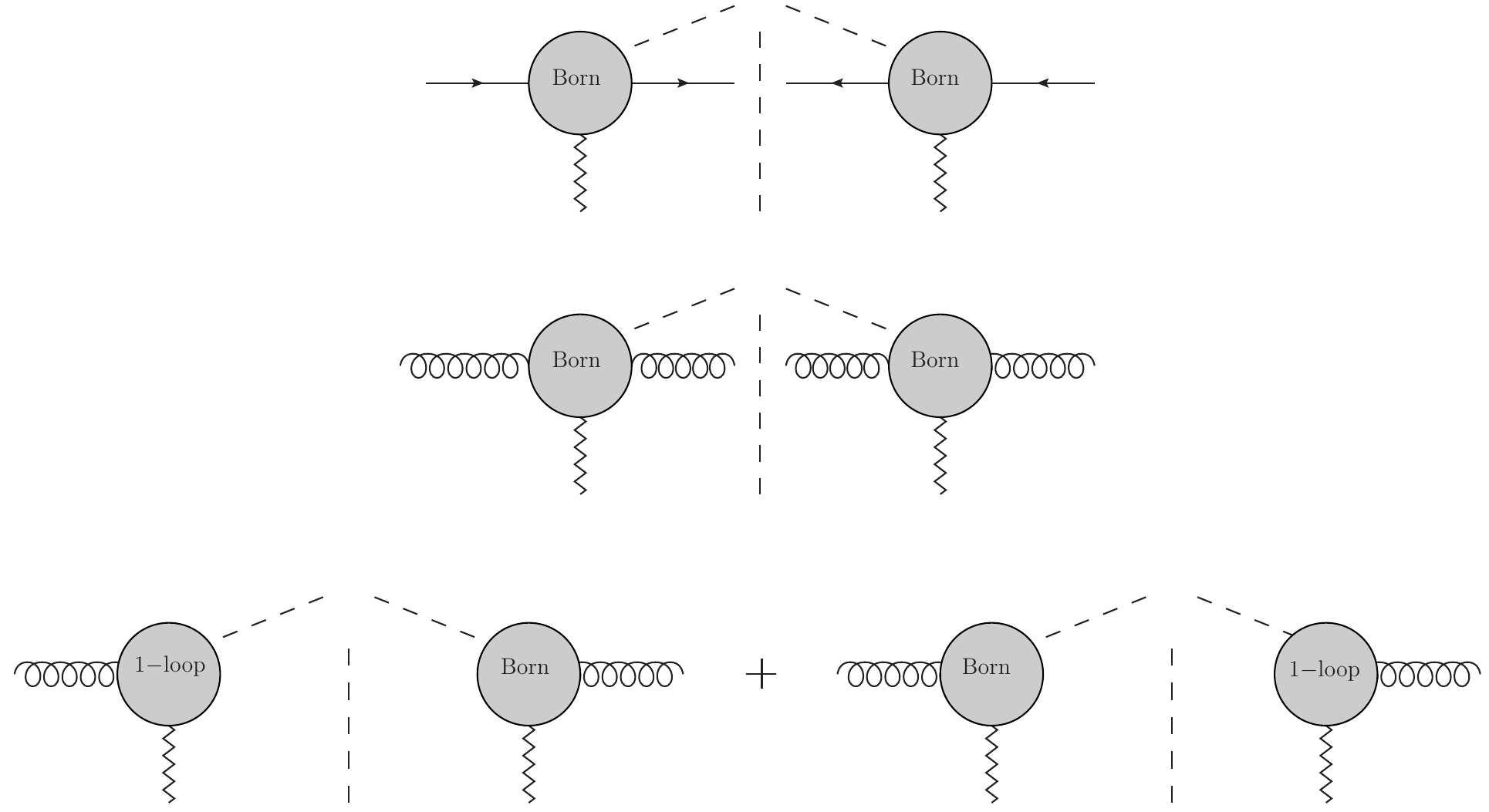}
  \end{center}
  \caption{Schematic description of the calculation of the impact factor at the NLO. In the first figure from above the hard process is initiated by a quark, and an additional quark is produced through the cut. Note that this quark ``crosses'' the cut to indicate that we will integrate on its kinematic variables.}
  \label{ProgramHiggsNLO}
\end{figure}
The program for calculating the NLO Higgs impact factor is the following: 
\begin{itemize}
    \item \textbf{Real corrections} \\
    At NLO the process can be initiated either by a gluon or by a quark extracted from the proton. Moreover, the Higgs must be accompanied by an additional parton (see the first two diagrams from the top of Fig.~\ref{ProgramHiggsNLO}). These corrections will be calculated in section~\ref{sec:real}, separating the quark-initiated case (subsection~\ref{ssec:quark}) and the gluon-initiated one (subsection~\ref{ssec:gluon}).
    \item \textbf{Virtual corrections} \\
    Another NLO order correction is obtained when we take one of the two effective vertices $\Gamma$ that appear in the definition of the impact factor at one loop and the other at Born level (see diagrams in the last line of Fig.~\ref{ProgramHiggsNLO}). This correction is trivial to compute once the 1-loop correction to the vertex~(\ref{gRHBorn}) has been extracted. This problem will be addressed in the section~\ref{sec:virtual}. 
    \item \textbf{Projection onto the eigenfunctions of the LO BFKL kernel and cancellation of divergences} \\
    In section~\ref{sec:projection} we perform the convolution with the PDFs for the previously calculated contributions and show that, after (i) carrying out the UV renormalization of the strong coupling, (ii) introducing the counterterms associated with the PDFs, (iii) performing the projection on the eigenfunctions of the LO BFKL kernel (more details in section~\ref{sec:projection}), the final result is free from any kind of divergence.
    
\end{itemize}

\section{NLO impact factor: Real corrections}
\label{sec:real}

In this section, we compute real corrections to the Higgs impact factor. At NLO both initial-state gluon and quark can contribute: if the process is initiated by a quark (gluon), then the intermediate state $\{f\}$ will contain the Higgs and a quark (gluon).
The Sudakov decomposition of the momentum of the produced parton reads
\begin{equation}
    p_p = z_p k_1 + \frac{\vec{p}_p^{\; 2}}{z_p s} k_2 + p_{p \perp} \; , \hspace{0.5 cm} p_p^2 = 0 \; , 
\end{equation}
where the subscript $p$ is equal to $q$ ($g$) in the quark (gluon) case. We have then
\begin{equation}
     s_{pR} = (p_p+p_H)^2 = \frac{z_p (z_H+z_p) m_H^2 + (z_p \vec{p}_H - z_H \vec{p}_p)^2}{z_H z_p} \; ,
     \label{SpR}
\end{equation}
\begin{equation}
    \frac{d s_{pR}}{2 \pi} d \rho_{pH} = \delta (1 - z_p - z_H) \delta^{(2)} (\vec{p}_p + \vec{p}_H - \vec{q} \; ) \frac{d z_p d z_H}{z_p z_H} \frac{d^{D-2} p_p d^{D-2} p_H}{2 (2 \pi)^{D-1}} \; .
\end{equation}
The integration over the kinematic variables of the produced parton is trivial due to the
presence of the delta functions and results in the constraints
\begin{equation}
    z_p = 1 - z_H \; , \hspace{0.6 cm} \vec{p}_p \equiv \vec{q} - \vec{p}_H \; .
\end{equation}
The integration over $z_H$ and $\vec p_H$ is not performed, since our target is an impact factor differential in the kinematic variables of the Higgs.

In the gluon case we will use the Sudakov decompositions of the polarization vectors
of the initial-state gluon with momentum $k_1$ and of the produced gluon with momentum $p_g$,
\begin{equation}
   \varepsilon (k_1) =  \varepsilon_{\perp} (k_1) \;, \hspace{0.6 cm}
   \varepsilon (p_g) = - \frac{(p_{g , \perp} \cdot \varepsilon_{\perp} (p_g))}{(p_g \cdot k_2)} k_2 + \varepsilon_{\perp} (p_g) \; ,
   \label{Pol12}
  \end{equation}
which encode the gauge condition $\varepsilon\cdot k_2=0$ for both gluons. 

We observe that, when calculating NLO real corrections, the $s_0$-dependent factor in the definition~(\ref{Int:Eq:ImpactproNext}) can be taken at the lowest order in the perturbative expansion and therefore put equal to one. Moreover, in the quark case there is no rapidity divergence for $s_{pR}\to \infty$, therefore the regulator $s_\Lambda$ can be sent to infinity in the argument of the theta function and the second term in Eq.~(\ref{Int:Eq:ImpactproNext}) (the so-called ``BFKL counterterm'') can be omitted.

\subsection{Quark-initiated contribution}
\label{ssec:quark}

\begin{figure}
\begin{center}
\includegraphics[scale=0.50]{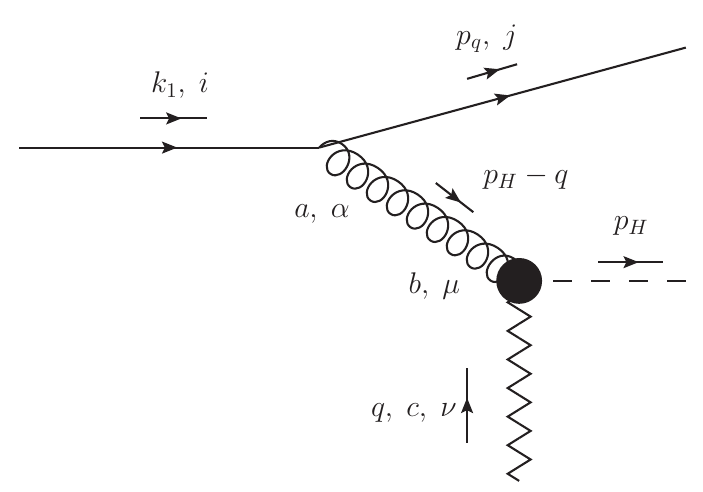}
\end{center}
\caption{Feynman diagram contributing to the $q \;R \rightarrow q \; H$ amplitude.}
\label{QuarkDiagr}
\end{figure}

In the case of incoming quark, the contribution to the impact factor reads
\begin{equation}
    d \Phi_{q q}^{\{H q \}} (\vec{q}) = \langle cc^{\prime} | \hat{\cal P}_{0} | 0 \rangle \frac{1}{2 N} \sum_{\substack{i, j \\ \lambda,\lambda'}}\int \frac{d s_{q R}}{2 \pi} d \rho_{\{H q \}} \Gamma_{\{ H q \} q}^{c(0)} (q) \left( \Gamma_{\{ H q \} q}^{c'(0)} (q) \right)^{*}  \; .
    \label{QuarkImpDef}
\end{equation}
where color and spin states have been averaged for the initial-state quark and summed over for 
the produced one. There is only one Feynman diagram to consider, shown in Fig.~\ref{QuarkDiagr}, leading to
\begin{equation}
  \Gamma_{\{ H q \} q}^{c(0)} = - g g_H t_{i j}^c \frac{1-z_H}{(\vec{q}-\vec{p}_H)^2} \bar{u} (p_q) \left( \frac{\slashed{k}_2}{s} (p_H-q) \cdot q - \slashed{q} \left( (p_H-q) \cdot \frac{k_2}{s} \right) \right) u(k_1) \; .
  \label{QuarkAmpBorn}
\end{equation}
Substituting Eq.~(\ref{QuarkAmpBorn}) into Eq.~(\ref{QuarkImpDef}), one easily gets the quark initiated contribution to the impact factor,
\begin{gather}
     \frac{d \Phi_{q q}^{\{H q \}} (z_H, \vec{p}_H, \vec{q})}{d z_H d^2 \vec{p}_H} = \frac{\sqrt{N^2-1}}{16 N (2 \pi)^{D-1}} \frac{g^2 g_H^2}{[(\vec{q}-\vec{p}_H)^2]^2} \nonumber \\
    \times \left[ \frac{4 (1-z_H) \left[(\vec{q}-\vec{p}_H) \cdot \vec{q} \; \right]^2 + z_H^2 \vec{q}^{\; 2} (\vec{q} - \vec{p}_H)^2}{z_H} \right] \; .
    \label{QuarkConImpacFin}
\end{gather}
We observe that this contribution to the impact factor vanishes in the limit $\vec q \to 0$, as required by gauge invariance.

\subsection{Gluon-initiated contribution}
\label{ssec:gluon}

\begin{figure}
  \begin{center}
  \includegraphics[scale=0.50]{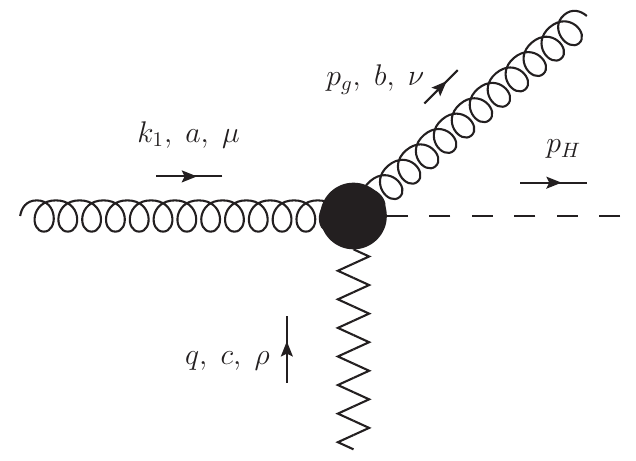} \hspace{1.5 cm}
  \includegraphics[scale=0.50]{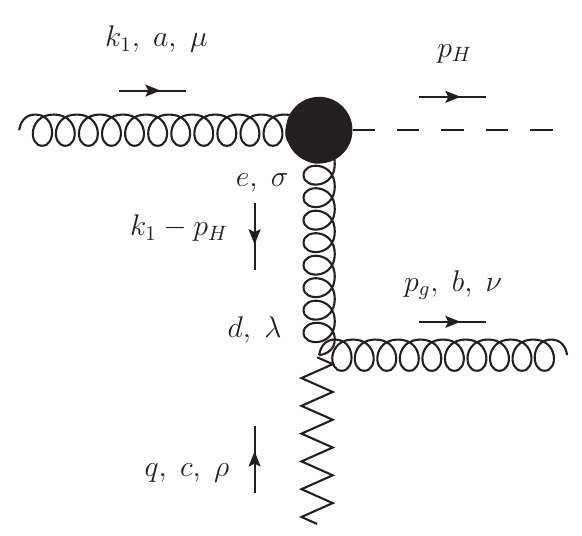} \vspace{1.0 cm} \\
  \includegraphics[scale=0.55]{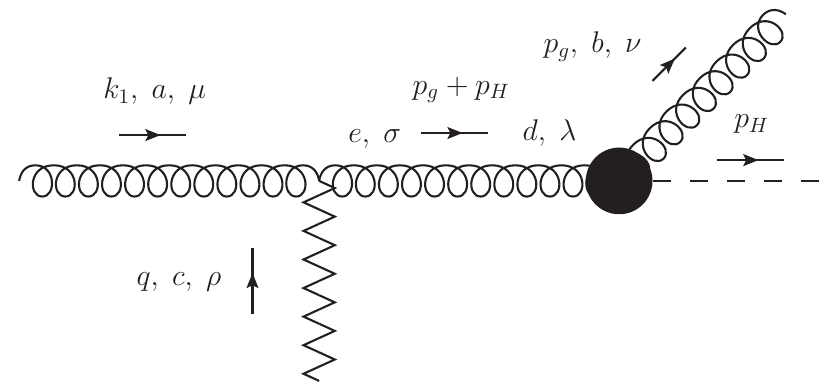} \hspace{0.3 cm}
  \includegraphics[scale=0.55]{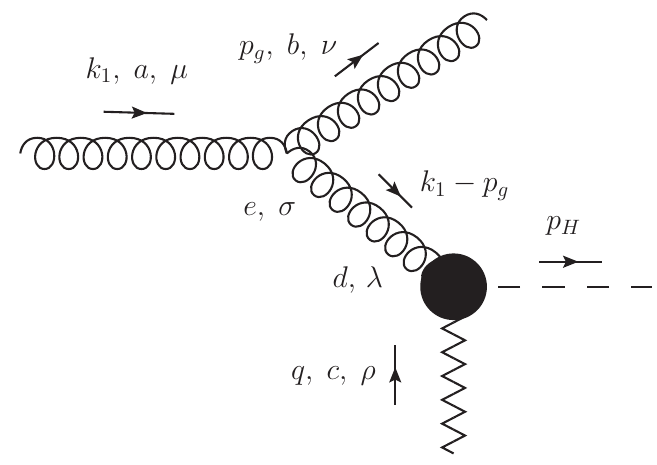}
  \end{center}
  \caption{Feynman diagrams contributing to the $g R \rightarrow g H$ amplitude. In the text we label them from (1) to (4) starting from the top left.}
  \label{GluonDiagrams}
\end{figure}

In the case of gluon in the initial state, the NLO real corrections, which are given, up to the BFKL counterterm, by the first term in Eq.~(\ref{Int:Eq:ImpactproNext}), take the following form
\begin{equation}
    d \Phi_{g g}^{\{H g \}} (\vec{q}) =  \frac{\braket{c c'|\mathcal{P}|0}}{2(1-\epsilon) (N^2-1)} \sum_{\substack{a, b \\ \lambda,\lambda'}}\int \frac{d s_{g R}}{2 \pi} d \rho_{\{H g \}}  \Gamma_{\{ H g \} g}^{a b c(0)} (q) \left( \Gamma_{\{ H g \} g}^{a b c' (0)} (q) \right)^{*} \theta (s_{\Lambda}-s_{gR})\; ,
    \label{GluonIniImp}
\end{equation}
where color and polarization states have been averaged for the initial-state gluon and summed over for the produced one.
The Feynman diagrams contributing to the amplitude $\Gamma_{\{ H g \} g}^{a b c(B)}$ of the $g R \rightarrow g H$ process are shown in Fig.~\ref{GluonDiagrams}, and give
\begin{equation}
\Gamma_1 = - i g g_H f^{abc} \varepsilon_{\mu} (k_1) \varepsilon_{\nu}^{*} (p_g)  \frac{2-z_H}{2}  g^{\mu \nu} \; ,
\end{equation}
\begin{equation}
\Gamma_2 = \frac{-i g g_H f^{abc} \varepsilon_{\mu} (k_1) \varepsilon_{\nu}^{*} (p_g) }{2[(z_H-1)m_H^2 - \vec{p}_H^{\; 2}]} \left[ (1-z_H) (m_H^2 + \vec{p}_H^{\; 2})g^{\mu \nu} - 2 z_H p_H^{\mu} (p_H^{\nu} - z_H k_1^{\nu}) \right] \; , 
\end{equation}
\begin{equation*}
\Gamma_3 = \frac{i g g_H f^{abc}\varepsilon_{\mu} (k_1) \varepsilon_{\nu}^{*} (p_g)}{2 \left[(1-z_H) m_H^2 + (\vec{p}_H-z_H \vec{q} \; )^2 \right]}
\end{equation*}
\begin{equation}
    \times \left[ g^{\mu \nu} \left( (1-z_H)^2 m_H^2 + (\vec{p}_H-z_H \vec{q} \; )^2 \right) +2 z_H (1-z_H)^2 p_H^{\mu} p_H^{\nu} -2 z_H^2 (1-z_H) p_g^{\mu} p_H^{\nu} )\right] \; ,
\end{equation}
\begin{equation}
\Gamma_4 = -i g g_H f^{abc} \varepsilon_{\mu} (k_1) \varepsilon_{\nu}^{*} (p_g) \frac{(1-z_H)}{(\vec{q}-\vec{p}_H)^2} \left[ - g^{\mu \nu} (\vec{q} - \vec{p}_H) \cdot \vec{q} - z_H (p_H^{\mu} k_1^{\nu} + p_g^{\mu} p_H^{\nu}) \right] \; ,
\end{equation}
which sum up to
\begin{equation*}
\Gamma_{\{ H g \} g}^{a b c(0)} = i g g_H f^{abc} \varepsilon_{\mu} (k_1) \varepsilon_{\nu}^{*} (p_g) \left[ \frac{\left[ (1-z_H) (m_H^2 + \vec{p}_H^{\; 2})g^{\mu \nu} - 2 z_H p_H^{\mu} (p_H^{\nu} - z_H k_1^{\nu}) \right] }{2[(1-z_H) m_H^2 + \vec{p}_H^{\; 2}]} \right.
\end{equation*}
\begin{equation*}
    + \left. \frac{\left[ g^{\mu \nu} \left( (1-z_H)^2 m_H^2 + (\vec{p}_H-z_H \vec{q} \; )^2 \right) +2 z_H (1-z_H)^2 p_H^{\mu} p_H^{\nu} -2 z_H^2 (1-z_H) p_g^{\mu} p_H^{\nu} )\right]}{2 \left[(1-z_H) m_H^2 + (\vec{p}_H-z_H \vec{q} \; )^2 \right]} \right. 
\end{equation*}
\begin{equation}
    \left. -\frac{2-z_H}{2}  g^{\mu \nu} + \frac{(1-z_H)\left[ g^{\mu \nu} (\vec{q} - \vec{p}_H) \cdot \vec{q} + z_H (p_H^{\mu} k_1^{\nu} + p_g^{\mu} p_H^{\nu}) \right]}{(\vec{q}-\vec{p}_H)^2}  \right] \; .
\end{equation}
Using Eqs.~(\ref{Pol12}), we can decouple longitudinal and transverse degrees of freedom and get
\begin{equation*}
\Gamma_{\{ H g \} g}^{a b c(0)} = -i g g_H f^{abc} \varepsilon_{\mu \perp} (k_1) \varepsilon_{\nu \perp}^{*} (p_g) \left[ \frac{2 z_H p_{H \perp}^{\mu} p_{H \perp}^{\nu} - z_H (1-z_H) m_H^2 g^{\mu \nu}}{2 \left[ (1-z_H) m_H^2 + \vec{p}_H^{\;2} \right]} \right.
\end{equation*}
\begin{equation}
    \left. - \frac{2 z_H \Delta_{\perp}^{\mu} \Delta_{\perp}^{\nu} - z_H (1-z_H) m_H^2 g^{\mu \nu}}{2 \left[ (1-z_H) m_H^2 + \vec{\Delta}^2 \right]}  + \frac{z_H (q_{\perp}^{\mu} r_{\perp}^{\nu} - (1-z_H) r_{\perp}^{\mu} q_{\perp}^{\nu})-(1-z_H) \vec{r} \cdot \vec{q} \; g^{\mu \nu}}{\vec{r}^{\; 2}} \right] \; ,
    \label{FinFormGluAmp}
\end{equation}
where we have defined, similarly to~\cite{Hentschinski:2020tbi}, 
\begin{equation}
    \Delta_{\perp}^{\mu} \equiv  p_{H,\perp}^{\mu} - z_H q_{\perp}^{\mu} \; , \hspace{1 cm} r_{\perp} \equiv q_{\perp}^{\mu} - p_{H,\perp}^{\mu} \; .
\end{equation}
We can finally plug this expression into Eq.~(\ref{GluonIniImp}) and get the gluon-initiated contribution to the impact factor,
\begin{equation*}
    \frac{d \Phi_{g g}^{\{H g \}} (z_H, \vec{p}_H, \vec{q})}{d z_H d^{2} p_H} = \frac{g^2 g_H^2 C_A}{8 (2 \pi)^{D-1}(1-\epsilon) \sqrt{N^2-1}} \left \{ \frac{2}{z_H (1-z_H)} \right.
\end{equation*}
\begin{equation*}
     \left. \left[ 2 z_H^2 + \frac{(1-z_H)z_H m_H^2 (\vec{q} \cdot \vec{r}) [z_H^2 - 2 (1-z_H) \epsilon]+2 z_H^3 (\vec{p}_H \cdot \vec{r}) (\vec{p}_H \cdot \vec{q})}{\vec{r}^{\; 2} \left[ (1-z_H) m_H^2 + \vec{p}_H^{\; 2} \right]} - \frac{2 z_H^2 (1-z_H) m_H^2}{\left[ (1-z_H) m_H^2 + \vec{p}_H^{\; 2}  \right]}  \right. \right. 
\end{equation*}
\begin{equation*}
    -\frac{(1-z_H)z_H m_H^2 (\vec{q} \cdot \vec{r}) [z_H^2 - 2 (1-z_H) \epsilon]+2 z_H^3 (\vec{\Delta} \cdot \vec{r}) (\vec{\Delta} \cdot \vec{q})}{\vec{r}^{\; 2} \left[ (1-z_H) m_H^2 + \vec{\Delta}^{ 2} \right]} - \frac{2 z_H^2 (1-z_H) m_H^2}{\left[ (1-z_H) m_H^2 + \vec{\Delta}^{2}  \right]}
\end{equation*}
\begin{equation*}
   \left. + \frac{(1-\epsilon) z_H^2 (1-z_H)^2 m_H^4}{2} \left( \frac{1}{\left[ (1-z_H) m_H^2 + \Delta^{2}  \right]} + \frac{1}{\left[ (1-z_H) m_H^2 + \vec{p}_H^{\; 2}  \right]}  \right)^2 \right.
\end{equation*}
\begin{equation*}
   \left. - \frac{2 z_H^2 (\vec{p}_H \cdot \vec{\Delta})^2 - 2 \epsilon (1-z_H)^2 z_H^2 m_H^4}{\left[ (1-z_H) m_H^2 + \vec{p}_H^{\; 2}  \right] \left[ (1-z_H) m_H^2 + \Delta^{2}  \right]} \right]
\end{equation*}
\begin{equation}
     \left. + \frac{2 \vec{q}^{\; 2}}{\vec{r}^{\; 2}} \left[ \frac{z_H}{1-z_H} + z_H (1-z_H) + 2 (1-\epsilon) \frac{(1-z_H)}{z_H} \frac{(\vec{q} \cdot \vec{r})^2}{\vec{q}^{\; 2} \vec{r}^{\; 2}} \right] \right \} \theta \left( s_{\Lambda} - \frac{(1-z_H) m_H^2 + \vec{\Delta}^2}{z_H (1-z_H)} \right) \; .
     \label{GluonImp}
\end{equation}
We notice that this expression is compatible with gauge invariance, since it vanishes for $\vec q \to 0$. We have presented our result for $\Phi_{g g}^{\{H g \}}$ in a form similar to that given in Eq.~(46) of Ref.~\cite{Hentschinski:2020tbi} to facilitate the comparison. One can see that the expression in Ref.~\cite{Hentschinski:2020tbi} agrees with ours\footnote{We stress that in Ref.~\cite{Hentschinski:2020tbi} authors work in $D=4+2 \epsilon$, while we work in $D=4-2 \epsilon$}, except for:
\begin{itemize}
    \item two terms which are proportional to $z_H (1-z_H)^2$ instead of $z_H^2 (1-z_H)$; 
    \item two terms, proportional to $z_H^3$, which have a sign different from ours.
\end{itemize}
These little discrepancies are due to misprints in Ref.~\cite{Hentschinski:2020tbi}, as privately communicated to us by the authors of that paper.

\section{NLO impact factor: Virtual corrections}
\label{sec:virtual}

In this section, we compute the contribution to the impact factor coming from virtual corrections. The basic ingredient we need is the 1-loop correction to the high-energy vertex in Eq.~(\ref{gRHBorn}), that we indicate as\footnote{We remind that this is the effective vertex for the coupling of the initial-state gluon with the final-state Higgs and a $t$-channel Reggeized gluon, which is an object in the octet color representation with negative signature. Up to the NLO, the effective vertex takes contribution from the exchange in the $t$-channel of either one or two gluons; in the latter case, the projection onto the octet color representation with negative signature should be taken; in our calculation this projection is automatic in presence of an initial-state gluon, since the Higgs is a color singlet state.}
\begin{equation}
    \Gamma_{\{ H\} g}^{ac(1)} (q) =  \Gamma_{\{ H\} g}^{ac(0)} (q) \left[ 1 + \delta_{{\rm{NLO}}} \right] \; .
\end{equation}
Thereafter, the virtual contribution to the impact factor can be calculated as
\begin{equation*}
    \Phi_{gg}^{\{ H \}(1)} (\vec{q}; s_0 ) = \frac{\langle
cc^{\prime} | \hat{\cal P}_{0} | 0 \rangle}{2 (1-\epsilon) (N^2-1)}
\end{equation*}
\begin{equation}
   \times \sum_{a, \lambda } \int \frac{d s_{gR}}{2 \pi} d \rho_H  \Gamma_{\{ H \} g}^{ac(0)} (q) \left( \Gamma_{\{ H \} g}^{ac'(0)} (q) \right)^{*} \left[ \omega^{(1)}(-\vec{q}^{\; 2}) \ln \left( \frac{s_0}{\vec{q}^{\; 2}} \right) + \delta_{{\rm{NLO}}} + \delta_{{\rm{NLO}}}^{*} \right] \; ,
\label{HiggsVirtImp}
\end{equation}
where the $s_0$-dependent term comes from the expansion of $(s_0/\vec q^{\,2})^{\omega^{(1)}(-\vec q^{\,2})}$ in~(\ref{Int:Eq:ImpactproNext}). 
\begin{figure}
  \begin{center}
  \includegraphics[scale=0.50]{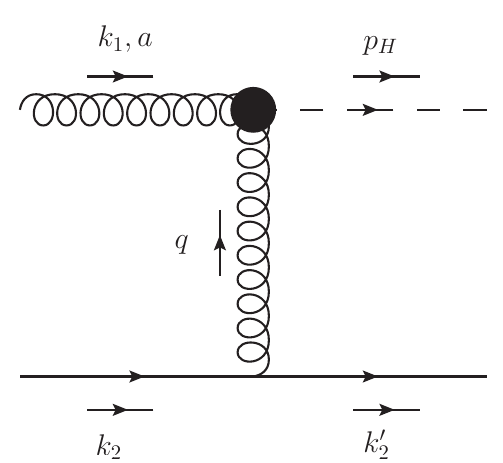} \hspace{1.5 cm}
  \end{center}
  \caption{$g q \rightarrow H q$ amplitude.}
  \label{LOgqHqAmp}
\end{figure}

The general strategy to find a NLO high-energy effective vertex is to compute, in the 
NLO and in the high-energy limit, a suitable amplitude and to compare it with the expected Regge form, written in terms of the needed effective vertex and of another known one. For our
purposes, we consider the diffusion of a gluon off a quark to produce a Higgs plus a quark, whose amplitude ${\cal A}_{g q \rightarrow H q}$, for the octet colour state and the negative signature in the $t$-channel has the following Reggeized form (see Fig.~\ref{LOgqHqAmp}):
$$
{\cal A}_{g q \rightarrow H q}^{(8,-)} = \Gamma_{ \{H\} g}^{ac} \frac{s}{t}\left[ \left( \frac{s}{-t} \right)^{\omega(t)} + \left(
\frac{-s}{-t} \right)^{\omega(t)} \right]\Gamma_{qq}^{c} \approx
\Gamma_{\{H\} g}^{ac(0)} \frac{2s}{t} \Gamma_{qq}^{c(0)}
$$
\begin{equation}
+ \Gamma_{\{H\} g}^{ac(0)} \frac{s}{t}\omega^{(1)}(t)
\left[ \ln\left( \frac{s}{-t} \right) + \ln\left(
\frac{-s}{-t} \right) \right]\Gamma_{q q}^{c(0)} + \Gamma_{\{H\} g}^{ac(0)} \frac{2s}{t}\Gamma_{q q}^{c(1)} +
\Gamma_{\{H\} g}^{ac(1)}\frac{2s}{t} \Gamma_{qq}^{c(0)} \; ,
\label{ReggeFormEx1}
\end{equation}
where 
\begin{equation}
    \Gamma_{qq}^{c(0)} = g t_{ji}^c \bar{u} (k_2-q) \frac{\slashed{k}_1}{s} u(k_2)
\end{equation} 
is the LO quark-quark-Reggeon effective vertex and $\Gamma_{qq}^{c(1)}$ its 1-loop correction,
known since long~\cite{Fadin:1993qb,Fadin:1995km}, and $t=q^2=-\vec q^{\,2}$.

Let us now split the 1-loop contributions to this amplitude into three pieces, related to
2-gluon (2g) or 1-gluon (1g) exchange or self-energy (se) diagrams in the $t$-channel:
$$
{\cal A}_{g q \rightarrow H q}^{(2g)(8,-)(1)} + {\cal A}
_{g q \rightarrow H q}^{(se)(1)} + {\cal A}_{g q \rightarrow H q}^{(1g)(1)} = \left\{ \Gamma_{\{H\} g}^{(2g)ac(1)}\frac{2s}{t}\Gamma_{qq}^{c(0)} + \Gamma_{\{H\} g}^{ac(0)}
\frac{2s}{t}\Gamma_{q q}^{(2g)c(1)} \right.
$$
$$
\left. + \Gamma_{\{H\} g}^{c(0)}\frac{s}{t}\omega^{(1)}(t)
\left[ \ln\left( \frac{s}{-t} \right) + \ln\left( \frac{-s}{-t}
\right) \right]\Gamma_{q q}^{c(0)} \right\}
$$
\begin{equation}
+ \left\{ \Gamma_{\{H\} g}^{(se)ac(1)}\frac{2s}{t}\Gamma
_{qq}^{c(0)} +  \Gamma_{\{H\} g}^{ac(0)}\frac{2s}{t}
\Gamma_{qq}^{(se)c(1)} \right\} +  \left\{ \Gamma_{\{H\} g}
^{(1g)ac(1)}\frac{2s}{t}\Gamma_{qq}^{c(0)} + \Gamma_{\{H\} g}
^{c(0)}\frac{2s}{t}\Gamma_{qq}^{(1g)c(1)} \right\},
\label{ReggeFormEx2}
\end{equation}
where the self-energy diagrams and 1-gluon exchange diagrams are automatically in the $8^-$
color representation.

We stress that we will use a unique regulator $\epsilon \equiv \epsilon_{\rm UV} \equiv \epsilon_{\rm IR}$ for both ultraviolet (UV) and infrared (IR) divergences. As a consequences, scaleless integrals such as
\begin{equation}
   \int \frac{d^D k}{i (2 \pi)^D} \frac{1}{k^2 (k+k_1)^2} 
\end{equation}
are equal to zero due to the exact cancellation between IR- and UV-divergence. In our case of massless partons, this implies that the contribution from the renormalization of the external quark and gluon lines is absent in dimensional regularization.

The 1-gluon-exchange contribution $\Gamma_{g H}^{(1g)ac(1)}$ can be calculated in a straightforward way by taking the radiative corrections to the amplitude of Higgs production in the collision of an on-shell gluon with an off-shell one (the gluon in the $t$-channel) having momentum $q$, colour index $c$ and ``nonsense'' polarization vector $-k_2/s$. In a similar manner one could calculate the 1-gluon-exchange contribution to the quark-quark-Reggeon effective vertex; in this case, the $t$-channel off-shell gluon must be taken with ``nonsense'' polarization vector $-k_1/s$. This is the consequence of the Gribov trick on the $t$-channel gluon propagator, valid in the high-energy limit.
The 1-gluon-exchange contribution does not include self-energy corrections to the $t$-channel gluon, which must be calculated separately and assigned with weight equal to one half to both $\Gamma_{\{H\}g}^{(se)ac(1)}$ and $\Gamma_{qq}^{(se)c(1)}$. 

For the 2-gluon exchange contributions we have the relation
\begin{equation*}
\Gamma_{\{H\} g}^{(2g)ac(1)}\frac{2s}{t}\Gamma_{qq}^{c(0)} +
\Gamma_{\{H\} g}^{ac(0)}\frac{2s}{t}\Gamma_{qq}^{(2g)c(1)} 
\end{equation*}
\begin{equation}
={\cal A}_{g q \rightarrow H q}^{(2g)(8,-)(1)} - \Gamma
_{\{H\} g}^{ac(0)}\frac{s}{t}\omega^{(1)}(t)\left[ \ln\left(
\frac{s}{-t} \right) + \ln\left( \frac{-s}{-t}
\right) \right]\Gamma_{qq}^{c(0)},
\label{FromExt2Gluon}
\end{equation}
which shows that we need to know the correction
$\Gamma_{qq}^{(2g)c(1)}$. This correction has been obtained in Ref.~\cite{Fadin:2001ap} and reads
\begin{equation}\label{311} \Gamma_{qq}^{(2g)c(1)} =\delta_{qq}^{(2g)}(t)
\Gamma_{qq}^{c(0)}\;, 
\end{equation}
with
$$
\delta_{qq}^{(2g)}(t) = \frac{1}{2}\omega^{(1)}(t)\left[ -\frac{1}{\epsilon}
+ \psi(1) + \psi(1+\epsilon) - 2\psi(1-\epsilon) \right]
$$
\begin{equation}
= g^2N\frac{\Gamma(2+\epsilon)}{(4\pi)^{2-\epsilon}}\frac{1}
{\epsilon}\left( -t \right)^{-\epsilon}\left( -\frac{1}{\epsilon} + 1 -
\epsilon + 4\epsilon\psi^\prime(1) \right) + {\cal O}(\epsilon).
\end{equation}

\subsection{The 1-loop correction: 1-gluon exchange and self-energy diagrams}
\label{ssec:virtual1Gluon}

\begin{figure}
  \begin{center}
  \includegraphics[scale=0.50]{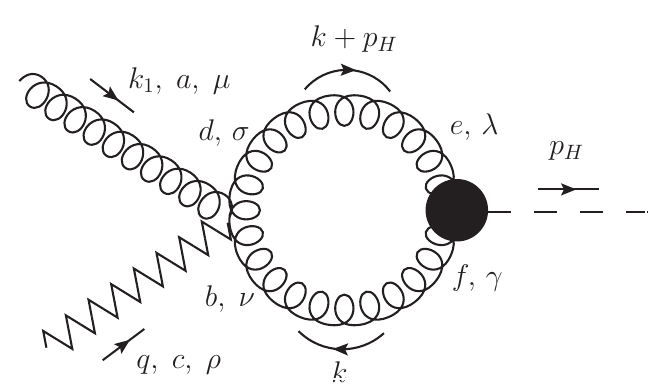} \hspace{1.5 cm}
  \includegraphics[scale=0.50]{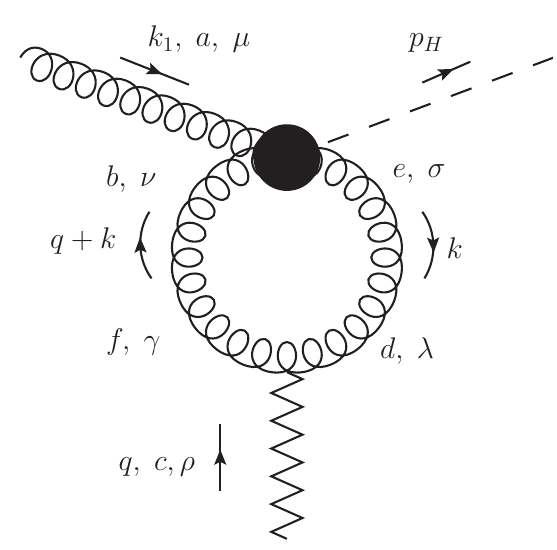} \vspace{1.0 cm} \\ \hspace{1.2 cm}
  \includegraphics[scale=0.53]{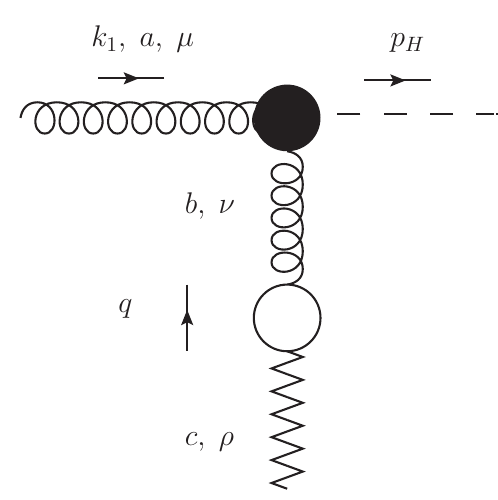} \hspace{1.0 cm} 
  \includegraphics[scale=0.50]{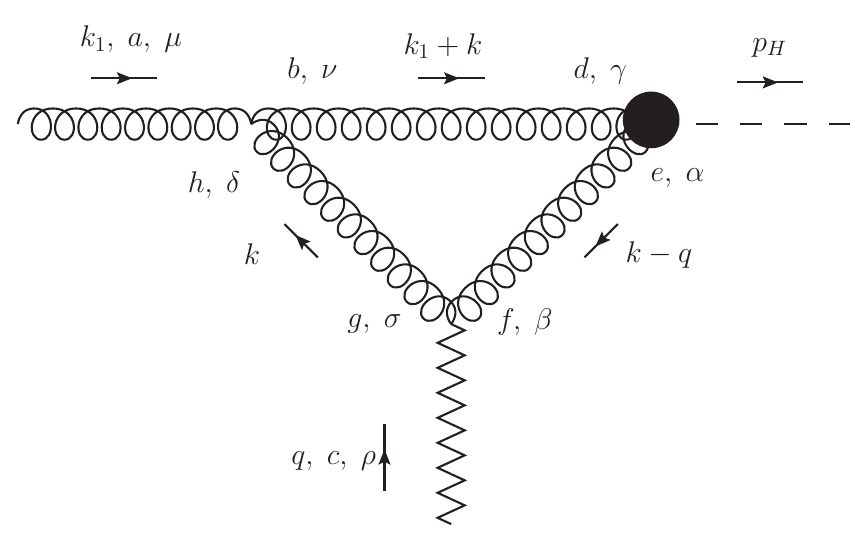} \\
  \end{center}
  \caption{1-gluon in the $t$-channel exchange diagrams. In the text we label them from (1) to (4) starting from the top left.}
  \label{VirtualDiagrams}
\end{figure}
There are four non-vanishing diagrams with 1-gluon exchange in the $t$-channel, including the one with self-energy corrections to the exchanged gluon; they are shown in Fig.~\ref{VirtualDiagrams}, where, the blob in the third diagram means the sum of gluon, ghost and quark loop contributions. 

Let's start by considering the first diagram; it gives
\begin{equation}
    \Gamma_{1,V}^{(1g)} = \frac{1}{2} i g^2 g_H \epsilon_{\mu} (k_1) \frac{k_{2,\rho}}{s} \int \frac{d^D k}{(2 \pi)^D} \frac{\Gamma_{abcb}^{\mu \rho \nu \sigma} H_{\sigma \nu} (-k-p_H,k)}{k^2 (k+p_H)^2} \; ,
\end{equation}
where 
\begin{equation*}
     \Gamma_{abcd}^{\mu \rho \nu \sigma} = \left[ (T^g)_{ab} (T^g)_{cd} (g^{\mu \rho} g^{\nu \sigma} - g^{\mu \sigma} g^{\nu \rho}) + (T^g)_{ac} (T^g)_{bd} (g^{\mu \nu} g^{\rho \sigma} - g^{\mu \sigma} g^{\nu \rho} ) \right. 
\end{equation*}
\begin{equation}
   \left. + (T^g)_{ad} (T^g)_{cb} (g^{\mu \rho} g^{\nu \sigma} - g^{\mu \nu} g^{\sigma \rho} ) \right]\;, \;\;\;\;\;  (T^a)_{bc}=-i f_{abc} \; ,
\end{equation}
and $H_{\sigma\nu}$ is defined in~(\ref{ggH}).  
After a trivial computation, we obtain
\begin{equation}
    \Gamma_{1,V}^{(1g)} =\Gamma_{\{H\} g}^{ac(0)} N g^2 \frac{(D-2)}{4(D-1)}  B_0(m_H^2) \equiv \Gamma_{\{H\} g}^{ac(0)} \delta_{1,V}^{(1g)} \; .
    \label{DeltaBubble}
\end{equation}
The definition and the value of the integral $B_0(m_H^2)$, together with all the other integrals that appear in the calculation of the virtual corrections, can be found in the Appendix~\ref{AppendixA2}.

The second diagram gives the following contribution
\begin{equation}
    \Gamma_{2,V}^{(1g)} = \frac{-i}{2} \epsilon_{\mu} (k_1) \frac{k_{2,\rho}}{s} N \delta^{ac} g^2 g_H \int \frac{d^D k}{(2 \pi)^D} \frac{V^{\mu \nu \sigma}(-k_1,-k-q,k) A^{\rho}_{\; \sigma \nu}(-k,k+q)}{k^2 (q+k)^2} \; ,
\end{equation}
where
\begin{equation}
      A^{\mu \rho \nu} (p,q) = g^{\nu \rho} (q-p)^{\mu} + g^{\rho \mu} (2p+q)^{\nu} -g^{\mu \nu} (p+2q)^{\rho}
  \end{equation}
and $V^{\mu\nu\sigma}$ is defined in~(\ref{gggH}). We find
\begin{equation}
       \Gamma_{2,V}^{(1g)} = \Gamma_{\{H\} g}^{ac(0)} \left[ -\frac{3}{2} N g^2 B_0 (-\vec{q}^{\; 2}) \right] \equiv \Gamma_{\{H\} g}^{ac(0)} \delta_{2,V}^{(1g)} \; .
       \label{DeltaFish}
  \end{equation}
  
The correction due to the Reggeon self-energy is universal and reads (see, {\it e.g.}, Ref.~\cite{Fadin:2001dc}) 
\begin{equation}
    \Gamma_{3,V}^{(1g)} = \Gamma_{\{H\} g}^{ac(0)} \left[ \omega^{(1)}(t) \frac{(5-3\epsilon)N-2(1-\epsilon)n_f}{4(1-2\epsilon)(3-2\epsilon)N} \right] \equiv \Gamma_{\{H\} g}^{ac(0)} \delta_{3,V}^{(1g)} \; .
    \label{DeltaSelf}
\end{equation}

The last correction is the one associated with the ``triangular'' diagram, 
\begin{equation*}
    \Gamma_{4,V}^{(1g)} = g^2 N \epsilon_{\mu} (k_1) \frac{k_{2,\rho}}{s} \delta^{ac} g_H
\end{equation*}
\begin{equation}
    \times \int \frac{d^D k}{i (2 \pi)^D} \frac{A_{\:\: \sigma}^{\mu \; \nu}(-k,k+k_1) A^{\; \sigma \rho \beta}(-q,q-k) H_{\nu \beta}(-k_1-k,k-q)}{k^2 (k+k_1)^2 (k-q)^2} \; .
\end{equation}
This correction is less trivial and we made use of the \textsc{FeynCalc}~\cite{Mertig:1991ca,Shtabovenko:2016cp} package to perform some analytic steps; up to terms power-suppressed in $s$, we end up with
\begin{equation*}
   \Gamma_{4,V}^{(1g)}  \equiv \Gamma_{H g}^{ac(0)} \delta_{4,V}^{(1g)} = \Gamma_{H g}^{ac(0)} \frac{N g^2}{4 (D-2) (1-D)
   \left(m_H^2+\vec{q}^{\; 2}\right)^2}  
   \end{equation*}
   \begin{equation*}
   \times \Bigg \{ 2 (D-1) \left[ \left(-5 (D-2) m_H^4-4(3 D-8) m_H^2
   \vec{q}^{\; 2}+(16-7 D) (\vec{q}_2^{\; 2})^2 \right) B_0(-\vec{q}^{\; 2}) \right. 
   \end{equation*}
   \begin{equation*}
   \left. +2 (D-2) m_H^2 \left(2
   m_H^4 +3 m_H^2 \vec{q}^{\; 2} + (\vec{q}^{\; 2})^2\right)
   C_0\left(0,-\vec{q}^{\; 2},m_H^2\right)\right] + \left( -D(D-2) (\vec{q}^{\; 2})^2 \right.
   \end{equation*}
   \begin{equation}
   + \left. (D (D (4 D-35)+92)-60)
   m_H^4-2 (D (-2 (D-9) D-41)+24) m_H^2 \vec{q}^{\; 2}\right)
   B_0\left(m_H^2\right)\Bigg\} \; .
   \label{DeltaTri}
\end{equation}

\subsection{The 1-loop correction: 2-gluon exchange diagrams}
\label{ssec:virtual2Gluon}

\begin{figure}
  \begin{center}
  \includegraphics[scale=0.60]{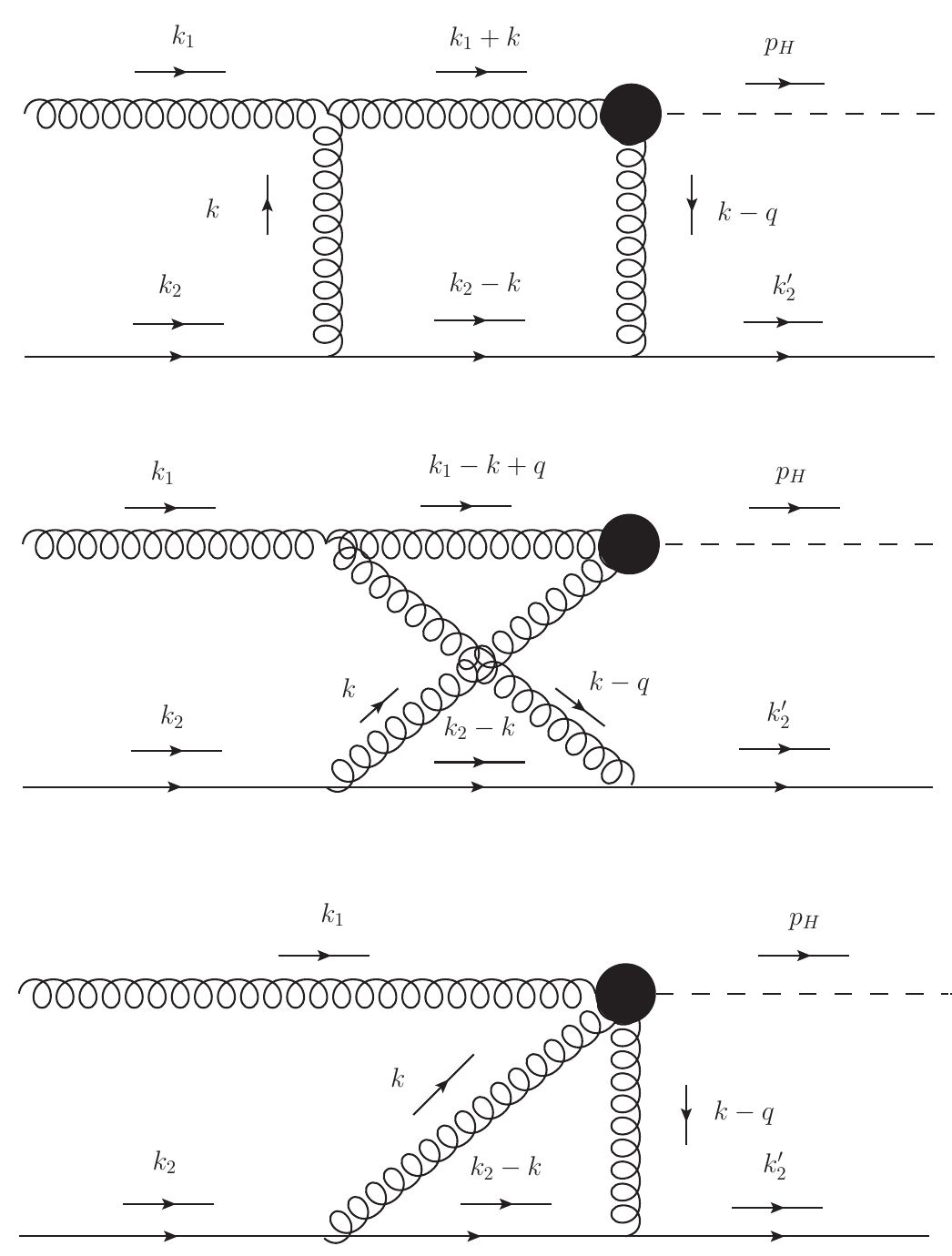} \hspace{1.5 cm}
  \end{center}
  \caption{Feynman diagrams contributing to the $g R \rightarrow g H$ amplitude.}
  \label{2GluonTchannel}
\end{figure}
As already mentioned, to extract this contribution we need to compute ${\cal A}_{g q \rightarrow H q}^{(2g)(8,-)(1)}$ and use Eq.~(\ref{FromExt2Gluon}). There are three diagrams contributing to this amplitude (see Fig.~\ref{2GluonTchannel}). If all vertices involved in these diagrams were from QCD, one could use the Gribov trick~(\ref{Gribov}) on both gluons in the $t$-channel and adopt the powerful technique explained in Refs.~\cite{Fadin:1999qc,Fadin:2001dc} to simplify the calculation. Instead, it is interesting to note that, the use of an effective Lagrangian containing an operator of dimension 5, requires a modification of the technique introduced in renormalizable theories for extracting the high-energy behaviour of amplitudes. In particular, for diagrams where a single gluon connects the two rapidity regions, the trick can be safely applied, while for diagrams with two gluons exchanged in the $t$-channel, the transverse part of the metric tensor should be kept in any gluon propagator directly connected to the effective vertex. For those propagators, one should use the modified Gribov prescription, 
\begin{equation}
g^{\mu \nu} = g_{\perp\perp}^{\mu \nu} +
2 \frac{k_1^{\mu} k_2^{\nu} + k_2^{\mu} k_1^{\nu}}{s} 
\;\;\to\;\;
g_{\perp\perp}^{\mu \nu} + 2 \frac{k_2^{\mu} k_1^{\nu}}{s} \; .
\end{equation}

We begin by computing the first diagram; the second one can be obtained from the first by analytic continuation from the $s$-channel to the $u$-channel, namely
\begin{equation}
    {\cal A}_{g q \rightarrow H q , \;2 }^{(2g)(8,-)(1)} (s) = - {\cal A}_{g q \rightarrow H q , \;1 }^{(2g)(8,-)(1)} (-s) \; .
\end{equation}
The amplitude relative to the first diagram is
\begin{equation*}
    {\cal A}_{g q \rightarrow H q , \;1 }^{(2g)(8,-)(1)} = \epsilon_{\mu}(k_1) g^3 g_H f^{abc} (t^b t^c)_{ji} \int \frac{d^D k}{(2 \pi)^{D}} \frac{A^{\mu \rho \nu}(-k,k+k_1) H_{\nu}^{\; \sigma} (-k-k_1, k-q)}{k^2(k+k_1)^2(k_2-k)^2 (k-q)^2}
\end{equation*}
\begin{equation}
    \times g_{\sigma \xi} g_{\rho \gamma} \bar{u} (k_2-q) \gamma^{\xi} (\slashed{k}_2 - \slashed{k}) \gamma^{\gamma} u(k_2) \; .
\end{equation}
Observing that 
\begin{equation*}
    f^{a b c} (t^b t^c)_{ji} = \frac{1}{2} f^{a b c} (t^b t^c - t^c t^b)_{ji} = \frac{i}{2} C_A t_{ji}^a = \frac{i}{2} C_A \delta^{ac} t_{ji}^c 
\end{equation*}
and then using \textsc{FeynCalc}~\cite{Mertig:1991ca,Shtabovenko:2016cp}, we obtain
\begin{equation*}
    {\cal A}_{g q \rightarrow \{H\} q , \;1 }^{(2g)(8,-)(1)} + {\cal A}_{g q \rightarrow H q , \;2 }^{(2g)(8,-)(1)} = \Gamma_{H g}^{ac(0)}  \left[ \frac{g^2 N}{(-4)} \left( \frac{2 \vec{q}^{\; 2} \left((5 D-12) m_H^2+(D-2)
   \vec{q}^{\; 2}\right) B_0\left(m_H^2\right)}{(D-2)
   \left(m_H^2+\vec{q}^{\; 2}\right)^2} \right. \right. 
   \end{equation*}
   \begin{equation*}
   \left. 
   +\frac{4 \left((D-2) m_H^4 + (D-2) m_H^2
   \vec{q}^{\; 2}+(2 D-5) (\vec{q}^{\; 2})^2 \right) B_0(-\vec{q}^{\; 2})}{(D-2)
   \left(m_H^2+\vec{q}^{\; 2}\right)^2}+2 s C_0\left(m_H^2,0,-s\right) \right. 
   \end{equation*}
   \begin{equation*}
   \left. -2 s
   C_0\left(m_H^2,0,s\right)-\frac{4 \left(m_H^4-m_H^2
   \vec{q}^{\; 2}-\vec{q}^{\; 2}\right)
   C_0\left(m_H^2,0,-\vec{q}^{\; 2}\right)}{m_H^2+\vec{q}^{\; 2}} +4 \vec{q}^{\; 2}
   C_0(0,0,-\vec{q}^{\; 2}) \right. \vspace{0.1cm}
   \end{equation*}
   \begin{equation*}
   \left. -2 s C_0(0,0,-s)+2 s C_0(0,0,s) +2 \vec{q}^{\; 2} s D_0\left(m_H^2,0,0,0;-\vec{q}^{\; 2},-s\right) \right. 
   \end{equation*}
   \begin{equation}
    -2 \vec{q}^{\; 2} s
   D_0\left(m_H^2,0,0,0;-\vec{q}^{\; 2},s\right) \Bigg ) \Bigg ] \frac{2s}{t} \Gamma_{qq}^{c(0)} \; .
\end{equation}

The third diagram gives a much simpler contribution:
\begin{equation}
   {\cal A}_{g q \rightarrow H q , \;3 }^{(2g)(8,-)(1)} = \Gamma_{\{H\} g}^{ac(0)} \left[ N g^2 B_0(-\vec{q}^{\; 2};0,0) \right] \frac{2 s}{t} \Gamma_{q q}^{c(0)} \; .
\end{equation}

At this point, using Eq.~(\ref{FromExt2Gluon}), we are able to extract the contribution to the $gRH$ vertex from the exchange of two gluons in the $t$-channel:
\begin{equation}
    \delta_{Hg}^{(2g)} = \frac{{\cal A}_{g q \rightarrow H q , \;1 }^{(2g)(8,-)(1)}+{\cal A}_{g q \rightarrow H q , \;2 }^{(2g)(8,-)(1)}+{\cal A}_{g q \rightarrow H q , \;3 }^{(2g)(8,-)(1)}}{\Gamma_{\{H\} g}^{ac(0)}\frac{2s}{t}\Gamma_{qq}^{c(0)}} -\delta_{qq}^{(2g)} - \frac{\omega^{(1)}(t)}{2} \left[ \ln\left(
\frac{s}{-t} \right) + \ln\left( \frac{-s}{-t}
\right) \right] \; .
\end{equation}

\subsection{Full 1-loop correction to the impact factor}
\label{ssec:virtual1full}

Adding $\delta_{Hg}^{(2g)}$ to $\delta_{1,V}^{(1g)}$, $\delta_{2,V}^{(1g)}$, $\delta_{3,V}^{(1g)}$, and $\delta_{4,V}^{(1g)}$ as defined in Eqs.~(\ref{DeltaBubble}), (\ref{DeltaFish}), (\ref{DeltaSelf}), and (\ref{DeltaTri}), respectively, we obtain
\begin{equation}
    \delta_{\rm NLO} \simeq \frac{\bar{\alpha}_s}{4 \pi} \left( \frac{\vec{q}^{\; 2}}{\mu^2} \right)^{-\epsilon} \left \{  - \frac{C_A}{\epsilon^2} + \frac{11 C_A -2 n_f}{6 \epsilon} - \frac{5 n_f}{9} + C_A \left( 2 {\rm{Li}}_2 \left( 1 + \frac{m_H^2}{\vec{q}^{\; 2}} \right) + \frac{\pi^2}{3} + \frac{67}{18} \right) \right \} \: , 
    \label{FullvertexCorrect}
\end{equation}
where we have kept only terms non-vanishing in the limit $\epsilon \to 0$ and have used
\begin{equation}
    \bar{\alpha}_s = \frac{g^2 \Gamma (1+\epsilon) \mu^{-2 \epsilon}}{(4 \pi)^{1-\epsilon}} \; , \hspace{0.5 cm} C_A = N \; .
\end{equation}
Substituting~(\ref{FullvertexCorrect}) into Eq.~(\ref{HiggsVirtImp}), we finally get the virtual correction to the Higgs impact factor:
\begin{equation*}
\frac{d \Phi_{gg}^{\{ H \}(1)} (z_H, \vec{p}_H, \vec{q}; s_0 )}{d z_H d^2 \vec{p}_H} = \frac{d \Phi_{gg}^{\{ H \}(0)} (z_H, \vec{p}_H,\vec{q})}{d z_H d^2 \vec{p}_H} \; \frac{\bar{\alpha}_s}{2 \pi} \left( \frac{\vec{q}^{\; 2}}{\mu^2}  \right)^{- \epsilon}  \left[  - \frac{C_A}{\epsilon^2} \right.
\end{equation*}
\begin{equation}
      \left. + \frac{11 C_A - 2n_f}{6 \epsilon} - \frac{C_A}{\epsilon} \ln \left( \frac{\vec{q}^{\; 2}}{s_0} \right) - \frac{5 n_f}{9} + C_A \left( 2\;\Re e \left( {\rm{Li}}_2 \left( 1 + \frac{m_H^2}{\vec{q}^{\; 2}} \right)\right) + \frac{\pi^2}{3} + \frac{67}{18} \right) + 11 \right] \; .
      \label{VirtualPartIMF}
\end{equation}
where $g_H$ has to be taken at leading order, while the last term equal to $11$ takes into account its next-to-leading contribution (see Eq.~(\ref{gH})). The result in Eq.~(\ref{VirtualPartIMF}) can be also obtained from Ref.~\cite{Schmidt:1997wr}, by taking the high-energy limit and using crossing symmetry. We used this alternative strategy as an independent check, finding perfect agreement.\footnote{To perform the comparison, we first used crossing symmetry to obtain from Ref.~\cite{Schmidt:1997wr} the NLO amplitude for the  Higgs plus quark production in the collision of a gluon with a quark, then we took the high-energy limit of this amplitude and confronted with the BFKL-factorized form of the same amplitude, taking advantage of the known expression for the NLO quark-quark-Reggeon vertex.}
The comparison with Ref.~\cite{Hentschinski:2020tbi}, whose virtual part contribution is based on Ref.~\cite{Nefedov:2019mrg}, is not immediate, since their calculation is based on the Lipatov effective action and on a different definition of the impact factor. This implies, for instance, that in their purely virtual result, Eq.~(43) of Ref.~\cite{Hentschinski:2020tbi}, the rapidity regulator is still present, which cancels when one combines the impact factor with their definition of the unintegrated gluon distribution\footnote{We recall that they considered the single-forward Higgs boson production.}.

\section{Projection onto the eigenfunctions of the BFKL kernel}
\label{sec:projection}

In this section we will carry out the projection of the Higgs impact factor onto the eigenfunctions of the LO BFKL kernel. There are two main reasons for performing this procedure. 

First, being our impact factor differential in the Higgs kinematic variables,
IR divergences do not appear in the real corrections and, therefore, it is not possible to observe their cancellation in the final result for the impact factor, as it would happen for a fully inclusive {\em hadronic} one (see Ref.~\cite{Fadin:1999qc}). The cancellation must of course be observed at the level of the cross section: the projection onto the LO BFKL eigenfunctions is an effective way to anticipate the integrations needed to get the cross section and, therefore, the projected impact factor, when all counterterms are taken into account, turns to be IR- and UV-finite. This procedure has been already successfully applied in 
Refs.~\cite{Ivanov:2012ms,Ivanov:2012iv}. The second reason is more practical: moving from the momentum representation to the representation in terms of the LO BFKL kernel eigenfunctions makes the numeric implementation somewhat simpler.

To understand the idea behind the projection, it is enough to consider the {\em partonic} amplitude in the LL,
\[
\Im m_{s}\left(({\cal A}^{(0)})_{AB}^{AB}\right) = \frac{s}{\left( 2\pi \right)^{D-2}}
\int \frac{d^{D-2}q_1}{\vec{q}_{1}^{\:2}}
\int \frac{d^{D-2}q_2}{\vec{q}_{2}^{\:2}}
\]
\begin{equation*}
\times \Phi _{A A}^{\left(0\right) }
\left( \vec{q}_{1};s_{0}\right)\int_{\delta -i\infty}^{\delta+i\infty}
\frac{d\omega }{2\pi i}\left[ \left( \frac{s}{s_{0}}\right)^{\omega }
G_{\omega }^{\left( 0\right) }\left( \vec{q}_{1},\vec{q}_{2}\right) 
\right] \Phi _{B B}^{\left( 0 \right) }\left( -\vec{q}_{2};s_{0}\right) \;,
\end{equation*}
and the spectral representation for the BFKL Green's function (\ref{sol}),
\begin{equation}
    G_{\omega }^{\left( 0\right) }\left( \vec{q}_{1},\vec{q}_{2}\right) = \sum_{n=-\infty}^\infty \int_{-\infty}^{+\infty} d \nu \frac{\phi_{\nu}^{n} (\vec{q}_1^{\; 2}) \phi_{\nu}^{n *}(\vec{q}_2^{\; 2})}{\omega - \frac{\alpha_s C_A}{\pi} \chi(n, \nu)} \; .
\end{equation}
If we use it, then, integrations over transverse momenta decouple and each impact factor can be separately projected onto the eigenfunctions of the BFKL kernel, so that
\begin{equation}
     \int \frac{d^{2-2\epsilon} q}{\pi \sqrt{2}} (\vec{q}^{\; 2})^{i \nu - \frac{3}{2}} e^{i n \phi} \Phi _{A A}^{\left(0\right)} (\vec{q} \;) \equiv 
     \Phi^{(0)}_{A A}(n,\nu) \; .
\label{ProjectionDef}
\end{equation}
and similarly for the $\Phi_{BB}$. The extension of the procedure to the NLL is straightforward (see the next chapter). Since we work in $D = 4 - 2\epsilon$, we rely again on the ``continuation'' of the LO BFKL kernel eigenfunctions to non-integer dimensions,
\[
(\vec q^{\:2})^\gamma e^{i n \phi} \ \rightarrow (\vec q^{\:2})^{\gamma -\frac{n}{2}} (\vec q \cdot \vec l)^n\;, \;\;\;\;\; \gamma\equiv i\nu-
\frac{1}{2} \; .
\]

As already discussed in subsection~\ref{ssec:LO_IF}, to get the Higgs impact factor for a proton-initiated process, we must take the convolution with the initial-state PDFs. This 
brings along initial-state collinear singularities, which must be canceled by suitable counterterms. In the rest of this section we will construct the $(n,\nu)$-projection
of all contributions to the hadronic Higgs impact factor, including the PDF counterterms, and will check the explicit cancellation of all IR divergences. The residual UV divergence will
be taken care of by the renormalization of the QCD coupling. To perform the $(n,\nu)$-projection, according to Eq.~(\ref{ProjectionDef}), we will make use of the integrals computed in Appendix~\ref{AppendixA3}.

\subsection{Projection of the LO impact factor}
\label{ssec:projection_LO}

Recalling the expression of the LO impact factor,
\begin{equation}
    \frac{d \Phi_{PP}^{ \{ H \}(0)} (x_H, \vec{p}_H, \vec{q})}{d x_H d^2 \vec{p}_H} = \int_{x_H}^1 \frac{d z_H}{z_H} f_g \left( \frac{x_H}{z_H} \right) \frac{d \Phi_{gg}^{ \{H \}(0)} (z_H, \vec{p}_H, \vec{q})}{d z_H d^2 \vec{p}_H} = \frac{g_H^2 \vec{q}^{\; 2} f_g (x_H) \delta^{(2)} ( \vec{q} - \vec{p}_H) }{8 (1-\epsilon) \sqrt{N^2-1}} \: ,
    \tag{\ref{Factorization}}
\end{equation}
we get immediately from Eq.~(\ref{ProjectionDef}) its projected counterpart,
\begin{equation}
    \frac{d \Phi_{ PP }^{\{H \}(0)} (x_H, \vec{p}_H, n, \nu)}{d x_H d^2 \vec{p}_H}  = \frac{g_H^2}{8 (1-\epsilon) \sqrt{N^2-1}} \frac{(\vec{p}_H^{\; 2})^{i \nu - \frac{1}{2}} e^{i n \phi_H}}{\pi \sqrt{2}} f_g (x_H)  \; .
\end{equation}
Here $\phi_H$ is the azimuthal angle of the vector $\vec p_H$ counted from the fixed direction in the transverse space. The projected LO impact factor is the starting point for the 
calculation of the NLO contribution to the projected impact factor from the gluon PDF and QCD coupling counterterms. From now on, we will omit the apex $^{(1)}$, since all contributions
to the projected impact factor are NLO.

\subsection{Projection of gluon PDF and coupling counterterms}
\label{ssec:projection_gluon_PDF}

Taking into account the running of $\alpha_s$,
\begin{equation}
    \alpha_s (\mu^2) = \alpha_s (\mu_R^2) \left[ 1 + \frac{\alpha_s (\mu_R^2)}{4 \pi} \left( \frac{11 C_A}{3} - \frac{2 n_f}{3} \right) \left( - \frac{1}{\epsilon} - \ln (4 \pi e^{-\gamma_E}) + \ln \left( \frac{\mu_R^2}{\mu^2} \right) \right) \right] \; ,
\end{equation}
and the running of the gluon PDF,
\begin{equation*}
    f_g (x, \mu) = f_g (x, \mu_F) - \frac{\alpha_s(\mu_F)}{2 \pi} \left( - \frac{1}{\epsilon} - \ln (4 \pi e^{- \gamma_E}) + \ln \left( \frac{\mu_F^2}{\mu^2} \right) \right) 
\end{equation*}
\begin{equation}
    \times \int_x^1 \frac{dz}{z} \left[ P_{gq} (z) \sum_{a=q \bar{q}} f_a \left( \frac{x}{z} , \mu_F \right) + P_{gg} (z) f_g \left( \frac{x}{z} , \mu_F \right) \right] \; ,
\end{equation}
where
\begin{equation}
    P_{gq} (z) = C_F \frac{1+(1-z)^2}{z} \; ,
\end{equation}
\begin{equation}
    P_{gg} (z) = 2 C_A \left( \frac{z}{(1-z)_{+}} + \frac{(1-z)}{z} + z (1-z) \right) + \frac{11 C_A - 2 n_f}{6} \delta (1-z) \; ,
\end{equation}
with plus prescription defined as
\begin{equation}
    \int_a^1 dx \frac{F(x)}{(1-x)_+} = \int_a^1 dx \frac{F(x)-F(1)}{(1-x)} - \int_0^a dx \frac{F(1)}{1-x} \; ,
\end{equation}
we obtain the following projected counterterms,
 \begin{equation*}
    \frac{d \Phi_{PP}^{ \{H \}} (x_H, \vec{p}_H, n, \nu)}{d x_H d^2 \vec{p}_H} \bigg|_{{\rm{coupling \; c.t.}}} = \frac{d \Phi_{PP}^{ \{H \}(0)} (x_H, \vec{p}_H, n, \nu)}{d x_H d^2 \vec{p}_H} \frac{\bar{\alpha}_s}{2 \pi} \left( \frac{\vec{p}_H^{\; 2}}{\mu^2} \right)^{-\epsilon}  
 \end{equation*}
 \begin{equation*}
    \times \left( \frac{11 C_A}{3} - \frac{2 n_f}{3} \right) \left( - \frac{1}{\epsilon} + \ln \left( \frac{\mu_R^2}{\vec{p}_H^{\; 2}}\right) \right) 
 \end{equation*}
 \begin{equation}
     \equiv \frac{d \Phi_{PP, {\rm{div}}}^{ \{H \}} (x_H, \vec{p}_H, n, \nu)}{d x_H d^2 \vec{p}_H} \bigg|_{{\rm{coupling \; c.t.}}} + \frac{d \Phi_{PP, {\rm{fin}}}^{ \{H \}} (x_H, \vec{p}_H, n, \nu)}{d x_H d^2 \vec{p}_H} \bigg|_{{\rm{coupling \; c.t.}}} \; ,
     \label{CoupliforDiv}
 \end{equation}
 \begin{equation*}
    \frac{d \Phi_{PP}^{ \{H \}}(x_H, \vec{p}_H, n, \nu)}{d x_H d^2 \vec{p}_H}\bigg|_{{\rm{P_{qg} \; c.t.}}} = - \frac{1}{f_g (x_H)} \frac{d \Phi_{PP}^{ \{H \}(0)} (x_H, \vec{p}_H, n, \nu)}{d x_H d^2 \vec{p}_H} \frac{\bar{\alpha}_s}{2 \pi} \left( \frac{\vec{p}_H^{\; 2}}{\mu^2} \right)^{-\epsilon} 
\end{equation*}
\begin{equation*}
    \times \left( - \frac{1}{\epsilon} + \ln \left( \frac{\mu_F^2}{\vec{p}_H^{\; 2}}\right) \right) \int_{x_H}^1 \frac{dz_H}{z_H} \left[ P_{gq} (z_H) \sum_{a=q \bar{q}} f_a \left( \frac{x_H}{z_H} , \mu_F \right) \right] 
\end{equation*}
\begin{equation}
    \equiv \frac{d \Phi_{PP, {\rm{div}}}^{ \{H \}}(x_H, \vec{p}_H, n, \nu)}{d x_H d^2 \vec{p}_H}\bigg|_{{\rm{P_{qg} \; c.t.}}} + \frac{d \Phi_{PP,{\rm{fin}}}^{ \{H \}}(x_H, \vec{p}_H, n, \nu)}{d x_H d^2 \vec{p}_H}\bigg|_{{\rm{P_{qg} \; c.t.}}} 
\label{PqgforDiv}
\end{equation}
and
\begin{equation*}
    \frac{d \Phi_{PP}^{ \{H \}}(x_H, \vec{p}_H, n, \nu)}{d x_H d^2 \vec{p}_H}\bigg|_{{\rm{P_{gg} \; c.t.}}} = - \frac{1}{f_g (x_H)} \frac{d \Phi_{PP}^{ \{H \}(0)} (x_H, \vec{p}_H, n, \nu)}{d x_H d^2 \vec{p}_H} \frac{\bar{\alpha}_s}{2 \pi} \left( \frac{\vec{p}_H^{\; 2}}{\mu^2} \right)^{-\epsilon} 
\end{equation*}
\begin{equation*}
    \times \left( - \frac{1}{\epsilon} + \ln \left( \frac{\mu_F^2}{\vec{p}_H^{\; 2}}\right) \right) \int_{x_H}^1 \frac{dz_H}{z_H} \left[ P_{gg} (z_H) f_g \left( \frac{x_H}{z_H} , \mu_F \right) \right] 
\end{equation*}
\begin{equation}
    \equiv \frac{d \Phi_{PP, {\rm{div}}}^{ \{H \}}(x_H, \vec{p}_H, n, \nu)}{d x_H d^2 \vec{p}_H}\bigg|_{{\rm{P_{gg} \; c.t.}}} + \frac{d \Phi_{PP,{\rm{fin}}}^{ \{H \}}(x_H, \vec{p}_H, n, \nu)}{d x_H d^2 \vec{p}_H}\bigg|_{{\rm{P_{gg} \; c.t.}}} \; .
    \label{PggforDiv}
\end{equation}
With obvious notation, we have implicitly defined divergent (``div'') and finite (``fin'') part of each contribution. We will keep adopting this notation in the following.

\subsection{Projection of high-rapidity real gluon contribution and BFKL counterterm}
\label{ssec:projection_BFKL}

The contribution from a real gluon emission has a divergence for $z_H \to 1$ or $ (z_g \to 0)$ at any value of the gluon momenta $\vec{q}-\vec{p}_H$. This divergence is regulated by the parameter $s_{\Lambda}$ and, in the final result, cancelled by the BFKL counterterm appearing in the definition of the NLO impact factor. In this section we make the cancellation of $s_{\Lambda}$ explicit and give the $(n,\nu)$-projection of the high-rapidity part of the real gluon production impact factor combined with the BFKL counterterm.

First of all, let's take the convolution of the impact factor for real gluon production, given in~(\ref{GluonImp}), with the gluon PDF and rewrite it in an equivalent form, by adding and subtracting three terms, for later convenience:
\[
\frac{d \Phi^{\{Hg\}}_{PP}(x_H, \vec{p}_H,\vec q;s_0)}{d x_H d^2 p_H}
= \frac{d \tilde{\Phi}^{\{Hg\}}_{PP}(x_H, \vec{p}_H,n,\vec q;s_0)}{d x_H d^2 p_H}
\]
\[
+ \frac{d \Phi^{\{Hg\} (1-x_H)}_{PP}(x_H, \vec{p}_H,\vec q;s_0)}{d x_H d^2 p_H}
+ \frac{d \Phi^{\{Hg\}{\rm plus}}_{PP}(x_H, \vec{p}_H,\vec q;s_0)}{d x_H d^2 p_H}
\]
\begin{equation}
+ \int_{x_H}^1 d z_H f_g (x_H)  \frac{d \Phi_{gg}^{ \{H g \}} (z_H, \vec{p}_H, \vec{q} ; s_0)}{d z_H d^2 \vec{p}_H} \bigg|_{z_H \rightarrow 1} \; ,
\label{subtractions}
\end{equation}
where
\[
    \frac{d \tilde{\Phi}^{\{Hg\}}_{PP}(x_H, \vec{p}_H,\vec q;s_0)}{d x_H d^2 p_H}
    =\int_{x_H}^1 \frac{d z_H}{z_H} f_g \left( \frac{x_H}{z_H} \right) \frac{d \Phi_{gg}^{ \{ H g \}}(z_H, \vec{p}_H, \vec{q};s_0)}{d z_H d^2 \vec{p}_H} 
\]
\begin{equation}
    - \int_{x_H}^1 d z_H f_g \left( \frac{x_H}{z_H} \right) \left[ \frac{d \Phi_{gg}^{ \{ H g \}}(z_H, \vec{p}_H, \vec{q};s_0)}{d z_H d^2 \vec{p}_H} \right]_{z_H=1} \;,
\end{equation}
\begin{equation}
\frac{d \Phi^{\{Hg\} (1-x_H)}_{PP}(x_H, \vec{p}_H,\vec q;s_0)}{d x_H d^2 p_H}
=\int_{0}^{x_H} d z_H f_g (x_H) \frac{d \Phi_{gg}^{ \{ H g \}}(z_H, \vec{p}_H, \vec{q};s_0)}{d z_H d^2 \vec{p}_H} \Bigg |_{z_H=1} 
\end{equation}
and
\[
\frac{d \Phi^{\{Hg\}{\rm plus}}_{PP}(x_H, \vec{p}_H,\vec q;s_0)}{d x_H d^2 p_H}
= - \int_{0}^{x_H} d z_H f_g (x_H) \frac{d \Phi_{gg}^{ \{ H g \}}(z_H, \vec{p}_H, \vec{q};s_0)}{d z_H d^2 \vec{p}_H} \Bigg |_{z_H=1}
\]
\begin{equation}
    + \int_{x_H}^1 d z_H \left( f_g \left( \frac{x_H}{z_H} \right) - f_g (x_H) \right) \left[\frac{d \Phi_{gg}^{ \{ H g \}}(z_H, \vec{p}_H, \vec{q};s_0)}{d z_H d^2 \vec{p}_H} \right]_{z_H=1} \; .
\end{equation}
The pieces $d\tilde{\Phi}$, $d\Phi^{\{Hg\}(1-x_H)}$, and $d\Phi^{\{Hg\}{\rm plus}}$ are free from the divergence for $z_H\to 1$ and therefore in their expressions the limit $s_{\Lambda} \rightarrow \infty$ can be safely taken, which means that $\theta(s_{\Lambda} - s_{PR})$ can be set to one.
The projection of these terms will be considered in subsection~\ref{ssec:projection_gluon}.
The last term in Eq.~(\ref{subtractions}) can be easily calculated, since
\begin{equation*}
    \frac{d \Phi_{gg}^{ \{H g \}} (z_H, \vec{p}_H, \vec{q} ; s_0)}{d z_H d^2 \vec{p}_H} \bigg|_{z_H \rightarrow 1} = \frac{\braket{cc'|\hat{\mathcal{P}}|0}}{2(1-\epsilon)(N^2-1)} 
\end{equation*}
\begin{equation*}
   \times \left[ \sum_{\{ f \}} \int \frac{d s_{PR} d\rho_f}{2 \pi} \Gamma^c_{P \{ f \}} \left( \Gamma^c_{P \{ f \}} \right)^{*} \theta \left( s_{\Lambda} - s_{PR} \right) \right]_{z_H \rightarrow 1}
\end{equation*}
\begin{equation}
    = \frac{g^2 g_H^2 C_A}{4 (1-\epsilon) \sqrt{N^2-1} (2 \pi)^{D-1} } \frac{\vec{q}^{\; 2}}{(\vec{q}- \vec{p}_H)^2} \frac{1}{(1-z_H)} \theta \left( s_{\Lambda} - \frac{(\vec{q}-\vec{p}_H)^2}{(1-z_H)} \right) \; .
\end{equation}
We get
\begin{equation*}
    \int_{x_H}^1 d z_H f_g (x_H)  \frac{d \Phi_{gg}^{ \{H g \}} (z_H, \vec{p}_H, \vec{q} ; s_0)}{d z_H d^2 \vec{p}_H} \bigg|_{z_H \rightarrow 1}
\end{equation*}
\begin{equation*}
=    \frac{g^2 g_H^2 C_A}{4 (1-\epsilon) \sqrt{N^2-1} (2 \pi)^{D-1} } \frac{\vec{q}^{\; 2}}{(\vec{q}- \vec{p}_H)^2} \int_{x_H}^1 d z_H \frac{1}{(1-z_H)} f_g (x_H) \theta \left( s_{\Lambda} - \frac{(\vec{q}-\vec{p}_H)^2}{(1-z_H)} \right) 
\end{equation*}
\begin{equation}
    = \frac{g^2 g_H^2 C_A}{4 (1-\epsilon) \sqrt{N^2-1} (2 \pi)^{D-1} } \frac{\vec{q}^{\; 2}}{(\vec{q}- \vec{p}_H)^2} f_g (x_H) \left[ \ln (1-x_H) - \frac{1}{2} \ln \left( \frac{\left[(\vec{q}-\vec{p}_H)^2 \right]^2}{s_{\Lambda}^2} \right) \right] \; .
\label{Phi_g_rap}
\end{equation}

Let's consider now the BFKL counterterm
\begin{equation}
   \frac{d \Phi^{\rm{BFKL \ c.t.}}_{gg}(z_H, \vec{p}_H, \vec q;s_0)}{d z_H d^2 p_H} = - \frac{1}{2} \int d^{D-2} q' \frac{ \vec{q}^{\; 2}}{\vec{q}^{\; '2}}  \frac{d \Phi_{gg}^{ \{H \} (0)} (\vec{q} \; ' )}{d z_H d^2 p_H}  \mathcal{K}^{(0)}_r (\vec{q} \; ', \vec{q} \; ) \ln \left( \frac{s_{\Lambda}^2}{(\vec{q} \; ' - \vec{q} \; )^2 s_0} \right) \; .
\end{equation}
Using Eq.~(\ref{LOHiggsImp2}) and Eq.~(\ref{Int:Eq:RealBornKert0}), we find
\begin{equation}
    \frac{d \Phi^{\rm{BFKL \ c.t.}}_{gg}(z_H, \vec{p}_H, \vec q;s_0)}{d z_H d^2 p_H} \!=\! \frac{-g^2 g_H^2 C_A}{8 (2 \pi)^{D-1} (1-\epsilon) \sqrt{N^2-1}} \frac{\vec{q}^{\; 2}}{(\vec{q}-\vec{p}_H)^2} \ln \left( \frac{s_{\Lambda}^2}{(\vec{q} - \vec{p}_H )^2 s_0}  \right) \delta (1-z_H)
\end{equation}
and, after convolution with the gluon PDF, we get
\[
 \frac{d \Phi^{{\rm{BFKL\ c.t.}}}_{PP}(x_H,\vec p_H,\vec q;s_0)}{d x_H d^2 p_H} = \int_{x_H}^1 \frac{d z_H}{z_H} f_g \left( \frac{x_H}{z_H} \right) \frac{d \Phi_{gg}^{{\rm{BFKL\ c.t.}}} (z_H, \vec{p}_H, \vec{q})}{d z_H d^2 \vec{p}_H} 
 \]
 \begin{equation}
= - \frac{g^2 g_H^2 C_A}{8 (2 \pi)^{D-1} \sqrt{N^2-1}} \frac{\vec{q}^{\; 2}}{(\vec{q}-\vec{p}_H)^2} \frac{f_g (x_H)}{(1-\epsilon)} \ln \left( \frac{s_{\Lambda}^2}{(\vec{q} - \vec{p}_H )^2 s_0}  \right) \; .
 \label{BFKLct}
\end{equation}
When we combine the last term of Eq.~(\ref{subtractions}), given in~(\ref{Phi_g_rap}), with the BFKL counterterm, given in~(\ref{BFKLct}), we obtain  
\begin{equation}
 \frac{d \Phi^{{\rm{BFKL}}}_{PP}(x_H, \vec{p}_H, \vec q;s_0)}{d x_H d^2 p_H} \equiv \frac{g^2 g_H^2 C_A}{4 (2 \pi)^{D-1} (1-\epsilon) \sqrt{N^2-1}} \frac{\vec{q}^{\; 2}}{(\vec{q}-\vec{p}_H)^2} f_g (x_H) \ln \left( \frac{(1-x_H) \sqrt{s_0}}{|\vec{q} - \vec{p}_H|}  \right) \,.
 \label{CountFin}
\end{equation}
Note that this term is finite as far as the high-energy divergence is concerned. The remaining divergences can be isolated after the projection, 
\begin{equation*}
   \frac{d \Phi^{{\rm{BFKL}}}_{PP}(x_H, \vec{p}_H,n,\nu;s_0)}{d x_H d^2 p_H} = \frac{d \Phi_{PP}^{\{ H \}(0)} (x_H, \vec{p}_H, n, \nu)}{d x_H d^2 \vec{p}_H} \frac{\bar{\alpha}_s}{2 \pi} \left( \frac{\vec{p}_H^{\; 2}}{\mu^2} \right)^{-\epsilon} \left \{ \frac{C_A}{\epsilon^2} + \frac{C_A}{\epsilon} \ln \left( \frac{\vec{p}_H^{\; 2}}{s_0} \right)  \right.
\end{equation*}
\begin{equation*}
    \left. - 2 \frac{C_A}{\epsilon} \ln (1-x_H) + C_A \left[ \ln \left( \frac{\vec{p}_H^{\; 2}}{s_0 (1-x_H)^2} \right) \left( 2  \gamma_E + \psi \left(\frac{1}{2} + \frac{n}{2} - i \nu \right) \right. \right. \right.
\end{equation*}
\begin{equation*}
    \left. + \psi \left( \frac{1}{2} + \frac{n}{2} + i \nu \right) \right) - 2 \gamma_E^2 - \zeta (2) - \frac{1}{2}\left(\psi' \left(\frac{1}{2} + \frac{n}{2} - i \nu\right) - \psi' \left(\frac{1}{2} + \frac{n}{2} + i \nu\right)\right) 
\end{equation*}
\begin{equation*}
    - 2 \gamma_E \left( \psi \left(\frac{1}{2} + \frac{n}{2} - i \nu\right) + \psi \left(\frac{1}{2} + \frac{n}{2} + i \nu\right) \right)
\end{equation*}
\begin{equation*}
- \left.\left. \frac{1}{2}  \left(\psi \left(\frac{1}{2} + \frac{n}{2} - i \nu \right) + \psi \left( \frac{1}{2} + \frac{n}{2} + i \nu \right) \right)^2 \right] \right \}
\end{equation*}
\begin{equation}
    \equiv \frac{d \Phi^{{\rm{BFKL}}}_{PP, {\rm{div}}}(x_H, \vec{p}_H,n,\nu;s_0)}{d x_H d^2 p_H} + \frac{d \Phi^{{\rm{BFKL}}}_{PP, {\rm{fin}}}(x_H, \vec{p}_H,n,\nu;s_0)}{d x_H d^2 p_H} \; .
    \label{BFKLforDiv}
\end{equation}

\subsection{Projection of virtual and real quark contributions}
\label{ssec:projection_quark}

Since the virtual contribution is proportional to the LO impact factor, the convolution with the gluon PDF and the $(n,\nu)$-projection are trivial and give
\begin{equation*}
\frac{d \Phi_{PP}^{\{ H \}(1)} (x_H, \vec{p}_H, n, \nu; s_0 )}{d x_H d^2 \vec{p}_H} = \frac{d \Phi_{PP}^{\{ H \}(0)} (x_H, \vec{p}_H,n, \nu)}{d x_H d^2 \vec{p}_H} \; \frac{\bar{\alpha}_s}{2 \pi} \left( \frac{\vec{p}_H^{\; 2}}{\mu^2}  \right)^{- \epsilon}  \left[  - \frac{C_A}{\epsilon^2} \right.
\end{equation*}
\begin{equation*}
      \left. + \frac{11 C_A - 2n_f}{6 \epsilon} - \frac{C_A}{\epsilon} \ln \left( \frac{\vec{p}_H^{\; 2}}{s_0} \right) - \frac{5 n_f}{9} + C_A \left( 2\;\Re e \left({\rm{Li}}_2 \left( 1 + \frac{m_H^2}{\vec{p}_H^{\; 2}} \right)\right) + \frac{\pi^2}{3} + \frac{67}{18} \right) + 11 \right] 
\end{equation*}
\begin{equation}
    \equiv \frac{d \Phi_{PP,{\rm{div}}}^{\{ H \}(1)} (x_H, \vec{p}_H, n, \nu; s_0 )}{d x_H d^2 \vec{p}_H} + \frac{d \Phi_{PP,{\rm{fin}}}^{\{ H \}(1)} (x_H, \vec{p}_H, n, \nu)}{d x_H d^2 \vec{p}_H} \; . \vspace{0.3 cm}
    \label{VirforDiv}
\end{equation}

We recall that the real quark contribution is
\begin{equation*}
    \frac{d \Phi_{q q}^{\{H q \}} (z_H, \vec{p}_H, \vec{q})}{d z_H d^2 \vec{p}_H} = \frac{\sqrt{N^2-1}}{16 N (2 \pi)^{D-1}} \frac{g^2 g_H^2}{[(\vec{q}-\vec{p}_H)^2]^2}
\end{equation*}
\begin{equation}
\times    \left[ \frac{4 (1-{z}_H) \left[(\vec{q}-\vec{p}_H) \cdot \vec{q} \; \right]^2 + z_H^2 \vec{q}^{\; 2} (\vec{q} - \vec{p}_H)^2}{z_H} \right] \; .
\tag{\ref{QuarkConImpacFin}}
\end{equation}
Using
\[
(\vec{q}-\vec{p}_H) \cdot \vec{q} = \frac{1}{2}\bigl[\vec q^{\:2} + (\vec{q}-\vec{p}_H)^2 - \vec p_{H}^{\:2} \bigl]\;,
\]
this contribution can be projected using the integral $I_1(\gamma_1, \gamma_2, n, \nu)$, defined in~(\ref{FirstMasterIntegral}), and suitably choosing $\gamma_1$ and $\gamma_2$ for the different terms. The projected quark contribution gives 
    \begin{equation*}
    \frac{d \Phi_{q q}^{\{H q \}} (z_H, \vec{p}_H, n, \nu)}{d z_H d^2 \vec{p}_H} = \frac{\sqrt{N^2-1}}{16 N (2 \pi)^{D-1}} g^2 g_H^2 \left \{ \left(z_H +2\, \frac{1-z_H}{z_H}\right) I_1(-1 , 1 , n , \nu) \right.
    \end{equation*}
\begin{equation*}    
    + \frac{1-z_H}{z_H} \bigg [ (\vec p_H^{\;2})^2 I_1(0, 2, n , \nu)  \bigg.
\end{equation*}
\begin{equation*}
     \left. \left. -2 \vec{p}_H^{\; 2} \bigg( I_1(0,1, n, \nu) + I_1(-1 , 2, n, \nu)\bigg)
    +I_1(-2,2, n, \nu)  \right] \right \} \; .
\end{equation*}
If we replace $I_1(\gamma_1, \gamma_2, n, \nu)$ by its explicit expression~(\ref{FirstMasterIntegral}), perform a partial $\epsilon$-expansion and take the convolution with the quark PDFs, we obtain
\begin{equation*}
    \frac{d \Phi_{P P}^{\{H q \}} (x_H, \vec{p}_H, n, \nu)}{d x_H d^2 \vec{p}_H} = \frac{1}{f_g(x_H)} \frac{d \Phi_{PP}^{\{H\}(0)}(x_H, \vec{p}_H, n , \nu)}{d x_H d^2 p_H} \frac{\bar{\alpha}_s}{2 \pi} \left( \frac{\vec{p}_H^{\; 2}}{\mu^2} \right)^{-\epsilon}     
\end{equation*}
\begin{equation*}
   \times \int_{x_H}^1 \frac{d z_H}{z_H} \sum_{a=q \bar{q}} f_a \left( \frac{x_H}{z_H}, \mu_F \right) \left \{ - \frac{1}{\epsilon} C_F \left( \frac{1+(1-z_H)^2}{z_H} \right) + (1-\gamma_E) C_F \left( \frac{1+(1-z_H)^2}{z_H} \right)  \right.
\end{equation*}
\begin{equation*}
    \left. \left. + \frac{C_F}{z_H} \bigg [ 4 (z_H-1)  - \frac{(1+n)(z_H-1)}{(\frac{1}{2} + \frac{n}{2} - i \nu)} - \frac{(1+(1-z_H)^2)}{(-\frac{3}{2} + \frac{n}{2} + i \nu)} - \frac{3+n(z_H-1)+z_H(z_H-3)}{(-\frac{1}{2} + \frac{n}{2} + i \nu)} \right. \right.
\end{equation*}
\begin{equation*}
    \left. \left.- (1 + (1-z_H)^2) \left( H_{-1/2+n/2-i \nu} + \psi (-\frac{3}{2}+\frac{n}{2}+i \nu) \right) \right] \right \} 
\end{equation*}
\begin{equation}
    \equiv \frac{d \Phi_{P P, {\rm{div}}}^{\{H q \}} (x_H, \vec{p}_H, n, \nu)}{d x_H d^2 \vec{p}_H} + \frac{d \Phi_{P P, {\rm{fin}}}^{\{H q \}} (x_H, \vec{p}_H, n, \nu)}{d x_H d^2 \vec{p}_H} \; .
    \label{QuarkforDiv}
\end{equation}
We observe that the $\epsilon$-singularity is cancelled when we combine this object with the counterterm containing $P_{g q} (z_H)$, which appears in~(\ref{PqgforDiv}). We emphasize that the limits $n \to 1$ or $n\to 3$, $\nu \rightarrow 0$ are safe from divergences.

\subsection{Projection of the real gluon contribution}
\label{ssec:projection_gluon}

In this subsection we discuss the projection of all terms appearing in~(\ref{subtractions}), but the last, which was already treated in subsection~\ref{ssec:projection_BFKL}.

We start with the terms labeled ``plus'', which, after $(n,\nu)$-projection, gives
\begin{equation*}
     \frac{d \Phi_{P P}^{\{H g\} {\rm{plus}}} (x_H, \vec{p}_H, n, \nu; s_0)}{d x_H d^2 \vec{p}_H} = - \frac{1}{f_g(x_H)} \frac{d \Phi_{P P}^{\{H \}(0)} (x_H, \vec{p}_H, n, \nu)}{d x_H d^2 \vec{p}_H} \frac{\bar{\alpha}_s}{2 \pi} \left( \frac{\vec{p}_H^{\; 2}}{\mu^2} \right)^{-\epsilon} 
\end{equation*}
\begin{equation*}
    \times \int_{x_H}^1 \frac{d z_H}{z_H} f_g \left( \frac{x_H}{z_H} \right) 2 C_A \frac{z_H}{(1-z_H)_{+}}  \left[ \frac{1}{\epsilon}  + \left( H_{-1/2+n/2 - i \nu} + H_{-1/2+n/2+i \nu} \right) \right]
\end{equation*}
\begin{equation}
    \equiv \frac{d \Phi_{P P, {\rm{div}}}^{\{H g\} {\rm{plus}}} (x_H, \vec{p}_H, n, \nu)}{d x_H d^2 \vec{p}_H} + \frac{d \Phi_{P P, {\rm{fin}}}^{\{H g\} {\rm{plus}}} (x_H, \vec{p}_H, n, \nu)}{d x_H d^2 \vec{p}_H} \; .
    \label{PlusforDiv}
\end{equation}
The divergence in this term is cancelled by the term containing the plus prescription in $P_{gg}(z_H)$, which appears in~(\ref{PggforDiv}).

Then, we project the term labeled ``$(1-x_H)$'' and find
\begin{equation*}
    \frac{d \Phi_{P P}^{\{H g\} (1-x_H)} (x_H, \vec{p}_H, n, \nu)}{d x_H d^2 \vec{p}_H} = \frac{d \Phi_{P P}^{\{H \}(0)} (x_H, \vec{p}_H, n, \nu)}{d x_H d^2 \vec{p}_H} \frac{\bar{\alpha}_s}{2 \pi} \left( \frac{\vec{p}_H^{\; 2}}{\mu^2} \right)^{-\epsilon} 
\end{equation*}
\begin{equation*}
     \times 2 C_A \ln (1-x_H) \left[ \frac{1}{\epsilon} + \left( H_{-1/2 + n/2 - i \nu} + H_{-1/2 + n/2 + i \nu} \right) \right] 
\end{equation*}
\begin{equation}
    \equiv \frac{d \Phi_{P P, \rm{div}}^{\{H g\} (1-x_H)} (x_H, \vec{p}_H, n, \nu)}{d x_H d^2 \vec{p}_H} + \frac{d \Phi_{P P, \rm{fin}}^{\{H g\} (1-x_H)} (x_H, \vec{p}_H, n, \nu)}{d x_H d^2 \vec{p}_H} \; .
    \label{(1-xH)forDiv}
\end{equation}
The divergence cancels with an analogous term present in $d \Phi_{PP}^{{\rm{BFKL}}}$, given in~(\ref{BFKLforDiv}). \\

We are left with the first term in Eq.~(\ref{subtractions}), {\it i.e.} 
\begin{equation}
\frac{d \tilde{\Phi}^{\{Hg\}}_{PP}(x_H, \vec{p}_H,\vec q;s_0)}{d x_H d^2 p_H}
\end{equation}
\[
=\int_{x_H}^1 \frac{d z_H}{z_H} f_g \left( \frac{x_H}{z_H} \right) \left[ \frac{d \Phi_{gg}^{ \{ H g \}}(z_H, \vec{p}_H, \vec{q})}{d z_H d^2 \vec{p}_H} - z_H \frac{d \Phi_{gg}^{ \{ H g \}}(z_H, \vec{p}_H, \vec{q})}{d z_H d^2 \vec{p}_H} \Bigg |_{z_H=1} \right] \; .
\]
The effect of this subtraction is to remove the first term of the last line in Eq.~(\ref{GluonImp}), which in fact represents the only true divergence for $z_H \rightarrow 1$, and we obtain
\begin{equation*}
    \frac{d \Phi_{gg}^{ \{ H g \}}(z_H, \vec{p}_H, \vec{q})}{d z_H d^2 \vec{p}_H} - z_H \frac{d \Phi_{gg}^{ \{ H g \}}(z_H, \vec{p}_H, \vec{q};s_0)}{d z_H d^2 \vec{p}_H} \Bigg |_{z_H=1}
    = \frac{g^2 g_H^2 C_A}{8 (2 \pi)^{D-1}(1-\epsilon) \sqrt{N^2-1}}
\end{equation*}
\begin{equation*}
    \times \left \{ \frac{2}{z_H (1-z_H)} \left. \left[ 2 z_H^2 + \frac{(1-z_H)z_H m_H^2 (\vec{q} \cdot \vec{r}) [z_H^2 - 2 (1-z_H) \epsilon]+2 z_H^3 (\vec{p}_H \cdot \vec{r}) (\vec{p}_H \cdot \vec{q})}{\vec{r}^{\; 2} \left[ (1-z_H) m_H^2 + \vec{p}_H^{\; 2} \right]} \right. \right.  \right.
\end{equation*}
\begin{equation*}
    - \frac{2 z_H^2 (1-z_H) m_H^2}{\left[ (1-z_H) m_H^2 + \vec{p}_H^{\; 2}  \right]}  -\frac{(1-z_H)z_H m_H^2 (\vec{q} \cdot \vec{r}) [z_H^2 - 2 (1-z_H) \epsilon]+2 z_H^3 (\vec{\Delta} \cdot \vec{r}) (\vec{\Delta} \cdot \vec{q})}{\vec{r}^{\; 2} \left[ (1-z_H) m_H^2 + \vec{\Delta}^{ 2} \right]} 
\end{equation*}
\begin{equation*}
   \left. - \frac{2 z_H^2 (1-z_H) m_H^2}{\left[ (1-z_H) m_H^2 + \vec{\Delta}^{2}  \right]} + \frac{(1-\epsilon) z_H^2 (1-z_H)^2 m_H^4}{2} \left( \frac{1}{\left[ (1-z_H) m_H^2 + \Delta^{2}  \right]} \right. \right.
\end{equation*}
\begin{equation*}
   \left. \left. + \frac{1}{\left[ (1-z_H) m_H^2 + \vec{p}_H^{\; 2}  \right]}  \right)^2 - \frac{2 z_H^2 (\vec{p}_H \cdot \vec{\Delta})^2 - 2 \epsilon (1-z_H)^2 z_H^2 m_H^4}{\left[ (1-z_H) m_H^2 + \vec{p}_H^{\; 2}  \right] \left[ (1-z_H) m_H^2 + \Delta^{2}  \right]} \right]
\end{equation*}
\begin{equation*}
     \left. + \frac{2 \vec{q}^{\; 2}}{\vec{r}^{\; 2}} \left[ z_H (1-z_H) + 2 (1-\epsilon) \frac{(1-z_H)}{z_H} \frac{(\vec{q} \cdot \vec{r})^2}{\vec{q}^{\; 2} \vec{r}^{\; 2}} \right] \right \}
\end{equation*}
\begin{equation}
    \equiv \frac{d \Phi_{gg}^{ \{ H g \}{\rm{coll}}}(z_H, \vec{p}_H, \vec{q})}{d z_H d^2 \vec{p}_H} + \frac{d \Phi_{gg}^{ \{ H g \} (1-z_H)}(z_H, \vec{p}_H, \vec{q})}{d z_H d^2 \vec{p}_H} + \frac{d \Phi_{gg}^{ \{ H g \}{\rm{rest}}}(z_H, \vec{p}_H, \vec{q})}{d z_H d^2 \vec{p}_H} \; .
\end{equation}
In the last equality, we split this term in three contributions: 1) $d \Phi_{g g}^{\{H g \}{\rm coll}}$ contains the pure collinear divergence remained, 2) $d \Phi_{g g}^{\{H g \}(1-z_H)}$ contains terms that taken alone are singular for $z_H \rightarrow 1$, but when combined are safe from divergences, 3) $d \Phi_{g g}^{\{H g \}{\rm rest}}$ is the rest. 

\vspace{0.2 cm}

\textit{Collinear term}

\begin{equation*}
\frac{d \Phi_{g g}^{\{H g \}{\rm{coll}}} (z_H, \vec{p}_H, \vec{q})}{d z_H d^2 \vec{p}_H} = \frac{g^2 g_H^2 C_A}{8 (2 \pi)^{D-1}(1-\epsilon)  \sqrt{N^2-1}} \frac{2 \vec{q}^{\; 2}}{\vec{r}^{\; 2}} 
\end{equation*}
\begin{equation}
\times \left[ z_H (1-z_H) + 2 (1-\epsilon) \frac{(1-z_H)}{z_H} \frac{(\vec{q} \cdot \vec{r})^2}{\vec{q}^{\; 2} \vec{r}^{\; 2}} \right] \; .
\end{equation}
This term can be projected and taken in convolution with the gluon PDF in a way quite analogous to the quark case; we find
\begin{equation*}
    \frac{d \Phi_{P P}^{\{H g \} {\rm{coll}}} (x_H, \vec{p}_H, n, \nu)}{d x_H d^2 \vec{p}_H} \equiv \frac{1}{f_g(x_H)} \frac{d \Phi_{P P}^{\{ H \}(0)}(x_H, \vec{p}_H, n , \nu)}{d x_H d^2 \vec{p}_H} \frac{\bar{\alpha}_s}{2 \pi} \left( \frac{\vec{p}_H^{\; 2}}{\mu^2} \right)^{-\epsilon} \int_{x_H}^1 \frac{d z_H}{z_H} f_g \left( \frac{x_H}{z_H} \right)
\end{equation*}
\begin{equation*}
 \times \left \{ - \frac{1}{\epsilon} 2 \; C_A \left( z_H (1-z_H) + \frac{(1-z_H)}{z_H} \right) - 2 \gamma_E  C_A \left( z_H (1-z_H) + \frac{(1-z_H)}{z_H} \right) \right.
\end{equation*}
\begin{equation*}
    \left. - \frac{2 C_A (1-z_H)}{z_H} \left[ 1 + \gamma_E + \gamma_E z_H^2 -  \frac{1+n}{2 (\frac{1}{2} + \frac{n}{2} - i \nu)} + \frac{1+z_H^2}{\left( -\frac{3}{2} + \frac{n}{2} + i \nu \right)} + \frac{3-n+2 z_H^2}{2(-\frac{1}{2} + \frac{n}{2} + i \nu)} \right. \right.
\end{equation*}
\begin{equation*}
    \left. \left. + (1 + z_H^2) \left( \psi \left(\frac{1}{2}+\frac{n}{2}-i \nu \right)  + \psi \left( -\frac{3}{2}+\frac{n}{2}+i \nu \right) \right) \right] \right \} 
\end{equation*}
\begin{equation}
   \equiv \frac{d \Phi_{P P, {\rm{div}}}^{\{H g \} {\rm{coll}}} (x_H, \vec{p}_H, n, \nu)}{d x_H d^2 \vec{p}_H} + \frac{d \Phi_{P P, {\rm{fin}}}^{\{H g \}{\rm{coll}}} (x_H, \vec{p}_H, n, \nu)}{d x_H d^2 \vec{p}_H} \; .
   \label{CollforDiv}
\end{equation}
We observe that the $\epsilon$-singularity is cancelled by a terms contained in the counterterm with $P_{g g} (z_H)$, which is given in~(\ref{PggforDiv}). As in the quark case, the above expression is safe from divergences in $n=1$ or $n=3$, $\nu=0$. 

\vspace{0.2 cm}

$(1-z_H)$-\textit{term}

\begin{equation*}
 \frac{d \Phi_{g g}^{\{H g \}(1-z_H)} (z_H, \vec{p}_H, \vec{q})}{d z_H d^2 \vec{p}_H} = \frac{g^2 g_H^2 C_A}{4 (2 \pi)^{D-1}(1-\epsilon)  \sqrt{N^2-1}}  \frac{1}{z_H (1-z_H)}\Biggl[ 2 z_H^2 \Biggr.
\end{equation*}
\begin{equation}
    \left. + \frac{2 z_H^3 (\vec{p}_H \cdot \vec{r}) (\vec{p}_H \cdot \vec{q})}{\vec{r}^{\; 2} \left[ (1-z_H) m_H^2 + \vec{p}_H^{\; 2} \right]} 
    -\frac{2 z_H^3 (\vec{\Delta} \cdot \vec{r}) (\vec{\Delta} \cdot \vec{q})}{\vec{r}^{\; 2} \left[ (1-z_H) m_H^2 + \vec{\Delta}^{ 2} \right]} 
    \right.
\end{equation}
\begin{equation*}
    \left. - \frac{2 z_H^2 (\vec{p}_H \cdot \vec{\Delta})^2}{\left[ (1-z_H) m_H^2 + \vec{p}_H^{\; 2}  \right] \left[ (1-z_H) m_H^2 + \Delta^{2}  \right]} \right] \; . 
\end{equation*}
Here, and in the rest of the calculation, we will use the following formula, 
\begin{equation}
    \vec{q} \cdot \vec{p}_H  = (\vec{q}^{\; 2})^{\frac{1}{2}} (\vec{p}_H^{\; 2})^{\frac{1}{2}} \cos( \phi - \phi_H ) = (\vec{q}^{\; 2})^{\frac{1}{2}} (\vec{p}_H^{\; 2})^{\frac{1}{2}} \left( \frac{e^{i(\phi-\phi_H)} + e^{-i(\phi-\phi_H)}}{2} \right)
    \label{angle trick} \; .
\end{equation}
Let us first observe that the term in square bracket gives zero in the limit $z_H \rightarrow 1$, so that there is no singularity in this limit. Now, using~(\ref{angle trick}), we perform the projection and convolution with the gluon PDF and obtain, up to vanishing terms in the $\epsilon\to 0$ limit 
\begin{equation*}
    \frac{d \Phi_{P P}^{\{H g \}(1-z_H)} (x_H, \vec{p}_H, n, \nu)}{d x_H d^2 \vec{p}_H} = \frac{1}{f_g(x_H)} \frac{d \Phi_{PP}^{ \{ H \}(0)} (x_H, \vec{p}_H, n, \nu)}{d x_H d^2 \vec{p}_H} \frac{2 \sqrt{2} C_A}{(\vec{p}_H^{\; 2})^{i \nu - \frac{1}{2}} e^{i n \phi_H}}  \frac{\alpha_s}{2 \pi}  
\end{equation*}
\begin{equation*}
    \times \int_{x_H}^1 \frac{d z_H}{z_H} f_g \left( \frac{x_H}{z_H} \right)  \frac{1}{z_H (1-z_H)} \left \{ \frac{2 z_H^3}{\left[ (1-z_H) m_H^2 + \vec{p}_H^{\; 2} \right]} \left[ \frac{\vec{p}_H^{\; 2}}{4} \left( e^{-2 i \phi_H} I_1(-1,1,n+2, \nu) \right. \right. \right.
\end{equation*}
\begin{equation*}
    \left. \left. \left. + e^{2 i \phi_H} I_1(-1,1,n-2, \nu) + 2 I_1(-1,1,n, \nu) \right) - \frac{(\vec{p}_H^{\; 2})^{\frac{3}{2}}}{2} \left( e^{- i \phi_H} I_1 \left(-\frac{1}{2},1,n+1, \nu \right) \right. \right. \right. 
\end{equation*}
\begin{equation*}
    \left. \left. \left.  + e^{ i \phi_H} I_1 \left( -\frac{1}{2},1,n-1, \nu \right) \right) \right] - \frac{2 z_H^2}{\left[ (1-z_H) m_H^2 + \vec{p}_H^{\; 2} \right]} \bigg[ (\vec{p}_H^{\; 2})^{2} I_3 (0,1,n,\nu)  \right.
\end{equation*}
\begin{equation*}
    \left. + \frac{z_H^2 \vec{p}_H^{\; 2}}{4} \left( e^{-2 i \phi_H} I_3(-1,1,n+2, \nu) + e^{2 i \phi_H} I_3(-1,1,n-2, \nu) + 2 I_3(-1,1,n, \nu) \right) \right. 
\end{equation*}
\begin{equation*}
    \left. - z_H (\vec{p}^{\;2}_H)^{\frac{3}{2}} \left( e^{-i \phi_H} I_3 \left(- \frac{1}{2}, 1, n+1, \nu \right) + e^{i \phi_H} I_3 \left(- \frac{1}{2}, 1, n-1, \nu \right) \right) \right] 
\end{equation*}
\begin{equation*}
    - 2 z_H^3 \left[ \frac{(\vec{p}_H^{\; 2})^{\frac{1}{2}}}{2} \left( e^{-i \phi_H} I_3 \left( -\frac{1}{2}, 1, n+1, \nu \right) + e^{i \phi_H} I_3 \left( -\frac{1}{2}, 1, n-1, \nu \right) \right) - z_H I_3 (-1,1,n, \nu) \right]
\end{equation*}
\begin{equation*}
    -2 z_H^3 (1-z_H) \left[ \frac{(\vec{p}^{\;2}_H)^{\frac{1}{2}}}{2} (1+z_H) \left( e^{-i \phi_H} I_2 \left( -\frac{3}{2},n+1,\nu \right) + e^{i \phi_H} I_2 \left( -\frac{3}{2},n-1,\nu \right) \right) -z_H \right.  
\end{equation*}
\begin{equation*}
   \left. \left. \times I_2 (-2,n,\nu) - \frac{\vec{p}_H^{\; 2}}{4} \left( e^{-2 i\phi_H} I_2 ( -1,n+2,\nu ) + e^{2 i \phi_H} I_2 ( -1,n-2,\nu ) + 2 I_2 (-1, n, \nu) \right) \right] \right \} \; .
   \label{(1-z_H)ForFin1}
\end{equation*}
Setting $\epsilon = 0$ and using the limits~(\ref{First Limit}), (\ref{Second Limit}), it is easy to show that the result is safe from $z_H \rightarrow 1$ divergences. In this result, there are combinations of integrals, which are safe from $\epsilon$-divergences, even if single integrals, taken alone, are divergent. In particular, we note that if we use Eq.~(\ref{FirstMasterIntegral}) for the integrals of the type $I_1$, the first five terms vanish up to ${\cal O}(\epsilon)$. Moreover, if we apply the replacement~(\ref{TrickAsy}), we see that the ``asymptotic'' counterparts of the last six integrals cancel completely and therefore any $I_2$ integral in the previous expression can be replaced by $I_{2, {\rm{reg}}}$. The final form for this contribution is
\begin{equation*}
    \frac{d \Phi_{P P}^{\{H g \}(1-z_H)} (x_H, \vec{p}_H, n, \nu)}{d x_H d^2 \vec{p}_H} = \frac{1}{f_g(x_H)} \frac{d \Phi_{P P}^{ \{ H \}} (x_H, \vec{p}_H, n, \nu)}{d x_H d^2 \vec{p}_H} \frac{2 \sqrt{2} C_A}{(\vec{p}_H^{\; 2})^{i \nu - \frac{1}{2}} e^{i n \phi_H}}  \frac{\alpha_s}{2 \pi}  
\end{equation*}
\begin{equation*}
    \times \int_{x_H}^1 \frac{d z_H}{z_H} f_g \left( \frac{x_H}{z_H} \right)  \frac{1}{z_H (1-z_H)} \left \{ - \frac{2 z_H^2}{\left[ (1-z_H) m_H^2 + \vec{p}_H^{\; 2} \right]} \bigg[ (\vec{p}_H^{\; 2})^{2} I_3 (0,1,n,\nu)  \right.
\end{equation*}
\begin{equation*}
    \left. + \frac{z_H^2 \vec{p}_H^{\; 2}}{4} \left( e^{-2 i \phi_H} I_3(-1,1,n+2, \nu) + e^{2 i \phi_H} I_3(-1,1,n-2, \nu) + 2 I_3(-1,1,n, \nu) \right) \right. 
\end{equation*}
\begin{equation*}
    \left. - z_H (\vec{p}^{\;2}_H)^{\frac{3}{2}} \left( e^{-i \phi_H} I_3 \left(- \frac{1}{2}, 1, n+1, \nu \right) + e^{i \phi_H} I_3 \left(- \frac{1}{2}, 1, n-1, \nu \right) \right) \right] 
\end{equation*}
\begin{equation*}
    - 2 z_H^3 \left[ \frac{(\vec{p}_H^{\; 2})^{\frac{1}{2}}}{2} \left( e^{-i \phi_H} I_3 \left( -\frac{1}{2}, 1, n+1, \nu \right) + e^{i \phi_H} I_3 \left( -\frac{1}{2}, 1, n-1, \nu \right) \right) - z_H I_3 (-1,1,n, \nu) \right]
\end{equation*}
\begin{equation*}
    -2 z_H^3 (1-z_H) \left[ \frac{(\vec{p}^{\;2}_H)^{\frac{1}{2}}}{2} (1+z_H) \left( e^{-i \phi_H} I_{2,\rm{reg}} \left( -\frac{3}{2},n+1,\nu \right) + e^{i \phi_H} I_{2,\rm{reg}} \left( -\frac{3}{2},n-1,\nu \right) \right) \right.  
\end{equation*}
\begin{equation}
\begin{split}
   \left. \left. -z_H I_{2,\rm{reg}} (-2,\nu, n) - \frac{\vec{p}_H^{\; 2}}{4} \left( e^{-2i  \phi_H} I_{2,\rm{reg}} ( -1,n+2,\nu ) \right. \right. \right. \\
   \left. \left. \left. + e^{2i \phi_H} I_{2,\rm{reg}} ( -1,n-2,\nu ) + 2 I_{2,\rm{reg}} (-1, n, \nu) \right) \right. \Bigg] \right. \Bigg \} \;. 
   \label{(1-z_H)ForFin}
\end{split}
\end{equation}

\vspace{0.5 cm} 

\textit{Rest term}

\begin{equation*}
    \frac{d \Phi_{g g}^{\{H g \}{\rm{rest}}} (z_H, \vec{p}_H, \vec{q})}{d z_H d^2 \vec{p}_H} \equiv \frac{g^2 g_H^2 C_A}{4 (2 \pi)^{D-1}(1-\epsilon) \sqrt{N^2-1}} \left \{ \frac{ m_H^2 (\vec{q} \cdot \vec{r}) [z_H^2 - 2 (1-z_H) \epsilon]}{\vec{r}^{\; 2} \left[ (1-z_H) m_H^2 + \vec{p}_H^{\; 2} \right]} \right.
\end{equation*}
\begin{equation*}
    \left.  + \frac{(1-\epsilon) z_H (1-z_H) m_H^4}{2} \left( \frac{1}{\left[ (1-z_H) m_H^2 + \Delta^{2}  \right]} + \frac{1}{\left[ (1-z_H) m_H^2 + \vec{p}_H^{\; 2}  \right]}  \right)^2 \right.
\end{equation*}
\begin{equation*}
  + \frac{ 2 \epsilon (1-z_H) z_H m_H^4}{\left[ (1-z_H) m_H^2 + \vec{p}_H^{\; 2}  \right] \left[ (1-z_H) m_H^2 + \Delta^{2}  \right]} - 2 z_H m_H^2 \left( \frac{1}{\left[ (1-z_H) m_H^2 + \vec{p}_H^{\; 2}  \right]} \right.
\end{equation*}
\begin{equation}
    \left. \left.  + \frac{1}{\left[ (1-z_H) m_H^2 + \vec{\Delta}^{2}  \right]} \right) -\frac{m_H^2 (\vec{q} \cdot \vec{r}) [z_H^2 - 2 (1-z_H) \epsilon]}{\vec{r}^{\; 2} \left[ (1-z_H) m_H^2 + \vec{\Delta}^{ 2} \right]} \right \} \; .
\end{equation}
We perform the projection and convolution with gluon PDF and, finally, we find, up to terms ${\cal O}(\epsilon)$,
\begin{equation*}
    \frac{d \Phi_{P P}^{\{H g \}{\rm{rest}}} (x_H, \vec{p}_H, n, \nu)}{d x_H d^2 \vec{p}_H} =  \frac{1}{f_g(x_H)} \frac{d \Phi_{P P}^{ \{H \}(0)} (x_H, \vec{p}_H, n, \nu)}{d x_H d^2 \vec{p}_H} \frac{ \sqrt{2} C_A}{(\vec{p}_H^{\; 2})^{i \nu - \frac{1}{2}} e^{i n \phi_H}}  \frac{\alpha_s}{2 \pi}  \int_{x_H}^1 \frac{d z_H}{z_H} f_g \left( \frac{x_H}{z_H} \right)
\end{equation*}
\begin{equation*}
   \times \left \{ \frac{2 m_H^2 \left[ z_H^2 - 2 (1-z_H) \epsilon \right]}{[(1-z_H) m_H^2 + \vec{p}_H^{\; 2}]} \left[ I_1(-1,1,n,\nu) - \frac{(\vec{p}_H^{\; 2})^{\frac{1}{2}}}{2} \left( e^{-i \phi_H} I_1 \left( -\frac{1}{2}, 1, n+1, \nu \right) \right. \right. \right.  
\end{equation*}
\begin{equation*}
 \left. \left. + e^{i \phi_H} I_1 \left( -\frac{1}{2}, 1, n-1, \nu \right) \right) \right] - 2 m_H^2 [z_H^2 - 2(1-z_H) \epsilon] \left. \Bigg[ I_2 \left( -1, n, \nu \right) \right. 
\end{equation*}
\begin{equation*}
- \frac{(\vec{p}_H^{\; 2})^{\frac{1}{2}}}{2} \left( e^{-i \phi_H} I_2 \left(-\frac{1}{2}, n+1, \nu \right) + e^{i \phi_H} I_2 \left( -\frac{1}{2},n-1,\nu \right) \right) \Bigg] - 4 z_H m_H^2 I_3 (0,1,n, \nu) 
\end{equation*}
\begin{equation*}
    + (1-\epsilon) z_H (1-z_H) m_H^4 \left[ I_3(0,2,n, \nu) + \left(2+\frac{4\epsilon}{1-\epsilon}\right) \frac{I_3 (0,1,n,\nu)}{[(1-z_H) m_H^2 + \vec{p}_H^{\; 2}]}\right] .
    \label{RestForFin1}
\end{equation*}
Using the explicit result for the integral $I_1$ and applying the same ideas as befor, after 
the $\epsilon$-expansion we end up with 
\begin{equation*}
    \frac{d \Phi_{P P}^{\{H g \}{\rm{rest}}} (x_H, \vec{p}_H, n, \nu)}{d x_H d^2 \vec{p}_H} =  \frac{1}{f_g(x_H)} \frac{d \Phi_{P P}^{ \{H \}(0)} (x_H, \vec{p}_H, n, \nu)}{d x_H d^2 \vec{p}_H} \frac{ \sqrt{2} C_A}{(\vec{p}_H^{\; 2})^{i \nu - \frac{1}{2}} e^{i n \phi_H}}  \frac{\alpha_s}{2 \pi}  \int_{x_H}^1 \frac{d z_H}{z_H} f_g \left( \frac{x_H}{z_H} \right)
\end{equation*}
\begin{equation*}
   \times \left \{ \frac{2 m_H^2 z_H^2 }{[(1-z_H) m_H^2 + \vec{p}_H^{\; 2}]} \left[ \frac{(\vec{p}_H^{\; 2})^{i \nu -\frac{1}{2} - \epsilon} e^{i n \phi_H}}{2 \sqrt{2}} \left( \frac{1}{\left( \frac{1}{2} + \frac{n}{2} - i \nu \right)} - \frac{1}{\left(-\frac{1}{2} + \frac{n}{2} + i \nu \right)} \right) \right] \right.  
\end{equation*}
\begin{equation*}
  - 2 m_H^2 z_H^2 \Bigg[ I_{2, \rm{reg}} \left( -1, n, \nu \right) \Biggr.
  \end{equation*}
  \begin{equation*}
      \left. - \frac{(\vec{p}_H^{\; 2})^{\frac{1}{2}}}{2} \left( e^{-i \phi_H} I_{2, \rm{reg}} \left( -\frac{1}{2}, n+1, \nu \right) + e^{i \phi_H} I_{2, \rm{reg}} \left( -\frac{1}{2}, n-1,\nu \right) \right) \Bigg] \right.
\end{equation*}
\begin{equation}
- 4 z_H m_H^2 I_3 (0,1,n, \nu) + z_H (1-z_H) m_H^4 \left[ I_3(0,2,n, \nu) + \frac{2 I_3 (0,1,n,\nu)}{[(1-z_H) m_H^2 + \vec{p}_H^{\; 2}]} \right] .
    \label{RestForFin}
\end{equation}

\subsection{Final result}
\label{ssec:projection_final}

In this subsection we present the final result. First of all, we observe that if we sum all the $\epsilon$-divergent contributions in Eqs.~(\ref{CoupliforDiv})-(\ref{PggforDiv}),~(\ref{BFKLforDiv}),~(\ref{VirforDiv}),~(\ref{QuarkforDiv})-(\ref{(1-xH)forDiv}),~(\ref{CollforDiv}), we get that they cancel completely. The finite parts in the same equations, together with the contributions in Eqs.~(\ref{(1-z_H)ForFin}), (\ref{RestForFin}) represent the final result. Setting $\epsilon=0$, we can cast the final result in the sum of three terms,
\begin{equation*}
    \frac{d \Phi_{P P}^{\{H \}, \rm{NLO}} (x_H, \vec{p}_H, n, \nu)}{d x_H d^2 \vec{p}_H} = \frac{d \Phi_{P P, 1}^{\{H \}, \rm{NLO}} (x_H, \vec{p}_H, n, \nu; s_0)}{d x_H d^2 \vec{p}_H} 
\end{equation*}
\begin{equation}
    + \frac{d \Phi_{P P, 2}^{\{H \}, \rm{NLO}} (x_H, \vec{p}_H, n, \nu)}{d x_H d^2 \vec{p}_H} + \frac{d \Phi_{P P, 3}^{\{H \}, \rm{NLO}} (x_H, \vec{p}_H, n, \nu)}{d x_H d^2 \vec{p}_H} \; ,
\end{equation}
where the first term is given by the sum of all contributions purely proportional to $f_g(x_H)$, {\it i.e.}
\begin{equation*}
    \frac{d \Phi_{P P, 1}^{\{H \}, \rm{NLO}} (x_H, \vec{p}_H, n, \nu; s_0)}{d x_H d^2 \vec{p}_H} \equiv \frac{d \Phi^{{\rm{BFKL}}}_{PP, {\rm{fin}}}(x_H, \vec{p}_H,n,\nu;s_0)}{d x_H d^2 p_H} 
\end{equation*}
\begin{equation}
    + \frac{d \Phi_{PP, {\rm{fin}}}^{ \{H \}} (x_H, \vec{p}_H, n, \nu)}{d x_H d^2 \vec{p}_H} \bigg|_{{\rm{coupling \; c.t.}}} + \frac{d \Phi_{PP,{\rm{fin}}}^{\{ H \}(1)} (x_H, \vec{p}_H, n, \nu)}{d x_H d^2 \vec{p}_H} \; ;
\end{equation}
the second is the sum of all contributions that are taken in convolution with $\sum_a f_a(x_H/z_H)$, {\it i.e.}
\begin{equation*}
    \frac{d \Phi_{P P, 2}^{\{H \}, \rm{NLO}} (x_H, \vec{p}_H, n, \nu; s_0)}{d x_H d^2 \vec{p}_H} \equiv \frac{d \Phi_{PP,{\rm{fin}}}^{ \{H \}}(x_H, \vec{p}_H, n, \nu)}{d x_H d^2 \vec{p}_H}\bigg|_{{\rm{P_{qg} \; c.t.}}} + \frac{d \Phi_{P P, {\rm{fin}}}^{\{H q \}} (x_H, \vec{p}_H, n, \nu)}{d x_H d^2 \vec{p}_H} \; ;
\end{equation*}
the third is the sum of all contributions that are taken in convolution with $f_g(x_H/z_H)$, {\it i.e.}
\begin{equation*}
    \frac{d \Phi_{P P, 3}^{\{H \}, \rm{NLO}} (x_H, \vec{p}_H, n, \nu; s_0)}{d x_H d^2 \vec{p}_H} \equiv \frac{d \Phi_{PP,{\rm{fin}}}^{ \{H \}}(x_H, \vec{p}_H, n, \nu)}{d x_H d^2 \vec{p}_H} \bigg|_{{\rm{P_{gg} \; c.t.}}} + \frac{d \Phi_{P P, {\rm{fin}}}^{\{H g\} {\rm{plus}}} (x_H, \vec{p}_H, n, \nu)}{d x_H d^2 \vec{p}_H} 
\end{equation*}
\begin{equation*}
    + \frac{d \Phi_{P P, {\rm{fin}}}^{\{H g\}(1-x_H)} (x_H, \vec{p}_H, n, \nu)}{d x_H d^2 \vec{p}_H} + \frac{d \Phi_{P P, {\rm{fin}}}^{\{H g\}{\rm{coll}}} (x_H, \vec{p}_H, n, \nu)}{d x_H d^2 \vec{p}_H} 
\end{equation*}
\begin{equation}
    + \frac{d \Phi_{P P}^{\{H g\}(1-z_H)} (x_H, \vec{p}_H, n, \nu)}{d x_H d^2 \vec{p}_H}
    + \frac{d \Phi_{P P}^{\{H g\}{\rm{rest}}} (x_H, \vec{p}_H, n, \nu)}{d x_H d^2 \vec{p}_H} \;.
\end{equation}

\section{Summary and outlook}

We calculated the full NLO correction to the impact factor for the production of a Higgs boson emitted by a proton in the forward rapidity region. 
Its analytic expression was obtained both in the momentum and in the Mellin representations. The latter is particularly relevant to clearly observe a complete cancellation of NLO singularities, and it is useful for future numeric studies. We relied on the large top-mass limit approximation, thus we employed the gluon-Higgs effective field theory. We have found that the Gribov trick (see Eq.~(\ref{Gribov})) cannot be applied to both the $t$-channel gluon legs connected to the effective gluon-Higgs vertex. This prevents the use of the technique outlined in Refs.~\cite{Fadin:1999qc,Fadin:2001dc} to simplify calculations. Formal studies on the generalization of that procedure to non-QCD vertices are underway~\cite{GribovNext}. As a first step forward towards phenomenology, we plan to extend the analysis on high-energy resummed distributions for the inclusive Higgs-plus-jet hadroproduction done in Ref.~\cite{Celiberto:2020tmb}. Our forward-Higgs NLO impact factor is a key ingredient to conduct a precise matching between BFKL and fixed-order approach. \\ 

From a more formal perspective, there are different natural developments of this work. One is the calculation of the Higgs impact factor {\it via} gluon fusion in the central-rapidity region. At the LO level it takes the form of a doubly off-shell coefficient function~\cite{Pasechnik:2006du}, where the top-quark loop connects two incoming Reggeized gluons with the outgoing scalar-boson line. Moving to NLO, the computation of contribution due to real emissions is not complicated, while technical issues are expected to emerge from the extraction of the vertex at 1-loop accuracy, due to the presence of an additional scale in the off-shellness of the incoming Reggeon. Once calculated, this doubly off-shell impact factor will be employed in the description of the central inclusive Higgs production in a pure high-energy factorization scheme, given as a $\kappa_T$-convolution between two UGDs describing the incoming protons and the aforementioned $g^*g^*H$ vertex. Another interesting development, that is under development is the inclusion of top-mass corrections in the forward case. Even in this case, real corrections are quite straightforward, while the biggest problems emerge from virtual corrections that require two-loop calculations.
\chapter{BFKL phenomenology}
\label{Chap:Pheno}

\begin{flushright}
\emph{\textit{Fermi’s inclination toward concrete questions verifiable
by direct \\ experiment was due, at least in part, to his desire to check \\
the soundness of his work by Nature, the infallible judge. \\ Emilio Segrè~\cite{Segrè:1987en}}}
\end{flushright}

In this chapter, we propose a number of new semi-hard reactions as testfield of BFKL dynamics. We focus on a class of partially inclusive processes featuring a forward-plus-backward two-particle final-state configuration. We investigate processes that can be studied at the LHC in proton-proton collisions, in particular, reactions involving, in the final state jets, Higgs bosons and identified heavy-flavored hadrons. This will require the use of a hybrid approach in which both collinear and BFKL ingredients are encoded, to which we will refer as \textit{hybrid high-energy/collinear factorization}. In the previous chapter, we have already seen a specific example by considering the Higgs impact factor. In the case of jets, in addition to a PDF for the description of the initial state in the impact factor, it is necessary to consider a jet algorithm for the final state, while, in the case of production of identified hadrons, it is necessary to make a further convolution with a fragmentation function (FF). The impact factors of the forward jets were calculated in~\cite{Bartels:2001ge,Bartels:2002yj,Caporale:2012ih}, while the impact factors of forward hadrons in~\cite{Ivanov:2012iv}. For our analyses, we use the expressions present in \cite{Caporale:2012ih,Ivanov:2012iv}, where the jet algorithm used is the \textit{small-cone} one~\cite{Furman:1981kf,Aversa:1988vb}. \\

The chapter contains four sections. In the first section, we introduce the hybrid factorization and construct the generic cross section within the NLLA. In the second, we describe the processes under investigation. In particular, we give details about the framework in which we study heavy-flavor production, the so-called \textit{variable-flavor number scheme} (VFNS). We also give details concerning the production of the Higgs in association with a jet or a charmed hadron. In the third section, we present numerical results, while in the fourth we summarize and discuss future perspectives. The material of this chapter is based on Refs.~\cite{Celiberto:2020tmb,Celiberto:2021dzy,Celiberto:2021fdp,Celiberto:2022dyf,Celiberto:2022zdg}, where the numerical elaboration of all the considered observables is done by making use of the {\tt JETHAD} modular work package~\cite{Celiberto:2020wpk,Celiberto:2022rfj}.
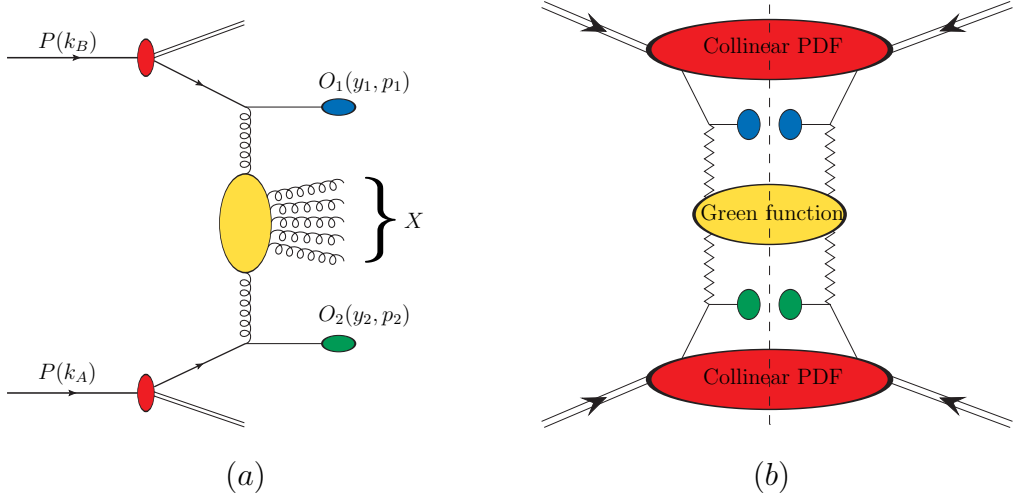
\begin{figure}
\begin{picture}(400,200)
\put(30,20){\includegraphics[width=0.3\textwidth]{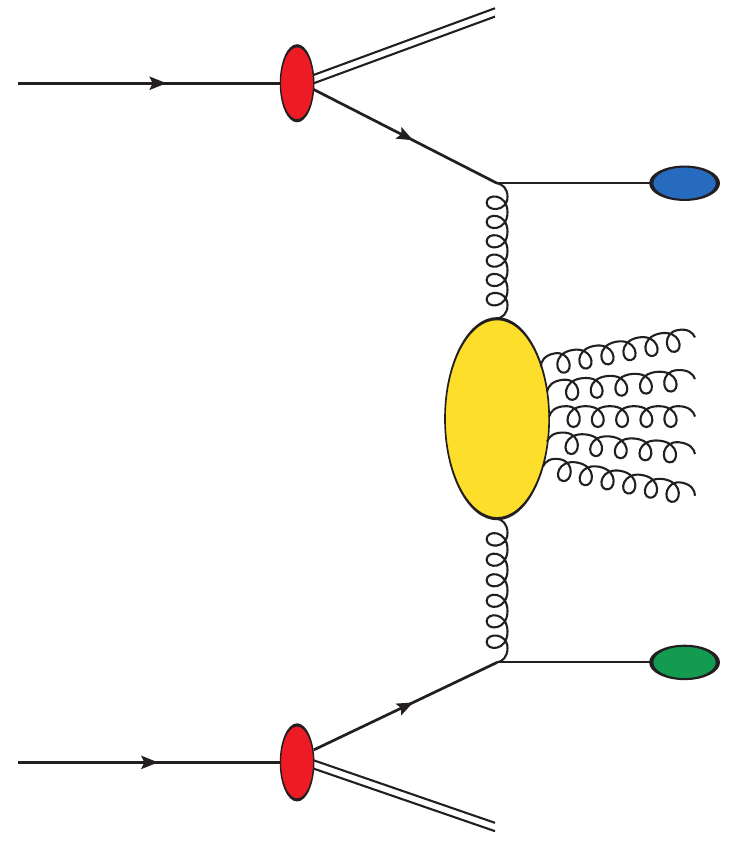}}
\put(230,20){\includegraphics[width=0.4\textwidth]{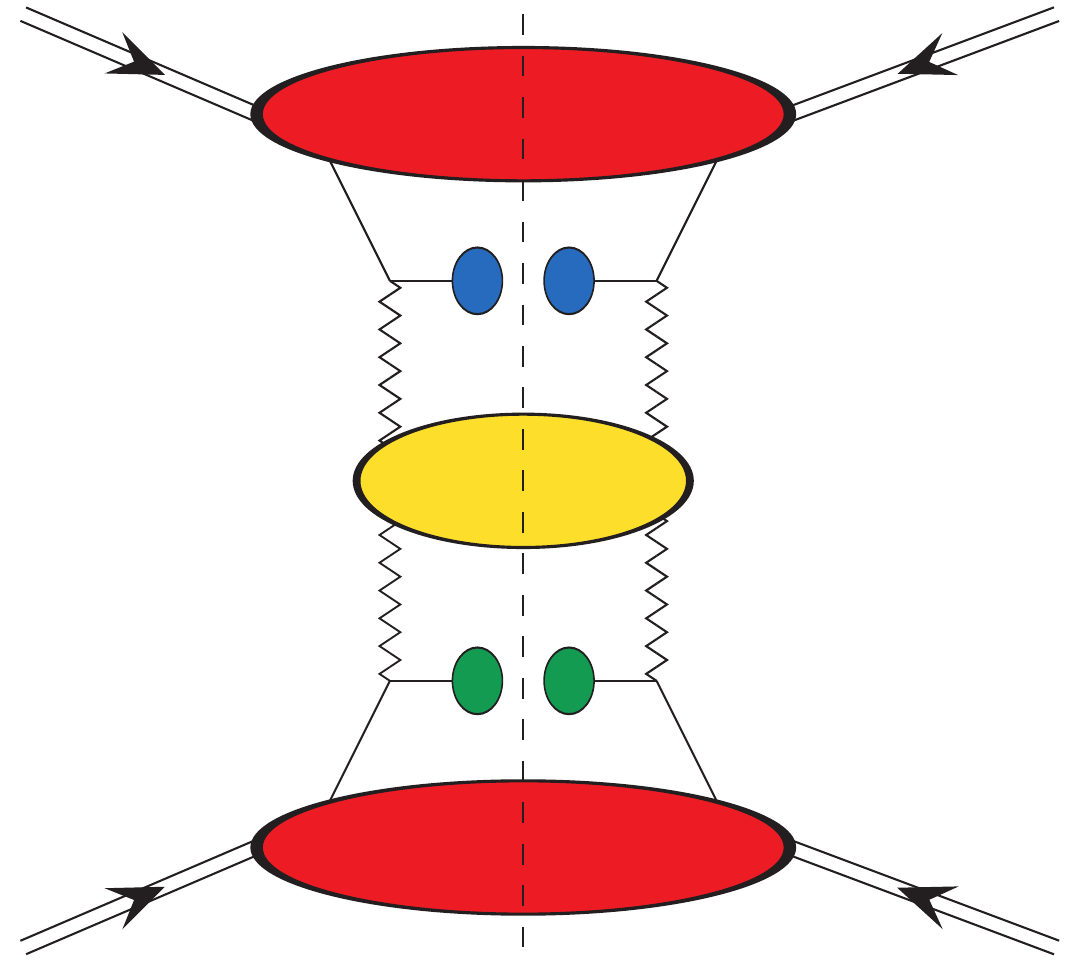}}
\put(115,0){$(a)$}
\put(312.5,0){$(b)$}
\put(45,40){\scalebox{0.7}{$P (k_A)$}}
\put(45,165){\scalebox{0.7}{$P (k_B)$}}
\put(150,150){\scalebox{0.7}{$O_1 (y_1, p_1)$}}
\put(150,62){\scalebox{0.7}{$O_2 (y_2, p_2)$}}
\put(165,92){\scalebox{2.7}{$\}$}}
\put(182,97){\scalebox{0.7}{$X$}}
\put(293,100){\scalebox{0.7}{Green function}}
\put(294,38){\scalebox{0.7}{Collinear PDF}}
\put(294,163){\scalebox{0.7}{Collinear PDF}}
\end{picture}
\caption{$(a)$ Picture of the generic inclusive forward/backward reaction. $(b)$ Schematic representation of the hybrid collinear/high-energy factorization.}
\label{Pheno:Fig:ForBack}
\end{figure}
\section{Theoretical framework}
We want to study the high-energy behavior of interesting observables for the following type of semi-hard reactions:
\begin{equation}
\label{Pheno:Eq:ProtoReac}
    P(k_A) + P(k_B) \longrightarrow O_1 (y_1, p_1) +  X + O_2 (y_2, p_2) \; .
\end{equation}
Here, two objects ($O_1$ and $O_2$) are emitted in proton-proton ($P$) collisions with large transverse momenta, $|\vec{p}_1|$ and $|\vec{p}_2| $, and wide separation in rapidity, $\Delta Y = y_1 - y_2$, together with the radiation of an undetected system, $X$. A schematic representation of the process is given in Fig.~\ref{Pheno:Fig:ForBack}. The hybrid factorization that we use in this section was firstly introduced in the context of Mueller-Navelet jets~\cite{Mueller:1986ey} (two jets separated in rapidity) and then successfully extended and investigated at the next-to-leading order~\cite{Bartels:2001ge,Bartels:2002yj,Caporale:2012ih,Colferai:2010wu,Ducloue:2013hia,Ducloue:2013bva,Caporale:2014gpa,Celiberto:2015yba,Celiberto:2016ygs,Celiberto:2017ius,Celiberto:2022gji}. \\

We begin by observing that the BFKL factorization allows us to construct cross sections as convolutions of two impact factors as defined in Eq.~(\ref{Int:Eq:ImpactFactSingGen}), and the Green's function for the scattering of two Reggeons. In the case of hadroproductions, this is sufficient to describe the parton collision only, but when we want to move on to the description of the whole physical process, we have to introduce some non-perturbative information on the distribution of partons inside protons\footnote{In the case of identified hadrons in the final state, also the hadronization phase must be taken into account.}. This is done exactly the way it was shown in the chapter \ref{Chap:HiggsImp} for the Higgs impact factor. The hybrid factorization is schematically depicted in Fig.~\ref{Pheno:Fig:ForBack}. \\

At this point, a clarification is in order. In literature, hybrid factorization can have slightly different meanings. For instance, recently a general hybrid $k_T$-factorization formula, in which one initial-state parton is light-like, and its associated PDF only depends on the longitudinal momentum fraction, while the other is space-like and its associated PDF also depends on $k_T$, has been established at NLO~\cite{vanHameren:2022mtk}, using the auxiliary parton method. Through this chapter, although our impact factors are obtained considering the collisions of a parton extracted from a PDF (on-shell) and a Reggeon that comes from the BFKL Green's function (off-shell), we never introduce a $k_T$-dependent PDF. This latter object is necessary, for example, if we want to describe single-forward productions. The formalism on which we rely can be adapted to such a situation considering the convolution of a single forward impact factor with an unintegrated gluon distribution (UGD). In this case, however, leaving out the PDF in the impact factor, this would look more like a pure $k_T$-factorization, in which the impact factor and the unintegrated gluon distribution are linked only through the factorization in the transverse momentum. \\

Another important clarification concerns the use of the optical theorem. As we know, in the forward elastic case, the BFKL approach allows us to construct the imaginary part of the amplitude of an elastic process, $A + B \longrightarrow A + B$. Through the optical theorem, Eq.~(\ref{Int:Eq:HighEneOpThe}), we can immediately obtain the total cross section of the process $A+B \rightarrow X$. Since in this case we want to study partially inclusive processes of the type (\ref{Pheno:Eq:ProtoReac}), we actually need to use one of the generalizations of the optical theorem\footnote{The generalization we need is due to A.~H.~Mueller and it can be found in~\cite{Collins:1977rt}.} and adjust the imaginary amplitude constructed via the BFKL approach accordingly. For this purpose, it is necessary to consider differential impact factors in which the kinematical variables of the tagged particles are left unintegrated and where the sum over all possible intermediate states in the $s$-channel cut is restricted to the final states compatible with the process~(\ref{Pheno:Eq:ProtoReac}).
\subsection{Next-to-leading order BFKL cross section}
In this subsection, we build the differential cross section of the generic process, (\ref{Pheno:Eq:ProtoReac}), within the NLLA.
We recall the representation introduced in Eq.~(\ref{Int:MomRepr}), which allows us to represent the BFKL cross section as
\begin{equation}
\label{Pheno:Eq:Diffcross1}
\frac{d\hat{\sigma}_{AB}(z_1, \vec{p}_1, z_2, \vec{p}_2 ,s)}{d z_1 d^2\vec{p}_1 d z_2 d^2 \vec{p}_2} = \frac{1}{(2\pi)^2} \int_{\delta-i\infty}^{\delta+i\infty} \frac{d\omega}{2\pi i} \left(\frac{\hat{s}}{s_0}\right)^{\omega} \braket{\frac{d\Phi_{AA}}{\vec{q}_A^{\; 2}} |\hat{G}_{\omega}|\frac{d\Phi_{BB}}{\vec{q}_B^{\; 2}}} \; ,
\end{equation}
where $z_1, z_2$ are the longitudinal Sudakov fractions with respect to the proton momenta $k_A$ of the two detected particles. 
In this representation, the Green's function solving the BFKL equation is
\begin{equation}
\label{Pheno:Eq:BFKLGreenSol}
\hat{G}_{\omega} =\left(\omega-\hat{K}\right)^{-1}.
\end{equation}
The NLO kernel can be expanded in the strong coupling,
\begin{equation}
\label{Pheno:Eq:expKernel}
\hat{K} = \bar{\alpha}_{s} \hat{K}^{0} + \bar{\alpha}_{s}^{2} \hat{K}^{1} ,
\end{equation}
where $\bar{\alpha}_{s}$ is defined as in Eq.~(\ref{Int:Eq:AlphaStrongBar}), $\hat{K}^{0}$ is the BFKL kernel in the LLA and $\hat{K}^{1}$ represents the NLLA correction.
To determine a cross section with next-to-leading accuracy we need an approximate solution of Eq.~(\ref{Pheno:Eq:BFKLGreenSol}). With the required accuracy, this solution is 
\begin{equation}
\label{Pheno:Eq:ExpGreen}
\hat{G}_{\omega} = \left(\omega-\bar{\alpha}_s \hat{K}^0 \right)^{-1}+\left(\omega-\bar{\alpha}_s \hat{K}^0 \right)^{-1} \left( \bar{\alpha}_s^2 \hat{K}^1 \right) \left(\omega-\bar{\alpha}_s \hat{K}^0 \right)^{-1} + \mathcal{O} \left[\left(\bar{\alpha}_s^2 \hat{K}^1 \right)^{2} \right] \; .
\end{equation}  
In this representation the eigenfunctions of the LO BFKL kernel are denoted by the set of kets $\{ \ket{n,\nu} \; : \; \nu \in {\rm I\!R} \; , \; n \in \mathbb{Z} \}$, defined by the relation
\begin{equation}
\braket{\vec{q} \; |n,\nu} = \frac{1}{\pi \sqrt{2}} \left(\vec{q}^{ \; 2}\right)^{i\nu-\frac{1}{2}}e^{in\theta} \; .
\end{equation}
The action of the LLA kernel is
\begin{equation}
\label{BFKLeigenvalue}
\hat{K}^0 \ket{n,\nu} = \chi \left(n,\nu\right)\ket{n,\nu}, 
\end{equation}
where the function $\chi \left(n,\nu\right)$ is the Lipatov characteristic function defined in Eq.~(\ref{Int:Eq:CharacteristicFunction}).
For the basis introduced, the orthonormality conditions
\begin{equation}
\braket {n',\nu'|n,\nu} = \int \frac{d^2 \vec{q}_A }{2\pi^2} \left( \vec{q}_A^{\; 2} \right)^{i\nu-i\nu'-1} e^{i(n-n') \theta_A} = \delta \left( \nu-\nu' \right) \delta_{nn'}
\end{equation} 
hold and the identity operator, with respect to this basis, can be written as
\begin{equation}
\label{Identity}
\hat{1} = \sum_{n=-\infty}^{\infty} \int_{-\infty}^{+\infty} d\nu \ket{n,\nu} \bra{n,\nu}.
\end{equation}
The action of the full NLA BFKL kernel on these functions may be expressed as follows:
\begin{equation*}
    \hat{K} \ket{n,\nu} = \bar{\alpha}_s \left( \mu_R \right) \chi \left(n,\nu \right) \ket{n,\nu} + \bar{\alpha}_s^2 \left( \mu_R \right) \Big( \chi^{(1)}\left( n,\nu \right) + \frac{\beta_0}{4N_c} \chi \left(n,\nu \right) \ln \left(\mu_R^2\right) \Big) \ket{n,\nu}
\end{equation*}
\begin{equation}
\label{NLLAKER}
+ \bar{\alpha}_s^2 \left(\mu_R\right) \frac{\beta_0}{4N_c} \chi \left(n,\nu \right) \Big(i\frac{\partial}{\partial \nu} \Big) \ket{n, \nu} ,
\end{equation}
where $\mu_R$ is the renormalization scale of the QCD coupling, while
\begin{equation}
\label{BETA}
\beta_0 = \frac{11}{3}N_c-\frac{2}{3}n_f 
\end{equation}
is the first coefficient of the QCD $\beta$-function. The first term of Eq.~(\ref{NLLAKER}) represents the action of LLA kernel, while the second term and the third ones stand for the diagonal and non-diagonal parts of the NLA kernel. For the calculation of $ \chi^{(1)} \left(n,\nu \right)$ see~\cite{Kotikov:2000pm,Kotikov:2002ab}. Here, only the final expression is presented:
\begin{equation}
\chi^{(1)}(n, \nu)=-\frac{\beta_0}{8N_c} \left( \chi^2(n, \nu)-\frac{10}{3} \chi(n, \nu)-i \chi'(n, \nu)\right)+ \bar{\chi}(n, \nu),
\end{equation}
where
\begin{equation*}
\bar{\chi}(n, \nu) =  - \frac{1}{4} \left[ \frac{\pi^2-4}{3} \chi(n, \nu)-6 \zeta(3) - \chi''(n, \nu)+2 \phi(n, \nu) + 2 \phi(n, -\nu) \right.
\end{equation*}
\begin{equation}
    \left. + \frac{\pi^2 \sinh(\pi \nu)}{2 \nu \cosh^2(\pi \nu)} \left( \left( 3+ \left(1+\frac{n_f}{N_c^3} \right) \frac{11+12 \nu^2}{16(1+\nu^2)} \right) \delta_{n0} -\left( 1+\frac{n_f}{N_c^3} \right) \frac{1+4\nu^2}{32(1+\nu^2)} \delta_{n2} \right) \right] \; , \vspace{0.2 cm}
\end{equation}
\begin{equation*}
\phi(n, \nu) = - \int_0^1 dx \frac{x^{-1/2+i\nu+n/2}}{1+x} \left[ \frac{1}{2} \left( \psi' \left( \frac{n+1}{2} \right) - \zeta(2) \right) + {\rm{Li}}_2(x) + {\rm{Li}}_2(-x) \right. 
\end{equation*}
\begin{equation*}
    \left. + \ln x \left( \psi(n+1) - \psi(1) + \ln(1+x) + \sum_{k=1}^{\infty} \frac{(-x)^k}{k+n} \right) \sum_{k=1}^{\infty} \frac{x^k}{(k+n)^2}(1-(-1)^k) \right]  
\end{equation*}
\begin{equation*}
    = \sum_{k=0}^{\infty} \frac{(-1)^{k+1}}{k+(n+1)/2+i\nu} \left[\psi'(k+n+1)-\psi'(k+1)+(-1)^{k+1}(\beta'(k+n+1)+\beta'(k+1)) \right. 
\end{equation*}
\begin{equation}
    \left. -\frac{1}{k+(n+1)/2+i\nu} (\psi(k+n+1)-\psi(k+1)) \right],
\end{equation}
and
\begin{equation}
\beta'(z) = \frac{1}{4} \left[ \psi' \left( \frac{z+1}{2} \right) -\psi' \left( \frac{z}{2} \right) \right], \quad {\rm{Li}}_2(x) = - \int_0^x dt \frac{\ln (1-t)}{t} \; .
\end{equation}
Here $\chi'(n, \nu)=d\chi(n, \nu)/d\nu$ and $\chi''(n, \nu)=d^2\chi(n, \nu)/d\nu^2$. \\
Starting from the expression (\ref{Pheno:Eq:Diffcross1}) for the differential cross section and inserting two identity operators in the form of Eq.~(\ref{Identity}), one obtains
\begin{equation*}
    \frac{d\hat{\sigma}_{AB}(z_1, \vec{p}_1, z_2, \vec{p}_2 ,s)}{ d z_1 d^2 \vec{p}_1 d z_2 d^2\vec{p}_2} = \frac{1}{(2\pi)^2} \sum_{n=-\infty}^{\infty} \int d\nu \sum_{n'=-\infty}^{\infty} \int d\nu'
\end{equation*}
\begin{equation}
\label{Diff.cross.sec2}
\times \int_{\delta-i\infty}^{\delta+i\infty} \frac{d\omega}{2\pi i} \left(\frac{\hat{s}}{s_0}\right)^{\omega} \braket{\frac{d\Phi_{AA}}{\vec{q}_A^{\; 2}} |n,\nu}  \braket{n, \nu |\hat{G}_{\omega}|n',\nu'} \braket{n',\nu'|\frac{d\Phi_{BB}}{\vec{q}_B^{\; 2}}}.
\end{equation}
In Eq.~(\ref{Diff.cross.sec2}), we have the projection of the impact factors onto the LO order BFKL eigenfunctions
\begin{equation}
\label{wavefunc1}
\frac{d\Phi_{AA} \left( z_1, \vec{p}_1, n, \nu \right)}{d z_1 d^2 \vec{p}_1} = \braket{ \frac{d\Phi_{AA}}{\vec{q}_A^{ \; 2}} | n, \nu} = \int d^2 \vec{q}_A \frac{1}{\vec{q}_A^{\; 2}} \frac{d\Phi_{AA} \left(z_1, \vec{p}_1, \vec{q}_A  \right)}{d z_1 d^2 \vec{p}_1 } \frac{1}{\pi \sqrt{2}} \left( \vec{q}_A^{ \; 2} \right)^{i\nu-\frac{1}{2}}e^{in \theta_A},
\end{equation}
\begin{equation}
\label{wavefunc2}
\frac{d\Phi_{BB}^* (z_2, \vec{p}_2, n, \nu)}{d z_2 d^2 \vec{p}_2} = \braket{ n', \nu'| \frac{d\Phi_{BB}}{\vec{q}_B^{ \; 2}} } = \int d^2 \vec{q}_B  \frac{1}{\vec{q}_B^{\; 2}} \frac{d\Phi_{BB} \left( z_2, \vec{p}_2, -\vec{q}_B \right)}{d z_2 d^2 \vec{p}_2} \frac{1}{\pi \sqrt{2}} \left( \vec{q}_B^{\; 2} \right)^{-i \nu' -\frac{1}{2}}e^{-i n' \theta_B} \; ,
\end{equation}
and the matrix element of the BFKL Green's function, which is obtained by substituting the expansion for $\hat{G}_{\omega}$ (Eq.~(\ref{Pheno:Eq:ExpGreen})) in the braket $\braket{n, \nu|\hat{G}_{\omega}|n', \nu'}$:
\begin{equation}
\braket{n, \nu|\hat{G}_{\omega}|n', \nu'} = \bra{n, \nu} \left[ \left(\omega-\bar{\alpha}_s \hat{K}^0 \right)^{-1}+\left(\omega-\bar{\alpha}_s \hat{K}^0 \right)^{-1} \left( \bar{\alpha}_s^2 \hat{K}^1 \right) \left(\omega-\bar{\alpha}_s \hat{K}^0 \right)^{-1} \right]\ket{n', \nu'}.
\end{equation}
Since one knows how the operators $\hat{K}_0$, $\hat{K}_1$ act on the states $\ket{n, \nu}$, this final expression can be easily calculated: 
\begin{equation*}
\braket{n, \nu |\hat{G}_{\omega}|n',\nu'} = \delta_{nn'} \left[ \delta \left( \nu-\nu' \right) \left( \frac{1}{ \omega -\bar{\alpha}_s \left(\mu_R \right) \chi \left( n,\nu \right)} \right. \right. 
\end{equation*}
\begin{equation*}
    \left. \left. + \frac{\bar{\alpha}_s^2 (\mu_R) \left( \bar{\chi} \left(n,\nu\right)+\frac{\beta_0}{8N_c}\left(-\chi^2\left(n,\nu\right)+ \frac{10}{3} \chi \left(n,\nu\right) + 2\chi \left(n, \nu \right) \ln \mu_R^2 +i\frac{\partial}{\partial \nu} \chi \left(n,\nu \right) \right) \right)}{ \left( \omega - \bar{\alpha}_s \left(\mu_R \right) \chi \left( n,\nu \right) \right)^2} \right) \right.
\end{equation*}
\begin{equation}
\left. + \frac{\frac{\beta_0}{4N_c}\bar{\alpha}_s^2\left(\mu_R\right)\chi \left(n,\nu'\right)}{\left( \omega -\bar{\alpha}_s \left(\mu_R \right) \chi \left( n,\nu \right) \right) \left( \omega -\bar{\alpha}_s \left(\mu_R \right) \chi \left( n,\nu' \right)\right)} \left(i \frac{d}{d\nu'} \delta \left( \nu - \nu' \right) \right) \right].
\label{matrixelementG}   
\end{equation}
The projected impact factors can be written as
\begin{equation}
\frac{d\Phi_{AA} \left(z_1, \vec{p}_1, n, \nu \right)}{d z_1 d^2 \vec{p}_1}  = \alpha_s \left(\mu_R \right) \left[ \tilde{c}_1 \left( z_1, \vec{p}_1, n, \nu \right) + \alpha_s (\mu_R) \tilde{c}_1^{(1)} \left( z_1, \vec{p}_1, n, \nu \right) \right] \; ,
\label{Pheno:Eq:ExpProImp1}
\end{equation}
\begin{equation}
\begin{split}
\frac{d\Phi_{BB}^{*} \left( z_2, \vec{p}_2 , n', \nu' \right)}{d z_2 d^2 \vec{p}_2 } = \alpha_s \left( \mu_R \right) \left[ \tilde{c}_2 \left( z_2, \vec{p}_2, n', \nu' \right) + \alpha_s (\mu_R) \tilde{c}_2^{(1)} \left( z_1, \vec{p}_1, n', \nu' \right) \right] \; .
\end{split}
\label{Pheno:Eq:ExpProImp2}
\end{equation}
Inserting the expression for the matrix element, (\ref{matrixelementG}), together with the expressions for the impact factors projected onto the eigenfunctions of the kernel, (\ref{Pheno:Eq:ExpProImp1}, \ref{Pheno:Eq:ExpProImp2}), in Eq.~(\ref{Diff.cross.sec2}), one obtains
\begin{equation*}
\hspace{-0.3 cm} \frac{d\hat{\sigma}_{AB}(z_1, \vec{p}_1, z_2, \vec{p}_2, s)}{ d z_1 d^2 \vec{p}_1 d z_2 d^2 \vec{p}_2 } = \frac{1}{(2\pi)^2} \sum_{n=-\infty}^{\infty} \hspace{-0.05 cm} \int \hspace{-0.1 cm} \frac{d \nu}{2 \pi i} \hspace{-0.05 cm} \int  d\nu' \hspace{-0.15 cm} \int_{\delta-i\infty}^{\delta+i\infty} \hspace{-0.65 cm} d \omega \left(\frac{\hat{s}}{s_0}\right)^{\omega} \hspace{-0.1 cm} \alpha_s^2(\mu_R) \tilde{c}_1(z_1, \vec{p}_1, n, \nu) \tilde{c}_2( z_2, \vec{p}_2 , n', \nu')   
\end{equation*}
\begin{equation*}
    \times  \left[ \delta \left( \nu-\nu' \right) \left( \frac{1}{ \omega -\bar{\alpha}_s \left(\mu_R \right) \chi \left( n,\nu \right)} \left( 1 + \alpha_s (\mu_R) \left( \frac{\tilde{c}_1^{(1)} \left( z_1, \vec{p}_1, n, \nu \right) }{\tilde{c}_1 \left( z_1, \vec{p}_1, n, \nu \right)} + \frac{\tilde{c}_2^{(1)} \left( z_1, \vec{p}_1, n', \nu' \right)}{\tilde{c}_2 \left( z_1, \vec{p}_1, n', \nu' \right)} \right) \right) \right. \right. 
\end{equation*}
\begin{equation*}
    \left. \left. + \frac{\bar{\alpha}_s^2 \left( \bar{\chi} \left(n,\nu\right)+\frac{\beta_0}{8N_c}\left(-\chi^2\left(n,\nu\right) + \frac{10}{3} \chi \left(n,\nu\right) + 2\chi \left(n, \nu \right) \ln \mu_R^2 +i\frac{\partial}{\partial \nu} \chi \left(n,\nu \right) \right) \right)}{ \left( \omega -\bar{\alpha}_s \left(\mu_R \right) \chi \left( n,\nu \right) \right)^2} \right) \right. 
\end{equation*}
\begin{equation}
    \left. + \frac{\frac{\beta_0}{4N_c}\bar{\alpha}_s^2\left(\mu_R\right)\chi \left(n,\nu'\right)}{\left( \omega -\bar{\alpha}_s \left(\mu_R \right) \chi \left( n,\nu \right) \right) \left( \omega -\bar{\alpha}_s \left(\mu_R \right) \chi \left( n,\nu' \right)\right)} \left(i \frac{d}{d\nu} \delta \left( \nu - \nu' \right) \right) \right].
    \label{long}
\end{equation}
In Eq.~(\ref{long}) there are three terms. The first term is
\begin{equation}
\begin{split}
& \frac{1}{(2\pi)^2} \sum_{n=-\infty}^{\infty} \int d\nu \int_{\delta-i\infty}^{\delta+i\infty} \frac{d\omega}{2\pi i} \left(\frac{\hat{s}}{s_0}\right)^{\omega} \alpha_s^2(\mu_R) \tilde{c}_1 (z_1, \vec{p}_1, n, \nu) \tilde{c}_2 (z_2, \vec{p}_2, n, \nu) \\ & \times \frac{1}{\omega-\bar{\alpha}_s \left(\mu_R \right) \chi \left( n,\nu \right)} \left( 1 + \alpha_s (\mu_R) \left( \frac{\tilde{c}_1^{(1)} \left( z_1, \vec{p}_1, n, \nu \right) }{\tilde{c}_1 \left( z_1, \vec{p}_1, n, \nu \right)} + \frac{\tilde{c}_2^{(1)} \left( z_1, \vec{p}_1, n', \nu' \right)}{\tilde{c}_2 \left( z_1, \vec{p}_1, n', \nu' \right)} \right) \right) \; ,
\end{split}
\end{equation}
that, using the residue theorem, becomes
\begin{equation}
\begin{split}
 \frac{1}{(2\pi)^2} & \sum_{n=-\infty}^{\infty} \int d\nu \; \alpha_s^2(\mu_R) \tilde{c}_1(z_1, \vec{p}_1, n, \nu) \tilde{c}_2( z_2, \vec{p}_2, n, \nu) \left(\frac{\hat{s}}{s_0}\right)^{\bar{\alpha}_s \chi(n, \nu)} \\ & \times \left( 1 + \alpha_s (\mu_R) \left( \frac{\tilde{c}_1^{(1)} \left( z_1, \vec{p}_1, n, \nu \right) }{\tilde{c}_1 \left( z_1, \vec{p}_1, n, \nu \right)} + \frac{\tilde{c}_2^{(1)} \left( z_1, \vec{p}_1, n', \nu' \right)}{\tilde{c}_2 \left( z_1, \vec{p}_1, n', \nu' \right)} \right) \right) \; ,
\end{split}
\end{equation}
The second term is
\begin{equation}
\begin{split}
& \frac{1}{(2\pi)^2} \sum_{n=-\infty}^{\infty} \int d\nu \int_{\delta-i\infty}^{\delta+i\infty} \frac{d\omega}{2\pi i} \left(\frac{\hat{s}}{s_0}\right)^{\omega} \alpha_s^2(\mu_R) \tilde{c}_1 (z_1, \vec{p}_1, n, \nu) \tilde{c}_2 (z_2, \vec{p}_2 , n, \nu) \\ & \times \frac{\bar{\alpha}_s^2 \left[ \bar{\chi}(n, \nu) + \frac{\beta_0}{8N_c} \left( -\chi(n, \nu)^2 + \frac{10}{3} \chi(n, \nu) + 2 \chi(n, \nu) \ln \mu_R^2 + i\chi'(n, \nu) \right) \right]}{\left( \omega -\bar{\alpha}_s \left(\mu_R \right) \chi \left( n,\nu \right) \right)^2} \; ,
\end{split}
\end{equation}
that, using the residue theorem, becomes
\begin{equation}
\begin{split}
& \frac{1}{(2\pi)^2} \sum_{n=-\infty}^{\infty} \int d\nu \; \alpha_s^2(\mu_R) \tilde{c}_1(z_1, \vec{p}_1, n, \nu ) \tilde{c}_2(z_2, \vec{p}_2,  n, \nu) \bar{\alpha}_s^2 ( \mu_R ) \left(\frac{\hat{s}}{s_0}\right)^{\bar{\alpha}_s (\mu_R) \chi(n, \nu)} \ln \left( \frac{\hat{s}}{s_0} \right) \\ & \times \left[ \bar{\chi}(n, \nu) + \frac{\beta_0}{8N_c} \left( -\chi(n, \nu)^2 + \frac{10}{3} \chi(n, \nu) + 2 \chi(n, \nu) \ln \mu_R^2 + i\chi'(n, \nu) \right) \right] \; .
\end{split}
\end{equation}
The third term is
\begin{equation}
\begin{split}
& \frac{1}{(2\pi)^2} \sum_{n=-\infty}^{\infty} \int d\nu \int d\nu' \int_{\delta-i\infty}^{\delta+i\infty} \frac{d\omega}{2\pi i} \left(\frac{\hat{s}}{s_0}\right)^{\omega} \alpha_s^2(\mu_R) \tilde{c}_1( z_1, \vec{p}_1, n, \nu ) \tilde{c}_2( z_2, \vec{p}_2, n, \nu ') \\ & \times \frac{\beta_0}{4N_C} \bar{\alpha}_s^2(\mu_R) \frac{\chi(n, \nu')}{(\omega-\bar{\alpha}_s \left(\mu_R \right) \chi \left( n,\nu \right))(\omega-\bar{\alpha}_s \left(\mu_R \right) \chi \left( n,\nu' \right))}\left( i\frac{d}{d\nu'} \delta(\nu-\nu') \right).
\end{split}
\end{equation}
Let us focus the attention only on the integration in $\nu$ and $\nu'$. One can use the following symmetrization:
\begin{equation*}
\int d\nu \int d\nu' \frac{\tilde{c}_1 (z_1, \vec{p}_1, n, \nu ) \tilde{c}_2 (z_2, \vec{p}_2 , n, \nu') \chi(n, \nu')}{(\omega-\bar{\alpha}_s \left(\mu_R \right) \chi \left( n,\nu \right))(\omega-\bar{\alpha}_s \left(\mu_R \right) \chi \left( n,\nu' \right))} i \frac{d}{d\nu'} \delta(\nu-\nu') 
\end{equation*}
\begin{equation*}
    = \frac{1}{2} \int d\nu \int d\nu' \frac{ \tilde{c}_1(z_1, \vec{p_1}, n, \nu) \tilde{c}_2(z_2, \vec{p}_2, n, \nu') \chi(n, \nu')}{(\omega-\bar{\alpha}_s \left(\mu_R \right) \chi \left( n,\nu \right))(\omega-\bar{\alpha}_s \left(\mu_R \right) \chi \left( n,\nu' \right))} i \frac{d}{d\nu'} \delta(\nu-\nu') 
\end{equation*}
\begin{equation}
    + \frac{ \tilde{c}_1 (z_1, \vec{p}_1, n, \nu') \tilde{c}_2 (z_2, \vec{p}_2, n, \nu) \chi(n, \nu)}{(\omega-\bar{\alpha}_s \left(\mu_R \right) \chi \left( n,\nu' \right))(\omega-\bar{\alpha}_s \left(\mu_R \right) \chi \left( n,\nu \right))} i \frac{d}{d\nu} \delta(\nu'-\nu) \; ,
\end{equation}
and, thereafter, the following properties:
\begin{equation*}
i \frac{d}{d\nu} \delta(\nu'-\nu) = -i \frac{d}{d\nu'} \delta(\nu'-\nu),
\end{equation*}
\begin{equation}
\int d\nu' \left( \frac{d}{d\nu'} \delta(\nu'-\nu) \right) f(\nu') = -f'(\nu')\big|_{\nu'=\nu} \; . 
\end{equation}
The resulting form for the term considered is
\begin{equation*}
- \frac{i}{2} \int d\nu \left[ \frac{ \tilde{c}_1( z_1, \vec{p}_1, n, \nu ) \tilde{c}'_2 ( z_2, \vec{p}_2, n, \nu) - \tilde{c}'_1 (z_1, \vec{p}_1, n, \nu) \tilde{c}_2(z_2, \vec{p}_2, n, \nu)}{(\omega-\bar{\alpha}_s \left(\mu_R \right) \chi \left( n,\nu \right))^2} \chi \left( n, \nu \right) \right.
\end{equation*}
\begin{equation}
\left. + \frac{\tilde{c}_1(z_1, \vec{p}_1, n, \nu) \tilde{c}_2(z_2, \vec{p}_2, n, \nu)}{(\omega-\bar{\alpha}_s \left(\mu_R \right) \chi \left( n,\nu \right))^2} \chi'(n, \nu) \right],
\end{equation}
that, using 
\begin{equation}
\begin{split}
\tilde{c}'_1 (& z_1, \vec{p}_1, n, \nu) \tilde{c}_2 (z_2, \vec{p}_2, n, \nu) - \tilde{c}_1 (z_1, \vec{p}_1, n, \nu) \tilde{c}'_2 (z_2, \vec{p}_2, n, \nu) \\ & = \tilde{c}_1 (z_1, \vec{p}_1, n, \nu) \tilde{c}_2 (z_2, \vec{p}_2, n, \nu) \frac{d}{d\nu} \ln \frac{ \tilde{c}_1 (z_1, \vec{p}_1, n, \nu)}{ \tilde{c}_2(z_2, \vec{p}_2, n, \nu)} \; ,
\end{split}
\end{equation}
can be put in the following form:
\begin{equation*}
- \frac{i}{2} \int d\nu \left[ \frac{- \tilde{c}_1 (z_1, \vec{p}_1, n, \nu) \tilde{c}_2 (z_2, \vec{p}_2, n, \nu) \frac{d}{d\nu} \ln \frac{ \tilde{c}_1(z_1, \vec{p}_1, n, \nu)}{ \tilde{c}_2 (z_2, \vec{p}_2, n, \nu)}}{(\omega-\bar{\alpha}_s \left(\mu_R \right) \chi \left( n,\nu \right))^2} \chi \left( n, \nu \right) \right.
\end{equation*}
\begin{equation}
\left. + \frac{\tilde{c}_1 (z_1, \vec{p}_1, n, \nu) \tilde{c}_2 (z_2, \vec{p}_2, n, \nu)}{(\omega-\bar{\alpha}_s \left(\mu_R \right) \chi \left( n,\nu \right))^2} \chi'(n, \nu) \right] \; .
\end{equation}
Now, one can go back to the complete expression for the third term and use the residue theorem to obtain
\begin{equation}
\begin{split}
\frac{1}{(2 \pi)^2} \sum_{n = -\infty}^{\infty} & \int d \nu \; \alpha_s^2 (\mu_R) \tilde{c}_1 (z_1, \vec{p}_1, n, \nu) \tilde{c}_2 (z_2, \vec{p}_2, n, \nu) \left( \frac{\hat{s}}{s_0} \right)^{\bar{\alpha}_s (\mu_R) \chi(n, \nu)} \ln \left( \frac{\hat{s}}{s_0} \right) \\ & \times \left[ \bar{\alpha}_s^2 (\mu_R) \frac{\beta_0}{8N_c} \chi(n, \nu) \left(i \frac{d}{d\nu} \ln \left( \frac{\tilde{c}_1 (z_1, \vec{p}_1, n, \nu)}{\tilde{c}_2 (z_2, \vec{p}_2, n, \nu)} \right)- i \frac{\chi'(n, \nu)}{\chi(n, \nu)} \right) \right] \; .
\end{split}
\end{equation}
Summing the three terms, the result is 
\begin{equation*}
\frac{d\hat{\sigma}_{AB}(z_1, \vec{p}_1, z_2, \vec{p}_2 ,s)}{d z_1 d^2\vec{p}_1 d z_2 d^2\vec{p}_2 } = \frac{1}{(2\pi)^2} \sum_{n=-\infty}^{\infty} \int d\nu \left(\frac{\hat{s}}{s_0}\right)^{\bar{\alpha}_s(\mu_R)\chi(n,\nu)} \hspace{-0.3 cm} \alpha_s^2(\mu_R) \tilde{c}_1(z_1, \vec{p}_1, n, \nu) \tilde{c}_2(z_2, \vec{p}_2, n, \nu)
\end{equation*}
\begin{equation*}
     \times \left[1 + \alpha_s (\mu_R) \left( \frac{\tilde{c}_1^{(1)} \left( z_1, \vec{p}_1, n, \nu \right) }{\tilde{c}_1 \left( z_1, \vec{p}_1, n, \nu \right)} + \frac{\tilde{c}_2^{(1)} \left( z_1, \vec{p}_1, n, \nu \right)}{\tilde{c}_2 \left( z_1, \vec{p}_1, n, \nu \right)} \right)  +  \bar{\alpha}_s^2 (\mu_R) \ln \left(\frac{\hat{s}}{s_0}\right) \right. 
\end{equation*}
\begin{equation}
\label{Diff.cross.sec.3}
 \left. \times \left\{ \bar{\chi}(n, \nu)+\frac{\beta_0}{8N_c} \chi(n, \nu) \left( -\chi(n, \nu)+\frac{10}{3}+ 2 \ln \mu_R^2 +i \frac{d}{d\nu} \ln \frac{\tilde{c}_1( z_1, \vec{p}_1, n, \nu)}{\tilde{c}_2( z_2, \vec{p}_2, n, \nu)} \right) \right\} \right] .
\end{equation}
The expression for the cross section given in Eq.~(\ref{Diff.cross.sec.3}) is valid both in the LLA and in the NLLA. However, it is not the only possible one. Actually, several NLA-equivalent expressions can be adopted; we use the \textit{exponential representation},  
\begin{equation*}
\frac{d\hat{\sigma}_{AB}(z_1, \vec{p}_1, z_2, \vec{p}_2 ,s)}{d z_1 d^2\vec{p}_1 d z_2 d^2\vec{p}_2 } = \frac{1}{(2\pi)^2} \sum_{n = -\infty}^{\infty} \int d\nu \left(\frac{\hat{s}}{s_0}\right)^{\bar{\alpha}_s\left(\mu_R\right)\chi\left(n,\nu\right)} 
\end{equation*}
\begin{equation*}
    \times \left(\frac{\hat{s}}{s_0}\right)^{\bar{\alpha}_s^2\left(\mu_R\right)\left(\bar{\chi}\left(n,\nu \right)+\frac{\beta_0}{8N_c}\chi\left(n,\nu\right)\left(-\chi \left(n,\nu\right)+\frac{10}{3}+2\ln\frac{\mu_R^2}{\sqrt{s_1 s_2}} \right)\right)} \alpha_s^2 (\mu_R) \tilde{c}_1 (z_1, \vec{p}_1, n, \nu) \tilde{c}_2 (z_2, \vec{p}_2, n, \nu) \vspace{0.1 cm}
\end{equation*}
\begin{equation}
\label{expoform}
     \times \left[ 1 + \alpha_s (\mu_R) \left( \frac{\tilde{c}_1^{(1)} \left( z_1, \vec{p}_1, n, \nu \right) }{\tilde{c}_1 \left( z_1, \vec{p}_1, n, \nu \right)} + \frac{\tilde{c}_2^{(1)} \left( z_1, \vec{p}_1, n, \nu \right)}{\tilde{c}_2 \left( z_1, \vec{p}_1, n, \nu \right)} \right) + \bar{\alpha}_s^2 ( \mu_R ) \ln \left(\frac{\hat{s}}{s_0} \right) \frac{\beta_0}{4N_c} \chi(n, \nu) f(\nu) \right],
\end{equation}
where
\begin{equation}
2 f(\nu) = i \frac{d}{d\nu} \ln \left( \frac{\tilde{c}_1 (z_1, \vec{p}_1, n, \nu)}{ \tilde{c}_2 (z_2, \vec{p}_2, n, \nu)} \right) + 2 \ln \sqrt{s_1 s_2} \; ,
\end{equation}
and $s_1,s_2$ denote here the hard scales which enter the impact factors $\tilde{c}_{1,2}$. We stress that the difference between the two representations is beyond the NLLA~\cite{Caporale:2014gpa}, as it can be easily checked. 
\subsection{From the partonic to the hadronic cross-section}
What we saw in the previous subsection is correct for our \textit{partonic} cross section. In this subsection, we want to understand how to pass to the \textit{hadronic} cross section. In order to implement the hybrid factorization, we perform the following manipulation:
\begin{equation}
    \left( \frac{x_1 x_2 s}{s_0} \right)^{ \bar{\alpha}_s \left(\mu_R\right) \chi \left(n,\nu\right)} =  \left( \frac{z_1 z_2 s}{s_0} \right)^{ \bar{\alpha}_s \left(\mu_R\right) \chi \left(n,\nu\right)} \left( \frac{x_1}{z_1} \right)^{\bar{\alpha}_s \left(\mu_R\right) \chi \left(n,\nu\right)} \left( \frac{x_2}{z_2} \right)^{\bar{\alpha}_s \left( \mu_R \right) \chi \left(n, \nu \right)} \; ,
\end{equation}
and reabsorb the second (third) factor on the right hand side in the definition of the upper (lower) hadronic impact factor\footnote{This means that we include it when convoluting partonic impact factor with PDFs.}. The impact of these corrective factors is seen at the sub-leading order\footnote{Note that $x_i$ and $z_i$ are fractions of longitudinal momenta of particles that are in the same fragmentation region and, therefore, correction proportional to $\ln (x_i/z_i)$ are not large.}. In all next-to-leading terms, corrective factors produce sub-sub-leading effects and hence we can immediately perform the replacement $x_1 x_2 \rightarrow z_1 z_2$. Keeping in mind that the impact factors have been re-defined, we obtain
\begin{equation*}
\frac{d\hat{\sigma}_{AB}(z_1, \vec{p}_1, z_2, \vec{p}_2 ,s)}{d z_1 d^2\vec{p}_1 d z_2 d^2\vec{p}_2 } = \frac{1}{(2\pi)^2} \sum_{n = -\infty}^{\infty} \int d\nu \left(\frac{z_1 z_2 s}{s_0}\right)^{\bar{\alpha}_s \left(\mu_R\right) \chi \left(n,\nu\right)} 
\end{equation*}
\begin{equation*}
    \times \left(\frac{z_1 z_2 s}{s_0}\right)^{\bar{\alpha}_s^2 \left( \mu_R \right) \left( \bar{\chi} \left(n,\nu \right) + \frac{\beta_0}{8N_c} \chi \left(n,\nu\right) \left(-\chi \left(n,\nu\right) + \frac{10}{3} + 2 \ln\frac{\mu_R^2}{\sqrt{s_1 s_2}} \right) \right)}  \alpha_s^2 ( \mu_R ) \tilde{c}_1 (z_1, \vec{p}_1, n, \nu) \tilde{c}_2 (z_2, \vec{p}_2, n, \nu) \vspace{0.1 cm}
\end{equation*}
\begin{equation}
\label{expoform2}
    \times \hspace{-0.05 cm} \left[ 1 \hspace{-0.1 cm} + \hspace{-0.05 cm} \alpha_s (\mu_R) \hspace{-0.05 cm} \left( \frac{\tilde{c}_1^{(1)} \left( z_1, \vec{p}_1, n, \nu \right) }{\tilde{c}_1 \left( z_1, \vec{p}_1, n, \nu \right)} \hspace{-0.05 cm} + \hspace{-0.05 cm} \frac{\tilde{c}_2^{(1)} \left( z_1, \vec{p}_1, n, \nu \right)}{\tilde{c}_2 \left( z_1, \vec{p}_1, n, \nu \right)} \right) \hspace{-0.1 cm} + \hspace{-0.05 cm} \bar{\alpha}_s^2 ( \mu_R ) \ln \left( \hspace{-0.05 cm} \frac{z_1 z_2 s}{s_0} \hspace{-0.05 cm} \right) \hspace{-0.05 cm} \frac{\beta_0}{4N_c} \chi(n, \nu) f(\nu) \right] .
\end{equation}
If one calculates the impact factors at the next-to-leading order, the expression of the cross section is that of Eq.~(\ref{expoform2}). 
Nonetheless, when the next-to-leading impact factors are unknown, some universal corrections may still be included.
Indeed, if the impact factors were calculated up to the next-to-leading order, the dependence on the arbitrary scale, $s_0$, and on the renormalization scale\footnote{In hadronic impact factors we also have the dependence on the factorization scale $\mu_F$. The argument can be extended also to this latter scale.}, $\mu_R$, would be next-to-next-to-leading. We can therefore add, to a prediction in which only the Green's function is known within the NLLA, some next-to-leading universal terms, coming from impact factors, which make the dependence on the scales sub-sub-leading. We adopt this scheme in the section devoted to Higgs production because the numerical implementation of the full next-to-leading Higgs impact factor is not yet available.
\subsection{Kinematics}
For the tagged particles we introduce the standard Sudakov decomposition, using as light-cone basis the momenta $k_A$ and $k_B$ of the collinding protons,
\begin{equation*}
    p_1 = z_1 k_A + \frac{m_1^{2} + \vec{p}_1^{\; 2}}{z_1 s} k_B + p_{1 \perp} \; , \hspace{0.5 cm} p_2 = z_2 k_A + \frac{m_2^{2} + \vec{p}_2^{\; 2}}{z_1 s} k_B + p_{2 \perp} \; ,
\end{equation*}
with $s = (k_A+k_B)^2 = 2 k_A \cdot k_B = 
 4 E_{k_A} E_{k_B}$, choosing the momenta of protons as
 \begin{equation*}
     k_A = E_{k_A} (1, \vec{0}, 1) \; , \hspace{0.5 cm} k_B = E_{k_B} (1, \vec{0}, -1) \; . 
 \end{equation*}
Now, we easily find that the rapidities of the two tagged particles are
\begin{equation}
    y_1 = \ln \left( \frac{2 z_1 E_{k_A}}{m_{1 \perp}} \right) \; , \hspace{0.5 cm}  y_2 = -\ln \left( \frac{2 z_2 E_{k_B}}{m_{2 \perp}} \right) \; ,
\end{equation}
where $m_{i \perp} = \sqrt{m_i^2 + \vec{p}_i^{\; 2}}$ denotes the transverse mass of the $i_{th}$-particle. The rapidity difference is given by
\begin{equation}
    \Delta Y = y_1 - y_2 = \ln \left( \frac{z_1 z_2 s}{m_{1 \perp} m_{2 \perp}} \right)
\end{equation}
For the semi-hard kinematics we have the requirement
\begin{equation}
    \frac{e^{\Delta Y}}{z_1 z_2} = \frac{s}{m_{1 \perp} m_{2 \perp}} \gg 1 \; .
\end{equation}
In what follows, we will need a cross section differential in the rapidities of the tagged
particles. For this reason we adopt the change of variable
\begin{equation*}
    d z_1 d z_2 = \frac{e^{\Delta Y} m_{1 \perp} m_{2 \perp}}{s} d y_1 d y_2 \; .
\end{equation*}
 If we set the scale $s_0=m_{1 \perp} m_{2 \perp}$, we can write the cross-section as
 \begin{equation*}
\frac{d\hat{\sigma}_{AB}(y_1, \vec{p}_1, y_2, \vec{p}_2 ,s)}{d y_1 d |\vec{p}_1| d \phi_1 d y_2 d |\vec{p}_2| d \phi_2 } = \frac{1}{(2\pi)^2} \frac{e^{\Delta Y} |\vec{p}_1| |\vec{p}_2| m_{1 \perp} m_{2 \perp} }{s} \sum_{n = -\infty}^{\infty} \int d\nu e^{\Delta Y \bar{\alpha}_s \left(\mu_R\right) \chi \left(n,\nu\right)}
\end{equation*}
\begin{equation*}
    \times e^{\Delta Y \bar{\alpha}_s^2 \left( \mu_R \right) \left( \bar{\chi} \left(n,\nu \right) + \frac{\beta_0}{8N_c} \chi \left(n,\nu\right) \left(-\chi \left(n,\nu\right) + \frac{10}{3} + 2 \ln\frac{\mu_R^2}{\sqrt{s_1 s_2}} \right) \right)}  \alpha_s^2 (\mu_R) \tilde{c}_1 (z_1, \vec{p}_1, n, \nu) \tilde{c}_2 (z_2, \vec{p}_2, n, \nu) \vspace{0.1 cm}
\end{equation*}
\begin{equation}
\label{Int:Eq:expoform3}
     \times \left[ 1 + \alpha_s (\mu_R) \left( \frac{\tilde{c}_1^{(1)} \left( z_1, \vec{p}_1, n, \nu \right) }{\tilde{c}_1 \left( z_1, \vec{p}_1, n, \nu \right)} + \frac{\tilde{c}_2^{(1)} \left( z_1, \vec{p}_1, n, \nu \right)}{\tilde{c}_2 \left( z_1, \vec{p}_1, n, \nu \right)} \right) + \bar{\alpha}_s^2 (\mu_R) \Delta Y \frac{\beta_0}{4N_c} \chi(n, \nu) f(\nu) \right] \; .
\end{equation}
The coefficients defining the impact factors projected in the $(n,\nu)$-space have the form
\begin{equation}
    \tilde{c}_1 (z_1, \vec{p}_1, n, \nu) = e^{\phi_1} c_1 (z_1, |\vec{p}_1|, n, \nu) \; , \hspace{0.5 cm} \tilde{c}_2 (z_2, \vec{p}_2, n, \nu) = e^{-\phi_2-\pi} c_2 (z_2, | \vec{p}_2|, n, \nu) \; .
    \label{Pheno:Eq:ImpExpPhi}
\end{equation}
Hence, the cross section (\ref{Int:Eq:expoform3}) can be re-expressed as
\begin{equation}
\frac{d\hat{\sigma}_{AB}(y_1, \vec{p}_1, y_2, \vec{p}_2 ,s)}{d y_1 d |\vec{p}_1| d \phi_1 d y_2 d |\vec{p}_2| d \phi_2 } = \frac{1}{(2\pi)^2} \left[ \mathcal{C}_0 + 2 \sum_{n = 1}^{\infty} 
\cos \left( n \varphi \right) \mathcal{C}_n \right] \; ,
\label{Pheno:Eq:CrossAzmExp}
\end{equation}
where
\begin{equation*}
\mathcal{C}_n = \int_0^{2 \pi} d \phi_1 \int_0^{2 \pi} d \phi_2 \cos (n \varphi ) \frac{d\hat{\sigma}_{AB}(y_1, \vec{p}_1, y_2, \vec{p}_2 ,s)}{d y_1 d |\vec{p}_1| d \phi_1 d y_2 d |\vec{p}_2| d \phi_2 } = \frac{e^{\Delta Y}}{s} \int d\nu e^{\Delta Y \bar{\alpha}_s \left(\mu_R\right) \chi \left(n,\nu\right)}  
\end{equation*}
\begin{equation*}
  \times e^{\Delta Y \bar{\alpha}_s^2 \left( \mu_R \right) \left( \bar{\chi} \left(n,\nu \right) + \frac{\beta_0}{8N_c} \chi \left(n,\nu\right) \left(-\chi \left(n,\nu\right) + \frac{10}{3} + 2 \ln \frac{\mu_R^2}{\sqrt{s_1 s_2}} \right) \right)}  \alpha_s^2 (\mu_R) c_1 (z_1, \vec{p}_1, n, \nu) c_2 (z_2, \vec{p}_2, n, \nu)
\end{equation*}
\begin{equation}
     \times \left[ 1 + \alpha_s (\mu_R) \left( \frac{ c_1^{(1)} \left( z_1, \vec{p}_1, n, \nu \right) }{c_1 \left( z_1, \vec{p}_1, n, \nu \right)} + \frac{c_2^{(1)} \left( z_1, \vec{p}_1, n, \nu \right)}{c_2 \left( z_1, \vec{p}_1, n, \nu \right)} \right) + \bar{\alpha}_s^2 (\mu_R) \Delta Y \frac{\beta_0}{4N_c} \chi(n, \nu) f(\nu) \right] \;
\label{Pheno:Eq:UnitCn}
\end{equation}
and $\varphi = \phi_1-\phi_2-\pi$. Eq.~(\ref{Pheno:Eq:CrossAzmExp}) represents the final form of our fully differential cross-section and it is the starting point for the construction of all observables under investigation in this thesis. For illustration purposes, we have constructed the cross section using a single $\mu_R$ scale throughout. This is not the only possible choice for this scale and also completely asymmetrical choices can be made. More in general, this is true for all scales involved. The extension is trivial and, since we will consider several processes, we refer the reader to the original works for details~\cite{Celiberto:2020tmb,Celiberto:2021dzy,Celiberto:2021fdp,Celiberto:2022dyf,Celiberto:2022zdg}.
\section{Processes under investigation}
\subsection{Heavy flavor production in a VFNS}
As it is well known, the treatment of the $c$ and $b$ quarks in pQCD is rather delicate. When we are in a perturbative regime, we have $ Q^2 \gg \Lambda_{\rm{QCD}}^2$, where $Q$ is the typical hard scale of the process and $\Lambda_{\rm{QCD}}$ is the QCD mass scale. For light quarks, this is sufficient to treat them as massless partons and to assume that they are always present in the initial state when introducing parton distribution functions. Instead, the presence in the initial state and the way one must treat the mass of an heavy quark ($Q = c,b$) depends on kinematical conditions. For this reason there are two main schemes for the treatment of an heavy quark:
\begin{itemize}
    \item \textbf{Fixed-Flavor Number Scheme (FFNS)}\footnote{For more details see, \emph{e.g.}, Ref.~\cite{Alekhin:2009ni} and references therein.}, where the heavy quark is always treated as a massive
    particle and never as a massless parton irrespective of the value of the scale $Q$. In this scheme, the heavy quark PDF is ignored and the number of active flavours is always kept fixed. This scheme takes into account heavy-quark mass effects in coefficient functions, but does not resum logarithmically enhanced terms of the form $\ln (Q^2/m_Q^2)$.
    \item \textbf{Zero-Mass Variable-Flavor Number scheme (ZM-VFNS)}\footnote{See Refs.~\cite{Mele:1990cw,Cacciari:1993mq,Buza:1996wv,Bierenbaum:2009mv,Cacciari:1993mq,Binnewies:1997xq}.}, where all heavy-quark mass effects are ignored. These corresponds to neglect powers of the ratio $m_Q^2/Q^2$. On the other hand, since the heavy quark is treated here as a massless parton, the scheme allows us to resum potentially large logarithms of the type $\ln (Q^2/m_Q^2)$ into parton distribution functions and fragmentation functions. In this scheme, the heavy quark is present in the initial state above a fixed threshold.
\end{itemize}
It is clear that the two approaches work in complementary regimes. The ZM-VFNS is applicable when $Q^2 \gg m_Q^2$ holds. In these kinematic conditions, power mass corrections are suppressed and DGLAP-type logarithms are large. On the other hand, the FFNS, which allows us to include all finite quark mass corrections, is only accurate in the region $Q \lesssim m_Q$. It is clear that there is no exact natural separation between these two approaches and it is reasonable to think of a more complete description.  This is provided by the so-called \textbf{General–Mass Variable Flavour Number scheme (GM–VFNS)}\footnote{See Refs.~\cite{Kramer:2000hn,Forte:2010ta,Blumlein:2018jfm,Aivazis:1993pi,Thorne:1997ga}.}, which combines the advantage of the massive and massless calculations by means of an interpolated scheme which is valid for any value of the scale $Q$, and that matches the FFN and ZM–VFN schemes at small and large values of $Q$, respectively. \\

Currently, a complete next-to-leading order description of the impact factors contributing to the production of heavy quarks in adroproduction channels is not available. For example, the impact factor(s) for producing the $J/ \Psi$ have been computed in Ref.~\cite{Boussarie:2017oae}, using two different approaches for the description of the charmonium state: 1) \textbf{Color evaporation model (CEM)}~\cite{Fritzsch:1977ay,Halzen:1977rs}, 2) \textbf{Non-relativistic QCD (NRQCD) formalism}~\cite{Caswell:1985ui,Thacker:1990bm,Bodwin:1994jh}. Even limiting ourselves to the first case, or only to the color octet production mechanism in the NRQCD formalism, the corrections are not known. At the level of gluon channel only, the diagrams to be calculated would be the same ones to be calculated for the next-to-leading order heavy-quark pair impact factor, shown at LO in the Appendix \ref{AppendixB} and originally computed in Ref.~\cite{Bolognino:2019yls}. This impact factor has been applied to the study of inclusive production of a heavy-light dijet system in Ref.~\cite{Bolognino:2021mrc}. \\

The remaining part of this section is dedicated to the description of semi-hard reactions involving the production of heavy bound states in the ZM-VFNS, leaving the inclusion of mass effects for future projects. In this way, the final state transition from partons to hadrons is described by a usual fragmentation function which is suitably convoluted with the partonic impact factor and with the parton distribution function. The leading order expression of this impact factor, projected onto the eigenfunctions of the LO BFKL kernel, is
\begin{equation*}
c_{1}(n , \nu,|\vec{p}_1 |, z_1) 
= 2 \sqrt{\frac{C_F}{C_A}}
(| \vec{p}_1 |^2)^{i\nu-1/2}\,\int_{z_1}^1\frac{d \zeta}{\zeta}
\left( \frac{\zeta}{z_1} \right)
^{2 i\nu-1} 
\end{equation*}
\begin{equation}
 \times \left[\frac{C_A}{C_F}f_g(\zeta)D_g^{ \mathcal{Q} } \left( \frac{z_1}{\zeta} \right)
 +\sum_{\alpha=q,\bar q}f_{\alpha}(\zeta) D_{\alpha}^{\mathcal{Q}}\left(\frac{z_1}{\zeta}\right)\right] \; ,
 \label{Pheno:Eq:LOImpHadro}
\end{equation}
while the next-to-leading corrections can be found in Ref.~\cite{Ivanov:2012iv}. Within the same accuracy, the LO impact factor for producing a jet is
\begin{equation}
 c_2 (n, \nu, | \vec{p}_2 |, z_2) = 2 \sqrt{\frac{C_F}{C_A}}
 (| \vec{p}_2 |^2)^{i\nu-1/2} \, \left(\frac{C_A}{C_F} f_g( z_2 )
 +\sum_{\beta=q,\bar q} f_{\beta}( z_2 ) \right) \; ,
\end{equation}
while the next-to-leading corrections can be found in Ref.~\cite{Caporale:2012ih}. \\

Using the aforementioned scheme, here, we analyze the following semi-hard reactions:
\begin{itemize}
    \item $P(k_A) + P(k_B) \longrightarrow \Lambda_c^{\pm} (y_1, p_1) +  X + \Lambda_c^{\pm} (y_2, p_2)$ ,
    \item $P(k_A) + P(k_B) \longrightarrow H_b (y_1, p_1) +  X + \text{jet} (y_2, p_2)$ ,
    \item $P(k_A) + P(k_B) \longrightarrow H_b (y_1, p_1) +  X + H_b (y_2, p_2)$ ,
    \item $P(k_A) + P(k_B) \longrightarrow J/ \psi (y_1, p_1) +  X + \text{jet} (y_2, p_2)$ ,
\end{itemize}
 where we are inclusive on the baryon charge of the $\Lambda_c$ particle, $H_b$ stands for a generic bottomed flavored hadron\footnote{In our analysis we are inclusive on the production of all species of $b$-hadrons whose lowest Fock state contains either a $b$ or $\bar b$ quark, but not both. Therefore, bottomed quarkonia are not considered. Furthermore, we ignore $B_c$ mesons since their production rate
is estimated to be at most 0.1\% of $b$-hadrons (see, \emph{e.g.}, Refs.\cite{LHCb:2014iah,LHCb:2016qpe}). Our choice is in line with the $b$-hadron FF determination of Ref.\cite{Kramer:2018vde}.}. \\

Collinear PDFs are calculated via the {\tt MMHT14} NLO PDF set~\cite{Harland-Lang:2014zoa} as provided by {\tt LHAPDFv6.2.1}
interpolator~\cite{Buckley:2014ana}, and a two-loop running coupling with $\alpha_s\left(M_Z\right)=0.11707$ and a
dynamic-flavor threshold is chosen. \\

We describe jet emissions at NLO perturbative accuracy in terms of a reconstruction algorithm calculated within the ``small-cone'' approximation (SCA)~\cite{Furman:1981kf,Aversa:1988vb}, \emph{i.e.} for a small-jet cone aperture in the rapidity/azimuthal angle plane. More in particular, we adopt the version derived in Ref.~\cite{Ivanov:2012ms}, which is \emph{infrared-safe} up to NLO perturbative and well-suited for numerical computations, with a cone-jet function selection~\cite{Colferai:2015zfa} and for the jet-cone radius fixed at ${\cal R}_J = 0.5$, as usually done in recent experimental analyses at CMS~\cite{Khachatryan:2016udy}. The expression for the NLO jet vertex can be obtained, \emph{e.g.}, by combining Eq.~(36) of Ref.~\cite{Caporale:2012ih} with Eqs.~(4.19)-(4.20) of Ref.~\cite{Colferai:2015zfa}.
\subsubsection{$\Lambda_c$ and $b$-hadrons fragmentation functions in VFNS}
 We described the parton fragmentation into $\Lambda_c$ baryons in terms of the novel {\tt KKSS19} NLO FF set~\cite{Kniehl:2020szu} (see also Refs.~\cite{Kniehl:2005de,Kniehl:2012ti}), whose native implementation was directly linked to {\tt JETHAD}. This parameterization mainly relies on a description \emph{\`a la} Bowler~\cite{Bowler:1981sb} for $c$ and $b$ quark/antiquark flavors. Technical details on the fitting procedure are presented in Section IV of Ref.~\cite{Kniehl:2020szu}. \\

We depicted the parton fragmentation to $b$-hadrons by the hand of the {\tt KKSS07} NLO~FFs, that were originally extracted from data of inclusive $B$-meson emissions in $e^+e^-$~annihilation~\cite{Kniehl:2008zza}.
In this parametrization the $b$ flavor has its starting scale at $\mu_0 = 4.5 \text{ GeV} \simeq m_b$ and is portrayed by a simple, three-parameter power-like \emph{Ansatz}~\cite{Kartvelishvili:1985ac}
\begin{equation}
\label{KKSS19_FF_power}
 D^{H_b}(x, \mu_0) = {\cal N} x^a (1-x)^b \;,
\end{equation}
whereas gluon and lighter quark (including $c$) FFs are generated
through DGLAP evolution and vanish at $\mu_F = \mu_0$.
Following Ref.~\cite{Kramer:2018vde}, we obtained the $b$-hadron FFs from the $B$-meson ones by simply removing the branching fraction for the $b \to B^\pm$ transition, which was assumed as $f_u = f_d = 0.397$ (see also Ref.~\cite{Kniehl:2008zza}). \\

We also show comparisons with predictions for lighter-hadron species ($\Lambda$ hyperons, pions, kaons, and protons), these latter are described in terms {\tt AKK08} NLO FFs~\cite{Albino:2008fy}, which are the closest in technology to {\tt KKSS19}.
\subsubsection{Quarkonium fragmentation functions}
We build our NLO collinear FF sets for the \emph{direct} $\JPsi$ or $\Yps$ meson production by taking, as a starting point, the recent work done in Ref.~\cite{Zheng:2019dfk}. There, a NLO calculation was performed for the heavy-quark FF depicting the transition $c \to \JPsi$ or the $b \to \Yps$ one, where $c$ ($b$) indistinctly refer to the charm (bottom) quark and its antiquark. It essentially relies on the NRQCD factorization formalism taken with NLO accuracy (see, \emph{e.g.}, Refs.~\cite{Thacker:1990bm,Bodwin:1994jh,QuarkoniumWorkingGroup:2004kpm,Brambilla:2010cs,Pineda:2011dg,Brambilla:2020ojz} and references therein), which allows us to write the FF function of a parton $i$ fragmenting into a heavy quarkonium $\cal Q$ with longitudinal fraction $z$ as
\begin{equation}
 \label{FF_NRQCD}
 D^{\cal Q}_i(z, \mu_F) = \sum_{[n]} {\cal D}^{\cal Q}_{i}(z, \mu_F, [n]) \langle {\cal O}^{\cal Q}([n]) \rangle \;.
\end{equation}
In Eq.~(\ref{FF_NRQCD}), ${\cal D}_{i}(z, \mu_F, [n])$ denotes the perturbative short-distance coefficient containing terms proportional to $\ln (\mu_F/m_{\cal Q})$ (to be resummed via DGLAP evolution), $\langle {\cal O}^{\cal Q}([n]) \rangle$ stands for the non-perturbative NRQCD LDME, and $[n] \equiv \,^{2S+1}L_J^{(c)}$ represents the quarkonium quantum numbers in the spectroscopic notation (see, \emph{e.g.}, Ref.~\cite{Bugge:1986xw}), the $(c)$ superscript identifying the color state, singlet (1) or octet (8).
Limiting ourselves to a spin-triplet (vector) and color-singlet quarkonium state, $^3S_1^{(1)}$, the analytic form of the initial-scale FF depicting the constituent heavy-quark to quarkonium transition, $Q \to {\cal Q}$ (we refer to $c \to \JPsi$ here\footnote{In the original paper, Ref.~\cite{Celiberto:2022dyf}, we consider the $b \to \Upsilon$} as well), reads (for details on its derivation, see Sections~II~and~III of Ref.~\cite{Zheng:2019dfk})
\begin{equation}
 \label{FF_Q-to-onium}
 D^{\cal Q}_Q(z, \mu_F \equiv \mu_0) 
 = D^{\cal Q, {\rm LO}}_Q(z)
 + \frac{\alpha_s^3(3m_Q)}{m_Q^3} \, |{\cal R}_{\cal Q}(0)|^2 \, \Gamma_Q^{\cal Q, {\rm NLO}}(z) \;,
\end{equation}
with $m_c = 1.5$~GeV, and the NRQCD radial wave-function at the origin of the quarkonium state set to~$|{\cal R}_{\JPsi}(0)|^2 = 0.810$~GeV$^3$, according to potential-model calculations (Ref.~\cite{Eichten:1994gt} and references therein).
The expression for the LO initial-scale FF was originally calculated in Ref.~\cite{Braaten:1993mp} and reads
\begin{equation}
 \label{FF_Q-to-onium_LO}
 D^{\cal Q, {\rm LO}}_Q(z) = 
 \frac{\alpha_s^2(3m_Q)}{m_Q^3} \, \frac{8z(1-z)^2}{27\pi (2-z)^6} \, |{\cal R}_{\cal Q}(0)|^2 \, (5z^4 - 32z^3 + 72z^2 -32z + 16) \;,
\end{equation}
and the polynomial function $\Gamma_Q^{\cal Q, {\rm NLO}}(z)$ entering the expression for the NLO-FF correction is
\begin{eqnarray}
 \label{FF_Gamma_JPsi}
 \Gamma_Q^{\JPsi, {\rm NLO}}(z) 
 &=& - 9.01726z^{10} + 18.22777z^9 + 16.11858z^8 - 82.54936z^7 
 \nonumber \\
 &+& 106.57565z^6 - 72.30107z^5 + 28.85798z^4 - 6.70607z^3
 \nonumber \\
 &+& 0.84950z^2 - 0.05376z - 0.00205
\end{eqnarray}
Coefficients of $z$-powers in Eqs.~(\ref{FF_Gamma_JPsi}) are obtained via a polynomial fit to the numerically-calculated NLO FFs.
Starting from $\mu_F \equiv \mu_0 = 3 m_Q$, in Ref.~\cite{Zheng:2019dfk} a DGLAP-evolved formula for the $D^{\cal Q}_Q(z, \mu_F)$ function was derived and then applied to phenomenological studies of $\JPsi$ production via $e^+ e^-$ single inclusive annihilation (SIA). \\

As pointed out in Ref.~\cite{Braaten:1994xb}, both $(c \to \JPsi)$ and $(g \to \JPsi)$ fragmentation channels are similar in size. The relative weight of the heavy-quark and gluon contributions is also driven by the size of the hard scattering producing these partons. Therefore, the number of large $p_T$-gluons emitted could be of the same order, if not larger, than the heavy-quark one.
Moreover, in a hadroproduction process such as the one considered in our study, the gluon FF is enhanced by the collinear convolution at LO with the corresponding gluon PDF (see Eq.~(\ref{Pheno:Eq:LOImpHadro})). Thus we expect, in our case, a stronger sensitivity on the gluon-fragmentation channel with respect to the case of a SIA-like reaction.
Therefore, we include in our analysis also the contribution coming from the gluon fragmentation. The gluon to vector-quarkonium LO fragmentation mechanism starts at $\alpha_s^3$, namely at the same order of the NLO correction to the heavy-quark FF in Eq.~(\ref{FF_Q-to-onium}). The $(g \to {^3S_1^{(1)}} g g)$ fragmentation function was computed in Ref.~\cite{Braaten:1993rw} and reads
\begin{equation*}
    D_g^{\cal Q} (z, 2 m_Q) = \frac{5}{36 (2\pi)^2} \alpha_s^3(2 m_Q) \frac{|{\cal R}_{\cal Q}(0)|^2}{m_{\cal Q}^3} \int_0^z d \xi \int_{(\xi+z^2)/2z}^{(1+\xi)/2} d \tau \frac{1}{(1-\tau)^2 (\tau-\xi)^2 (\tau^2-\xi)^2} 
\end{equation*}
\begin{equation}
    \sum_{i=1}^{2} z^i \left[ f^{(g)}_i (\xi, \tau) + g^{(g)}_i (\xi, \tau) \frac{1+\xi-2 \tau}{2 (\tau-\xi) \sqrt{\tau^2-\xi}} \ln \left( \frac{\tau - \xi + \sqrt{\tau^2-\xi}}{\tau - \xi - \sqrt{\tau^2-\xi}} \right) \right] \; ,
\label{FF_g-to-onium}
\end{equation}
with the six $f^{(g)}_i$ and $g^{(g)}_i$ functions being given in Eqs.~(4)-(9) of Ref.~\cite{Braaten:1995cj}. We stress that the mechanism considered here is the \emph{direct} production of the quarkonium from the parent gluon. Another contribution to $\JPsi$-meson production in high-energy processes, not considered in our analysis, is the production of a $P$-wave charmonium state $\chi_c$, followed by its radiative decay $\chi_c \to \JPsi + \gamma$ (see Refs.~\cite{Braaten:1993mp,Yuan:1994hn}). \\

\begin{figure*}[!t]
\centering
\includegraphics[width=0.45\textwidth]{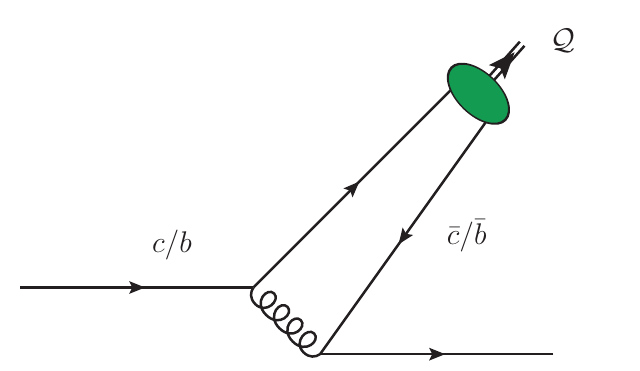}
\hspace{0.50cm}
\includegraphics[width=0.45\textwidth]{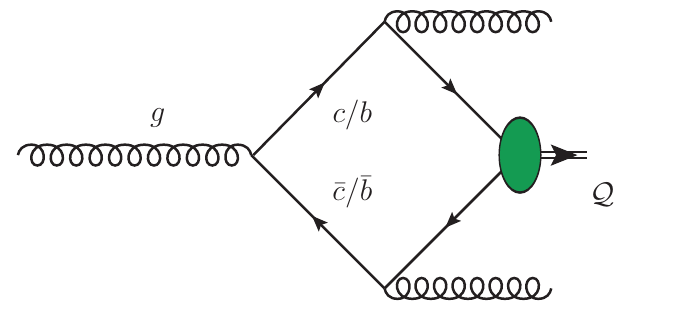}

\caption{Left: one of the leading diagrams contributing to the heavy-quark fragmentation to a $^3S_1^{(1)}$ vector quarkonium $\cal Q$ at order $\alpha_s^2$.
Right: one of the leading diagrams contributing to the gluon fragmentation to $^3S_1^{(1)}$ vector quarkonium $\cal Q$ at order $\alpha_s^3$.
The green blob denotes the corresponding non-perturbative NRQCD LDME.}
\label{fig:FF_diagrams}
\end{figure*}

The different initial energy scales at which quarks and gluons FF are taken is due to the production mechanism itself. The heavy-quark fragmentation involves at least three heavy quarks in the final state (Fig.~\ref{fig:FF_diagrams}, left panel), thus the running coupling in Eqs.~(\ref{FF_Q-to-onium}) and~(\ref{FF_Q-to-onium_LO}) is calculated at $\mu_R = 3m_Q$. Conversely, the gluon fragmentation involves two heavy quarks only (Fig.~\ref{fig:FF_diagrams}, right panel), and this explains why the running coupling in Eq.~(\ref{FF_g-to-onium}) is taken at $\mu_R = 2m_Q$. \\

Starting from the initial-scale FFs in Eqs.~(\ref{FF_Q-to-onium}, \ref{FF_g-to-onium}), we generate the corresponding functions for all parton species. This is a required step in order to perform analyses by means of our high-energy VFNS treatment. For a given quarkonium, $\JPsi$ or $\Yps$, we set the corresponding constituent heavy (anti-)quark FF, $D^{\JPsi}_c \equiv D^{\JPsi}_{\bar c}$ or $D^{\Yps}_b \equiv D^{\Yps}_{\bar b}$, to be equal to the parameterization given in Eq.~(\ref{FF_Q-to-onium}) at the initial scale $\mu_0 = 3 m_Q$ and we set the gluon FF, $D^{\JPsi}_g$ or $D^{\Yps}_g$, to be equal to the parameterization given in Eq.~(\ref{FF_g-to-onium}) at the initial scale $\mu_0 = 2 m_Q$. Then, we compute the DGLAP-evolved functions via the {\tt APFEL++} library~\cite{Bertone:2013vaa,Carrazza:2014gfa,Bertone:2017gds}, thus getting a {\tt LHAPDF} set of FFs that embodies all parton flavors. From now, according to names of Authors of Ref.~\cite{Zheng:2019dfk}, we will refer to these sets as the $\JPsi$ {\tt ZCW19}$^{+}$ NLO FF parameterizations\footnote{In Ref.~\cite{Celiberto:2022dyf} we named {\tt ZCW19} the set which does not include the gluon FF at the initial scale $\mu_0 = 2 m_Q$. In that set, also the gluon channel is purely generated by the evolution.}.

\subsection{Higgs plus jet/charmed-hadron}
In Ref.~\cite{Celiberto:2020tmb,Celiberto:2022zdg} the following semi-hard reactions
\begin{itemize}
    \item $P(k_A) + P(k_B) \longrightarrow H (y_1, p_1) + X + \text{jet} (y_2, p_2)$ ,
    \item $P(k_A) + P(k_B) \longrightarrow H (y_1, p_1) +  X + \Lambda_c (y_2, p_2)$ ,
\end{itemize}
have been proposed as testified of BFKL dynamics. The setup of the PDFs, FFs and the jet reconstruction algorithm is identical to the one described in the previous subsection. As the implementation of the next-to-leading order level impact factor for the production of a forward Higgs boson from a colliding proton is not yet available, we use the leading order one (including finite top-mass contributions), which reads~\cite{Celiberto:2020tmb}
\begin{equation}
\label{c-higgs}
c_H(\kappa,\nu, |\vec p_1|, z_1) = \frac{1}{v^2} \frac{|\mathcal{H}_f(\vec p_1^{\: 2})|^2}
{128\pi^{3}\sqrt{2(N^2_{c}-1)}}
\left( \vec p_1^{\: 2} \right)^ {i\nu + 1/2} f_g(z_1,\mu_{F_1}) \; ,
\end{equation}
where~\cite{DelDuca:2003ba}
\begin{equation*}
    \mathcal{H} (\vec p_1^{\: 2}) = \frac{4 m_t^2}{m_{H \perp}^2}
\end{equation*}
\begin{equation}
\label{IF_F}
 \times \left\{ \hspace{-0.05cm}
 \left( \frac{1}{2} - \frac{2 m_t^2}{m_{H \perp}^2} \right) \hspace{-0.05 cm} \left[ \Delta_h(\upsilon_2)^2 \hspace{-0.10 cm} - \hspace{-0.05cm} \Delta_h(\upsilon_1)^2 \right]
 \hspace{-0.05 cm} + \hspace{-0.05 cm} \left( \frac{2 \vec p_1^{\: 2}}{m_{H \perp}^2} \right)
 \left[ \sqrt{\upsilon_1} \, \Delta_h(\upsilon_1) - \sqrt{\upsilon_2} \, \Delta_h(\upsilon_2) \right] \hspace{-0.05cm} + \hspace{-0.05cm} 2
  \right\} \hspace{-0.05 cm} ,
\end{equation}
with $m_t = 173.21$ GeV the top-quark mass, $\upsilon_1 = 1
- 4 m_t^2/m_H^2$, $\upsilon_2 = 1 + 4 m_t^2/\vec p_H^{\: 2}$, the root
$\sqrt{\upsilon_1} = i \sqrt{|\upsilon_1|}$ holding for negative values of $\upsilon_1$.
Moreover one has
\begin{equation}
 \label{IF_W}
 \Delta_h(\upsilon) = \left\{
 \begin{aligned}
  &- 2 i \arcsin \frac{1}{\sqrt{1 - \upsilon}} \; , 
  &\upsilon < 0 \; ; \\ 
  &\ln \frac{1 + \sqrt{\upsilon}}{1 - \sqrt{\upsilon}} - i \pi \; ,
  &0 < \upsilon < 1 \; ; \\ 
  &\ln \frac{1 + \sqrt{\upsilon}}{\sqrt{\upsilon} - 1} \; ,
  &\upsilon > 1 \; . 
 \end{aligned}
 \right.
\end{equation}
In our study we consider a partial NLO implementation that includes the ``universal''  contributions to the Higgs impact factor, proportional to the corresponding LO impact factor. These terms are obtained on the basis of a renormalization group analysis, namely \emph{via} the requirement of stability at NLO under variations of energy scales. Thus we have
\begin{equation*}
c_H^{(1)}(\kappa,\nu,|\vec p_1|, z_1) \,=\,
c_H(\kappa, \nu, |\vec p_1|, z_1) \left\{ \frac{\beta_0}{2 \pi} 
\left(\ln \frac{\mu_{R_1}}{|\vec p_1|} + \frac{5}{6} \right) 
+ \chi\left(\kappa,\nu\right) \ln \left( \frac{\sqrt{s_0}}{m_{H \perp}} \right)
+ \frac{\beta_0}{2 \pi} \ln \frac{\mu_{R_1}}{m_{H \perp}}
\right.
\end{equation*}
\begin{equation}
\left.
\hspace{0.5 cm} - \; \frac{1}{\pi f_g(z_1,\mu_{F_1})} \ln \left( \frac{\mu_{F_1}}{m_{H \perp}} \right) \int_{z_1}^1\frac{d \xi}{\xi}
\left[P_{gg}(\xi) f_g \left(\frac{z_1}{\xi}, \mu_{F_1} \right) + \sum_{a={q,\bar q}} P_{ga}(\xi)
f_a \left(\frac{z_1}{\xi},\mu_{F_1}\right) \right]
\right\} \; .
\label{cH1}
\end{equation}

\section{Numerical results}
The numerical elaboration of all the considered observables is done by making use of the {\tt JETHAD} modular work package~\cite{Celiberto:2020wpk}. The sensitivity of our results on scale variation is assessed by letting $\mu_R$ and $\mu_F$ to be around their \emph{natural} values, up to a factor ranging from 1/2 to two. More specifically, we set\footnote{In case of jet or light hadrons the transverse mass coincide with the $| \vec{p} \; |$.}
\begin{equation}
    \mu_R = \mu_F = \mu_N = \sqrt{m_{1 \perp} m_{2 \perp}} \; ,
\label{Pheno:Eq:GeomMean}
\end{equation}
and the $C_{\mu}$ parameter entering plots as the ratio $\mu_{R,F}/\mu_N$ for all the consider processes, except the ones involving the production of a Higgs boson. In these latter cases, we use the geometrical mean in Eq.~(\ref{Pheno:Eq:GeomMean}) only the exponential factor in Eq.~(\ref{Pheno:Eq:UnitCn}), while the scale associated to each impact factor is chosen as 
\begin{equation*}
    \mu_{R_i} = \mu_{F_i} = C_{\mu} \mu_{N_i} =  C_{\mu} m_{i \perp} \; .
\end{equation*}
Error bands in our figures embody the combined effect of scale variation and phase-space multi-dimensional integration, the latter being steadily kept below 1\% by the {\tt JETHAD} integrators. All calculations of our observables is done in the $\MSb$ scheme. The kinematical cuts are always shown directly in plots, for more detailed discussion on these choices we refer to~\cite{Celiberto:2020tmb,Celiberto:2021dzy,Celiberto:2021fdp,Celiberto:2022dyf,Celiberto:2022zdg}. Furthermore, Refs.~\cite{Celiberto:2020tmb,Celiberto:2021dzy,Celiberto:2021fdp,Celiberto:2022dyf,Celiberto:2022zdg} contain for all the processes mentioned above, long and detailed studies, both at natural scales and at scales obtained through the \textit{Brodsky-Lepage-Mackenzie} (BLM) \textit{optimization} procedure~\cite{Brodsky:1996sg,Brodsky:1997sd,Brodsky:1998kn,Brodsky:2002ka}, which prescribes that the optimal scale value is the one that cancels the non-conformal $\beta_0$-terms in the considered observable. In the following, we want to provide a review of the observables that can be studied, focusing on the possibility of carrying out studies at natural scales. The reason for this choice is that, although the application of the BLM method led to a significant improvement of the agreement between predictions for azimuthal correlations of the two Mueller-Navelet jets and CMS data~\cite{Khachatryan:2016udy}, the scale values found, much higher than the natural ones, generally bring to a substantial reduction of cross sections (observed for the first time in inclusive light charged dihadron emissions~\cite{Celiberto:2016hae,Celiberto:2017ptm}). This issue clearly hampers the possibility of doing precision studies.
\subsection{Azimuthal-angle coefficients}
The \emph{azimuthal-angle coefficients}, are obtained by integrating coefficients ${\cal C}_n$ in Eq.~(\ref{Pheno:Eq:UnitCn}) over the phase space of the outgoing particles, at fixed values of their mutual rapidity separation, $\DY$. One has
    \begin{equation}
     \label{Cn_int}
     C_n(\DY, s) =
     \int_{p^{\rm{min}}_1}^{p^{\rm max}_1} d | \vec{p}_1|
     \int_{p^{\rm{min}}_2}^{{p^{\rm max}_2}} d |\vec{p}_2|
     \int_{y^{\rm{min}}_1}^{y^{\rm{max}}_1} d y_1
     \int_{y^{\rm{min}}_2}^{y^{\rm{max}}_2} d y_2
     \, \delta \left( y_1 - y_2 - \DY \right)
     \, {\cal C}_n 
     \, .
    \end{equation}
We investigate the $\DY$-behavior of the $\varphi$-summed cross section (or $\DY$-distribution), $C_0(\DY, s)$, of the azimu\-thal-correlation moments, $R_{n0}(\DY, s) = C_{n}/C_{0} \equiv \langle \cos n \varphi \rangle$, and of their ratios, $R_{nm} = C_{n}/C_{m}$~\cite{Vera:2006un,Vera:2007kn}.
\subsubsection{$\DY$-distribution}
\begin{figure}
\centering
  \hspace{-0.6 cm} \includegraphics[scale=0.48,clip]{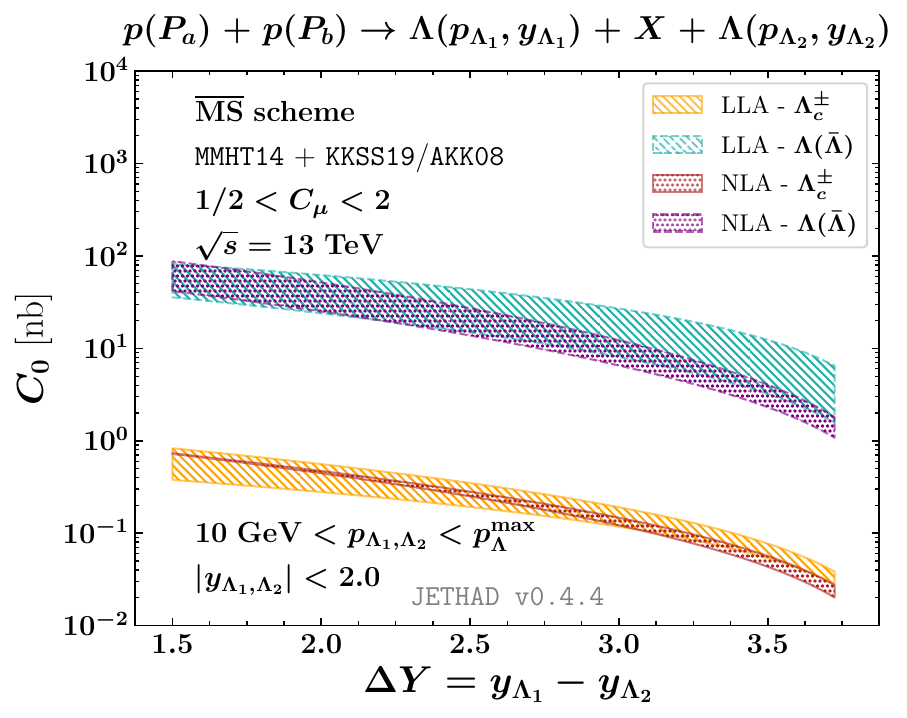}
   \hspace{0.6 cm}
   \includegraphics[scale=0.48,clip]{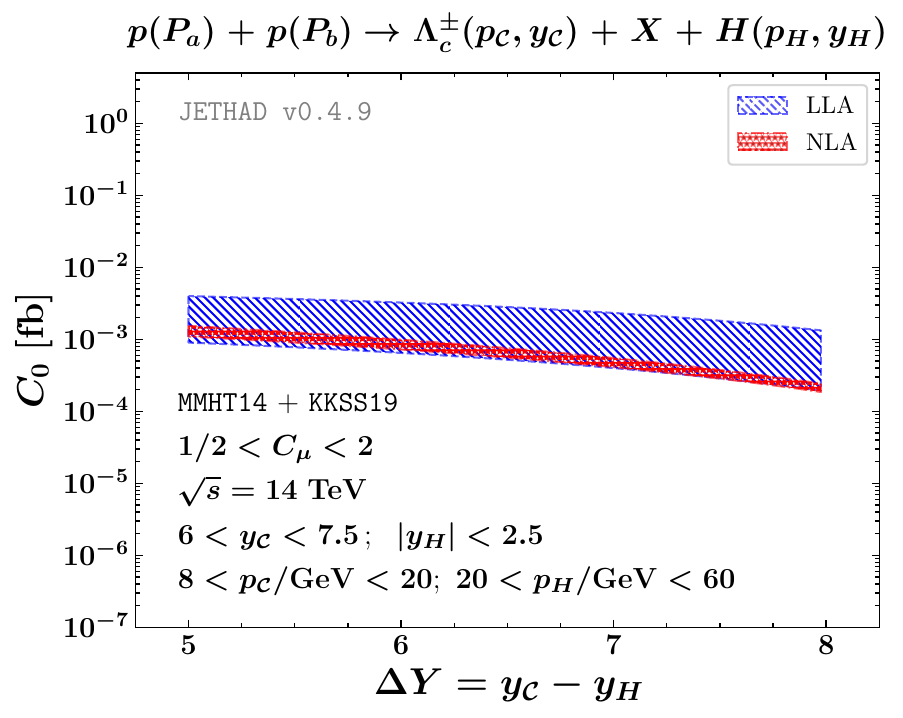}
   \vspace{0.15cm}
   \includegraphics[scale=0.48,clip]{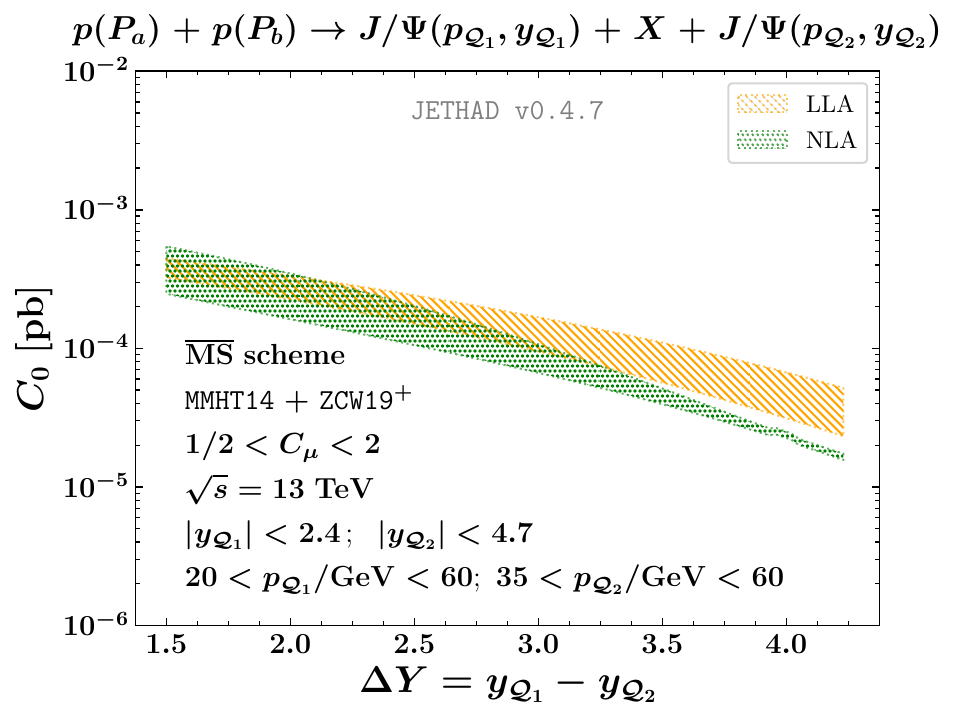}
   \hspace{0.15cm}
   \includegraphics[scale=0.48,clip]{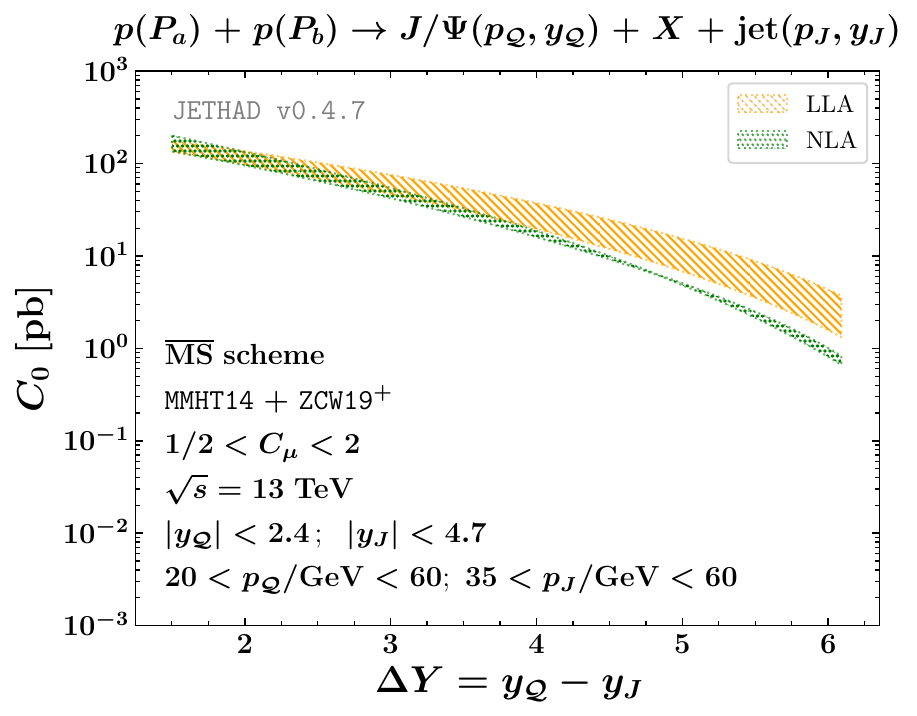}
   \vspace{0.15cm}
\caption{Behavior of the $\varphi$-summed cross section, $C_0$, as a function of $\Delta Y$, in the double $\Lambda_c$ channel (upper left panel), in the $\Lambda_c$ plus Higgs channel (upper right panel), in the double $\JPsi$ (lower left panel) and in the $\JPsi$ plus jet channel (lower right panel). $\sqrt{s}=14$ for the $\Lambda_c$ plus Higgs channel and $\sqrt{s}=14$ in the remaining cases. Predictions for $\Lambda_c$ emissions are compared with configurations where $\Lambda$ hyperons are detected. Text boxes inside panels exhibit final-state kinematic cuts.}
\label{fig:C0DeltaY}
\end{figure}

In Fig.~\ref{fig:C0DeltaY}, we show the $\DY$-behavior of $C_0$ for a series of different semi-hard reactions. The common feature is the downtrend exhibited by the cross section at increasing $\Delta Y$. It emerges as the interplay of two competing effects. On the one hand, the pure high-energy evolution leads to the well-known growth with energy of partonic cross sections. On the other hand, collinear parton distributions and fragmentation functions quench hadronic cross sections when $\DY$ increases. All the plots show a promising partial stabilization of the high-energy series. \\

Upper left panel of Fig.~\ref{fig:C0DeltaY} shows the $\Delta Y$-dependence of the $\varphi$-summed cross section in
the double $\Lambda_c$ channel, together with corresponding predictions for detection of $\Lambda$ hyperons, at NLLA. We note that NLA bands are almost nested (except for large values of $\Delta Y$, for which they overlap but are not fully nested) inside LLA ones and they are generally narrower in the $\Lambda_c$ case. Since this analysis is performed around natural scales and without implementing any BLM optization, we claim that this is a clear effect of a (partially) reached stability of the high-energy series, for both hadron emissions. The fact that the stabilization effect is stronger for heavy species was corroborated in Ref.~\cite{Celiberto:2021dzy}, observing that, while predictions for hyperons lose almost one order of magnitude when passing from natural scales to the expanded BLM ones, results for $\Lambda_c$ baryons are much more stable, with the NLA band becoming even wider in the BLM case. Moreover, in Ref.~\cite{Celiberto:2021dzy} it was proposed that the stabilization effect is connected to the smooth and non-decreasing with $\mu_F$ FFs behavior of the $\Lambda_c$ fragmentation functions. Subsequently, a similar behavior was observed for other hadronic species. \\

Another manifestation of stabilizing effects is shown in the upper right panel of Fig.~\ref{fig:C0DeltaY}, where we consider the $\Lambda_c + H$ channel. We perform phenomenological study by imposing realistic kinematic cuts of forthcoming experimental analyses at the LHC. In particular, we allow the charmed-hadron transverse momentum to be in the range 8~GeV~$< p_{\cal C} <$~20~GeV and its rapidity in the ultraforward rapidity window $6 < y_{\cal C} < 7.5$. This choice is in line with nominal acceptances of FPF detector plans~\cite{Anchordoqui:2021ghd,Feng:2022inv}. We observe that the NLA predictions are systematically contained inside LLA ones and the width of uncertainty bands considerably decreases when moving to the higher order. It should be specified that, even if this stabilization effect is very strong in the latter case, especially if one considers into the high rapidity-difference between the two detected objects, the analysis is only partial next-to-leading, due to the presence of the Higgs impact factor, whose complete next-to-leading implementation is not yet available\footnote{Nonetheless, we include all of the universal NLO terms presented in the previous section.}. \\

Lastly, in lower panels of Fig.~\ref{fig:C0DeltaY} we study, with the full NLLA, the $\DY$-behavior of the cross section in the double $J/\psi$ and $J/\psi$-plus-jet channels.
Predictions are obtained by making use of the {\tt ZCW19$^+$} set. We note that values of $C_0$ are everywhere larger than 0.5 pb in the $J/\psi$-plus-jet channel. This leads to a very promising statistics, although being substantially lower than the one for heavy-baryon and heavy-light meson emissions~\cite{Celiberto:2021dzy,Celiberto:2021fdp}. 

\subsubsection{Azimu\-thal-correlation moments and their ratios}
\begin{figure}
\centering
 \hspace{-0.6 cm}  \includegraphics[scale=0.48,clip]{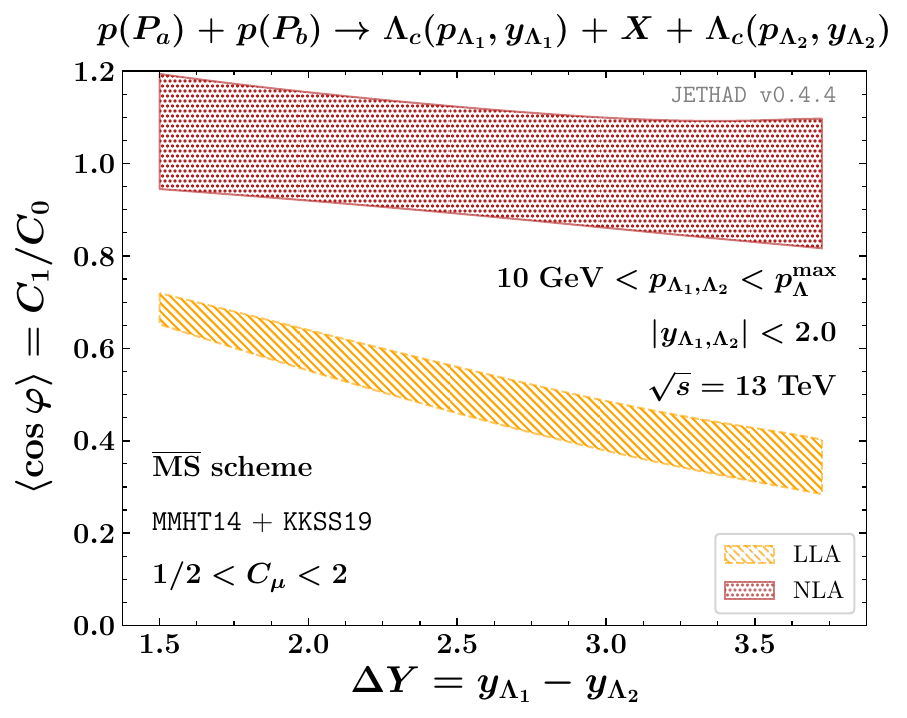}
   \hspace{0.3 cm}
   \includegraphics[scale=0.48,clip]{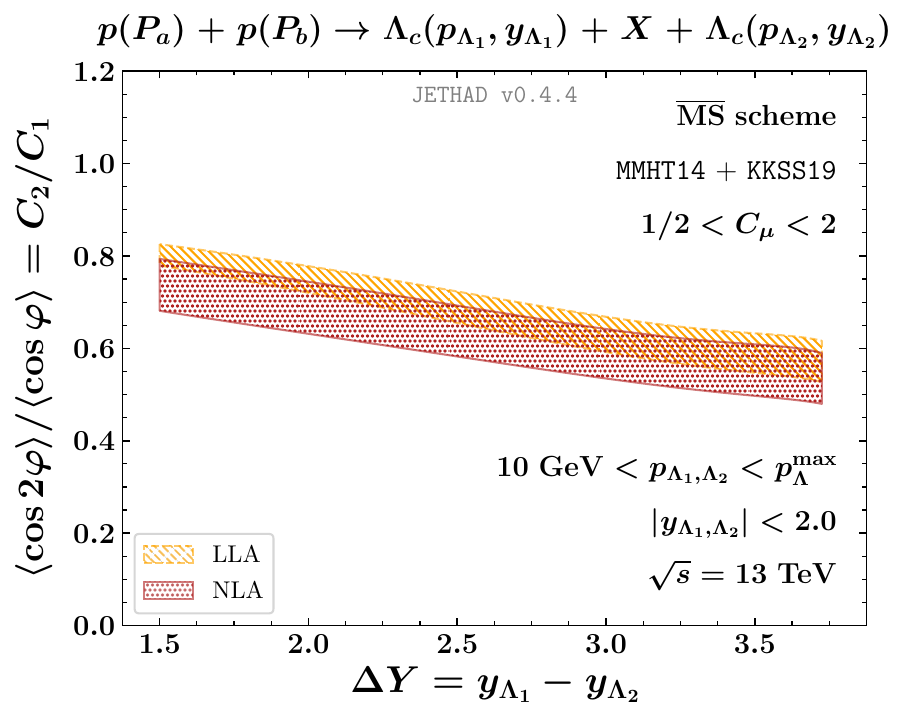}
   \vspace{0.15cm}
   \includegraphics[scale=0.48,clip]{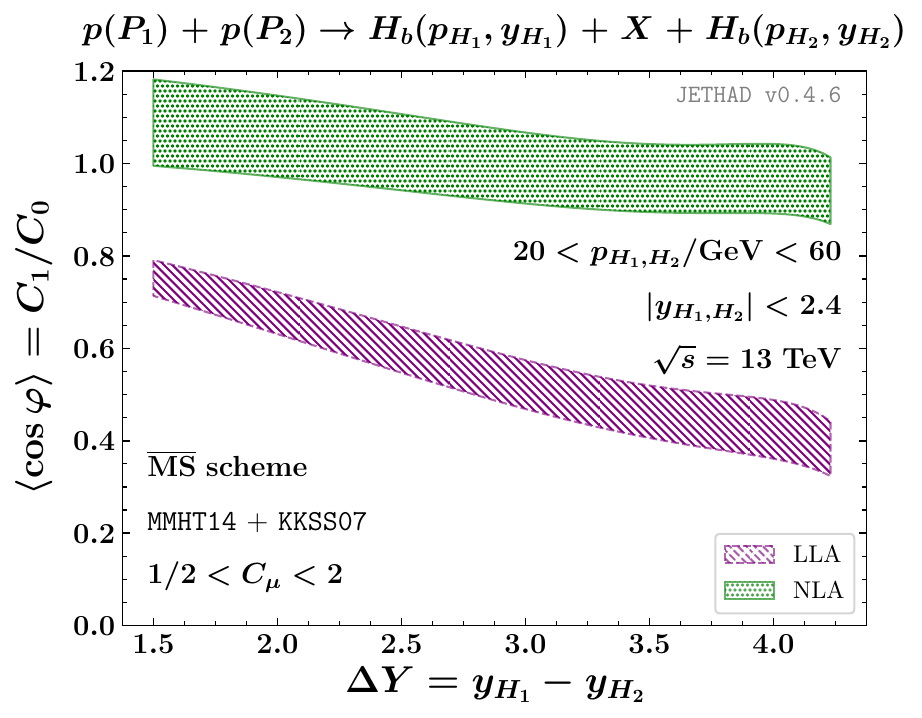}
   \hspace{0.15cm}
   \includegraphics[scale=0.48,clip]{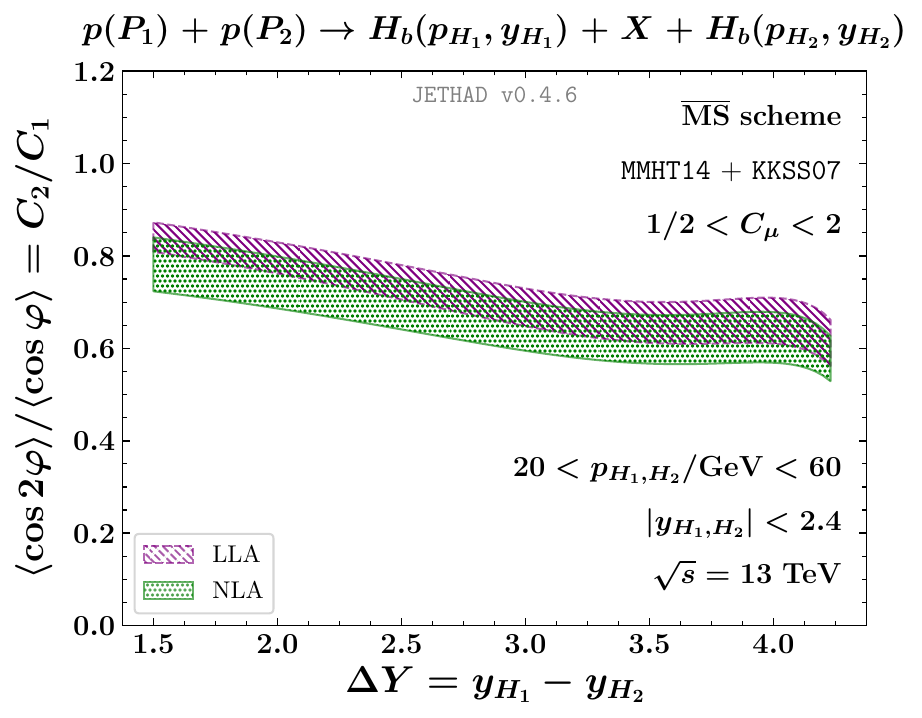}
   \vspace{0.15cm}
\caption{Behavior of azimuthal-correlation moments, $R_{nm} \equiv C_{n}/C_{m}$, as functions of $\Delta Y$, in the double $\Lambda_c$ channel and in the double $H_b$ channel, at natural scales, and for $\sqrt{s} = 13$ TeV. Text boxes inside panels exhibit final-state kinematic cuts.}
\label{fig:RnmDelY}
\end{figure}

Azimu\-thal-correlation moments and their ratio are widely recognized as very sensitive to the BFKL dynamics. It is also well known that, in the Mueller-Navelet channel, the instabilities under higher order corrections and scale variations prevents any realistic analysis around natural values\footnote{At least for the pure azimuthal correlation moments.}. In Fig.~\ref{fig:RnmDelY}, predictions for the $R_{10}, R_{21}$ azimuthal ratios in the double $\Lambda_c$ (upper panels) and in the double $H_b$ (lower panels) channels, at natural scales, are presented. The downtrend of all these ratios when $\Delta Y$ grows is a well know signal of the onset of high-energy dynamics. Larger rapidity distances heighten the weight of undetected gluons, thus leading to a decorrelation pattern in the azimuthal plane, which is more pronounced in pure LLA series. \\

Although, instabilities rising at natural scales are strong, they are milder than the ones observed in the Mueller-Navelet dijet channel. In the presented plots, the value of the $R_{10}$ moment exceeds one in the small-$\Delta Y$ region. This unphysical effect is fairly explained by the fact that contributions which are power-suppressed in energy and are not included in our BFKL treatment start to become relevant in those kinematic ranges, thus worsening the accuracy of our predictions. \\ 

The situation improves, when one considers ratio of azimuthal correlation moments. As firstly observed in Refs.~\cite{Vera:2006un,Vera:2007kn}, the $R_{21}$ ratios exhibits a fair stability under NLA corrections for both both final-state channels. 

\subsection{Azimuthal distribution}
 The \emph{azimuthal distribution} of the two tagged particles, as a function of $\varphi$ and at fixed values of $\DY$, is 
\begin{equation}
     \label{azimuthaldistribution}
     \frac{d \sigma_{pp} (\varphi, \DY, s)}{\sigma_{pp} d \varphi} = \frac{1}{\pi} \left\{ \frac{1}{2} + \sum_{n=1}^{\infty} \cos(n \varphi) \langle \cos(n \varphi) \rangle \right\}
     \equiv \frac{1}{\pi} \left\{ \frac{1}{2} + \sum_{n=1}^{\infty} \cos(n \varphi) R_{n0} \right\} \; .
\end{equation}

Proposed for the first time in the context for Mueller--Navelet studies~\cite{Ducloue:2013hia,Marquet:2007xx}, this distribution represents one of the most directly accessible observables in experimental analyses. 
Indeed, experimental measurements hardly cover the whole azimuthal-angle plane due to limitations of the apparatus. Therefore, distributions differential on the azimuthal-angle difference, $\varphi$, could be easier compared with data. \\

\begin{figure}
\centering
   \includegraphics[scale=0.48,clip]{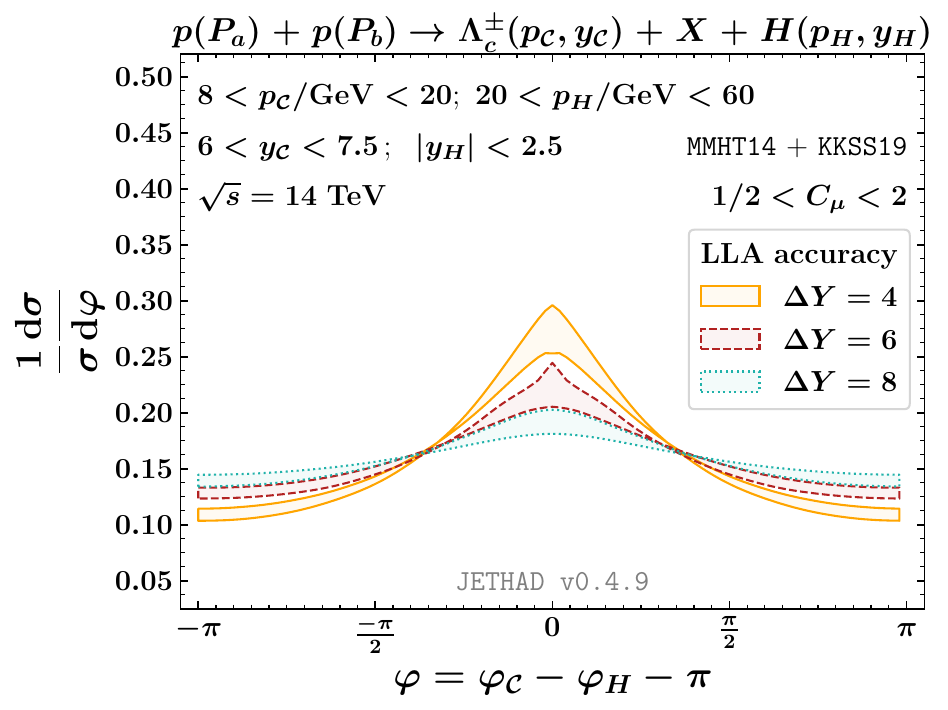}
   \hspace{0.15cm}
   \includegraphics[scale=0.48,clip]{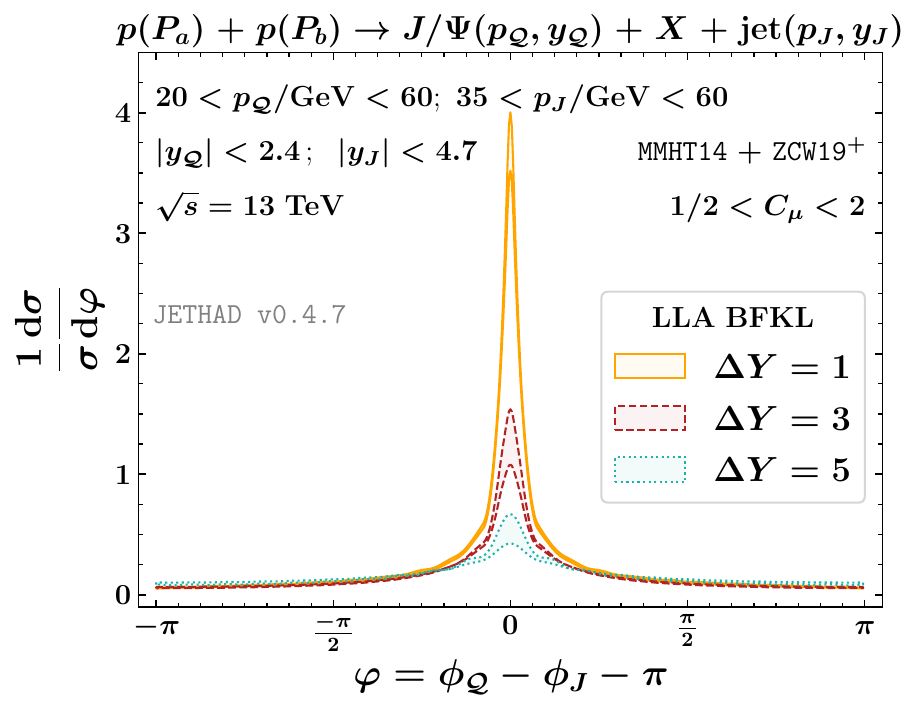}
   \vspace{0.15cm}
   \includegraphics[scale=0.48,clip]{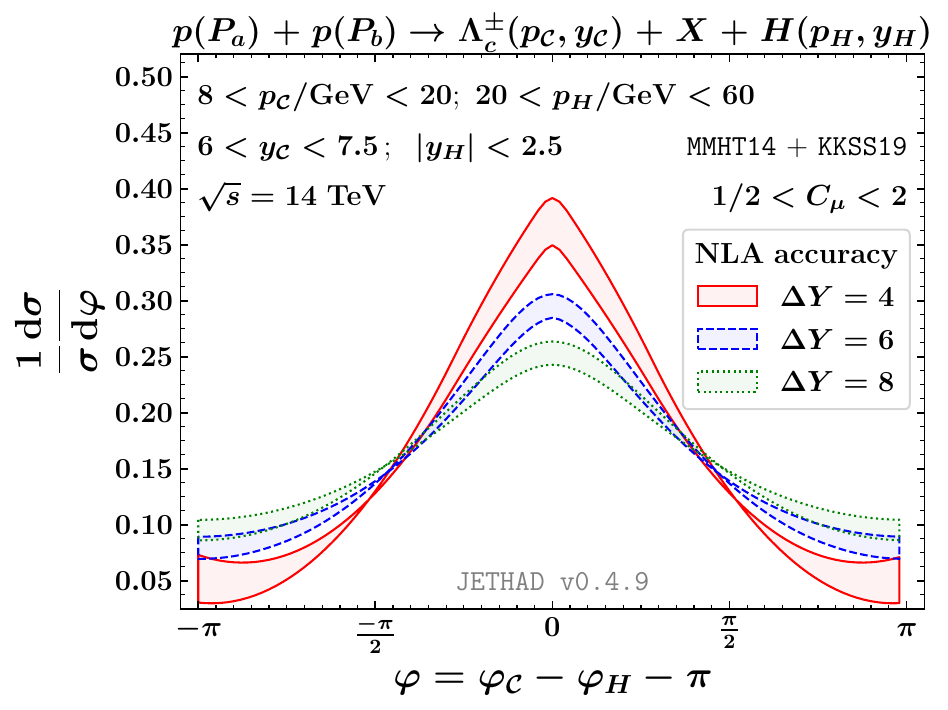}
   \hspace{0.15cm}
   \includegraphics[scale=0.48,clip]{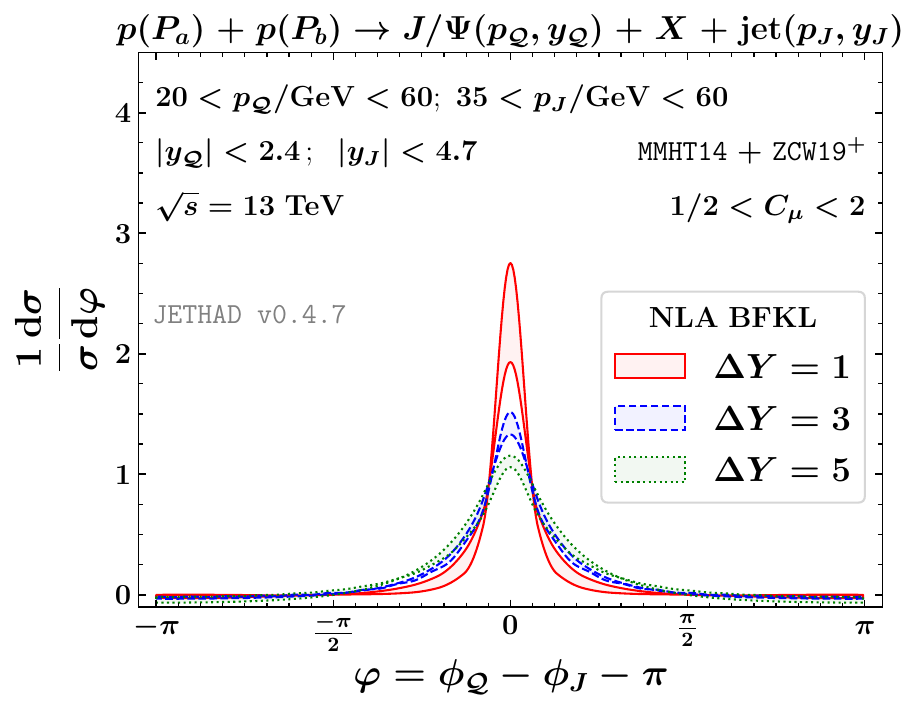}
   \vspace{0.15cm}
\caption{LLA (upper panels) and NLA (lower panels) predictions for the $\varphi$-distribution in the $\Lambda_c$~$+$~$H$ (left) and in the $\JPsi$~$+$~jet (right) channel, at $\sqrt{s} = 14$ TeV and $\sqrt{s} = 13$ TeV, respectively, and for three distinct values of $\DY$.
Text boxes inside panels exhibit final-state kinematic cuts.}
\label{fig:PhiDist}
\end{figure}

In Fig.~\ref{fig:PhiDist}, we present predictions for the azimuthal distribution in the $\Lambda_c + H$ (left panels) and in the $\JPsi$~$+$~jet (right panels) channels, for three distinct values of the rapidity interval. LLA predictions are given in the upper panels, while NLA are found in the lower ones. In all cases the height of the peak visibly diminishes when $\DY$ grows, while the distribution width slightly widens. Again, at large values of $\DY$ the weight of gluons strongly ordered in rapidity predicted by BFKL increases, thus bringing to a reduction of  the azimuthal correlation between the two detected objects, so that the number of back-to-back events lowers. \\

Focusing on patterns for the $\varphi$-distribution in the $\JPsi$~$+$~jet channel, we observe that at $\DY = 1$ the LLA peak is much more pronounced than the corresponding NLA one, at $\DY = 3$ the LLA and NLA peaks are similar in height, and at $\DY = 5$ the LLA peak is beyond the NLA one.
This behavior as a straightforward explanation. The small-$\DY$ range stays at the limit of applicability of the BFKL resummation, since the low values of the partonic center-of-mass energies reduces the phase space for secondary-gluon emissions. This leads to a stronger discrepancy between LLA and NLA results. Conversely, in the moderate-$\DY$ regime the high-energy series shows a fair stability when NLA corrections are switched on. Finally, the large-$\DY$ territory is very sensitive to the BFKL dynamics, this reflecting in a stronger weight of the re-correlation effects, predicted by the NLA resummation, over pure LLA results. \\

The same pattern is observed in the $\Lambda_c + H$ channel. The only difference is that, from the lower $\Delta Y$ value, the LLA predictions are below the NLLA ones. This is due to the fact that in this analysis we are considering ultra-forward kinematics, in fact, the first value of the rapidity difference is four. At these values, re-correlation effects, predicted by the NLA resummation, are already very strong.

\subsection{Transverse-momentum distributions}
Cross sections and azimuthal-angle correlations differential in the final-state rapidity interval, $\DY$, are excellent testing grounds for the high-energy resummation. However, in order to probe regimes where other resummation dynamics are also relevant, more differential distributions in the $p_T$-spectrum are needed. Indeed, when the measured transverse momenta range in wider windows, other regions that are contiguous to the strict semi-hard one get probed. On one hand, when the transverse momenta are very large or their mutual distance is large, the weight of DGLAP-type logarithms as well as \emph{threshold} contaminations~\cite{Bonciani:2003nt,deFlorian:2005fzc,Muselli:2017bad} grows, thus making the description by our formalism inadequate. On the other hand, in the very low-$p_T$ limit a pure high-energy treatment would also fail since large transverse-momentum logarithms entering the perturbative series are systematically neglected by BFKL. Therefore, in this section, rather than claiming to be able to describe the whole spectrum in $p_T$ of the observables, we want to show that there exist kinematic windows in which high-energy resummation is relevant and in which therefore it must be taken into account in a serious precision program. 
\subsubsection{Single differential $p_T$-distribution}
\begin{figure}
\centering
   \includegraphics[scale=0.48,clip]{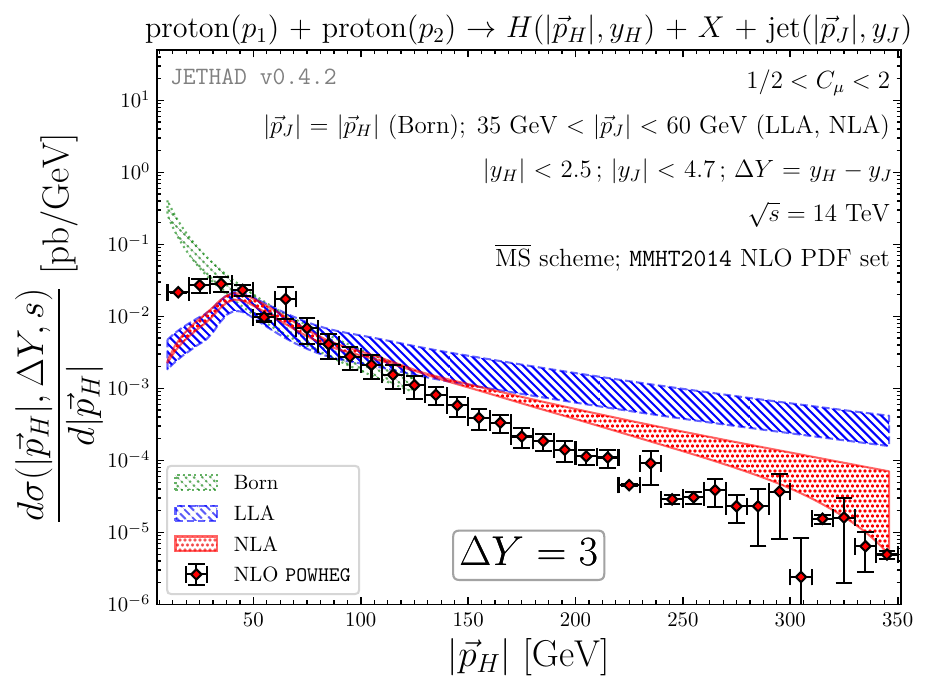}
   \hspace{0.15cm}
   \includegraphics[scale=0.48,clip]{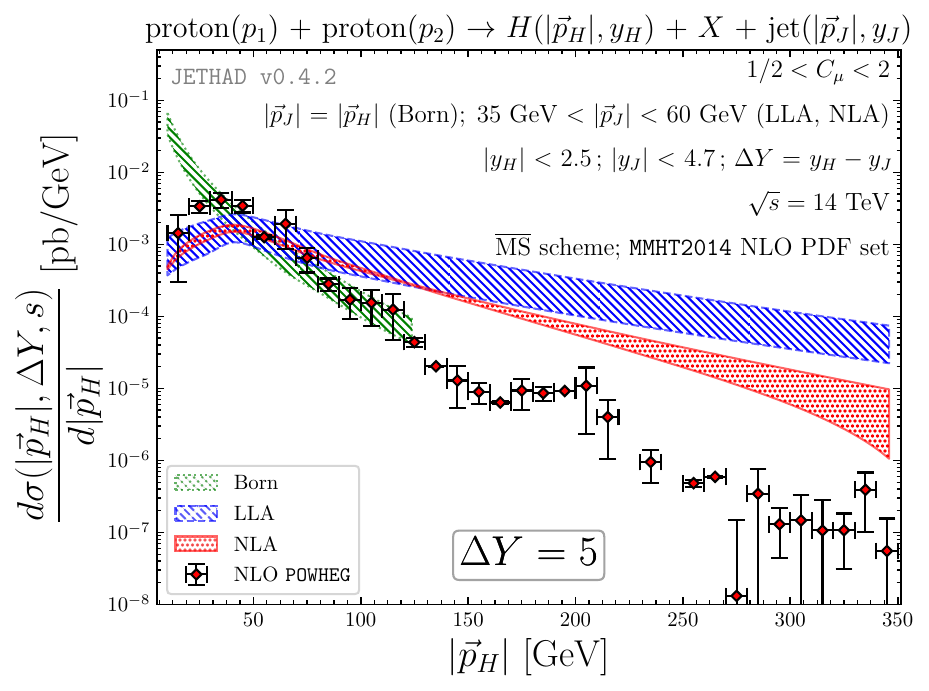}
   \vspace{0.15cm}
   \includegraphics[scale=0.48,clip]{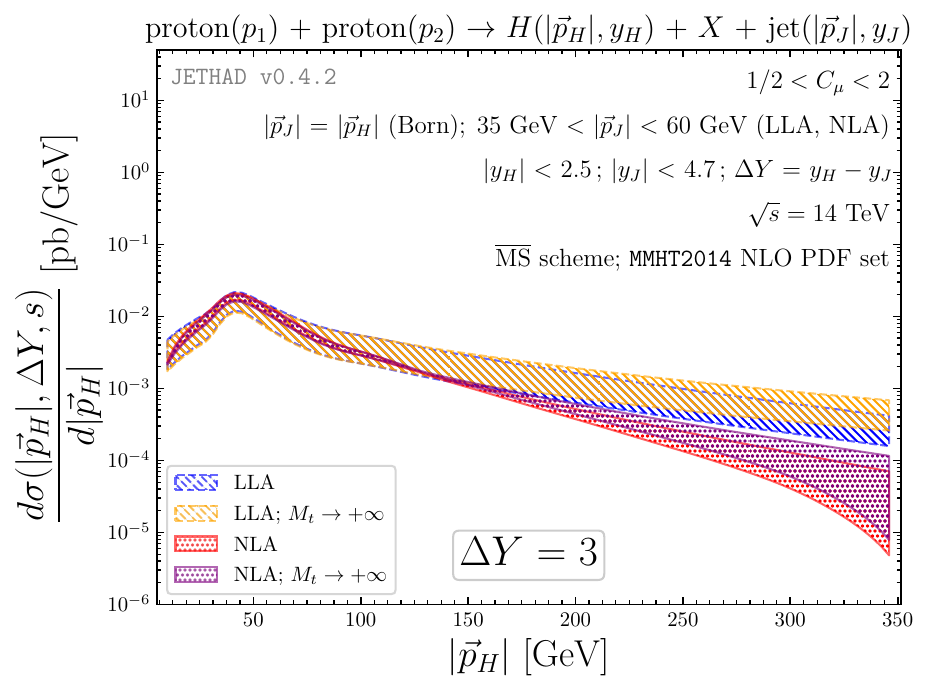}
   \hspace{0.15cm}
   \includegraphics[scale=0.48,clip]{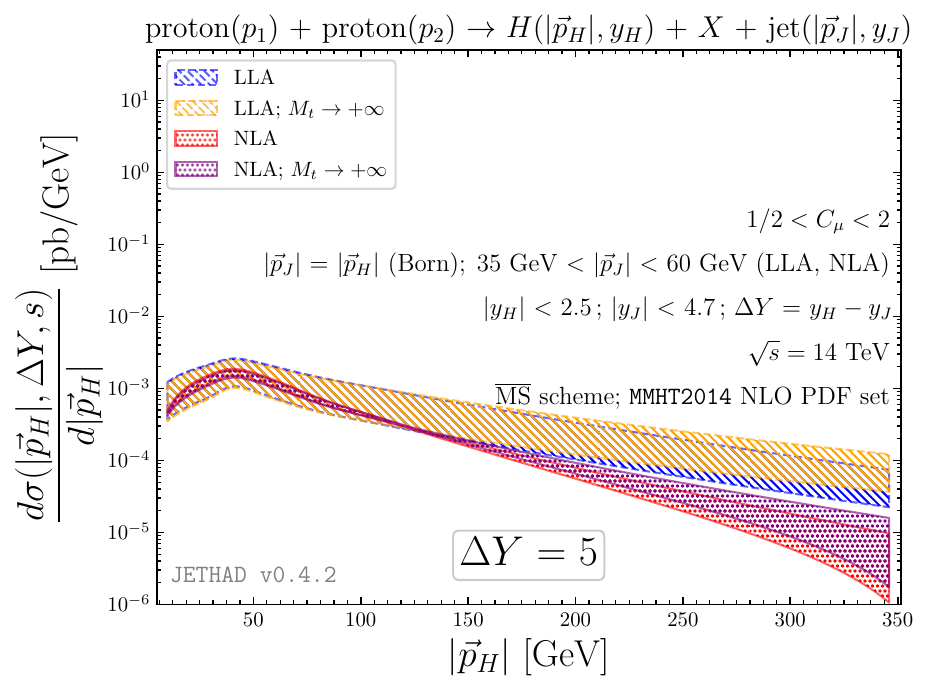}
   \vspace{0.15cm}
\caption{$p_T$-dependence of the cross section for the inclusive Higgs-jet hadroproduction for $\sqrt{s} = 14 \; {\rm{TeV}}$ and for $ \Delta Y = 3, 5$. In the top panels, resumed predictions are compared to the fixed order result obtained through the {\tt POWHEG} method. In the bottom panels, predictions in the standard case and in the large top-mass limit are compared. Text boxes inside panels exhibit final-state kinematic cuts.}
\label{fig:Y5-pT}
\end{figure}

We start by studying the \emph{transverse-momentum distribution} of one particle at fixed values of $\DY$, \textit{i.e.}
    \begin{equation}
     \label{pT_distribution}
     \frac{ d \sigma_{pp}(|\vec p_1|, \DY, s)}{ d |\vec p_1| d \DY} =
     \int_{p^{\rm min}_2}^{{p^{\rm max}_2}} d |\vec{p}_2|
     \int_{y^{\rm{min}}_Q}^{y^{\rm{max}}_Q} d y_1
     \int_{y^{\rm{min}}_J}^{y^{\rm{max}}_J} d y_2
     \, \delta \left( y_1 - y_2 - \DY \right)
     \, {\cal C}_0 .
    \end{equation}
We investigate this distribution in the Higgs~$+$~jet channel. Our calculation in the Born limit at $\Delta Y = $ 3 (left panel of
Fig.~\ref{fig:Y5-pT}) is in fair agreement with the corresponding pattern in Ref.~\cite{DelDuca:2003ba} (solid line in the left panel of Fig.~2), up to a factor two, due to the fact that we restricted $\Delta Y$ to be positive, which means that the Higgs particle is always more forward than the jet\footnote{Note that in Ref.~\cite{DelDuca:2003ba} the Higgs mass is a free parameter. We compare our result with the corresponding one at $M_H =$ 120 GeV.}. In our study, this calculation cannot exceed a given upper cut-off in the $|\vec p_H|$-range, say around 125 GeV. This is due to our choice for the final-state kinematic ranges, where consistency with experimental cuts in the rapidities of the detected objects would lead to $x_J > 1$ for sufficiently large jet transverse momenta. \\

Both the LLA (blue) and the NLA (red) series in upper panels of Fig.~\ref{fig:Y5-pT} show a peak (not present in the Born case) at $|\vec p_H|$ around 40 GeV for the two values of $\Delta Y$, and a decreasing behavior at large $|\vec p_H|$. For the sake of simplicity, we distinguish three kinematic subregions. The low-$|\vec p_H|$ region, {\emph{i.e.}} $|\vec p_H| < 10$ GeV, has been excluded from our analysis, since it is dominated by large transverse-momentum logarithms, which call for the corresponding all-order resummation, not accounted by our formalism. To the intermediate-$|\vec p_H|$ region the set of configurations where $|\vec p_H|$ is of the same order of $|\vec p_J|$, which ranges from 35 to 60 GeV, corresponds. It is essentially the peak region plus the first part of the decreasing tail, where NLA bands are totally nested inside the LLA ones. Here, the impressive stability of the perturbative series unambiguously confirms the validity of our description at the hand of the BFKL resummation. Finally, in the large-$|\vec p_H|$ region represented by the long tail, NLA distributions decouple from LLA ones and exhibit an increasing sensitivity to scale variation. Here, DGLAP-type logarithms together with \emph{threshold} effects start to become relevant, thus spoiling the convergence of the high-energy series. In Fig.~\ref{fig:Y5-pT} we present also the $p_H$-distributions at $\Delta Y=3$ and$~5$, as obtained by a fixed-order NLO calculation through the POWHEG method~\cite{Nason:2004rx,Frixione:2007vw,Alioli:2010xd}, by suitably adapting the subroutines dedicated to the inclusive Higgs plus jet final state~\cite{Campbell:2012am,Hamilton:2012rf}. It is interesting to observe that, both at $\Delta Y=3$ and $\Delta Y=5$, the NLO fixed-order prediction is systematically lower than the LLA- and NLA-BFKL ones and this is more evident at the larger $\Delta Y$, where the effect of resummation is expected to be more important. This observation provides with an interesting window for discrimination between fixed-order and high-energy-resummed approaches. \\

Finally, in lower panels of Fig.~(\ref{fig:Y5-pT}), we compare the $p_T$-distributions presented above with the corresponding ones obtained in the large top-mass limit, $M_t \to + \infty$. In Ref.~\cite{Celiberto:2020tmb}, it was noted that, when this limit is taken, cross sections become at most $5 \div 7 \, \%$ larger, whereas the effect on azimuthal correlations is very small or negligible. We do not show figures related with this comparison, since the bands related to the large top-mass limit are hardly distinguishable from the ones with physical top mass. The impact on the $p_H$-distribution is also quite small in the $|\vec p_H| \sim |\vec p_J|$ range, while it become more manifest when the value of $|\vec p_H|$ increases. \\

All these considerations brace the message that an exhaustive study of the $|\vec p_H|$-distribution would rely on a unified formalism where distinct resummations are concurrently embodied. In particular, the impact of the BFKL resummation could depend on the delicate interplay among the Higgs transverse mass, the Higgs transverse momentum and the jet transverse momentum entering, in logarithmic form, the expressions of partial NLO corrections to impact factors. Future studies including full higher-order corrections will allow us to further gauge the stability of our calculations. 
\subsubsection{Double differential $p_T$-distribution}
\begin{figure}
\centering
   \includegraphics[scale=0.37,clip]{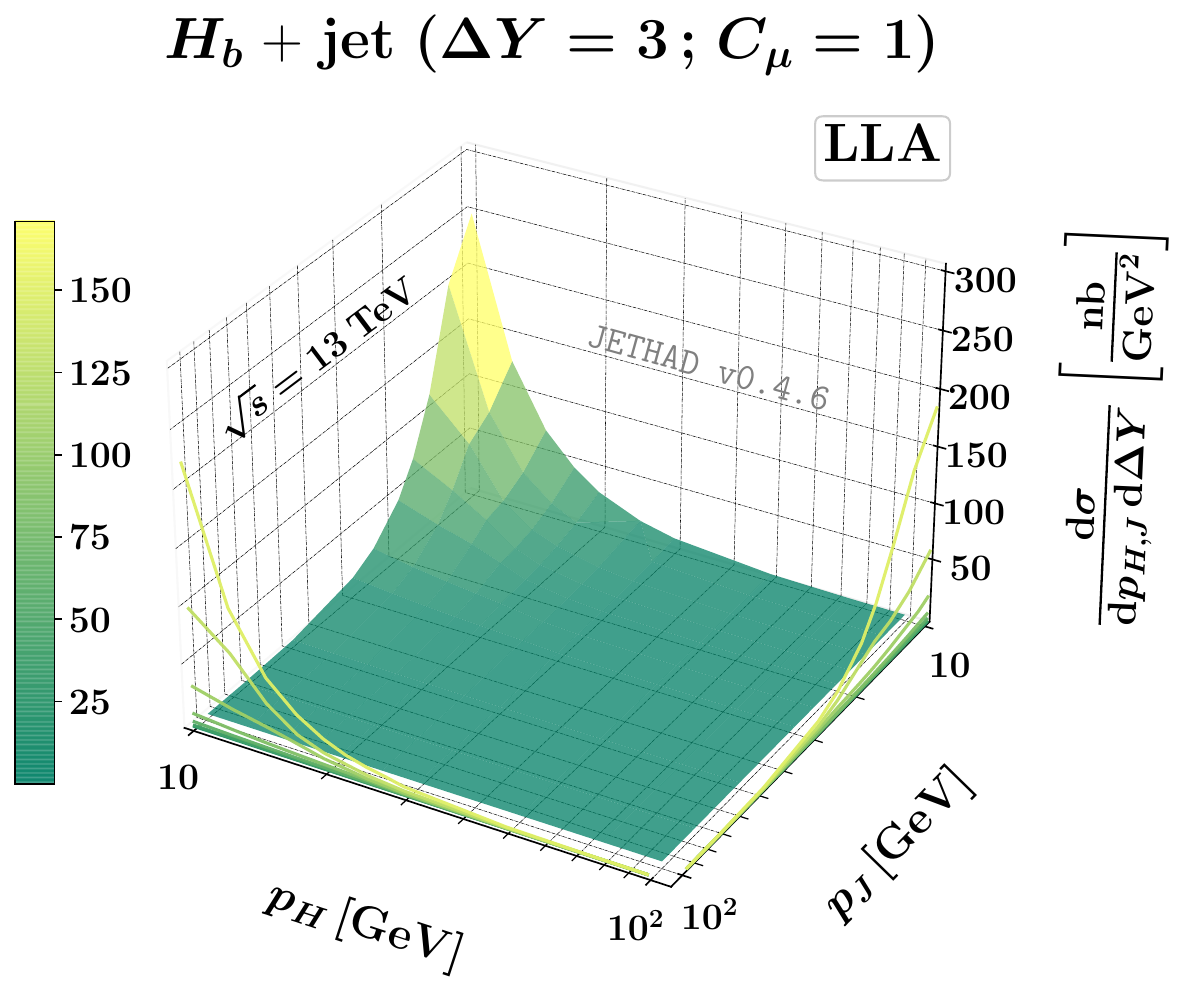}
   \hspace{0.25cm}
   \includegraphics[scale=0.37,clip]{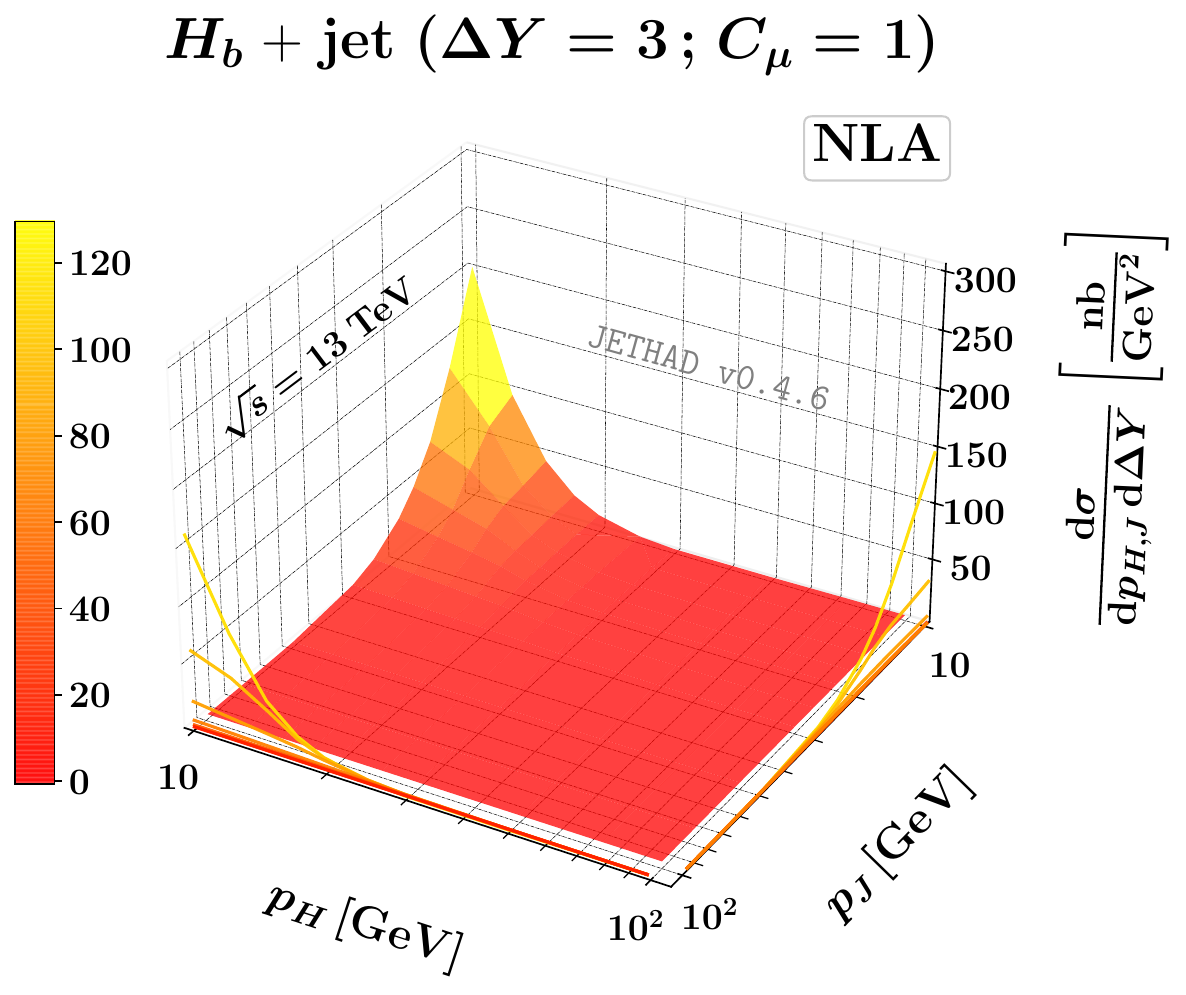}
   \vspace{0.15cm}
   \includegraphics[scale=0.37,clip]{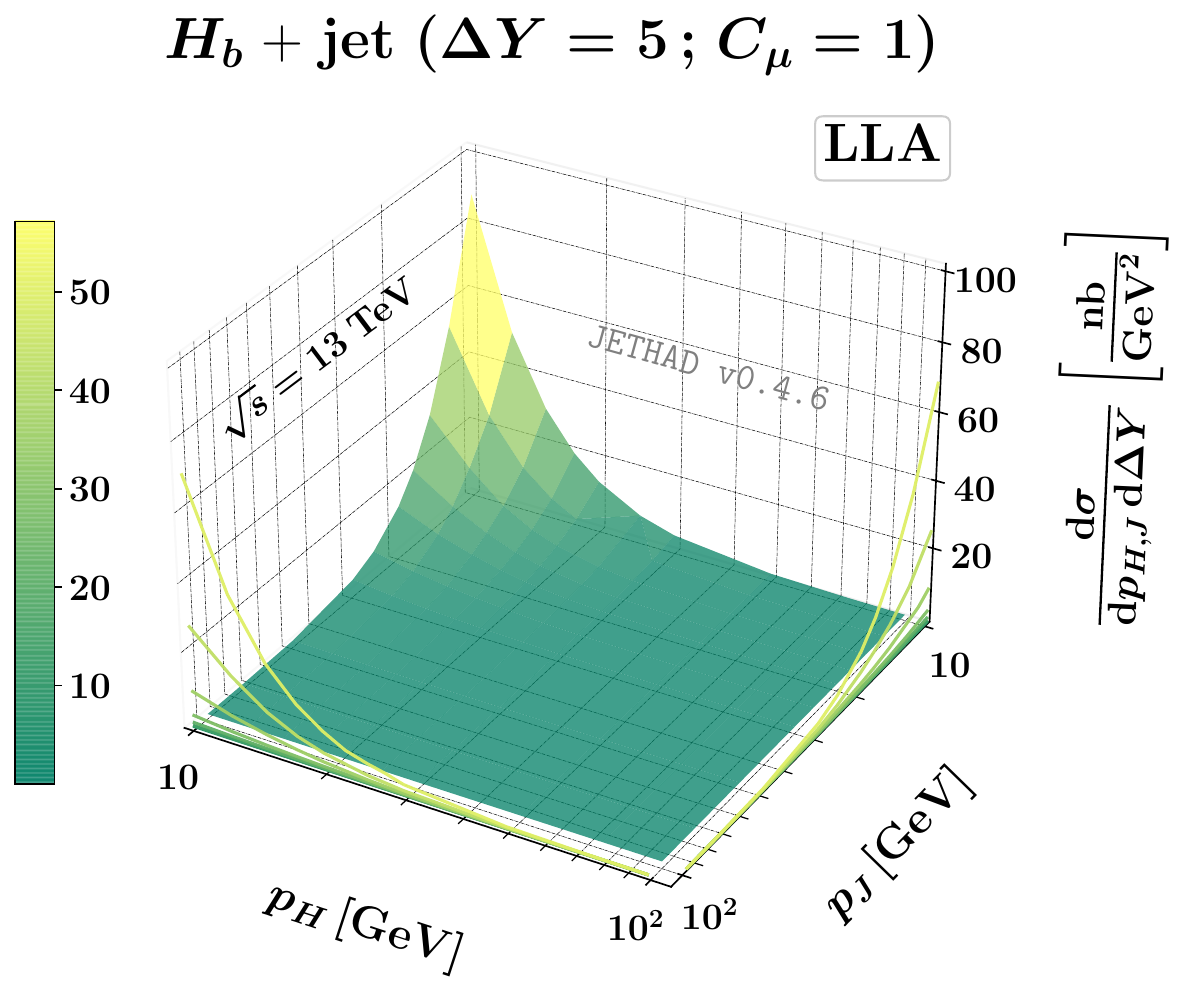}
   \hspace{0.25cm}
   \includegraphics[scale=0.37,clip]{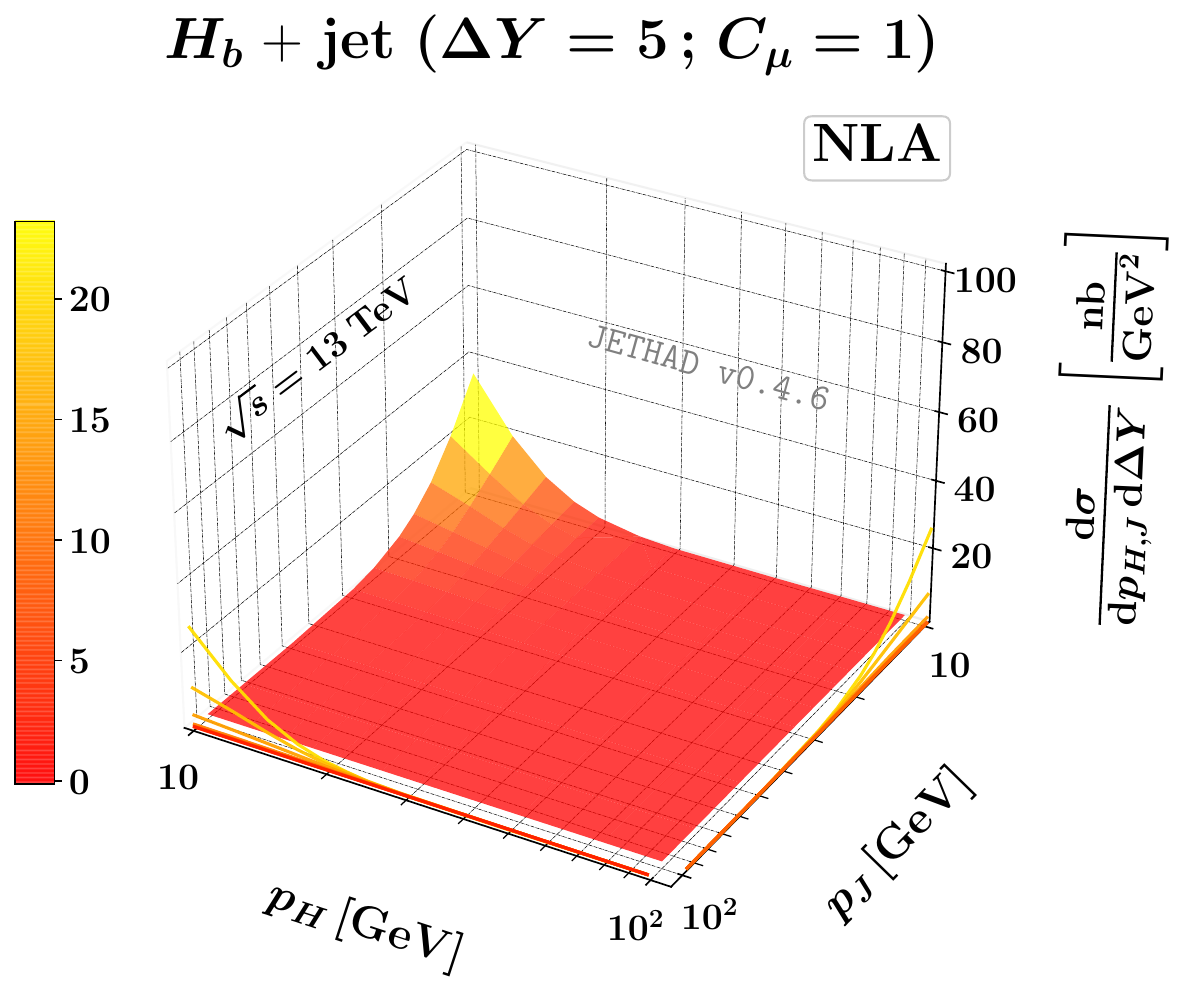}
   \vspace{0.15cm}
\caption{Double differential $p_T$-distribution for the $H_b$~$+$~jet channel at $\DY=3$ (top) and $\DY=5$ (bottom), $\sqrt{s} = 13$ TeV, and in the LLA (left) and NLA (right) resummation accuracy.}
\label{fig:Y3-2pT0}
\end{figure}

Lastly, we also investigate the \emph{double differential $p_T$-distribution} at fixed values of $\DY$,
    \begin{equation}
     \label{DoublepT_distribution}
     \frac{ d \sigma_{pp}(|\vec p_1|, |\vec p_2|, \DY, s)}{ d |\vec p_1| d |\vec p_2| d \DY} =
     \int_{y^{\rm{min}}_1}^{y^{\rm{max}}_1} d y_1
     \int_{y^{\rm{min}}_2}^{y^{\rm{max}}_2} d y_2
     \, \delta \left( y_1 - y_2 - \DY \right)
     \, {\cal C}_0 .
    \end{equation}
In Ref.~\cite{Celiberto:2021fjf} this study was proposed, without pretension of catching all the dominant features of this observable by the hand of our hybrid factorization, but rather to set the ground for futures studies where the interplay of different resummations (among all BFKL, Transverse momentum, and threshold one) can be deeply investigated. Results for our distributions in the $H_b$~$+$~jet channel at $\DY=3$ and $5$ are presented in Fig.~\ref{fig:Y3-2pT0}. In this analysis no BLM scale optimization is employed. We note that predictions fall off very fast when the two observable transverse momenta, $|\vec p_H|$ and $|\vec p_J|$, become larger or when their mutual distance grows. As generally predicted by the BFKL dynamics, LLA predictions (left panels) are always larger than NLA ones (right panels). Furthermore, we do not observe any peak, which could be present in the low-$p_T$ region, namely where Transverse momentum-resummation effects are dominant and that it is excluded from our analysis.

\section{Summary and outlook}

In this chapter, we have proposed a series of new semi-hard reactions that can be studied at the LHC, presenting a whole series of observables useful for detecting high-energy effects. We believe that such effects should be included in the precision programs of the LHC and of future accelerators capable of reaching even higher center-of-mass energies. \\

There are several interesting future developments on the side of forward/backward phenomenology. The analysis presented here in the case of heavy-flavor production only partially covers the $p_T$-spectrum. In fact, as already said, in the region of small-$p_T$ of the produced heavy-hadron a FFNS description is more reliable. The construction of predictions within this scheme (or in a full GM-VFNS), necessarily requires finding the NLO corrections to impact factors for the production of massive quarks ($Q$). The impact factor for the $Q \rightarrow Q$ transition is already known at next-to-leading order~\cite{Chachamis:2013bwa}. Nevertheless, the production of heavy-flavor involves other channels that are even dominanting. Among them, the contributions calculated at LO in the references~\cite{Boussarie:2017oae,Bolognino:2019yls} are necessary to study the productions of quarkonia or heavy-light mesons. Another interesting channel in the context of forward/backward productions has been proposed in Ref.~\cite{Golec-Biernat:2018kem}, \textit{i.e.} forward Drell-Yan and backward jet production. In this case, a full NLL description, requires the calculation of the next-to-leading impact factor for the forward Drell-Yan production. \\

In the case of Higgs-plus-jet production, it is already possible to perform a full next-to-leading logarithmic analysis. Using the next-to-leading order result, for example obtained through {\tt POWHEG}, it is also possible to implement a matching procedure to obtain a full next-to-leading order prediction supplemented by a complete resummation of high-energy logarithms. Through such a procedure it will be on the one hand clearer to distinguish the high-energy effects from those of the fixed order dynamics and on the other hand to provide valid predictions in a broader kinematic regime. \\ 

\newpage
\thispagestyle{empty}
\vspace*{\fill}
    \begin{center}
      { \Huge \textbf{Part II}} \vspace{0.3 cm} \\
      { \Huge \textbf{Beyond the NLLA}}
    \end{center}
\vspace*{\fill}
\addcontentsline{toc}{chapter}{Beyond the NLLA}
\chapter{Lipatov vertex in QCD with higher $\epsilon$-accuracy}
\label{Chap:Lipatov}

\begin{flushright}
\emph{\textit{How can it be that mathematics, being after all a product of \\ human thought which is independent of experience, is so \\ admirably appropriate to the objects of reality? \\ Albert Einstein~\cite{Einstein:1921ge}}}
\end{flushright}

In this chapter, we deal with the computation of the Lipatov vertex with higher accuracy in the dimensional regularization parameter $\epsilon = (D-4)/2$. There are a number of reasons motivating the need of the NLO Lipatov effective vertex with higher $\epsilon$-accuracy: first, it is the building block of the next-to-NLO contribution to the BFKL kernel from the production of one gluon in the collision of two Reggeons; second, it enters the expression of the impact factors for the Reggeon-gluon transition~\cite{Bartels:2003jq}, which appear in the derivation of the bootstrap conditions for inelastic amplitudes; these discontinuities are needed in the derivation of the BFKL equation in the NNLLA; third, the discontinuities of multiple gluon production amplitudes in the MRK can be used~\cite{Fadin:2014yxa} for a simple demonstration of violation of the ABDK-BDS (Anastasiou-Bern-Dixon-Kosower — Bern-Dixon-Smirnov) ansatz~\cite{Anastasiou:2003kj,Bern:2005iz} for amplitudes with maximal helicity violation in Yang-Mills theories with maximal super-symmetry ($\mathcal{N}=4$ SYM) in the planar limit and for the calculations of the remainder functions to this ansatz. In the planar maximally supersymmetric $ \mathcal{N} = 4$ Yang-Mills theory, the Lipatov vertex, within accuracy $\epsilon^2$, has been computed in~\cite{DelDuca:2009ae}. \\

The chapter contains four sections. In the first, we briefly discuss the formulation of the BFKL approach in the NNLLA. In the second, we review the derivation of the Lipatov vertex at one-loop. In the third, we compute the fundamental integrals appearing in the vertex up to the required accuracy. In the third and last section, we summarize and discuss future perspectives. The material of this chapter is based on Refs.~\cite{Fadin:2016wso,Fadin:2023roz}.
\section{BFKL beyond NLLA}
\label{BFKL beyond NLLA}
Extending BFKL beyond the NLLA is an extremely complex task. Let's summarize what we have seen in the previous chapters. \\

The $s$-channel discontinuities of the processes $A + B \rightarrow A' + B'$ are presented as the convolutions $ \Phi_{A A'} \; \otimes \; G \; \otimes \; \Phi_{B B'}$, where the impact factors $\Phi_{A A'}$ and
$ \Phi_{B B'}$ describe transitions $A \rightarrow A'$ and $B \rightarrow B'$ due to interactions with Reggeized gluons and $G$ is the Green’s function for two interacting Reggeized gluons. The Mellin transform of the Green function, satisfy the BFKL equation~(\ref{Int:Eq:BFKLOPEform})
\begin{equation*}
    \omega \hat{G}_{\omega} = 1 + \hat{ \mathcal{K} } \hat{G}_{\omega} \; .
\end{equation*}
Energy dependence of scattering amplitudes is determined by the BFKL kernel, which is
universal (process independent). \\
In the LLA, up to the required accuracy,
\begin{equation}
   \hat{\mathcal{K}} = \hat{\omega}_1 + \hat{\mathcal{K}}_{G}^{(0)} \;  
\end{equation}
where $\hat{\omega}_1$ is the one-loop Regge trajectory and $\hat{\mathcal{K}}_{G}^{(0)}$ is the part of the kernel related to real production of one gluon. \\
In the NLLA, we can still rely on the previous program of computations, but the kernel, up to the required accuracy, is
\begin{equation}
    \hat{\mathcal{K}} = \hat{\omega}_1 + \hat{\omega}_2 + \hat{\mathcal{K}}_{G}^{(0)} + \hat{\mathcal{K}}_{G}^{(1)} + \hat{\mathcal{K}}_{GG}^{(0)} + \hat{\mathcal{K}}_{Q \bar{Q}}^{(0)} \; ,  
\end{equation}
where, in the ``virtual" part, we have included the two-loop corrections to the Regge trajectory, while,  in the real particle production part, we have included, together with the
contribution from the one-gluon production in the RR collisions (this time calculated at one-loop
accuracy), contributions from the two-gluon and quark-antiquark productions (these latter computed at Born level). \\

One might think that this scheme is applicable in the NNLLA as well. In this case it
would be sufficient to calculate three-loop corrections to the trajectory, two-loop corrections to $\hat{\mathcal{K}}_G$, one-loop corrections to $\hat{\mathcal{K}}_{Q \bar{Q}}$ and $\hat{\mathcal{K}}_{GG}$ and to find in the Born approximation two new contributions, $\hat{\mathcal{K}}_{Q \bar{Q} G}$ and $\hat{\mathcal{K}}_{G G G}$. Unfortunately, the scheme based on the forms (\ref{Int:Eq:ReggeizedAmp}) and (\ref{Int:Eq:FinMulInelastic}) does not provide a full resummation in the NNLLA. The reason is the need to take account of the contributions of Regge cuts and the imaginary parts of the amplitudes in the unitarity conditions (\ref{Unitary}). \\

The main difficult in the extention of BFKL beyond the NLLA is the appearance of three-Reggeon cut which invalidates the forms (\ref{Int:Eq:ReggeizedAmp}) in the NNLLA. This was first shown in~\cite{DelDuca:2001gu} when considering the non-logarithmic terms in two-loop amplitudes for elastic scattering. A detailed consideration of the terms responsible for the breaking of the pole-Regge form in two- and three-loop amplitudes was performed in~\cite{DelDuca:2013ara,DelDuca:2013dsa,DelDuca:2014cya}. In~\cite{Fadin:2016wso,Fadin:2017nka,Caron-Huot:2017fxr} it was shown that the observed violation of the pole-Regge form can be explained by the contributions of the three-Reggeon cuts. A procedure for disentangling the Regge cut and Regge pole in QCD in all orders of perturbation theory has been suggested in~\cite{Falcioni:2021dgr}. 
Moreover, in deriving BFKL in the LLA, we have not emphasized much that the form of the inelastic amplitudes in Eq.~(\ref{Int:Eq:FinMulInelastic}) is strictly valid only for the real part of the amplitude. The reason is that only the real part is really needed in the LLA and in the NLLA. Indeed, the imaginary parts are suppressed by a power of $\ln s_i$ with respect to the real ones. In the LLA, no logarithm can be ``lost" and hence only the real part is important. In the NLLA, in principle, the products of imaginary and real parts in the unitarity relations (\ref{Unitary}) are important, but they cancel due to summation of contributions complex conjugated to each other. In the NNLLA, we can ``lose" two large logarithms and hence imaginary parts should be taken into account. At computational level, this is not very complicated since we need imaginary parts just in the main approximation. Nonetheless, this complicates the derivation of the BFKL equation and deprives it of its universality (see Ref.~\cite{Fadin:2016wso} for details). \\

In the following, as already mentioned, we focus our attention just on the one-loop Lipatov vertex, extending its knowledge in the $\epsilon$-expansion up to the required accuracy for NNLL formulation of BFKL. In fact, the NNLLA formulation of BFKL requires not only two and three-loop calculations, but also  higher $\epsilon$-accuracy of the one-loop results. Thus, since the NLO RRG vertex has the singularity $1/\epsilon^2$, it must be known in general up to terms of order $\epsilon^2$ to ensure the accuracy  $\epsilon^0$ in the part of the kernel containing the product of two RRG vertices. Of course,  in the region of small transverse momentum $\vec p$ of the produced gluon, the accuracy must be higher ($\epsilon^3$). Fortunately, in this region the vertex was obtained in~\cite{Fadin:1996yv} exactly in $\epsilon$.

\section{Review of the Lipatov vertex at one-loop}
\label{sec:Rew}
\subsection{The gluon production amplitude}
\label{subsec:gluonproductionamplitude}
The RRG vertex can be obtained from a generic $A + B \rightarrow A' + g + B'$ amplitude taken in MRK and with the gluon emitted in the central kinematical region. In this case we follow the construction in~\cite{Fadin:2000yp} and consider gluon-gluon collision. We will use the denotations $p_A$ and $p_B$ ($p_{A'}$ and $p_{B'}$) for the momenta of the incoming (outgoing) gluons, $p$ and $e(p)$ for momentum and polarization vector of the produced gluon; $q_1= p_A - p_{A'}$ and $q_2= p_{B'} - p_B$ are the momentum transfers, so that $p=q_1-q_2$. The denotations are the same as in~\cite{Fadin:2000yp}, except that we use $p$ for the ``central" outgoing gluon, instead of $k$. The MRK kinematics is defined by the relations
\begin{equation}
s \gg s_1 , s_2 \gg |t_1| \sim |t_2|\;,
\label{2}
\end{equation}
where
\begin{equation}
s=(p_A+p_B)^2\;, \;\;\;\;\; s_1=(p_{A'}+p)^2\;, \;\;\;\;\; 
s_2=(p_{B'}+p)^2\;, 
\;\;\;\;\; t_{1,2}=q_{1,2}^2\;.
\label{MRKMin}
\end{equation}
In terms of the parameters of the Sudakov decomposition
\begin{equation}
p = \beta_p \, p_A + \alpha_p \, p_B + p_\perp \;, \;\;\;\;\;\;\;\;\;\;
q_i = \beta_i \, p_A + \alpha_i \, p_B + {q_i}_{\perp} \;, 
\label{4}
\end{equation}
the relations~(\ref{2}) give
\[
1 \gg \beta_p \approx \beta_1 \gg -\alpha_1 \simeq \frac{\vec q_1^{\:2}}{s} \;,
\;\;\;\;\;\;\;\;\;\;
1 \gg \alpha_p \approx -\alpha_2 \gg \beta_2 \simeq\frac{\vec q_2^{\:2}}{s} \;,
\]
\begin{equation}
s_1 \approx s \, \alpha_p \;, \;\;\;\;\; s_2 \approx s \, \beta_p \;, \;\;\;\;\;
\vec{p}^{\; 2} = - p_\perp^2 \approx \frac{s_1 s_2}{s}\;.
\label{5}
\end{equation} 
Here and below, the vector sign is used for the components of the momenta transverse to the plane of the momenta of the initial particles $p_A$ and $p_B$. \\

To extract the RRG vertex, we can restrict ourselves to amplitudes with conservation of the helicities of the scattered gluons. The form of this amplitude is well known:
\begin{equation}
A_{2\rightarrow 3} = 2s \, g^3\, T_{A'A}^{c_1} \, \frac{1}{t_1} \, T_{c_2c_1}^{d} 
\, \frac{1}{t_2}\,T_{B'B}^{c_2} \, e^*_\mu(p) \, {\cal A}^\mu(q_2,q_1)\;,
\label{6}
\end{equation} 
where $T_{bc}^a$ are matrix elements of the colour group generator 
in the adjoint representation and the amplitude
${\cal A}^\mu$ in the Born approximation is equal to $C^\mu(q_2,q_1)$: 
\begin{equation}
{\cal A}_{\rm Born}^\mu = C^\mu(q_2,q_1) = -{q_1}_{\perp}^\mu -{q_2}_{\perp}^\mu
+\frac{p_A^\mu}{s_1}\,(\vec p^{\; 2} - 2\vec q_1^{\:2})
-\frac{p_B^\mu}{s_2}\,(\vec p^{\; 2} - 2\vec q_2^{\:2}) \; .
\label{7}
\end{equation}
It was shown in~\cite{Fadin:1993rc} that at one-loop order the amplitude can be presented
 in its gauge invariant form,
\begin{equation}
{\cal A}^\mu = C^\mu(q_2,q_1) (1+\overline g^2 r_C) + {\cal P}^\mu \overline g^2
\, 2 t_1 t_2 \, r_{\cal P}\;,
\label{26}
\end{equation}
where the terms proportional to $p^\mu$ were obviously omitted and 
\begin{equation*}
    \bar{g} = \frac{N g^2 \Gamma (1-\epsilon)}{(4 \pi)^{2+\epsilon}} \; ,
\end{equation*}
\begin{equation*}
    \mathcal{P}^{\mu} = \frac{p_A^{\mu}}{s_1} - \frac{p_B^{\mu}}{s_2} \; ,
\end{equation*}
\[
r_C = \left\{t_1 t_2 \left(r_{_+}+\frac{{\cal F}_5}{2}\right)+t_2{\cal I}_{4B} 
+ \frac{\Gamma^2(\epsilon)}{\Gamma(2\epsilon)} (\vec q_1^{\:2})^\epsilon
\left[ -\frac{1}{2}\ln\left(\frac{s_1 (-s_1)}{t_1^2}\right)
+2 \psi(\epsilon) - \psi(2\epsilon)
\right.\right.
\] 
\[
- \psi(1-\epsilon)+\frac{1}{2\epsilon(1+2\epsilon)(3+2\epsilon)}\left(-3(1+\epsilon)
-\frac{\epsilon^2}{1+\epsilon} + \frac{t_1(3+14\epsilon+8\epsilon^2)
-t_2(3+3\epsilon+\epsilon^2)}{t_1-t_2} \right. 
\]
\[
\left.\left.
+ \frac{\vec p^{\; 2} \, t_1 \, \epsilon}{(t_1-t_2)^3}
\biggl((2+\epsilon)t_2-\epsilon\,t_1\biggr)\right]\right\}
+ \biggl\{A \longleftrightarrow B \biggr\}\;,
\]
\[
r_{\cal P} = \left\{(\vec q_1 \cdot \vec q_2) r_{_+} - \frac{t_1+t_2}{4}{\cal F}_5
-\frac{{\cal I}_{4B}}{2} 
+ \frac{1}{2}\frac{\Gamma^2(\epsilon)}{\Gamma(2\epsilon)} 
(\vec q_1^{\:2})^{\epsilon-1}
\left( -\frac{1}{2}\ln\left(\frac{s_1(-s_1)s_2(-s_2)}{s(-s)t_1^2}\right)
+ \psi(1)
\right.\right.
\]
\[
+\psi(\epsilon) - \psi(1-\epsilon) - \psi(2\epsilon)
+\frac{t_1}{t_2\,(1+2\epsilon)(3+2\epsilon)}
\left[ \frac{t_2}{t_1-t_2} (11+7\epsilon) \right.
\]
\begin{equation}
\left.\left.\left.
+ \frac{\vec p^{\; 2}}{(t_1-t_2)^3}\biggl(t_2(t_1+t_2)-\epsilon\,t_1(t_1-t_2)\biggr)
+ \frac{(\vec p^{\; 2})^2}{(t_1-t_2)^3}\biggl((2+\epsilon)t_2-\epsilon\,t_1\biggr)
\right]\right)\right\} + \biggl\{ A \longleftrightarrow B \biggr\} .
\label{27}
\end{equation}
The expression for the function $r_+$ appearing in the definition of ${\cal A}^\mu$ is
\begin{equation}
r_{_+} = \frac{f_{_-} (\vec q_1^{\:2}-\vec q_2^{\:2})-f_{_+} \vec p^{\; 2}}
{8 (\vec q_1^{\:2} \vec q_2^{\:2} - (\vec q_1 \vec q_2)^2)} \;,
\end{equation}
with
\[
f_{_-} \equiv \left[-t_1 {\cal F}_5 - {\cal I}_{4B}
- \frac{\Gamma^2(\epsilon)}{\Gamma(2\epsilon)}(\vec q_2^{\: 2})^{\epsilon-1}
\left(\frac{1}{2} \ln\left(\frac{s_1 (-s_1) s_2(-s_2)}{s(-s) t_2^2}\right)
\right.\right.
\]
\begin{equation}
\biggl. \biggl.
+ \psi(1-\epsilon) - \psi(\epsilon) + \psi(2 \epsilon) - \psi(1) \biggr)\biggr]
- \biggl[ A \longleftrightarrow B \biggr] \;, 
\label{21}
\end{equation}
\[
 f_{_+} \equiv \left\{-t_1 {\cal F}_5 
-{\cal I}_{4B}-\frac{\Gamma^2(\epsilon)}{\Gamma(2\epsilon)}
\left[(\vec q_2^{\: 2})^{\epsilon-1}
\left(\frac{1}{2} \ln\left(\frac{s_1 (-s_1) s_2(-s_2)}{s(-s) t_2^2}\right)
\right.\right.\right.
\]
\[
\biggl. 
+ \psi(1-\epsilon) - \psi(\epsilon) + \psi(2 \epsilon) - \psi(1) \biggr)
\]
\begin{equation}
\left.\left.
-(\vec p^{\; 2})^{\epsilon-1}
\left(\frac{1}{2} \ln\left(\frac{s_1 (-s_1) s_2(-s_2)}{s(-s) (\vec p^{\; 2})^2}
\right) + \psi(1-\epsilon) - \psi(\epsilon)\right)\right]\right\}
+ \biggl\{ A \longleftrightarrow B \biggr\} \; , 
\label{22}
\end{equation}
and
\begin{equation}
    {\cal F}_5 = {\cal I}_5 - {\cal L}_3 - \frac{1}{2} 
\ln\left(\frac{s (-s) (\vec p^{\; 2})^2}{s_1 (-s_1) s_2 (-s_2)}\right) {\cal I}_3\;.
\end{equation}
The total amplitude is given in terms of five structures $ \mathcal{I}_3,  \; \mathcal{L}_3, \; \mathcal{I}_{4B}, \; \mathcal{I}_{4A}, \; \mathcal{I}_{5}$, defined as~\cite{Fadin:2000yp}
\begin{equation}
    \mathcal{I}_3 = \int \frac{d^{2+2\epsilon} k}{\pi^{1+\epsilon}\Gamma(1-\epsilon)} \frac{1}{\vec{k}^{\, 2}  (\vec{k}-\vec{q}_1)^2 (\vec{k}-\vec{q}_2)^2} \; ,
    \label{I3cal}
\end{equation}
\begin{equation}
    \mathcal{L}_3 = \int \frac{d^{2+2\epsilon} k}{\pi^{1+\epsilon} \Gamma(1-\epsilon)} \frac{1}{\vec{k}^{\, 2}  (\vec{k}-\vec{q}_1)^2 (\vec{k}-\vec{q}_2)^2} \left[\ln \left( \frac{(\vec{k}-\vec{q}_1)^2(\vec{k}-\vec{q}_2)^2}{\vec{k}^{\, 2}  (\vec{q}_1-\vec{q}_2)^{\; 2}} \right) \right] \; ,
    \label{L3cal}
\end{equation}
\begin{equation*}
\mathcal{I}_{4B} = \int_0^1 \frac{dx}{x} \int \frac{d^{2+2\epsilon} k}{\pi^{1 + \epsilon} \Gamma (1-\epsilon)} 
\end{equation*} 
\begin{equation}
   \times \left[ \frac{1-x}{\left( x \vec{k}^2 + (1-x) (\vec{k}-\vec{q}_1)^2 \right) (\vec{k}-(1-x)(\vec{q}_1-\vec{q}_2))^2} - \frac{1}{(\vec{k}-\vec{q}_{1})^2 (\vec{k}-(\vec{q}_1-\vec{q}_2))^2} \right] \; ,
\label{I4Bcal}
\end{equation}
\begin{equation*}
    \mathcal{I}_5 = \int_0^1 \frac{d x}{1-x} \int \frac{ d^{2+2\epsilon} k}{\pi^{1+\epsilon} \Gamma(1-\epsilon)} \frac{1}{\vec{k}^{\, 2} [(1-x) \vec{k}^{\, 2}  + x (\vec{k}-\vec{q}_1)^2]}
\end{equation*}
\begin{equation}
   \times \left[ \frac{x^2}{(\vec{k}-x(\vec{q}_1-\vec{q}_2))^2} - \frac{1}{(\vec{k}-\vec{q}_1+\vec{q}_2)^2} \right] 
    \label{I5cal}
\end{equation}
and $\mathcal{I}_{4A}= \mathcal{I}_{4B} (\vec{q}_1 \leftrightarrow -\vec{q_2})$.

\subsection{The Lipatov vertex}
\label{subsec: Lipatov vertex}
Imposing general requirements of
analyticity, unitarity and crossing symmetry, the production amplitude must take the Regge the form~\cite{Fadin:1993rc,Bartels:1980}
\[
8\, g^3 \, {\cal A}^\mu = \Gamma(t_1;\vec p^{\; 2}) \Gamma(t_2;\vec p^{\; 2})
\left\{
\left[\left(\frac{s_1}{\vec p^{\; 2}}\right)^{\omega_1-\omega_2}
+\left(\frac{-s_1}{\vec p^{\; 2}}\right)^{\omega_1-\omega_2}\right]
\left[\left(\frac{s}{\vec p^{\; 2}}\right)^{\omega_2}
+\left(\frac{-s}{\vec p^{\; 2}}\right)^{\omega_2}\right]\, R^\mu \right.
\]
\begin{equation}
\left.
+ \left[\left(\frac{s_2}{\vec p^{\; 2}}\right)^{\omega_2-\omega_1}
+\left(\frac{-s_2}{\vec p^{\; 2}}\right)^{\omega_2-\omega_1}\right]
\left[\left(\frac{s}{\vec p^{\; 2}}\right)^{\omega_1}
+\left(\frac{-s}{\vec p^{\; 2}}\right)^{\omega_1}\right]\, L^\mu 
\right\}\;,
\label{28}
\end{equation}
where $\omega_i = \omega(t_i)$ (we have chosen $\vec p^{\; 2}$ as the scale of energy), $\Gamma(t_i;\vec p^{\; 2})$ are the helicity conserving gluon-gluon-Reggeon vertices, $R^\mu$ and $L^\mu$ are the right and left 
RRG vertices, depending on $\vec q_1$ and $\vec q_2$. They are real 
in all physical channels, as well as $\Gamma(t_i; \vec p^{\; 2})$. 

In the one-loop approximation we have 
\begin{equation}
\omega(t) = \omega^{(1)}(t) = - \overline g^2 \frac{\Gamma^2(\epsilon)}
{\Gamma(2\epsilon)} (\vec q^{\:2})^\epsilon\;,
\label{29}
\end{equation}
\begin{equation}
\Gamma(t;\vec p^{\; 2}) =  g \biggl(1+\Gamma^{(1)}(t;\vec p^{\; 2}) \biggr)\;,
\label{30}
\end{equation}
where
\[
\Gamma^{(1)}(t;\vec p^{\; 2}) = \overline g^2 \frac{\Gamma^2(\epsilon)}
{\Gamma(2\epsilon)} (\vec q^{\:2})^\epsilon \left[\psi(\epsilon)
- \frac{1}{2} \psi(1) - \frac{1}{2} \psi(1-\epsilon) \right.
\]
\begin{equation}
\left.
+\frac{9 (1+\epsilon)^2+2}{4(1+\epsilon) (1+2\epsilon) (3+2\epsilon)}
-\frac{1}{2} \ln\left(\frac{\vec p^{\; 2}}{\vec q^{\:2}}\right)\right] \;.
\label{31}
\end{equation}
In the same approximation we obtain from~(\ref{28})
\[
2 g \left\{{\cal A}^\mu - C^\mu(q_2,q_1)\left[\Gamma^{(1)}(t_1;\vec p^{\; 2})
+\Gamma^{(1)}(t_2;\vec p^{\; 2})
+\frac{\omega_1}{2}\ln\left(\frac{s_1(-s_1)}{(\vec p^{\; 2})^2}\right)
+\frac{\omega_2}{2}\ln\left(\frac{s_2(-s_2)}{(\vec p^{\; 2})^2}\right) \right.\right.
\]
\begin{equation}
\left.\left.
+\frac{\omega_1+\omega_2}{4}\ln\left(\frac{s(-s)(\vec p^{\; 2})^2}{s_1(-s_1)s_2(-s_2)}
\right)\right]\right\}
= R^\mu + L^\mu + (R^\mu - L^\mu) \frac{\omega_1-\omega_2}{4} 
\ln\left(\frac{s_1(-s_1)s_2(-s_2)}{s(-s)(\vec p^{\; 2})^2}\right) .
\label{32}
\end{equation}
Since in all physical channels
\begin{equation}
    \ln\left(\frac{s_1(-s_1)s_2(-s_2)}{s(-s)(\vec p^{\; 2})^2}\right) = i \pi \;,
\end{equation}
the combinations $R^{\mu} + L^{\mu}$ and $R^{\mu} - L^{\mu}$ are respectively related to the real and imaginary parts of the production amplitude. Using this we find that
\[
R^\mu + L^\mu = 2g\left\{C^\mu(q_2,q_1)+
\overline g ^2\left(\frac{(\vec q_1 \cdot \vec q_2)C^\mu(q_2,q_1)+
2\vec q_1^{\:2}\vec q_2^{\:2}{\cal P}^\mu}{2(\vec q_1^{\:2}\vec q_2^{\:2}
-(\vec q_1 \cdot \vec q_2)^2)}\left[
\frac{\vec q_1^{\:2}\vec q_2^{\:2}\vec{p}^{\; 2}}{2}
({\cal I}_5-{\cal L}_3)
\right.\right.\right.
\]
\[
\left.
-\vec q_2^{\:2}(\vec q_1 \cdot \vec{p})\,{\cal I}_{4B}
+\frac{\Gamma^2(\epsilon)}{\Gamma(2\epsilon)}
\left(\frac{(\vec{p}^{\; 2})^{\epsilon}}{2}(\vec q_1 \cdot \vec q_2)\left(
\psi(\epsilon)-\psi(1-\epsilon)\right)+(\vec q_1^{\:2})^{\epsilon}
(\vec q_2 \cdot \vec{p})\left(\ln \left(\frac{\vec p^{\; 2}}
{\vec q_1^{\:2}}\right)-\psi(1)\right.\right.\right.
\]
\[
\left.\left.\left. -\psi(\epsilon) +\psi(1-\epsilon)
+\psi(2\epsilon)\phantom{\frac{1}{1}}\!\!\!
\right)\right)\right]+C^\mu(q_2,q_1)\left[-\frac{\vec q_2^{\:2}}{2}{\cal I}_{4B}+
\frac{\Gamma^2(\epsilon)}{2\Gamma(2\epsilon)}\left(
\frac{(\vec{p}^{\; 2})^\epsilon}{2}(\psi(\epsilon)-\psi(1-\epsilon))
\right.\right.
\]
\[
+(\vec q_1^{\:2})^{\epsilon}\left[\psi(\epsilon)
-\psi(2\epsilon)+\frac{1}{2(1+2\epsilon)(3+2\epsilon)}\left(\frac
{\vec q_1^{\:2}+\vec q_2^{\:2}}{\vec q_1^{\:2}-\vec q_2^{\:2}}
(11+7\epsilon)+2\epsilon\frac
{\vec p^{\; 2}\vec q_1^{\:2}}{(\vec q_1^{\:2}-\vec q_2^{\:2})^2} \right.\right.
\]
\[
\left.\left.\left.\left.
-4\frac{\vec p^{\; 2}\vec q_1^{\:2}\vec q_2^{\:2}}
{(\vec q_1^{\:2}-\vec q_2^{\:2})^3}\right)\right]\right)\right]
+{\cal P}^\mu\frac{\Gamma^2(\epsilon)(\vec q_1^{\:2})^{\epsilon}}
{\Gamma(2\epsilon)(1+2\epsilon)(3+2\epsilon)}
\left[\frac{\vec q_1^{\:2}\vec q_2^{\:2}}{\vec q_1^{\:2}-\vec q_2^{\:2}}
(11+7\epsilon)+\epsilon \vec p^{\; 2}\vec q_1^{\:2}\frac
{\vec q_1^{\:2}-\vec p^{\; 2}}{(\vec q_1^{\:2}-\vec q_2^{\:2})^2}
\right. 
\]
\begin{equation}
\left.\left.\left.
-\vec p^{\; 2}\vec q_1^{\:2}\vec q_2^{\:2}
\frac{\vec q_1^{\:2}+\vec q_2^{\:2}-2\vec p^{\; 2}}
{(\vec q_1^{\:2}-\vec q_2^{\:2})^3}\right]\right)
-\overline g ^2\left(A\longleftrightarrow B\right)\right\}  
\label{Rmu+Lmu}
\end{equation}
and
\[
R^\mu - L^\mu =  \frac{2g \overline g^2 }{\omega_1-\omega_2}
\left\{-C^\mu(q_2,q_1)\frac{\Gamma^2(\epsilon)}
{\Gamma(2\epsilon)}(\vec p^{\; 2})^{\epsilon}
+\frac{(\vec q_1 \cdot \vec q_2)C^\mu(q_2,q_1)+2\vec q_1^{\:2}
\vec q_2^{\:2}{\cal P}^{\mu}}{\vec q_1^{\:2} \vec q_2^{\:2}
-(\vec q_1 \cdot \vec q_2)^2}
\right.
\]
\begin{equation}
\left.\times \left[\vec q_1^{\:2}\vec q_2^{\:2}\vec p^{\; 2}
{\cal I}_3+\frac{\Gamma^2(\epsilon)}{\Gamma(2\epsilon)}
\left((\vec q_1^{\:2})^{\epsilon}(\vec q_2 \cdot \vec p) 
-(\vec q_2^{\:2})^{\epsilon}(\vec q_1 \cdot \vec p) 
-(\vec p^{\; 2})^{\epsilon}(\vec q_1 \cdot \vec q_2)\right)\right]\right\}\;.
\label{Rmu-Lmu}
\end{equation} 
It is clear that, in order to know the Lipatov vertex at a certain order in the $\epsilon$-expansion, we must calculate the integrals $\mathcal{I}_3$, $\mathcal{I}_{4A}$, $\mathcal{I}_{4B}$, and $\mathcal{I}_5-\mathcal{L}_3$ with the same accuracy.  
\\

In the region of small momenta of the central emitted gluon (the {\em soft region}, from now on) the vertex must be known to all orders in $\epsilon$. The soft limit of integrals entering the vertex are computed in appendix~\ref{AppendixC4}. By using Eqs.~(\ref{I3calSoft}),~(\ref{I4BcalSoft}) and (\ref{I5cal-L3calSoft}), we easily find
\begin{equation}
R^\mu + L^\mu = 2g C^\mu(q_2,q_1)\left(1+\overline g^2
\frac{\Gamma^2(\epsilon)}{2\Gamma(2\epsilon)}(\vec p^{\; 2})^{\epsilon}
\left[\psi(\epsilon)-\psi(1-\epsilon)\right]\right), 
\label{48b}   
\end{equation}
\begin{equation}
R^\mu - L^\mu = - \frac{2g \overline g^2 }{\omega_1-\omega_2}
C^\mu(q_2,q_1)\frac{\Gamma^2(\epsilon)}
{\Gamma(2\epsilon)}(\vec p^{\; 2})^{\epsilon} \; . 
\label{36c}   
\end{equation}
The last two equation confirm the result in~\cite{Fadin:1996yv,DelDuca:1998cx}.

\section{Fundamental integrals for the one-loop RRG vertex}
\label{sec:Fund}

In this section we present the result at $\epsilon^2$-accuracy of the integrals in the transverse momentum space entering the one-loop RRG vertex, expanding expressions
already available in the literature and, whenever possible, cross-checking them with an alternative calculation. 

\subsection{$\mathcal{I}_3$: Bern-Dixon-Kosower (BDK) method}
\label{subsec:I31}

We start from the integral
\[
\mathcal{I}_3' \equiv \pi^{1+\epsilon} \Gamma(1-\epsilon)\ \mathcal{I}_3 
\]
\begin{equation}
= \int d^{2+2\epsilon}k \frac{1}{\vec{k}^{2} (\vec{k}-\vec{q}_{1})^{2} (\vec{k}-\vec{q}_{2})^{2}} = \int d^{2+2\epsilon}k_{E} \frac{1}{k_{E}^{2} (k_{E}-q_{1E})^{2} (k_{E}-q_{2E})^{2}} \; ,
\label{I3prime_def}
\end{equation}
where $k_{E}, q_{1E}, q_{2E}$ are Euclidean vectors in dimension $2+2\epsilon$ and $q_{1E}^{\;2}, q_{2E}^{\;2}, (q_{1E}-q_{2E})^2 \neq 0$. We can relate this Euclidean integral to a corresponding Minkowskian integral by Wick rotation:
\begin{equation}
  \int d^{2+2 \epsilon}k_{E} \frac{1}{k_{E}^{2} (k_{E}-q_{1E})^{2} (k_{E}-q_{2E})^{2}}
  = i \int d^{2+2 \epsilon}k \frac{1}{k^{2} (k-q_1)^{2} (k-q_2)^{2}}\;,
\end{equation}
where $k = (i k_{E}^0, k_{E}^1)$, $q_1 = (i q_{1E}^0, q_{1E}^1)$, $q_2
= (i q_{2E}^0, q_{2E}^1)$. We note that
\begin{equation}
\hspace{-0.1 cm}  q_1^{\;2} = - q_{1E}^{\;2}=-\vec q_1^{\;2}, \hspace{0.4 cm} q_2^{\;2} = - q_{2E}^{\;2}=-\vec q_2^{\;2}\;,
  \hspace{0.4 cm}
  (q_1-q_2)^2 = -(q_{1E}-q_{2E})^2 =- (\vec q_1- \vec q_2)^{\;2} = - \vec{p}^{\; 2} .
\end{equation}
Hence, we need a triangular integral with three massive external legs. This
result to all order in $\epsilon$-expansion can be found in~\cite{Bern_1994},
taking care to apply the replacement ($ \epsilon \rightarrow -\epsilon + 1$),
because they calculated the integral in $d^{4 - 2 \epsilon}k$, instead we need it
in $d^{2 + 2 \epsilon}k$. One finds that
\begin{equation}
\mathcal{I}_3' = \pi^{1+\epsilon} \alpha_1 \alpha_2 \alpha_3 \left( -\frac{1}{2} \frac{\Gamma(2-\epsilon) \Gamma^2(\epsilon)}{\Gamma(2 \epsilon -1)} \right) \frac{\hat{\Delta}_3^{1/2-\epsilon}}{(1-\epsilon)^2} \left[ f \left( \delta_1 \right) + f \left( \delta_2 \right) + f \left( \delta_3 \right) + c \right] \;,
\end{equation} 
where
\begin{equation}
  \alpha_1 = \sqrt{\frac{\vec{q}_{2}^{\; 2}}{\vec{q}_{1}^{\;2} (\vec{q}_{1}-\vec{q}_{2})^2}}\;,\hspace{0.5 cm}
  \alpha_2 = \sqrt{\frac{(\vec{q}_{1}-\vec{q}_{2})^2}{\vec{q}_{1}^{\; 2} \vec{q}_{2}^{\;2}}}\;, \hspace{0.5 cm}
  \alpha_3 = \sqrt{\frac{\vec{q}_{1}^{\; 2}}{\vec{q}_{2}^{\; 2} (\vec{q}_{1}-\vec{q}_{2})^2}}\;,
\end{equation}
\begin{equation}
  \gamma_1 = -\alpha_1 + \alpha_2 + \alpha_3\;, \hspace{0.5 cm}
  \gamma_2 = \alpha_1 - \alpha_2 + \alpha_3 \;, \hspace{0.5 cm}
  \gamma_3 = \alpha_1 + \alpha_2 - \alpha_3 \;,
\end{equation}
\begin{equation}
  \delta_1 = \frac{\gamma_1}{\sqrt{\hat{\Delta}_3}}\;, \hspace{0.5 cm}
  \delta_2 = \frac{\gamma_2}{\sqrt{\hat{\Delta}_3}}\;, \hspace{0.5 cm}
  \delta_3 = \frac{\gamma_3}{\sqrt{\hat{\Delta}_3}} \;,
\end{equation}
\begin{equation}
  \hat{\Delta}_3 = - \alpha_1^2 - \alpha_2^2 - \alpha_3^2 + 2 \alpha_1 \alpha_2
  + 2 \alpha_2 \alpha_3 + 2 \alpha_1 \alpha_3  \;,
\end{equation}
\begin{equation}
  c = -2 \pi (1-\epsilon) \frac{\Gamma(2 \epsilon-1)}{\Gamma^2(\epsilon)}
  \;,
\end{equation}
\begin{equation}
\begin{split}
f \left( \delta \right) = & \frac{1}{i} \left[ \left( \frac{1+i\delta}{1-i\delta} \right)^{1-\epsilon} \; _2 F_1 \left(2 - 2 \epsilon, 1 - \epsilon, 2 - \epsilon; -\left( \frac{1+i\delta}{1-i\delta} \right) \right) \right. \\ &
   \left. - \left( \frac{1-i\delta}{1+i\delta} \right)^{1-\epsilon} \; _2 F_1 \left(2 - 2 \epsilon, 1 - \epsilon, 2 - \epsilon; -\left( \frac{1-i\delta}{1+i\delta} \right) \right) \right] \;. 
\end{split}
\end{equation}
Using the expansion (\ref{HyperExp1}) and the definitions
\begin{equation}
    z_i = - \left( \frac{1+ i \delta_i}{1-i \delta_i} \right) \; ,
\end{equation}
we find 
\begin{equation*}
    \mathcal{I}_3' = \pi^{1+\epsilon} \Gamma (1-\epsilon) \alpha_1 \alpha_2 \alpha_3 \hat{\Delta}_3^{1/2-\epsilon} \frac{ \Gamma^2(1+\epsilon)}{\Gamma(1+2 \epsilon)} \frac{ (1-2 \epsilon) }{(1-\epsilon)} \Bigg \{ \frac{\alpha_1 + \alpha_2 + \alpha_3}{\sqrt{\hat{\Delta}_3} \epsilon} + \Bigg( \pi + \sum_{i=1}^{3} \Bigg[ \frac{i \; z_i}{1-z_i} 
\end{equation*}
\begin{equation*}
    \times \left( 1 - \ln \left( -z_i \right) + \frac{1+z_i}{z_i} \ln \left( 1-z_i \right) \right) - (z_i \rightarrow z_{i}^{-1}) \Bigg] \Bigg) + \left( \pi + \sum_{i = 1}^{3} \left[ \frac{ i \; z_i}{1-z_i} \right. \right.
\end{equation*}
\begin{equation*}
    \left. \left. \times \left( 2 + \left( \frac{1+z_i}{z_i} \right) \ln \left( 1-z_i \right) \ln \left( (1-z_i)e \right) + \frac{1}{2} \ln^2 \left(-z_i \right) - \frac{1-z_i}{z_i} {\rm{Li}}_2 \left(z_i \right) \right. \right. \right.
\end{equation*}
\begin{equation*}
    \left. \left. \left. - \ln \left( -z_i \right) \left( 1 + \frac{1+z_i}{z_i} \ln \left( 1-z_i \right) \right) \right) - (z_i \rightarrow z_{i}^{-1}) \right] \right) \epsilon + \bigg( \pi (2 + \zeta (2)) 
\end{equation*}
\begin{equation*}
    \left. + \sum_{i}^{3} \left[ \frac{i \; z_i}{1-z_i} \left( 4 + \ln (1-z_i) \bigg( \frac{2(1+z_i)}{z_i} + \frac{2(1-z_i)}{z_i} \zeta (2) + \frac{\ln (1-z_i)}{3 z_i} \bigl( 3 (1+z_i) \right. \right. \right. \bigr.
\end{equation*}
\begin{equation*}
       \bigl. \left. \left. \left. + 2 (1+z_i) \ln (1-z_i) -3 (1-z_i) \ln z_i \right) \bigr) - \frac{(1-z_i)}{z_i} (1+2 \ln (1-z_i)) {\rm{Li}}_2 (z_i) - \frac{2 (1-z_i)}{z_i} \right. \right. 
\end{equation*}
\begin{equation*}
      \left. \left. \left. \times \left( {\rm{Li}}_3 (1-z_i) + \frac{{\rm{Li}}_3 (z_i)}{2} - \zeta(3) \right) - \frac{\ln^3 (-z_i)}{6} + \frac{\ln^2 (-z_i)}{2} \left( 1 + \left( \frac{1+z_i}{z_i} \right) \ln (1-z_i)  \right)  \right. \right. \right.
\end{equation*}
\begin{equation}
  \left. \left. \left. \hspace{-0.15 cm} - \ln (-z_i) \left( 2 + \left( \frac{1+z_i}{z_i} \right) \ln \left( 1-z_i \right) \ln \left( (1-z_i)e \right) - \frac{1-z_i}{z_i} {\rm{Li}}_2 \left(z_i \right) \right) \right) - (z_i \rightarrow z_{i}^{-1}) \phantom{\frac{1}{1}}\!\!
    \right]  \right) \epsilon^2 \Bigg \} \; .  
\end{equation}
Please note that $\delta_i \rightarrow - \delta_i$ is equivalent to $z_i \rightarrow z_i^{-1}$. We also observe that $\alpha_2 = \alpha_1 (\vec{q}_2 \leftrightarrow \vec{q}_1-\vec{q}_2)$ and $\alpha_3 = \alpha_1 (\vec{q}_1 \leftrightarrow \vec{q}_2)$. From this we realize that the second and the third term of the summation can be obtained from the first by two simple substitutions. We can rewrite:
\begin{equation}
    \begin{split}
        & \mathcal{I}_3' = \pi^{1+\epsilon} \Gamma (1-\epsilon) \alpha_1 \alpha_2 \alpha_3 \hat{\Delta}_3^{1/2-\epsilon} \frac{ \Gamma^2(1+\epsilon)}{\Gamma(1+2 \epsilon)} \frac{ (1-2 \epsilon) }{(1-\epsilon)} \hat{\mathcal{S}} \left \{ \frac{\alpha_1}{\sqrt{\hat{\Delta}_3} \epsilon} + \Bigg( \hspace{-0.05 cm} \pi \hspace{-0.05 cm} + \hspace{-0.05 cm} \Bigg[ \frac{i \; z_1}{1-z_1} \left( 1 - \ln \left( -z_1 \right) \right. \right. \\ & \left. \left. + \frac{1+z_1}{z_1} \ln \left( 1-z_1 \right) \hspace{-0.05 cm} \right) \hspace{-0.05 cm} - \hspace{-0.05 cm} (z_1 \rightarrow z_{1}^{-1}) \Bigg] \hspace{-0.05 cm} \Bigg) \hspace{-0.05 cm} + \hspace{-0.05 cm} \left( \hspace{-0.05 cm} \pi \hspace{-0.1 cm} + \left[ \frac{ i \; z_1}{1-z_1} \hspace{-0.05 cm} \left( \hspace{-0.05 cm} 2 + \hspace{-0.05 cm} \left( \hspace{-0.05 cm} \frac{1+z_1}{z_1} \hspace{-0.05 cm} \right) \hspace{-0.05 cm} \ln \left( 1-z_1 \right) \ln \left( (1-z_1)e \right) \right. \right. \right. \right. \\ & \left. \left. \left. \left. + \frac{1}{2} \ln^2 \left(-z_1 \right) - \frac{1-z_1}{z_1} {\rm{Li}}_2 \left(z_1 \right) - \ln \left( -z_1 \right) \left( 1 + \frac{1+z_1}{z_1} \ln \left( 1-z_1 \right) \right) \right) - (z_1 \rightarrow z_{1}^{-1}) \right]  \right) \epsilon  \right. \\ &  + \left( \pi (2 + \zeta (2)) + \left[ \frac{i \; z_1}{1-z_1} \left( 4 + \ln (1-z_1) \left( \frac{2(1+z_1)}{z_1} + \frac{2(1-z_1)}{z_1} \zeta (2) + \frac{\ln (1-z_1)}{3 z_1}  \right. \right. \right. \right. \\ & \left. \left. \left. \times \bigg ( 3 (1+z_1) + 2 (1+z_1) \ln (1-z_1)  -3 (1-z_1) \ln z_1 \bigg ) \bigg ) - \frac{(1-z_1)}{z_1} (1+2 \ln (1-z_1)) {\rm{Li}}_2 (z_1) \right. \right. \right. \\ & \left. \left. \left. \hspace{-0.3 cm} \left. - \frac{2 (1-z_1)}{z_1} \hspace{-0.05 cm} \left( \hspace{-0.05 cm} {\rm{Li}}_3 (1-z_1) \hspace{-0.05 cm} + \hspace{-0.05 cm} \frac{{\rm{Li}}_3 (z_1)}{2} \hspace{-0.05 cm} - \hspace{-0.05 cm} \zeta(3) \hspace{-0.05 cm} \right) \hspace{-0.05 cm} - \hspace{-0.05 cm} \frac{\ln^3 (-z_1)}{6} + \frac{\ln^2 (-z_1)}{2} \left( \hspace{-0.05 cm} 1 \hspace{-0.05 cm} + \hspace{-0.05 cm} \left( \frac{1+z_1}{z_1} \right) \ln (1-z_1)  \right) \right. \right. \right. \right. \\ &  \left. - \ln (-z_1) \left( 2 + \left( \hspace{-0.05 cm} \frac{1+z_1}{z_1} \hspace{-0.05 cm} \right)  \ln \left( 1-z_1 \right) \ln \left( (1-z_1)e \right) - \frac{1-z_1}{z_1} {\rm{Li}}_2 \left(z_1 \right) \hspace{-0.1 cm} \right) \hspace{-0.05 cm} \right) \left. \hspace{-0.1 cm} - ( z_1 \rightarrow z_1^{-1} ) \bigg]  \bigg) \epsilon^2 \right \} ,
    \end{split}
    \label{I3BDKe^2}
\end{equation}
where the operator $\hat{S}$ acts on a generic function  $f(\vec{q}_{1}^{\; 2}, \vec{q}_{2}^{\; 2}, \vec{p}^{\; 2} )$ as
\begin{equation}
\hat{S} \left \{ f \left( \vec{q}_{1}^{\; 2}, \vec{q}_{2}^{\; 2}, \vec{p}^{\; 2} \right) \right \} = f \left( \vec{q}_{1}^{\; 2}, \vec{q}_{2}^{\; 2}, \vec{p}^{\; 2} \right) + f \left( \vec{q}_{1}^{\; 2}, \vec{p}^{\; 2}, \vec{q}_2^{\; 2} \right) + f \left( \vec{q}_{2}^{\; 2}, \vec{q}_{1}^{\; 2}, \vec{p}^{\; 2} \right) \; .
\end{equation}
 
\subsection{$\mathcal{I}_3$: Alternative calculation}
\label{subsec:I32}
We verify the previous result through a different calculation procedure. We introduce the Feynman parametrization to get
\begin{equation}
    \mathcal{I}_3 = (1-\epsilon) \int_0^1 dx \int_0^1 dy \frac{y^{\epsilon-1}}{(A-By)^{2-\epsilon}} \;,
\end{equation}
where $A=x \vec{q}_1^{\; 2} + (1-x) \vec{q}_2^{\; 2}$ and $B=A-x(1-x) \vec{p}^{\; 2}$. Now, it is convenient to perform the decomposition
\begin{equation*}
    \int_0^1 dx \int_0^1 dy \frac{y^{\epsilon-1}}{(A-By)^{2-\epsilon}} = \int_0^1 dy \frac{y^{\epsilon}}{(A-By)^{-\epsilon}}
\left[\frac{1/A^2}{y} +\frac{B/A^2}{A-By}+\frac{B/A}{(A-By)^2}\right]
\end{equation*}
\begin{equation*}
    = \int_0^1 dy \left[ \frac{y^{\epsilon-1}}{A^{2-\epsilon}} \left( 1 - \frac{B}{A} y \right)^{\epsilon} + \frac{B y^{\epsilon}}{A^2 (A-By)^{1-\epsilon}} + \frac{B y^{\epsilon}}{A (A-By)^{2-\epsilon}} \right]\;.
\end{equation*}
After this, we can safely expand $\left( 1 - \frac{B}{A} y \right)^{\epsilon}$ in the first term and $y^{\epsilon}$ in the second and third terms, getting
\begin{equation*}
    \mathcal{I}_3 = (1-\epsilon) \int_0^1 dx \int_0^1 dy \left \{ \frac{y^{\epsilon-1}}{A^{2-\epsilon}} + \frac{B}{A^2 (A-By)^{1-\epsilon}} + \frac{B}{A (A-By)^{2-\epsilon}} + \epsilon \left( \frac{\ln \left( 1- \frac{B}{A} y \right)}{A^2 y} \right. \right.
\end{equation*}
\begin{equation*}
    \left. \left. + \frac{B \ln y}{A^2 (A-By)} + \frac{B \ln y}{A (A-By)^2} \right) + \epsilon^2 \left[ \frac{\ln (1- \frac{B}{A} y)}{y A^2} \left( \frac{\ln (1- \frac{B}{A} y)}{2} + \ln A + \ln y \right) \right. \right.
\end{equation*}
\begin{equation*}
        \left. \left. + \frac{B \ln y}{A (A-By)}  \left( \frac{\ln y}{2} + \ln (A-By) \right) \left( \frac{1}{A} + \frac{1}{A-By} \right) \right] \right \} + {\cal O}(\epsilon^3) \;.
\end{equation*}
Integrating over $y$, we obtain
\begin{equation*}
\mathcal{I}_3 = \int_0^1 dx \left \{ \frac{(2-3 \epsilon)}{\epsilon} \frac{1}{A^{2-\epsilon}} - \frac{(1- \epsilon)}{\epsilon} \frac{(A-B)^{\epsilon}}{A^{2}} + \frac{1}{A(A-B)^{1-\epsilon}} + \epsilon \left( \frac{1}{A^2} \ln \left( \frac{A-B}{A} \right) \right. \right.
\end{equation*}
\begin{equation*}
    \left. \left. - \frac{2}{A^2} {\rm{Li}}_2 \left( \frac{B}{A} \right) \right) \right \} + 2 \epsilon^2 \int_{0}^1  \frac{dx}{A^2} \left[ 2 \; {\rm{Li}}_2 \left( \frac{B}{A} \right) + \; {\rm{S}}_{1,2} \left( \frac{B}{A} \right) \right. 
\end{equation*}
\begin{equation}
   \left. + {\rm{Li}}_3 \left( \frac{B}{A} \right) + \frac{\ln A}{2} \left( \ln \left( \frac{A-B}{A} \right) -2 \; {\rm{Li}}_2 \left( \frac{B}{A} \right)  \right) + \frac{1}{4} \ln^2 \left( \frac{A-B}{A} \right) \right] + {\cal O}(\epsilon^3)  \;.
\label{I3final}
\end{equation}

In this form, all divergences are contained in the first three terms, which we can promptly compute, to get
\begin{equation*}
     \int_0^1 dx \frac{(2-3 \epsilon)}{\epsilon A^{2-\epsilon} } = \frac{2}{ab} \left[ \frac{1}{\epsilon} + \frac{a \ln b - b \ln a}{a-b} - \frac{1}{2} + \frac{\epsilon}{2} \left( \frac{a \ln^2 b - b \ln^2 a}{a-b} -\frac{a \ln b - b \ln a}{a-b}  -1 \right) \right. 
\end{equation*}
\begin{equation}
   \left. + \frac{\epsilon^2}{6} \left( - 3 - 3 \frac{a \ln b - b \ln a}{a-b} - \frac{3}{2}  \frac{a \ln^2 b - b \ln^2 a}{a-b} + \frac{a \ln^3 b - b \ln^3 a}{a-b} \right) \right] + {\cal O}(\epsilon^3) \; , \vspace{0.25 cm} \\
\label{I3_div_1}
\end{equation}

\begin{equation*}
  - \frac{(1- \epsilon)}{\epsilon} \int_0^1 dx \frac{(A-B)^{\epsilon}}{A^{2}} = - \frac{c^{\epsilon}}{ab} \left \{ \frac{1}{\epsilon} -1 - \frac{a+b}{a-b} \ln \left( \frac{a}{b} \right)-\epsilon \left[ \zeta(2) + \frac{a+b}{a-b} \left( {\rm{Li}}_2 \left( 1- \frac{a}{b} \right) \right. \right. \right.
\end{equation*}
\begin{equation*}
    \left. \left. \left. - {\rm{Li}}_2 \left( 1- \frac{b}{a} \right) - \ln \left( \frac{a}{b} \right) \right) \right] + \epsilon^2 \left[ \zeta(2) - \frac{a+b}{a-b} \ln \left( \frac{a}{b} \right) \left( \zeta(2) + \ln \left( \frac{a}{b} \right) \left( -\frac{1}{2}- \ln \left( 1 - \frac{a}{b} \right) \right. \right. \right. \right. 
\end{equation*}
\begin{equation}
  \left. \left. \left. \left. \hspace{-0.5 cm} + \frac{1}{6} \ln \left( \frac{a}{b} \right) \right) \right) - \frac{4 b}{a-b} \zeta (3) + 2 \frac{a+b}{a-b} \left( {\rm{Li}}_2 \left( 1- \frac{a}{b} \right) + 2 {\rm{Li}}_3 \left( 1- \frac{a}{b} \right) + {\rm{Li}}_3 \left( \frac{a}{b} \right) \right) \right] \right \} \ + {\cal O}(\epsilon^3)  , \vspace{0.25 cm} \\
\label{I3_div_2}
\end{equation}

\begin{equation*}
    \int_0^1 dx \frac{1}{A(A-B)^{1-\epsilon}} = \frac{1}{abc} \left \{ \frac{a+b}{\epsilon} + (a+b) \ln c - (a-b) \ln \left( \frac{a}{b} \right) + \epsilon \left[ \frac{(a+b) \ln^2 c}{2}  \right. \right.
\end{equation*}
\begin{equation*}
   \left. - (a-b) \ln c \ln \left( \frac{a}{b} \right) -(a+b) \zeta (2) - (a-b) \left( {\rm{Li}}_2 \left( 1 - \frac{a}{b} \right) - {\rm{Li}}_2 \left( 1 - \frac{b}{a} \right) \right) \right ] + \epsilon^2 \left[ \frac{a+b}{6} \ln^3 c \right. 
\end{equation*}
\begin{equation*}
  \left. \left. \left. - \frac{a-b}{2} \ln \left( \frac{a}{b} \right) \ln^2 c - \ln c \bigg ( (a+b) \zeta(2) + (a-b) \left( {\rm{Li}}_2 \left( 1 - \frac{a}{b} \right) - {\rm{Li}}_2 \left( 1 - \frac{b}{a} \right) \right) \right) + 4 b \zeta (3) \right. \right.
\end{equation*}
\begin{equation*}
     \left. \left. \hspace{-0.15 cm} - (a-b) \left( \ln \left( \frac{a}{b} \right) \left( \zeta (2) -\ln \left( 1 - \frac{a}{b} \right) \ln \left( \frac{a}{b} \right) + \frac{1}{6} \ln^2 \left( \frac{a}{b} \right) \right) - 4 {\rm{Li}}_3 \left( 1 - \frac{a}{b} \right) - 2 {\rm{Li}}_3 \left( \frac{a}{b} \right)   \right)  \right] \right \} 
\end{equation*}
\begin{equation}
    + {\cal O}(\epsilon^3) \; ,
\label{I3_div_3}
\end{equation}
where $a= \vec{q}_1^{\; 2}$, $b= \vec{q}_2^{\; 2}$, $c=\vec{p}^{\; 2}$.
Plugging into (\ref{I3final}) the results given in (\ref{I3_div_1})-(\ref{I3_div_3}), one gets 
an expression for $\mathcal{I}_3$, valid up to the order $\epsilon^2$, in terms of finite 
one-dimensional integrals. A quick numerical comparison shows that this expression is perfectly equivalent to the one obtained in the previous section. \\

By limiting the accuracy to the order $\epsilon$, we can express the result in a very compact form. We calculate the two residual one-dimensional integrals,
\begin{equation}
    \int_0^1 dx \frac{1}{A^2} \ln \left( \frac{A-B}{A} \right) = - \frac{1}{2ab} \left[ 2 - \ln \left( \frac{c^2}{a b} \right) + \frac{a+b}{a-b} \ln \left( \frac{a}{b} \right) \right] \; ,
\end{equation}
\begin{equation*}
    \int_0^1 dx \int_0^1 dy \frac{\ln \left( 1 - \frac{B}{A} y \right)}{y A^2} = \frac{1}{2 a b} \left[ 2 -2 \zeta (2) - \ln \left( \frac{c^2}{ab} \right) + \frac{a+b}{a-b} \ln \left( \frac{a}{b} \right) + \frac{(a-b)^2 - c (a+b)}{c(a-b)} \right.
\end{equation*}
\begin{equation}
  \left.  \times \left( {\rm{Li}}_2 \left( 1 - \frac{a}{b} \right) - {\rm{Li}}_2 \left( 1 - \frac{b}{a} \right) + \frac{1}{2} \ln \left( \frac{c^2}{ab} \right) \ln \left( \frac{a}{b} \right) \right) \right] + \frac{(a-b)^2 -2 c(a+b) +c^2}{2 a b c} I_{a,b,c} \; ,
\end{equation}
where
\begin{equation}
    I_{a,b,c} = \int_0^1 dx \ \frac{1}{a x + b (1-x) - c x (1-x)} \ln \left( \frac{a x + b (1-x)}{c x(1-x)} \right) \; .
\label{FadinGorba}
\end{equation}
Various properties and representations of the integral~(\ref{FadinGorba}), together with its explicit value for $I_{\vec{q}_1^{\; 2}, \vec{q}_2^{\; 2}\vec{p}^{\; 2}}$, are given in appendix~\ref{AppendixC2}. Combining everything we find
\begin{equation*}
    \mathcal{I}_3 = \frac{\Gamma^2 (1+\epsilon)}{\epsilon \Gamma (1+2 \epsilon)} \left[ (\vec{p}^{\; 2})^{\epsilon} \left( \frac{1}{\vec{p}^{ \; 2} \vec{q}_1^{\; 2}} + \frac{1}{\vec{p}^{\; 2} \vec{q}_2^{\; 2}} - \frac{1}{\vec{q}_1^{ \; 2} \vec{q}_2^{\; 2}} \right) \right.
\end{equation*}
\begin{equation*}
    \left. + (\vec{q}_1^{\; 2})^{\epsilon} \left( \frac{1}{ \vec{q}_1^{\; 2} \vec{q}_2^{\; 2}} + \frac{1}{\vec{q}_1^{\; 2} \vec{p}^{\; 2}} - \frac{1}{ \vec{q}_2^{\; 2} \vec{p}^{\; 2} } \right) + (\vec{q}_2^{\; 2})^{\epsilon} \left( \frac{1}{ \vec{q}_2^{\; 2} \vec{q}_1^{\; 2}} + \frac{1}{\vec{q}_2^{\; 2} \vec{p}^{\; 2}} - \frac{1}{ \vec{q}_1^{\; 2} \vec{p}^{\; 2} } \right) \right.
\end{equation*}
\begin{equation}
    + \frac{\epsilon^2}{\vec{q}_1^{\; 2} \vec{q}_2^{\; 2} \vec{p}^{\;2}} ((\vec{p}^{\;2})^2+(\vec{q}_1^{\; 2})^2+(\vec{q}_2^{\; 2})^2-2 \vec{q}_1^{\; 2} \vec{q}_2^{\; 2}-2 \vec{q}_1^{\; 2} \vec{p}^{\; 2} -2 \vec{q}_2^{\; 2} \vec{p}^{\; 2} ) I_{\vec{q}_1^{\; 2}, \vec{q}_2^{\; 2}, \vec{p}^{\; 2}} \; ,
\end{equation}
or 
\begin{equation*}
    \mathcal{I}_3 = \frac{\Gamma^2 (1+\epsilon)}{\epsilon \Gamma (1+2 \epsilon)}
\end{equation*}    
\begin{equation}
   \times \hat{\mathcal{S}} \left \{ (\vec{q}_2^{\; 2})^{\epsilon} \left( \frac{1}{ \vec{q}_2^{\; 2} \vec{q}_1^{\; 2}} + \frac{1}{\vec{q}_2^{\; 2} \vec{p}^{\; 2}} - \frac{1}{ \vec{q}_1^{\; 2} \vec{p}^{\; 2} } \right)
    + \frac{\epsilon^2}{\vec{q}_1^{\; 2} \vec{q}_2^{\; 2} \vec{p}^{\; 2}} ((\vec{q}_2^{\; 2})^2- \vec{q}_1^{\; 2} \vec{q}_2^{\; 2}- \vec{q}_2^{\; 2} \vec{p}^{\; 2} ) I_{\vec{q}_1^{\; 2}, \vec{q}_2^{\; 2}, \vec{p}^{\; 2}} \right \} \; .
\end{equation}
This result, after multiplication by the factor $\pi^{1+\epsilon} \Gamma(1-\epsilon)$ (see the definition~(\ref{I3prime_def})), is equivalent to (\ref{I3BDKe^2}) at the order $\epsilon$.

\subsection{$\mathcal{I}_{4 B}$ and $\mathcal{I}_{4 A}$: BDK method}
\label{subsec:I41}
The integral that has to be evaluated is
\begin{equation*}
\mathcal{I}_{4B} = \int_0^1 \frac{dx}{x} \int \frac{d^{D-2} k}{\pi^{1 + \epsilon} \Gamma (1-\epsilon)} 
\end{equation*}
\begin{equation}
   \times \left[ \frac{1-x}{\left( x \vec{k}^2 + (1-x) (\vec{k}-\vec{q}_1)^2 \right) (\vec{k}-(1-x)(\vec{q}_1-\vec{q}_2))^2} - \frac{1}{(\vec{k}-\vec{q}_{1})^2 (\vec{k}-(\vec{q}_1-\vec{q}_2))^2} \right] \; ,
\tag{\ref{I4Bcal}}
\end{equation}
Let's start from the integral $I_{4B}$ defined as 
\begin{equation}
  I_{4B} = \frac{1}{i} \int d^{D}k \frac{1}{(k^{2} + i \varepsilon) [(k+q_1)^{2}+i\varepsilon]
    [(k+q_2)^{2}+i\varepsilon] [(k-p_B)^2 + i \varepsilon]}\;.
\label{I4B}
\end{equation}
Using the BDK method and working in the Euclidean region ($s,s_1,s_2,t_1,t_2<0$) the result for this integral is~\cite{Bern_1994}
\begin{equation}
\begin{split}
I_{4B} = & \frac{\pi^{2+\epsilon}}{s_2 t_2} \frac{\Gamma(1-\epsilon) \Gamma^2(1+\epsilon)}{\Gamma(1+2 \epsilon)} \frac{2}{\epsilon^2} \left[ (-\alpha_4 (\alpha_1 - \alpha_5))^{-\epsilon} \; _2F_1 \left( \epsilon, \epsilon, 1+\epsilon; \frac{\alpha_1 \alpha_4 + \alpha_5 \alpha_3 - \alpha_5 \alpha_4}{\alpha_4 (\alpha_1 - \alpha_5)} \right) \right. \\ & \left. +(-\alpha_5 (\alpha_3 - \alpha_4))^{-\epsilon} \; _2F_1 \left( \epsilon, \epsilon, 1+\epsilon; \frac{\alpha_1 \alpha_4 + \alpha_5 \alpha_3 - \alpha_5 \alpha_4}{\alpha_5 (\alpha_3 - \alpha_4)} \right) \right. \\ & \left. -((\alpha_1-\alpha_5) (\alpha_3 - \alpha_4))^{-\epsilon} \; _2F_1 \left( \epsilon, \epsilon, 1+\epsilon; -\frac{\alpha_1 \alpha_4 + \alpha_5 \alpha_3 - \alpha_5 \alpha_4}{(\alpha_1-\alpha_5) (\alpha_3 - \alpha_4)} \right) \right]\;,
\end{split}
\label{C}
\end{equation}
where
\begin{equation*}
  \alpha_1 = \sqrt{-\frac{s_1 s_2}{s t_2 t_1}}\;, \hspace{0.5 cm}
  \alpha_2 = \sqrt{-\frac{s_2 t_2}{s s_1 t_1}}\;, \hspace{0.5 cm}
  \alpha_3 = \sqrt{-\frac{s t_2}{s_2 s_1 t_1}}\;, 
\end{equation*}
\begin{equation}
  \alpha_4 = \sqrt{-\frac{s t_1}{s_1 s_2 t_2}}\;, \hspace{0.5 cm}
  \alpha_5 = \sqrt{-\frac{s_1 t_1}{s s_2 t_2}} \;.
  \label{alpha}
\end{equation}
We then have
\begin{equation}
\begin{split}
I_{4B} = & \frac{\pi^{2+\epsilon}}{s_2 t_2} \frac{\Gamma(1-\epsilon) \Gamma^2(1+\epsilon)}{\Gamma(1+2 \epsilon)} \frac{2}{\epsilon^2} \left[ \frac{(-s_2)^{\epsilon} (-t_2)^{\epsilon}}{(s_2-t_1)^{\epsilon}} \; _2F_1 \left( \epsilon, \epsilon, 1+\epsilon; 1 - \frac{(-t_2)}{s_2-t_1} \right) \right. \\ & \left. +\frac{(-s_2)^{\epsilon} (-t_2)^{\epsilon}}{(t_2-t_1)^{\epsilon}} \; _2F_1 \left( \epsilon, \epsilon, 1+\epsilon; 1 - \frac{s_2}{t_1-t_2} \right) \right. \\ & \left. - \frac{(-s_2)^{\epsilon} (-t_2)^{\epsilon} (-t_1)^{\epsilon}}{(s_2-t_1)^{\epsilon} (t_2-t_1)^{\epsilon}} \; _2F_1 \left( \epsilon, \epsilon, 1+\epsilon; 1 - \frac{s_2 t_2}{(s_2-t_1)(t_2-t_1)} \right)  \right]\; .
\end{split}
\label{C1}
\end{equation} 
Using~(\ref{HyperProp1}), this result can be put in the following simpler form:
\begin{equation} 
I_{4B} = \frac{\pi^{2+\epsilon}}{s_2 t_2} \frac{\Gamma(1-\epsilon) \Gamma^2(1+\epsilon)}{\Gamma(1+2 \epsilon)} \frac{2}{\epsilon^2} \left[ (-s_2)^{\epsilon}\; _2F_1 \left(1, \epsilon, 1+\epsilon; 1 - \frac{s_2-t_1}{(-t_2)} \right) \right.
\label{C1bis}
\end{equation} 
\[
\left. + (-t_2)^{\epsilon} \; _2F_1 \left(1, \epsilon, 1+\epsilon; 1 - \frac{t_1-t_2}{s_2} \right)
- (-t_1)^{\epsilon} 
\; _2F_1 \left(1, \epsilon, 1+\epsilon; 1 - \frac{(s_2-t_1)(t_1-t_2)}{s_2(-t_2)} \right)  \right]\; .
\]
We re-derive this result in appendix \ref{AppendixC2}. \\
For the first term in the square roots in~(\ref{C}) we have
\begin{equation}
\left(-t_2\right)^\epsilon\left(-s_2\right)^\epsilon\left(s_2-t_1\right)^{-\epsilon}{ }_2 F_1\left(\epsilon, \epsilon ; 1+\epsilon ; 1- \frac{\left(-t_2\right)}{s_2-t_1}\right) \; .
\end{equation}
Analytically continuing the result in the region of positive $s_2$ by the replacement
\begin{equation}
    (-s_2) \rightarrow e^{-i \pi } s_2 \;,
\end{equation}
using the MRK approximation to neglect $(-t_2)/(s_2-t_1)$ with respect to one
in the argument of the hypergeometric function and recalling that
\begin{equation}
    { }_2 F_1(\epsilon, \epsilon , 1+\epsilon ; 1)=\Gamma(1-\epsilon) \Gamma(1+\epsilon)=\frac{\pi \epsilon}{\sin (\pi \epsilon)} \; ,
\end{equation}
we obtain
\begin{equation}
\left(-t_2\right)^\epsilon e^{-i \pi \epsilon} \Gamma(1-\epsilon) \Gamma(1+\epsilon)=\left(-t_2\right)^\epsilon\left[\cos (\pi \epsilon) \frac{\pi \epsilon}{\sin (\pi \epsilon)}-i \pi \epsilon\right] \; .
\end{equation}
The second term,
\begin{equation}
    \left(\frac{-t_2 s_2}{t_1-t_2} \right)^\epsilon{ }_2 F_1\left(\epsilon, \epsilon , 1+\epsilon ; 1- \frac{s_2}{t_1-t_2} \right) \;,
\end{equation}
with the help of~(\ref{HyperProp1}) and~(\ref{HyperExp2}) gives, assuming $t_1 > t_2$, 
\begin{equation}
    \left(-t_2\right)^\epsilon{ }_2 F_1\left(1, \epsilon , 1+\epsilon ; 1 - \frac{ t_1-t_2 }{s_2} \right)=\left(-t_2\right)^\epsilon\left(1-\epsilon \ln \left( \frac{t_1-t_2 }{s_2} \right)-\sum_{n=2}^{\infty}(-\epsilon)^n \zeta(n) \right ) \; .
\end{equation}
The third term,
\begin{equation}
    -\left( \frac{t_1 t_2}{t_1-t_2} \right)^\epsilon{ }_2 F_1\left(\epsilon, \epsilon , 1+\epsilon ; 1+ \frac{t_2}{t_1-t_2} \right) \;,
\end{equation}
using~(\ref{HyperProp1}) and~(\ref{HyperExp2}), again assuming $t_1 > t_2$, becomes
\begin{equation}
  -\left(-t_1\right)^\epsilon{ }_2 F_1\left(1, \epsilon , 1+\epsilon ; \frac{t_1}{t_2} \right)=-\left(-t_1\right)^\epsilon\left(1-\epsilon \ln \left(1- \frac{t_1}{t_2} \right) - \sum_{n=2}^{\infty}(-\epsilon)^n {\rm{Li}}_n\left( \frac{t_1}{ t_2} \right)\right) \; .  
\end{equation}
We finally obtain
\begin{gather}
    I_{4B} = \frac{\pi^{2+\epsilon}}{s_2 t_2} \frac{\Gamma(1-\epsilon) \Gamma^2(1+\epsilon)}{\Gamma(1+2 \epsilon)} \frac{2}{\epsilon^2} \left(-t_1\right)^\epsilon \left[ \left(\frac{t_2}{t_1}\right)^\epsilon \bigg ( e^{-i \pi \epsilon} \Gamma(1-\epsilon) \Gamma(1+\epsilon) \right. \nonumber \\
    \left. \left. + 1  -\epsilon \ln \left( \frac{ t_1-t_2 }{s_2} \right) -\sum_{n=2}^{\infty}(-\epsilon)^n \zeta(n) \right ) - 1 + \epsilon \ln \left(1- \frac{t_1}{t_2} \right) + \sum_{n=2}^{\infty}(-\epsilon)^n {\rm{Li}}_n\left( \frac{t_1}{ t_2} \right) \right ] \; .
    \label{I4B_BDK}
\end{gather}
Furthermore, starting from the integral in eq.~(\ref{I4B}), using the standard Sudakov decomposition for the 4-momentum $k$, one finds that, in the multi-Regge kinematics (see appendix~\ref{AppendixC3}),
\begin{equation}
I_{4B}=- \frac{\pi^{2+\epsilon} \Gamma (1-\epsilon)}{s_2} \left[ \frac{\Gamma^2 (\epsilon)}{\Gamma (2 \epsilon)} (-t_2)^{\epsilon -1} \left( \ln \left( \frac{- s_2}{- t_2} \right) + \psi (1-\epsilon) - 2 \psi (\epsilon) + \psi (2 \epsilon)  \right) + \mathcal{I}_{4B} \right] \;,
\label{I4B-I4Bcal}
\end{equation}
where $\mathcal{I}_{4B}$ is exactly the integral defined in eq.~(\ref{I4Bcal}). From this relation we can hence derive an expression for $\mathcal{I}_{4B}$ that, after some manipulations, can be cast in the following form:
\begin{gather}
    \mathcal{I}_{4 B}=\frac{\Gamma^2(1+\epsilon)}{\Gamma(1+2 \epsilon)} \frac{2}{\epsilon^2} \frac{\left(-t_1\right)^\epsilon}{-t_2}\left[\left(\frac{t_2}{t_1}\right)^\epsilon\left(-\frac{1}{2}+\frac{\pi \epsilon}{\sin (\pi \epsilon)} \cos (\pi \epsilon)-\epsilon \ln \left(1- \frac{t_1}{t_2} \right)\right.\right. \nonumber \\ 
    \left.\left.+\sum_{n=2}^{\infty} \epsilon^n \zeta(n)\left(1-(-1)^n\left(2^{n-1}-1\right)\right)\right)-1+\epsilon \ln \left(1- \frac{t_1}{t_2} \right)+\sum_{n=2}^{\infty}(-\epsilon)^n {\rm{Li}}_n\left( \frac{t_1}{t_2} \right)\right] \; .
\label{I4BcalFin}
\end{gather} 
The result remains valid for $t_2 > t_1$. In particular, truncating the summation at $n=4$ gives the desired result. \\
This result is completely equivalent to the one obtained by calculating directly $\mathcal{I}_{4B}$ introducing Feynman parameters (see the next subsection). The integral $\mathcal{I}_{4A}$ is defined similarly to $\mathcal{I}_{4B}$ in eq.~(\ref{I4Bcal}), up to the replacement $q_1 \rightarrow -q_2$, implying that
\begin{equation}
    \mathcal{I}_{4A} = \mathcal{I}_{4B} (t_1 \leftrightarrow t_2) \; .
\end{equation}

\subsection{$\mathcal{I}_{4 B}$ and $\mathcal{I}_{4 A}$: Alternative calculation}
\label{subsec:I42}
Starting from the expression
\begin{equation*}
\mathcal{I}_{4B} = \int_0^1 \frac{dx}{x} \int \frac{d^{D-2} k}{\pi^{1 + \epsilon} \Gamma (1-\epsilon)} 
\end{equation*} 
\begin{equation*}
   \times \left[ \frac{1-x}{\left( x \vec{k}^2 + (1-x) (\vec{k}-\vec{q}_1)^2 \right) (\vec{k}-(1-x)(\vec{q}_1-\vec{q}_2))^2} - \frac{1}{(\vec{k}-\vec{q}_{1})^2 (\vec{k}-(\vec{q}_1-\vec{q}_2))^2} \right] \; ,
\end{equation*}
introducing the Feynman parametrization and performing the integration over $k$, we obtain
\begin{gather}
\mathcal{I}_{4 B}=\int_0^1 \frac{d x}{x}\left[ \int_0^1 d z \frac{(1-x)}{[z(1-x)(a x+b(1-x)(1-z))]^{1-\epsilon}}-\int_0^1 d z \frac{1}{[bz(1-z)]^{1-\epsilon}}\right] \; . 
\end{gather}
Defining
\begin{equation}
  a = \vec{q}_1^{\; 2} = - t_1 \; , \hspace{0.5 cm} b = \vec{q}_2^{\; 2} = - t_2 \; , \hspace{0.5 cm} c = 1 - \frac{a}{b} \; ,
\end{equation}
we can write $\mathcal{I}_{4 B} \equiv b^{\epsilon-1} F(c)$, with 
\begin{equation}
    F(c) = \int_0^1 \frac{dx}{x} \left[ \int_0^1 \frac{dz}{z} \frac{z^{\epsilon} (1-x)^{\epsilon}}{ (1-x c- z(1-x) 
 )^{1-\epsilon}} - \int_0^1 dz \; z^{\epsilon-1} (1-z)^{\epsilon-1} \right] \; . 
\end{equation}
Performing the transformation $y=(1-x)z$ in the first term, we get
\begin{equation}
    F (c) = \int_0^1 \frac{dx}{x} \left[ \int_0^{1-x} \frac{d y}{y} \frac{y^{\epsilon}}{ (1-x c- y 
 )^{1-\epsilon}} - \int_0^1 dz \; z^{\epsilon-1} (1-z)^{\epsilon-1} \right] \; .
\end{equation}
It is very simple to compute the integral for $c=0$; we find
\begin{gather}
    F (0) = \int_0^1 \frac{dx}{x} \left[ \int_0^{1-x} d y \; y^{\epsilon-1} (1-y)^{\epsilon-1} - \int_0^1 dz \; z^{\epsilon-1} (1-z)^{\epsilon-1} \right] \nonumber \\
    = - \int_0^1 \frac{dx}{x} \int_{1-x}^1 dz \; z^{\epsilon-1} (1-z)^{\epsilon-1} = - \int_{0}^1 dz \; z^{\epsilon-1} (1-z)^{\epsilon-1} \int_{1-z}^1 \frac{dx}{x} \nonumber \\ \hspace{1.0 cm} = \int_{0}^1 dz \; z^{\epsilon-1} (1-z)^{\epsilon-1} \ln (1-z) = \frac{d}{d \delta} \left[ \int_{0}^1 dz \; z^{\epsilon-1} (1-z)^{\delta-1} \right]_{\delta=\epsilon}  \; .
\end{gather}
The function $F(c)$ in $c=0$ is then
\begin{equation}
    F(0) = \frac{\Gamma^2 (\epsilon)}{\Gamma (2 \epsilon)} \left( \psi (\epsilon) - \psi (2 \epsilon) \right) \; .
\end{equation}
We now compute the derivative with respect to $c$ of the function $F$ and get
\begin{gather}
F^{\prime}(c)=(1-\epsilon) \int_0^1 dx \int_0^{1-x} \frac{d y}{y} y^\epsilon(1-x c-y)^{\epsilon-2} = (1-\epsilon) \int_0^1 \frac{d y}{y} y^\epsilon \int_0^{1-y} \hspace{-0.2 cm} d x (1-x c-y)^{\epsilon-2} \nonumber \\ =\frac{1}{c} \int_0^1 \frac{d y}{y} y^\epsilon\left(((1-c)(1-y))^{\epsilon-1}-(1-y)^{\epsilon-1}\right) =\frac{1}{c} \frac{\Gamma^2(\epsilon)}{\Gamma(2 \epsilon)}\left((1-c)^{\epsilon-1}-1\right) \; .
\end{gather}
Having this information, we can write
\begin{equation}
    F (c) = F (0) + \int_0^c dx \; F'(x) = \frac{ \Gamma^2 (\epsilon)}{\Gamma (2 \epsilon)} \left[ \psi (\epsilon) - \psi (2 \epsilon) + \int_0^c dx \left( \frac{(1-x)^{\epsilon-1} -1}{x} \right) \right] 
\end{equation}
The integral on the right-hand side can be computed to all orders in $\epsilon$:
\begin{gather}
    \int_0^c dx \left( \frac{(1-x)^{\epsilon-1} -1}{x} \right) = \int_0^c dx \left[ (1-x)^{\epsilon} \left( \frac{1}{1-x} + \frac{1}{x} \right) - \frac{1}{x} \right] \nonumber \\ = \frac{1}{\epsilon} \left[ 1- (1-c)^{\epsilon}  \right] + \int_0^c \frac{dx}{x} ((1-x)^{\epsilon}-1) = \sum_{n=1}^{\infty} \left( \frac{\epsilon^n}{n!} \int_0^1 \frac{dx}{x} \ln^n (1-cx) -\epsilon^{n-1} \frac{\ln^n (1-c)}{n!}  \right).
\end{gather}
Using
\begin{equation}
    \frac{1}{n!} \int_0^1 \frac{dx}{x} \ln^n (1-c x) = (-1)^n \;  {\rm{S}}_{1,n} (c) \; ,
\end{equation}
we finally find
\begin{equation}
    \int_0^c dx \left( \frac{(1-x)^{\epsilon-1} -1}{x} \right) = \sum_{n=1}^{\infty} \epsilon^{n-1} \left( -\frac{\ln^n (1-c)}{n!} + \epsilon \; (-1)^n \;  {\rm{S}}_{1,n} (c) \right) \; .
\end{equation}
The final result for $\mathcal{I}_{4B}$ is then 
\begin{equation}
    \mathcal{I}_{4 B} = \frac{ \Gamma^2 (\epsilon)}{\Gamma (2 \epsilon)} (-t_2)^{\epsilon-1} \left[ \psi (\epsilon) - \psi (2 \epsilon) + \sum_{n=1}^{\infty} \epsilon^{n-1} \left( -\frac{\ln^n \left( t_1 / t_2 \right)}{n!} + \epsilon \; (-1)^n \;  {\rm{S}}_{1,n} \left( 1 - \frac{t_1}{t_2} \right) \right) \right] \; .
    \label{CalI4BDerTrick}
\end{equation}
The expression~(\ref{CalI4BDerTrick}) gives an alternative representation of $\mathcal{I}_{4 B}$, which is completely equivalent to (\ref{I4BcalFin}). Again, the complete $\epsilon^2$ result is obtained truncating the series at $n=4$. \\
We can achieve an alternative representation in terms of the hypergeometric function. In fact, we note that
\begin{gather} 
\int_0^c \frac{d x}{x}\left((1-x)^\epsilon-1\right)=\left((1-c)^\epsilon-1\right) \ln c+\epsilon\left(\int_0^1 \frac{d y}{y} \ln (1-y) y^\epsilon-\int_0^{1-c} \frac{d y}{y} \ln (1-y) y^\epsilon\right) \nonumber \\ =\left((1-c)^\epsilon-1\right) \ln c+\psi(1)-\psi(1+\epsilon)-\frac{(1-c)^\epsilon}{\epsilon}\bigg( { }_2 F_1(1, \epsilon, 1+\epsilon ; 1-c)-1+\epsilon \ln c \bigg) .
\label{HyperReprIntInCal4}
\end{gather}
The last equality can be proved by using the explicit result for the first integral, \textit{i.e.} 
\begin{equation} 
\int_0^1 \frac{d y}{y} \ln (1-y) y^{\epsilon} = \left. \frac{d}{d \delta} \int_0^1 \frac{d y}{y}(1-y)^\delta y^\epsilon\right|_{\delta=0} = \left. \frac{d}{d \delta} \frac{\Gamma(1+\delta) \Gamma(\epsilon)}{\Gamma(1+\delta+\epsilon)}\right|_{\delta=0}= \frac{1}{\epsilon}(\psi(1)-\psi(1+\epsilon)) \; ,
\end{equation}
and the following expression for the second, in terms of the hypergeometric function,
\begin{gather}
{ }_2 F_1(1, \epsilon, 1+\epsilon ; 1-c)=\epsilon \int_0^1 \frac{d y}{y(1-y(1-c))} y^\epsilon=\epsilon \int_0^1 y^\epsilon d \ln \frac{y}{(1-y(1-c))} \nonumber \\ =-\epsilon \ln c -\epsilon^2 \int_0^1 d y y^{\epsilon-1} \ln y+\epsilon^2 \int_0^1 d y y^{\epsilon-1} \ln (1-y(1-c)) \nonumber \\  = -\epsilon \ln c+1+\epsilon^2(1-c)^{-\epsilon} \int_0^{1-c} d y y^{\epsilon-1} \ln (1-y) . 
\end{gather}
Using (\ref{HyperReprIntInCal4}), we obtain
\begin{equation}
    \mathcal{I}_{4 B}=\frac{\Gamma^2(\epsilon)}{\Gamma(2 \epsilon)} b^{\epsilon-1}\left(\frac{1}{2 \epsilon}+\psi(1)-\psi(1+2 \epsilon) - \ln c -\frac{(1-c)^\epsilon}{\epsilon}{ }_2 F_1(1, \epsilon, 1+\epsilon ; 1-c)\right) \; ,
\end{equation}
which is completely equivalent to both (\ref{I4BcalFin}) and (\ref{CalI4BDerTrick}). 
\subsection{Combination $\mathcal{I}_{5} - \mathcal{L}_3$}
\label{subsec:I5I3}
Let's start by considering the pentagonal integral $I_5$, defined as
\begin{equation}
  I_{5} = \frac{1}{i} \int d^{D}k \frac{1}{(k^{2} + i \varepsilon) [(k+q_1)^{2}+i\varepsilon]
    [(k+q_2)^{2}+i\varepsilon] [(k+p_A)^2 + i \varepsilon]
    [(k-p_B)^2 + i \varepsilon]}
  \;.
  \label{I5}
\end{equation}
In multi-Regge kinematics (and in $D=4+2 \epsilon$), it is given by the following combination:
\begin{equation}
I_5 = \frac{\pi^{2+\epsilon} \Gamma (1-\epsilon)}{s} \left[ \ln \left( \frac{(-s) (\vec{q}_1-\vec{q}_2)^2}{(-s_1)(-s_2)} \right) \mathcal{I}_3 + \mathcal{L}_3 -\mathcal{I}_5 \right]
\label{I5inTransver}
\end{equation}
Inverting this relation, we obtain
\begin{equation}
 \mathcal{I}_5 - \mathcal{L}_3 = \ln \left( \frac{(-s) (\vec{q}_1-\vec{q}_2)^2}{(-s_1)(-s_2)} \right) \mathcal{I}_3 - \frac{s}{\pi^{2+\epsilon} \Gamma (1-\epsilon)} I_5 \; .
\end{equation}
Hence, the integral $I_{5}$ can be used (together with $\mathcal{I}_{3}$) to calculate
$\mathcal{I}_{5} - \mathcal{L}_3$ (this combination is the one appearing in the Lipatov vertex~\cite{Fadin:2000yp}). \\
We start by working in the Euclidean region, defined by $s,s_1,s_2,t_1,t_2 <0$, the analytical continuation to the physical region is obtained according to the prescriptions
\begin{equation}
    (-s) \rightarrow e^{-i \pi } s \; , \hspace{1 cm} (-s_1) \rightarrow e^{-i \pi } s_1 \; , \hspace{1 cm} (-s_2) \rightarrow e^{-i \pi} s_2 \; .
\end{equation}
We use the quantities $\alpha_i$ defined before, which we recall here:
\begin{equation*}
  \alpha_1 = \sqrt{-\frac{s_1 s_2}{s t_2 t_1}}\;, \hspace{0.5 cm}
  \alpha_2 = \sqrt{-\frac{s_2 t_2}{s s_1 t_1}}\;, \hspace{0.5 cm}
  \alpha_3 = \sqrt{-\frac{s t_2}{s_2 s_1 t_1}}\;, 
\end{equation*}
\begin{equation}
  \alpha_4 = \sqrt{-\frac{s t_1}{s_1 s_2 t_2}}\;, \hspace{0.5 cm}
  \alpha_5 = \sqrt{-\frac{s_1 t_1}{s s_2 t_2}} \;.
\tag{\ref{alpha}}
\end{equation}
Let's define the reduced integral $\hat{I}_5$ in this way:
\begin{equation}
I_5= -\pi^{D/2} \alpha_1 \alpha_2 \alpha_3 \alpha_4 \alpha_5 \hat{I}_5 \;.
\end{equation}
In Ref.~\cite{Bern_1994} it is shown that this integral is given by the
recursive relation
\begin{equation}
  \hat{I}_5 = \frac{1}{2} \left[ \sum_{i=1}^{5} \gamma_i \hat{I}_4^{(i)} -
    2 \epsilon \Delta_5 \hat{I}_5^{(D = 6+2 \epsilon)} \right]\;,
    \label{IterativeRelI5}
\end{equation}
where $\Delta_5$ is the following quantity:
\begin{equation}
  \Delta_5 = \sum_{i=1}^5 (\alpha_i^2 - 2 \alpha_i \alpha_{i+1}
  + 2 \alpha_i \alpha_{i+2}) \;.
\end{equation} 
The $\gamma_i$'s are
\begin{eqnarray}
\gamma_1 &=& \alpha_1 -\alpha_2 + \alpha_3 + \alpha_4 - \alpha_5 \;, \\
\gamma_2 &=& -\alpha_1 +\alpha_2 - \alpha_3 + \alpha_4 + \alpha_5 \;, \\
\gamma_3 &=& \alpha_1 - \alpha_2 + \alpha_3 - \alpha_4 + \alpha_5 \;, \\
\gamma_4 &=& \alpha_1 + \alpha_2 - \alpha_3 + \alpha_4 - \alpha_5 \;, \\
\gamma_5 &=& -\alpha_1 + \alpha_2 + \alpha_3 - \alpha_4 + \alpha_5 \;. 
\end{eqnarray}
$\hat{I}_4^{(i)}$ are the reduced version of $I_4^{(i)}$ (apart for a trivial $\pi^{2+\epsilon}$), \textit{i.e.} 
\begin{equation}
    \hat{I}_4^{(i)} = \frac{\alpha_1 \alpha_2 \alpha_3 \alpha_4 \alpha_5}{\alpha_i} I_4^{(i)} \;.
\end{equation}
\subsubsection{Box integrals part}
\label{subsubsec:Boxes}
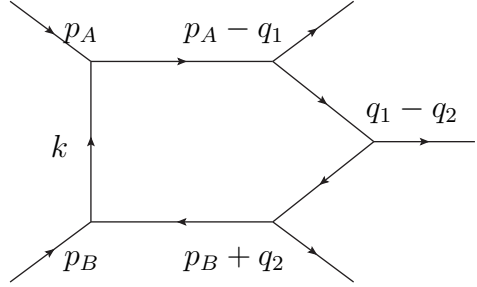
\begin{figure}
\begin{picture}(400,120)
\put(145,0){\includegraphics[width=0.4\textwidth]{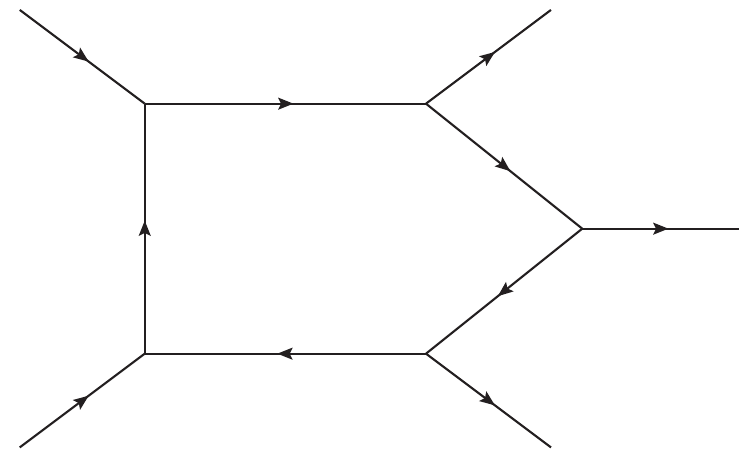}}
\put(165,50){$k$}
\put(170,95){$p_A$}
\put(215,95){$p_A-q_1$}
\put(170,09){$p_B$}
\put(215,09){$p_B+q_2$}
\put(283,65){$q_1-q_2$}
\end{picture}
\caption{Schematic representation of the pentagonal integral. The arrows denote the direction of the momenta.}
\label{Fig:Pentagon}
\end{figure}
The integrals $I_4^{(i)}$ are box integrals with all external legs on-shell,
but one, which is off-shell. They can be obtained starting from $I_5$ (see Fig.~\ref{Fig:Pentagon}) in this
way (apart from a factor $\pi^{2+\epsilon}$):
\begin{itemize}
\item $I_4^{(1)}$ \\
  Making the propagator between $p_A$ and $p_B$ ``collapse'', so that the two
  on-shell legs associated with $p_A$ and $p_B$ become a unique off-shell
  external leg with (incoming) momenta $p_A+p_B$.
\item $I_4^{(2)}$ \\
  Making the propagator between $p_A$ and $p_{A'} = p_A - q_1$ ``collapse'',
  so that the two on-shell legs associated with $p_A$ and $p_{A'}$ become a
  unique off-shell external leg with (incoming) momenta $q_1$.    
\item $I_4^{(3)}$ \\
  Making the propagator between $p_{A'} = p_A - q_1$ and $q_1 - q_2$
  ``collapse'', so that the two on-shell legs associated with $p_{A'}$ and
  $q_1 - q_2$ become a unique off-shell external leg with (outgoing) momenta
  $p_A - q_2$.   
\item $I_4^{(4)}$ \\
  Making the propagator between $q_1-q_2$ and $p_{B'} = p_B + q_2$ ``collapse'',
  so that the two on-shell legs associated with $q_1-q_2$ and $p_{B'}$ become
  a unique off-shell external leg with (outgoing) momenta $p_B + q_1$.   
\item $I_4^{(5)}$ \\
  Making the propagator between $p_{B}$ and $p_{B'}$ ``collapse'', so that the
  two on-shell legs associated with $p_{B}$ and $p_{B'}$ become a unique
  off-shell external leg with (outgoing) momenta $q_2$. 
\end{itemize} 
The results for the reduced version of these integrals are
\begin{equation}
\begin{split}
\hat{I}_{4}^{(1)} (s_1,s_2,s) = & \frac{2}{\epsilon^2} r_{\Gamma} \left[ (-\alpha_3 (\alpha_5 - \alpha_4))^{-\epsilon} \; _2F_1 \left( \epsilon, \epsilon, 1+\epsilon; \frac{\alpha_5 \alpha_3 + \alpha_2 \alpha_4 - \alpha_4 \alpha_3}{\alpha_3 (\alpha_5 - \alpha_4)} \right) \right. \\ & \left. +(-\alpha_4 (\alpha_2 - \alpha_3))^{-\epsilon} \; _2F_1 \left( \epsilon, \epsilon, 1+\epsilon; \frac{\alpha_5 \alpha_3 + \alpha_2 \alpha_4 - \alpha_4 \alpha_3}{\alpha_4 (\alpha_2 - \alpha_3)} \right) \right. \\ & \left. -((\alpha_5-\alpha_4) (\alpha_2 - \alpha_3))^{-\epsilon} \; _2F_1 \left( \epsilon, \epsilon, 1+\epsilon; -\frac{\alpha_5 \alpha_3 + \alpha_2 \alpha_4 - \alpha_4 \alpha_3}{(\alpha_5-\alpha_4) (\alpha_2 - \alpha_3)} \right) \right] \;,
\end{split}
\end{equation}
\begin{equation}
\begin{split}
\hat{I}_{4}^{(2)} (s_2,t_2,t_1) = & \frac{2}{\epsilon^2} r_{\Gamma} \left[ (-\alpha_4 (\alpha_1 - \alpha_5))^{-\epsilon} \; _2F_1 \left( \epsilon, \epsilon, 1+\epsilon; \frac{\alpha_1 \alpha_4 + \alpha_3 \alpha_5 - \alpha_4 \alpha_5}{\alpha_4 (\alpha_1 - \alpha_5)} \right) \right. \\ & \left. + (-\alpha_5 (\alpha_3 - \alpha_4))^{-\epsilon} \; _2F_1 \left( \epsilon, \epsilon, 1+\epsilon; \frac{\alpha_1 \alpha_4 + \alpha_5 \alpha_3 - \alpha_5 \alpha_4}{\alpha_5 (\alpha_3 - \alpha_4)} \right) \right. \\ & \left. -((\alpha_1-\alpha_5) (\alpha_3 - \alpha_4))^{-\epsilon} \; _2F_1 \left( \epsilon, \epsilon, 1+\epsilon; -\frac{\alpha_1 \alpha_4 + \alpha_5 \alpha_3 - \alpha_5 \alpha_4}{(\alpha_1-\alpha_5) (\alpha_3 - \alpha_4)} \right) \right] \; ,
\end{split}
\end{equation}
\begin{equation}
\begin{split}
\hat{I}_{4}^{(3)} (s,s_1,t_2) = &  \frac{2}{\epsilon^2} r_{\Gamma} \left[ (-\alpha_5 (\alpha_2 - \alpha_1))^{-\epsilon} \; _2F_1 \left( \epsilon, \epsilon, 1+\epsilon; \frac{\alpha_2 \alpha_5 + \alpha_1 \alpha_4 - \alpha_1 \alpha_5}{\alpha_5 (\alpha_2 - \alpha_1)} \right) \right. \\ & \left. +(-\alpha_1 (\alpha_4 - \alpha_5))^{-\epsilon} \; _2F_1 \left( \epsilon, \epsilon, 1+\epsilon; \frac{\alpha_2 \alpha_5 + \alpha_1 \alpha_4 - \alpha_1 \alpha_5}{\alpha_1 (\alpha_4 - \alpha_5)} \right) \right. \\ & \left. -((\alpha_2-\alpha_1) (\alpha_4 - \alpha_5))^{-\epsilon} \; _2F_1 \left( \epsilon, \epsilon, 1+\epsilon; -\frac{\alpha_2 \alpha_5 + \alpha_1 \alpha_4 - \alpha_1 \alpha_5}{(\alpha_2-\alpha_1) (\alpha_4 - \alpha_5)} \right) \right]\;,
\end{split}
\end{equation}
\begin{equation}
\begin{split}
\hat{I}_{4}^{(4)} (s,s_2,t_1) = & \frac{2}{\epsilon^2} r_{\Gamma} \left[ (-\alpha_1 (\alpha_3 - \alpha_2))^{-\epsilon} \; _2F_1 \left( \epsilon, \epsilon, 1+\epsilon; \frac{\alpha_1 \alpha_3 + \alpha_2 \alpha_5 - \alpha_2 \alpha_1}{\alpha_1 (\alpha_3 - \alpha_2)} \right) \right. \\ & \left. +(-\alpha_2 (\alpha_5 - \alpha_1))^{-\epsilon} \; _2F_1 \left( \epsilon, \epsilon, 1+\epsilon; \frac{\alpha_1 \alpha_3 + \alpha_2 \alpha_5 - \alpha_2 \alpha_1}{\alpha_2 (\alpha_5 - \alpha_1)} \right) \right. \\ & \left. -((\alpha_3-\alpha_2) (\alpha_5 - \alpha_1))^{-\epsilon} \; _2F_1 \left( \epsilon, \epsilon, 1+\epsilon; -\frac{\alpha_1 \alpha_3 + \alpha_2 \alpha_5 - \alpha_2 \alpha_1}{(\alpha_3-\alpha_2) (\alpha_5 - \alpha_1)} \right) \right]\;,
\end{split}
\end{equation}
\begin{equation}
\begin{split}
\hat{I}_{4}^{(5)} (s_1,t_1,t_2) = & \frac{2}{\epsilon^2} r_{\Gamma} \left[ (-\alpha_3 (\alpha_1 - \alpha_2))^{-\epsilon} \; _2F_1 \left( \epsilon, \epsilon, 1+\epsilon; \frac{\alpha_1 \alpha_3 + \alpha_2 \alpha_4 - \alpha_2 \alpha_3}{\alpha_3 (\alpha_1 - \alpha_2)} \right) \right. \\ & \left. +(-\alpha_2 (\alpha_4 - \alpha_3))^{-\epsilon} \; _2F_1 \left( \epsilon, \epsilon, 1+\epsilon; \frac{\alpha_1 \alpha_3 + \alpha_2 \alpha_4 - \alpha_2 \alpha_3}{\alpha_2 (\alpha_4 - \alpha_3)} \right) \right. \\ & \left. -((\alpha_1-\alpha_2) (\alpha_4 - \alpha_3))^{-\epsilon} \; _2F_1 \left( \epsilon, \epsilon, 1+\epsilon; -\frac{\alpha_1 \alpha_3 + \alpha_2 \alpha_4 - \alpha_2 \alpha_3}{(\alpha_1-\alpha_2) (\alpha_4 - \alpha_3)} \right) \right]\; ,
\end{split}
\end{equation}
where
\begin{equation}
    r_{\Gamma} = \frac{\Gamma(1-\epsilon) \Gamma^2(1+\epsilon)}{\Gamma(1+2 \epsilon)} \; .
\end{equation}
It is clear that with appropriate exchanges of the invariants
$s,s_1,s_2,t_1,t_2$ they can be obtained from each other. In particular, as
pointed out in~\cite{Bern_1994}, starting from one of these integrals the
others can be obtained by cyclic permutations of the $\alpha_i$'s. 

By expanding these results, in multi-Regge kinematics, we have
\begin{equation}
   \hat{I}_{4}^{(1)} (s_1,s_2,s) \simeq \frac{ \Gamma(1-\epsilon) \Gamma^2(1+\epsilon)}{\Gamma(1+2 \epsilon)} \frac{2}{\epsilon^2} \left( \frac{(-s_1)(-s_2)}{(-s)} \right)^{\epsilon}  \left \{ 1 + \sum_{n=1}^{\infty} \epsilon^{2n} \;  2 \left( 1 - \frac{1}{2^{2 n -1}} \right) \zeta (2 n) \right \} \; ,
\end{equation}
\begin{gather}
    \hat{I}_{4}^{(2)} (s_2,t_2,t_1) =  \frac{\Gamma(1-\epsilon) \Gamma^2(1+\epsilon)}{\Gamma(1+2 \epsilon)} \frac{2}{\epsilon^2} \left(-t_1\right)^\epsilon \left[ \left(\frac{t_2}{t_1}\right)^\epsilon \bigg ( e^{-i \pi \epsilon} \Gamma(1-\epsilon) \Gamma(1+\epsilon) \right. \nonumber \\
    \left. \left. + 1  -\epsilon \ln \left( \frac{ t_1-t_2 }{s_2} \right) -\sum_{n=2}^{\infty}(-\epsilon)^n \zeta(n) \right ) - 1 + \epsilon \ln \left(1- \frac{t_1}{t_2} \right) + \sum_{n=2}^{\infty}(-\epsilon)^n {\rm{Li}}_n\left( \frac{t_1}{ t_2} \right) \right ] \; .
\end{gather}
\begin{equation}
\hat{I}_{4}^{(3)} (s,s_1,t_2) \simeq \frac{ \Gamma(1-\epsilon) \Gamma^2(1+\epsilon)}{\Gamma(1+2 \epsilon)} \frac{2}{\epsilon^2} (-t_2)^{\epsilon} \left \{ 1 + \epsilon \ln \left( \frac{-s}{-s_1} \right) - \sum_{n=2}^{\infty} (-\epsilon)^{n} \zeta(n) \right \} \; ,
\end{equation}
\begin{equation}
\hat{I}_{4}^{(4)} (s,s_2,t_1) \simeq \frac{ \Gamma(1-\epsilon) \Gamma^2(1+\epsilon)}{\Gamma(1+2 \epsilon)} \frac{2}{\epsilon^2} (-t_1)^{\epsilon} \left \{ 1 + \epsilon \ln \left( \frac{-s}{-s_2} \right) - \sum_{n=2}^{\infty} (-\epsilon)^{n} \zeta(n) \right \} \; ,
\end{equation}
\begin{gather}
    \hat{I}_{4}^{(5)} (s_1,t_1,t_2) =  \frac{\Gamma(1-\epsilon) \Gamma^2(1+\epsilon)}{\Gamma(1+2 \epsilon)} \frac{2}{\epsilon^2} \left(-t_2\right)^\epsilon \left[ \left(\frac{t_1}{t_2}\right)^\epsilon \bigg ( e^{-i \pi \epsilon} \Gamma(1-\epsilon) \Gamma(1+\epsilon) \right. \nonumber \\
    \left. \left. + 1  -\epsilon \ln \left( \frac{ t_2-t_1 }{s_1} \right) -\sum_{n=2}^{\infty}(-\epsilon)^n \zeta(n) \right ) - 1 + \epsilon \ln \left(1- \frac{t_2}{t_1} \right) + \sum_{n=2}^{\infty}(-\epsilon)^n {\rm{Li}}_n\left( \frac{t_2}{t_1} \right) \right ] \; .
\end{gather} 
Again, truncating the summation at $n=4$ gives the desired result. The integral $\hat{I}_4^{(2)}$, up to trivial factors, coincides with~(\ref{I4B_BDK}), as can be easily verified. Since we have computed the integral $\mathcal{I}_{4B}$ through two independent methods, the integral $\hat{I}_4^{(2)}$ can be cross-checked. We can use the technique explained in appendix~\ref{AppendixC3} to obtain the relation (\ref{I4B-I4Bcal}) and then, by using the knowledge of $\mathcal{I}_{4B}$ integral from the alternative method explained in section~\ref{subsec:I42}, obtain an alternative expression for $\hat{I}_4^{(2)}$. This automatically verifies $\hat{I}_4^{(5)}$, that is obtained from $\hat{I}_4^{(2)}$ by the substitutions $s_2 \rightarrow s_1$ and $t_2 \leftrightarrow t_1$. The technique explained in appendix~\ref{AppendixC3} can be also applied to $\hat{I}_4^{(1)}$, $\hat{I}_4^{(3)}$, $\hat{I}_4^{(4)}$ in order to verify also the results. These latter cases are really much simpler with respect to the ones computed explicitly in appendix~\ref{AppendixC3}. We can also verify these integrals using direct Feynman technique, as explained in \ref{AppendixC2} for $I_{4B}$. 

\subsubsection{Del Duca, Duhr, Glover, Smirnov result for $\hat{I}_5^{D = 6 + 2 \epsilon}$ part}
\label{subsubsec:DDGS}
As it can be seen from eq.~(\ref{IterativeRelI5}), beyond the constant order in the $\epsilon$-expansion, the pentagon in dimension $D=4+2 \epsilon$ takes a contribution from the pentagon integral in dimension $D=6 + 2 \epsilon$. The latter pentagon integral is finite and hence does not contribute to the divergent and finite parts in~(\ref{IterativeRelI5}). It starts to contribute at the order $\epsilon$, and hence, if one wants to obtain the pentagon in $D=4+2 \epsilon$ up to $\epsilon^2$ accuracy, the first two non-trivial orders of the pentagon in $D=6 + 2 \epsilon$ must be computed. There are different results in literature for the 6-dimensional pentagon integral~\cite{DelDuca:2009ac,Kniehl:2010aj,Kozlov:2015kol,Syrrakos:2020kba}; among these, the one obtained by Del Duca, Duhr, Glover, Smirnov (DDGS) is the most suitable for our aims, as it is calculated in multi-Regge kinematics and its $\epsilon$-expansion up to the desired order is given. In~\cite{DelDuca:2009ac} the integral is computed by two independent methods: 1) negative dimension approach and 2) Mellin-Barnes technique. The final result can be expressed in terms of transcendental double sums that, following Ref.~\cite{DelDuca:2009ac}, we denote by $\mathcal{M}$-functions; their definition is given in appendix~\ref{AppendixA}. Adapting their notation to ours, we define
\begin{equation}
    x_1 \equiv \frac{s t_1}{s_1 s_2} = \frac{t_1}{\vec{p}^{ \; 2}}  \; , \hspace{0.5 cm} x_2 = \frac{s t_2}{s_1 s_2} = \frac{t_2}{\vec{p}^{ \; 2}} \; , \hspace{0.5 cm} - \vec{p}^{\; 2} = \frac{(-s_1)(-s_2)}{(-s)} \; , \hspace{0.5 cm} y_1 \equiv \frac{1}{x_2} \; , \hspace{0.5 cm} y_2 = \frac{x_1}{x_2} \; .
\end{equation}
In Ref.~\cite{DelDuca:2009ac} the three regions contributing to the pentagonal are identified as
\begin{itemize}
    \item \textbf{Region I} : $\sqrt{x_1} + \sqrt{x_2} < 1$ 
    \item \textbf{Region II (a)} : $-\sqrt{x_1} + \sqrt{x_2} > 1$ 
    \item \textbf{Region II (b)} : $\sqrt{x_1} - \sqrt{x_2} > 1$ 
\end{itemize}
The solution in the \textbf{Region II (a)} is
\begin{equation}
    \hat{I}_{5}^{(D=6+2 \epsilon)} = -  \frac{ \Gamma(1-\epsilon) \Gamma^2(1+\epsilon)}{\Gamma(1+2 \epsilon)} \sqrt{- \frac{s_1 s_2 t_1}{s t_2}} \left( \frac{(-s_1)(-s_2)}{(-s)} \right)^{\epsilon} \mathcal{I}_{DDGS}^{II(a)} (\vec{p}^{\; 2}, t_1, t_2) \; ,
    \label{DDGSsol}
\end{equation}
where
\begin{equation}
   \mathcal{I}_{DDGS}^{II(a)} (\vec{p}^{\; 2}, t_1, t_2) = i_0^{II(a)} (y_1, y_2) - \epsilon \; i_1^{II(a)} (y_1, y_2) + \mathcal{O} (\epsilon^2)\;,
\end{equation}
with 
\begin{align}
   & i^{(IIa)}_0 (y_1,y_2)=
(-8 \ln y_1-4 \ln y_2) \cM\big(0,0,(1,1);-y_1, y_2\big)-4 \ln y_2 \cM\big((1,1),0,0;-y_1, y_2\big) \nonumber \\
+&\,18 \cM\big(0,0,(1,2);-y_1, y_2\big)+18 \cM\big(0,0,(2,1);-y_1, y_2\big)-24 \cM\big(0,0,(1,1,1);-y_1, y_2\big) \nonumber \\
+&\,8 \cM\big(0,1,(1,1);-y_1, y_2\big)+16 \cM\big(1,0,(1,1);-y_1, y_2\big)-8 \cM\big((1,1),0,1;-y_1, y_2\big) \nonumber \\
+&\,8 \cM\big((1,1),1,0;-y_1, y_2\big)-\cM\big(0,0,0;-y_1, y_2\big) \Big(\frac{\pi ^2 \ln y_1}{3}+\frac{\ln^2 y_1 \ln y_2}{2}+\frac{\pi ^2 \ln y_2}{2}-2 \zeta(3)\Big) \nonumber \\
-&\,\cM\big(0,0,1;-y_1, y_2\big) \Big(2 \ln y_1 \ln y_2+\ln^2 y_1+\frac{5 \pi ^2}{3}\Big)+(6 \ln y_1+3 \ln y_2) \cM\big(0,0,2;-y_1, y_2\big) \nonumber \\
+&\,\Big(2 \ln y_1 \ln y_2+\frac{2 \pi ^2}{3}\Big) \cM\big(1,0,0;-y_1, y_2\big)+(4 \ln y_1+4 \ln y_2) \cM\big(1,0,1;-y_1, y_2\big) \nonumber \\
+&\,4 \ln y_1 \cM\big(0,1,1;-y_1, y_2\big)-4 \ln y_1 \cM\big(1,1,0;-y_1, y_2\big)+\Big(\ln^2 y_1+\pi ^2\Big) \cM\big(0,1,0;-y_1, y_2\big) \nonumber \\
+&\,\ln y_2 \cM\big(2,0,0;-y_1, y_2\big)-12 \cM\big(0,0,3;-y_1, y_2\big)-6 \cM\big(0,1,2;-y_1, y_2\big) \nonumber \\
-&\,12 \cM\big(1,0,2;-y_1, y_2\big)-8 \cM\big(1,1,1;-y_1, y_2\big)+2 \cM\big(2,0,1;-y_1, y_2\big) \nonumber \\
-&\,2 \cM\big(2,1,0;-y_1, y_2\big),
\end{align}
\begin{align}
& i^{(IIa)}_1 (y_1,y_2)=
\cM\big(0,0,(1,1);-y_1, y_2\big) \Big(4 \ln y_1 \ln y_2-4 \ln^2 y_1+2 \ln^2 y_2+4 \pi ^2\Big) \nonumber \\
+&\,(2 \ln^2 y_2-4 \ln y_1 \ln y_2) \cM\big((1,1),0,0;-y_1, y_2\big)+(8 \ln y_1-12 \ln y_2) \cM\big(1,0,(1,1);-y_1, y_2\big) \nonumber \\
+&\,(4 \ln y_2-8 \ln y_1) \cM\big((1,1),0,1;-y_1, y_2\big)+(8 \ln y_1-4 \ln y_2) \cM\big((1,1),1,0;-y_1, y_2\big) \nonumber \\
-&\,15 \ln y_2 \cM\big(0,0,(1,2);-y_1, y_2\big)-15 \ln y_2 \cM\big(0,0,(2,1);-y_1, y_2\big) \nonumber \\
+&\,20 \ln y_2 \cM\big(0,0,(1,1,1);-y_1, y_2\big)-4 \ln y_2 \cM\big(0,1,(1,1);-y_1, y_2\big) \nonumber \\
-&\,\ln y_2 \cM\big((1,2),0,0;-y_1, y_2\big)-\ln y_2 \cM\big((2,1),0,0;-y_1, y_2\big) \nonumber \\ +&\,4 \ln y_2 \cM\big((1,1,1),0,0;-y_1, y_2\big)+32 \cM\big(0,0,(1,3);-y_1, y_2\big)+36 \cM\big(0,0,(2,2);-y_1, y_2\big) \nonumber \\
+&\,32 \cM\big(0,0,(3,1);-y_1, y_2\big)-48 \cM\big(0,0,(1,1,2);-y_1, y_2\big)-48 \cM\big(0,0,(1,2,1);-y_1, y_2\big) \nonumber \\
-&\,48 \cM\big(0,0,(2,1,1);-y_1, y_2\big)+64 \cM\big(0,0,(1,1,1,1);-y_1, y_2\big)+12 \cM\big(0,1,(1,2);-y_1, y_2\big) \nonumber \\
+ & 12 \cM\big(0,1,(2,1);-y_1, y_2\big)-16 \cM\big(0,1,(1,1,1);-y_1, y_2\big)+18 \cM\big(1,0,(1,2);-y_1, y_2\big) \nonumber \\
+&\,18 \cM\big(1,0,(2,1);-y_1, y_2\big)-24 \cM\big(1,0,(1,1,1);-y_1, y_2\big)+8 \cM\big(1,1,(1,1);-y_1, y_2\big) \nonumber \\
-&\,2 \cM\big((1,2),0,1;-y_1, y_2\big)+2 \cM\big((1,2),1,0;-y_1, y_2\big)-2 \cM\big((2,1),0,1;-y_1, y_2\big) \nonumber \\
+&\,2 \cM\big((2,1),1,0;-y_1, y_2\big)+8 \cM\big((1,1,1),0,1;-y_1, y_2\big)-8 \cM\big((1,1,1),1,0;-y_1, y_2\big) \nonumber \\
+&\,\cM\big(0,0,1;-y_1, y_2\big) \Big(\ln y_1 \ln^2 y_2-\frac{\pi ^2 \ln y_1}{3}-\frac{\ln^2 y_1 \ln y_2}{2}-\frac{2 \ln^3 y_1}{3}+\frac{3 \pi ^2 \ln y_2}{2}-6 \zeta(3)\Big) \nonumber \\
+&\,\cM\big(0,1,0;-y_1, y_2\big) \Big(\frac{\pi ^2 \ln y_1}{3}-\frac{\ln^2 y_1 \ln y_2}{2}+\frac{2 \ln^3 y_1}{3}-\frac{\pi ^2 \ln y_2}{2}+2 \zeta(3)\Big) \nonumber \\
+&\,\cM\big(1,0,0;-y_1, y_2\big) \Big(-\ln y_1 \,\ln^2 y_2+\frac{\pi ^2 \ln y_1}{3}+\frac{3 \ln^2 y_1 \ln y_2}{2}-\frac{\pi ^2 \ln y_2}{2}+2 \zeta(3)\Big) \nonumber \\
+&\,\cM\big(0,0,2;-y_1, y_2\big) \Big(-3 \ln y_1 \ln y_2+3 \ln^2 y_1-\frac{3 \ln^2 y_2}{2}-3 \pi ^2\Big) \nonumber \\
+&\,\cM\big(0,1,1;-y_1, y_2\big) \Big(-2 \ln y_1 \ln y_2+2 \ln^2 y_1-\frac{4 \pi ^2}{3}\Big) \nonumber \\
+&\,\cM\big(1,1,0;-y_1, y_2\big) \Big(2 \ln y_1 \ln y_2-3 \ln^2 y_1+\frac{\pi ^2}{3}\Big)+(\ln y_2-2 \ln y_1) \cM\big(2,1,0;-y_1, y_2\big) \nonumber \\
+&\,\cM\big(2,0,0;-y_1, y_2\big) \Big(\ln y_1 \ln y_2-\frac{\ln^2 y_2}{2}\Big)+(9 \ln y_2-6 \ln y_1) \cM\big(1,0,2;-y_1, y_2\big) \nonumber \\
+&\,(4 \ln y_2-4 \ln y_1) \cM\big(1,1,1;-y_1, y_2\big)+(2 \ln y_1-\ln y_2) \cM\big(2,0,1;-y_1, y_2\big) \nonumber \\
+&\,\cM\big(0,0,0;-y_1, y_2\big) \Big(\frac{\ln^2 y_1 \ln^2 y_2}{4}-\frac{\pi ^2 \ln^2 y_1}{6}-\frac{\ln^3 y_1 \ln y_2}{3}-2 \ln y_2 \zeta(3)+\frac{\pi ^2 \ln^2 y_2}{4}+\frac{2 \pi ^4}{15}\Big) \nonumber \\
+&\,\Big(3 \ln^2 y_1-2 \ln^2 y_2-\pi ^2\Big) \cM\big(1,0,1;-y_1, y_2\big)+10 \ln y_2 \cM\big(0,0,3;-y_1, y_2\big) \nonumber \\
+&\,3 \ln y_2 \cM\big(0,1,2;-y_1, y_2\big)-20 \cM\big(0,0,4;-y_1, y_2\big)-8 \cM\big(0,1,3;-y_1, y_2\big) \nonumber \\
-&\,12 \cM\big(1,0,3;-y_1, y_2\big)-6 \cM\big(1,1,2;-y_1, y_2\big).
\end{align}
The solution in the \textbf{Region II (b)} is obtained by the replacement
\begin{equation}
   \mathcal{I}_{DDGS}^{II(a)} (\vec{p}^{\; 2}, t_1, t_2) \rightarrow \mathcal{I}_{DDGS}^{II(b)} (\vec{p}^{\; 2}, t_1, t_2) = \frac{t_2}{t_1} \mathcal{I}_{DDGS}^{II(a)} (\vec{p}^{\; 2}, t_2, t_1) 
\end{equation}
in eq.~(\ref{DDGSsol}). \\
The solution in the \textbf{Region I} can be written as  
\begin{equation}
    \hat{I}_{5}^{(6+2 \epsilon)} = -  \frac{ \Gamma(1-\epsilon) \Gamma^2(1+\epsilon)}{\Gamma(1+2 \epsilon)} \sqrt{- \frac{s t_1 t_2}{s_1 s_2}} \left( \frac{(-s_1)(-s_2)}{(-s)} \right)^{\epsilon} \mathcal{I}_{DDGS}^{I} (\vec{p}^{\; 2}, t_1, t_2) \; ,
\end{equation}
where $\mathcal{I}_{DDGS}^{I} (\vec{p}^{\; 2}, t_1, t_2)$ can be obtained from $\mathcal{I}_{DDGS}^{II(a)} (\vec{p}^{\; 2}, t_1, t_2)$ by analytical continuation, according to the prescription $y_1 \rightarrow 1/y_1$.

\section{Summary and outlook}

\label{Conclusions}

We have calculated at the NLO the Reggeon-Reggeon-gluon (also called ``Lipatov") effective vertex in QCD with accuracy up to the order $\epsilon^2$, with $\epsilon=(D-4)/2$ and $D$ the space-time dimension. The NLO Lipatov effective vertex can be expressed in terms of a few integrals (triangle, boxes, pentagon), which we obtained at the required accuracy in a two-fold way: 1) taking their expressions, known at arbitrary $\epsilon$, from the literature~\cite{Bern_1994,DelDuca:2009ac} and expanding them to the required order; 2) calculating them from scratch\footnote{With the exception of the part of the pentagon integral in $D = 4+2 \epsilon$ which depends on the same integral in $D = 6+2 \epsilon$. Nevertheless, in Ref.~\cite{DelDuca:2009ac} this contribution was computed through two independent methods. In the case of $\mathcal{I}_3$ we have used an independent method valid just up to the order $\epsilon^2$.}, by an independent method. The purpose of the latter calculation is not only cross-checking, but also providing with an alternative, though equivalent, expression for the integrals, which could turn out to be more convenient for the uses of the NLO Lipatov effective vertex. For instance, the result up to order $\epsilon$ of the integral $\mathcal{I}_3$, calculated in section~\ref{subsec:I32}, is very compact compared to the result for the same integral computed in~\ref{subsec:I31}. It contains the structure $I_{\vec{q}_1^{\;2},\vec{q}_2^{\;2},\vec{p}^{\;2}}$ which has been extensively used in the BFKL literature (see \textit{e.g.}~\cite{Ioffe:2010zz}).  \\

The integrals $\mathcal{I}_{4B}$, $\mathcal{I}_{4A}$ and $\hat{I}_{4}^{(i)}$ have been written as expansion to all orders in $\epsilon$, in terms of polylogarithms. The integral $\mathcal{I}_3$ has been expanded up to the order $\epsilon^2$, but can be expanded to higher orders, if needed. The residual and most complicated term is the one related to the pentagonal integral in dimension $6-2\epsilon$. Restricting to the order $\epsilon^2$ in MRK, it is given in section \ref{subsubsec:DDGS}. The knowledge of the one-loop Lipatov vertex in QCD at any successive order in the $\epsilon$-expansion is completely reduced to the computation of this integral with higher $\epsilon$-accuracy. \\
\newpage
\thispagestyle{empty}
\addcontentsline{toc}{chapter}{Beyond the linear regime: saturation in the Shockwave formalism}
\vspace*{\fill}
    \begin{center}
      { \Huge \textbf{Part III}} \vspace{0.3 cm} \\
      { \Huge \textbf{Beyond the linear regime: Saturation in the Shockwave formalism}}
    \end{center}
\vspace*{\fill}
\chapter{The Shockwave formalism}
\begin{flushright}
\emph{\textit{Physics is a very difficult thing; ...Unless you think that \\ physics is the most important thing in your life, \\ you should not do it. It takes passion, precision, patience. \\
S. Ting }}
\end{flushright}

In this chapter, the Shockwave formalism is introduced following~\cite{Balitsky:2001re,Boussarie:2016txb}. This approach, or equivalently his CGC formulation\footnote{CGC can be used also in the scattering of two dense systems.}, applies when one consider the scattering of a dilute projectile on a dense target with high center-of-mass energy. \\
The chapter is organized in four sections. In the first section, we introduce the general picture of saturation. In the second section, we build the effective Lagrangian of the Shockwave approach. In the third, for pedagogical purposes, we derive the propagator of a quark crossing the Shockwave. Finally, in the last section, we derive the famous B-JIMWLK evolution equation for the dipole operator.
\section{Saturation picture}
The BFKL equation has gained much of its fame for being able to predict the rapid growth of the $\gamma^{*} p$ cross section at increasing center-of-mass energy $s$. In the large $s$ limit, which in this context corresponds to the small Bjorken-$x$ limit, the $k_{t}$-factorization is the most suitable scheme to investigate QCD processes. \\
If we consider the total cross section of Deep-Inelastic Scattering (DIS) through Eq.~(\ref{Int:Eq:TotCross}), we have
\begin{equation}
\sigma_{\gamma^{*} P}(x) = \frac{1}{(2 \pi)^{D-2}} \hspace{-0.1 cm} \int \hspace{-0.1 cm} \frac{d^{D-2} k_{\perp}}{\vec{k}^{2}} \frac{\Phi_{\gamma^{*} \gamma^{*}} (\vec{k} \; )}{\vec{k}^{2}} \hspace{-0.1 cm} \int \hspace{-0.1 cm} \frac{d^{D-2} k_{\perp}'}{\vec{k}^{'2}} \frac{\Phi_{P P}(-\vec{k}')}{\vec{k}^{'2}} \hspace{-0.1 cm} \int_{\delta-i\infty}^{\delta+i\infty} \hspace{-0.05 cm} \frac{d\omega}{2 \pi i} \left( \frac{1}{x} \right)^{\omega}  G_{\omega}(\vec{q}_A, \vec{q}_B) ,
\label{Int:Eq:TotCrossDIS}
\end{equation}  
where $ x = \frac{Q^2}{s} $. Obviously, it is important to make a clarification. Two impact factors appear in Eq.~(\ref{Int:Eq:TotCrossDIS}), one describing the $\gamma^{*}$-$\gamma^{*}$ transition and the other describing the proton-proton transition. The second, clearly, contains non-perturbative information and cannot be defined and treated like the parton impact factors that we introduced in previous sections. Suppose we model it on the basis of reasonable physical arguments and define its convolution in the transverse momentum $\vec{k}'$ with the Green's function as $\mathcal{F}(x, \vec{k})$. We then have
\begin{equation}
\label{Int:Eq:RiseOfCrossSection}
    \sigma_{\gamma^{*} P}(x) = \Phi_{\gamma^{*} \gamma^{*}} (\vec{k} \; ) \otimes_{ \vec{k} } \mathcal{F} (x, \vec{k}) \; ,
\end{equation}
where $\otimes_{ \vec{k} }$ means convolution in the transverse momenta $\vec{k}$. The object we have denoted by $\mathcal{F} (x, \vec{k})$ is known as \textit{unintegrated gluon density} and it is defined in such a way that the gluon collinear parton distribution function, $f_g$, is given by
\begin{equation*}
    f_g (x, Q^2) = \int \frac{d^2 k}{\pi \vec{k}^{\; 2}} \mathcal{F} (x, \vec{k}) \theta (Q^2 - \vec{k}^{\; 2}) \; .
\end{equation*}
From Eq.~(\ref{Int:Eq:RiseOfCrossSection}) it is clear that the growth of cross sections that BFKL predicts is 
\begin{equation*}
    \sigma_{\gamma^{*} P}(x) \sim \left( \frac{1}{x} \right)^{\omega_0} = \left( \frac{s}{Q^2} \right)^{\omega_0} \; .
\end{equation*}
This growth violates the Froissart bound~\cite{Froissart:1961ux},
\begin{equation*}
    \sigma_{tot} < c \ln^2 s,
\end{equation*}
where $c$ is a constant. The violation cannot be removed by calculations of radiative corrections to a fixed $NNN...NL$ order and it is physically interpretable as an infinite growth of the UGD at low-$x$. The target appears to become a denser and denser gluon medium, until at some point it becomes infinitely dense. \\
It is clear that this scenario is incomplete and that some \textit{saturation} effects must come into play in order to slow down this growth. The first idea to incorporate these effects was introduced by Gribov, Levin and Ryskin (GLR)~\cite{Gribov:1984tu} and it is based on a non-linear modification of the BFKL equation to incorporate recombination effects. 
This approach makes use of the double-log approximation, \textit{i.e.} resummation of $\alpha_s \ln( s) \ln (Q^2) $ terms.\\
The most modern approach to saturation physics is based on a hierarchy of equations, known under the name of Balitsky-JIMWLK (B-JIMWLK) equations. They were derived, in the so-called Shockwave approach, by Balitsky~\cite{Balitsky:2001re,Balitsky:1995ub,Balitsky:1998kc,Balitsky:1998ya} and, in the so-called Color Glass Condensate (CGC) approach, by Jalilian-Marian, Iancu, McLerran, Weigert, Leonidov and Kovner (JIMWLK)~\cite{Jalilian-Marian:1997qno,Jalilian-Marian:1997jhx,Jalilian-Marian:1997ubg,Jalilian-Marian:1998tzv,Kovner:2000pt,Weigert:2000gi,Iancu:2000hn,Iancu:2001ad,Ferreiro:2001qy}. 
In the large $N$-limit, a closed equation, which is known as Balitsky-Kovchegov (BK), is obtained. This large-$N$ truncation was recovered by Kovchegov using Mueller’s dipole formalism~\cite{Kovchegov:1999yj,Kovchegov:1999ua}. It can also be shown that, in the dipole operator
case and in the double logarithmic limit, the GLR equation is recovered. \\
In saturation physics, a new dimensional scale is introduced, the so called \textit{saturation scale},
\begin{equation}
\label{Int:Eq:SaturationScale}
    Q_s^2 (x) \equiv (A x^{-1})^{\frac{1}{3}} \Lambda_{\rm{QCD}}^2 \; ,
\end{equation}
where $A$ is the mass number of the target. It is estimated that the CGC formalism (or equivalently Balitsky’s shockwave formalism) must be applied instead of the BFKL formalism for  
\begin{equation*}
    Q^2 < Q_s^2 \; .
\end{equation*}
Eq.~(\ref{Int:Eq:SaturationScale}) contains in itself extremely important information. The saturation scale grows not only with the inverse of the Bjorken-$x$, but also with the mass number of the target. Saturation effects are therefore emphasized in large nuclei. This is crucial, because it means that if very low values of $x$ have to be reached for a proton to perceive these effects, for a large enough nucleus, these effects will be seen at larger $x$.
\section{Boosted gluonic field and effective Lagrangian}
\subsection{Notation and conventions}
We introduce the dimensionless Sudakov vectors $n_1$ and $n_2$, defined as
\begin{equation}
    n_1 \equiv \frac{1}{\sqrt{2}} (1, 0_{\perp}, 1), \hspace{0.2 cm}  n_2 \equiv \frac{1}{\sqrt{2}} (1, 0_{\perp}, -1) \; , \hspace{0.2 cm} n_1^{+} = n_2^{-} = n_1 \cdot n_2 = 1 \; .
\end{equation}
With these definitions, for any four-vector $p$, we can introduce the components
\begin{equation*}
    p^{+} = p_{-} \equiv (p \cdot n_2) = \frac{1}{\sqrt{2}} (p^0 + p^3) \; , \hspace{1 cm} p_{+} = p^{-} \equiv (p \cdot n_1) = \frac{1}{\sqrt{2}} (p^0 - p^3) \; ,
\end{equation*}
\begin{equation}
    p = p^{+} n_1 + p^{-} n_2 + p_{ \perp } \; .
\end{equation}
A generic scalar product can therefore be written as
\begin{equation}
    (p \cdot k) = p^{\mu} k_{\mu} = p^{+} k^{-} + p^{-} k^{+} + (p_{\perp} \cdot k_{\perp}) \equiv p_{+} k_{-} + p_{-} k_{+} - (\Vec{p} \cdot \Vec{k}) \; .
\end{equation}
We use the same notation for Dirac matrices, and denote
\begin{equation}
    \gamma^{+} = \gamma_{-} = \gamma^{\mu} n_{2, \mu} \; , \hspace{1 cm } \gamma^{-} = \gamma_{+} = \gamma^{\mu} n_{1, \mu} \; .
\end{equation}
Throughout the chapter we also use the notation
\begin{equation}
    \gamma^{\mu} p_{\mu} = \hat{p} \; .
\end{equation}
In the Regge limit, both the projectile and the target possess large components, respectively along the $n_1$ and along $n_2$, of the order
\begin{equation}
    p_{p}^{+} \sim p_{t}^{-} \sim \sqrt{\frac{s}{2}} \; .
\end{equation}
Again, we work in a $k_t$-formalism, therefore the scattering amplitudes will be expressed as convolutions in transverse momenta between a projectile impact factor and a target impact factor. The gluons exchanged in the $t$-channel, describing the interaction between projectile and Shockwave field, are once again eikonal and carry an effective polarization proportional to $n_2$. Once again, we work in dimension $D = d + 2 = 4 + 2 \epsilon $. \\

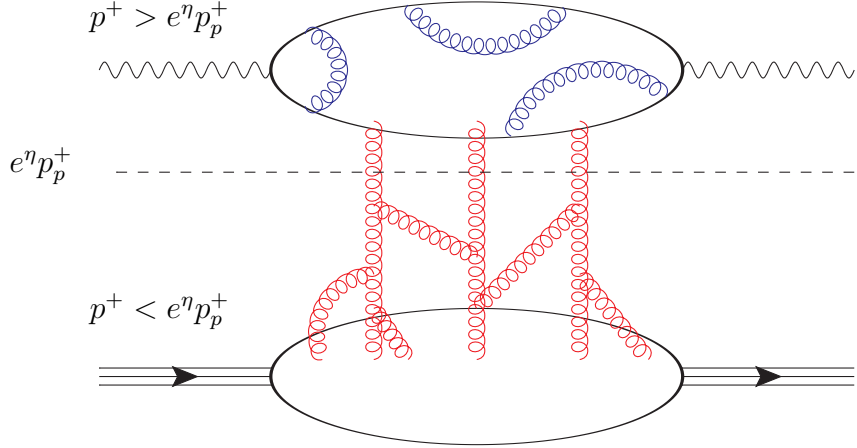
\begin{figure}
\begin{picture}(400,180)
\put(80,0){\includegraphics[scale=0.40]{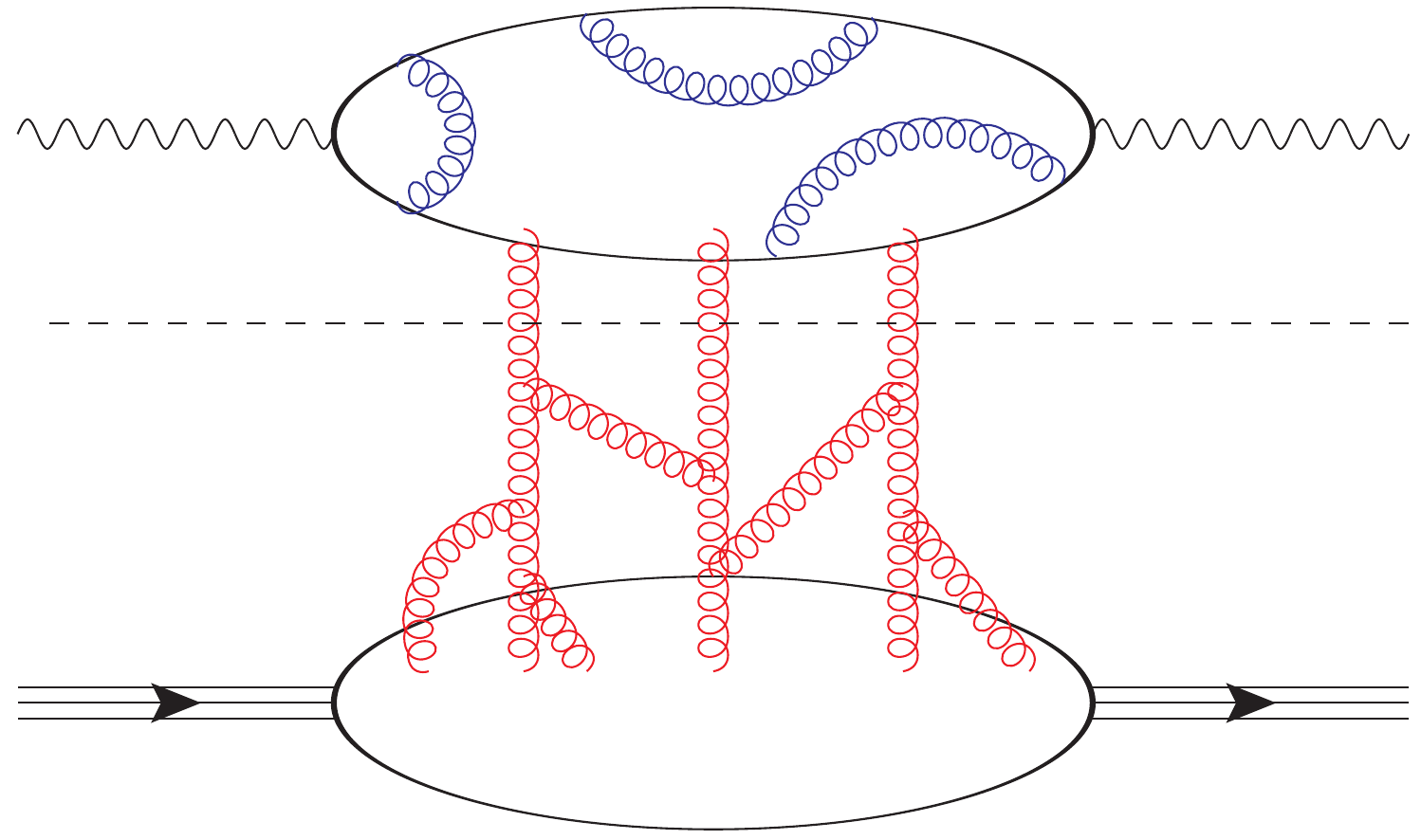}}
\put(80,160){$p^{+} > e^{\eta} p_p^{+}$}
\put(50,105){$e^{\eta} p_p^{+}$}
\put(80,50){$p^{+} < e^{\eta} p_p^{+}$}
\end{picture}
\caption{Schematic representation of the separation in the rapidity space between quantum ``fast" modes (blue) and classical ``slow" modes (red).}
\label{Int:Fig:ShockwaveFact}
\end{figure}

The fundamental idea of the Shockwave approach is to immediately exploit the Regge kinematics in such a way as to formulate the high-energy scattering problem in terms of the relevant degrees of freedom. In particular, since in Regge kinematics all the transverse momenta are of the same order of magnitude, but colliding particles strongly differ in rapidity, it is natural to factorize in the rapidity space (see Fig.~\ref{Int:Fig:ShockwaveFact}). This type of factorization is achieved by introducing a rapidity cut-off, $e^{\eta} p_{p}^{+}$, which separates the ``slow" modes from the ``fast" modes. The interpretation of these modes as ``fast" and ``slow" is thought in the rest reference frame of the target. Gluons with $+$-momentum component below the cut-off act as an external \textit{classical} field. The interaction between projectile and target is through these ``slow" modes and, as we will see, it is described in terms of a Wilson-line operator. The B-JIMWLK evolution equations are the result of quantum corrections, due to ``fast" gluons, to the Wilson line. Practically, in the spirit of the renormalization group, it is the dependence of the relevant matrix elements on the cut-off $\eta$ which determines the high-energy behavior of the original amplitude.  \\

One could reasonably ask why slow gluons can be treated classically. The answer lies in the fact that, in the saturation regime $Q^2 < Q_s^2 (x)$, we are dealing with weakly coupled QCD, \textit{i.e.}
\begin{equation}
    \alpha_s (Q_s^2) \ll 1 \; .
\end{equation}
As it is well known, in the small-coupling regime, field theories are dominated by classical solutions. In this observation lies one of the most interesting aspects of saturation approaches. In fact, like other semi-classical approaches, they are able to describe interesting phenomena, which occur for small coupling, but for which approaches based on perturbation theory only are inadequate. 
\subsection{Boosted gluonic field}
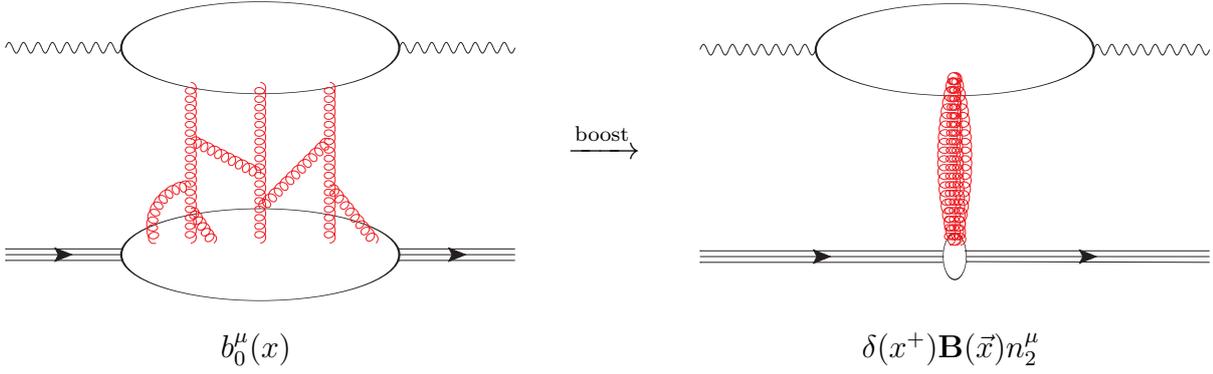
\begin{figure}
\begin{picture}(400,145)
\put(0,20){\includegraphics[scale=0.27]{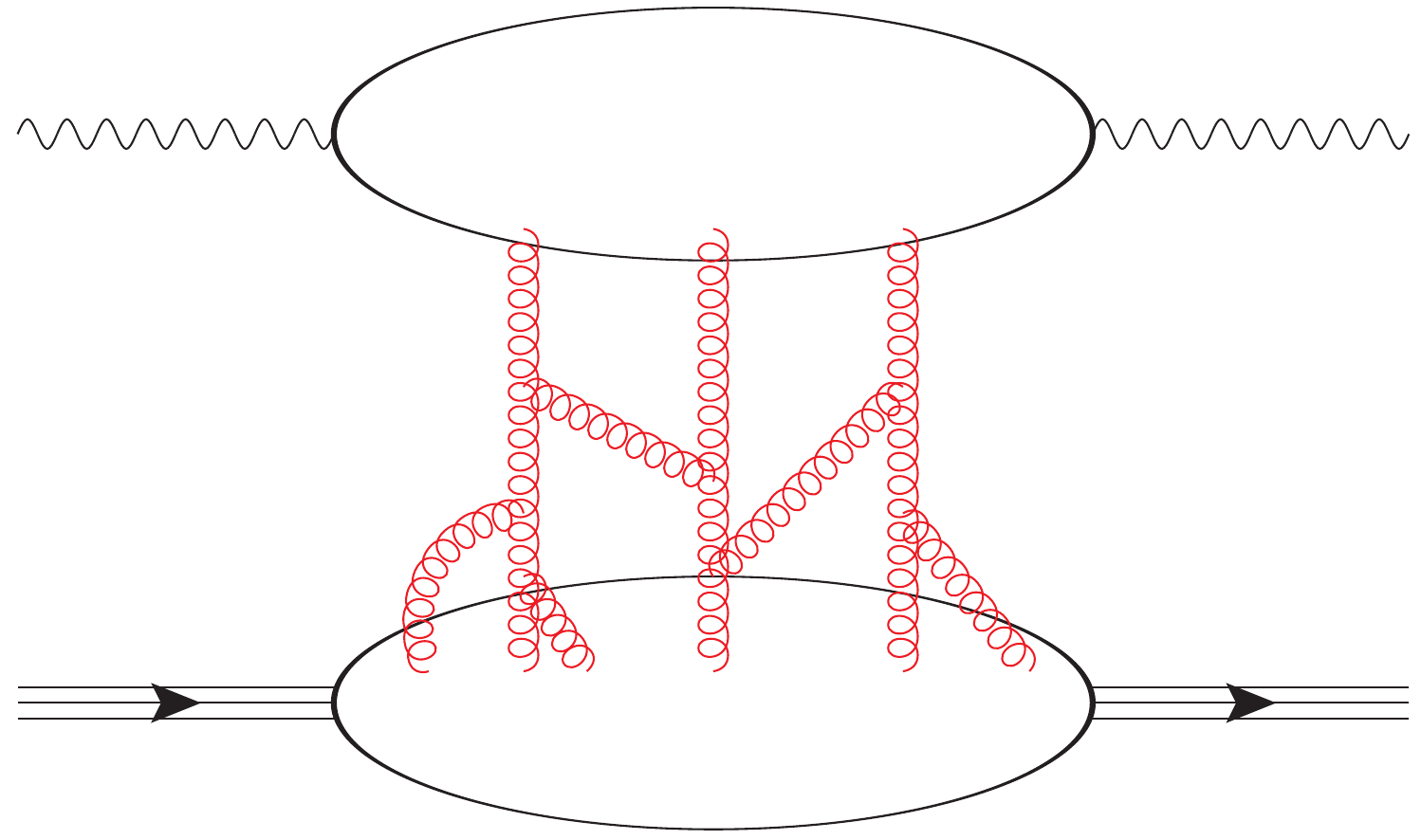}}
\put(83,0){$b_0^{\mu} (x)$}
\put(213,75){$\xrightarrow{{\rm{boost}}}$}
\put(260,28){\includegraphics[scale=0.27]{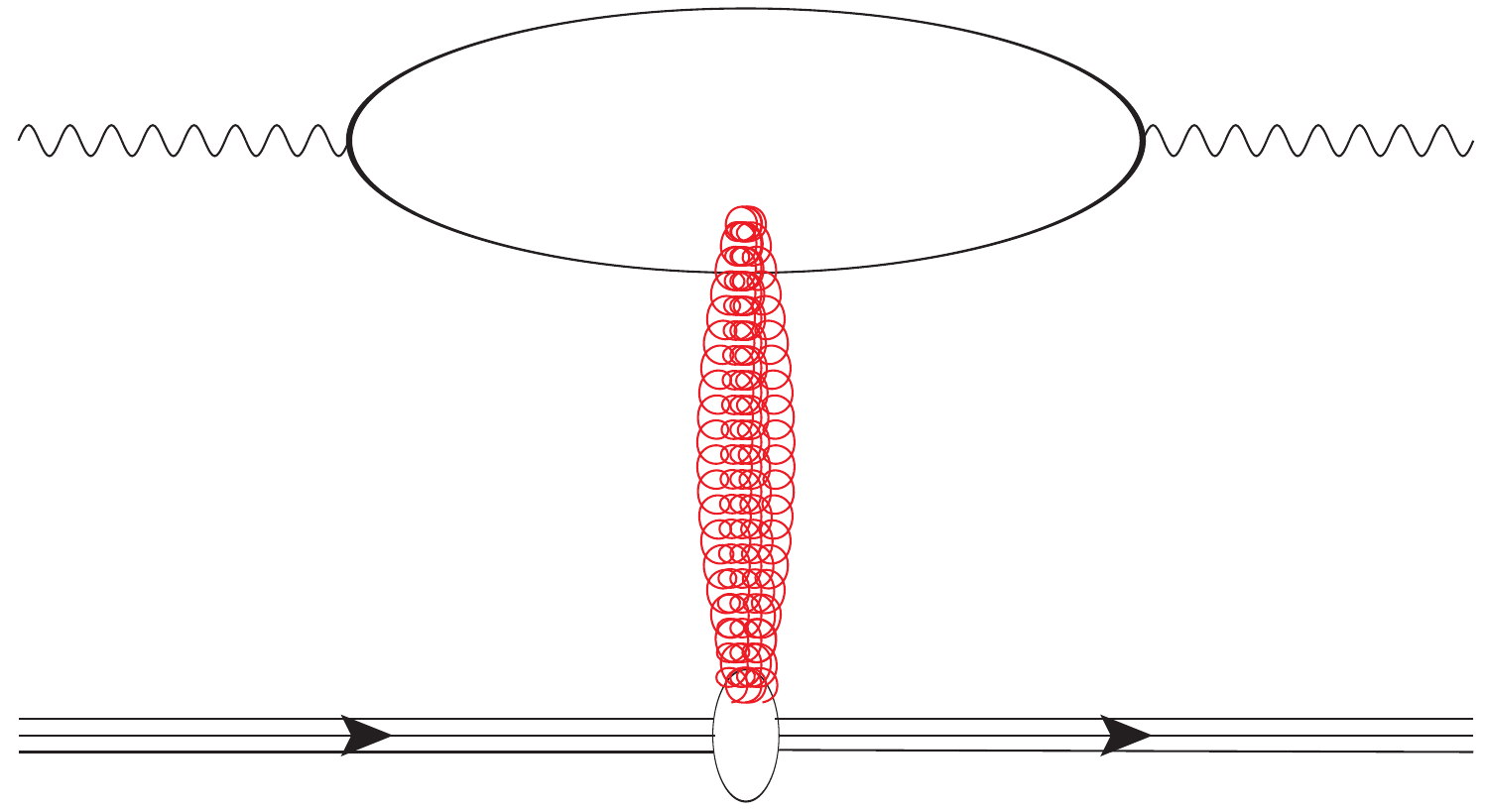}}
\put(323,0){$ \delta (x^{+}) \mathbf{B}(\vec{x}) n_2^{\mu}$}
\end{picture}
\caption{Schematic representation of the Shockwave approximation.}
\label{Int:Fig:ShockwaveBoost}
\end{figure}

Let's consider a gluon field $b_0 (\mu)$ in the target rest frame. Let's now perform a boost, with velocity $\beta$, along the $z^+$ axis to move to the projectile rest frame. The coordinate transformation is
\begin{equation}
    (x^{+}, x^{-}, \vec{x} \; ) \equiv \left( \frac{z^{+}}{\Lambda}, \Lambda z^{-}, \vec{x} \right) \; , \hspace{0.5 cm} \text{with} \hspace{0.5 cm} \Lambda = \sqrt{\frac{1+\beta}{1-\beta}} \; .
\end{equation}
In the new frame, the gluonic field reads
\begin{equation}
 \begin{aligned} 
b^{+}\left(x^{+}, x^{-}, \vec{x} \; \right) & =\frac{1}{\Lambda} b_0^{+}\left(\Lambda x^{+}, \frac{x^{-}}{\Lambda}, \vec{x}\right) \; , \\ b^{-}\left(x^{+}, x^{-}, \vec{x} \; \right) & =\Lambda b_0^{-}\left(\Lambda x^{+}, \frac{x^{-}}{\Lambda}, \vec{x}\right) \; , \\ b^{\; i}\left(x^{+}, x^{-}, \vec{x} \; \right) & =b_0^i\left(\Lambda x^{+}, \frac{x^{-}}{\Lambda}, \vec{x}\right) \; .
\end{aligned}   
\end{equation}
Assuming that the field vanishes to infinity, and remembering that we are making a very large boost, the $+$ and $i$ components of the field vanish. Hence, up to $\Lambda^{-1}$ corrections, we get that
\begin{equation}
 \begin{aligned} 
b^{+}\left(x^{+}, x^{-}, \vec{x} \; \right) & = 0 \; , \\ b^{-}\left(x^{+}, x^{-}, \vec{x} \; \right) & =\Lambda b_0^{-}\left(\Lambda x^{+}, 0, \vec{x}\right) \; , \\ b^{\; i} \left(x^{+}, x^{-}, \vec{x} \; \right) & = 0 \; ,
\end{aligned}   
\end{equation}
where we note that, in this approximation, the surviving component of the field is always evaluated in $x^{-} = 0$. Exploiting that, for any integrable function $F$,
\begin{equation}
    \lim_{\Lambda \rightarrow \infty} \Lambda F (\Lambda x) \propto \delta (x) \; ,
\end{equation}
we can write
\begin{equation}
    b^{\mu} \left( x^{+}, x^{-}, \vec{x} \; \right) = b^{-} \left( x^{+}, \vec{x} \; \right) n_2^{ \mu } \equiv \delta ( x^{+} ) \mathbf{B} (\vec{x} \; ) n_2^{\mu} \; .
\end{equation}
The last equation expresses precisely the idea of re-formulating the problem in terms of the relevant degrees of freedom. From the point of view of the projectile, the interaction takes place with an eikonal field ($n_2^{\mu}$), well localized in the $+$-component ($ \delta (x^{+})$) and whose shape is described by a function depending on transverse coordinates only ($\mathbf{B} (\vec{x} \; )$). The effect of this transformation is depicted in Fig.~(\ref{Int:Fig:ShockwaveBoost}). \\

Let's consider the collision of our projectile with a large momentum $p_{p}$ along $p^{+}$ on a target with a large momentum $p_{t}$ along $p^{-}$ and with mass $m_t$. The energy of the target in the projectile's frame is
\begin{equation}
\label{Int:Eq:EnegyTar}
    E = \frac{m_t}{\sqrt{1-\beta^{2}}} = \frac{p_t^{+}+p_t^{-}}{\sqrt{2}} \sim \frac{p_t^{-}}{\sqrt{2}} \; .
\end{equation}
From Eq.~(\ref{Int:Eq:EnegyTar}) it is simple to see that
\begin{equation*}
    \Lambda \sim \sqrt{\frac{s}{m_t^2}} \; ,
\end{equation*}
when $p_p^{+}, p_t^{-} \sim \sqrt{\frac{s}{2}}$. Thus, the above approximation (neglecting $\Lambda^{-1}$ corrections) is equivalent to neglect terms suppresed by $\frac{1}{\sqrt{s}}$ factor.
\subsection{Effective Lagrangian}
We now start from the QCD Lagrangian,
\begin{equation}
\begin{aligned} \mathcal{L}= & -\frac{1}{4} \mathcal{F}_{a \mu \nu} \mathcal{F}^{a \mu \nu}+i \bar{\psi} \hat{D} \psi =\mathcal{L}_{\text {free }} + \mathcal{L}_{\text {int}} \; ,
\end{aligned}
\end{equation}
where we split the free and the interaction part; this latter reads
\begin{equation}
   \mathcal{L}_{\text{int}} = -g f_{a b c}\left(\partial_\mu \mathcal{A}_\nu^a\right)\left(\mathcal{A}^{\mu b} \mathcal{A}^{\nu c}\right)-\frac{1}{4} g^2 f_{a b c} f_{a d e}\left(\mathcal{A}_\mu^a \mathcal{A}_\nu^b \mathcal{A}^{\mu d} \mathcal{A}^{\nu e}\right) +i \bar{\psi}\left(-i g t^a \hat{\mathcal{A}}^a\right) \psi \; .
\end{equation}
In this interaction part, we write the total gluon field, $\mathcal{A}$, as the combination of the external field, $b$, and the usual gluon field $A$.  The former describes the effective interaction of the projectile with the small-$x$ gluons from the target, while the latter contributes to the quantum evolution of the system. We then have,
\begin{equation}
    \mathcal{A}_{\mu}^{a} = A_{\mu}^{a} + b_{\mu}^{a} \; .
\end{equation}
Using the fact that $b^{\mu} b_{\mu} \propto n_2^2 =0$, we obtain
\begin{equation}
\begin{aligned} \mathcal{L}_{\text{int}} = & -g f_{a b c}\left[\left(\left(A^b \cdot \partial\right) A^a+\left(b^b \cdot \partial\right) A^a\right) \cdot\left(A^c+b^c\right)+\left(\left(A^b \cdot \partial\right) b^a+\left(b^b \cdot \partial\right) b^a\right) \cdot A^c\right] \\ & -\frac{1}{4} g^2 f_{a b c} f_{a d e}\left[\left(\left(A^a \cdot A^d\right)+\left(b^d \cdot A^a\right)\right)\left(\left(A^b \cdot A^e\right)+\left(b^e \cdot A^b\right)+\left(b^b \cdot A^e\right)\right)\right. \\ & \left.+\left(b^a \cdot A^d\right)\left(\left(A^b \cdot A^e\right)+\left(b^e \cdot A^b\right)+\left(b^b \cdot A^e\right)\right)\right] +i \bar{\psi}\left[-i g t^a\left(\hat{A}^a+\hat{b}^a\right)\right] \psi .
\end{aligned}
\label{Int:Eq:InteractionPartL1}
\end{equation}
In the lightcone gauge $A \cdot n_2 = 0$, 
\begin{equation*}
  A \cdot b \propto A \cdot n_2 = 0  
\end{equation*}
holds, and hence, Eq.~(\ref{Int:Eq:InteractionPartL1}) can be simplified. Therefore, in this case, unlike the formulation of the BFKL approach presented in the previous sections, it is essential to choose an axial gauge. In the aforementioned gauge, we then have 
\begin{gather}
    \mathcal{L}_{\text{int}} =  -g f_{a b c}\left(A^b \cdot \partial\right)\left(A^a \cdot A^c\right)-\frac{1}{4} g^2 f_{a b c} f_{a d e}\left[\left(A^a \cdot A^d\right)\left(A^b \cdot A^e\right)\right] +i \bar{\psi}\left[-i g t^a \hat{A}^a\right] \psi \nonumber \\  -g f_{a b c}\left(b^b \cdot \partial\right)\left(A^a \cdot A^c\right)+g t^a\left(\bar{\psi} \hat{b}^a \psi\right).
\end{gather}
The first three terms describe the usual interactions of QCD, while the last two terms describe the interaction between the internal field and the Shockwave field. We denote this part of the Lagrangian as
\begin{equation}
    \mathcal{L}_{\text {int }}^{\mathcal{S}} = -g f_{a c b} b^{-c} g_{\alpha \beta}\left[A_\alpha^a \frac{\partial A_\beta^b}{\partial x^{-}}\right]+g\left(\bar{\psi} t^a \hat{b}^a \psi\right).
\end{equation}
\section{Quark propagator through the shockwave field}
To give an explicit example of how Feynman rules describing the effective interaction can be found, let us re-derive the one for a quark propagator crossing the Shockwave. 
\subsection{Two gluons interaction}
We start by considering the propagator interacting twice with the external field while propagating from a point $z_0$ (at negative light-cone time) to a point $z_3$ (at positive light-cone time). We have\footnote{$b^{-}$ contains the $SU(3)$ matrix in the fundamental representation.}
\begin{equation}
\begin{aligned}
& G\left(z_3, z_0\right)|_{z_3^{+}>0>z_0^{+}} \\ = & \int d^D z_2 d^D z_1 G_0\left(z_{32}\right)\left[i g b^{-}\left(z_2\right) \gamma^{+}\right] G_0\left(z_{21}\right)\left[i g b^{-}\left(z_1\right) \gamma^{+}\right] G_0\left(z_{10}\right) \\ = & \int d^D z_2 d^D z_1\left[i g b^{-}\left(z_2\right) i g b^{-}\left(z_1\right)\right] \int \frac{d^D p_3}{(2 \pi)^D} \frac{d^D p_2}{(2 \pi)^D} \frac{d^D p_1}{(2 \pi)^D} G_0\left(p_3\right) \gamma^{+} G_0\left(p_2\right) \gamma^{+} G_0\left(p_1\right) \\ & \times \exp \left[-i\left(p_3 \cdot z_3\right)+i\left(p_3-p_2\right) \cdot z_2+i\left(p_2-p_1\right) \cdot z_1+i\left(p_1 \cdot z_0\right)\right]  \; ,
\end{aligned}   
\end{equation}
where we denote the free quark propagator by $G_0$. As explained before, the field $b^{-}(z)$ has no dependencies on $z^{-}$; this means that we can immediately integrate over $z_{1}^{-}$, $z_{2}^{-}$. These integrations produce two Dirac deltas which impose $p_3^{+} = p_2^{+} = p_1^{+} \equiv p^{+}$:
\begin{equation}
\label{Int:Eq:QuarkProp1}
\begin{aligned}
& \left.G\left(z_3, z_0\right)\right|_{z_3^{+}>0>z_0^{+}}= (2 \pi)^2 \int d z_2^{+} d z_1^{+} d^d \vec{z}_2 d^d \vec{z}_1\left[i g b^{-}\left(z_2^{+}, \vec{z}_2\right) i g b^{-}\left(z_1^{+}, \vec{z}_1\right)\right] \\ & \times \int d p^{+} \int \frac{d^d \vec{p}_3}{(2 \pi)^D} \frac{d^d \vec{p}_2}{(2 \pi)^D} \frac{d^d \vec{p}_1}{(2 \pi)^D} \exp \left[i\left(\vec{p}_3 \cdot \vec{z}_{32}\right)+i\left(\vec{p}_2 \cdot \vec{z}_{21}\right)+i\left(\vec{p}_1 \cdot \vec{z}_{10}\right)-i p^{+} z_{30}^{-}\right] \\ & \times \int d p_3^{-} \frac{i\left(p^{+} \gamma^{-}+\hat{p}_{3 \perp}\right)}{2 p^{+}\left(p_3^{-}-\frac{\vec{p}_3^{\; 2}-i 0}{2 p^{+}}\right)} \exp \left[-i p_3^{-} z_{32}^{+}\right] \int d p_2^{-} \gamma^{+} \frac{i\left(p^{+} \gamma^{-}+\hat{p}_{2 \perp}\right)}{2 p^{+}\left(p_2^{-} -\frac{\vec{p}_2^{\; 2}-i 0}{2 p^{+}}\right)} \gamma^{+} \exp \left[-i p_2^{-} z_{21}^{+}\right] \\ & \times \int d p_1^{-} \frac{i\left(p^{+} \gamma^{-}+\hat{p}_{1 \perp}\right)}{2 p^{+}\left(p_1^{-}-\frac{\vec{p}_1^{\; 2}-i 0}{2 p^{+}}\right)} \exp \left[-i p_1^{-} z_{10}^{+}\right] \; .
\end{aligned}
\end{equation}
In Eq.~(\ref{Int:Eq:QuarkProp1}), we have already removed the $\gamma^{+}$ component in all quarks propagators exploiting $\gamma^{+} \gamma^{+} = 0$. The integration over the three variables $p_i^{-}$ can be preformed using Jordan's lemma. Due to the conditions imposed by theta functions, only the region $p^{+} >0$ gives a non-zero contribution:
\begin{equation}
\label{Int:Eq:QuarkProp2}
    \begin{aligned}
   & \left.G\left(z_3, z_0\right)\right|_{z_3^{+}>0>z_0^{+}}= (2 \pi)^5 \int d z_2^{+} d z_1^{+} d^d \vec{z}_2 d^d \vec{z}_1\left[i g b^{-}\left(z_2^{+}, \vec{z}_2\right) i g b^{-}\left(z_1^{+}, \vec{z}_1\right)\right] \\ & \times \int d p^{+} \theta\left(p^{+}\right) \theta\left(z_{32}^{+}\right) \theta\left(z_{21}^{+}\right) \theta\left(z_{10}^{+}\right) \exp \left[-i p^{+} z_{30}^{-}\right] \\ & \times \int \frac{d^d \vec{p}_3}{(2 \pi)^D} \frac{\left(p^{+} \gamma^{-}+\hat{p}_{3 \perp}\right)}{2 p^{+}} \exp \left[-i \frac{z_{32}^{+}}{2 p^{+}}\left\{\left(\vec{p}_3-\frac{p^{+}}{z_{32}^{+}} \vec{z}_{32}\right)^2-\left(\frac{p^{+}}{z_{32}^{+}}\right)^2 \vec{z}_{32}^{\; 2}-i 0\right\}\right] \\ & \times \int \frac{d^d \vec{p}_2}{(2 \pi)^D} \gamma^{+} \exp \left[-i \frac{z_{21}^{+}}{2 p^{+}}\left(\vec{p}_2^{\; 2}-i 0\right)+i\left(\vec{p}_2 \cdot \vec{z}_{21}\right)\right] \\ & \times \int \frac{d^d \vec{p}_1}{(2 \pi)^D} \frac{\left(p^{+} \gamma^{-}+\hat{p}_{1 \perp}\right)}{2 p^{+}} \exp \left[-i \frac{z_{10}^{+}}{2 p^{+}}\left\{\left(\vec{p}_1-\frac{p^{+}}{z_{10}^{+}} \vec{z}_{10}\right)^2-\left(\frac{p^{+}}{z_{10}^{+}}\right)^2 \vec{z}_{10}^{\; 2}-i 0\right\}\right].
    \end{aligned}
\end{equation}
Up to corrections of order $\Lambda^{-1} \sim s^{-1/2}$, $b(z_{i}^{+}, \vec{z}_{i}) \propto \delta(z_{i}^{+})$. This allows us to approximate the Gaussian integral over $\vec{p}_2$ by setting $z_{12}^{+}=0$ and getting therefore a $\delta (\vec{z}_{12})$. This is important, because it tells us that the interaction with the shockwave field occurs at a single transverse coordinate. We note that the remaining Gaussian integrals are always convergent since $\frac{z_{10}^{+}}{p^{+}}, \frac{z_{32}^{+}}{p^{+}}~>~0$. Observing that we can clearly rescale the pole prescription $i0$ by using an arbitrary positive quantity, we can make the following manipulation
\begin{equation}
    \begin{aligned}
    & -i \frac{z_{32}^{+}}{2 p^{+}}\left[\left(\vec{p}_3-\frac{p^{+}}{z_{32}^{+}} \vec{z}_{32}\right)^2-\left(\frac{p^{+}}{z_{32}^{+}}\right)^2 \vec{z}_{31}^{\; 2}-i 0\right] \\ = & -i \frac{z_{32}^{+}}{2 p^{+}}\left[\left\{\left(\vec{p}_3-\frac{p^{+}}{z_{32}^{+}} \vec{z}_{32}\right)^2-i 0\right\}-\left(\frac{p^{+}}{z_{32}^{+}}\right)^2\left(\vec{z}_{31}^{\; 2}+i 0\right)\right] \\ = & -i \frac{z_{32}^{+}}{2 p^{+}}(1-i 0)\left(\vec{p}_3-\frac{p^{+}}{z_{32}^{+}} \vec{z}_{32}\right)^2+i \frac{p^{+}}{2 z_{32}^{+}}\left(\vec{z}_{31}^{\; 2}+i 0\right) \; .
    \end{aligned}
\end{equation}
An analogous manipulation can be done for the exponent depending on $p_1$, and then both integrations can be performed to get
\begin{equation}
\begin{aligned}
& \left.G\left(z_3, z_0\right)\right|_{z_3^{+}>0>z_0^{+}}= \int d z_2^{+} d z_1^{+} d^d \vec{z}_2 d^d \vec{z}_1\left[i g b^{-}\left(z_2^{+}, \vec{z}_2\right) i g b^{-}\left(z_1^{+}, \vec{z}_1\right)\right] \frac{\delta\left(\vec{z}_{21}\right)}{4(2 \pi)^{2 D-3}} \\ & \times\left(\gamma^{-}+\frac{\hat{z}_{31 \perp}}{z_{32}^{+}}\right) \gamma^{+}\left(\gamma^{-}+\frac{\hat{z}_{10 \perp}}{z_{10}^{+}}\right) \int d p^{+}\left(\frac{-2 i \pi p^{+}}{z_{32}^{+}}\right)^{\frac{d}{2}}\left(\frac{-2 i \pi p^{+}}{z_{10}^{+}}\right)^{\frac{d}{2}} \theta\left(p^{+}\right) \theta\left(z_{32}^{+}\right) \\ & \times \theta\left(z_{21}^{+}\right) \theta\left(z_{10}^{+}\right) \exp \left[-i p^{+} z_{30}^{-}+i \frac{p^{+}}{2 z_{32}^{+}}\left(\vec{z}_{31}^{\; 2}+i 0\right)+i \frac{p^{+}}{2 z_{10}^{+}}\left(\vec{z}_{10}^{\; 2}+i 0\right)\right] \\ & =  i \frac{\Gamma(D-1)}{4(2 \pi)^{D-1}} \int d z_2^{+} d z_1^{+} d^d \vec{z}_1\left[i g b^{-}\left(z_2^{+}, \vec{z}_1\right) i g b^{-}\left(z_1^{+}, \vec{z}_1\right)\right] \\ & \times \frac{\left(z_3^{+} \gamma^{-}+\hat{z}_{31 \perp}\right) \gamma^{+}\left(-z_0^{+} \gamma^{-}+\hat{z}_{10 \perp}\right)}{\left(-z_3^{+} z_0^{+}\right)^{\frac{D}{2}}} \frac{\theta\left(z_{32}^{+}\right) \theta\left(z_{21}^{+}\right) \theta\left(z_{10}^{+}\right)}{\left(-z_{30}^{-}+\frac{\vec{z}_{31}^{\; 2}+i 0}{2 z_3^{+}}-\frac{\vec{z}_{10}^{\; 2}+i 0}{2 z_0^{+}}\right)^{D-1}} .
\end{aligned}
\end{equation}
Now, it is useful to define  
\begin{equation}
\label{Int:Eq:WilsonLine2Ord}
    U_{\vec{z}}^{(2)}=(i g)^2 \int d z_2^{+} d z_1^{+} \theta\left(z_{32}^{+}\right) \theta\left(z_{21}^{+}\right) \theta\left(z_{10}^{+}\right) b^{-}\left(z_2^{+}, \vec{z}\right) b^{-}\left(z_1^{+}, \vec{z}\right) \; ,
\end{equation}
and re-write the previous expression as
\begin{equation}
    \left.G\left(z_3, z_0\right)\right|_{z_3^{+}>0>z_0^{+}}=i \frac{\Gamma(D-1)}{4(2 \pi)^{D-1}} \int d^d \vec{z}_1 U_{\vec{z}_1}^{(2)} \frac{\left(z_3^{+} \gamma^{-}+\hat{z}_{31 \perp}\right) \gamma^{+}\left(-z_0^{+} \gamma^{-}+\hat{z}_{10 \perp}\right)}{\left(-z_3^{+} z_0^{+}\right)^{\frac{D}{2}}\left(-z_{30}^{-}+\frac{\vec{z}_{31}^{\; 2}}{2 z_3^{+}}-\frac{\vec{z}_{10}^{\; 2}}{2 z_0^{+}}+i 0\right)^{D-1}} \; .
\end{equation}
The physical interpretation of this expression is not immediate. We therefore introduce a representation which, as we shall see, is completely equivalent:
\begin{equation}
\label{Int:Eq:QuarkPropPhysiInt}
    \left.\tilde{G}\left(z_3, z_0\right)\right|_{z_3^{+}>0>z_0^{+}} \equiv \int d^D z_1 \delta\left(z_1^{+}\right) G_0\left(z_{31}\right) \gamma^{+} G_0\left(z_{10}\right) \theta\left(z_3^{+}\right) \theta\left(-z_0^{+}\right) U_{\vec{z}_1}^{(2)} \; .
\end{equation}
From Eq.~(\ref{Int:Eq:QuarkPropPhysiInt}), the physical picture is clear: first the quark propagates from $z_0$ to $z_1$, then
it interacts instantly at $z_{1}^{+} = 0$ with the external field, then it propagates again from $z_1$ to $z_3$. Let's now prove that the expression we gave is completely equivalent. We first integrate over $z_1^{+}$ to get
\begin{equation}
\left.\tilde{G}\left(z_3, z_0\right)\right|_{z_3^{+}>0>z_0^{+}}=\left[\frac{\Gamma\left(\frac{D}{2}\right)}{2 \pi^{\frac{D}{2}}}\right]^2 \int d z_1^{-} d^d \vec{z}_1 U_{\vec{z}_1}^{(2)} \frac{\left(z_3^{+} \gamma^{-}+\hat{z}_{31 \perp}\right) \gamma^{+}\left(-z_0^{+} \gamma^{-}+\hat{z}_{10 \perp}\right)}{\left(-2 z_3^{+} z_{31}^{-}+\vec{z}_{31}^{\; 2}+i 0\right)^{\frac{D}{2}}\left(2 z_0^{+} z_{10}^{-}+\vec{z}_{10}^{\; 2}+i 0\right)^{\frac{D}{2}}} \; ,
\end{equation}
and then we introduce the Schwinger representation for the denominators,
\begin{equation}
    \frac{1}{(A \pm i 0)^n}=\frac{(\mp i)^n}{\Gamma(n)} \int_0^{+\infty} d \alpha\left(\alpha^{n-1}\right) e^{\pm i \alpha(A \pm i 0)} \; .
\end{equation}
Performing the integration over $z_{1}^{-}$ and then over one Schwinger parameter, we find
\begin{equation}
\label{Int:Eq:ExpressForInductionProof}
    \begin{aligned}
    \left.\tilde{G}\left(z_3, z_0\right)\right|_{z_3^{+}>0>z_0^{+}}= & -\frac{(-i)^D}{4 \pi^{D-1}} \int d^d \vec{z}_1\left(-\frac{z_3^{+}}{z_0^{+}}\right)^{\frac{d}{2}} \frac{\left(z_3^{+} \gamma^{-}+\hat{z}_{31 \perp}\right) \gamma^{+}\left(-z_0^{+} \gamma^{-}+\hat{z}_{10 \perp}\right)}{z_0^{+}} \\ & \times U_{\vec{z}_1}^{(2)} \int_0^{+\infty} d \alpha_1\left(\alpha_1\right)^d \exp \left[i \alpha_1\left(-2 z_3^{+} z_{30}^{-}+\vec{z}_{31}^{\; 2}-\frac{z_3^{+}}{z_0^{+}} \vec{z}_{10}^{\; 2}+i 0\right)\right] .
    \end{aligned}
\end{equation}
Integration with respect to the Schwinger parameter now gives
\begin{equation}
    \begin{aligned}
    \left.\tilde{G}\left(z_3, z_0\right)\right|_{z_3^{+}>0>z_0^{+}}= & \frac{i \Gamma(D-1)}{4(2 \pi)^{D-1}} \int d^d \vec{z}_1 \frac{\left(z_3^{+} \gamma^{-}+\hat{z}_{31 \perp}\right) \gamma^{+}\left(-z_0^{+} \gamma^{-}+\hat{z}_{10 \perp}\right)}{\left(-z_3^{+} z_0^{+}\right)^{\frac{D}{2}}} \\ & \times \frac{\theta\left(z_3^{+}\right) \theta\left(-z_0^{+}\right) U_{z_1}^{(2)}}{\left(-z_{30}^{-}+\frac{ \vec{z}_{31}^{\; 2}}{2 z_3^{+}}-\frac{ \vec{z}_{10}^{\; 2}}{2 z_0^{+}}+i 0\right)^{D-1}} .
    \end{aligned}
\end{equation}
We have proved that
\begin{equation}
   \left.G\left(z_3, z_0\right)\right|_{z_3^{+}>0>z_0^{+}}=\left.\tilde{G}\left(z_3, z_0\right)\right|_{z_3^{+}>0>z_0^{+}}. 
\end{equation}
\subsection{The appearance of the Wilson line}
We would like to generalize what we saw previously and resum all possible interactions of the quark line with the external field $b^{\mu}$. A great hint in this direction is given by form of $U_{\vec{z}}^{(2)}$ in Eq.~(\ref{Int:Eq:WilsonLine2Ord}). It is easy to see that it is the $(ig)^2$ term in the $(ig)$-expansion of the Wilson line (in the fundamental representation)
\begin{equation}
\label{Int:Eq:WilsonLine}
    \left[z_3^{+}, z_0^{+}\right]_{\vec{z}}=\mathcal{P} \exp \left[i g \int_{z_0^{+}}^{z_3^{+}} d z^{+} b^{-}\left(z^{+}, \vec{z}\right)\right] \; ,
\end{equation}
where $\mathcal{P}$ means time-ordered product. We want to establish that the generalization we seek is obtained by replacing $U_{\vec{z}}^{(2)}$ with the full Wilson line defined in Eq.~(\ref{Int:Eq:WilsonLine}), \textit{i.e.}
\begin{equation}
\label{Int:Eq:FullPropShock}
    \left.G\left(z_3, z_0\right)\right|_{z_3^{+}>0>z_0^{+}}= \int d^D z_1 \delta\left(z_1^{+}\right) G_0\left(z_{31}\right) \gamma^{+} G_0\left(z_{10}\right)\left[z_3^{+}, z_0^{+}\right]_{\vec{z}_1}.
\end{equation}
We prove this by mathematical induction. We already proved that (\ref{Int:Eq:FullPropShock}) is correct for $n=2$, where $n$ is the number of interactions, and for $n=1$ is trivial. Let's prove the $n=0$ case by setting the Wilson line to the identity in Eq.~(\ref{Int:Eq:ExpressForInductionProof}), we get
\begin{equation}
    \begin{aligned}
    \left.G^{(0)}\left(z_3, z_0\right)\right|_{z_3^{+}>0>z_0^{+}}= & -\frac{(-i)^D}{4 \pi^{D-1}} \int d^d \vec{z}_1\left(-\frac{z_3^{+}}{z_0^{+}}\right)^{\frac{d}{2}} \frac{\left(z_3^{+} \gamma^{-}+\hat{z}_{31 \perp}\right) \gamma^{+}\left(-z_0^{+} \gamma^{-}+\hat{z}_{10 \perp}\right)}{z_0^{+}} \\ & \times \int_0^{+\infty} d \alpha_1\left(\alpha_1\right)^d \exp \left[i \alpha_1\left(-2 z_3^{+} z_{30}^{-}+\vec{z}_{31}^{\; 2}-\frac{z_3^{+}}{z_0^{+}} \vec{z}_{10}^{\; 2}+i 0\right)\right] \; .
    \end{aligned}
\end{equation}
A shift in the variable $\vec{z}_1$ allows us to have a Gaussian integral and hence to obtain
\begin{equation}
    \begin{aligned}
    \left.G^{(0)}\left(z_3, z_0\right)\right|_{z_3^{+}>0>z_0^{+}}= & -\frac{(-i)^D}{4 \pi^{D-1} z_0^{+}}\left(-\frac{z_3^{+}}{z_0^{+}}\right)^{\frac{a}{2}} \int_0^{+\infty} d \alpha_1\left(\alpha_1\right)^{\frac{d}{2}} \\ & {\left[-2 z_3^{+} z_0^{+}\left(\gamma^{-}\right)-2 \frac{z_3^{+} z_0^{+}}{z_{30}^{+}}\left(\hat{z}_{30 \perp}\right)-\frac{z_3^{+} z_0^{+} \vec{z}_{30}^{\; 2}}{\left(z_{30}^{+}\right)^2}\left(\gamma^{+}\right) + i \frac{d}{2} \frac{z_0^{+}}{\alpha_1 z_{30}^{+}}\left(\gamma^{+}\right) \right] } \\ & \times \exp \left[i \alpha_1\left(-2 z_3^{+} z_{30}^{-}+\frac{z_3^{+}}{z_{30}^{+}} \vec{z}_{30}^{\; 2}+i 0\right)\right]\left(\frac{-i \pi z_0^{+}}{z_{30}^{+}}\right)^{\frac{d}{2}} \; .    
    \end{aligned}
\end{equation}
After the last integration we recover the usual free quark propagator in coordinates space:
\begin{equation}
    \begin{aligned}
    \left.G^{(0)}\left(z_3, z_0\right)\right|_{z_3^{+}>0>z_0^{+}} & =\frac{i \Gamma\left(\frac{D}{2}\right)}{2 \pi^{\frac{D}{2}}} \frac{\hat{z}_{30}}{\left(-z_{30}^{\; 2}+i 0\right)^{\frac{D}{2}}} =G_0\left(z_3, z_0\right) .
    \end{aligned}
\end{equation}
To complete our proof, we only need to show that, assuming the result holds for $n$ interactions, the case $n+1$ follows. If for $n$ we have
\begin{equation}
    \begin{aligned}
    \left.G^{(n)}\left(z_3, z_0\right)\right|_{z_3^{+}>0>z_0^{+}}= \int d^D z_1 \delta\left(z_1^{+}\right) G_0\left(z_{31}\right) \gamma^{+} G_0\left(z_{10}\right)\left[z_3^{+}, z_0^{+}\right]_{\vec{z}_1}^{(n)} \; ,
    \end{aligned}
\end{equation}
then
\begin{equation}
    \begin{aligned}
    & \left.G^{(n+1)}\left(z_3, z_0\right)\right|_{z_3^{+}>0>z_0^{+}} = \int d^D z_2 G_0\left(z_{32}\right)\left[i g \gamma^{+} b^{-}\left(z_2\right)\right] G^{(n)}\left(z_2, z_0\right) \\ = & (i g) \int d^D z_2 d^D z_1 b^{-}\left(z_2\right) \delta\left(z_1^{+}\right)\left[G_0\left(z_{32}\right) \gamma^{+} G_0\left(z_{21}\right) \gamma^{+} G_0\left(z_{10}\right)\right] \theta\left(z_2^{+}\right)\left[z_2^{+}, z_0^{+}\right]_{\vec{z}_1}^{(n)} .
    \end{aligned}
\end{equation}
The integration is performed as explained before and we get
\begin{equation}
    \begin{aligned}
    \left.G^{(n+1)}\left(z_3, z_0\right)\right|_{z_3^{+}>0>z_0^{+}}= &  \int d^D z_1 \delta\left(z_1^{+}\right)\left[G_0\left(z_{31}\right) \gamma^{+} G_0\left(z_{10}\right)\right] \\ & \times(i g) \int d z_2^{+} \theta\left(z_{32}^{+}\right) \theta\left(z_2^{+}\right) \theta\left(-z_0^{+}\right) b^{-}\left(z_2^{+}, \vec{z}_1\right)\left[z_2^{+}, z_0^{+}\right]_{\vec{z}_1}^{(n)} \; .
    \end{aligned}
\end{equation}
This last expression can be immediately re-written as
\begin{equation}
    \left.G^{(n+1)}\left(z_3, z_0\right)\right|_{z_3^{+}>0>z_0^{+}}=  \int d^D z_1 \delta\left(z_1^{+}\right)\left[G_0\left(z_{31}\right) \gamma^{+} G_0\left(z_{10}\right)\right] \theta\left(z_3^{+}\right) \theta\left(-z_0^{+}\right)\left[z_3^{+}, z_0^{+}\right]_{\vec{z}_1}^{(n+1)} \; .
\end{equation}
This completes our proof. \\ 
We want to make one last observation. Since the interaction is localized in light-cone time, we can have two situations:
\begin{itemize}
     \item The integration boundaries $z_{0}^{+}$ and $z_3^{+}$ have the same sign. In this case the Wilson line reduces to the identity, the Shockwave is not crossed and we have a normal quark propagator.
     \item The integration boundaries $z_{0}^{+}$ and $z_3^{+}$ possess opposite sign. In this case the Wilson line is nontrivial and the Shockwave is crossed.  It is clear that in this case, only the point $z^{+}=0$ of the entire integration domain contributes and that we can therefore extend the integration from $-\infty$ to $+\infty$. 
\end{itemize}
We therefore define
\begin{equation}
    \begin{aligned} 
    U_{\vec{z}} & \equiv[-\infty,+\infty]_{\vec{z}} =\mathcal{P} \exp \left[i g \int_{-\infty}^{+\infty} d z^{+} b^{-}\left(z^{+}, \vec{z}\right)\right] \; ,
    \end{aligned}
\end{equation}
and its Fourier transform
\begin{equation}
  U\left(p_{\perp}\right) \equiv \int d^d z_{\perp} e^{i\left(p_{\perp} \cdot z_{\perp}\right)} U_{\vec{z}} \; .
\end{equation}
In defining the Feynman rules, we use these latter definitions. We avoid to report all of them here, they can be found in Ref.~\cite{Boussarie:2016txb}, where the notation is identical to the one used in the part of this thesis devoted to the Shockwave approach.

\section{B-JIMWLK evoultion for the dipole operator}
In this section, we want to study quantum corrections to the previous picture. As already anticipated, we will study the effect of a change in the cut-off $\eta$. We consider the external field at rapidity $\eta_1 = \eta + \Delta \eta$, where $\Delta \eta$ should be intended as a small deviation from $\eta$. We separate it as
\begin{equation}
  b_{\eta + \Delta \eta}^{-} (z) =   b_{\eta}^{-} (z) + b_{ \Delta \eta}^{-} (z) \; ,
\end{equation}
where 
\begin{equation}
\begin{aligned} b_{\Delta \eta}^{-}(z) & \equiv \int \frac{d^D p}{(2 \pi)^D} e^{-i(p \cdot z)} b^{-}(p) \theta\left(e^{\eta+\Delta \eta} p_p^{+}-p^{+}\right) \theta\left(p^{+}-e^\eta p_p^{+}\right) 
\end{aligned}
\end{equation}
contains gluons with $+$-momenta between $e^{\eta} p_p^{+}$ and $e^{\eta+\Delta \eta} p_p^{+}$. Let's consider now the Wilson line at rapidity $\eta_1$ for the propagation between the lightcone time $x^{+}$ and the lightcone time $y^{+}$. From Eq.~(\ref{Int:Eq:WilsonLine}), it can be written as
\begin{equation}
    \left[x^{+}, y^{+}\right]_{ \vec{z} }^{\eta_1}=\sum_N \int_{x^{+}}^{y^{+}} \hspace{-0.3 cm} d z_1^{+}(i g) b_{\eta_1}^{-}\left(z_1^{+}, \vec{z}\right) \int_{x^{+}}^{z_1^{+}} \hspace{-0.3 cm} d z_2^{+}(i g) b_{\eta_1}^{-}\left(z_2^{+}, \vec{z}\right) \ldots \int_{x^{+}}^{z_{N-1}} \hspace{-0.3 cm} d z_N^{+}(i g) b_{\eta_1}^{-}\left(z_N^{+}, \vec{z} \; \right) \; .
\end{equation}
Now, we perform a perturbative expansion in the small parameter $i g \Delta \eta$ to show how the Wilson line at rapidity $\eta$ is modified by the small perturbation. Using that $[z_{i}^{+}, z_{j}^{+}]_{\vec{z}}^{\eta} = 1$ if $z_{i}^{+} z_{j}^{+} > 0$, we find
\begin{equation}
    \begin{aligned} 
    & {\left[x^{+}, y^{+}\right]_{\vec{z}}^{\eta+\Delta \eta}=}  {\left[x^{+}, y^{+}\right]_{\vec{z}}^\eta+(i g) \int_{x^{+}}^y d z_1^{+}\left[x^{+}, z_1^{+}\right]_{\vec{z}}^\eta b_{\Delta \eta}^{-}\left(z_1^{+}, \vec{z}\right)\left[z_1^{+}, y^{+}\right]_{\vec{z}}^\eta } \\ & +(i g)^2 \int_{x^{+}}^{y^{+}} d z_1^{+} d z_2^{+}\left[x^{+}, z_1^{+}\right]_{\vec{z}}^\eta b_{\Delta \eta}\left(z_1^{+}, \vec{z}\right)\left[z_1^{+}, z_2^{+}\right]_{\vec{z}}^\eta b_{\Delta \eta}\left(z_2^{+}, \vec{z}\right)\left[z_2^{+}, y^{+}\right]_{\vec{z}}^\eta \theta\left(z_{21}^{+}\right) .
    \end{aligned}
\label{Int:Eq:ExpFluctEta}
\end{equation}
We study two eikonal lines propagating through the shockwave at rapidity $\eta_1 = \eta + \Delta$ and with a color singlet interaction with the external field. They interact ``classically" with the part of the field generated by $b_{\eta}^{-}$. However, this \textit{classical propagation} is modified by the quantum fluctuations generated by the field $b_{\Delta \eta}^{-}$. These quantum fluctuations lead to the evolution equation. \\
In the spirit of the expansion (\ref{Int:Eq:ExpFluctEta}), taking into account these corrections is equivalent to considering the two lines that propagate in the part of the field generated by $b_{\eta}^{-}$ with the possibility of emitting and reabsorbing a gluon which has a $+$-component of momenta strictly between $e^{\eta} p_p^{+}$ and $e^{\eta+\Delta \eta} p_p^{+}$ (see Figs.~\ref{Int:Fig:RealBJIMWLK}, \ref{Int:Fig:VirtualBJIMWLK}). \\
The propagator of this gluon is built from $b_{\Delta \eta}^{-}$ and reads
\begin{equation}
\label{Int:Eq:GluonPropEta}
    \begin{aligned}
    \left(G_{\Delta \eta}^{\mu \nu}\right)^{a b}= & (-i)^d \int \frac{d^d \vec{z}_3}{(2 \pi)^{d+1}}\left(U_{\vec{z}_3}\right)^{a b} \int_{e^\eta}^{e^{\eta+\Delta \eta}} d p^{+} \frac{\left(p^{+}\right)^{d-1}}{2\left(-z_2^{+} z_1^{+}\right)^{\frac{d}{2}+1}} \\ & \times \exp \left[-i p^{+}\left\{z_{21}^{-}-\frac{\vec{z}_{23}^{\; 2}}{2 z_2^{+}}+\frac{\vec{z}_{31}^{\; 2}}{2 z_1^{+}}-i 0\right\}\right]\left(\vec{z}_{23} \cdot \vec{z}_{31}\right)\left(n_2^\mu n_2^\nu\right) \; ,
    \end{aligned}
\end{equation}
where we want to remark the restriction on the $+$-component. \\

Before proceeding, we want to make a small digression to compare BFKL and Shockwave approach. Obviously, the physical picture is the same, there is a big rapidity gap between two interacting objects, and hence, there is phase space for gluon radiation strongly ordered in rapidity. In both cases, we are interested in resuming the large logarithmic corrections generated by these kinematic conditions. However, it is important to note that the way we want to construct the evolution is slightly different. For BFKL we had considered two lines of quarks strongly separated in rapidity which, precisely because of this separation, exchanged an infinite ladder of gluons. The effective polarization of gluons attached to the upper line  was $n_2^{\mu}$, while the one of gluons attached to the lower line was $n_1^{\mu}$. In the present case, the two incoming lines represent the whole projectile; their longitudinal components are both of order $p_p^{+}$, \textit{i.e.} they are in the same fragmentation region and there is no rapidity gap between them. The separation in rapidity is between this projectile and the gluons produced by the entire $b_{\eta_1}^{\mu}$ gluon field. This way of deriving evolution is closer in spirit to that of the Muller dipole, but, as we shall see, in the correct limit is equivalent to BFKL. \\

Let's move on to the calculation of corrections produced by the field $b_{\Delta \eta}^{-}$; they lead to ten Feynman diagrams. In four of these diagrams, the additional gluon crosses the Shockwave (see Fig.~\ref{Int:Fig:RealBJIMWLK}), and thus, this class of diagrams involves an additional adjoint Wilson line $ \left( U_{\vec{z}_3}^{\eta} \right)^{ab}$. We refer to this part as \textit{double dipole contribution}. In the remaining six, the Shockwave is crossed only by the eikonal lines (see Fig.~\ref{Int:Fig:VirtualBJIMWLK}), we refer to this part as \textit{dipole contribution}. \\ 

\begin{figure}
\begin{picture}(417,300)
\put(27,172){\includegraphics[scale=0.45]{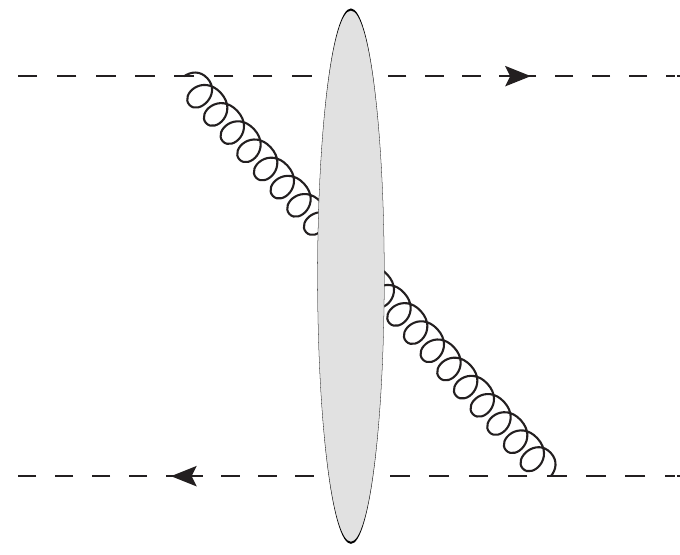}}
\put(95, 157){(a)}
\put(98, 189){$\vec{z}_{2}$}
\put(98, 237){$\vec{z}_{3}$}
\put(98, 270){$\vec{z}_{1}$}
\put(65, 282){$z_{1}^{+}$}
\put(140, 178){$z_{2}^{+}$}
\put(10, 170){$ -\infty^{+}$}
\put(170, 170){$ \infty^{+}$}
\put(270,172){\includegraphics[scale=0.45]{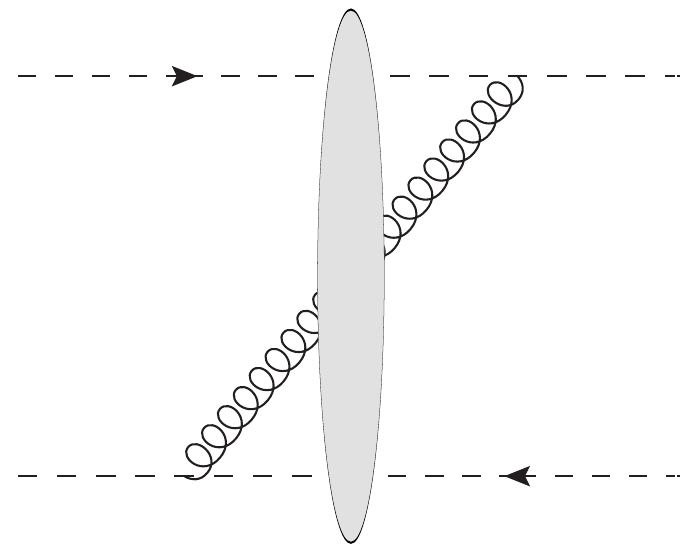}}
\put(338, 157){(b)}
\put(341, 189){$\vec{z}_{2}$}
\put(341, 232){$\vec{z}_{3}$}
\put(341, 270){$\vec{z}_{1}$}
\put(305, 178){$z_{1}^{+}$}
\put(380, 282){$z_{2}^{+}$}
\put(255, 170){$ -\infty^{+}$}
\put(410, 170){$ \infty^{+}$}
\put(27,20){\includegraphics[scale=0.45]{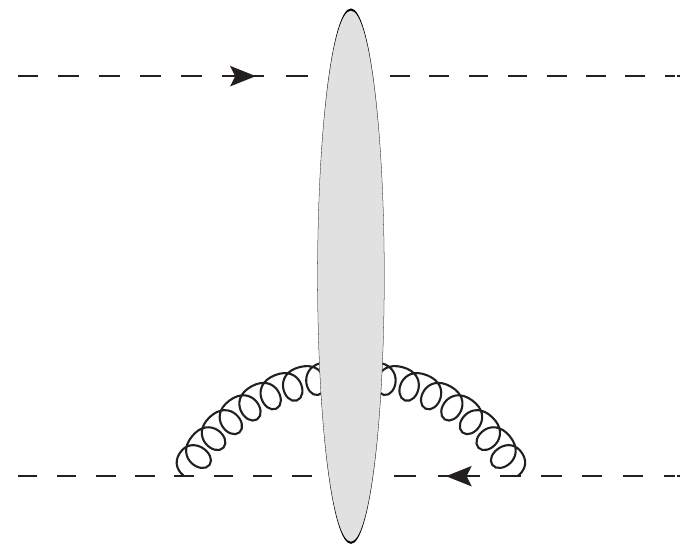}}
\put(95, 05){(c)}
\put(98, 35){$\vec{z}_{2}$}
\put(98, 57){$\vec{z}_{3}$}
\put(98, 120){$\vec{z}_{1}$}
\put(62, 25){$z_{1}^{+}$}
\put(137, 25){$z_{2}^{+}$}
\put(10, 05){$ -\infty^{+}$}
\put(170, 05){$ \infty^{+}$}
\put(270,20){\includegraphics[scale=0.45]{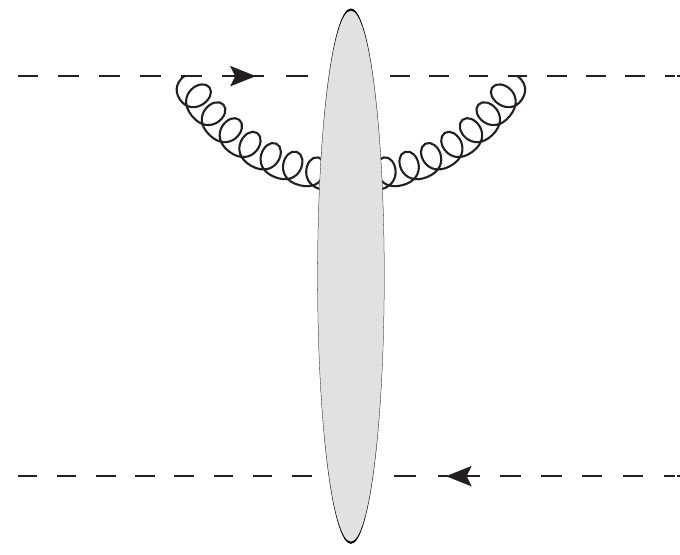}}
\put(338, 05){(d)}
\put(341, 35){$\vec{z}_{2}$}
\put(341, 98){$\vec{z}_{3}$}
\put(341, 120){$\vec{z}_{1}$}
\put(305, 132){$z_{1}^{+}$}
\put(380, 132){$z_{2}^{+}$}
\put(255, 05){$ -\infty^{+}$}
\put(410, 05){$ \infty^{+}$}
\end{picture}
\caption{Double dipole contribution to the evolution. The Shockwave is always crossed at $z_3^{+} = 0$.}
\label{Int:Fig:RealBJIMWLK}
\end{figure}
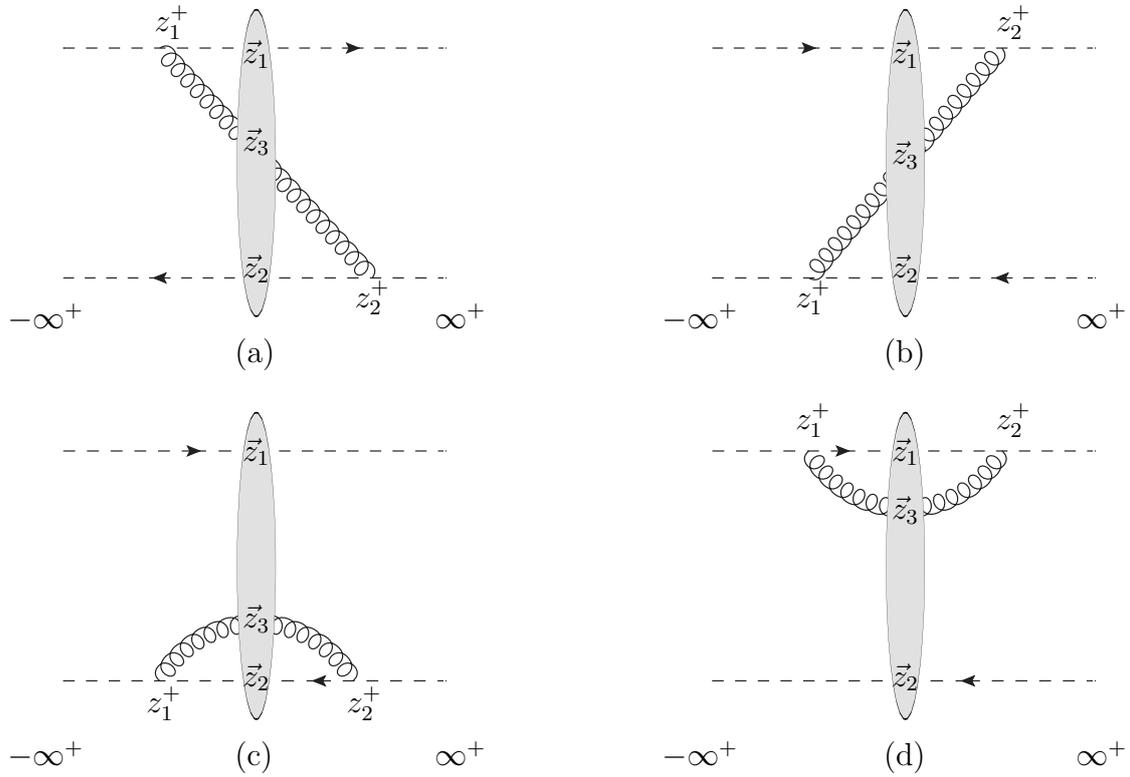
\begin{figure}
\begin{picture}(417,205)
\put(17,102){\includegraphics[scale=0.35]{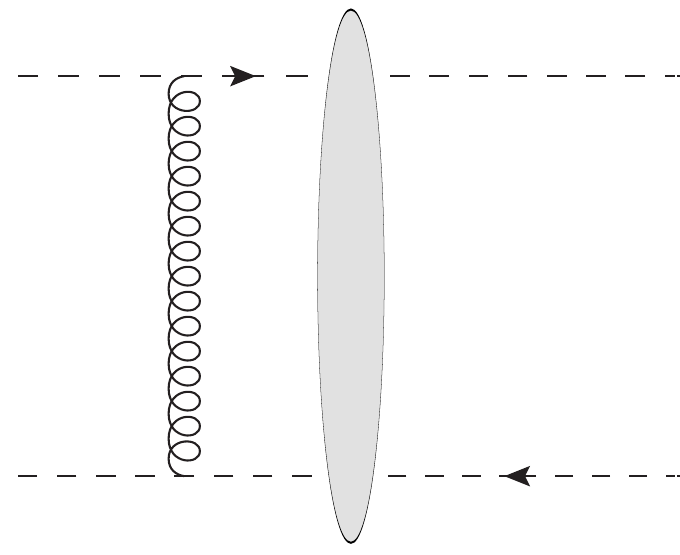}}
\put(167,102){\includegraphics[scale=0.35]{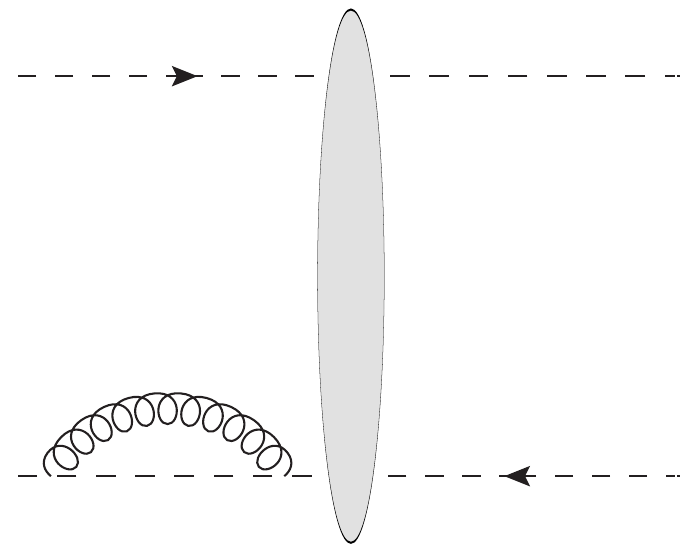}}
\put(317,102){\includegraphics[scale=0.35]{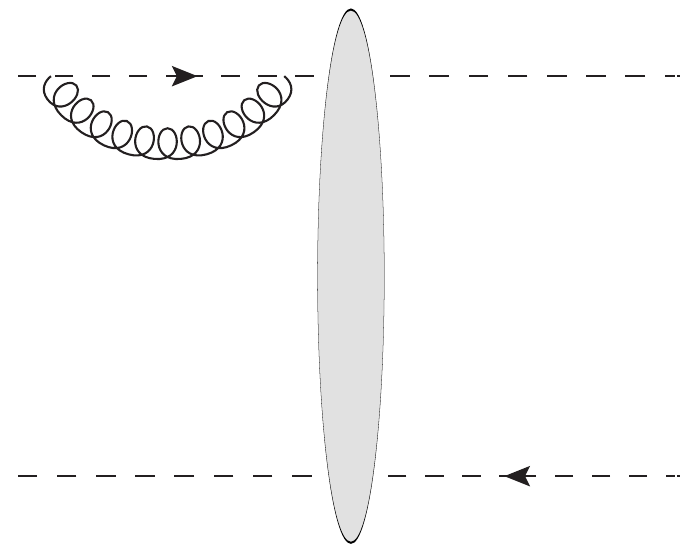}}
\put(17,0){\includegraphics[scale=0.35]{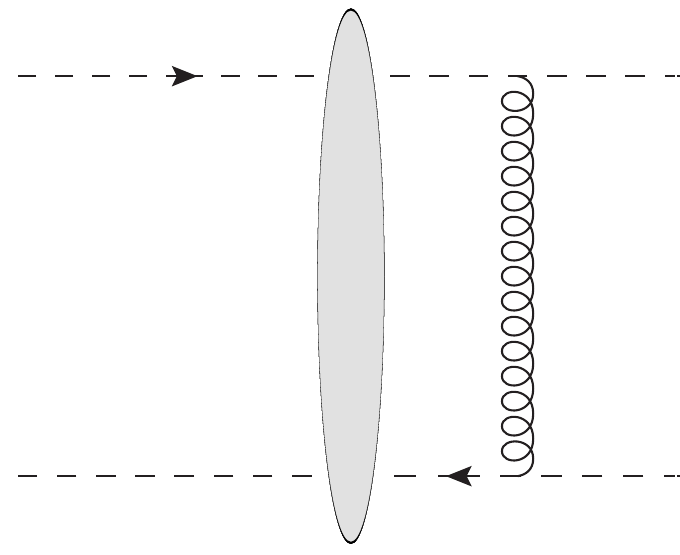}}
\put(167,0){\includegraphics[scale=0.35]{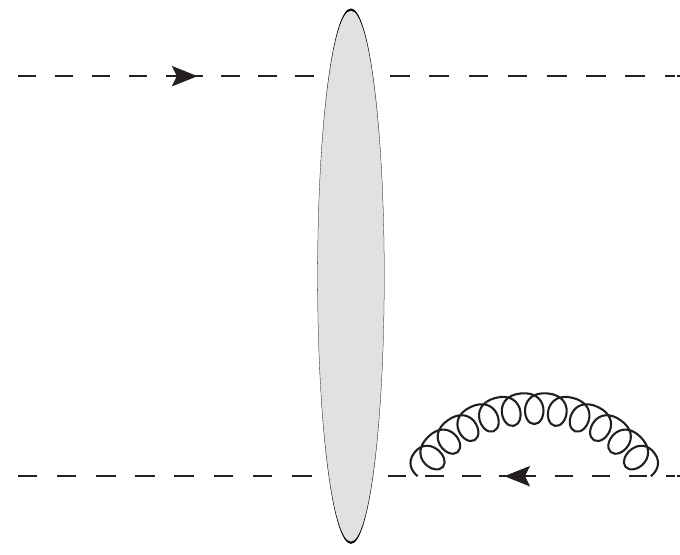}}
\put(317,0){\includegraphics[scale=0.35]{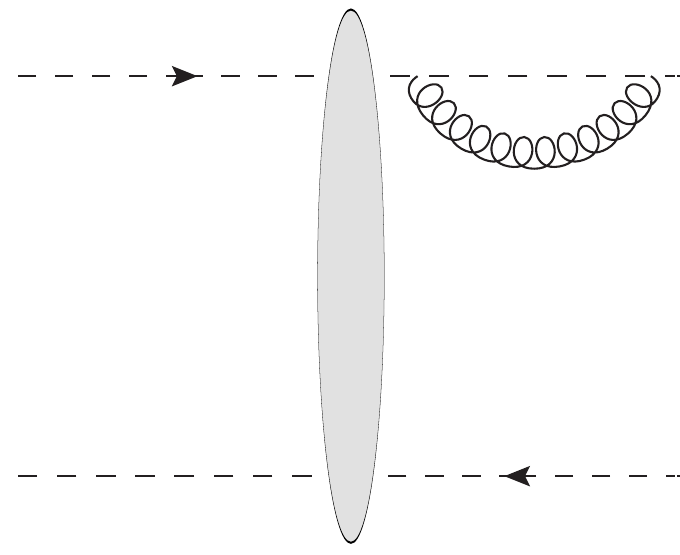}}
\end{picture}
\caption{Dipole contribution to the evolution. The Shockwave is not crossed and hence only single dipole cntributions are generated.}
\label{Int:Fig:VirtualBJIMWLK}
\end{figure}
We start by considering the diagram (a) in Fig.~\ref{Int:Fig:RealBJIMWLK}, it reads\footnote{$\operatorname{Tr}$ means trace with respect to the fundamental representation and it is needed in order to have a singlet exchange.}
\begin{equation*}
\begin{aligned} 
& I_{(a)} = (i g)(-i g) \mu^{-2 \epsilon} \int_{-\infty}^0 d z_1^{+} \int_0^{+\infty} d z_2^{+} \\ & \times \operatorname{Tr}\left(\left[+\infty, z_1^{+}\right]_{\vec{z}_1}^\eta t^b\left[z_1^{+},-\infty\right]_{\vec{z}_1}^\eta\left[-\infty, z_2^{+}\right]_{\vec{z}_2}^\eta t^a\left[z_2^{+},+\infty\right]_{\vec{z}_2}^\eta\right)\left(G_{\Delta \eta}\right)^{a b} 
\end{aligned}
\end{equation*}
\begin{equation}
\begin{aligned}
 & = \frac{(-i)^d g^2 \mu^{-2 \epsilon}}{2(2 \pi)^{d+1}} \operatorname{Tr}\left(U_{\vec{z}_1}^\eta t^b U_{\vec{z}_2}^{\eta \dagger} t^a\right) \int d^d \vec{z}_3\left(U_{\vec{z}_3}^\eta\right)^{a b} \int_{e^\eta}^{e^{\eta+\Delta \eta}} d p^{+}\left(p^{+}\right)^{d-1} \\ & \times \int_{-\infty}^0 \frac{d z_1^{+}}{\left(-z_1^{+}\right)^{\frac{d}{2}+1}} \int_0^{+\infty} \frac{d z_2^{+}}{\left(z_2^{+}\right)^{\frac{d}{2}+1}} \exp \left[-i p^{+}\left\{z_{21}^{-}-\frac{\vec{z}_{23}^{\; 2}}{2 z_2^{+}}+\frac{\vec{z}_{31}^{\; 2}}{2 z_1^{+}}-i 0\right\}\right]\left(\vec{z}_{23} \cdot \vec{z}_{31}\right) \\ = & \frac{g^2 \mu^{-2 \epsilon} 2^d\left[\Gamma\left(\frac{d}{2}\right)\right]^2}{2(2 \pi)^{d+1}}  \int  d^d \vec{z}_3 \frac{\left(\vec{z}_{23} \cdot \vec{z}_{31}\right)}{\left(\vec{z}_{23}^{\; 2}\right)^{\frac{d}{2}}\left(\vec{z}_{31}^{\; 2}\right)^{\frac{d}{2}}}\left[\operatorname{Tr}\left(U_{\vec{z}_1}^\eta t^b U_{\vec{z}_2}^{\eta \dagger} t^a\right)\left(U_{\vec{z}_3}^\eta\right)^{a b}\right] \int_{e^\eta}^{e^{n+\Delta_\eta}} \hspace{-0.6 cm}d p^{+} \frac{e^{-i p^{+}\left(z_{21}^{-}-i 0\right)}}{p^{+}} \; .
\end{aligned}
\end{equation}
This implies 
\begin{equation}
    I_{(a)} \sim \frac{g^2 \mu^{-2 \epsilon} 2^d\left[\Gamma\left(\frac{d}{2}\right)\right]^2}{2(2 \pi)^{d+1}} \int d^d \vec{z}_3 \frac{\left(\vec{z}_{23} \cdot \vec{z}_{31}\right)}{\left(\vec{z}_{23}^{\; 2}\right)^{\frac{d}{2}}\left(\vec{z}_{31}^{\; 2}\right)^{\frac{d}{2}}}\left[\operatorname{Tr}\left(U_{\vec{z}_1}^\eta t^b U_{\vec{z}_2}^{\eta \dagger} t^a\right)\left(U_{\vec{z}_3}^\eta\right)^{a b}\right](\Delta \eta) \; .
\end{equation}
The diagram (c) in Fig.~\ref{Int:Fig:RealBJIMWLK} is computed similarly and gives
\begin{equation}
   I_{(c)} \sim \frac{g^2 \mu^{-2 \epsilon}\left[\Gamma\left(\frac{d}{2}\right)\right]^2 2^d}{2(2 \pi)^{d+1}} \int d^d \vec{z}_3 \operatorname{Tr}\left(t^a U_{\vec{z}_2}^{\eta \dagger} t^b U_{\vec{z}_1}^\eta\right)\left(U_{\vec{z}_3}^\eta\right)^{b a} \frac{1}{\left(\vec{z}_{32}^{\; 2} +i 0\right)^{d-1}}(\Delta \eta) \; .
\end{equation}
Finally, diagram (b) and (d) are obtained from (a) and (c) respectively, by the substitutions $ \vec{z}_1 \leftrightarrow \vec{z}_2 $ and $ig \leftrightarrow -ig$. We have
\begin{equation}
    I_{(b)} \sim \frac{g^2 \mu^{-2 \epsilon} 2^d\left[\Gamma\left(\frac{d}{2}\right)\right]^2}{2(2 \pi)^{d+1}} \int d^d \vec{z}_3 \frac{\left(\vec{z}_{23} \cdot \vec{z}_{31}\right)}{\left(\vec{z}_{23}^{\; 2} \right)^{\frac{d}{2}}\left(\vec{z}_{31}^{\; 2} \right)^{\frac{d}{2}}}\left[\operatorname{Tr}\left(U_{\vec{z}_1}^\eta t^b U_{\vec{z}_2}^{\eta \dagger} t^a\right)\left(U_{\vec{z}_3}^\eta\right)^{a b}\right](\Delta \eta) \; ,
\end{equation}
\begin{equation}
    I_{(d)} \sim \frac{g^2 \mu^{-2 \epsilon}\left[\Gamma\left(\frac{d}{2}\right)\right]^2 2^d}{2(2 \pi)^{d+1}} \int d^d \vec{z}_3 \operatorname{Tr}\left(t^a U_{\vec{z}_2}^{\eta \dagger} t^b U_{\vec{z}_1}^\eta\right)\left(U_{\vec{z}_3}^\eta\right)^{b a} \frac{1}{\left(\vec{z}_{31}^{\; 2}\right)^{d-1}}(\Delta \eta) \; .
\end{equation}
We denote the sum of all these contributions by
\begin{equation}
    \begin{aligned}
     I_R \equiv & \frac{\alpha_s \mu^{-2 \epsilon} \Delta \eta\left[\Gamma\left(\frac{d}{2}\right)\right]^2}{\pi^d} \int d^d \vec{z}_3 \operatorname{Tr}\left(t^a U_{\vec{z}_1}^\eta t^b U_{\vec{z}_2}^{\eta \dagger}\right)\left(U_{\vec{z}_3}^\eta\right)^{a b} \\ & \times\left[\frac{2\left(\vec{z}_{23} \cdot \vec{z}_{31}\right)}{\left(\vec{z}_{23}^{\; 2}\right)^{\frac{d}{2}}\left(\vec{z}_{31}^{\; 2}\right)^{\frac{d}{2}}}+\frac{1}{\left(\vec{z}_{23}^{\; 2} \right)^{d-1}}+\frac{1}{\left(\vec{z}_{31}^{\; 2}\right)^{d-1}}\right] \; .
     \end{aligned}
\end{equation}
Now, we should compute the six diagrams without crossing. Instead of a direct computation, there is an elegant trick that allows to obtain directly the sum of all six contributions. Let's denote the sum of all these contributions as $I_{V}$. First, we observe that if we set all Wilson lines to the identity\footnote{From a physical point of view it clearly means that there is no interaction between the eikonal lines and the target.}, we must have
\begin{equation}
    \left( I_R + I_V \right) \big |_{U_{\vec{z}_{1,2,3}}^{\eta} \rightarrow 1} = 0 \; \implies   I_R \big |_{U_{\vec{z}_{1,2,3}}^{\eta} \rightarrow 1} = - I_V \big |_{U_{\vec{z}_{1,2,3}}^{\eta} \rightarrow 1} \; .
\end{equation}
Indeed, we are computing the linear term in the $\Delta \eta$ expansion of the dipole operator. When every Wilson line is set to one, any dependence on $\eta$ disappears so this term in the expansion in exactly zero. Moreover, it is clear that 
\begin{equation}
    I_V \equiv C_V \operatorname{Tr}\left(U_{\vec{z}_1}^\eta U_{\vec{z}_2}^{\eta \dagger}\right) \implies I_V \big |_{U_{\vec{z}_{1,2,3}}^{\eta} \rightarrow 1} = C_V N_c  \; .
\end{equation}
Since 
\begin{equation}
    I_R \big |_{U_{\vec{z}_{1,2,3}}^{\eta}} = \frac{\alpha_s \mu^{-2 \epsilon} \Delta \eta\left[\Gamma\left(\frac{d}{2}\right)\right]^2}{\pi^d} \frac{N_c-1}{2} \int d^d \vec{z}_3  \left[\frac{2\left(\vec{z}_{23} \cdot \vec{z}_{31}\right)}{\left(\vec{z}_{23}^{\; 2} \right)^{\frac{d}{2}}\left(\vec{z}_{31}^{\; 2} \right)^{\frac{d}{2}}}+\frac{1}{\left(\vec{z}_{23}^{\; 2} \right)^{d-1}}+\frac{1}{\left(\vec{z}_{31}^{\; 2} \right)^{d-1}}\right] \; ,
\end{equation}
we immediately get
\begin{equation}
    C_V = - \frac{\alpha_s \mu^{-2 \epsilon} \Delta \eta\left[\Gamma\left(\frac{d}{2}\right)\right]^2}{\pi^d} \frac{N_c-1}{2 N_c} \int d^d \vec{z}_3  \left[\frac{2\left(\vec{z}_{23} \cdot \vec{z}_{31}\right)}{\left(\vec{z}_{23}^{\; 2} \right)^{\frac{d}{2}}\left(\vec{z}_{31}^{\; 2} \right)^{\frac{d}{2}}}+\frac{1}{\left(\vec{z}_{23}^{\; 2} \right)^{d-1}}+\frac{1}{\left(\vec{z}_{31}^{\; 2} \right)^{d-1}}\right] \; ,
\end{equation}
and hence the full dipole contribution, which reads
\begin{equation*}
    I_V = -\frac{\alpha_s \mu^{-2 \epsilon} \Delta \eta\left[\Gamma\left(\frac{d}{2}\right)\right]^2}{\pi^d} \frac{N_c^2-1}{2 N_c} 
\end{equation*}
\begin{equation}
    \times \int d^d \vec{z}_3\left[\frac{2\left(\vec{z}_{23} \cdot \vec{z}_{31}\right)}{\left(\vec{z}_{23}^{\; 2}\right)^{\frac{d}{2}}\left(\vec{z}_{31}^{\; 2}\right)^{\frac{d}{2}}}+\frac{1}{\left(\vec{z}_{23}^{\; 2} \right)^{d-1}}+\frac{1}{\left(\vec{z}_{31}^{\; 2}\right)^{d-1}}\right] \operatorname{Tr}\left(U_{\vec{z}_1}^\eta U_{\vec{z}_2}^{\eta \dagger}\right) \; .
\end{equation}
Considering together the dipole and double dipole contributions and taking the limit $\Delta \eta \rightarrow 0$, we get
\begin{equation}
\label{BJIMWLKTrace}
\begin{aligned} 
\frac{\partial \operatorname{Tr}\left(U_{\vec{z}_1}^\eta U_{\bar{z}_2}^{\eta \dagger}\right)}{\partial \eta}= & \frac{\alpha_s \mu^{-2 \epsilon}\left[\Gamma\left(\frac{d}{2}\right)\right]^2}{\pi^d} \int d^d \vec{z}_3\left[\operatorname{Tr}\left(U_{\vec{z}_1}^\eta t^b U_{\vec{z}_2}^{\eta \dagger} t^a\right)\left(U_{\vec{z}_3}\right)^{a b}-\frac{N_c^2-1}{2 N_c} \operatorname{Tr}\left(U_{\vec{z}_1}^\eta U_{\vec{z}_2}^{\eta \dagger}\right)\right] \\ & \times\left[\frac{2\left(\vec{z}_{23} \cdot \vec{z}_{31}\right)}{\left(\vec{z}_{23}^{\; 2} \right)^{\frac{d}{2}}\left(\vec{z}_{31}^{\; 2} \right)^{\frac{d}{2}}}+\frac{1}{\left(\vec{z}_{23}^{\; 2}\right)^{d-1}}+\frac{1}{\left(\vec{z}_{31}^{\; 2} \right)^{d-1}}\right] .
\end{aligned}
\end{equation}
We introduce now the dipole operators
\begin{equation}
    \mathcal{U}_{i j}^\eta=\frac{1}{N_c} \operatorname{Tr}\left(U_{\vec{z}_i}^\eta U_{\vec{z}_j}^{\eta \dagger}\right) - 1 \; ,
\end{equation}
Using the relation
\begin{equation}
    \left(U_{\vec{z}_3}^\eta\right)^{a b}=2 \operatorname{Tr}\left(t^a U_{\vec{z}_3}^\eta t^b U_{\vec{z}_3}^{\eta \dagger}\right)
\end{equation}
and the Fierz identity
\begin{equation}
    \left(t_{i j}^a\right)\left(t_{k l}^a\right)=\frac{1}{2} \delta_{i l} \delta_{j k}-\frac{1}{2 N_c} \delta_{i j} \delta_{k l} \; ,
\end{equation}
we can immediately find
\begin{equation}
\begin{aligned} & \operatorname{Tr}\left(U_{\vec{z}_1}^\eta t^b U_{\vec{z}_2}^{\eta \dagger} t^a\right)\left(U_{\vec{z}_3}^\eta\right)^{a b}-\frac{N_c^2-1}{2 N_c} \operatorname{Tr}\left(U_{\vec{z}_1}^\eta U_{\vec{z}_2}^{\eta \dagger}\right) \\ = & 2 \operatorname{Tr}\left(t^a U_{\vec{z}_3}^\eta t^b U_{\vec{z}_3}^{\eta \dagger}\right) \operatorname{Tr}\left(t^a U_{\vec{z}_1}^\eta t^b U_{\bar{z}_2}^{\eta \dagger}\right)-\frac{N_c^2-1}{2 N_c} \operatorname{Tr}\left(U_{\vec{z}_1}^\eta U_{\vec{z}_2}^{\eta \dagger}\right) \\ = & 2\left[\frac{1}{2} \delta_{i n} \delta_{j m}-\frac{1}{2 N_c} \delta_{i j} \delta_{m n}\right]\left[\frac{1}{2} \delta_{k q} \delta_{p l}-\frac{1}{2 N_c} \delta_{k l} \delta_{p q}\right] \\ & \times\left(U_{\vec{z}_3}^\eta\right)_{j k}\left(U_{\vec{z}_3}^{\eta \dagger}\right)_{l i}\left(U_{\vec{z}_1}^\eta\right)_{n p}\left(U_{\vec{z}_2}^{\eta \dagger}\right)_{q m}-\frac{N_c^2-1}{2 N_c} \operatorname{Tr}\left(U_{\vec{z}_1}^\eta U_{\vec{z}_2}^{\eta \dagger}\right) \\ = & \frac{1}{2}\left[\operatorname{Tr}\left(U_{\vec{z}_1}^\eta U_{\vec{z}_3}^{\eta \dagger}\right) \operatorname{Tr}\left(U_{\vec{z}_3}^\eta U_{\vec{z}_2}^{\eta \dagger}\right)-N_c \operatorname{Tr}\left(U_{\vec{z}_1}^\eta U_{\vec{z}_2}^{\eta \dagger}\right)\right] \\ = & \frac{N_c^2}{2}\left[\mathcal{U}_{13}^\eta+\mathcal{U}_{32}^\eta-\mathcal{U}_{12}^\eta+\mathcal{U}_{13}^\eta \mathcal{U}_{32}^\eta\right] .
\end{aligned}
\end{equation}
Therefore, Eq.~(\ref{BJIMWLKTrace}) in terms of the dipole operator takes the form 
\begin{equation}
\label{BJIMWLKDipole}
\begin{aligned}
\frac{\partial \mathcal{U}_{12}^\eta}{\partial \eta}= & \frac{\alpha_s N_c\left[\Gamma\left(\frac{d}{2}\right)\right]^2\left(\mu^2\right)^{1-\frac{d}{2}}}{2 \pi^d} \\ &
\times \int d^d \vec{z}_3\left[\mathcal{U}_{13}^\eta+\mathcal{U}_{32}^\eta-\mathcal{U}_{12}^\eta+\mathcal{U}_{13}^\eta \mathcal{U}_{32}^\eta\right]\left[\frac{2\left(\vec{z}_{23} \cdot \vec{z}_{31}\right)}{\left(\vec{z}_{23}^{\; 2}\right)^{\frac{d}{2}}\left(\vec{z}_{31}^{\; 2} \right)^{\frac{d}{2}}}+\frac{1}{\left(\vec{z}_{23}^{\; 2} \right)^{d-1}}+\frac{1}{\left(\vec{z}_{31}^{\; 2} \right)^{d-1}}\right] \; ,
\end{aligned}
\end{equation}
which is the final evolution equation. Taking $D=4$, one recovers the usual B-JIMWLK equation for the dipole operator
\begin{equation}
    \frac{\partial \mathcal{U}_{12}^\eta}{\partial \eta}=\frac{\alpha_s N_c}{2 \pi^2} \int d^2 \vec{z}_3\left(\frac{\vec{z}_{12}^{\; 2}}{\vec{z}_{23}^{\; 2} \vec{z}_{31}^{\; 2}}\right)\left[\mathcal{U}_{13}^\eta+\mathcal{U}_{32}^\eta-\mathcal{U}_{12}^\eta-\mathcal{U}_{13}^\eta \mathcal{U}_{32}^\eta\right] \; .
\label{Int:Eq:B-JIMWLKD2}
\end{equation}
The operators constructed from the Wilson lines lead us to the physical scattering amplitudes when we consider the matrix elements between two ($in$ and $out$) proton states. In this sense, the evolution equation for the dipole operator involves a double dipole operator.
This second operator, involving four Wilson lines, in general, cannot be described as a product of two dipole operators. For the latter a second evolution equation has to be written. This equation will depend on operators with higher number of Wilson lines. We thus obtain an infinite cascade of equations known as \textit{Balitsky hierarchy}. We can recover the closed non-linear BK equation only in the large-$N_c$ approximation ('t Hooft limit). In this case the double dipole operator becomes a product of dipoles. It is well known that the next-to-leading corrections in the large $N_c$ approximation are of the order $1/N_c^2$ and hence, naively, we would expect corrections to BK dynamics to be of order $ 11 \% $. However, saturation effects tend to play an important role in suppressing these corrections~\cite{Kovchegov:2012mbw}. Finally, we observe that if we completely neglect the non-linear term (dilute approximation) in Eq.~(\ref{Int:Eq:B-JIMWLKD2}) we obtain the colour-dipole version of the BFKL equation~(see \textit{e.g.}~\cite{Ioffe:2010zz}).
\chapter{Inclusive diffractive di-hadron production}
\begin{flushright}
\emph{\textit{Another roof, another proof. \\ Paul Erdős~\cite{Baker:1990tp}}}
\end{flushright}

In this chapter, we consider again an hybrid high-energy/collinear factorization for the description of the semi-diffractive di-hadron production. This time, we rely on the Shockwave formalism introduced in the previous chapter and on results obtained in Ref.~\cite{Boussarie:2016ogo}. The cancellation of divergences is explicitly shown, and the finite parts of the NLO differential cross-sections are found. We work in arbitrary kinematics such that both photoproduction and leptoproduction are considered. The results obtained, are usable to detect saturation effects, at both the future Electron-Ion-Collider (EIC) or already at LHC, using Ultra Peripheral Collisions (UPC). \\

The chapter contains six sections. In the first section, we set-up the general framework and we present the program of computations. In the second, we show counterterms coming from renormalization of FFs. In the third and fourth section we extract divergences from real and virtual corrections, respectively. In the fifth chapter we give all additional finite terms needed to construct the NLO cross section. Lastly, in the sixth section we summarize and discuss possible developments. The material of this chapter is based on Ref.~\cite{Fucilla:2022wcg}.

\section{Theoretical framework}
\label{sec: framework}
\subsection{Hybrid collinear/high-energy factorization}
\begin{figure}
   \centering
    \begin{minipage}{.5\textwidth}
        \centering
        \includegraphics[scale=0.36]{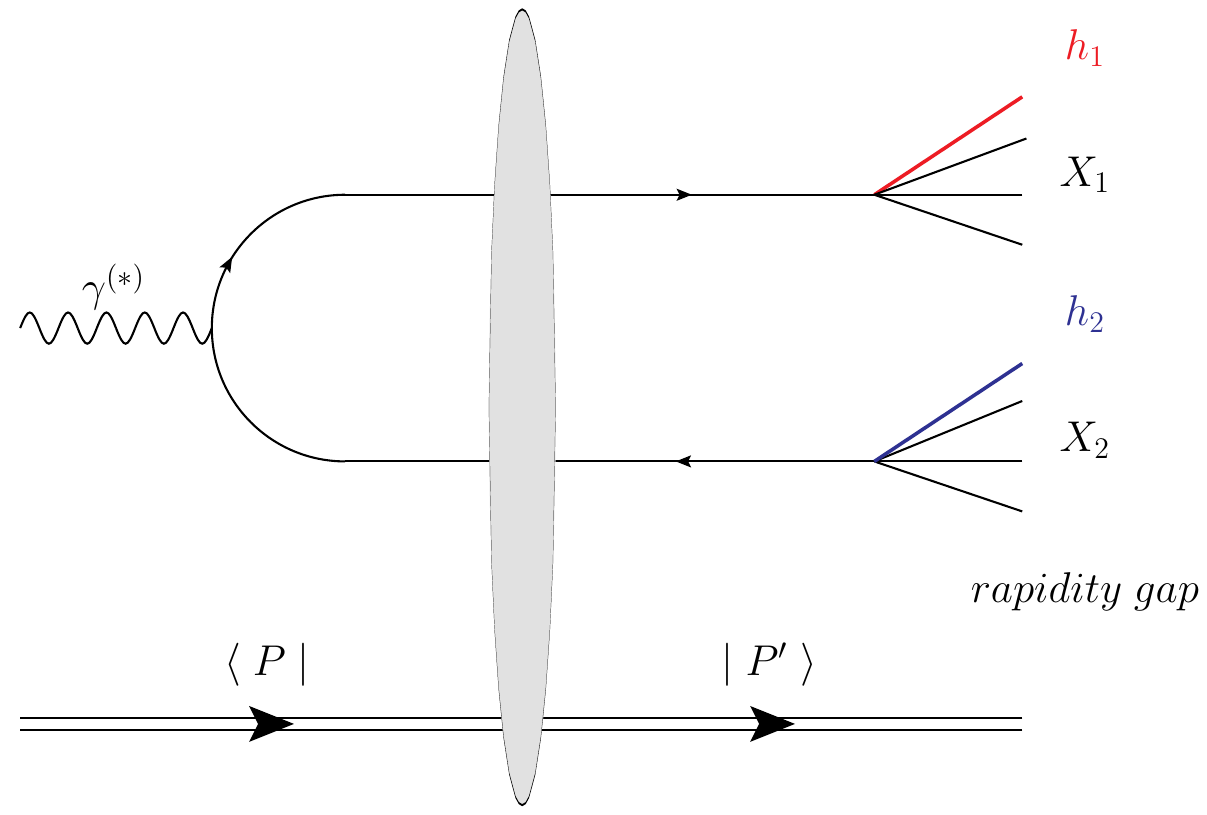}
    \end{minipage}%
    \begin{minipage}{0.5\textwidth}
        \centering
        \includegraphics[scale=0.36]{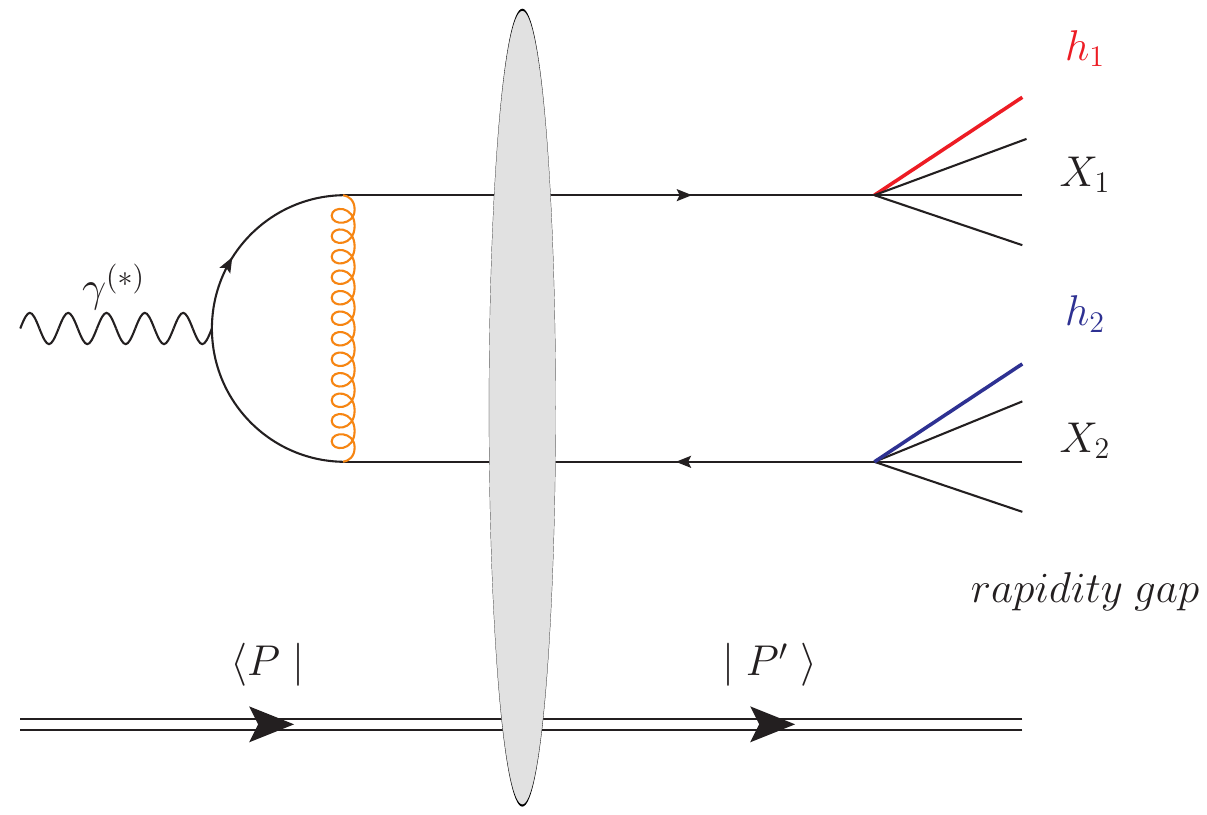}
    \end{minipage}
    \caption{Left: Amplitude of the process (\ref{process}) at LO. Right: An example of diagram contributing to the amplitude of the process (\ref{process}) at NLO. The grey blob symbolizes the QCD shockwave. The double line symbolizes the target, which remains intact in the figure, but could just as well break. The quark and antiquark fragment into the systems $(h_1 X_1)$ and $(h_2 X_2)$. The two tagged hadrons $h_1$ and $h_2$ are drawn in red and blue.}
    \label{fig:process}
\end{figure}

We want to perform a full NLO computation  of the semi-inclusive diffractive di-hadron production in the high-energy limit: 
\begin{equation}
\label{process}
    \gamma^{(*)}(p_\gamma) + P(p_0) \rightarrow h_1(p_{h_1}) + h_2(p_{h_2})  + X + P'(p'_0) \; ,
\end{equation}
where $P$ is a nucleon or a nucleus target, generically called proton in the following. The initial photon plays the role of a probe (also named projectile). Our computation applies both to the photoproduction case (including ultraperipheral collisions) and to the electroproduction case (e.g. at EIC). A gap in rapidity is assumed between the outgoing nucleon/nucleus and the diffractive system $(X h_1 h_2)$. This is illustrated by Fig.~\ref{fig:process}.

We will be working in a combination of collinear factorization and small-$x$ factorization, more precisely in the shockwave formalism for the latter. 

\subsubsection*{Kinematics}
We work in a reference frame such that the target moves ultra-relativistically and such that 
$s = (p_\gamma + p_0)^2 \sim 2 p_\gamma^+ p_0^- \gg \Lambda_{\text{QCD}}^2$, $s$ also being larger than any other scale.  Particles on the projectile side are moving in the $n_1$ (\textit{i.e.} $+$) direction while particles on the target side have a large component along $n_2$ (\textit{i.e.} $-$) direction.

We will use a kinematics such that the photon with virtuality $Q$ is forward, and thus it does not carry any transverse momentum\footnote{Any transverse momentum in Euclidean space will be denoted with an arrow, while a $\perp$ index will be used in Minkowski space.}:
\begin{equation}
\vec{p}_{\gamma}=0,\quad p_{\gamma}^{\mu}=p_{\gamma}^{+}n_{1}^{\mu}+\frac{p_{\gamma}^{2}}{2p_{\gamma}^{+}}n_{2}^{\mu},\quad-p_{\gamma}^{2}\equiv Q^{2}\geq 0. \label{photonk}
\end{equation}
We will denote its transverse polarization $\varepsilon_T$. Its longitudinal polarization vector reads
\begin{equation}
\varepsilon_{L}^{\alpha}=\frac{1}{\sqrt{-p_{\gamma}^{2}}}\left(  p_{\gamma
}^{+}n_{1}^{\alpha}-\frac{p_{\gamma}^{2}}{2p_{\gamma}^{+}}n_{2}^{\alpha
}\right)  ,\quad\varepsilon_{L}^{+}=\frac{p_{\gamma}^{+}}{Q},\quad
\varepsilon_{L}^{-}=\frac{Q}{2p_{\gamma}^{+}}.
\end{equation}
We write the momentum of the produced hadrons as
\beqa
\label{ph}
p^\mu_{h_i}=p^+_{h_i} n_1^\mu + \frac{m_{h_i}^2 + \vec{p}_{h_i}^{\,2}}{2 p^+_{h_i}} n_2^\mu + p^\mu_{h_i\perp}\quad (i=1,2) \,.
\eqa
The momenta of the fragmenting quark of virtuality $p_q^2$ reads
\beqa
\label{pq}
p^\mu_q=p^+_q n_1^\mu + \frac{p_q^2+\vec{p}_{q}^{\,2}}{2 p^+_{q}} n_2^\mu + p^\mu_{q\perp}\,
\eqa
and similarly for an antiquark of virtuality $p_{\bar{q}}^2$ ,
\beqa
\label{pqbar}
p^\mu_{\bar{q}}=p^+_{\bar{q}} n_1^\mu + \frac{p^2_{\bar{q}}+\vec{p}_{\bar{q}}^{\,2}}{2 p^+_{\bar{q}}} n_2^\mu + p^\mu_{\bar{q}\perp}\,.
\eqa
From now, we will use the notation $p_{ij}=p_i-p_j.$

\subsubsection*{Collinear factorization}

The kinematical region considered here is such that $\vec{p}_{h_1}^{\,2} \sim \vec{p}_{h_2}^{\,2} \gg \Lambda_{\text{QCD}}^2$. The hadron momenta are the hard scale, making the use of perturbative QCD and collinear factorization possible. 
The constraint $\vec{p}^{\, 2} \gg \vec{p}_{h_{1,2}}^{\,2} $, with $\vec{p}$ the relative transverse momentum of the two hadrons has also been considered. This means that the two produced hadrons have a large enough separation angle (or, in other words, a large enough invariant mass) so that it will not be necessary to consider the di-hadron unpolarized fragmentation functions: each hadron, typically pion, can be produced by two well-separated fragmentation cascades.
The quark and antiquark in the hard part after collinear factorization will be treated as on-shell particles. For further use, we introduce the longitudinal momentum fraction $x_q$ and $x_{\bar{q}}$ as
\beqa
\label{xq-xqbar}
p_q^+ = x_q p^+_{\gamma} \quad \hbox{ and } \quad p_{\bar{q}}^+ = x_{\bar{q}} p^+_{\gamma}\,.
\eqa
Similarly, we denote
\beqa
\label{xh}
p_{h_i}^+ = x_{h_i} p^+_{\gamma} \,.
\eqa

\subsubsection*{Shockwave approach}
The small-$x$ factorization applies here and the scattering amplitude is the convolution of the projectile impact factor and the non-perturbative matrix element of operators from the Wilson lines operators on the target states. 
One of such operators is the dipole operator, which in the fundamental representation of $SU(N_c)$ takes the form:
\begin{equation}
\left[\operatorname{Tr} \left(U_1 U_2^\dag\right)-N_c\right]\left(\vec{p_1},\vec{p}_2\right) = \int d^d \vec{z}_{1} d^d \vec{z}_{2\perp} e^{- i \vec{p}_1 \cdot \vec{z}_1} e^{- i \vec{p}_2 \cdot \vec{z}_2} \left[\operatorname{Tr} \left(U_{\vec{z}_1} U_{\vec{z}_2}^\dag\right)-N_c\right]\,,
\end{equation}
where
$\vec{z}_{1,2}$ are the transverse positions of the $q,\bar{q}$ coming from the photon and $\vec{p}_{1,2}$ their respective transverse momentums kicks from the shockwave.

The proton matrix element should be parameterized. This can be done  through a generic function $F$, following the definition of Ref.~\cite{Boussarie:2016ogo} 
\begin{eqnarray}
\left\langle P^{\prime}\left(p_{0}^{\prime}\right)\left|T\left(\operatorname{Tr}\left(U_{\frac{z_{\perp}}{2}} U_{-\frac{z_{\perp}}{2}}^{\dagger}\right)-N_{c}\right)\right| P\left(p_{0}\right)\right\rangle
& \equiv & 2 \pi \delta\left(p_{00^{\prime}}^{-}\right) F_{p_{0 \perp} p_{0 \perp}^{\prime}}\left(z_{\perp}\right) \nonumber \\
   & \equiv & 2 \pi \delta\left(p_{00^{\prime}}^{-}\right) F\left(z_{\perp}\right) 
\end{eqnarray}
and its Fourier Transform (FT) is
\begin{equation}
\label{eq: FT F}
\int d^{d} z_{\perp} e^{i\left(z_{\perp} \cdot p_{\perp}\right)} F\left(z_{\perp}\right) \equiv \mathbf{F}\left(p_{\perp}\right).
\end{equation}

Similar definitions exist for the double dipole operator and its action on proton states, as can be seen with eqs.~(5.3) and (5.6) in \cite{Boussarie:2016ogo}, with
\begin{eqnarray}
    && \left\langle P^{\prime}\left(p_0^{\prime}\right)\left|\left(\operatorname{Tr}\left(U_{\frac{z}{2}} U_x^{\dagger}\right) \operatorname{Tr}\left(U_x U_{-\frac{z}{2}}^{\dagger}\right)-N_c \operatorname{Tr}\left(U_{\frac{z}{2}} U_{-\frac{z}{2}}^{\dagger}\right)\right)\right| P\left(p_0\right)\right\rangle  \nonumber \\
     &&\equiv 2 \pi \delta\left(p_{00^{\prime}}^{-}\right) \tilde{F}_{p_{0 \perp} p_{0 \perp}^{\prime}}\left(z_{\perp}, x_{\perp}\right) \equiv 2 \pi \delta\left(p_{00^{\prime}}^{-}\right) \tilde{F}\left(z_{\perp}, x_{\perp}\right)
\end{eqnarray}
and its FT is
\begin{equation}
\int d^d z_{\perp} d^d x_{\perp} e^{i\left(p_{\perp} \cdot x_{\perp}\right)+i\left(z_{\perp} \cdot q_{\perp}\right)} \tilde{F}\left(z_{\perp}, x_{\perp}\right) \equiv \tilde{\mathbf{F}}\left(q_{\perp}, p_{\perp}\right).
\end{equation}

In this chapter, dimensional regularization will be used with $D=2 + d$, where $d= 2+2\epsilon$ is the transverse dimension.

\subsection{LO order}
\begin{figure}
\begin{picture}(430,70)
\put(100,1){\includegraphics[scale=0.35]{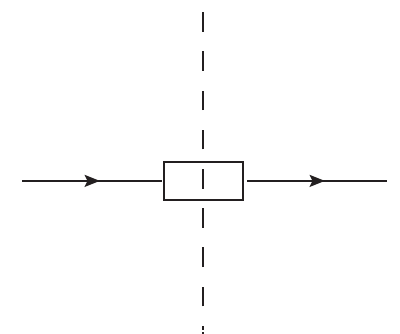}}
\put(182,24){$\equiv$}
\put(200,0){\includegraphics[scale=0.35]{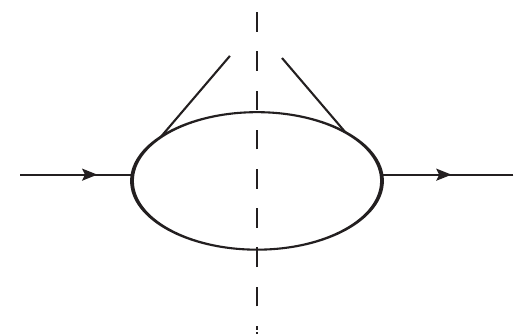}}
\put(225,45){$h$}
\end{picture} 
 \caption{Graphical convention for the fragmentation function of a parton (here a quark for illustration) to a hadron $h$ plus spectators. In the rest of this article, we will use the left-hand side of this drawing.}
    \label{fig:fragmentation}
\end{figure}
QCD collinear factorization stipulates that the total cross section, at leading twist and LO, reads, see ref~\cite{Collins:2011zzd} (chap.~12) and Ref.~\cite{Altarelli:1979kv}  
\begin{equation}
\label{eq: coll facto}
\frac{d \sigma_{0JI}^{h_1 h_2}}{d x_{h_1} d x_{h_2}}= \sum_{q} \int_{x_{h_1}}^1 \frac{d x_q}{x_q} \int_{x_{h_2}}^1 \frac{d x_{\bar{q}}}{x_{\bar{q}}} D_q^{h_1}\left(\frac{x_{h_1}}{x_q},\mu_F\right) D_{\bar{q}}^{h_2}\left(\frac{x_{h_2}}{x_{\bar{q}}}, \mu_F\right) \frac{d\hat{\sigma}_{JI}}{d x_q d x_{\bar{q}}} + (h_1 \leftrightarrow h_2) \; ,
\end{equation}
where $q$ specifies the quark flavor types ($q=u,d,s,c,b$), and $J,I=L,T$ specify the photon polarization since we deal here with a modulus square amplitude ($J$ labels the photon polarization in the complex conjugated amplitude and $I$ in the amplitude). Here  $x_q$ and $x_{\bar{q}}$ are the longitudinal fractions of the photon momentum carried by the fragmenting partons, $x_ {h_1,h_2}$ are the longitudinal fraction of the photon momentum carried by the produced hadrons, $\mu_F$ is the factorization scale, $D_{q(\bar{q})}^{h}$ denotes the quark (antiquark) Fragmentation Function (FF) and $d\hat{\sigma}$ is the partonic cross section, i.e. the cross section for the subprocess 
\begin{equation}
\label{partonic_LO}
     \gamma^{(*)}(p_\gamma) + P(p_0) \rightarrow q (p_q) + \bar{q}(p_{\bar{q}})  + P'(p'_0)\,.
\end{equation}
The graphical convention used in the present article for any fragmentation function is given in Fig.~\ref{fig:fragmentation}.

All detailed computations will be done considering only the first term in \eqref{eq: coll facto}, remembering the second term is just simply obtained by the replacement $h_1 \leftrightarrow h_2$. 

\begin{figure}
\centering
 \includegraphics[scale=0.35]{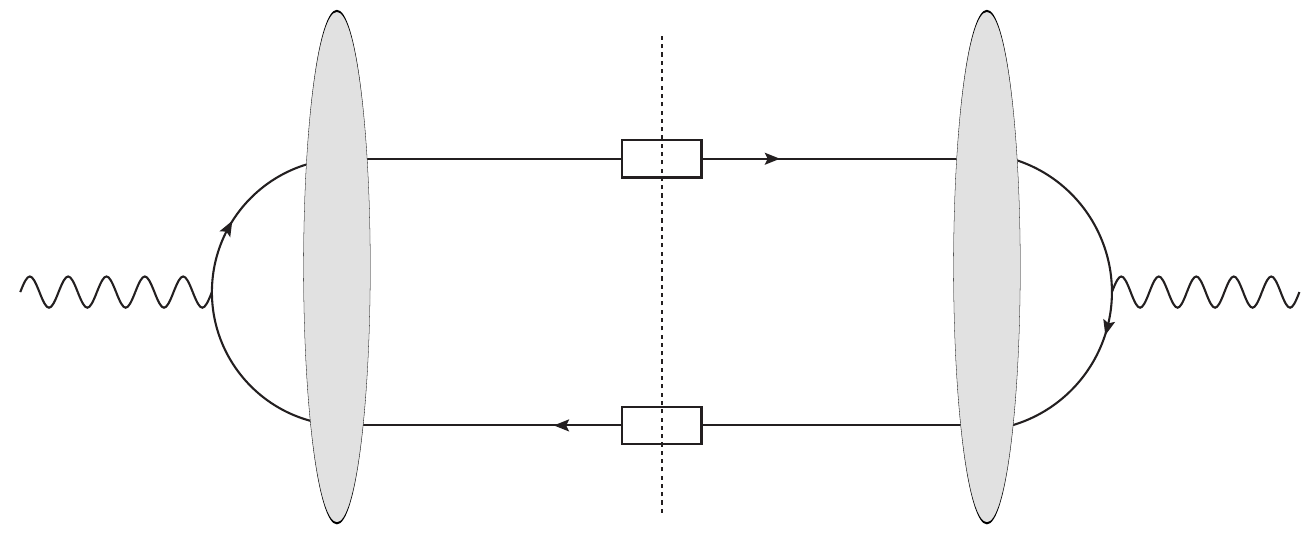}
 \caption{Diagram of the LO process at cross section level. The blob is the shockwave (we do not draw the coupling with the target for clarity) and the squares the FFs, see Fig.~\ref{fig:fragmentation}. The dashed line is to represent the integration over phase space.}
    \label{fig:LO}
\end{figure}

The partonic cross section (\ref{partonic_LO}) has been computed in the shockwave framework. The structure of the result for the whole process \ref{process} at LO is illustrated in Fig.~\ref{fig:LO}.

Collinear factorization means that the produced hadrons should fly collinearly to the fragmenting partons. This means here that the following constraints should be fulfilled 
\beqa
\label{constraint-collinear-q}
p^+_q &=& \frac{x_q}{x_{h_1}} p^+_{h_1},    \quad \vec{p}_q = \frac{x_q}{x_{h_1}} \vec{p}_{h_1}\,,
\\
\label{constraint-collinear-qbar}
p^+_{\bar{q}} &=& \frac{x_{\bar{q}}}{x_{h_2}} p^+_{h_2}, \quad \vec{p}_{\bar{q}}= \frac{x_{\bar{q}}}{x_{h_2}}\vec{p}_{h_2}\,.
\eqa

Since the photon in the initial state can appear with different polarizations, we construct the density matrix 
\begin{equation}
\label{eq:density_matrix}
d\sigma_{JI}=
\begin{pmatrix}
d\sigma_{LL} & d\sigma_{LT}\\
d\sigma_{TL} & d\sigma_{TT}
\end{pmatrix}
,\qquad d\sigma_{TL}=d\sigma_{LT}^{\ast}.
\end{equation}
Each element of this matrix has a LO contribution $d\sigma_{0}$.
This Born order result, see Eq.~(5.14) of Ref.~\cite{Boussarie:2016ogo}, has the following structure:
\begin{align}  
\label{dsigma0}
d\sigma_{0JI}  & =  \frac{\alpha_{\mathrm{em}}Q_{q}^{2}}{\left(2\pi\right)^{4\left(d-1\right)}N_{c}}\frac{\left(p_{0}^{-}\right)^{2}}{2x_q x_{\bar{q}} s^{2}}d x_q d x_{\bar{q}} d^{d}p_{q\perp}d^{d}p_{\bar{q}\perp}\delta\left(1-x_q-x_{\bar{q}} \right)\left(\varepsilon_{I\beta}\varepsilon_{J\gamma}^\ast\right)\nonumber \\
& \quad  \times  \int d^{d}p_{1\perp}d^{d}p_{2\perp}d^{d}p_{1^{\prime}\perp}d^{d}p_{2^{\prime}\perp}\delta\left(p_{q1\perp}+p_{\bar{q}2\perp}\right)\delta\left(p_{11^{\prime}\perp}+p_{22^{\prime}\perp}\right)\nonumber \\
& \quad  \times  \sum_{\lambda_q,\lambda_{\bar{q}}}\Phi_{0}^{\beta}\left(p_{1\perp},\, p_{2\perp}\right)\Phi_{0}^{\gamma*}\left(p_{1^{\prime}\perp},\, p_{2^{\prime}\perp}\right)\mathbf{F}\left(\frac{p_{12\perp}}{2}\right)\mathbf{F^{*}}\left(\frac{p_{1^{\prime}2^{\prime}\perp}}{2}\right) .
\end{align}
Using the explicit expressions of the product $\Phi_{0}^{\beta}\Phi_{0}^{\gamma*}$, see Eq.~(5.18-20) of Ref.~\cite{Boussarie:2016ogo}, as well as  Eq.~\eqref{eq: coll facto}, 
the LO cross sections are obtained and read for $LL$

\begin{eqnarray}
\label{eq:LL-LO}
&&  \hspace{-1cm}  \frac{d \sigma_{0 L L}^{h_1 h_2}}{d x_{h_1}d x_{h_2}d^d p_{h_1 \perp}d^d p_{h_2 \perp}}  = \frac{4 \alpha_{\mathrm{em}} Q^2 }{(2\pi)^{4(d-1)}N_c} \sum_{q} \int_{x_{h_1}}^1 d x_q \int_{x_{h_2}}^1 d x_{\bar{q}} \;  x_q x_{\bar{q}} \left(\frac{x_q}{x_{h_1}}\right)^d \left(\frac{x_{\bar{q}}}{x_{h_2}}\right)^d \nonumber  \\
& \times &  \delta (1-x_q - x_{\bar{q}})  Q_q^2 D_q^{h_1}\left(\frac{x_{h_1}}{x_q} \right) D_{\bar{q}}^{h_2}\left(\frac{x_{h_2}}{x_{\bar{q}}}\right)  {\cal F}_{LL}  + (h_1 \leftrightarrow h_2)\,,
\end{eqnarray}
where
\begin{equation}
\label{F-LL}
{\cal F}_{LL} =  \left|\int d^{d} p_{2 \perp} \frac{\mathbf{F}\left(\frac{x_q}{2x_{h_1}}  p_{h_1\perp} + \frac{x_{\bar{q}} }{2 x_{h_2}}p_{h_2\perp} -p_{2\perp}\right)}{\left(\frac{x_{\bar{q}}}{x_{h_2}}\vec{p}_{h_2}- \vec{p}_{2}\right)^{2}+x_q x_{\bar{q}} Q^{2}} \right|^{2}   \,.
\end{equation}

This LO cross section can be written differently, using transverse momentum conservation, see Eq.~\eqref{dsigma0},
\begin{eqnarray}
\label{eq:LL-LO-minus}
&&  \hspace{-1cm}  \frac{d \sigma_{0 L L}^{h_1 h_2}}{d x_{h_1}d x_{h_2}d^d p_{h_1 \perp}d^d p_{h_2 \perp}}  = \frac{4 \alpha_{\mathrm{em}} Q^2 }{(2\pi)^{4(d-1)}N_c} \sum_{q} \int_{x_{h_1}}^1 d x_q \int_{x_{h_2}}^1 d x_{\bar{q}} \; x_q x_{\bar{q}} \left(\frac{x_q}{x_{h_1}}\right)^d \left(\frac{x_{\bar{q}}}{x_{h_2}}\right)^d \nonumber  \\
& \times &  \delta (1-x_q - x_{\bar{q}})  Q_q^2 D_q^{h_1}\left(\frac{x_{h_1}}{x_q} \right) D_{\bar{q}}^{h_2}\left(\frac{x_{h_2}}{x_{\bar{q}}}\right)   \tilde{\cal F}_{LL}  + (h_1 \leftrightarrow h_2)\,,
\end{eqnarray}
where
\begin{equation}
\label{Ftilde-LL}
{\cal \tilde{F}}_{LL} =  \left|\int d^{d} p_{1 \perp} \frac{\mathbf{F}\left(-\left(\frac{x_q}{2x_{h_1}}  p_{h_1\perp} + \frac{x_{\bar{q}} }{2 x_{h_2}}p_{h_2\perp} -p_{1\perp}\right)\right)}{\left(\frac{x_q}{x_{h_1}}\vec{p}_{h_1}- \vec{p}_{1}\right)^{2}+x_q x_{\bar{q}} Q^{2}} \right|^{2}   \,.
\end{equation}

Both forms can be used interchangeably in terms of NLO cross sections that are proportional to the LO cross sections, \textit{i.e.} when dealing with the soft, virtual, and counter-term contribution to the NLO cross sections. For the collinear quark-gluon contribution, Eq.~\eqref{eq:LL-LO} will be used, while, for the collinear anti-quark-gluon contribution, Eq.~\eqref{eq:LL-LO-minus} will be used. 

Similarly, the non-diagonal interference term $TL$ can be written in the two equivalent forms:
\begin{eqnarray}
\label{eq:TL-LO}
&&  \hspace{-1cm}  \frac{d \sigma_{0 T L}^{h_1 h_2}}{d x_{h_1}d x_{h_2}d^d p_{h_1 \perp}d^d p_{h_2 \perp}}  = \frac{2 \alpha_{\mathrm{em}} Q }{(2\pi)^{4(d-1)}N_c}  \sum_{q} \int_{x_{h_1}}^1 d x_q \int_{x_{h_2}}^1 d x_{\bar{q}}  \left(\frac{x_q}{x_{h_1}}\right)^d \left(\frac{x_{\bar{q}}}{x_{h_2}}\right)^d \nonumber  \\
    & \times &  (x_{\bar{q}}-x_q) \delta (1-x_q - x_{\bar{q}}) Q_q^2 D_q^{h_1}\left(\frac{x_{h_1}}{x_q}\right) D_{\bar{q}}^{h_2}\left(\frac{x_{h_2}}{x_{\bar{q}}}\right)  {\cal F}_{TL}  + (h_1 \leftrightarrow h_2)\,,
\end{eqnarray}
where
\begin{equation}
\begin{aligned}
{\cal F}_{TL} & =  \int d^d p_{2\perp} \frac{\mathbf{F}\left( \frac{x_q}{2 x_{h_1}} p_{h_1 \perp} + \frac{x_{\bar{q}}}{2 x_{h_2}} p_{h_2 \perp} - p_{2 \perp}
\right)}{ \left( \frac{x_{\bar{q}}}{x_{h_2}} \vec{p}_{h_2} -\vec{p}_2  \right)^2 + x_q x_{\bar{q}} Q^2} \\
&  \times \left[\int d^d p_{2'\perp} \frac{\mathbf{F}\left( \frac{x_q}{2 x_{h_1}} p_{h_1 \perp} + \frac{x_{\bar{q}}}{2 x_{h_2}} p_{h_2 \perp} - p_{2' \perp}
\right)}{ \left( \frac{x_{\bar{q}}}{x_{h_2}} \vec{p}_{h_2} -\vec{p}_{2'}  \right)^2 + x_q x_{\bar{q}} Q^2} \left( \frac{x_{\bar{q}}}{x_{h_2}} \vec{p}_{h_2}  - \vec{p}_{2'}\right) \cdot \vec{\varepsilon}_T  \right]^*
\end{aligned}
\end{equation}
and
\begin{eqnarray}
\label{eq:TL-LO-minus}
&&  \hspace{-1cm}  \frac{d \sigma_{0 T L}^{h_1 h_2}}{d x_{h_1}d x_{h_2}d^d p_{h_1 \perp}d^d p_{h_2 \perp}}  = \frac{2 \alpha_{\mathrm{em}} Q }{(2\pi)^{4(d-1)}N_c} \sum_{q} \int_{x_{h_1}}^1 d x_q \int_{x_{h_2}}^1 d x_{\bar{q}}  \left(\frac{x_q}{x_{h_1}}\right)^d \left(\frac{x_{\bar{q}}}{x_{h_2}}\right)^d \nonumber  \\
    & \times & (x_{\bar{q}}-x_q)  \delta (1-x_q - x_{\bar{q}})  Q_q^2 D_q^{h_1}\left(\frac{x_{h_1}}{x_q}\right) D_{\bar{q}}^{h_2}\left(\frac{x_{h_2}}{x_{\bar{q}}}\right)  {\cal \tilde{F}}_{TL}  + (h_1 \leftrightarrow h_2)\,,
\end{eqnarray}
where
\begin{equation}
\begin{aligned}
{\cal \tilde{F}}_{TL} & =  \int d^d p_{1\perp} \frac{\mathbf{F}\left( - \left(\frac{x_q}{2 x_{h_1}} p_{h_1 \perp} + \frac{x_{\bar{q}}}{2 x_{h_2}} p_{h_2 \perp} - p_{1 \perp}
\right)\right)}{ \left( \frac{x_q}{x_{h_1}} \vec{p}_{h_1} -\vec{p}_1  \right)^2 + x_q x_{\bar{q}} Q^2} \\
&  \times \left[\int d^d p_{1'\perp} \frac{\mathbf{F} \left( - \left(\frac{x_q}{2 x_{h_1}} p_{h_1 \perp} + \frac{x_{\bar{q}}}{2 x_{h_2}} p_{h_2 \perp} - p_{1' \perp}
\right) \right)}{ \left( \frac{x_q}{x_{h_1}}  \vec{p}_{h_1} -\vec{p}_{1'}  \right)^2 + x_q x_{\bar{q}} Q^2} \left( \frac{x_q}{x_{h_1}} p_{h_1 }  - p_{1'}\right) \cdot \varepsilon_{T}  \right]^*.
\end{aligned}
\end{equation}
The most complicated contribution $TT$ reads
\begin{eqnarray}
\label{eq:TT-LO}
&&  \hspace{-1cm}  \frac{d \sigma_{0 T T}^{h_1 h_2}}{d x_{h_1}d x_{h_2}d^d p_{h_1 \perp}d^d p_{h_2 \perp}}  = \frac{\alpha_{\mathrm{em}} }{(2\pi)^{4(d-1)}N_c} \sum_{q} \int_{x_{h_1}}^1 \frac{d x_q}{x_q} \int_{x_{h_2}}^1 \frac{d x_{\bar{q}} }{x_{\bar{q}}}  \left(\frac{x_q}{x_{h_1}}\right)^d \left(\frac{x_{\bar{q}}}{x_{h_2}}\right)^d \nonumber  \\
    & \times &  \delta (1-x_q - x_{\bar{q}})  Q_q^2 D_q^{h_1}\left(\frac{x_{h_1}}{x_q}\right) D_{\bar{q}}^{h_2}\left(\frac{x_{h_2}}{x_{\bar{q}}}\right)  {\cal F}_{TT}  + (h_1 \leftrightarrow h_2)\,,
\end{eqnarray}
where
\begin{equation}
\begin{aligned}
{\cal F}_{TT} 
& =   \left[ (x_{\bar{q}} -x_q )^2 g_{\perp}^{ri} g_\perp^{lk} - g_\perp^{rk} g_\perp^{li} + g_{\perp}^{rl} g_\perp^{ik} \right] \\
& \times  \int d^d p_{2\perp} \frac{\mathbf{F}\left( \frac{x_q}{2 x_{h_1}} p_{h_1 \perp} + \frac{x_{\bar{q}}}{2 x_{h_2}} p_{h_2 \perp} - p_{2 \perp}
\right)}{ \left( \frac{x_{\bar{q}}}{x_{h_2}} \vec{p}_{h_2} -\vec{p}_2  \right)^2 + x_q x_{\bar{q}} Q^2} \left( \frac{x_{\bar{q}}}{x_{h_2}} p_{h_2 }  - p_{2}\right)_{r}  \varepsilon_{T i}  \\
&  \times  \left[\int d^d p_{2'\perp} \frac{\mathbf{F}\left( \frac{x_q}{2 x_{h_1}} p_{h_1 \perp} + \frac{x_{\bar{q}}}{2 x_{h_2}} p_{h_2 \perp} - p_{2' \perp}
\right)}{ \left( \frac{x_{\bar{q}}}{x_{h_2}} \vec{p}_{h_2} -\vec{p}_{2'}  \right)^2 + x_q x_{\bar{q}} Q^2} \left( \frac{x_{\bar{q}}}{x_{h_2}} p_{h_2}  - p_{2'}\right)_l  \varepsilon_{T k} \right]^*
\end{aligned}
\end{equation}
or equivalently
\begin{eqnarray}
\label{eq:TT-LO-minus}
&&  \hspace{-1cm}  \frac{d \sigma_{0 T T}^{h_1 h_2}}{d x_{h_1}d x_{h_2}d^d p_{h_1 \perp}d^d p_{h_2 \perp}}  = \frac{\alpha_{\mathrm{em}} }{(2\pi)^{4(d-1)}N_c} \sum_{q} \int_{x_{h_1}}^1 \frac{d x_q}{x_q} \int_{x_{h_2}}^1 \frac{d x_{\bar{q}} }{x_{\bar{q}}}  \left(\frac{x_q}{x_{h_1}}\right)^d \left(\frac{x_{\bar{q}}}{x_{h_2}}\right)^d \nonumber  \\
    & \times &  \delta (1-x_q - x_{\bar{q}})  Q_q^2 D_q^{h_1}\left(\frac{x_{h_1}}{x_q}\right) D_{\bar{q}}^{h_2}\left(\frac{x_{h_2}}{x_{\bar{q}}}\right)  {\cal \tilde{F}}_{TT}  + (h_1 \leftrightarrow h_2)\,,
\end{eqnarray}
where 
\begin{equation}
\begin{aligned}
{\cal \tilde{F}}_{TT} 
& =   \left[ (x_{\bar{q}} -x_q )^2 g_{\perp}^{ri} g_\perp^{lk} - g_\perp^{rk} g_\perp^{li} + g_{\perp}^{rl} g_\perp^{ik} \right] \\
& \times  \int d^d p_{1\perp} \frac{\mathbf{F}\left( -\left( \frac{x_q}{2 x_{h_1}} p_{h_1 \perp} + \frac{x_{\bar{q}}}{2 x_{h_2}} p_{h_2 \perp} - p_{1 \perp}
\right) \right)}{ \left( \frac{x_q}{x_{h_1}} \vec{p}_{h_1} -\vec{p}_1 \right)^2 + x_q x_{\bar{q}} Q^2} \left( \frac{x_q}{x_{h_1}} p_{h_1 }  - p_{1}\right)_{r}  \varepsilon_{T i}  \\
&  \times  \left[\int d^d p_{1'\perp} \frac{\mathbf{F}\left( - \left(\frac{x_q}{2 x_{h_1}} p_{h_1 \perp} + \frac{x_{\bar{q}}}{2 x_{h_2}} p_{h_2 \perp} - p_{1' \perp}
\right) \right)}{ \left( \frac{x_q}{x_{h_1}} \vec{p}_{h_1} -\vec{p}_{1'}  \right)^2 + x_q x_{\bar{q}} Q^2} \left( \frac{x_q}{x_{h_1}} p_{h_1}  - p_{1'}\right)_l \varepsilon_{T k} \right]^*.
\end{aligned}
\end{equation}

Compared to the $LL$ cross section, the $TL$ cross section has the same form up to a factor of 
$$\frac{1}{Q} \frac{x_{\bar{q}}-x_q}{2 x_{\bar{q}} x_q } \left( \frac{x_{\bar{q}}}{x_{h_2}} \vec{p}_{h_2}  - \vec{p}_{2'}\right) \cdot \vec{\varepsilon}^{\,*}_T $$
or $$ \frac{1}{Q} \frac{x_{\bar{q}}-x_q}{2 x_{\bar{q}} x_q }  \left( \frac{x_q}{x_{h_1}} p_{h_1}  - p_{1'}\right) \cdot \varepsilon_{T}^{\,*} \,.$$
The $TT$ cross section differs from the $LL$ cross section by a factor of 
$$\frac{1}{Q^2} \frac{1}{4 x_q^2 x_{\bar{q}}^2 } \left[ (x_{\bar{q}} -x_q )^2 g_{\perp}^{ri} g_\perp^{lk} - g_\perp^{rk} g_\perp^{li} + g_{\perp}^{rl} g_\perp^{ik} \right] \left( \frac{x_q}{x_{h_1}} p_{h_1 }  - p_{1}\right)_{r}  \varepsilon_{T i}  \left( \frac{x_q}{x_{h_1}} p_{h_1}  - p_{1'}\right)_l \varepsilon_{T k}^{\,*} $$
or 
$$
\frac{1}{Q^2} \frac{1}{4 x_q^2 x_{\bar{q}}^2 } \left[ (x_{\bar{q}} -x_q )^2 g_{\perp}^{ri} g_\perp^{lk} - g_\perp^{rk} g_\perp^{li} + g_{\perp}^{rl} g_\perp^{ik} \right]
\left( \frac{x_{\bar{q}}}{x_{h_2}} p_{h_2 }  - p_{2}\right)_{r}  \varepsilon_{T i} \left( \frac{x_{\bar{q}}}{x_{h_2}} p_{h_2}  - p_{2'}\right)_l  \varepsilon_{T k}^{\,*} \,.
$$

The factors of $1/Q$ and $1/Q^2$ come from the photon polarization while the other modifications come from the expression of the squared of the impact factors.   
Those modifications and additional factors between $TL$ and $TT$ cross section with respect to $LL$ will remain true when going to NLO, for what concerns the extraction of divergences. This means that no additional detailed calculations are needed for those cases.  

\subsection{NLO computations in a nutshell}

\subsubsection*{Different types of contributions in the dipole picture}

\begin{figure}
\begin{picture}(430,470)

\put(180,470){\fbox{virtual contributions}}

\put(10,400){\includegraphics[scale=0.25]{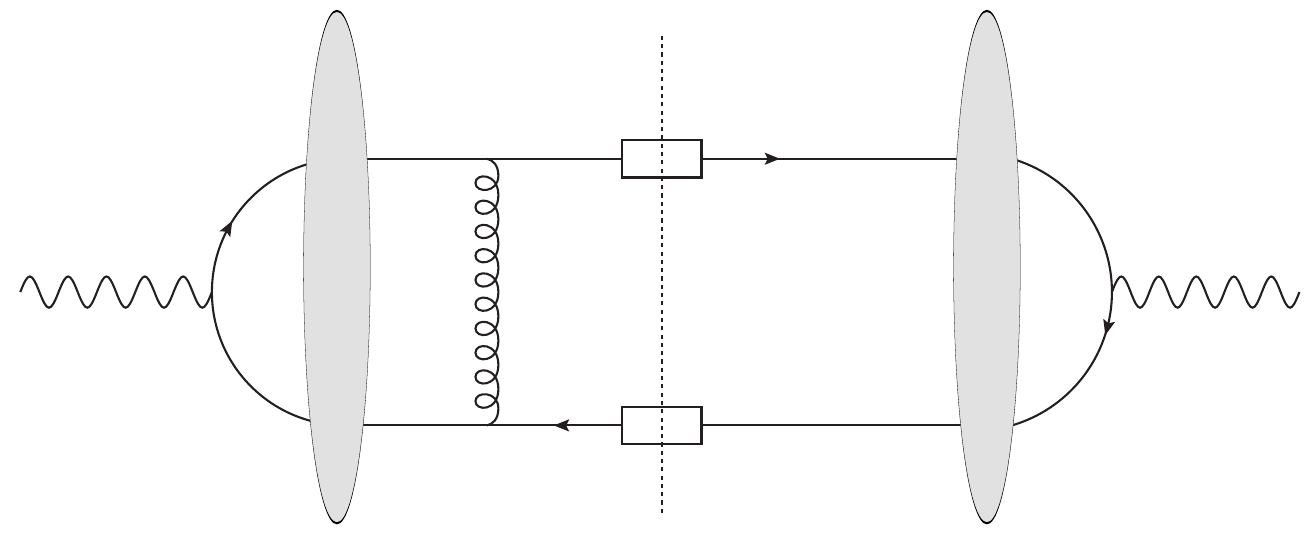}}
\put(180,425){$\Arrow{1cm}$}
\put(230,425){$d\sigma_{1IJ}$}
\put(280,425){dipole $\times$ dipole}

\put(10,300){\includegraphics[scale=0.25]{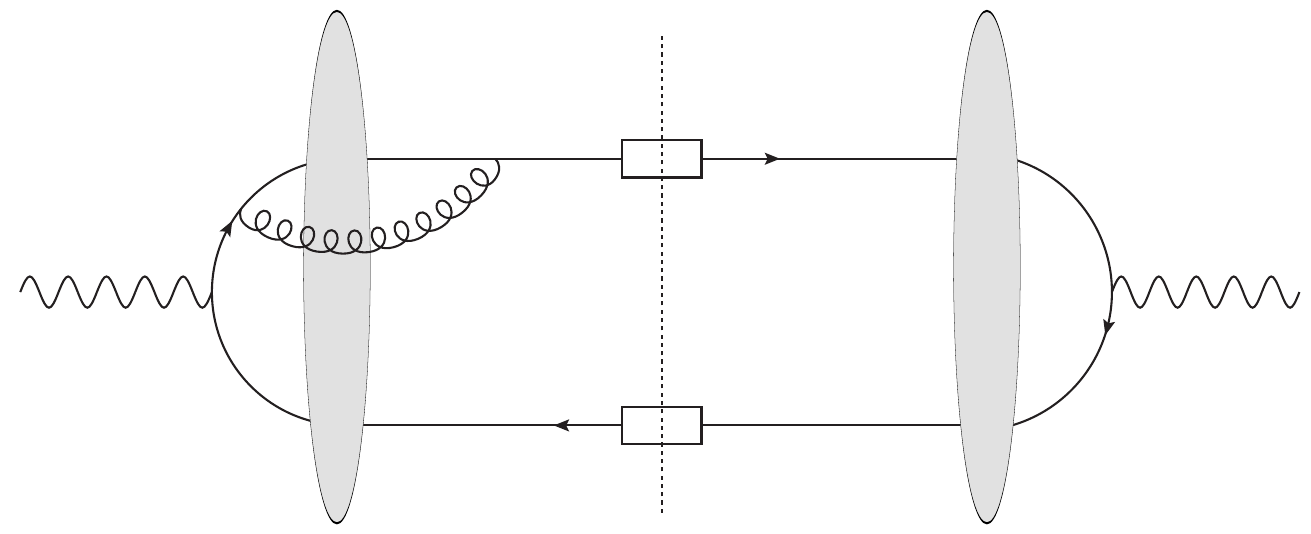}}
\put(180,350){\rotatebox{58}{$\Arrow{1.7cm}$}}
\put(180,325){$\Arrow{1cm}$}
\put(230,325){$d\sigma_{2IJ}$}
\put(280,325){double dipole $\times$ dipole}

\put(180,270){\fbox{real contributions}}

\put(10,200){\includegraphics[scale=0.25]{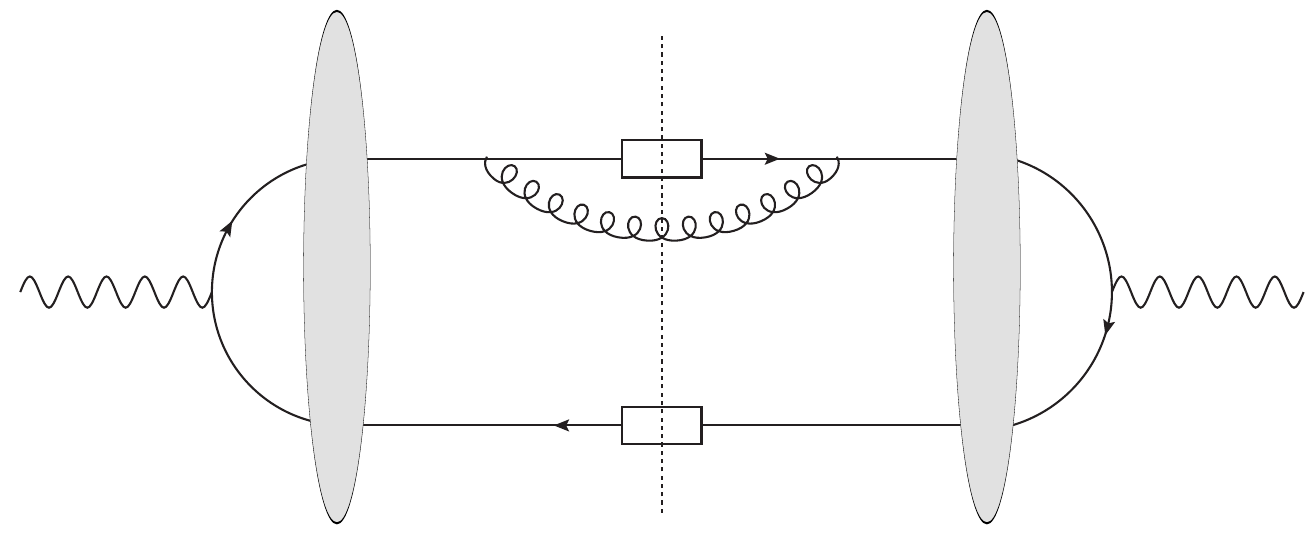}}
\put(180,225){$\Arrow{1cm}$}
\put(230,225){$d\sigma_{3IJ}$}
\put(280,225){dipole $\times$ dipole}

\put(10,100){\includegraphics[scale=0.25]{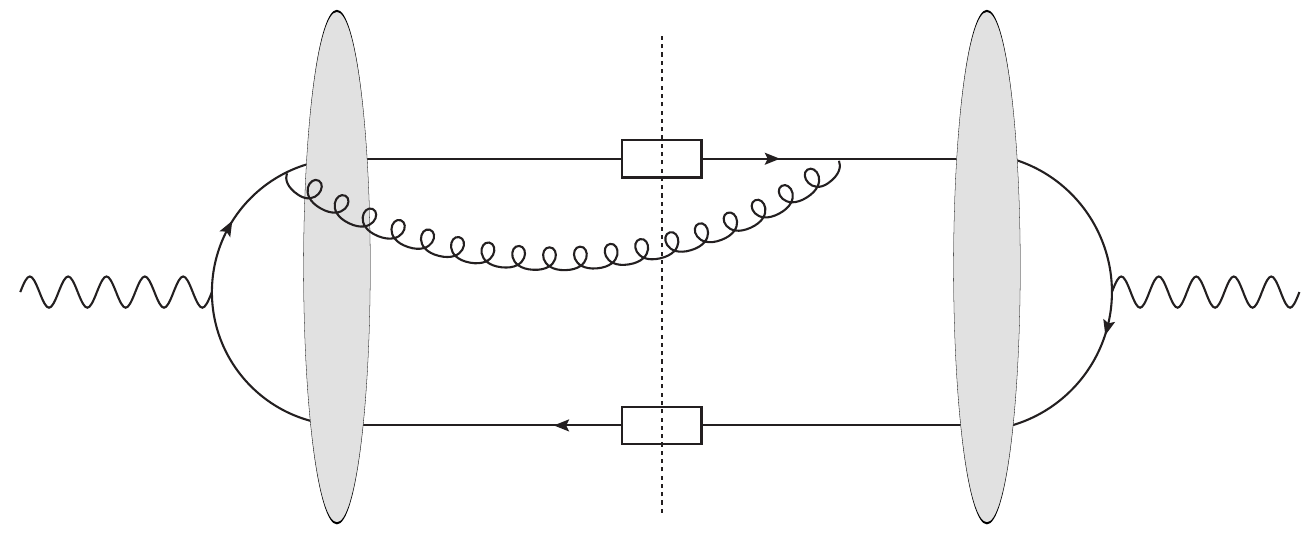}}
\put(180,125){$\Arrow{1cm}$}
\put(180,150){\rotatebox{58}{$\Arrow{1.7cm}$}}
\put(230,125){$d\sigma_{4IJ}$}
\put(280,125){double dipole $\times$ dipole}

\put(10,0){\includegraphics[scale=0.25]{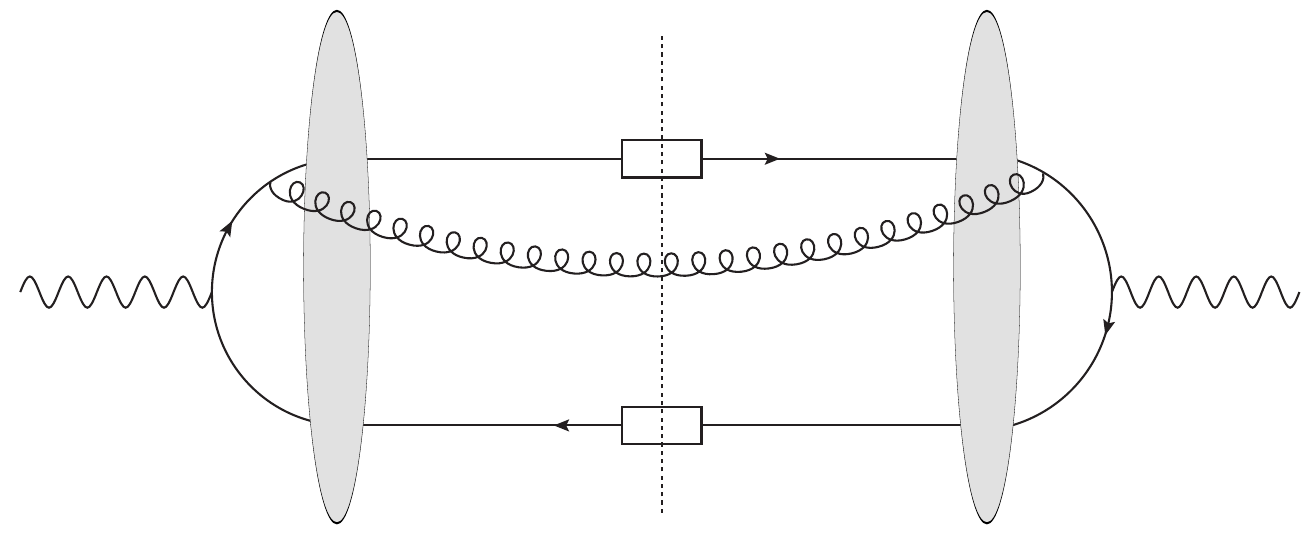}}
\put(180,25){$\Arrow{1cm}$}
\put(180,50){\rotatebox{58}{$\Arrow{1.7cm}$}}
\put(180,70){\rotatebox{75}{$\Arrow{4cm}$}}
\put(230,25){$d\sigma_{5IJ}$}
\put(280,25){double dipole $\times$ double dipole}

\end{picture}
\vspace{.1cm}
\caption{Illustration of the five kinds of  contributions to the NLO cross section from the dipole point of view. Arrows show to which combination of dipole structures each type of diagrams contributes.}
  \label{fig:sigma-NLO-dipole}
\end{figure}
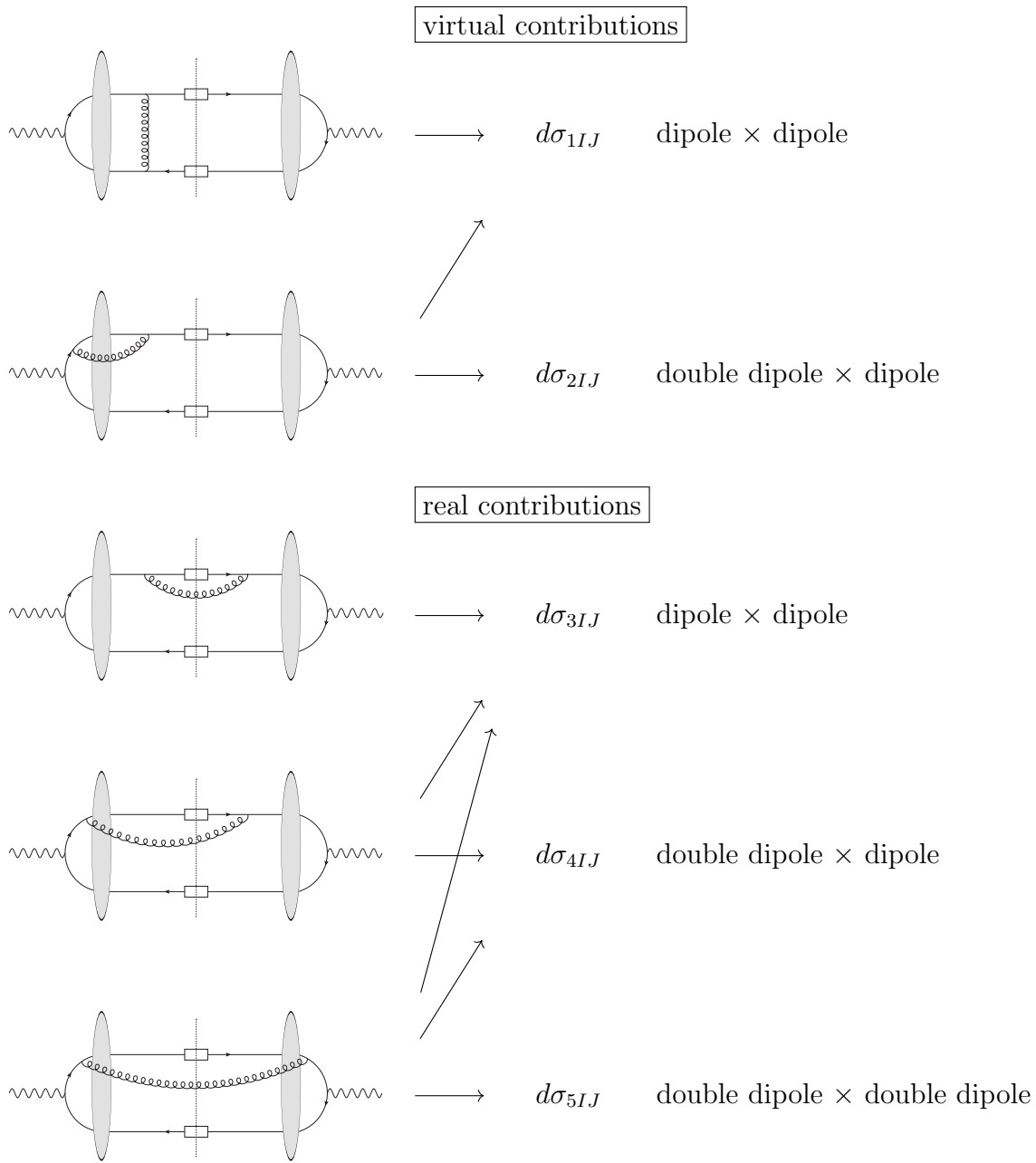

At NLO, since we rely on the shockwave approach, it is convenient to separate the various contributions from the dipole point of view, as illustrated in Fig.~\ref{fig:sigma-NLO-dipole}. In this figure, we exhibit a few examples of diagrams, either virtual or real, as a representative of each five classes of diagrams. There are indeed five classes of contributions from the dipole point of view, namely $d \sigma_{iJI}\, (i=1,\cdots 5)$, so that the NLO density matrix can be written as
\begin{equation}
d\sigma_{JI}=d\sigma_{0JI}+d\sigma_{1JI}+d\sigma_{2JI}+d\sigma_{3JI}+d\sigma_{4JI}+d\sigma_{5JI}. 
\label{sigmaNLO}
\end{equation}
Now, we will shortly discuss each of these five NLO corrections.

For the virtual diagrams, there are two classes of diagrams: the diagrams in which the virtual gluon does not cross the shockwave, thus contributing to $d \sigma_{1IJ}$, purely made of dipole $\times$ dipole terms; the diagrams in which the virtual gluon does cross the shockwave, contributing both to $d \sigma_{1IJ}$, made of dipole $\times$ dipole terms, as well as to $d \sigma_{2IJ}$, made of double dipole $\times$ dipole (and dipole $\times$ double-dipole) terms.

For the real diagrams, there are three classes of diagrams: the diagrams in which the real gluon does not cross the shockwave, thus contributing to $d \sigma_{3IJ}$, purely made of dipole $\times$ dipole terms; the diagrams in which the real gluon crosses  exactly once the shockwave, contributing both to $d \sigma_{3IJ}$, made of dipole $\times$ dipole terms as well as to $d \sigma_{4IJ}$, made of double dipole $\times$ dipole  (and dipole $\times$ double-dipole) terms; the diagrams in which the real gluon crosses  exactly twice the shockwave, contributing to $d \sigma_{3IJ}$, made of dipole $\times$ dipole terms, to $d \sigma_{4IJ}$, made of double dipole $\times$ dipole  (and dipole $\times$ double-dipole) terms, and to $d \sigma_{5IJ}$, made of double dipole $\times$ double dipole terms.

\subsubsection*{Overview of cancellation of divergences}
Before providing technical details, let us sketch the way the computation will be done, putting emphasis on the infrared (IR) sector.

When generically decomposing any on-shell parton momentum in the Sudakov basis as\footnote{Here $p^+$ is a large fixed momentum, eg $p_\gamma^+$ in our present case.}
\begin{equation}
\label{p-sudakov}
p^\mu = z p^+ n_1^\mu + \frac{\vec{p}^{\,2}}{2 z p^+} n_2^\mu + p_\perp^\mu\,,
\end{equation} 
in the IR sector, we face three kinds of divergences:
\begin{itemize}
    \item Rapidity:
$z$ goes to zero and $p_{\perp}$ arbitrary.
    \item Soft:
any component of the gluon momentum goes linearly to zero (obtained with both $z$ and $p_\perp = z \tilde{p}_\perp \sim z$ going to zero).
    \item Collinear:
parton's $p_\perp$ goes to zero, $z$ being arbitrary.
\end{itemize} 

Technically, since the $z$ integration is regulated through a lower cut-off (named $\alpha$), one should be careful with the fact that the appearance of 
$\ln \alpha$ may have originated from both rapidity or soft divergences.  

The calculation goes as follows. First, the rapidity divergences, appearing only in the virtual corrections in the present computation, are taken care of at the amplitude level by absorbing them in the shockwave through one step of B-JIMWLK evolution. This removes part of terms with $\ln \alpha$ related to pure rapidity divergences.

Next, at the level of cross section, we separate the soft divergent contribution from the non-soft divergent terms. Combining real and virtual contributions, these soft divergent terms will disappear as guaranteed by the Kinoshita-Lee-Nauenberg theorem.

Finally, the remaining type of divergences, which are of purely collinear nature, will be absorbed into the fragmentation functions through one step of the Dokshitzer-Gribov-Lipatov-Altarelli-Parisi (DGLAP) evolution equation~\cite{Gribov:1972ri, Lipatov:1974qm, Altarelli:1977zs, Dokshitzer:1977sg}.

\subsubsection*{Different fragmentation contributions to the NLO cross section}

\def\sca{.27}
\begin{figure}[htbp]
\begin{picture}(430,400)
\put(10,300){\includegraphics[scale=\sca]{FF_dihadron_LO_box.pdf}}
\put(90,370){NLO}
\put(188,328){$=$}
\put(202,300){\includegraphics[scale=\sca]{FF_dihadron_LO_box.pdf}}
\put(250,370){1-loop}
\put(380,328){+ c.c}
\put(282,275){(a)}
\put(0,200){+}
\put(10,170){\includegraphics[scale=\sca]{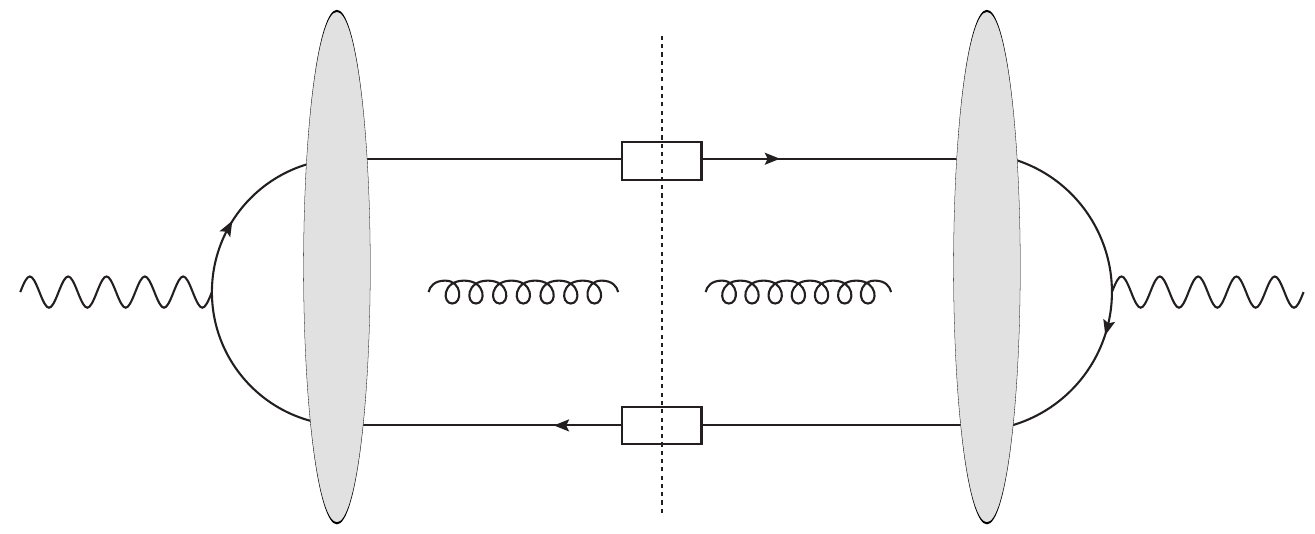}}
\put(188,198){+}
\put(202,170){\includegraphics[scale=\sca]{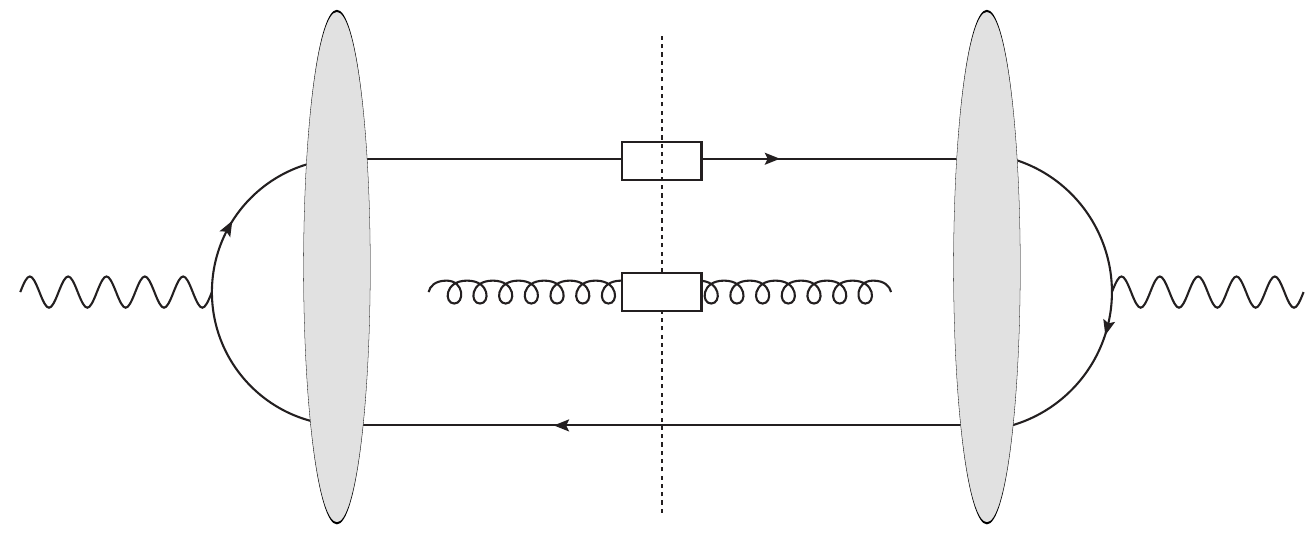}}
\put(82,145){(b)}
\put(282,145){(c)}
\put(0,70){+}
\put(10,40){\includegraphics[scale=\sca]{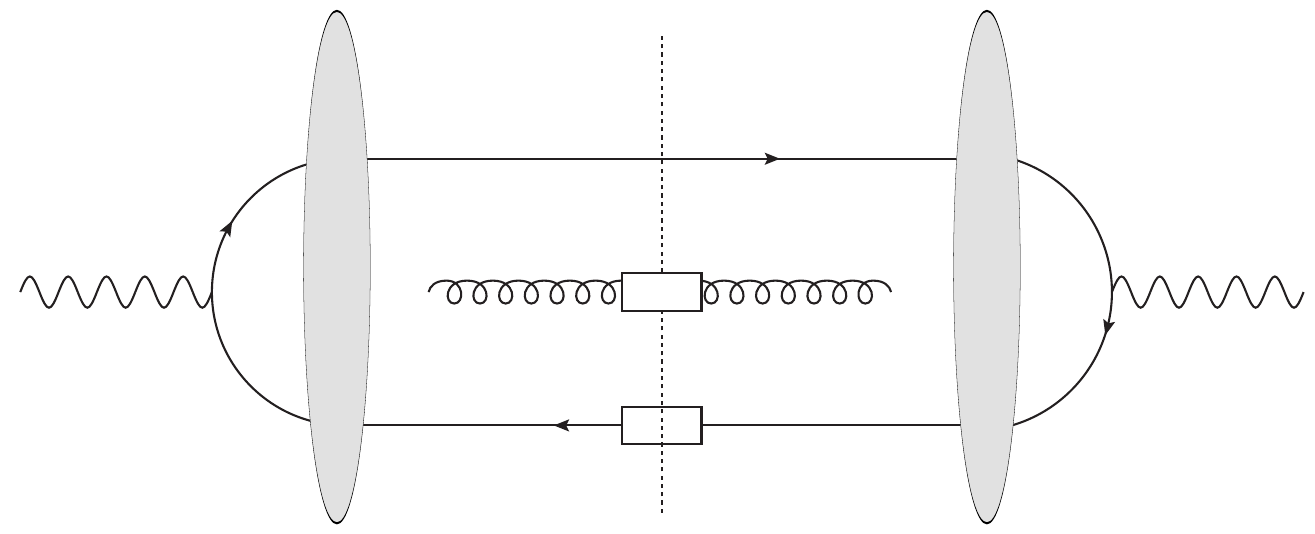}}
\put(188,68){+}
\put(205,40){\includegraphics[scale=\sca]{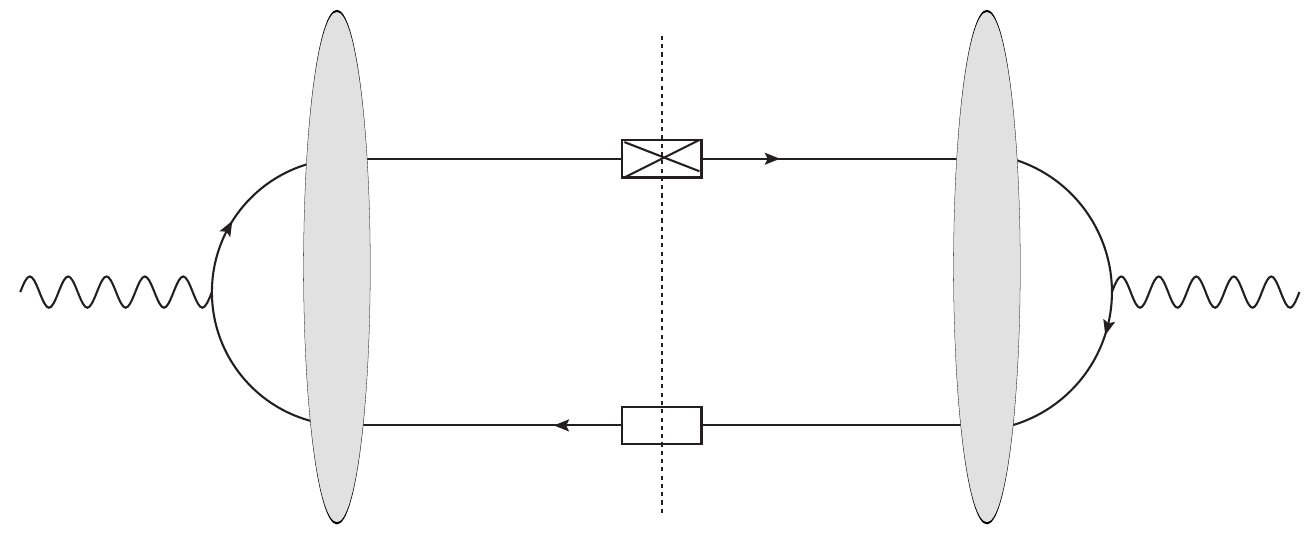}}
\put(197,68){\scalebox{2}{\Bigg ( }}
\put(380,68){$+ \, q \leftrightarrow \bar{q}$ \scalebox{2}{\Bigg )}}
\put(82,15){(d)}
\put(282,15){(e)}
\end{picture}
  \caption{The five kinds of  contributions to the NLO cross section.}
  \label{fig:sigma-NLO}
\end{figure}

At NLO, we have to deal with five kinds of contributions to the cross section, illustrated in Fig.~\ref{fig:sigma-NLO}:
\begin{enumerate}
    \item[(a)] $\gamma^{*} + P \rightarrow h_1 + h_2 + X + P$ cross section at one loop (i.e. virtual contributions),
    \item[(b)] $\gamma^{*} + P \rightarrow h_1 + h_2 + g + X + P$ cross section at Born level (i.e. real contributions),
    \item[(c)] $\gamma^{*} + P \rightarrow h_1 + h_2 + \bar{q} + X + P$ cross section at Born level (i.e. real contributions),
    \item[(d)] $\gamma^{*} + P \rightarrow h_1 + h_2 + q + X + P$ cross section at Born level (i.e. real contributions),
    \item[(e)] FFs counterterms,
\end{enumerate}
where $X$ denotes the remnants of the fragmentation.

\begin{figure}[htbp]
\begin{picture}(420,410)
\put(100,340){\includegraphics[scale=\sca]{FF_dihadron_NLOqqbarg_box.pdf}}
\put(180,315){(b)}
\put(0,240){=}
\put(10,210){\includegraphics[scale=\sca]{FF_dihadron_soft_1.pdf}}
\put(195,240){+}
\put(210,210){\includegraphics[scale=\sca]{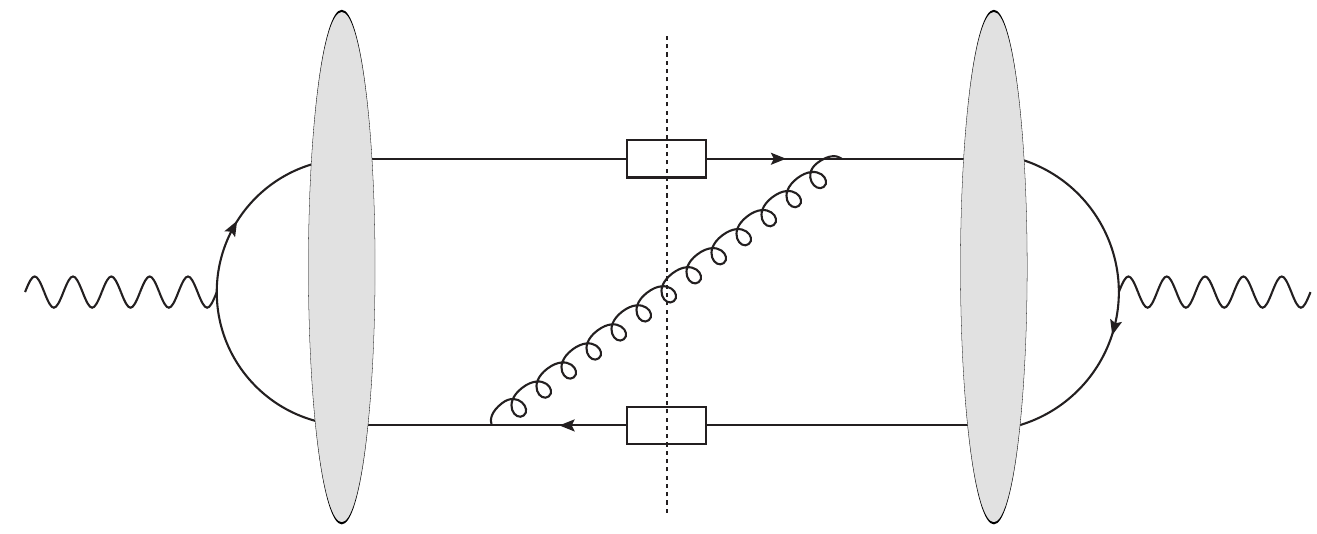}}
\put(92,185){(1)}
\put(292,185){(2)}
\put(0,112){+}
\put(10,80){\includegraphics[scale=\sca]{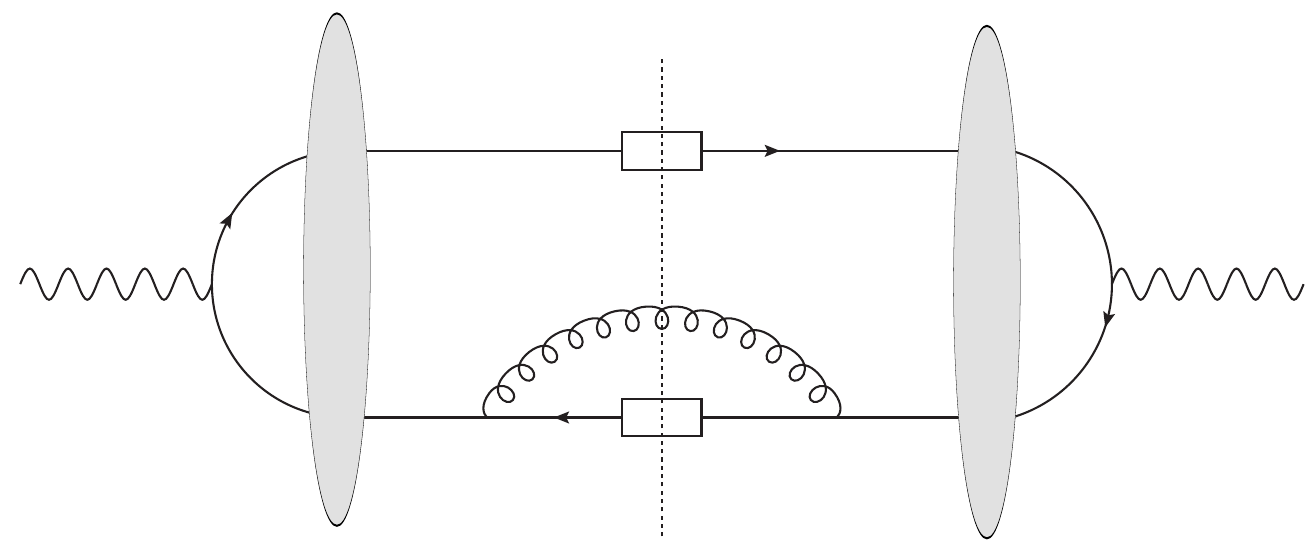}}
\put(195,112){+}
\put(210,80){\includegraphics[scale=\sca]{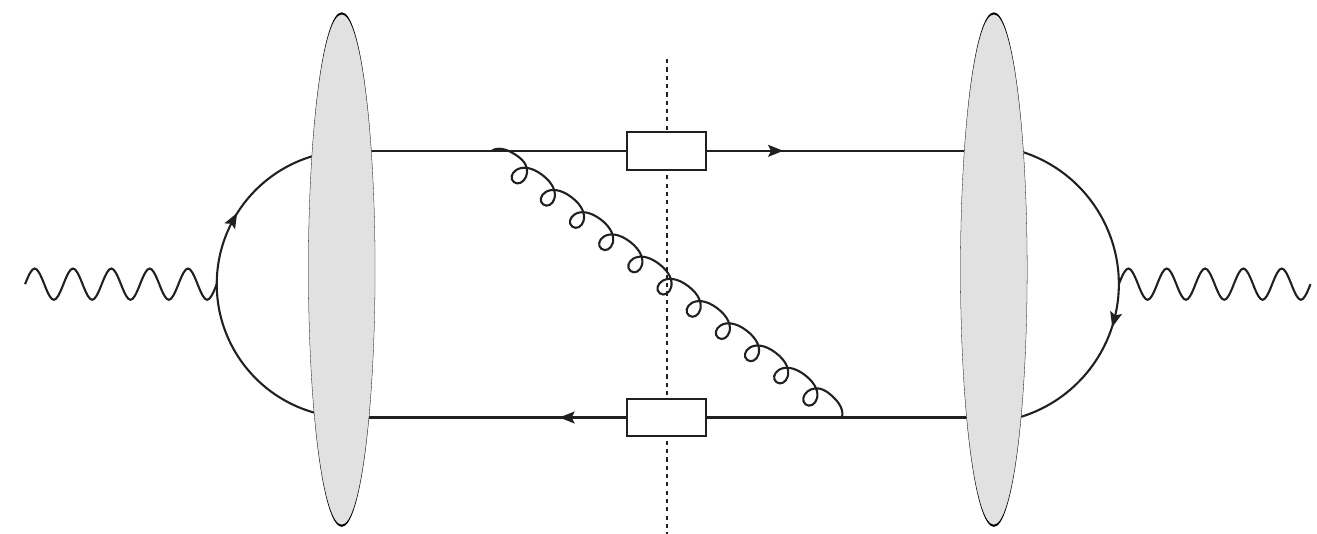}}
\put(92,45){(3)}
\put(292,45){(4)}
\put(0,0){+ finite contributions}
\end{picture}
\vspace{.5cm}
\caption{NLO cross section in the case of fragmentation from the quark and the antiquark. We explicitly isolate the diagrams which contain divergences.
  Diagram (1) contains a collinear divergence between the fragmenting quark and the gluon as well as a soft gluon divergence. Diagram (2) contains a soft gluon divergence. Diagram (3) contains a collinear divergence between the fragmenting antiquark and the gluon as well as a soft gluon divergence. Diagram (4) contains a soft gluon divergence. By ``finite terms", here, we mean all diagrams in which the gluon crosses the shockwave at least once.}
  \label{fig:NLO-b-div}
\end{figure}
Contributions (a) and (e) are easy to treat since (a) is simply the convolution of a known one-loop result with fragmentation functions, while (e) is obtained from the Born result when one renormalizes the fragmentation functions. We just split them into finite and divergent parts. 

For the real contributions (b), (c), (d), the treatment is less straightforward even if the partonic real corrections are also already known.

Contribution (b) is the most complicated one, it contains both soft and collinear divergences. When we square the amplitude contributing to (b), see Fig.~\ref{fig:NLO-b-div}, there is a series of finite contributions plus one, represented by the sum of contributions (1), (2), (3), (4) of Fig.~\ref{fig:NLO-b-div}, that contains all divergences, and that belongs only to the dipole-dipole contribution. 
We  add and subtract to the latter its soft limit to obtain the following structure
\begin{eqnarray}
\label{sigmatilde_q-qbar}
\tilde{\sigma}_{(b)div} &=&  \sum_{\lambda_q,\lambda_g, \lambda_{\bar{q}}}|A_{q\bar{q},sing-dipole}  |^2_{div}= \tilde{\sigma}_{(b)div,1} + \tilde{\sigma}_{(b)div,2} + \tilde{\sigma}_{(b)div,3} + \tilde{\sigma}_{(b)div,4} 
\nonumber \\
&=& \tilde{\sigma}_{(b)div}^{soft}
+ (\tilde{\sigma}_{(b)div,1}-\tilde{\sigma}_{(b)div,1}^{soft})
+
(\tilde{\sigma}_{(b)div,2}-\tilde{\sigma}_{(b)div,2}^{soft})
+
(\tilde{\sigma}_{(b)div,3}-\tilde{\sigma}_{(b)div,3}^{soft}) \nonumber \\
&&+
(\tilde{\sigma}_{(b)div,4}-\tilde{\sigma}_{(b)div,4}^{soft})
\end{eqnarray}
with the meaning that each of these four contributions is in one-to-one correspondence to the four diagrams (1), (2), (3), (4) in Fig.~\ref{fig:NLO-b-div}.
Among these various terms, the terms $(\tilde{\sigma}_{(b)div,1}-\tilde{\sigma}_{(b)div,1}^{soft})$ and $(\tilde{\sigma}_{(b)div,3}-\tilde{\sigma}_{(b)div,3}^{soft})$ are collinearly divergent, while the terms $(\tilde{\sigma}_{(b)div,2}-\tilde{\sigma}_{(b)div,2}^{soft})$ and $(\tilde{\sigma}_{(b)div,4}-\tilde{\sigma}_{(b)div,4}^{soft})$ are finite.

\begin{figure}
\begin{picture}(420,140)
\put(10,70){\includegraphics[scale=\sca]{diagram4a_box.pdf}} \put(195,100){=}
\put(210,67){\includegraphics[scale=\sca]{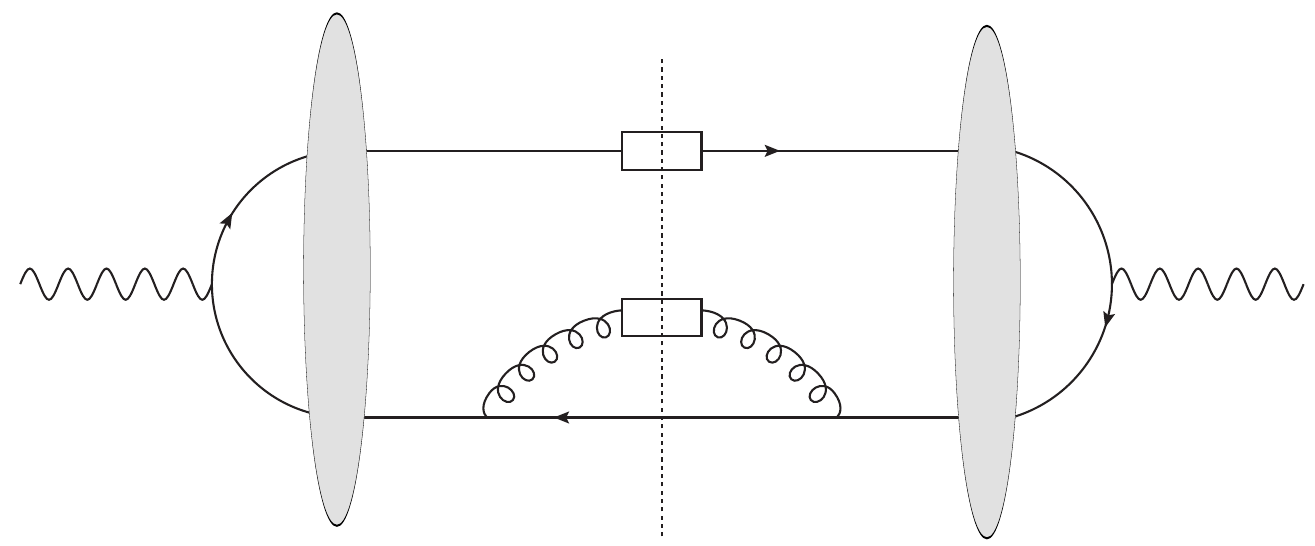}}
\put(200,0){+ finite contributions.}
\put(90,45){(c)}
\put(290,45){(3)}
\end{picture}

\caption{NLO cross section in the case of fragmentation from the gluon and the quark. We explicitly isolate the diagram which contains divergences, namely a collinear divergence between the fragmenting gluon and the antiquark.}
\label{fig:NLO-c-div}
\end{figure}
In Fig.~\ref{fig:NLO-c-div}, the contribution with fragmentation from quark and gluon is considered. Again, we have
\begin{eqnarray}
\label{sigmatilde_q-gluon}
\tilde{\sigma}_{(c)div} &=&  \sum_{\lambda_q,\lambda_g, \lambda_{\bar{q}}}|A_{qg,sing-dipole}  |^2_{div}= 
\tilde{\sigma}_{(c)div,3}   
\,.
\end{eqnarray}
Here, one does not encounter any soft divergence. The only divergence comes from the contribution $\tilde{\sigma}_{(c)div,3}$ which has a collinear divergence when the fragmenting gluon and the antiquark are collinear.

\begin{figure}
\begin{picture}(420,150)
\put(10,70){\includegraphics[scale=\sca]{diagram3a_box.pdf}} \put(195,100){=}
\put(210,70){\includegraphics[scale=\sca]{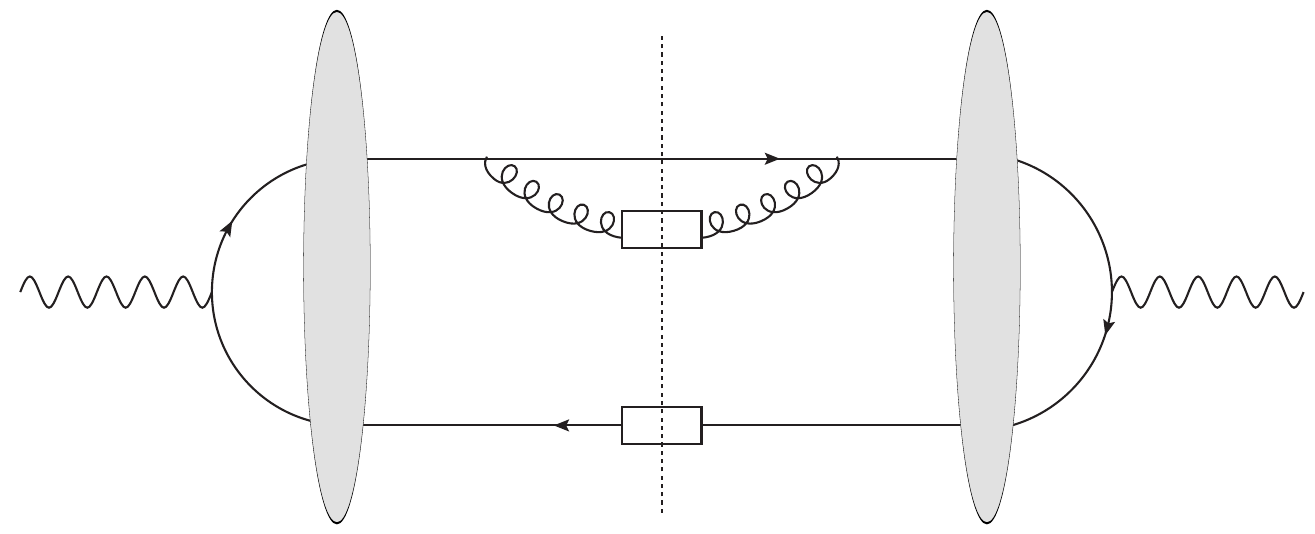}}
\put(200,0){+ finite contributions.}
\put(90,45){(d)}
\put(290,45){(1)}
\end{picture}
\vspace{.5cm}

\caption{NLO cross section in the case of fragmentation from the gluon and the antiquark. We explicitly isolate the diagram which contains divergences, namely a collinear divergence between the fragmenting gluon and the quark.}
\label{fig:NLO-d-div}
\end{figure}

The discussion for the fourth case, see Fig.~\ref{fig:NLO-d-div}, involving the fragmentation from antiquark and gluon goes along the same line: the only divergence comes from the contribution $\tilde{\sigma}_{(d)div,1}$ which has a collinear divergence when the fragmenting gluon and the quark are collinear.

\section{Counterterms from FFs renormalization and evolution}
\label{sec:CounterTerms}

The renormalization and evolution equations of FFs express the bare FFs in terms of dressed ones. In $\overline{\text{MS}}$ scheme, at factorization scale $\mu_F$, they take the form, following notations of Ref.~\cite{Ivanov:2012iv} 
\begin{equation}
\label{eq: FF evolution} 
\begin{aligned}
    D_{q}^{h}(x)& =D_{q}^{h}\left(x, \mu_{F}\right)-\frac{\alpha_{s}}{2 \pi}\left(\frac{1}{\hat{\epsilon}}+\ln \frac{\mu_{F}^{2}}{\mu^{2}}\right) \int_{x}^{1} \frac{d z}{z}\left[D_{q}^{h}\left(\frac{x}{z}, \mu_{F}\right) P_{q q}(z)+D_{g}^{h}\left(\frac{x}{z}, \mu_{F}\right) P_{gq}(z)\right], \\
    D_g^h (x) &= D_g^h (x, \mu_F) - \frac{\alpha_s}{2 \pi} \!\left( \frac{1}{\hat{\epsilon}}+\ln \frac{\mu_{F}^{2}}{\mu^{2}}\right)\! \!\int_{x}^{1} \!\frac{d z}{z} \! \left[ \sum_{q,\bar{q}} D_{q}^{h}\left(\frac{x}{z}, \mu_{F}\right) \! P_{qg}(z)+D_{g}^{h}\left(\frac{x}{z}, \mu_{F}\!\right)\! P_{gg}(z)\right]\!,
\end{aligned}
\end{equation}
where $\frac{1}{\hat{\epsilon}} = \frac{\Gamma (1- \epsilon)}{\epsilon (4 \pi )^\epsilon} \sim \frac{1}{\epsilon} + \gamma_E - \ln (4 \pi)$ and $\mu$ is the arbitrary parameter introduced by dimensional regularization. 
The LO splitting functions are given by  
\begin{eqnarray}
    P_{qq}(z) &=& C_F \left[ \frac{1 + z^2}{(1-z)_+} + \frac{3}{2} \delta (1-z) \right], \\
    P_{gq}(z) &=& C_F \frac{1 + (1-z)^2}{z}\,, \\
    P_{qg}(z) &=& T_R \left[z^2 + (1-z)^2 \right] \ \text{ with } \ T_R = \frac{1}{2} \,,\\\
    P_{gg}(z) &=& 2 C_A \left[ \frac{1}{(1-z)_+} + \frac{1}{z} - 2 + z(1-z) \right] + \left( \frac{11}{6}C_A - \frac{n_f}{3} \right) \delta (1-z)\, ,
\end{eqnarray}
where the + prescription is defined as 
\begin{equation}
\label{eq: plus prescription}
    \int_a^1 d \beta \frac{F(\beta)}{(1-\beta)_+} = \int_a^1 d \beta \frac{F(\beta)- F(1)}{(1-\beta)} - \int_0^{a} d\beta \frac{F(1)}{1-\beta} \,.\\ 
\end{equation}
The collinear counterterms due to the renormalization of the bare FFs are calculated by inserting Eq.~\eqref{eq: FF evolution} in the contributions~(\ref{eq:LL-LO}, \ref{eq:TL-LO}, \ref{eq:TT-LO}) to the LO cross section. This corresponds to the contribution  (e) in Fig. \ref{fig:sigma-NLO}. 
For the $LL$ cross section, this counterterm takes the form  
    \begin{align*}
    \allowdisplaybreaks
   & \frac{d \sigma_{L L}^{q\bar{q} \rightarrow h_1 h_2}}{d x_{h_1}d_{h_2}d^d p_{h_1 \perp}d^d p_{h_2 \perp}} \bigg |_{\text{ct}}  \\
   & = \frac{4 \alpha_{\mathrm{em}} Q^2 }{(2\pi)^{4(d-1)}N_c} \sum_{q} \int_{x_{h_1}}^1 \!\!\!\! d x_q \int_{x_{h_2}}^1 \!\!\!\! d x_{\bar{q}} \; x_q x_{\bar{q}}  \left( \frac{x_q}{x_{h_1}}\right)^d \! \left(\frac{x_{\bar{q}}}{x_{h_2}}\right)^d \! \delta (1-x_q - x_{\bar{q}}) \\
     & \times  {\cal F}_{LL} \left(- \frac{\alpha_s}{2\pi}\right) 
     \left(\frac{1}{\hat{\epsilon}}+\ln \frac{\mu_{F}^{2}}{\mu^{2}}\right)
      Q_q^2 \left\{ \int_{\frac{x_{h_1}}{x_q}}^1 \frac{d \beta_1}{\beta_1} \left[ P_{qq}(\beta_1) D_q^{h_1}\left(\frac{x_{h_1}}{\beta_1 x_q},\mu_F\right)D_{\bar{q}}^{h_2}\left(\frac{x_{h_2}}{x_{\bar{q}}},\mu_F\right)  \right. \right. \\
    & \left. + P_{gq}(\beta_1) D_g^{h_1}\left(\frac{x_{h_1}}{\beta_1 x_q}, \mu_F\right) D_{\bar{q}}^{h_2}\left(\frac{x_{h_2}}{x_{\bar{q}}},\mu_F\right) \right]\\
    & + \int_{\frac{x_{h_2}}{x_{\bar{q}}}}^1 \frac{d \beta_2}{\beta_2} \left[ P_{qq}(\beta_2) D_q^{h_1}\left(\frac{x_{h_1}}{x_q}, \mu_F\right)D_{\bar{q}}^{h_2}\left(\frac{x_{h_2}}{\beta_2 x_{\bar{q}}}, \mu_F\right) \right. \\
    & \left.  \left. + P_{gq}(\beta_2) D_q^{h_1}\left(\frac{x_{h_1}}{x_q}, \mu_F\right)D_{g}^{h_2}\left(\frac{x_{h_2}}{\beta_2 x_{\bar{q}}}, \mu_F\right) \right] \right\} + (h_1 \leftrightarrow h_2)  \\ 
    &= \frac{d \sigma_{L L}^{h_1 h_2}}{d x_{h_1}d_{h_2}d^d p_{h_1 \perp}d^d p_{h_2 \perp}} \bigg |_{\text{ct div}} +  \frac{d \sigma_{L L}^{h_1 h_2}}{d x_{h_1}d_{h_2}d^d p_{h_1 \perp}d^d p_{h_2 \perp}} \bigg |_{\text{ct fin}} .
    \numberthis[ct_LL]
   \end{align*}
The divergent part is the one containing $1/\hat{\epsilon}$ and the finite term is the one with $\ln (\mu_F^2/ \mu^2) $. The dependence on the arbitrary parameter $\mu$ disappears at the end when all finite terms are put together. 
We stress here that, for any separate term in the curly bracket, one can indifferently use $\mathcal{F}_{LL}$ or $\tilde{\mathcal{F}}_{LL}$. In particular, for the first two terms we can use $\mathcal{F}_{LL}$ and for the last two $\tilde{\mathcal{F}}_{LL}$. This simple observation is useful when observing the cancellation of divergences at the level of integrands. The same remark applies also for other transitions.  \\

For cross sections involving other combinations of polarizations, we have
\begin{equation}
   \allowdisplaybreaks
    \begin{aligned}
   & \frac{d \sigma_{TL}^{q\bar{q} \rightarrow h_1 h_2}}{d x_{h_1}d_{h_2}d^d p_{h_1 \perp}d^d p_{h_2 \perp}} \bigg |_{\text{ct}}  \\
   & = \frac{2 \alpha_{\mathrm{em}} Q}{(2\pi)^{4(d-1)}N_c} \sum_{q}  \int_{x_{h_1}}^1 \!\!\!\! d x_q \int_{x_{h_2}}^1 \!\!\!\! d x_{\bar{q}}  \left( \frac{x_q}{x_{h_1}}\right)^d \! \left(\frac{x_{\bar{q}}}{x_{h_2}}\right)^d \! (x_{\bar{q}}-x_q ) \; \delta (1-x_q - x_{\bar{q}}) \\
     & \times  {\cal F}_{TL} \left(- \frac{\alpha_s}{2\pi}\right)
     \left(\frac{1}{\hat{\epsilon}}+\ln \frac{\mu_{F}^{2}}{\mu^{2}}\right)
     Q_q^2 \left\{ \int_{\frac{x_{h_1}}{x_q}}^1 \frac{d \beta_1}{\beta_1} \left[ P_{qq} (\beta_1) D_q^{h_1}\left(\frac{x_{h_1}}{\beta_1 x_q},\mu_F\right)D_{\bar{q}}^{h_2}\left(\frac{x_{h_2}}{x_{\bar{q}}},\mu_F\right)  \right. \right. \\
    & \left. + P_{gq}(\beta_1) D_g^{h_1}\left(\frac{x_{h_1}}{\beta_1 x_q}, \mu_F\right) D_{\bar{q}}^{h_2}\left(\frac{x_{h_2}}{x_{\bar{q}}},\mu_F\right) \right] \\
    & + \int_{\frac{x_{h_2}}{x_{\bar{q}}}}^1 \frac{d \beta_2}{\beta_2} \left[ P_{qq}(\beta_2) D_q^{h_1}\left(\frac{x_{h_1}}{x_q}, \mu_F\right)D_{\bar{q}}^{h_2}\left(\frac{x_{h_2}}{\beta_2 x_{\bar{q}}}, \mu_F\right) \right.  \\
    & \left. \left. + P_{gq}(\beta_2) D_q^{h_1}\left(\frac{x_{h_1}}{x_q}, \mu_F\right)D_{g}^{h_2}\left(\frac{x_{h_2}}{\beta_2 x_{\bar{q}}}, \mu_F\right) \right]
    \right\} + (h_1 \leftrightarrow h_2) \\ 
    &= \frac{d \sigma_{T L}^{h_1 h_2}}{d x_{h_1}d_{h_2}d^d p_{h_1 \perp}d^d p_{h_2 \perp}} \bigg |_{\text{ct div}} +  \frac{d \sigma_{T L}^{h_1 h_2}}{d x_{h_1}d_{h_2}d^d p_{h_1 \perp}d^d p_{h_2 \perp}} \bigg |_{\text{ct fin }} ,
   \end{aligned}
\label{eq:ct TL }   
\end{equation}
and
\begin{align}
   \allowdisplaybreaks
   & \frac{d \sigma_{TT}^{q\bar{q} \rightarrow h_1 h_2}}{d x_{h_1}d_{h_2}d^d p_{h_1 \perp}d^d p_{h_2 \perp}} \bigg |_{\text{ct}} \nonumber \\
   & = \frac{ \alpha_{\mathrm{em}}  }{(2\pi)^{4(d-1)}N_c} \sum_{q} \int_{x_{h_1}}^1 \!\!\!\! \frac{d x_q}{x_q} \int_{x_{h_2}}^1 \!\!\!\! \frac{d x_{\bar{q}} }{x_{\bar{q}}}  \left( \frac{x_q}{x_{h_1}}\right)^d \! \left(\frac{x_{\bar{q}}}{x_{h_2}}\right)^d \! \delta (1-x_q - x_{\bar{q}}) \nonumber \\
     & \times  {\cal F}_{TT} \left(- \frac{\alpha_s}{2\pi}\right) 
     \left(\frac{1}{\hat{\epsilon}}+\ln \frac{\mu_{F}^{2}}{\mu^{2}}\right)
      Q_q^2 \left\{ \int_{\frac{x_{h_1}}{x_q}}^1 \frac{d \beta_1}{\beta_1} \left[  P_{qq}(\beta_1) D_q^{h_1}\left(\frac{x_{h_1}}{\beta_1 x_q},\mu_F\right)D_{\bar{q}}^{h_2}\left(\frac{x_{h_2}}{x_{\bar{q}}},\mu_F\right)  \right. \right. \nonumber \\
    & \left. + P_{gq}(\beta_1) D_g^{h_1}\left(\frac{x_{h_1}}{\beta_1 x_q}, \mu_F\right) D_{\bar{q}}^{h_2}\left(\frac{x_{h_2}}{x_{\bar{q}}},\mu_F\right) \right] \nonumber \\
    & + \int_{\frac{x_{h_2}}{x_{\bar{q}}}}^1 \frac{d \beta_2}{\beta_2} \left[ P_{qq}(\beta_2) D_q^{h_1}\left(\frac{x_{h_1}}{x_q}, \mu_F\right)D_{\bar{q}}^{h_2}\left(\frac{x_{h_2}}{\beta_2 x_{\bar{q}}}, \mu_F\right) \right. \nonumber \\
    & \left. \left.  + P_{gq}(\beta_2) D_q^{h_1}\left(\frac{x_{h_1}}{x_q}, \mu_F\right)D_{g}^{h_2}\left(\frac{x_{h_2}}{\beta_2 x_{\bar{q}}}, \mu_F\right) \right] \right\} + (h_1 \leftrightarrow h_2) \nonumber \\ 
    &= \frac{d \sigma_{TT}^{h_1 h_2}}{d x_{h_1}d_{h_2}d^d p_{h_1 \perp}d^d p_{h_2 \perp}} \bigg |_{\text{ct div}} +  \frac{d \sigma_{TT}^{h_1 h_2}}{d x_{h_1}d_{h_2}d^d p_{h_1 \perp}d^d p_{h_2 \perp}} \bigg |_{\text{ct fin }} .
   \label{eq:ct TT }
   \end{align}
The divergent parts are the ones containing $1/\hat{\epsilon}$ and the finite terms are the ones with $\ln (\mu_F^2/ \mu^2) $. The dependence on the arbitrary parameter $\mu$ disappears at the end when all finite terms are put together. 

\section{NLO cross section: Virtual corrections}
\label{sec: VirtualDiv}
Here, we compute 1-loop virtual corrections to the leading order cross section. 
For sake of comprehension, we report the  dipole-dipole virtual corrections to the $\gamma^{*} \rightarrow q \bar{q}$ cross section, as presented in (5.24) of \cite{Boussarie:2016ogo}, which we refer to as our partonic cross section. Adapting to our notation, we have
\begin{equation}
\allowdisplaybreaks
\label{eq: VirtBeg}
\begin{aligned}
d \hat{\sigma}_{1 L L}  &= \frac{\alpha_{s}}{2\pi} \frac{\Gamma(1-\epsilon)}{(4 \pi)^{\epsilon}} C_F \left( \frac{S_{V}+S_{V}^{*}}{2} \right) d \hat{\sigma}_{0 L L} \\
&  + \frac{\alpha_{s} Q^{2}}{4 \pi}\left(\frac{N_{c}^{2}-1}{N_{c}}\right) \frac{\alpha_{\mathrm{em}} Q_{q}^{2}}{(2 \pi)^{4} N_{c}} d x_q d x_{\bar{q}} d^{2} p_{q \perp} d^{2} p_{\bar{q} \perp} \delta(1-x_q-x_{\bar{q}}) \\
 & \times \int d^{2} p_{1 \perp} d^{2} p_{2 \perp} d^{2} p_{1 \perp}^{\prime} d^{2} p_{2 \perp}^{\prime} \delta\left(p_{q 1 \perp}+p_{\bar{q} 2 \perp}\right) \frac{\delta\left(p_{11^{\prime} \perp}+p_{22^{\prime} \perp}\right)}{\vec{p}_{q 1^{\prime}}^{\,2}+x_q x_{\bar{q}} Q^{2}} \mathbf{F}\left(\frac{p_{12 \perp}}{2}\right) \mathbf{F}^{*}\left(\frac{p_{1^{\prime} 2^{\prime} \perp}}{2}\right) \\
 &  \times {\left[\frac{6 x_q^{2} x_{\bar{q}}^{2}}{\vec{p}_{q 1}^{\,2}+x_q x_{\bar{q}} Q^{2}} \ln \left(\frac{x_q^{2} x_{\bar{q}}^{2} \mu^{4} Q^{2}}{\left(x_q \vec{p}_{\bar{q}}-x_{\bar{q}} \vec{p}_{q}\right)^{2}\left(\vec{p}_{q 1}^{\,2}+x_q x_{\bar{q}} Q^{2}\right)^{2}}\right)\right.} \\
& + \left.\frac{\left(p_{0}^{-}\right)^{2}}{s^{2} p_{\gamma}^{+}} \operatorname{tr}\left(\left(C_{\|}^{4}+C_{1 \|}^{5}+C_{1 \|}^{6}\right) \hat{p}_{\bar{q}} \gamma^{+} \hat{p}_{q}\right)\right]+ h.c . 
\end{aligned}
\end{equation} 
in which the divergences are inside the contribution
\begin{gather}   
\frac{S_V + S_V^*}{2}  =  \frac{1}{\epsilon} \Bigg [ - 4 \epsilon \ln (\alpha) \ln \left(\frac{x_q ^2 x_{\bar{q}}^2 \mu^{2}}{\left(x_q \vec{p}_{\bar{q}}-x_{\bar{q}} \vec{p}_{q}\right)^{2}}\right) + 4 \ln (\alpha) + 4 \epsilon \ln^2(\alpha)  - 2 \ln (x_q x_{\bar{q}})+ 3 \nonumber \\ 
+ 2 \epsilon \ln \left(\frac{x_q x_{\bar{q}} \mu^{2}}{\left(x_q \vec{p}_{\bar{q}}-x_{\bar{q}}\vec{p}_{q}\right)^{2}}\right) \ln (x_q x_{\bar{q}}) + \epsilon \ln^2 (x_q x_{\bar{q}})  - 3 \epsilon \ln \left(\frac{x_q x_{\bar{q}} \mu^{2}}{\left(x_q \vec{p}_{\bar{q}}-x_{\bar{q}} \vec{p}_{q}\right)^{2}}\right)  - \frac{\pi^2}{3}\epsilon + 6 \epsilon \Bigg ] , \, 
\label{SV}
\end{gather}
where $\alpha$ is an infra-red cut-off imposed on the longitudinal fraction of gluon momenta in order to regularize rapidity divergences. 
In Eq.~\eqref{eq: VirtBeg} the 
$C$ functions have been parametrized using (5.25) and (5.26) in Ref.~\cite{Boussarie:2016ogo} as
\begin{align}
\frac{(p_{0}^{-})^{2}}{s^{2}p_{\gamma}^{+}}tr(C_{||}^{4}\hat{p}_{\bar{q}}\gamma^{+}\hat{p}_{q})=\int_{0}^{x}dz\left[(\phi_4)_{LL}\right]_+ +(q\leftrightarrow\bar{q}) \, , \label{phi1LL}
\end{align}
and
\begin{align}
\frac{(p_{0}^{-})^{2}}{s^{2}p_{\gamma}^{+}}tr(C_{1||}^{n}\hat{p}_{\bar{q}}\gamma^{+}\hat{p}_{q})=\int_{0}^{x}dz\left[(\phi_n)_{LL}\right]_+|_{\vec{p}_3=\vec{0}} +(q\leftrightarrow\bar{q}) \, ,
\end{align}
where $n=5 \, \, \mathrm{or} \, \, 6$, and $(q\leftrightarrow\bar{q})$ stands for $p_{q}\leftrightarrow p_{\bar{q}},\,p_{1}^{(\prime)}\leftrightarrow p_{2}^{(\prime)},\,x_{q}\leftrightarrow
x_{\bar{q}}.$ The expressions for $(\phi_n)_{LL}$ are given in Appendix~\ref{AppendixD1}.

By using a factorization formula analogous to Eq.~\eqref{eq: coll facto} and the expression  \eqref{eq: VirtBeg}, as well as the collinear constraints (\ref{constraint-collinear-q}) and (\ref{constraint-collinear-qbar}), we obtain the dipole-dipole virtual corrections to the full cross section. We split it into divergent part,
\begin{equation}
\label{eq:virtual div LL}
    \begin{aligned}
& \frac{d \sigma_{1 L L}^{q\bar{q} \rightarrow  h_1 h_2}}{d x_{h_1}d_{h_2}d^d p_{h_1 \perp}d^d p_{h_2 \perp}}\bigg |_{\text{div}} \\
&=  \frac{4 \alpha_{\mathrm{em}} Q^2}{( 2 \pi)^{4(d-1)} N_c}  \sum_{q}   \int_{x_{h_1}}^1 d x_q  \int_{x_{h_2}}^1 d x_{\bar{q}} \;  x_q x_{\bar{q}} \left( \frac{x_q}{x_{h_1}}\right)^d \left( \frac{x_{\bar{q}}}{x_{h_2}}\right)^d  \delta(1-x_q-x_{\bar{q}})\\
& \times Q_q^2 D_q^{h_1}\left(\frac{x_{h_1}}{x_q}, \mu_F\right) D_{\bar{q}}^{h_2}\left(\frac{x_{h_2}}{x_{\bar{q}}}, \mu_F\right)  {\cal F}_{LL} \\
& \times \frac{\alpha_s}{2 \pi}C_F  \frac{1}{\hat{\epsilon}}  \left[ - 4 \epsilon \ln (\alpha) \ln \left(\frac{  \mu^{2}}{\left( \frac{\vec{p}_{h_2}}{x_{h_2}}- \frac{\vec{p}_{h_1}}{x_{h_1}} \right)^{2}}\right) + 4 \ln (\alpha)  \right. \\
& + \left.   4 \epsilon \ln^2(\alpha)   - 2 \ln (x_q x_{\bar{q}})+ 3 \right] + (h_1 \leftrightarrow h_2)
    \end{aligned}
\end{equation}
and finite part,
\begin{align*}
& \frac{d \sigma_{1 L L}^{q\bar{q} \rightarrow  h_1 h_2}}{d x_{h_1}d_{h_2}d^d p_{h_1 \perp}d^d p_{h_2 \perp}}\bigg |_{\text{fin}} \\
& =  \frac{4 \alpha_{\mathrm{em}} Q^2}{( 2 \pi)^{4(d-1)} N_c}   \sum_{q} \int_{x_{h_1}}^1 d x_q  \int_{x_{h_2}}^1 d x_{\bar{q}} \;   x_q x_{\bar{q}} \left( \frac{x_q}{x_{h_1}}\right)^d \left( \frac{x_{\bar{q}}}{x_{h_2}}\right)^d  \delta(1-x_q-x_{\bar{q}})\\
& \times  Q_q^2 D_q^{h_1}\left(\frac{x_{h_1}}{x_q}, \mu_F\right) D_{\bar{q}}^{h_2}\left(\frac{x_{h_2}}{x_{\bar{q}}}, \mu_F\right)  {\mathcal{F}}_{LL} \\
& \times \frac{\alpha_s}{2 \pi}C_F  \frac{1}{\hat{\epsilon}} \left[2 \epsilon \ln \left(\frac{ \mu^{2}}{ x_q x_{\bar{q}} \left( \frac{\vec{p}_{h_2}}{x_{h_2}}- \frac{\vec{p}_{h_1}}{x_{h_1}}\right)^{2}}\right)  \ln (x_q x_{\bar{q}}) \right. \\
& \left. + \epsilon \ln^2 (x_q x_{\bar{q}})  - 3 \epsilon \ln \left(\frac{ \mu^{2}}{ x_q x_{\bar{q}} \left( \frac{\vec{p}_{h_2}}{x_{h_2}}- \frac{\vec{p}_{h_1}}{x_{h_1}}\right)^{2}}\right) - \frac{\pi^2}{3}\epsilon + 6 \epsilon\right] \\
& + \frac{\alpha_s Q^2}{4\pi}\left(\frac{N_c^2-1}{N_c}\right) \frac{\alpha_{\mathrm{em}}}{(2 \pi)^{4} N_{c}} \sum_{q} \int_{x_{h_1}}^{1} \frac{d x_q}{x_q} \int_{x_{h_2}}^1  \frac{d x_{\bar{q}}}{x_{\bar{q}}}  \delta(1-x_q-x_{\bar{q}}) \\
& \times \left(\frac{x_{q}}{x_{h_1}}\right)^d \left(\frac{x_{\bar{q}}}{x_{h_2}}\right)^d  Q_{q}^{2} D_{q}^{h_1}\left(\frac{x_{h_1}}{x_q},\mu_F\right) D_{\bar{q}}^{h_2}\left(\frac{x_{h_2}}{x_{\bar{q}}} ,\mu_F\right)  \\ 
 & \times \int d^{d} p_{2 \perp} \frac{\mathbf{F}\left(\frac{x_q}{2x_{h_1}}  p_{h_1\perp} + \frac{x_{\bar{q}} }{2 x_{h_2}}p_{h_2\perp} -p_{2\perp}\right)}{\left(\frac{x_{\bar{q}}}{x_{h_2}}\vec{p}_{h_2}- \vec{p}_{2}\right)^{2}+x_q x_{\bar{q}} Q^{2}} \int d^{d} p_{2' \perp} \frac{\mathbf{F}^{*}\left(\frac{x_q}{2x_{h_1}}  p_{h_1\perp} + \frac{x_{\bar{q}} }{2 x_{h_2}}p_{h_2\perp} -p_{2'\perp}\right)}{\left(\frac{x_{\bar{q}}}{x_{h_2}}\vec{p}_{h_2}- \vec{p}_{2'} \right)^{2}+x_q x_{\bar{q}} Q^{2}} \\ & \times \left[ 6 x_q^{2} x_{\bar{q}}^{2}  
  \ln \left( \frac{\mu^{4} Q^{2}}{\left( \frac{\vec{p}_{h_2}}{x_{h_2}}- \frac{\vec{p}_{h_1}}{x_{h_1}}\right)^{2} \left(\left(\frac{x_{\bar{q}}}{x_{h_2}} \vec{p}_{h_2}- \vec{p}_2 \right)^{2}+x_q x_{\bar{q}} Q^{2}\right)^{2}}\right) +  \left(  \int_0^{x_q} dz \left[(\phi_4)_{LL}\right]_+   \right. \right.  \\
 & \left. \left. + \sum_{n=5,6} \left[(\phi_n)_{LL}\right]_+ \bigg |_{\vec{p}_3 = \vec{0}}  + (q \leftrightarrow \bar{q}) \right) \left(\left(\frac{x_{\bar{q}}}{x_{h_2}}\vec{p}_{h_2}-\vec{p}_{2}\right)^2 + x_q x_{\bar{q}}Q^2 \right) \right] + (h_1 \leftrightarrow h_2) \; .
\numberthis[virtual finite LL]
\end{align*}

For the $d \sigma_{TL}$ element of the matrix \eqref{eq:density_matrix}, we get from (5.28) of \cite{Boussarie:2016ogo}
    \begin{align*}
& \frac{d \sigma_{1 T L}^{q\bar{q} \rightarrow  h_1 h_2}}{d x_{h_1}d_{h_2}d^d p_{h_1 \perp}d^d p_{h_2 \perp}}\bigg |_{\text{div}} \\
&=  \frac{2 \alpha_{\mathrm{em}} Q}{( 2 \pi)^{4(d-1)} N_c}   \sum_{q}  \int_{x_{h_1}}^1 d x_q  \int_{x_{h_2}}^1 d x_{\bar{q}}   \left( \frac{x_q}{x_{h_1}}\right)^d \left( \frac{x_{\bar{q}}}{x_{h_2}}\right)^d (x_{\bar{q}}-x_q ) \\
& \times \delta(1-x_q-x_{\bar{q}}) Q_q^2 D_q^{h_1}\left(\frac{x_{h_1}}{x_q}, \mu_F\right) D_{\bar{q}}^{h_2}\left(\frac{x_{h_2}}{x_{\bar{q}}}, \mu_F\right)  {\cal F}_{TL} \\
& \times \frac{\alpha_s}{2 \pi}C_F  \frac{1}{\hat{\epsilon}}  \left[ - 4 \epsilon \ln (\alpha) \ln \left(\frac{  \mu^{2}}{\left( \frac{\vec{p}_{h_2}}{x_{h_2}}- \frac{\vec{p}_{h_1}}{x_{h_1}} \right)^{2}}\right) + 4 \ln (\alpha)   \right. \\
& \left. + 4 \epsilon \ln^2(\alpha)  - 2 \ln (x_q x_{\bar{q}})+ 3 \right] + (h_1 \leftrightarrow h_2) \; ,  
\numberthis[virtual div TL]
\end{align*}
and
{\allowdisplaybreaks
\begin{align*}
& \frac{d \sigma_{1T L}^{q\bar{q} \rightarrow h_1 h_2}}{d x_{h_1}d_{h_2}d^d p_{h_1 \perp}d^d p_{h_2 \perp}}\bigg |_{\text{fin}} \\
& =  \frac{2 \alpha_{\mathrm{em}} Q}{( 2 \pi)^{4(d-1)} N_c}   \sum_{q} \int_{x_{h_1}}^1 d x_q  \int_{x_{h_2}}^1 d x_{\bar{q}}  \left( \frac{x_q}{x_{h_1}}\right)^d \left( \frac{x_{\bar{q}}}{x_{h_2}}\right)^d (x_{\bar{q}}-x_q ) \\
& \times \delta(1-x_q-x_{\bar{q}})  Q_q^2 D_q^{h_1}\left(\frac{x_{h_1}}{x_q}, \mu_F\right) D_{\bar{q}}^{h_2}\left(\frac{x_{h_2}}{x_{\bar{q}}}, \mu_F\right)  {\cal F}_{TL} \\
& \times \frac{\alpha_s}{2 \pi}C_F  \frac{1}{\hat{\epsilon}} \left[2 \epsilon \ln \left(\frac{ \mu^{2}}{ x_q x_{\bar{q}} \left( \frac{\vec{p}_{h_2}}{x_{h_2}}- \frac{\vec{p}_{h_1}}{x_{h_1}}\right)^{2}}\right) \ln (x_q x_{\bar{q}}) + \epsilon \ln^2 (x_q x_{\bar{q}}) \right. \\
& \left. - 3 \epsilon \ln \left(\frac{ \mu^{2}}{ x_q x_{\bar{q}} \left( \frac{\vec{p}_{h_2}}{x_{h_2}}- \frac{\vec{p}_{h_1}}{x_{h_1}}\right)^{2}}\right)  - \frac{\pi^2}{3}\epsilon + 6 \epsilon\right] \\
& + \frac{\alpha_s Q}{4\pi}\left(\frac{N_c^2-1}{N_c}\right) \frac{\alpha_{\mathrm{em}}}{(2 \pi)^{4} N_{c}} \sum_{q}  \int_{x_{h_1}}^{1} \frac{d x_q}{x_q} \int_{x_{h_2}}^1  \frac{d x_{\bar{q}}}{x_q} \;  \delta(1-x_q-x_{\bar{q}})  \\
& \times \left(\frac{x_{q}}{x_{h_1}}\right)^d \left(\frac{x_{\bar{q}}}{x_{h_2}}\right)^d  Q_{q}^{2} D_{q}^{h_1}\left(\frac{x_{h_1}}{x_q},\mu_F\right) D_{\bar{q}}^{h_2}\left(\frac{x_{h_1}}{x_q},\mu_F\right)\\ 
& \times \int d^d p_{1 \perp} d^d p_{2 \perp} \mathbf{F}\left(\frac{p_{12\perp}}{2}\right) \delta \left( \frac{x_q}{x_{h_1}} p_{h_1 \perp} -p_{1\perp} + \frac{x_{\bar{q}}}{x_{h_2}} p_{h_2 \perp} - p_{2 \perp} \right) \\
& \times  \int d^d p_{1' \perp} d^d p_{2' \perp} \mathbf{F}^{*} \left(\frac{p_{1'2'\perp}}{2}\right) \delta \left( \frac{x_q}{x_{h_1}} p_{h_1 \perp} -p_{1'\perp} + \frac{x_{\bar{q}}}{x_{h_2}} p_{h_2 \perp} - p_{2' \perp} \right) \varepsilon_{Ti}^* \\
& \times   \left[ \frac{1}{\left(\frac{x_q}{x_{h_1}} \vec{p}_{h_1} - \vec{p}_1\right)^2 + x_q x_{\bar{q}}Q^2 }  \left( \int_0^{x_q} \left[(\phi_4^i)_{TL}\right]_+ + \sum_{n=5,6} \int_0^{x_q} \left[(\phi_n^i)_{TL}\right]_+ \bigg |_{\vec{p}_3 = \vec{0}} + (q \leftrightarrow \bar{q} ) \right)^\dag  \right. \\
& + \frac{3 x_q x_{\bar{q}} (x_{\bar{q}} - x_q) \left(\frac{x_q}{x_{h_1}} p_{h_1 \perp} - p_{1' \perp}\right)^i }{\left(\left(\frac{x_q}{x_{h_1}} \vec{p}_{h_1} - \vec{p}_1\right)^2 + x_q x_{\bar{q}}Q^2 \right) \left(\left(\frac{x_q}{x_{h_1}} \vec{p}_{h_1} - \vec{p}_{1'}\right)^2 + x_q x_{\bar{q}}Q^2 \right)} \\
& \times \left( \ln \left( \frac{ Q^2 \mu^8 \left(\frac{\vec{p}_{h_2}}{x_{h_2}} - \frac{\vec{p}_{h_1}}{x_{h_1}}\right)^{-4}}{x_q x_{\bar{q}} \left(\left(\frac{x_q}{x_{h_1}} \vec{p}_{h_1} - \vec{p}_1\right)^2 + x_q x_{\bar{q}}Q^2 \right) \left(\left(\frac{x_q}{x_{h_1}} \vec{p}_{h_1} - \vec{p}_{1'}\right)^2 + x_q x_{\bar{q}}Q^2 \right)} \right)\right. \\ 
& \left. \hspace{-0.1 cm} - \frac{x_q x_{\bar{q}} Q^2}{\left(\frac{x_q}{x_{h_1}} \vec{p}_{h_1} - \vec{p}_{1'} \right)^2} \ln \left( \frac{x_q x_{\bar{q}} Q^2}{\left(\frac{x_q}{x_{h_1}} \vec{p}_{h_1} - \vec{p}_{1'} \right)^2 + x_q x_{\bar{q}}Q^2 }\right) \hspace{-0.1 cm} \right) \hspace{-0.1 cm} + \frac{1}{2 x_q x_{\bar{q}} \left(\left(\frac{x_q}{x_{h_1}} \vec{p}_{h_1} - \vec{p}_{1'} \right)^2 + x_q x_{\bar{q}} Q^2 \right)} \\
& \times  \left.   \left( \int_0^{x_q} \left[(\phi_4^i)_{LT}\right]_+ + \sum_{n=5,6} \int_0^{x_q} \left[(\phi_n^i)_{LT}\right]_+ \bigg |_{\vec{p}_3 = \vec{0}}  +   (q \leftrightarrow \bar{q} ) \right)\right]  + (h_1 \leftrightarrow h_2) \,. \numberthis[virtual finite TL]
\end{align*}}
In the case of the $TT$ transition, the result was obtained in Ref.~\cite{Boussarie:2016ogo} Eq.~(5.35) and the divergent part is
{\allowdisplaybreaks
    \begin{align*}
& \frac{d \sigma_{1T T}^{q\bar{q} \rightarrow  h_1 h_2}}{d x_{h_1}d_{h_2}d^d p_{h_1 \perp}d^d p_{h_2 \perp}}\bigg |_{\text{div}} \\
&=  \frac{ \alpha_{\mathrm{em}} }{( 2 \pi)^{4(d-1)} N_c}  \sum_{q}  \int_{x_{h_1}}^1 \frac{d x_q}{x_q}  \int_{x_{h_2}}^1 \frac{d x_{\bar{q}} }{x_{\bar{q}}}   \left( \frac{x_q}{x_{h_1}}\right)^d \left( \frac{x_{\bar{q}}}{x_{h_2}}\right)^d \delta(1-x_q -x_{\bar{q}}) \\
& \times  Q_q^2 D_q^{h_1}\left(\frac{x_{h_1}}{x_q}, \mu_F\right) D_{\bar{q}}^{h_2}\left(\frac{x_{h_2}}{x_{\bar{q}}}, \mu_F\right)  {\cal F}_{TT} \\
& \times \frac{\alpha_s}{2 \pi}C_F  \frac{1}{\hat{\epsilon}}  \left[ - 4 \epsilon \ln (\alpha) \ln \left(\frac{  \mu^{2}}{\left( \frac{\vec{p}_{h_2}}{x_{h_2}}- \frac{\vec{p}_{h_1}}{x_{h_1}} \right)^{2}}\right) + 4 \ln (\alpha)  \right. \\
& \left.  + 4 \epsilon \ln^2(\alpha) - 2 \ln (x_q x_{\bar{q}})+ 3 \right] + (h_1 \leftrightarrow h_2)  \numberthis[virtual div TT]
    \end{align*}}
while the finite part reads 
{\allowdisplaybreaks
\begin{align*}
& \frac{d \sigma_{1T T}^{q\bar{q} \rightarrow h_1 h_2}}{d x_{h_1}d_{h_2}d^d p_{h_1 \perp}d^d p_{h_2 \perp}} \bigg |_{\text{fin}} \\
& =  \frac{\alpha_{\mathrm{em}} }{( 2 \pi)^{4(d-1)} N_c}   \sum_{q} \int_{x_{h_1}}^1 \frac{d x_q }{x_q} \int_{x_{h_2}}^1 \frac{d x_{\bar{q}}}{x_{\bar{q}}}  \left( \frac{x_q}{x_{h_1}}\right)^d \left( \frac{x_{\bar{q}}}{x_{h_2}}\right)^d  \delta(1-x_q -x_{\bar{q}}) \\
& \times  Q_q^2 D_q^{h_1}\left(\frac{x_{h_1}}{x_q}, \mu_F\right) D_{\bar{q}}^{h_2}\left(\frac{x_{h_2}}{x_{\bar{q}}}, \mu_F\right)  {\cal F}_{TT} \\
& \times \frac{\alpha_s}{2 \pi}C_F  \frac{1}{\hat{\epsilon}} \left[2 \epsilon\ln \left(\frac{ \mu^{2}}{ x_q x_{\bar{q}} \left( \frac{\vec{p}_{h_2}}{x_{h_2}}- \frac{\vec{p}_{h_1}}{x_{h_1}}\right)^{2}}\right)  \ln (x_q x_{\bar{q}}) + \epsilon \ln^2 (x_q x_{\bar{q}}) \right. \\
& \left. - 3 \epsilon \ln \left(\frac{ \mu^{2}}{ x_q x_{\bar{q}} \left( \frac{\vec{p}_{h_2}}{x_{h_2}}- \frac{\vec{p}_{h_1}}{x_{h_1}}\right)^{2}}\right) - \frac{\pi^2}{3}\epsilon + 6 \epsilon\right] \\
& + \frac{\alpha_s }{4 \pi} \left(\frac{N_c^2 -1 }{N_c}\right) \frac{\alpha_{\mathrm{em}}}{(2\pi)^4 N_c} \sum_{q} \int_{x_{h_1}}^1 \frac{ d x_q }{x_q} \int_{h_2}^1 \frac{ d x_{\bar{q}}}{x_{\bar{q}}} \; \delta (1-x_q -x_{\bar{q}} ) \\ 
& \times \left(\frac{x_q}{x_{h_1}}\right)^d \left(\frac{x_{\bar{q}}}{x_{h_2}}\right)^d   Q_q^2 D_q^{h_1}\left(\frac{x_{h_1}}{x_q}, \mu_F\right) D_{\bar{q}}^{h_2}\left(\frac{x_{h_2}}{x_{\bar{q}}}, \mu_F\right) \\
&\times  \int d^d p_{1 \perp} d^d p_{2 \perp} \mathbf{F}\left(\frac{p_{12\perp}}{2}\right) \delta \left( \frac{x_q}{x_{h_1}} p_{h_1 \perp} -p_{1\perp} + \frac{x_{\bar{q}}}{x_{h_2}} p_{h_2 \perp} - p_{2 \perp} \right) \\
& \times \int d^d p_{1' \perp} d^d p_{2' \perp} \mathbf{F}^{*} \left(\frac{p_{1'2'\perp}}{2}\right) \delta \left( \frac{x_q}{x_{h_1}} p_{h_1 \perp} -p_{1'\perp} + \frac{x_{\bar{q}}}{x_{h_2}} p_{h_2 \perp} - p_{2' \perp} \right) \\
& \times  \varepsilon_{T i} \varepsilon_{T k}^* \left\{ \frac{3}{2} \frac{\left(\frac{x_q}{x_{h_1}} p_{h_1 \perp}- p_{1 \perp}\right)_r \left(\frac{x_q}{x_{h_1}} p_{h_1 \perp}- p_{1'\perp }\right)_l}{\left(\left(\frac{x_q}{x_{h_1}} \vec{p}_{h_1}- \vec{p}_{1 }\right)^2 + x_q x_{\bar{q} } Q^2 \right) \left(\left(\frac{x_q}{x_{h_1}} \vec{p}_{h_1}- \vec{p}_{1'}\right)^2 + x_q x_{\bar{q} } Q^2 \right)} \right. \\ 
&  \times \left[(x_{\bar{q}} -x_q)^2 g_{\perp}^{ri} g_{\perp}^{lk} - g_{\perp}^{rk} g_{\perp}^{li} + g_{\perp}^{rl} g_{\perp}^{ik}\right] \\
& \times  \left[ \ln \left( \frac{\mu^4}{x_q x_{\bar{q}} \left(\frac{\vec{p}_{h_2}}{x_{h_2}} - \frac{\vec{p}_{h_1}}{x_{h_1}}\right)^2 \left(\left(\frac{x_q}{x_{h_1}} \vec{p}_{h_1}- \vec{p}_{1 }\right)^2 + x_q x_{\bar{q} } Q^2 \right)  }\right) \right. \\ 
& \left. - \frac{x_q x_{\bar{q}}Q^2}{\left(\frac{x_q}{x_{h_1}} \vec{p}_{h_1}- \vec{p}_{1 }\right)^2  } \ln \left( \frac{x_q x_{\bar{q}} Q^2 }{\left(\frac{x_q}{x_{h_1}} \vec{p}_{h_1}- \vec{p}_{1 }\right)^2 + x_q x_{\bar{q} } Q^2  }\right) \right] +  \frac{1}{\left(\left(\frac{x_q}{x_{h_1}} \vec{p}_{h_1}- \vec{p}_{1 '}\right)^2 + x_q x_{\bar{q} } Q^2 \right) x_q x_{\bar{q}}} \\ 
& \times  \left(\int_{0}^{x_q} \left[(\phi_4)^{ik}_{TT}\right]_+ + \sum_{n=5,6} \int_{0}^{x_q} \left[(\phi_n)^{ij}_{TT}\right]_+ \bigg |_{\vec{p}_3 = \vec{0} } + (q \leftrightarrow \bar{q}) \right)   \left. + h.c. \bigg  |_{1\leftrightarrow 1', i \leftrightarrow k} \right\}  \\
& + (h_1 \leftrightarrow h_2) \numberthis[virtual finite TT] \,.
\end{align*}}

\section{NLO cross section: Real corrections}
\label{sec: RealDiv}
In this section, we will discuss the real corrections. Since, as explained above, the calculation is almost completely identical in the $LL$, $TL$, and $TT$ cases (apart from factors that do not affect the general strategy), we will show the details of the $LL$ case only. For the others, we will just report the final results. \\
The dipole-dipole partonic cross section is given by Eq.~(6.6) of Ref.~\cite{Boussarie:2016ogo}: 
\begin{equation}
\begin{aligned}
    d \hat{\sigma}_{3JI} &  = \frac{\alpha_s}{\mu^{2\epsilon}} \left( \frac{N_c^2 -1}{N_c}\right) \frac{\alpha_{\mathrm{em}}Q_q^2}{(2\pi)^{4(d-1)}N_c} \frac{(p_0^-)^2}{s^2 x_q'x_{\bar{q}}'} \varepsilon_{I\alpha} \varepsilon_{J\beta}^*  d x_q' d x_{\bar{q}}'   \delta (1-x_q'-x_{\bar{q}}'-x_g) d^d p_{q\perp}  d^d p_{\bar{q}\perp} \\
    & \times  \frac{d x_g  d^d p_{g\perp}}{x_g (2\pi)^d} \int d^d p_{1\perp} d^d p_{2\perp} \mathbf{F} \left(\frac{p_{12\perp}}{2}\right)\delta (p_{q1\perp} + p_{\bar{q}2\perp} + p_{g\perp}) \\
    & \times  \int d^d p_{1'\perp} d^d p_{2'\perp} \mathbf{F}^*\left(\frac{p_{1'2'\perp}}{2}\right)  \delta (p_{q1'\perp} + p_{\bar{q}2'\perp} + p_{g\perp}) \\
   &  \times \Phi_3^\alpha (p_{1\perp}, p_{2\perp}) \Phi_3^{\beta*} (p_{1'\perp}, p_{2'\perp}) \, , 
\end{aligned}
\end{equation}
where we introduce shorthand notation by suppressing summation over helicities of partons
\begin{equation}
    \Phi_3^\alpha (p_{1\perp}, p_{2\perp}) \Phi_3^{\beta*} (p_{1'\perp}, p_{2'\perp}) \equiv \sum_{\lambda_q, \lambda_g, \lambda_{\bar{q}} } \Phi_3^\alpha (p_{1\perp}, p_{2\perp}) \Phi_3^{\beta*} (p_{1'\perp}, p_{2'\perp}) \; .
    \label{eq:ShortHand}
\end{equation}
The impact factor has the form $\Phi_3^\alpha = \Phi_4^\alpha |_{\vec{p}_3 = 0} + \Tilde{\Phi}_3^\alpha$. Only the square of $\Tilde{\Phi}_3^\alpha$ provides divergences in the cross section and it is given by (B.3) in Ref.~\cite{Boussarie:2016ogo}. The $LL$ contribution reads
\begin{equation}
\label{eq: div real impact factor}
\begin{aligned}
   &    \Tilde{\Phi}_3^+(\vec{p}_1, \vec{p}_2) \Tilde{\Phi}_3^{+*}(\vec{p}_{1'}, \vec{p}_{2'}) \\
   &= \frac{8 x_q' x_{\bar{q}}' (p_\gamma^+)^4 \left( d x_g^2 + 4 x_q' (x_q' + x_g) \right)}{\left(Q^2 + \frac{\vec{p}_{\bar{q}2}^{\,2}}{x_{\bar{q}}'(1-x_{\bar{q}}')} \right) \left(Q^2 + \frac{\vec{p}_{\bar{q}2'}^{\,2}}{x_{\bar{q}}'(1-x_{\bar{q}}')} \right) (x_q' \vec{p}_g -x_g \vec{p}_q)^2 } \\ 
   & - \frac{8 x_q' x_{\bar{q}}'(p_\gamma^+)^4 \left(2 x_g -d x_g^2 + 4 x_q' x_{\bar{q}}' \right) \left(x_q' \vec{p}_g - x_g \vec{p}_q \right) \cdot \left( x_{\bar{q}}' \vec{p}_g - x_g \vec{p}_{\bar{q}} \right)}{\left(Q^2 + \frac{\vec{p}_{\bar{q}2'}^{\,2}}{x_{\bar{q}}'(1-x_{\bar{q}}')} \right) \left(Q^2 + \frac{\vec{p}_{q1}^{\,2}}{x_q' (1-x_q')} \right)  \left(x_q' \vec{p}_g - x_g \vec{p}_q \right)^2  \left( x_{\bar{q}}' \vec{p}_g - x_g \vec{p}_{\bar{q}} \right)^2 }
   \\
   & +  \frac{8 x_q' x_{\bar{q}}' (p_\gamma^+)^4 \left( d x_g^2 + 4 x_{\bar{q}}' (x_{\bar{q}}' + x_g) \right)}{\left(Q^2 + \frac{\vec{p}_{q1}^{\,2}}{x_q' (1-x_q')} \right) \left(Q^2 + \frac{\vec{p}_{q1'}^{\,2}}{x_q' (1-x_q')} \right) (x_{\bar{q}}' \vec{p}_g -x_g \vec{p}_{\bar{q}})^2 } \\ 
   & - \frac{8 x_q' x_{\bar{q}}' (p_\gamma^+)^4 \left(2 x_g -d x_g^2 + 4 x_q' x_{\bar{q}}' \right) \left(x_q'  \vec{p}_g - x_g \vec{p}_q \right) \cdot \left( x_{\bar{q}}' \vec{p}_g - x_g \vec{p}_{\bar{q}} \right)}{\left(Q^2 + \frac{\vec{p}_{q1'}^{\,2}}{x_q' (1-x_q')} \right) \left(Q^2 + \frac{\vec{p}_{\bar{q}2}^{\,2}}{x_{\bar{q}}' (1-x_{\bar{q}}')} \right)  \left(x_q' \vec{p}_g - x_g \vec{p}_q \right)^2  \left( x_{\bar{q}}' \vec{p}_g - x_g \vec{p}_{\bar{q}} \right)^2 }\,.
\end{aligned}
\end{equation}
The $TL$ contribution is
{\allowdisplaybreaks
\begin{align*}
&  \tilde{\Phi}_3^{+}(\vec{p}_{1},\vec{p}_{2}) \tilde{\Phi}_3^{i*}(\vec{p}_{1'},\vec{p}_{2'}) \\*
& = \frac{4 x_q'\left(p_\gamma^{+}\right)^3}{\left(x_q' +x_g\right)\left(Q^2+\frac{\vec{p}_{\bar{q} 2'}^{\,2}}{x_{\bar{q}}'\left(1-x_{\bar{q}}'\right)}\right)\left(Q^2+\frac{\vec{p}_{q1}^{\,2}}{x_q'\left(1-x_q'\right)}\right)}\left(\frac{ \left(x_q' p_{g\perp} - x_g p_{q\perp}\right)_\mu \left(x_{\bar{q}}' p_{g\perp} - x_g p_{\bar{q}\perp}\right)_\nu}{\left(x_q' \vec{p}_g - x_g \vec{p}_q\right)^2 \left(x_{\bar{q}}' \vec{p}_g - x_g \vec{p}_{\bar{q}}\right)^2}\right) \\ 
& \times \left[x_g\left(4 x_{\bar{q}}' +x_g d-2\right)\left(p_{\bar{q} 2' \perp}^\mu g_{\perp}^{i \nu}-p_{\bar{q} 2' \perp}^\nu g_{\perp}^{\mu i}\right)-\left(2 x_{\bar{q}}'-1\right)\left(4 x_q' x_{\bar{q}}'+x_g\left(2-x_g d\right)\right) g_{\perp}^{\mu \nu} p_{\bar{q} 2' \perp}^i\right] \\
& -\frac{4 x_q ' \left(p_\gamma^{+}\right)^3\left(2 x_{\bar{q}}' -1\right)\left(x_g^2 d+4 x_q' \left(x_q' +x_g\right)\right) p_{\bar{q} 2' \perp}^i}{\left(x_q'+x_g\right)\left(Q^2+\frac{\vec{p}_{\bar{q} 2'}^{\,2}}{x_{\bar{q}}'\left(1-x_{\bar{q}}'\right)}\right)\left(Q^2+\frac{\vec{p}_{\bar{q} 2}^{\,2}}{x_{\bar{q}}'\left(1-x_{\bar{q}}'\right)}\right) \left(x_q' \vec{p}_g- x_g \vec{p}_q\right)^2}
+(q \leftrightarrow \bar{q}) \,,
\numberthis[impact_factor_TL]
\end{align*}}
and finally, the $TT$ contribution reads
{\allowdisplaybreaks
\begin{align*}
&  \tilde{\Phi}_3^i(\vec{p}_1,\vec{p}_2) \tilde{\Phi}_3^{k*}(\vec{p}_{1'},\vec{p}_{2'}) \\
& =\frac{-2\left(p_\gamma^{+}\right)^2}{\left(x_q'+x_g\right)\left(x_{\bar{q}}'+x_g\right)\left(Q^2+\frac{\vec{p}_{\bar{q} 2}^{\,2}}{x_{\bar{q}}'\left(1-x_{\bar{q}}'\right)}\right)\left(Q^2+\frac{\vec{p}_{q 1^{\prime}}^{\,2}}{x_q'\left(1-x_q'\right)}\right)} \\
& \times  \left(\frac{ \left(x_q' p_{g\perp} - x_g p_{q\perp}\right)_\mu \left(x_{\bar{q}}' p_{g\perp} - x_g p_{\bar{q}\perp}\right)_\nu}{\left(x_q' \vec{p}_g - x_g \vec{p}_q\right)^2  \left(x_{\bar{q}}' \vec{p}_g - x_g \vec{p}_{\bar{q}}\right)^2}\right) \left\{ x_g ((d-4)) x_g -2) \left[p_{q 1^{\prime} \perp}^\nu\left(p_{\bar{q} 2 \perp}^\mu g_{\perp}^{i k}+p_{\bar{q} 2 \perp}^k g_{\perp}^{\mu i}\right) \right. \right. \\
& \left. +g_{\perp}^{\mu \nu}\left(\left(\vec{p}_{q 1^{\prime}} \cdot \vec{p}_{\bar{q} 2}\right) g_{\perp}^{i k}+p_{q 1^{\prime} \perp}^i p_{\bar{q} 2 \perp}^k\right) -g_{\perp}^{\nu k} p_{q 1^{\prime} \perp}^i p_{\bar{q} 2 \perp}^\mu -g_{\perp}^{\mu i} g_{\perp}^{\nu k}\left(\vec{p}_{q 1^{\prime}} \cdot \vec{p}_{\bar{q} 2}\right) \right] -g_{\perp}^{\mu \nu} \\
& \times  \left[ \left(2x_q' -1 \right) \left(2 x_{\bar{q}}' - 1\right) p_{q1'\perp}^k p_{\bar{q}2\perp}^i \left( 4 x_q' x_{\bar{q}}' + x_g (2 - x_g d)\right)  + 4 x_q' x_{\bar{q}}' ((\vec{p}_{q1'} \cdot \vec{p}_{\bar{q}2})g_\perp^{ik} + p_{q1'\perp}^i p_{\bar{q}2\perp}^k  )\right] \\
& + \left( p_{q1'\perp}^\mu p_{\bar{q}2\perp}^\nu g_\perp^{ik} - p_{q1'\perp}^\mu p_{\bar{q}2\perp}^k g_\perp^{\nu i } - p_{q1'\perp}^i p_{\bar{q}2\perp}^\nu g_\perp^{\mu k } - g_\perp^{\mu k } g_\perp^{\nu i } (\vec{p}_{q1'} \cdot \vec{p}_{\bar{q}2} ) \right) \\ 
& \times x_g ((d-4)x_g + 2) + x_g (2x_{\bar{q}}' - 1 ) (x_g d + 4 x_q' -2 ) \left( g_\perp^{\mu k } p_{q1'\perp}^\nu - g_\perp^{\nu k} p_{q1'\perp}^\mu \right) p_{\bar{q}2\perp}^i \\
& \left.  + x_g (2 x_q' -1 ) p_{q1'\perp}^k (4 x_{\bar{q}}' + x_g d -2) \left( g_\perp^{\nu i } p_{\bar{q}2\perp}^\mu -g_\perp^{\mu i } p_{\bar{q}2\perp}^\nu \right) \right\} \\
& - \frac{2 x_q' (p_\gamma^+)^2 (x_g^2 d + 4x_q'(x_q'+ x_g)) \left( (\vec{p}_{\bar{q}2} \cdot \vec{p}_{\bar{q}2'}) g_\perp^{ik}-(1-2x_{\bar{q}}')^2 p_{\bar{q}2\perp}^i p_{\bar{q}2'\perp}^k + p_{\bar{q}2'\perp}^i p_{\bar{q}2\perp}^k\right) }{x_{\bar{q}}' (x_q' + x_g)^2 \left(Q^2 + \frac{\vec{p}_{\bar{q}2}^{\,2} }{x_{\bar{q}}' (1 - x_{\bar{q}}')} \right) \left(Q^2 + \frac{\vec{p}_{\bar{q}2'}^{\,2}}{x_{\bar{q}}' (1 - x_{\bar{q}}')} \right) \left(x_q' \vec{p}_{g} - x_g \vec{p}_q \right)^2 }    \\
& + (q \leftrightarrow \bar{q}) \numberthis[impact_factor_TT] \,.
\end{align*}} 
The divergent part of the $LL$ partonic cross section from real emission is given by
%
\begin{align*}
   \left. d \hat{\sigma}_{3LL}\right|_{div} &=  \frac{4 \alpha_{\mathrm{em}} Q^2}{(2\pi)^{4(d-1)}N_c} Q_q^2 d x_q' d x_{\bar{q}}' \delta(1-x_q'-x_{\bar{q}}' -x_g) d^d p_{q\perp} d^d p_{\bar{q}\perp}  \frac{\alpha_s C_F}{\mu^{2\epsilon}} \frac{d x_g}{x_g} \frac{d^d p_{g\perp}}{(2\pi)^d} \\
    & \times  \int d^d p_{1\perp} d^d p_{2\perp}  \delta (p_{q1\perp} + p_{\bar{q}2\perp} + p_{g\perp})  \; \mathbf{F} \left(\frac{p_{12\perp}}{2}\right)  \\
    & \times \int d^d p_{1'\perp} d^d p_{2'\perp} \delta (p_{q1'\perp} + p_{\bar{q}2'\perp} + p_{g\perp}) \; \mathbf{F}^*\left(\frac{p_{1'2'\perp}}{2}\right)   \\
    & \times  \left\{   \frac{ d x_g^2 + 4 x_q' (x_q' + x_g) }{\left(Q^2 + \frac{\vec{p}_{\bar{q}2}^{\,2}}{x_{\bar{q}}'(1-x_{\bar{q}}')} \right) \left(Q^2 + \frac{\vec{p}_{\bar{q}2'}^{\,2}}{x_{\bar{q}}'(1-x_{\bar{q}}')} \right) (x_q' \vec{p}_g -x_g \vec{p}_q)^2 } \right. \\ 
   & - \frac{ \left(2 x_g -d x_g^2 + 4 x_q' x_{\bar{q}}'\right) \left(x_q' \vec{p}_g - x_g \vec{p}_q \right) \cdot \left( x_{\bar{q}}' \vec{p}_g - x_g \vec{p}_{\bar{q}} \right)}{\left(Q^2 + \frac{\vec{p}_{\bar{q}2'}^{\,2}}{x_{\bar{q}}'(1-x_{\bar{q}}')} \right) \left(Q^2 + \frac{\vec{p}_{q1}^{\,2}}{x_q' (1-x_q')} \right)  \left(x_q' \vec{p}_g - x_g \vec{p}_q \right)^2  \left( x_{\bar{q}}' \vec{p}_g - x_g \vec{p}_{\bar{q}} \right)^2 }
   \\
   & +  \frac{ d x_g^2 + 4 x_{\bar{q}}' (x_{\bar{q}}' + x_g) }{\left(Q^2 + \frac{\vec{p}_{q1}^{\,2}}{x_q' (1-x_q')} \right) \left(Q^2 + \frac{\vec{p}_{q1'}^{\,2}}{x_q' (1-x_q')} \right) (x_{\bar{q}}' \vec{p}_g -x_g \vec{p}_{\bar{q}})^2 } \\ 
   & \left. - \frac{ \left(2 x_g -d x_g^2 + 4 x_q' x_{\bar{q}}'\right) \left(x_q' \vec{p}_g - x_g \vec{p}_q \right) \cdot \left( x_{\bar{q}}' \vec{p}_g - x_g \vec{p}_{\bar{q}} \right)}{\left(Q^2 + \frac{\vec{p}_{q1'}^{\,2}}{x_q' (1-x_q')} \right) \left(Q^2 + \frac{\vec{p}_{\bar{q}2}^{\,2}}{x_{\bar{q}}' (1-x_{\bar{q}}')} \right)  \left(x_q' \vec{p}_g - x_g \vec{p}_q \right)^2  \left( x_{\bar{q}}' \vec{p}_g - x_g \vec{p}_{\bar{q}} \right)^2 } \right \} \; .
   \\  \numberthis[real_div_LL]
\end{align*}

When two partons labeled $i$ and $j$ become collinear, the variable
\beqa
\vec{A}_{ij} = x_i \vec{p}_j - x_j \vec{p}_i
\eqa
vanishes. In the present case, 
the first term in the bracket of Eq.~\eqref{eq:real_div_LL} gives the collinear divergences ($\vec{A}_{qg}^2 \to 0$, i.e. quark-gluon channel)  and the third ($\vec{A}_{\bar{q}g}^2 \to 0$, i.e. antiquark-gluon channel). 

For  the $LT$, the relevant divergent squared impact factor is \eqref{eq:impact_factor_TL} and for the $TT$, it is \eqref{eq:impact_factor_TT}.
 
\subsection{Fragmentation from quark and anti-quark}
\label{sec:qqbarfrag}

As explained above, there are several contributions to the final cross section that contain divergences. In this section, we deal with extracting the soft and collinear divergences associated with the contribution (b) in Fig. \ref{fig:sigma-NLO}. This contribution corresponds to the situation in which the quark and the anti-quark fragment and there is an additional emission of a gluon with respect to the LO case. Below,
\begin{itemize}
    \item[\textbullet] We compute the collinear divergence of the diagram (1) of Fig. \ref{fig:NLO-b-div} and show that it is removed by the + prescription part of the first term of Eq. \eqref{eq:ct_LL}. 
    \item[\textbullet] Similarly, we calculate the collinear divergence of the diagram (3) of Fig. \ref{fig:NLO-b-div} and show that it is removed by the + prescription part of the third term of Eq. \eqref{eq:ct_LL}.
    \item[\textbullet] We extract the soft divergences of diagrams (1), (2), (3), (4) of Fig. \ref{fig:NLO-b-div} and discuss the complete cancellation of divergences of this contribution.
\end{itemize}
The calculations of the collinear divergences is done by Fourier transforming the $\mathbf{F} (\frac{p_{12\perp}}{2})$, as defined in \eqref{eq: FT F}, and by using the identity  (see eg Ref.~\cite{Chirilli:2012jd}):
\begin{equation}
\label{eq:expo}
\frac{1}{\mu^{2\epsilon}} \int d^{2+2 \epsilon} q_\perp e^{-i q_\perp \cdot r_\perp} \frac{1}{q_\perp^2} = \pi \left( \frac{4\pi}{\mu^2  r_\perp^2} \right)^{\epsilon} \Gamma (\epsilon) \,.
\end{equation}
We also have to change variables from the usual fraction of longitudinal photon momentum of the partons $(x_i',x_g)$ with $i= q,\bar{q}$, in the spirit of the definition \eqref{xq-xqbar}, as used in Eq.~\eqref{eq:real_div_LL}, to the variable of the fraction of longitudinal photon momentum of the parent parton $x_i$ and longitudinal fraction $\beta$  with respect to the parent parton and not with respect to the photon anymore:
\beqa
(x_i',x_g) \rightarrow (x_i,\beta) \qquad \hbox{ with } \qquad x_i' = \beta x_i \,.
\eqa
Changing variables is necessary to be able to compare with \eqref{eq:ct_LL}, \eqref{eq:ct TL }, and \eqref{eq:ct TT }. Note that to make notations lighter, we will remove the apex ' when one particle is a spectator, see Figs.~\ref{fig:kinematics_qg} and~\ref{fig:kinematics_qbarg}.

The Fourier transform of $\mathbf{F} \left(\frac{p_{12\perp}}{2}\right)$ is necessary in order to be able to integrate over the transverse momentum of the spectator parton, as it allows for the complete factorization of this momentum from this non-perturbative function. 

\subsubsection{Collinear contributions: $q$-$g$ splitting}
\label{sec:qqbarfragColl-qg}

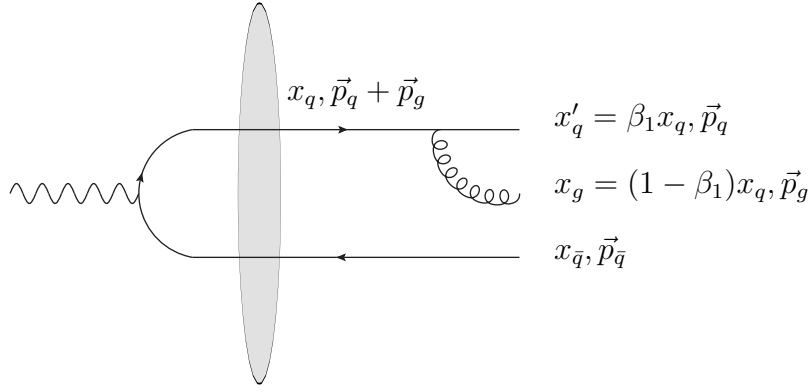
\begin{figure}
\begin{picture}(420,130)
\put(-50,0){\includegraphics[scale=0.50]{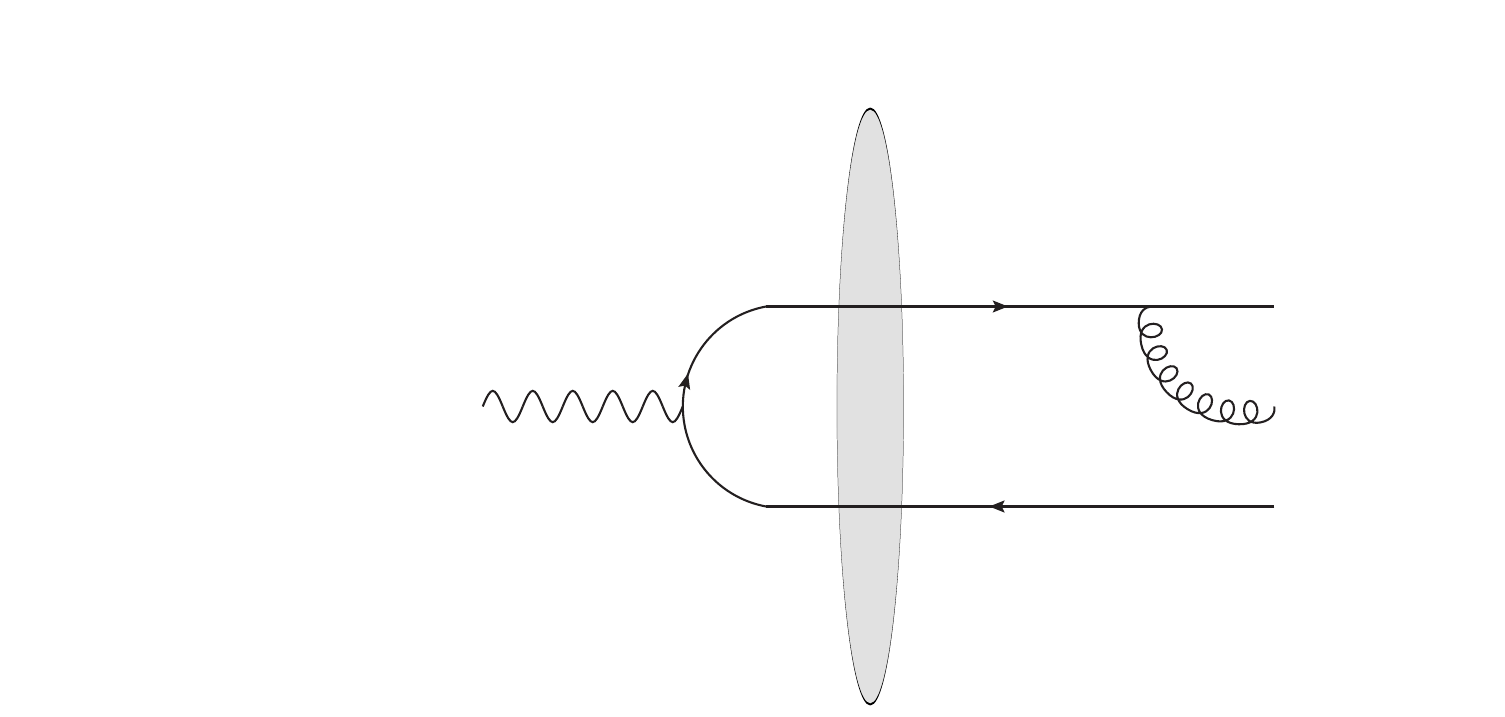}}
\put(170,110){$x_q, \vec{p}_q+\vec{p}_g$}
\put(270,100){$x'_q = \beta_1 x_q, \vec{p}_q$}
\put(270,75){$x_g=(1-\beta_1) x_q, \vec{p}_g$}
\put(270,50){$x_{\bar{q}}, \vec{p}_{\bar{q}}$}
\end{picture}
\caption{Kinematics for the $q-g$ splitting contribution. We indicate the longitudinal fraction of momentum carried by the partons as well as their transverse momenta.}
\label{fig:kinematics_qg}
\end{figure}
We use the kinematics illustrated in Fig.~\ref{fig:kinematics_qg}.
The term in \eqref{eq: div real impact factor} considered for this collinear contribution is the first one in the bracket:
\newpage
{\allowdisplaybreaks
\begin{align*}
 &  d \sigma_{3LL}^{q \bar{q} \rightarrow h_1 h_2}|_{\text{coll. qg}} \\
    & = d x_{h_1} d x_{h_2} \frac{4  \alpha_{\mathrm{em}} Q^2}{(2\pi)^{4(d-1)} N_c} \sum_{q}  \int_{x_{h_1}}^1 \frac{d x_q'}{x_q'} \int_{x_{h_2}}^1 \frac{d x_{\bar{q}}}{x_{\bar{q}}} \int_{\alpha}^1 \frac{d x_g}{x_g} \delta(1-x_q'-x_{\bar{q}}-x_g) \\
    & \times Q_q^2 D_q^{h_1}\left(\frac{x_{h_1}}{x_q'}, \mu_F\right) D_{\bar{q}}^{h_2}\left(\frac{x_{h_2}}{x_{\bar{q}}}, \mu_F\right) d^d p_{q\perp} d^d p_{\bar{q}\perp}   \frac{\alpha_s}{\mu^{2\epsilon}}C_F  \frac{d^d p_{g\perp}}{(2\pi)^d}   \\ 
     & \times \int d^d p_{1\perp}  d^d p_{2\perp}    \delta(p_{q1\perp} + p_{\bar{q}2\perp} + p_{g\perp}) \; \mathbf{F} \left(\frac{p_{12\perp}}{2}\right)  \\
     & \times \int d^d p_{1'\perp}   d^d p_{2'\perp} \delta(p_{q1'\perp} + p_{\bar{q}2'\perp}  + p_{g\perp}) \; \mathbf{F}^*\left(\frac{p_{1'2'\perp}}{2}\right)  \\
    & \times \frac{(d x_g^2 + 4 x_q' (x_q' + x_g)) x_{\bar{q}}^2 (1-x_{\bar{q}})^2}{\left( x_{\bar{q}} (1-x_{\bar{q}})Q^2 + \vec{p}_{\bar{q}2}^{\,2}\right) \left( x_{\bar{q}} (1-x_{\bar{q}})Q^2 + \vec{p}_{\bar{q}2'}^{\,2}\right) (x_q' \vec{p}_g - x_g \vec{p}_q)^2}  + (h_1 \leftrightarrow h_2) \\ 
    & = d x_{h_1} d x_{h_2} d^d p_{h_1 \perp} d^d p_{h_2\perp} \frac{4  \alpha_{\mathrm{em}} Q^2}{(2\pi)^{4(d-1)} N_c} \sum_{q} \int_{x_{h_1}}^1 \frac{d x_q'}{x_q'} \int_{\alpha}^1 \frac{d x_g}{x_g} \int_{x_{h_2}}^1 \frac{d x_{\bar{q}}}{x_{\bar{q}}} \delta(1-x_q'-x_{\bar{q}}-x_g) \\
    & \times   \left(\frac{x_q'}{x_{h_1}}\right)^d \left(\frac{x_{\bar{q}}}{x_{h_2}}\right)^d  Q_q^2 D_q^{h_1}\left(\frac{x_{h_1}}{x_q'}, \mu_F\right) D_{\bar{q}}^{h_2}\left(\frac{x_{h_2}}{x_{\bar{q}}}, \mu_F\right)  \frac{\alpha_s}{\mu^{2\epsilon}} C_F  \frac{d^d p_{g\perp}}{(2\pi)^d} \\
     & \times \int d^d p_{1\perp}  d^d p_{2\perp}    \delta\left(\frac{x_q'}{x_{h_1}} p_{h_1\perp} -p_{1\perp} + \frac{x_{\bar{q}}}{x_{h_2}} p_{h_2 \perp} -p_{2\perp} + p_{g\perp}\right)  \mathbf{F}\left(\frac{p_{12\perp}}{2}\right) \\
     & \times \int d^d p_{1'\perp}  d^d p_{2'\perp}    \delta \left(\frac{x_q'}{x_{h_1}} p_{h_1\perp} -p_{1'\perp} + \frac{x_{\bar{q}}}{x_{h_2}}p_{h_2 \perp} -p_{2'\perp} + p_{g\perp}\right)  \mathbf{F}^*\left(\frac{p_{1'2'\perp}}{2}\right) \\
     & \times \frac{(d x_g^2 + 4 x_q' (x_q' + x_g)) x_{\bar{q}}^2 (1-x_{\bar{q}})^2}{\left( x_{\bar{q}} (1-x_{\bar{q}})Q^2 \hspace{-0.1 cm} + \hspace{-0.05 cm} \left(\frac{x_{\bar{q}}}{x_{h_2}} \vec{p}_{h_2}-\vec{p}_{2} \hspace{-0.05 cm} \right)^2 \hspace{-0.05 cm} \right) \left( x_{\bar{q}} (1-x_{\bar{q}})Q^2 \hspace{-0.1 cm} + \hspace{-0.05 cm} \left(\frac{x_{\bar{q}}}{x_{h_2}} \vec{p}_{h_2}-\vec{p}_{2'} \hspace{-0.05 cm} \right)^2 \hspace{-0.05 cm} \right) \left(x_q' \vec{p}_g - x_g \frac{x_{q}'}{x_{h_1}}\vec{p}_{h_1} \hspace{-0.05 cm} \right)^2} \\
     & + (h_1 \leftrightarrow h_2)\\ 
      & = d x_{h_1} d x_{h_2} d^d p_{h_1 \perp} d^d p_{h_2\perp} \frac{4  \alpha_{\mathrm{em}} Q^2}{(2\pi)^{4(d-1)} N_c} \sum_{q} \int_{x_{h_1}}^1 \frac{d x_q'}{x_q'} \int_{\alpha}^1 \frac{d x_g}{x_g} \int_{x_{h_2}}^1 \frac{d x_{\bar{q}}}{x_{\bar{q}}} \delta(1-x_q'-x_{\bar{q}}-x_g) \\
    & \times   \left(\frac{x_q'}{x_{h_1}}\right)^d \left(\frac{x_{\bar{q}}}{x_{h_2}}\right)^d  Q_q^2 D_q^{h_1}\left(\frac{x_{h_1}}{x_q'}, \mu_F\right) D_{\bar{q}}^{h_2}\left(\frac{x_{h_2}}{x_{\bar{q}}}, \mu_F\right)  \frac{\alpha_s}{\mu^{2\epsilon}} C_F  \frac{d^d p_{g\perp}}{(2\pi)^d} \\
     & \times \int d^d p_{2\perp}   \;  \mathbf{F} \left(\frac{x_q'}{2 x_{h_1}} p_{h_1\perp} + \frac{x_{\bar{q}}}{2 x_{h_2}} p_{h_2 \perp} -p_{2\perp} + \frac{p_{g\perp}}{2}\right)  \\
     & \times \int  d^d p_{2'\perp}  \;   \mathbf{F}^* \left(\frac{x_q'}{2x_{h_1}} p_{h_1\perp} + \frac{x_{\bar{q}}}{ 2 x_{h_2}}p_{h_2 \perp} -p_{2'\perp} + \frac{p_{g\perp}}{2}\right)  \\
     & \times \hspace{-0.1 cm} \frac{(d x_g^2 + 4 x_q' (x_q' + x_g)) x_{\bar{q}}^2 (1-x_{\bar{q}})^2}{\left( x_{\bar{q}} (1-x_{\bar{q}})Q^2 + \hspace{-0.1 cm} \left(\frac{x_{\bar{q}}}{x_{h_2}} \vec{p}_{h_2}-\vec{p}_{2}\right)^2 \hspace{-0.05 cm} \right) \hspace{-0.1 cm} \left( x_{\bar{q}} (1-x_{\bar{q}})Q^2 \hspace{-0.1 cm} +\left(\frac{x_{\bar{q}}}{x_{h_2}} \vec{p}_{h_2}-\vec{p}_{2'}\right)^2 \hspace{-0.05 cm} \right) \hspace{-0.1 cm} \left(x_q' \vec{p}_g - x_g \frac{x_{q}'}{x_{h_1}}\vec{p}_{h_1} \hspace{-0.05 cm} \right)^2} \\
     & + (h_1 \leftrightarrow h_2) \; .  
\end{align*}}
After performing the change of variable
\begin{equation}
\begin{split}
        & x_q' = \beta_1 x_q \nonumber \\ 
        & x_g = (1-\beta_1) x_q 
\end{split}
\label{eq:Transbeta}
\end{equation}
and using the
Jacobian $d x_q' d x_g = x_q \,d x_q d \beta_1 $
we can rewrite the longitudinal integration in the symbolic form
    \begin{align*}
         & \int_{x_{h_1}}^1 \frac{d x_q'}{x_q'} \int_{x_{h_2}}^1 \frac{d x_{\bar{q}} }{x_{\bar{q}}} \int_{\alpha}^1 \frac{d x_g}{x_g} \delta(1-x_q'-x_{\bar{q}}-x_g) \\
        & = \int_{x_{h_1}}^1 \frac{d x_q'}{x_q'} \int_{\alpha}^1 \frac{d x_g}{x_g} \int_{- \infty}^{+\infty} \frac{d x_{\bar{q}}}{x_{\bar{q}}} \theta(x_{\bar{q}}-x_{h_2}) \theta(1- x_{\bar{q}}) \delta(1-x_q'-x_{\bar{q}}-x_g) \\
        &= \int_{x_{h_1}}^1 \frac{d x_q'}{x_q'} \int_{\alpha}^1 \frac{d x_g}{x_g}  \theta(1-x_q'-x_g-x_{h_2}) \theta(x_q'+x_g) \frac{1}{1-x_q'-x_g} \\
        &= \int_{x_{h_1}}^{1-x_{h_2}} \frac{d x_q}{x_q} \frac{1}{1-x_q} \int_{\frac{x_{h_1}}{x_q}}^{1-\frac{\alpha}{x_q}}\frac{d\beta_1}{ \beta_1  (1-\beta_1)} \,.\numberthis[change_variable_integral]
    \end{align*}
After this manipulation, we obtain
{\allowdisplaybreaks
\begin{align*}
& \frac{d \sigma_{3LL}^{q \bar{q} \rightarrow h_1 h_2}}{ d x_{h_1} d x_{h_2} d^d p_{h_1 \perp} d^d p_{h_2\perp} } \Bigg |_{\text{coll qg}} \\
&=  \frac{4  \alpha_{\mathrm{em}} Q^2}{(2\pi)^{4(d-1)} N_c} \sum_{q} \int_{x_{h_1}}^{1-x_{h_2}} d x_q x_q (1-x_q) \left(\frac{ x_q}{x_{h_1}}\right)^d \left(\frac{1-x_q}{x_{h_2}}\right)^d  \\
&\times \int_{\frac{x_{h_1}}{x_q}}^{1-\frac{\alpha}{x_q}}\frac{d\beta_1}{\beta_1}    Q_q^2 D_q^{h_1}\left(\frac{x_{h_1}}{\beta_1 x_q}, \mu_F\right) D_{\bar{q}}^{h_2}\left(\frac{x_{h_2}}{1-x_q}, \mu_F\right)  \\
 & \times \int   d^d p_{2\perp}  \int d^d z_{1\perp}   \frac{e^{i z_{1\perp}\cdot \left(\frac{\beta_1 x_q}{2 x_{h_1}} p_{h_1\perp} + \frac{1-x_q}{2 x_{h_2}} p_{h_2 \perp} -p_{2\perp} \right)} F(z_{1\perp})}{x_q (1-x_q)Q^2 +\left(\frac{1-x_q}{x_{h_2}} \vec{p}_{h_2}-\vec{p}_{2}\right)^2} \\
& \times \int   d^d p_{2'\perp}  \int d^d z_{2\perp}   \frac{e^{-i z_{2\perp}\cdot \left(\frac{\beta_1 x_q}{2 x_{h_1}} p_{h_1\perp} + \frac{1-x_q}{2 x_{h_2}} p_{h_2 \perp} -p_{2'\perp} \right)} F^*(z_{2\perp})}{x_q (1-x_q)Q^2 +\left(\frac{1-x_q}{x_{h_2}} \vec{p}_{h_2}-\vec{p}_{2'}\right)^2} \\
&  \times \frac{2(1+\beta_1^2)+ 2 \epsilon (1-\beta_1)^2 + 4 \epsilon (1+\beta_1^2)  \ln \beta_1  }{ 1-\beta_1}  \\
& \times e^{i \left(\frac{z_{1\perp}-z_{2\perp}}{2}\right)\cdot \frac{(1-\beta_1)x_q}{x_{h_1}} p_{h_1\perp}}   \frac{\alpha_s}{\mu^{2\epsilon}} C_F  \int \frac{d^d p_{g\perp}}{(2\pi)^d} \frac{e^{i \left(\frac{z_{1\perp}-z_{2\perp}}{2}\right)\cdot p_{g\perp}}}{\left(\vec{p}_g  \right)^2} + (h_1 \leftrightarrow h_2 ).
\numberthis[dsigmaLL-qg-collinear]
\end{align*} }

\noindent
The integral over $p_{g\perp}$ gives, using Eq.~\eqref{eq:expo},
\begin{align*}
\mu^{-2\epsilon}\int \frac{d^d p_{g\perp}}{(2\pi)^d} \frac{e^{i \left(\frac{z_{1\perp}-z_{2\perp}}{2}\right)\cdot p_{g\perp}}}{\left(\vec{p}_g  \right)^2} &  = \frac{1}{(2\pi)^d} \pi \mu^{-2\epsilon} \left[ \frac{\left( \frac{z_{1\perp}-z_{2\perp}}{2}\right)^2}{4\pi}\right]^{-\epsilon} \Gamma (\epsilon) \\*
& = \frac{1}{4\pi} \left( \frac{1}{\hat{\epsilon}} + \ln \left( \frac{c_0^2}{\left(\frac{z_{1\perp} - z_{2\perp}}{2}\right)^2 \mu^2}\right) \right) + O(\epsilon) \\
& = \frac{1}{4\pi} \frac{1}{\hat{\epsilon}} \left(\frac{c_0^2}{\left(\frac{z_{1\perp}-z_{2\perp}}{2}\right)^2 \mu^2}\right)^\epsilon + O(\epsilon)
\numberthis[integration-over-pgperp]
\end{align*}
where $c_0 = 2 e^{-\gamma_E}.$ This leads to
\begin{align*}
& \frac{d \sigma_{3LL}^{q \bar{q} \rightarrow h_1 h_2}}{ d x_{h_1} d x_{h_2} d^d p_{h_1 \perp} d^d p_{h_2\perp} } \Bigg |_{\text{coll qg.}} \\
& =  \frac{4  \alpha_{\mathrm{em}} Q^2}{(2\pi)^{4(d-1)} N_c}  \sum_{q} \int_{x_{h_1}}^{1} d x_q  \int_{x_{h_2}}^{1} d x_{\bar{q}} x_q x_{\bar{q}} \delta(1-x_q -x_{\bar{q}}) \left(\frac{x_q}{x_{h_1}}\right)^d \left(\frac{x_{\bar{q}}}{x_{h_2}}\right)^d \\
& \times \int   d^d p_{2\perp}  \int d^d z_{1\perp}   \frac{e^{i z_{1\perp}\cdot \left(\frac{x_q}{2 x_{h_1}} p_{h_1\perp} + \frac{1-x_q}{2 x_{h_2}} p_{h_2 \perp} -p_{2\perp} \right)} F(z_{1\perp})}{x_q (1-x_q)Q^2 +\left(\frac{1-x_q}{x_{h_2}} \vec{p}_{h_2}-\vec{p}_{2}\right)^2} \\
& \times \int   d^d p_{2'\perp}  \int d^d z_{2\perp}   \frac{e^{-i z_{2\perp}\cdot \left(\frac{x_q}{2 x_{h_1}} p_{h_1\perp} + \frac{1-x_q}{2 x_{h_2}} p_{h_2 \perp} -p_{2'\perp} \right)} F^*(z_{2\perp})}{x_q (1-x_q)Q^2 +\left(\frac{1-x_q}{x_{h_2}} \vec{p}_{h_2}-\vec{p}_{2'}\right)^2} \\
& \times  \int_{\frac{x_{h_1}}{x_q}}^{1-\frac{\alpha}{x_q}}\frac{d\beta_1}{\beta_1}  Q_q^2 D_q^{h_1}\left(\frac{x_{h_1}}{\beta_1 x_q}, \mu_F\right) D_{\bar{q}}^{h_2}\left(\frac{x_{h_2}}{1-x_q}, \mu_F\right) \\
& \times \frac{\alpha_s C_F}{2\pi} \left[ \frac{1}{\hat{\epsilon}} \left(\frac{c_0^2}{\left(\frac{z_{1\perp}-z_{2\perp}}{2}\right)^2 \mu^2}\right)^\epsilon \frac{1+ \beta_1^2}{1-\beta_1}  + \frac{(1-\beta_1)^2 + 2 (1+ \beta_1^2) \ln \beta_1 }{(1-\beta_1)} \right] + (h_1 \leftrightarrow h_2 ). \numberthis[coll_qg_qqbar_FF_1]
\end{align*}
Here in Eq.~\eqref{eq:coll_qg_qqbar_FF_1}
we have put back the integral in $x_{\bar{q}} $ using
\begin{equation}
 \label{eq: xq xbarq }
\begin{aligned}
\int_{x_{h_1}}^1 d x_q \int_{x_{h_2}}^1 d x_{\bar{q}}\, \delta(1-x_q-x_{\bar{q}}) & =  \int_{x_{h_1}}^1 d x_q \int_{-\infty}^{+\infty} d x_{\bar{q}} \, \delta(1-x_q-x_{\bar{q}}) \theta(1-x_{\bar{q}}) \theta(x_{\bar{q}}-x_{h_2}) \\
&= \int_{x_{h_1}}^{1-x_{h_2}} d x_q  
\end{aligned}
\end{equation}
in order to have the same form as in the LO cross section \eqref{eq:LL-LO}.
  
Now, to separate the collinear and soft contribution, we introduce the plus prescription, as defined in Eq.~\eqref{eq: plus prescription}, and after, we expand the factor $ \frac{1}{\epsilon} \left(\frac{c_0^2}{\left(\frac{z_{1\perp}-z_{2\perp}}{2}\right)^2 \mu^2}\right)^\epsilon$ within accuracy of order $\epsilon^0$, only in those terms whose integrand is  safe in the limit $\beta_1 \rightarrow 1$:   
\begin{align*}
\allowdisplaybreaks
& \frac{d \sigma_{3LL}^{q \bar{q} \rightarrow h_1 h_2}}{ d x_{h_1} d x_{h_2} d^d p_{h_1 \perp} d^d p_{h_2\perp} } \Bigg |_{\text{coll qg.}} \\
& =  \frac{4  \alpha_{\mathrm{em}} Q^2}{(2\pi)^{4(d-1)} N_c}  \sum_{q} \int_{x_{h_1}}^{1} d x_q  \int_{x_{h_2}}^{1} d x_{\bar{q}} \; x_q x_{\bar{q}} \delta(1-x_q -x_{\bar{q}}) \left(\frac{x_q}{x_{h_1}}\right)^d \left(\frac{x_{\bar{q}}}{x_{h_2}}\right)^d \\
& \times \int   d^d p_{2\perp}  \int d^d z_{1\perp}   \frac{e^{i z_{1\perp}\cdot \left(\frac{x_q}{2 x_{h_1}} p_{h_1\perp} + \frac{x_{\bar{q}}}{2 x_{h_2}} p_{h_2 \perp} -p_{2\perp} \right)} F(z_{1\perp})}{x_q x_{\bar{q}}Q^2 +\left(\frac{x_{\bar{q}}}{x_{h_2}} \vec{p}_{h_2}-\vec{p}_{2}\right)^2} \\
& \times \int   d^d p_{2'\perp}  \int d^d z_{2\perp}   \frac{e^{-i z_{2\perp}\cdot \left(\frac{x_q}{2 x_{h_1}} p_{h_1\perp} + \frac{x_{\bar{q}}}{2 x_{h_2}} p_{h_2 \perp} -p_{2'\perp} \right)} F^*(z_{2\perp})}{x_q x_{\bar{q}}Q^2 +\left(\frac{x_{\bar{q}}}{x_{h_2}} \vec{p}_{h_2}-\vec{p}_{2'}\right)^2} \\
& \times \left \{ \int_{\frac{x_{h_1}}{x_q}}^{1}\frac{d\beta_1}{\beta_1}  Q_q^2 D_q^{h_1}\left(\frac{x_{h_1}}{\beta_1 x_q}, \mu_F\right) D_{\bar{q}}^{h_2}\left(\frac{x_{h_2}}{x_{\bar{q}}}, \mu_F\right) \frac{\alpha_s C_F}{2\pi} \frac{1}{\hat{\epsilon}}  \frac{1+ \beta_1^2}{(1-\beta_1)_+} \right. \\
& + \int_{\frac{x_{h_1}}{x_q}}^{1-\frac{\alpha}{x_q}} d\beta_1  Q_q^2 D_q^{h_1}\left(\frac{x_{h_1}}{ x_q}, \mu_F\right) D_{\bar{q}}^{h_2}\left(\frac{x_{h_2}}{x_{\bar{q}}}, \mu_F\right) \frac{\alpha_s C_F}{2\pi} \frac{1}{\hat{\epsilon}} \left(\frac{c_0^2}{\left(\frac{z_{1\perp}-z_{2\perp}}{2}\right)^2 \mu^2}\right)^\epsilon \frac{2}{1-\beta_1}  \\
& - Q_q^2 D_q^{h_1}\left(\frac{x_{h_1}}{x_q},\mu_F\right) D_{\bar{q}}^{h_2}\left(\frac{x_{h_2}}{x_{\bar{q}}},\mu_F \right) \frac{\alpha_s C_F}{2\pi} \frac{1}{\hat{\epsilon}} 2 \ln \left(1-\frac{x_{h_1}}{x_q}\right) \\
& + \int_{\frac{x_{h_1}}{x_q}}^{1}\frac{d\beta_1}{\beta_1}  Q_q^2 D_q^{h_1}\left(\frac{x_{h_1}}{\beta_1 x_q}, \mu_F\right) D_{\bar{q}}^{h_2}\left(\frac{x_{h_2}}{x_{\bar{q}}}, \mu_F\right) \frac{\alpha_s C_F}{2\pi} \left[ \ln \left( \frac{c_0^2}{\left(\frac{z_{1\perp}-z_{2\perp}}{2}\right)^2 \mu^2} \right) \frac{1+ \beta_1^2}{(1-\beta_1)_+} \right. \\ & \left. 
+ \frac{(1-\beta_1)^2 + 2 (1+ \beta_1^2) \ln \beta_1 }{(1-\beta_1)} \right] - Q_q^2 D_q^{h_1}\left(\frac{x_{h_1}}{x_q},\mu_F\right) D_{\bar{q}}^{h_2}\left(\frac{x_{h_2}}{x_{\bar{q}}},\mu_F \right) \\
& \left. \times \frac{\alpha_s C_F}{2\pi} 2 \ln \left(1-\frac{x_{h_1}}{x_q}\right) \ln \left(\frac{c_0^2}{\left(\frac{z_{1\perp}-z_{2\perp}}{2}\right)^2 \mu^2}\right) \right \} + (h_1 \leftrightarrow h_2 ) \\
&= \frac{d \sigma_{3LL}^{q \bar{q} \rightarrow h_1 h_2}}{ d x_{h_1} d x_{h_2} d p_{h_1 \perp} d^d p_{h_2\perp} } \bigg |_{\text{coll. qg div}}     + \frac{d \sigma_{3LL}^{q \bar{q} \rightarrow h_1 h_2}}{ d x_{h_1} d x_{h_2} d p_{h_1 \perp} d^d p_{h_2\perp} } \bigg |_{\text{coll. qg fin}}   \numberthis[coll_qg_qqbar_FF]  \;,
    \end{align*} 
where the term denoted by the label ``coll. qg div" corresponds to the sum of the first three terms in the curly bracket, whereas the remaining terms in the curly bracket contribute to the term denoted ``coll. qg fin".

This gives the following expression for the divergent part: 
\begin{align*}
\allowdisplaybreaks
& \frac{d \sigma_{3LL}^{q \bar{q} \rightarrow h_1 h_2}}{ d x_{h_1} d x_{h_2} d p_{h_1 \perp} d^d p_{h_2\perp} } \bigg |_{\text{coll. qg div}}       \\
&=  \frac{4  \alpha_{\mathrm{em}} Q^2}{(2\pi)^{4(d-1)} N_c} \sum_{q} \int_{x_{h_1}}^{1} d x_q \int_{x_{h_2}}^1 d x_{\bar{q}} \; x_q x_{\bar{q}} \left(\frac{x_q}{x_{h_1}}\right)^d \left(\frac{x_{\bar{q}}}{x_{h_2}}\right)^d \delta(1-x_q-x_{\bar{q}}) \\*
& \times \int   d^d p_{2\perp}  \int d^d z_{1\perp}   \frac{e^{i z_{1\perp}\cdot \left(\frac{x_q}{2 x_{h_1}} p_{h_1\perp} + \frac{x_{\bar{q}}}{2 x_{h_2}} p_{h_2 \perp} -p_{2\perp} \right)} F(z_{1\perp})}{x_q x_{\bar{q}} Q^2 +\left(\frac{x_{\bar{q}}}{x_{h_2}} \vec{p}_{h_2}-\vec{p}_{2}\right)^2} \\
& \times \int   d^d p_{2'\perp}  \int d^d z_{2\perp}   \frac{e^{-i z_{2\perp}\cdot \left(\frac{x_q}{2 x_{h_1}} p_{h_1\perp} + \frac{x_{\bar{q}}}{2 x_{h_2}} p_{h_2 \perp} -p_{2'\perp} \right)} F^*(z_{2\perp})}{x_q x_{\bar{q}} Q^2 +\left(\frac{x_{\bar{q}}}{x_{h_2}} \vec{p}_{h_2}-\vec{p}_{2'}\right)^2} \\
& \times \frac{\alpha_s}{2\pi} \frac{1}{\hat{\epsilon}}  Q_q^2 \left[
\int_{\frac{x_{h_1}}{x_q}}^{1}\frac{d\beta_1}{\beta_1} C_F \frac{1+ \beta_1^2}{(1-\beta_1)_+} D_q^{h_1}\left(\frac{x_{h_1}}{\beta_1 x_q},\mu_F\right) D_{\bar{q}}^{h_2}\left(\frac{x_{h_2}}{x_{\bar{q}}},\mu_F \right) \right. \\*
& + \int_{\frac{x_{h_1}}{x_q}}^{1-\frac{\alpha}{x_q}} d \beta_1 C_F \frac{2}{1-\beta_1} \left(\frac{c_0^2}{\left(\frac{z_{1\perp}-z_{2\perp}}{2}\right)^2 \mu^2}\right)^\epsilon D_q^{h_1}\left(\frac{x_{h_1}}{x_q},\mu_F\right) D_{\bar{q}}^{h_2}\left(\frac{x_{h_2}}{x_{\bar{q}}},\mu_F \right) \\
& \left. - 2  C_F \ln \left(1-\frac{x_{h_1}}{x_q}\right)  D_q^{h_1}\left(\frac{x_{h_1}}{x_q},\mu_F\right) D_{\bar{q}}^{h_2}\left(\frac{x_{h_2}}{x_{\bar{q}}},\mu_F \right) \right] + (h_1 \leftrightarrow h_2 ) \,.  \numberthis[coll_div_q] 
\end{align*} 
The first term in the bracket cancels part of the first term in the bracket in Eq.~\eqref{eq:ct_LL}, \textit{i.e.} the part involving the $+$ prescription in the spitting function $P_{qq}$, and the remaining part of the $P_{qq}$ term is cancelled by an analogous contribution in the virtual part. The second term in Eq.~\eqref{eq:coll_div_q}  has to be removed to avoid double counting as it corresponds to the soft contribution and will be taken into account later in the chapter. The third and last term, in Eq.~\eqref{eq:coll_div_q}, will compensate an analogous term in the soft contribution.

The finite part for the $LL$ contribution takes the form 
\begin{align*}
&  \frac{d \sigma_{3LL}^{q \bar{q} \rightarrow h_1 h_2}}{ d x_{h_1} d x_{h_2} d p_{h_1 \perp} d^d p_{h_2\perp} } \bigg |_{\text{coll. qg fin}} \\
&= \frac{4  \alpha_{\mathrm{em}} Q^2}{(2\pi)^{4(d-1)} N_c} \sum_{q} \int_{x_{h_1}}^{1} d x_q  \int_{x_{h_2}}^{1} d x_{\bar{q}} \; x_q x_{\bar{q}} \delta(1-x_q -x_{\bar{q}}) \left(\frac{x_q}{x_{h_1}}\right)^d \left(\frac{x_{\bar{q}}}{x_{h_2}}\right)^d \\
& \times \int   d^d p_{2\perp}  \int d^d z_{1\perp}   \frac{e^{i z_{1\perp}\cdot \left(\frac{x_q}{2 x_{h_1}} p_{h_1\perp} + \frac{x_{\bar{q}}}{2 x_{h_2}} p_{h_2 \perp} -p_{2\perp} \right)} F(z_{1\perp})}{x_q x_{\bar{q}} Q^2 +\left(\frac{x_{\bar{q}}}{x_{h_2}} \vec{p}_{h_2}-\vec{p}_{2}\right)^2} \\
& \times \int   d^d p_{2'\perp}  \int d^d z_{2\perp}   \frac{e^{-i z_{2\perp}\cdot \left(\frac{x_q}{2 x_{h_1}} p_{h_1\perp} + \frac{x_{\bar{q}}}{2 x_{h_2}} p_{h_2\perp} -p_{2'\perp} \right)} F^*(z_{2\perp})}{x_q x_{\bar{q}} Q^2 +\left(\frac{x_{\bar{q}}}{x_{h_2}} \vec{p}_{h_2}-\vec{p}_{2'}\right)^2} \\
& \times  \frac{\alpha_s C_F}{2\pi}  \left \{ \int_{\frac{x_{h_1}}{x_q}}^{1} \frac{d\beta_1}{\beta_1}  Q_q^2 D_q^{h_1}\left(\frac{x_{h_1}}{\beta_1 x_q}, \mu_F\right) D_{\bar{q}}^{h_2}\left(\frac{x_{h_2}}{x_{\bar{q}}}, \mu_F\right) \right. \\
& \times  \left[ \ln \left( \frac{c_0^2}{\left(\frac{z_{1\perp} - z_{2\perp}}{2}\right)^2 \mu^2}\right)  \frac{1+ \beta_1^2}{(1-\beta_1)_+}  + \frac{(1-\beta_1)^2 + 2 (1+ \beta_1^2) \ln \beta_1 }{(1-\beta_1)} \right] \\
& \left. - 2 \ln \left( 1- \frac{x_{h_1}}{x_q}\right) \ln \left( \frac{c_0^2}{\left(\frac{z_{1\perp} - z_{2\perp}}{2}\right)^2 \mu^2}\right) D_{q}^{h_1} \left(\frac{x_{h_1}}{x_q},\mu_F\right) D_{\bar{q}}^{h_q} \left(\frac{x_{h_2}}{x_{\bar{q}}},\mu_F \right)\right\} + (h_1 \leftrightarrow h_2 )\,. \numberthis[collqg_LL_fin]
    \end{align*}

Similarly, one gets for the $TL$ case
\begin{align*}
& \frac{d \sigma_{3TL}^{q \bar{q} \rightarrow h_1 h_2}}{ d x_{h_1} d x_{h_2} d p_{h_1 \perp} d^d p_{h_2\perp} } \bigg |_{\text{coll. qg }} \\
& =  \frac{2  \alpha_{\mathrm{em}} Q}{(2\pi)^{4(d-1)} N_c} \sum_{q}  \int_{x_{h_1}}^{1} d x_q  \int_{x_{h_2}}^{1} d x_{\bar{q}} \; (x_{\bar{q}}-x_q)\delta(1-x_q -x_{\bar{q}}) \left(\frac{x_q}{x_{h_1}}\right)^d \left(\frac{x_{\bar{q}}}{x_{h_2}}\right)^d \\
& \times \int   d^d p_{2\perp}  \int d^d z_{1\perp}   \frac{e^{i z_{1\perp}\cdot \left(\frac{x_q}{2 x_{h_1}} p_{h_1\perp} + \frac{x_{\bar{q}}}{2 x_{h_2}} p_{h_2 \perp} -p_{2\perp} \right)} F(z_{1\perp})}{x_q x_{\bar{q}} Q^2 +\left(\frac{x_{\bar{q}}}{x_{h_2}} \vec{p}_{h_2}-\vec{p}_{2}\right)^2} \\
& \times \int   d^d p_{2'\perp}  \int d^d z_{2\perp}   \frac{e^{-i z_{2\perp}\cdot \left(\frac{x_q}{2 x_{h_1}} p_{h_1\perp} + \frac{x_{\bar{q}}}{2 x_{h_2}} p_{h_2 \perp} -p_{2'\perp} \right)} F^*(z_{2\perp})}{x_q x_{\bar{q}} Q^2 +\left(\frac{x_{\bar{q}}}{x_{h_2}} \vec{p}_{h_2}-\vec{p}_{2'}\right)^2}   \left( \frac{x_{\bar{q}}}{x_{h_2}} \vec{p}_{h_2}  - \vec{p}_{2'}\right) \cdot \vec{\varepsilon}^{\,*}_T  \\
& \times  \int_{\frac{x_{h_1}}{x_q}}^{1-\frac{\alpha}{x_q}}\frac{d\beta_1}{\beta_1} Q_q^2 D_q^{h_1}\left(\frac{x_{h_1}}{\beta_1 x_q}, \mu_F\right) D_{\bar{q}}^{h_2}\left(\frac{x_{h_2}}{x_{\bar{q}}}, \mu_F\right) \\
& \times \frac{\alpha_s C_F}{2\pi} \left[ \frac{1}{\hat{\epsilon}} \left(\frac{c_0^2}{\left(\frac{z_{1\perp}-z_{2\perp}}{2}\right)^2 \mu^2}\right)^\epsilon \frac{1+ \beta_1^2}{1-\beta_1}  + \frac{(1-\beta_1)^2 + 2 (1+ \beta_1^2) \ln \beta_1 }{(1-\beta_1)} \right] + (h_1 \leftrightarrow h_2 ) \\
&= \frac{d \sigma_{3TL}^{q \bar{q} \rightarrow h_1 h_2}}{ d x_{h_1} d x_{h_2} d p_{h_1 \perp} d^d p_{h_2\perp} } \bigg |_{\text{coll. qg div}}     + \frac{d \sigma_{3TL}^{q \bar{q} \rightarrow h_1 h_2}}{ d x_{h_1} d x_{h_2} d p_{h_1 \perp} d^d p_{h_2\perp} } \bigg |_{\text{coll. qg fin}}   ,  \numberthis
    \end{align*} %
where \newpage
\begin{align*}
& \frac{d \sigma_{3TL}^{q \bar{q} \rightarrow h_1 h_2}}{ d x_{h_1} d x_{h_2} d p_{h_1 \perp} d^d p_{h_2\perp} } \bigg |_{\text{coll. qg div}}       \\
&=  \frac{ 2 \alpha_{\mathrm{em}} Q}{(2\pi)^{4(d-1)} N_c} \sum_{q} \int_{x_{h_1}}^{1} d x_q \int_{x_{h_2}}^1 d x_{\bar{q}}  \; (x_{\bar{q}}-x_q) \left(\frac{x_q}{x_{h_1}}\right)^d \left(\frac{x_{\bar{q}}}{x_{h_2}}\right)^d \delta(1-x_q-x_{\bar{q}}) \\*
& \times \int   d^d p_{2\perp}  \int d^d z_{1\perp}   \frac{e^{i z_{1\perp}\cdot \left(\frac{x_q}{2 x_{h_1}} p_{h_1\perp} + \frac{x_{\bar{q}}}{2 x_{h_2}} p_{h_2 \perp} -p_{2\perp} \right)} F(z_{1\perp})}{x_q x_{\bar{q}} Q^2 +\left(\frac{x_{\bar{q}}}{x_{h_2}} \vec{p}_{h_2}-\vec{p}_{2}\right)^2} \\
& \times \int   d^d p_{2'\perp}  \int d^d z_{2\perp}   \frac{e^{-i z_{2\perp}\cdot \left(\frac{x_q}{2 x_{h_1}} p_{h_1\perp} + \frac{x_{\bar{q}}}{2 x_{h_2}} p_{h_2 \perp} -p_{2'\perp} \right)} F^*(z_{2\perp})}{x_q x_{\bar{q}} Q^2 +\left(\frac{x_{\bar{q}}}{x_{h_2}} \vec{p}_{h_2}-\vec{p}_{2'}\right)^2} \left( \frac{x_{\bar{q}}}{x_{h_2}} \vec{p}_{h_2}  - \vec{p}_{2'}\right) \cdot \vec{\varepsilon}^{\,*}_T \\
& \times \frac{\alpha_s}{2\pi} \frac{1}{\hat{\epsilon}}  Q_q^2 \left[
\int_{\frac{x_{h_1}}{x_q}}^{1}\frac{d\beta_1}{\beta_1} C_F \frac{1+ \beta_1^2}{(1-\beta_1)_+} D_q^{h_1}\left(\frac{x_{h_1}}{\beta_1 x_q},\mu_F\right) D_{\bar{q}}^{h_2}\left(\frac{x_{h_2}}{x_{\bar{q}}},\mu_F \right) \right. \\*
& + \int_{\frac{x_{h_1}}{x_q}}^{1-\frac{\alpha}{x_q}} d \beta_1 C_F \frac{2}{1-\beta_1} \left(\frac{c_0^2}{\left(\frac{z_{1\perp}-z_{2\perp}}{2}\right)^2 \mu^2}\right)^\epsilon D_q^{h_1}\left(\frac{x_{h_1}}{x_q},\mu_F\right) D_{\bar{q}}^{h_2}\left(\frac{x_{h_2}}{x_{\bar{q}}},\mu_F \right) \\
& \left. - 2  C_F \ln \left(1-\frac{x_{h_1}}{x_q}\right)  D_q^{h_1}\left(\frac{x_{h_1}}{x_q},\mu_F\right) D_{\bar{q}}^{h_2}\left(\frac{x_{h_2}}{x_{\bar{q}}},\mu_F \right) \right] + (h_1 \leftrightarrow h_2 ) \,,  
\end{align*}
and
\begin{align*}
&  \frac{d \sigma_{3TL}^{q \bar{q} \rightarrow h_1 h_2}}{ d x_{h_1} d x_{h_2} d p_{h_1 \perp} d^d p_{h_2\perp} } \bigg |_{\text{coll. qg fin}} \\
&= \frac{2 \alpha_{\mathrm{em}} Q}{(2\pi)^{4(d-1)} N_c} \sum_{q} \int_{x_{h_1}}^{1} d x_q  \int_{x_{h_2}}^{1} d x_{\bar{q}}  \; (x_{\bar{q}}-x_q) \delta(1-x_q -x_{\bar{q}}) \left(\frac{x_q}{x_{h_1}}\right)^d \left(\frac{x_{\bar{q}}}{x_{h_2}}\right)^d \\
& \times \int   d^d p_{2\perp}  \int d^d z_{1\perp}   \frac{e^{i z_{1\perp}\cdot \left(\frac{x_q}{2 x_{h_1}} p_{h_1\perp} + \frac{x_{\bar{q}}}{2 x_{h_2}} p_{h_2 \perp} -p_{2\perp} \right)} F(z_{1\perp})}{x_q x_{\bar{q}} Q^2 +\left(\frac{x_{\bar{q}}}{x_{h_2}} \vec{p}_{h_2}-\vec{p}_{2}\right)^2} \\
& \times \int   d^d p_{2'\perp}  \int d^d z_{2\perp}   \frac{e^{-i z_{2\perp}\cdot \left(\frac{x_q}{2 x_{h_1}} p_{h_1\perp} + \frac{x_{\bar{q}}}{2 x_{h_2}} p_{h_2\perp} -p_{2'\perp} \right)} F^*(z_{2\perp})}{x_q x_{\bar{q}} Q^2 +\left(\frac{x_{\bar{q}}}{x_{h_2}} \vec{p}_{h_2}-\vec{p}_{2'}\right)^2} \left( \frac{x_{\bar{q}}}{x_{h_2}} \vec{p}_{h_2}  - \vec{p}_{2'}\right) \cdot \vec{\varepsilon}^{\,*}_T\\
& \times  \frac{\alpha_s C_F}{2\pi}  \left \{ \int_{\frac{x_{h_1}}{x_q}}^{1} \frac{d\beta_1}{\beta_1}  Q_q^2 D_q^{h_1}\left(\frac{x_{h_1}}{\beta_1 x_q}, \mu_F\right) D_{\bar{q}}^{h_2}\left(\frac{x_{h_2}}{x_{\bar{q}}}, \mu_F\right) \right. \\
& \times  \left[ \ln \left( \frac{c_0^2}{\left(\frac{z_{1\perp} - z_{2\perp}}{2}\right)^2 \mu^2}\right)  \frac{1+ \beta_1^2}{(1-\beta_1)_+}  + \frac{(1-\beta_1)^2 + 2 (1+ \beta_1^2) \ln \beta_1 }{(1-\beta_1)} \right] \\
& \left. - 2 \ln \left( 1- \frac{x_{h_1}}{x_q}\right) \ln \left( \frac{c_0^2}{\left(\frac{z_{1\perp} - z_{2\perp}}{2}\right)^2 \mu^2}\right) D_{q}^{h_1} \left(\frac{x_{h_1}}{x_q},\mu_F\right) D_{\bar{q}}^{h_q} \left(\frac{x_{h_2}}{x_{\bar{q}}},\mu_F \right)\right\} + (h_1 \leftrightarrow h_2 )\,, \numberthis[TL_coll_qg_finite]
    \end{align*}%
and finally, one obtains for the $TT$ case
\begin{align*}
& \frac{d \sigma_{3TT}^{q \bar{q} \rightarrow h_1 h_2}}{ d x_{h_1} d x_{h_2} d p_{h_1 \perp} d^d p_{h_2\perp} } \bigg |_{\text{coll. qg }} \\*
& =  \frac{  \alpha_{\mathrm{em}} }{(2\pi)^{4(d-1)} N_c} \sum_{q} \int_{x_{h_1}}^{1} \frac{d x_q}{x_q}  \int_{x_{h_2}}^{1} \frac{d x_{\bar{q}}}{x_{\bar{q}}} \delta(1-x_q -x_{\bar{q}}) \left(\frac{x_q}{x_{h_1}}\right)^d \left(\frac{x_{\bar{q}}}{x_{h_2}}\right)^d \\
& \times \left[ (x_{\bar{q}} -x_q)^2 g_{\perp}^{ri}g_{\perp}^{lk} - g_{\perp}^{rk}g_{\perp}^{li} + g_{\perp}^{rl}g_{\perp}^{ik} \right] \\
& \times \int   d^d p_{2\perp}  \int d^d z_{1\perp}   \frac{e^{i z_{1\perp}\cdot \left(\frac{x_q}{2 x_{h_1}} p_{h_1\perp} + \frac{x_{\bar{q}}}{2 x_{h_2}} p_{h_2 \perp} -p_{2\perp} \right)} F(z_{1\perp})}{x_q (1-x_q)Q^2 +\left(\frac{x_{\bar{q}}}{x_{h_2}} \vec{p}_{h_2}-\vec{p}_{2}\right)^2} \left(\frac{x_{\bar{q}}}{x_{h_2}} p_{h_2} - p_{2}\right)_r \varepsilon_{T i} \\
& \times \int   d^d p_{2'\perp}  \int d^d z_{2\perp}   \frac{e^{-i z_{2\perp}\cdot \left(\frac{x_q}{2 x_{h_1}} p_{h_1\perp} + \frac{x_{\bar{q}}}{2 x_{h_2}} p_{h_2 \perp} -p_{2'\perp} \right)} F^*(z_{2\perp})}{x_q x_{\bar{q}} Q^2 +\left(\frac{x_{\bar{q}}}{x_{h_2}} \vec{p}_{h_2}-\vec{p}_{2'}\right)^2} \left(\frac{x_{\bar{q}}}{x_{h_2}} p_{h_2 } - p_{2'}\right)_l \varepsilon_{T k }^*  \\
& \times  \int_{\frac{x_{h_1}}{x_q}}^{1-\frac{\alpha}{x_q}}\frac{d\beta_1}{\beta_1}  Q_q^2 D_q^{h_1}\left(\frac{x_{h_1}}{\beta_1 x_q}, \mu_F\right) D_{\bar{q}}^{h_2}\left(\frac{x_{h_2}}{x_{\bar{q}}}, \mu_F\right) \\
& \times \frac{\alpha_s C_F}{2\pi} \left[ \frac{1}{\hat{\epsilon}} \left(\frac{c_0^2}{\left(\frac{z_{1\perp}-z_{2\perp}}{2}\right)^2 \mu^2}\right)^\epsilon \frac{1+ \beta_1^2}{1-\beta_1}  + \frac{(1-\beta_1)^2 + 2 (1+ \beta_1^2) \ln \beta_1 }{(1-\beta_1)} \right] + (h_1 \leftrightarrow h_2 ) \\
&= \frac{d \sigma_{3TT}^{q \bar{q} \rightarrow h_1 h_2}}{ d x_{h_1} d x_{h_2} d p_{h_1 \perp} d^d p_{h_2\perp} } \bigg |_{\text{coll. qg div}}     + \frac{d \sigma_{3TT}^{q \bar{q} \rightarrow h_1 h_2}}{ d x_{h_1} d x_{h_2} d p_{h_1 \perp} d^d p_{h_2\perp} } \bigg |_{\text{coll. qg fin}} \; ,  \numberthis   
    \end{align*}
where
{\allowdisplaybreaks
\begin{align*}
& \frac{d \sigma_{3TT}^{q \bar{q} \rightarrow h_1 h_2}}{ d x_{h_1} d x_{h_2} d p_{h_1 \perp} d^d p_{h_2\perp} } \bigg |_{\text{coll. qg div}}       \\
&=  \frac{  \alpha_{\mathrm{em}} }{(2\pi)^{4(d-1)} N_c} \sum_{q} \int_{x_{h_1}}^{1} \frac{d x_q}{x_q} \int_{x_{h_2}}^1 \frac{d x_{\bar{q}}}{x_{\bar{q}}} \left(\frac{x_q}{x_{h_1}}\right)^d \left(\frac{x_{\bar{q}}}{x_{h_2}}\right)^d \delta(1-x_q-x_{\bar{q}}) \\*
& \times \left[ (x_{\bar{q}} -x_q)^2 g_{\perp}^{ri}g_{\perp}^{lk} - g_{\perp}^{rk}g_{\perp}^{li} + g_{\perp}^{rl}g_{\perp}^{ik} \right] \\
& \times \int   d^d p_{2\perp}  \int d^d z_{1\perp}   \frac{e^{i z_{1\perp}\cdot \left(\frac{x_q}{2 x_{h_1}} p_{h_1\perp} + \frac{x_{\bar{q}}}{2 x_{h_2}} p_{h_2 \perp} -p_{2\perp} \right)} F(z_{1\perp})}{x_q x_{\bar{q}} Q^2 +\left(\frac{x_{\bar{q}}}{x_{h_2}} \vec{p}_{h_2}-\vec{p}_{2}\right)^2}  \left(\frac{x_{\bar{q}}}{x_{h_2}} p_{h_2 } - p_{2 }\right)_r \varepsilon_{T i}  \\
& \times \int   d^d p_{2'\perp}  \int d^d z_{2\perp}   \frac{e^{-i z_{2\perp}\cdot \left(\frac{x_q}{2 x_{h_1}} p_{h_1\perp} + \frac{x_{\bar{q}}}{2 x_{h_2}} p_{h_2 \perp} -p_{2'\perp} \right)} F^*(z_{2\perp})}{x_q x_{\bar{q}} Q^2 +\left(\frac{x_{\bar{q}}}{x_{h_2}} \vec{p}_{h_2}-\vec{p}_{2'}\right)^2} \left(\frac{x_{\bar{q}}}{x_{h_2}} p_{h_2} - p_{2'}\right)_l \varepsilon_{T k}^* \\
& \times \frac{\alpha_s}{2\pi} \frac{1}{\hat{\epsilon}}  Q_q^2 \left[
\int_{\frac{x_{h_1}}{x_q}}^{1}\frac{d\beta_1}{\beta_1} C_F \frac{1+ \beta_1^2}{(1-\beta_1)_+} D_q^{h_1}\left(\frac{x_{h_1}}{\beta_1 x_q},\mu_F\right) D_{\bar{q}}^{h_2}\left(\frac{x_{h_2}}{x_{\bar{q}}},\mu_F \right) \right. \\*
& + \int_{\frac{x_{h_1}}{x_q}}^{1-\frac{\alpha}{x_q}} d \beta_1 C_F \frac{2}{1-\beta_1} \left(\frac{c_0^2}{\left(\frac{z_{1\perp}-z_{2\perp}}{2}\right)^2 \mu^2}\right)^\epsilon D_q^{h_1}\left(\frac{x_{h_1}}{x_q},\mu_F\right) D_{\bar{q}}^{h_2}\left(\frac{x_{h_2}}{x_{\bar{q}}},\mu_F \right) \\
& \left. - 2  C_F \ln \left(1-\frac{x_{h_1}}{x_q}\right)  D_q^{h_1}\left(\frac{x_{h_1}}{x_q},\mu_F\right) D_{\bar{q}}^{h_2}\left(\frac{x_{h_2}}{x_{\bar{q}}},\mu_F \right) \right] + (h_1 \leftrightarrow h_2 ) \,,  
\end{align*} }
and 
{\allowdisplaybreaks
\begin{align*}
&  \frac{d \sigma_{3TT}^{q \bar{q} \rightarrow h_1 h_2}}{ d x_{h_1} d x_{h_2} d p_{h_1 \perp} d^d p_{h_2\perp} } \bigg |_{\text{coll. qg fin}} \\
&= \frac{ \alpha_{\mathrm{em}} }{(2\pi)^{4(d-1)} N_c} \sum_{q} \int_{x_{h_1}}^{1} \frac{d x_q}{x_q}  \int_{x_{h_2}}^{1} \frac{d x_{\bar{q}}}{x_{\bar{q}}} \delta(1-x_q -x_{\bar{q}}) \left(\frac{x_q}{x_{h_1}}\right)^d \left(\frac{x_{\bar{q}}}{x_{h_2}}\right)^d \\
& \times \left[ (x_{\bar{q}} -x_q)^2 g_{\perp}^{ri}g_{\perp}^{lk} - g_{\perp}^{rk}g_{\perp}^{li} + g_{\perp}^{rl}g_{\perp}^{ik} \right] \\
& \times \int   d^d p_{2\perp}  \int d^d z_{1\perp}   \frac{e^{i z_{1\perp}\cdot \left(\frac{x_q}{2 x_{h_1}} p_{h_1\perp} + \frac{x_{\bar{q}}}{2 x_{h_2}} p_{h_2 \perp} -p_{2\perp} \right)} F(z_{1\perp})}{x_q x_{\bar{q}} Q^2 +\left(\frac{x_{\bar{q}}}{x_{h_2}} \vec{p}_{h_2}-\vec{p}_{2}\right)^2} \left(\frac{x_{\bar{q}}}{x_{h_2}} p_{h_2} - p_{2}\right)_r \varepsilon_{T i} \\
& \times \int   d^d p_{2'\perp}  \int d^d z_{2\perp}   \frac{e^{-i z_{2\perp}\cdot \left(\frac{x_q}{2 x_{h_1}} p_{h_1\perp} + \frac{x_{\bar{q}}}{2 x_{h_2}} p_{h_2\perp} -p_{2'\perp} \right)} F^*(z_{2\perp})}{x_q x_{\bar{q}} Q^2 +\left(\frac{x_{\bar{q}}}{x_{h_2}} \vec{p}_{h_2}-\vec{p}_{2'}\right)^2} \left(\frac{x_{\bar{q}}}{x_{h_2}} p_{h_2} - p_{2'}\right)_l \varepsilon_{T k}^*\\
& \times  \frac{\alpha_s C_F}{2\pi}  \left \{ \int_{\frac{x_{h_1}}{x_q}}^{1} \frac{d\beta_1}{\beta_1}  Q_q^2 D_q^{h_1}\left(\frac{x_{h_1}}{\beta_1 x_q}, \mu_F\right) D_{\bar{q}}^{h_2}\left(\frac{x_{h_2}}{x_{\bar{q}}}, \mu_F\right) \right. \\
& \times  \left[ \ln \left( \frac{c_0^2}{\left(\frac{z_{1\perp} - z_{2\perp}}{2}\right)^2 \mu^2}\right)  \frac{1+ \beta_1^2}{(1-\beta_1)_+}  + \frac{(1-\beta_1)^2 + 2 (1+ \beta_1^2) \ln \beta_1 }{(1-\beta_1)} \right] \\
& \left. - 2 \ln \left( 1- \frac{x_{h_1}}{x_q}\right) \ln \left( \frac{c_0^2}{\left(\frac{z_{1\perp} - z_{2\perp}}{2}\right)^2 \mu^2}\right) D_{q}^{h_1} \left(\frac{x_{h_1}}{x_q},\mu_F\right) D_{\bar{q}}^{h_q} \left(\frac{x_{h_2}}{x_{\bar{q}}},\mu_F \right)\right\} + (h_1 \leftrightarrow h_2 )\,. \numberthis[collqg_TT_fin]
    \end{align*} }

\subsubsection{Collinear contributions: $\bar{q}$-$g$ splitting}
\label{sec:qqbarfragColl-qbarg}

Here the term in \eqref{eq: div real impact factor} to consider is the third one. The calculation proceeds in the same way as for the quark-gluon collinear contribution but this time the integration is performed over $p_{2 \perp}$ and $p_{2' \perp}$. To observe the cancellation of these collinear divergences, one has to use the different representations we gave for the LO cross section, as explained before, see eqs.~(\ref{eq:LL-LO-minus}, \ref{Ftilde-LL}) with respect to eqs.~(\ref{eq:LL-LO}, \ref{F-LL}).

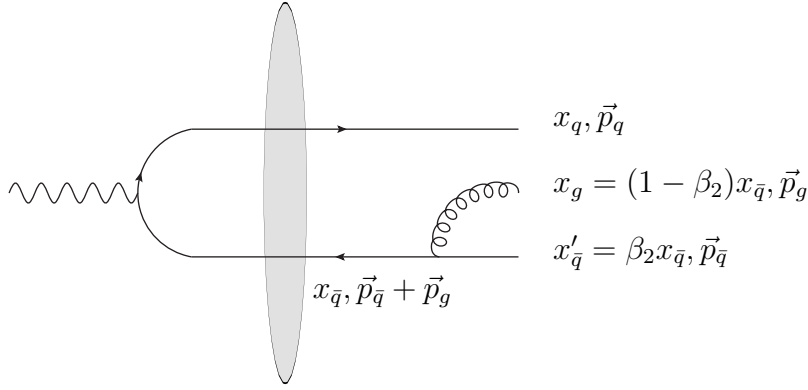
\begin{figure}
\begin{picture}(420,160)
\put(-50,0){\includegraphics[scale=0.5]{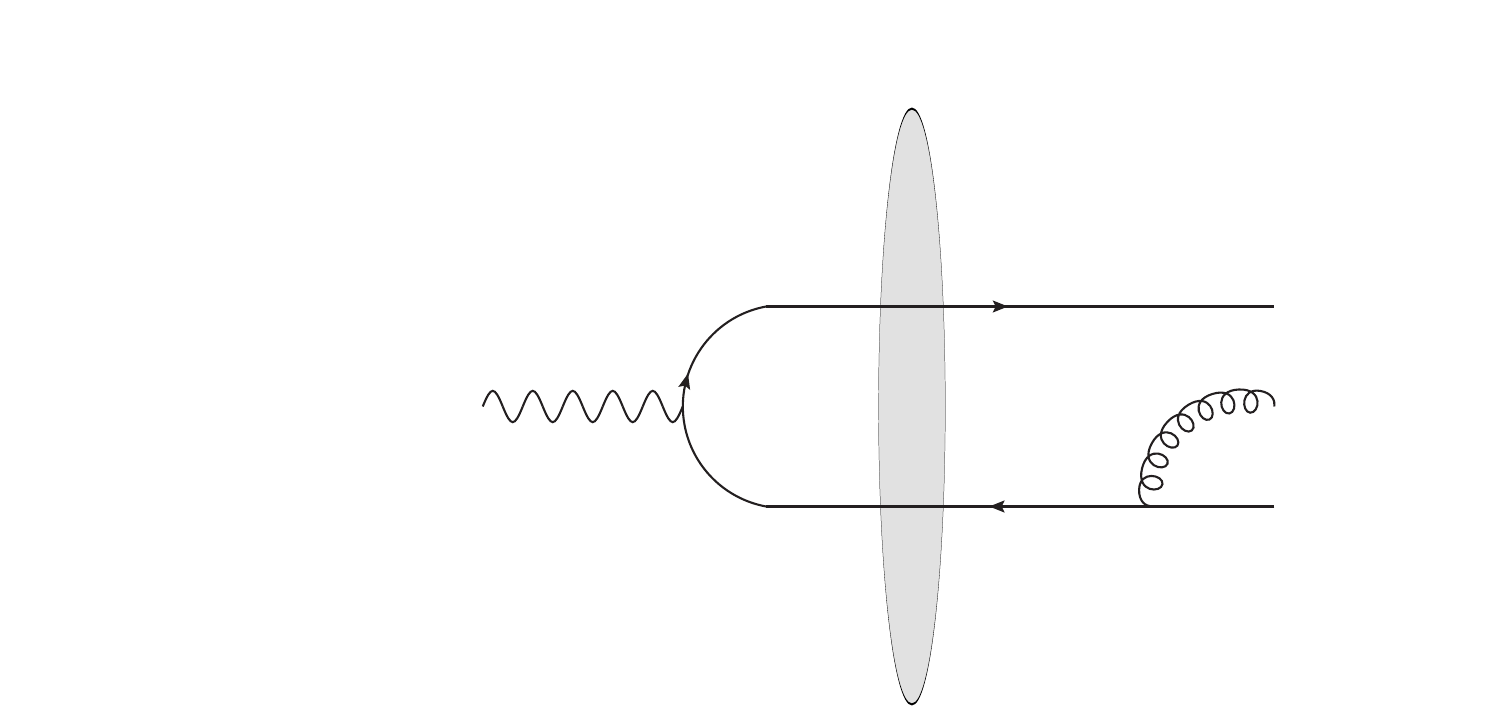}}
\put(180,35){$x_{\bar{q}}, \vec{p}_{\bar{q}}+\vec{p}_g$}
\put(270,100){$x_q, \vec{p}_q$}
\put(270,75){$x_g=(1-\beta_2)x_{\bar{q}}, \vec{p}_g$}
\put(270,50){$x'_{\bar{q}}=\beta_2 x_{\bar{q}}, \vec{p}_{\bar{q}}$}
\end{picture}
\caption{Kinematics for the $\bar{q}-g$ splitting contribution. We indicate the longitudinal fraction of momentum carried by the partons as well as their transverse momenta.}
\label{fig:kinematics_qbarg}
\end{figure}

This time, the change of variable to be done is
\begin{align*}
    x_{\bar{q}}' & = \beta_2 x_{\bar{q}} , \\
    x_g & = (1-\beta_2) x_{\bar{q}} \,.\numberthis[transBeta2]
\end{align*}
This kinematics is illustrated in Fig.~\ref{fig:kinematics_qbarg}.
The boundaries of integration for $(x_{\bar{q}},\beta_2)$ are calculated in the same spirit as in eq~\eqref{eq:change_variable_integral}. 

Thus, after changes of variable and integrations, the third term in \eqref{eq: div real impact factor} takes the form 
{ \allowdisplaybreaks
\begin{align*} 
& \frac{d \sigma_{3LL}^{q \bar{q} \rightarrow h_1 h_2}}{ d x_{h_1} d x_{h_2} d^d p_{h_1 \perp} d^d p_{h_2\perp} } \bigg |_{\text{coll. } \bar{q}g}  \\
& =  \frac{4  \alpha_{\mathrm{em}} Q^2}{(2\pi)^{4(d-1)} N_c} \sum_{q} \int_{x_{h_1}}^{1} d x_q  \int_{x_{h_2}}^{1} d x_{\bar{q}} \;  x_q x_{\bar{q}} \delta(1-x_q -x_{\bar{q}}) \left(\frac{x_q}{x_{h_1}}\right)^d \left(\frac{x_{\bar{q}}}{x_{h_2}}\right)^d \\
& \times \int   d^d p_{1\perp}  \int d^d z_{1\perp}   \frac{e^{i z_{1\perp}\cdot \left( -\frac{x_q}{2 x_{h_1}} p_{h_1\perp} - \frac{x_{\bar{q}}}{2 x_{h_2}} p_{h_2 \perp} + p_{1\perp} \right)} F(z_{1\perp})}{x_q x_{\bar{q}} Q^2 +\left(\frac{x_q}{x_{h_1}} \vec{p}_{h_1}-\vec{p}_{1}\right)^2} \\
& \times \int   d^d p_{1'\perp}  \int d^d z_{2\perp}   \frac{e^{-i z_{2\perp}\cdot \left(- \frac{x_q}{2 x_{h_1}} p_{h_1\perp} - \frac{x_{\bar{q}}}{2 x_{h_2}} p_{h_2 \perp} + p_{1'\perp} \right)} F^*(z_{2\perp})}{x_q x_{\bar{q}} Q^2 +\left(\frac{x_q}{x_{h_1}} \vec{p}_{h_1}-\vec{p}_{1'}\right)^2} \\
& \times  \int_{\frac{x_{h_2}}{x_{\bar{q}}}}^{1-\frac{\alpha}{x_{\bar{q}}}}\frac{d\beta_2}{\beta_2}  Q_q^2 D_q^{h_1}\left(\frac{x_{h_1}}{x_q}, \mu_F\right) D_{\bar{q}}^{h_2}\left(\frac{x_{h_2}}{\beta_2 x_{\bar{q}}}, \mu_F\right) \\
& \times \frac{\alpha_s C_F}{2\pi} \left[ \frac{1}{\hat{\epsilon}} \left(\frac{c_0^2}{\left(\frac{z_{2\perp}-z_{1\perp}}{2}\right)^2 \mu^2}\right)^\epsilon \frac{1+ \beta_2^2}{1-\beta_2}  + \frac{(1-\beta_2)^2 + 2 (1+ \beta_2^2) \ln \beta_2 }{(1-\beta_2)} \right] + (h_1 \leftrightarrow h_2 ) \\
& = \frac{d \sigma_{3LL}^{q \bar{q} \rightarrow h_1 h_2}}{ d x_{h_1} d x_{h_2} d p_{h_1 \perp} d^d p_{h_2\perp} } \bigg |_{\text{coll. } \bar{q}g  \text{ div}}     + \frac{d \sigma_{3LL}^{q \bar{q} \rightarrow h_1 h_2}}{ d x_{h_1} d x_{h_2} d p_{h_1 \perp} d^d p_{h_2\perp} } \bigg |_{\text{coll. } \bar{q}g \text{ fin}} \; . \numberthis[coll_qbar_g_qqbar_FF]
\end{align*} }

Adding the $+$ prescription and expanding up to $\epsilon^0$, just like for the collinear $qg$ contribution,  one gets the finite and divergent part of \eqref{eq:coll_qbar_g_qqbar_FF}. 
The divergent part is 
{\allowdisplaybreaks
\begin{align*}
&\frac{d \sigma_{3LL}^{q \bar{q} \rightarrow h_1 h_2}}{ d x_{h_1} d x_{h_2} d p_{h_1 \perp} d^d p_{h_2\perp} } \bigg |_{\text{coll. } \bar{q}g  \text{ div}} \\ 
&=  \frac{4  \alpha_{\mathrm{em}} Q^2}{(2\pi)^{4(d-1)} N_c} \sum_{q} \int_{x_{h_1}}^{1} d x_q \int_{x_{h_2}}^1 d x_{\bar{q}} \; x_q x_{\bar{q}} \left(\frac{x_q}{x_{h_1}}\right)^d \left(\frac{x_{\bar{q}}}{x_{h_2}}\right)^d \delta(1-x_q-x_{\bar{q}}) \\
& \times \int   d^d p_{1\perp}  \int d^d z_{1\perp}   \frac{e^{i z_{1\perp}\cdot \left( -\frac{x_q}{2 x_{h_1}} p_{h_1\perp} - \frac{x_{\bar{q}}}{2 x_{h_2}} p_{h_2 \perp} + p_{1\perp} \right)} F(z_{1\perp})}{x_q x_{\bar{q}} Q^2 +\left(\frac{x_q}{x_{h_1}} \vec{p}_{h_1}-\vec{p}_{1}\right)^2} \\
& \times \int   d^d p_{1'\perp}  \int d^d z_{2\perp}   \frac{e^{-i z_{2\perp}\cdot \left(- \frac{x_q}{2 x_{h_1}} p_{h_1\perp} - \frac{x_{\bar{q}}}{2 x_{h_2}} p_{h_2 \perp} + p_{1'\perp} \right)} F^*(z_{2\perp})}{x_q x_{\bar{q}} Q^2 +\left(\frac{x_q}{x_{h_1}} \vec{p}_{h_1}-\vec{p}_{1'}\right)^2} \\
& \times \frac{\alpha_s}{2\pi} \frac{1}{\hat{\epsilon}}  Q_q^2 \left[
\int_{\frac{x_{h_2}}{x_{\bar{q}}}}^{1}\frac{d\beta_2}{\beta_2} C_F \frac{1+ \beta_2^2}{(1-\beta_2)_+} D_q^{h_1}\left(\frac{x_{h_1}}{ x_q},\mu_F\right) D_{\bar{q}}^{h_2}\left(\frac{x_{h_2}}{\beta_2 x_{\bar{q}}},\mu_F \right) \right. \\
& + \int_{\frac{x_{h_2}}{x_{\bar{q}}}}^{1-\frac{\alpha}{x_q}} d \beta_2 C_F \frac{2}{1-\beta_2} \left(\frac{c_0^2}{\left(\frac{z_{1\perp}-z_{2\perp}}{2}\right)^2 \mu^2}\right)^\epsilon D_q^{h_1}\left(\frac{x_{h_1}}{x_q},\mu_F\right) D_{\bar{q}}^{h_2}\left(\frac{x_{h_2}}{x_{\bar{q}}},\mu_F \right) \\
& \left. - 2  C_F \ln \left(1-\frac{x_{h_2}}{x_{\bar{q}}}\right)  D_q^{h_1}\left(\frac{x_{h_1}}{x_q},\mu_F\right) D_{\bar{q}}^{h_2}\left(\frac{x_{h_2}}{x_{\bar{q}}},\mu_F \right) \right] + (h_1 \leftrightarrow h_2 ) \,.      \numberthis[coll_qbarg_div]
\end{align*} }

The first term cancels with the + prescription term in the second $P_{qq}$ in \eqref{eq:ct_LL}.
We have to remove the second term to avoid double counting with the soft contribution. The third term is to be removed by the soft contribution. 

The finite part is 
{\allowdisplaybreaks
\begin{align*}
&\frac{d \sigma_{3LL}^{q \bar{q} \rightarrow h_1 h_2}}{ d x_{h_1} d x_{h_2} d p_{h_1 \perp} d^d p_{h_2\perp} } \bigg |_{\text{coll. } \bar{q}g \text{ fin}} \\
&= \frac{4  \alpha_{\mathrm{em}} Q^2}{(2\pi)^{4(d-1)} N_c} \sum_{q} \int_{x_{h_1}}^{1} d x_q  \int_{x_{h_2}}^{1} d x_{\bar{q}} \;  x_q x_{\bar{q}} \delta(1-x_q -x_{\bar{q}}) \left(\frac{x_q}{x_{h_1}}\right)^d \left(\frac{x_{\bar{q}}}{x_{h_2}}\right)^d \\
& \times \int   d^d p_{1\perp}  \int d^d z_{1\perp}   \frac{e^{i z_{1\perp}\cdot \left( -\frac{x_q}{2 x_{h_1}} p_{h_1\perp} - \frac{x_{\bar{q}}}{2 x_{h_2}} p_{h_2 \perp} + p_{1\perp} \right)} F(z_{1\perp})}{x_q x_{\bar{q}} Q^2 +\left(\frac{x_q}{x_{h_1}} \vec{p}_{h_1}-\vec{p}_{1}\right)^2} \\
& \times \int   d^d p_{1'\perp}  \int d^d z_{2\perp}   \frac{e^{-i z_{2\perp}\cdot \left(- \frac{x_q}{2 x_{h_1}} p_{h_1\perp} - \frac{x_{\bar{q}}}{2 x_{h_2}} p_{h_2 \perp} + p_{1'\perp} \right)} F^*(z_{2\perp})}{x_q x_{\bar{q}} Q^2 +\left(\frac{x_q}{x_{h_1}} \vec{p}_{h_1}-\vec{p}_{1'}\right)^2} \\
& \times  \frac{\alpha_s C_F}{2\pi} \left\{ \int_{\frac{x_{h_2}}{x_{\bar{q}}}}^1\frac{d\beta_2}{\beta_2}  Q_q^2 D_q^{h_1}\left(\frac{x_{h_1}}{x_q}, \mu_F\right) D_{\bar{q}}^{h_2}\left(\frac{x_{h_2}}{\beta_2 x_{\bar{q}}}, \mu_F\right) \right. \\
& \times \left[ \ln \left( \frac{c_0^2}{\left(\frac{z_{1\perp} - z_{2\perp}}{2}\right)^2 \mu^2}\right)  \frac{1+ \beta_2^2}{(1-\beta_2)_+}  + \frac{(1-\beta_2)^2 + 2 (1+ \beta_2^2) \ln \beta_2}{(1-\beta_2)_+} \right] \\
& \left. - 2 \ln \left( 1- \frac{x_{h_2}}{x_{\bar{q}}}\right) \ln \left( \frac{c_0^2}{\left(\frac{z_{1\perp} - z_{2\perp}}{2}\right)^2 \mu^2}\right) D_{q}^{h_1} \left(\frac{x_{h_1}}{x_q},\mu_F\right) D_{\bar{q}}^{h_q} \left(\frac{x_{h_2}}{x_{\bar{q}}},\mu_F \right)\right\} + (h_1 \leftrightarrow h_2 )  \,. \numberthis[collqbarg_LL_fin]
\end{align*} }

For the $TL$ transition, 
{\allowdisplaybreaks
\begin{align*} 
 & \frac{d \sigma_{3TL}^{q \bar{q} \rightarrow h_1 h_2}}{ d x_{h_1} d x_{h_2} d p_{h_1 \perp} d^d p_{h_2\perp} } \bigg |_{\text{coll. } \bar{q}g}  \\*
 & =  \frac{2 \alpha_{\mathrm{em}} Q}{(2\pi)^{4(d-1)} N_c} \sum_{q} \int_{x_{h_1}}^{1} d x_q  \int_{x_{h_2}}^{1} d x_{\bar{q}} \; (x_{\bar{q}} -x_q) \delta(1-x_q -x_{\bar{q}}) \left(\frac{x_q}{x_{h_1}}\right)^d \left(\frac{x_{\bar{q}}}{x_{h_2}}\right)^d \\
& \times \int   d^d p_{1\perp}  \int d^d z_{1\perp}   \frac{e^{i z_{1\perp}\cdot \left( -\frac{x_q}{2 x_{h_1}} p_{h_1\perp} - \frac{x_{\bar{q}}}{2 x_{h_2}} p_{h_2 \perp} + p_{1\perp} \right)} F(z_{1\perp})}{x_q x_{\bar{q}} Q^2 +\left(\frac{x_q}{x_{h_1}} \vec{p}_{h_1}-\vec{p}_{1}\right)^2} \\
& \times \int   d^d p_{1'\perp}  \int d^d z_{2\perp}   \frac{e^{-i z_{2\perp}\cdot \left(- \frac{x_q}{2 x_{h_1}} p_{h_1\perp} - \frac{x_{\bar{q}}}{2 x_{h_2}} p_{h_2 \perp} + p_{1'\perp} \right)} F^*(z_{2\perp})}{x_q x_{\bar{q}} Q^2 +\left(\frac{x_q}{x_{h_1}} \vec{p}_{h_1}-\vec{p}_{1'}\right)^2} \left( \frac{x_q}{x_{h_1}} p_{h_1 }  - p_{1'}\right) \cdot \varepsilon^* _{T}   \\
& \times  \int_{\frac{x_{h_2}}{x_{\bar{q}}}}^{1-\frac{\alpha}{x_{\bar{q}}}}\frac{d\beta_2}{\beta_2}  Q_q^2 D_q^{h_1}\left(\frac{x_{h_1}}{x_q}, \mu_F\right) D_{\bar{q}}^{h_2}\left(\frac{x_{h_2}}{ \beta_2 x_{\bar{q}}}, \mu_F\right) \\
& \times \frac{\alpha_s C_F}{2\pi} \left[ \frac{1}{\hat{\epsilon}} \left(\frac{c_0^2}{\left(\frac{z_{2\perp}-z_{1\perp}}{2}\right)^2 \mu^2}\right)^\epsilon \frac{1+ \beta_2^2}{1-\beta_2}  + \frac{(1-\beta_2)^2 + 2 (1+ \beta_2^2) \ln \beta_2 }{(1-\beta_2)} \right] + (h_1 \leftrightarrow h_2 ) \\
&= \frac{d \sigma_{3TL}^{q \bar{q} \rightarrow h_1 h_2}}{ d x_{h_1} d x_{h_2} d p_{h_1 \perp} d^d p_{h_2\perp} } \bigg |_{\text{coll. } \bar{q}g  \text{ div}}     + \frac{d \sigma_{3TL}^{q \bar{q} \rightarrow h_1 h_2}}{ d x_{h_1} d x_{h_2} d p_{h_1 \perp} d^d p_{h_2\perp} } \bigg |_{\text{coll. } \bar{q}g \text{ fin}}      \numberthis
\end{align*}
}
where 
{\allowdisplaybreaks
\begin{align*}
&\frac{d \sigma_{3TL}^{q \bar{q} \rightarrow h_1 h_2}}{ d x_{h_1} d x_{h_2} d p_{h_1 \perp} d^d p_{h_2\perp} } \bigg |_{\text{coll. } \bar{q}g  \text{ div}} \\ 
&=  \frac{2  \alpha_{\mathrm{em}} Q}{(2\pi)^{4(d-1)} N_c} \sum_{q} \int_{x_{h_1}}^{1} d x_q \int_{x_{h_2}}^1 d x_{\bar{q}}  (x_{\bar{q}} -x_q)  \left(\frac{x_q}{x_{h_1}}\right)^d \left(\frac{x_{\bar{q}}}{x_{h_2}}\right)^d \delta(1-x_q-x_{\bar{q}}) \\
& \times \int   d^d p_{1\perp}  \int d^d z_{1\perp}   \frac{e^{i z_{1\perp}\cdot \left( -\frac{x_q}{2 x_{h_1}} p_{h_1\perp} - \frac{x_{\bar{q}}}{2 x_{h_2}} p_{h_2 \perp} + p_{1\perp} \right)} F(z_{1\perp})}{x_q x_{\bar{q}} Q^2 +\left(\frac{x_q}{x_{h_1}} \vec{p}_{h_1}-\vec{p}_{1}\right)^2} \\
& \times \int   d^d p_{1'\perp}  \int d^d z_{2\perp}   \frac{e^{-i z_{2\perp}\cdot \left(- \frac{x_q}{2 x_{h_1}} p_{h_1\perp} - \frac{x_{\bar{q}}}{2 x_{h_2}} p_{h_2 \perp} + p_{1'\perp} \right)} F^*(z_{2\perp})}{x_q x_{\bar{q}} Q^2 +\left(\frac{x_q}{x_{h_1}} \vec{p}_{h_1}-\vec{p}_{1'}\right)^2} \left( \frac{x_q}{x_{h_1}} p_{h_1}  - p_{1'}\right) \cdot \varepsilon^* _{T}  \\
& \times \frac{\alpha_s}{2\pi} \frac{1}{\hat{\epsilon}}  Q_q^2 \left[
\int_{\frac{x_{h_2}}{x_{\bar{q}}}}^{1}\frac{d\beta_2}{\beta_2} C_F \frac{1+ \beta_2^2}{(1-\beta_2)_+} D_q^{h_1}\left(\frac{x_{h_1}}{ x_q},\mu_F\right) D_{\bar{q}}^{h_2}\left(\frac{x_{h_2}}{\beta_2 x_{\bar{q}}},\mu_F \right) \right. \\
& + \int_{\frac{x_{h_2}}{x_{\bar{q}}}}^{1-\frac{\alpha}{x_{\bar{q}}}} d \beta_2 C_F \frac{2}{1-\beta_2} \left(\frac{c_0^2}{\left(\frac{z_{1\perp}-z_{2\perp}}{2}\right)^2 \mu^2}\right)^\epsilon D_q^{h_1}\left(\frac{x_{h_1}}{x_q},\mu_F\right) D_{\bar{q}}^{h_2}\left(\frac{x_{h_2}}{x_{\bar{q}}},\mu_F \right) \\
& \left. - 2  C_F \ln \left(1-\frac{x_{h_2}}{x_{\bar{q}}}\right)  D_q^{h_1}\left(\frac{x_{h_1}}{x_q},\mu_F\right) D_{\bar{q}}^{h_2}\left(\frac{x_{h_2}}{x_{\bar{q}}},\mu_F \right) \right] + (h_1 \leftrightarrow h_2 ) \,,   \numberthis
\end{align*} }

and 
\begin{align*}
&\frac{d \sigma_{3TL}^{q \bar{q} \rightarrow h_1 h_2}}{ d x_{h_1} d x_{h_2} d p_{h_1 \perp} d^d p_{h_2\perp} } \bigg |_{\text{coll. } \bar{q}g \text{ fin}} \\
&= \frac{2  \alpha_{\mathrm{em}} Q}{(2\pi)^{4(d-1)} N_c} \sum_{q} \int_{x_{h_1}}^{1} d x_q  \int_{x_{h_2}}^{1} d x_{\bar{q}} \;  (x_{\bar{q}} -x_q)  \delta(1-x_q -x_{\bar{q}}) \left(\frac{x_q}{x_{h_1}}\right)^d \left(\frac{x_{\bar{q}}}{x_{h_2}}\right)^d \\
& \times \int   d^d p_{1\perp}  \int d^d z_{1\perp}   \frac{e^{i z_{1\perp}\cdot \left( -\frac{x_q}{2 x_{h_1}} p_{h_1\perp} - \frac{x_{\bar{q}}}{2 x_{h_2}} p_{h_2 \perp} + p_{1\perp} \right)} F(z_{1\perp})}{x_q x_{\bar{q}} Q^2 +\left(\frac{x_q}{x_{h_1}} \vec{p}_{h_1}-\vec{p}_{1}\right)^2} \\
& \times \int   d^d p_{1'\perp}  \int d^d z_{2\perp}   \frac{e^{-i z_{2\perp}\cdot \left(- \frac{x_q}{2 x_{h_1}} p_{h_1\perp} - \frac{x_{\bar{q}}}{2 x_{h_2}} p_{h_2 \perp} + p_{1'\perp} \right)} F^*(z_{2\perp})}{x_q x_{\bar{q}} Q^2 +\left(\frac{x_q}{x_{h_1}} \vec{p}_{h_1}-\vec{p}_{1'}\right)^2} \left( \frac{x_q}{x_{h_1}} p_{h_1}  - p_{1'}\right) \cdot \varepsilon^* _{T}  \\
& \times \frac{\alpha_s C_F}{2\pi} \left\{ \int_{\frac{x_{h_2}}{x_{\bar{q}}}}^1 \frac{d\beta_2}{\beta_2}  Q_q^2 D_q^{h_1}\left(\frac{x_{h_1}}{x_q}, \mu_F\right) D_{\bar{q}}^{h_2}\left(\frac{x_{h_2}}{\beta_2 x_{\bar{q}}}, \mu_F\right) \right. \\
& \times  \left[ \ln \left( \frac{c_0^2}{\left(\frac{z_{1\perp} - z_{2\perp}}{2}\right)^2 \mu^2}\right)  \frac{1+ \beta_2^2}{(1-\beta_2)_+}  + \frac{(1-\beta_2)^2 + 2 (1+ \beta_2^2) \ln \beta_2}{(1-\beta_2)_+} \right] \\
& \left. - 2 \ln \left( 1- \frac{x_{h_2}}{x_{\bar{q}}}\right) \ln \left( \frac{c_0^2}{\left(\frac{z_{1\perp} - z_{2\perp}}{2}\right)^2 \mu^2}\right) D_{q}^{h_1} \left(\frac{x_{h_1}}{x_q},\mu_F\right) D_{\bar{q}}^{h_q} \left(\frac{x_{h_2}}{x_{\bar{q}}},\mu_F \right)\right\} + (h_1 \leftrightarrow h_2 )  \,. \numberthis[collqbarg_TL_fin]
\end{align*}

For the $TT$ case, we get 

{\allowdisplaybreaks
\begin{align*} 
& \frac{d \sigma_{3TT}^{q \bar{q} \rightarrow h_1 h_2}}{ d x_{h_1} d x_{h_2} d p_{h_1 \perp} d^d p_{h_2\perp} } \bigg |_{\text{coll. } \bar{q}g}  \\*
 & =  \frac{ \alpha_{\mathrm{em}} }{(2\pi)^{4(d-1)} N_c} \sum_{q} \int_{x_{h_1}}^{1} \frac{d x_q}{x_q}  \int_{x_{h_2}}^{1} \frac{d x_{\bar{q}} }{x_{\bar{q}}} \delta(1-x_q -x_{\bar{q}}) \left(\frac{x_q}{x_{h_1}}\right)^d \left(\frac{x_{\bar{q}}}{x_{h_2}}\right)^d  \\
 & \times \left[ (x_{\bar{q}} -x_q)^2 g_{\perp}^{ri}g_{\perp}^{lk} - g_{\perp}^{rk}g_{\perp}^{li} + g_{\perp}^{rl}g_{\perp}^{ik} \right] \\
& \times \int   d^d p_{1\perp}  \int d^d z_{1\perp}   \frac{e^{i z_{1\perp}\cdot \left( -\frac{x_q}{2 x_{h_1}} p_{h_1\perp} - \frac{x_{\bar{q}}}{2 x_{h_2}} p_{h_2 \perp} + p_{1\perp} \right)} F(z_{1\perp})}{x_q x_{\bar{q}} Q^2 +\left(\frac{x_q}{x_{h_1}} \vec{p}_{h_1}-\vec{p}_{1}\right)^2} \left(\frac{x_{q}}{x_{h_1}} p_{h_1} - p_{1}\right)_r \varepsilon_{T i}  \\
& \times \int   d^d p_{1'\perp}  \int d^d z_{2\perp}   \frac{e^{-i z_{2\perp}\cdot \left(- \frac{x_q}{2 x_{h_1}} p_{h_1\perp} - \frac{x_{\bar{q}}}{2 x_{h_2}} p_{h_2 \perp} + p_{1'\perp} \right)} F^*(z_{2\perp})}{x_q x_{\bar{q}} Q^2 +\left(\frac{x_q}{x_{h_1}} \vec{p}_{h_1}-\vec{p}_{1'}\right)^2}  \left(\frac{x_{q}}{x_{h_1}} p_{h_1} - p_{1'}\right)_l \varepsilon_{T k}^*  \\
& \times  \int_{\frac{x_{h_2}}{x_{\bar{q}}}}^{1-\frac{\alpha}{x_{\bar{q}}}}\frac{d\beta_2}{\beta_2}  Q_q^2 D_q^{h_1}\left(\frac{x_{h_1}}{x_q}, \mu_F\right) D_{\bar{q}}^{h_2}\left(\frac{x_{h_2}}{\beta_2 x_{\bar{q}}}, \mu_F\right)  \\
& \times \frac{\alpha_s C_F}{2\pi} \left[ \frac{1}{\hat{\epsilon}} \left(\frac{c_0^2}{\left(\frac{z_{2\perp}-z_{1\perp}}{2}\right)^2 \mu^2}\right)^\epsilon \frac{1+ \beta_2^2}{1-\beta_2}  + \frac{(1-\beta_2)^2 + 2 (1+ \beta_2^2)\ln \beta_2 }{(1-\beta_2)} \right] + (h_1 \leftrightarrow h_2 ) \\
&= \frac{d \sigma_{3TT}^{q \bar{q} \rightarrow h_1 h_2}}{ d x_{h_1} d x_{h_2} d p_{h_1 \perp} d^d p_{h_2\perp} } \bigg |_{\text{coll. } \bar{q}g  \text{ div}}     + \frac{d \sigma_{3TT}^{q \bar{q} \rightarrow h_1 h_2}}{ d x_{h_1} d x_{h_2} d p_{h_1 \perp} d^d p_{h_2\perp} } \bigg |_{\text{coll. } \bar{q}g \text{ fin}}  ,   \numberthis
\end{align*}}
where 
{\allowdisplaybreaks
\begin{align*}
&\frac{d \sigma_{3TT}^{q \bar{q} \rightarrow h_1 h_2}}{ d x_{h_1} d x_{h_2} d p_{h_1 \perp} d^d p_{h_2\perp} } \bigg |_{\text{coll. } \bar{q}g  \text{ div}} \\ 
&=  \frac{  \alpha_{\mathrm{em}} }{(2\pi)^{4(d-1)} N_c} \sum_{q} \int_{x_{h_1}}^{1} \frac{d x_q}{x_q}\int_{x_{h_2}}^1 \frac{d x_{\bar{q}} }{x_{\bar{q}}} \left(\frac{x_q}{x_{h_1}}\right)^d \left(\frac{x_{\bar{q}}}{x_{h_2}}\right)^d \delta(1-x_q-x_{\bar{q}}) \\
& \times \left[ (x_{\bar{q}} -x_q)^2 g_{\perp}^{ri}g_{\perp}^{lk} - g_{\perp}^{rk}g_{\perp}^{li} + g_{\perp}^{rl}g_{\perp}^{ik} \right] \\
& \times \int   d^d p_{1\perp}  \int d^d z_{1\perp}   \frac{e^{i z_{1\perp}\cdot \left( -\frac{x_q}{2 x_{h_1}} p_{h_1\perp} - \frac{x_{\bar{q}}}{2 x_{h_2}} p_{h_2 \perp} + p_{1\perp} \right)} F(z_{1\perp})}{x_q x_{\bar{q}} Q^2 +\left(\frac{x_q}{x_{h_1}} \vec{p}_{h_1}-\vec{p}_{1}\right)^2} \left(\frac{x_{q}}{x_{h_1}} p_{h_1} - p_{1}\right)_r \varepsilon_{T i}  \\
& \times \int   d^d p_{1'\perp}  \int d^d z_{2\perp}   \frac{e^{-i z_{2\perp}\cdot \left(- \frac{x_q}{2 x_{h_1}} p_{h_1\perp} - \frac{x_{\bar{q}}}{2 x_{h_2}} p_{h_2 \perp} + p_{1'\perp} \right)} F^*(z_{2\perp})}{x_q x_{\bar{q}} Q^2 +\left(\frac{x_q}{x_{h_1}} \vec{p}_{h_1}-\vec{p}_{1'}\right)^2}  \left(\frac{x_{q}}{x_{h_1}} p_{h_1} - p_{1'}\right)_l \varepsilon_{T k}^*  \\
& \times \frac{\alpha_s}{2\pi} \frac{1}{\hat{\epsilon}}  Q_q^2 \left[
\int_{\frac{x_{h_2}}{x_{\bar{q}}}}^{1}\frac{d\beta_2}{\beta_2} C_F \frac{1+ \beta_2^2}{(1-\beta_2)_+} D_q^{h_1}\left(\frac{x_{h_1}}{ x_q},\mu_F\right) D_{\bar{q}}^{h_2}\left(\frac{x_{h_2}}{\beta_2 x_{\bar{q}}},\mu_F \right) \right. \\
& + \int_{\frac{x_{h_2}}{x_{\bar{q}}}}^{1-\frac{\alpha}{x_{\bar{q}}}} d \beta_2 C_F \frac{2}{1-\beta_2} \left(\frac{c_0^2}{\left(\frac{z_{1\perp}-z_{2\perp}}{2}\right)^2 \mu^2}\right)^\epsilon D_q^{h_1}\left(\frac{x_{h_1}}{x_q},\mu_F\right) D_{\bar{q}}^{h_2}\left(\frac{x_{h_2}}{x_{\bar{q}}},\mu_F \right) \\
& \left. - 2  C_F \ln \left(1-\frac{x_{h_2}}{x_{\bar{q}}}\right)  D_q^{h_1}\left(\frac{x_{h_1}}{x_q},\mu_F\right) D_{\bar{q}}^{h_2}\left(\frac{x_{h_2}}{x_{\bar{q}}},\mu_F \right) \right] + (h_1 \leftrightarrow h_2 ) \,,   \numberthis
\end{align*} }
\newpage and 
{\allowdisplaybreaks
\begin{align*}
&\frac{d \sigma_{3TT}^{q \bar{q} \rightarrow h_1 h_2}}{ d x_{h_1} d x_{h_2} d p_{h_1 \perp} d^d p_{h_2\perp} } \bigg |_{\text{coll. } \bar{q}g \text{ fin}} \\
&= \frac{  \alpha_{\mathrm{em}} }{(2\pi)^{4(d-1)} N_c} \sum_{q} \int_{x_{h_1}}^{1} \frac{d x_q}{x_q}  \int_{x_{h_2}}^{1} \frac{d x_{\bar{q}}}{x_{\bar{q}}} \delta(1-x_q -x_{\bar{q}}) \left(\frac{x_q}{x_{h_1}}\right)^d \left(\frac{x_{\bar{q}}}{x_{h_2}}\right)^d \\
& \times \left[ (x_{\bar{q}} -x_q)^2 g_{\perp}^{ri}g_{\perp}^{lk} - g_{\perp}^{rk}g_{\perp}^{li} + g_{\perp}^{rl}g_{\perp}^{ik} \right] \\
& \times \int   d^d p_{1\perp}  \int d^d z_{1\perp}   \frac{e^{i z_{1\perp}\cdot \left( -\frac{x_q}{2 x_{h_1}} p_{h_1\perp} - \frac{x_{\bar{q}}}{2 x_{h_2}} p_{h_2 \perp} + p_{1\perp} \right)} F(z_{1\perp})}{x_q x_{\bar{q}} Q^2 +\left(\frac{x_q}{x_{h_1}} \vec{p}_{h_1}-\vec{p}_{1}\right)^2}  \left(\frac{x_{q}}{x_{h_1}} p_{h_1} - p_{1}\right)_r \varepsilon_{T i}  \\
& \times \int   d^d p_{1'\perp}  \int d^d z_{2\perp}   \frac{e^{-i z_{2\perp}\cdot \left(- \frac{x_q}{2 x_{h_1}} p_{h_1\perp} - \frac{x_{\bar{q}}}{2 x_{h_2}} p_{h_2 \perp} + p_{1'\perp} \right)} F^*(z_{2\perp})}{x_q x_{\bar{q}} Q^2 +\left(\frac{x_q}{x_{h_1}} \vec{p}_{h_1}-\vec{p}_{1'}\right)^2} \left(\frac{x_{q}}{x_{h_1}} p_{h_1} - p_{1'}\right)_l \varepsilon_{T k}^*   \\
& \times \frac{\alpha_s C_F}{2\pi} \left\{ \int_{\frac{x_{h_2}}{x_{\bar{q}}}}^{1}\frac{d\beta_2}{\beta_2}  Q_q^2 D_q^{h_1}\left(\frac{x_{h_1}}{x_q}, \mu_F\right) D_{\bar{q}}^{h_2}\left(\frac{x_{h_2}}{\beta_2 x_{\bar{q}}}, \mu_F\right) \right. \\
& \times  \left[ \ln \left( \frac{c_0^2}{\left(\frac{z_{1\perp} - z_{2\perp}}{2}\right)^2 \mu^2}\right)  \frac{1+ \beta_2^2}{(1-\beta_2)_+}  + \frac{(1-\beta_2)^2 + 2 (1+ \beta_2^2) \ln \beta_2}{(1-\beta_2)_+} \right] \\
& \left. - 2 \ln \left( 1- \frac{x_{h_2}}{x_{\bar{q}}}\right) \ln \left( \frac{c_0^2}{\left(\frac{z_{1\perp} - z_{2\perp}}{2}\right)^2 \mu^2}\right) D_{q}^{h_1} \left(\frac{x_{h_1}}{x_q},\mu_F\right) D_{\bar{q}}^{h_q} \left(\frac{x_{h_2}}{x_{\bar{q}}},\mu_F \right)\right\} + (h_1 \leftrightarrow h_2 )  \,. \numberthis[collqbarg_TT_fin]
\end{align*} }

\subsubsection{Soft contribution}
\label{sec: soft contribution}
To calculate the soft contribution of the divergent part of the real emission cross section, the soft limit of \eqref{eq:real_div_LL} is taken by setting  $\vec{p}_g = x_g \vec{u}$, where $|\vec{u}| \sim |\vec{p}_{h}|,$ which extracts the divergence on $x_g$.
{\allowdisplaybreaks
\begin{align*}
& \frac{d \sigma_{3LL}^{q \bar{q} \rightarrow h_1 h_2 }}{d x_{h_1}d x_{h_2} d^d p_{h_1\perp} d p_{h_2 \perp}}\Bigg |_{\text{soft}} \\ 
&=  \frac{4  \alpha_{\mathrm{em}} Q^2}{(2\pi)^{4(d-1)} N_c} \sum_{q} \int_{x_{h_1}}^1 \frac{d x_q'}{x_q'} \int_{x_{h_2}}^1 \frac{d x_{\bar{q}}'}{ x_{\bar{q}}'}  Q_q^2 D_q^{h_1}\left(\frac{x_{h_1}}{x_q'},\mu_F\right)D_{\bar{q}}^{h_2}\left(\frac{x_{h_2}}{x_{\bar{q}}'}, \mu_F\right) \\
& \times \left(\frac{x_q'}{x_{h_1}}\right)^d \left(\frac{x_{\bar{q}}'}{x_{h_2}}\right)^d \int_{\alpha}^1 \frac{d x_g}{x_g^{3-d}} \delta(1-x_q'-x_{\bar{q}}'-x_g) \frac{\alpha_s C_F}{\mu^{2\epsilon}} \int \frac{d^d \vec{u}}{(2\pi)^d} \\
& \times \int d^d p_{1\perp} d^d p_{2\perp} \; \mathbf{F}\left(\frac{p_{12\perp}}{2}\right) \delta\left(\frac{x_q'}{x_{h_1}} p_{h_1\perp} - p_{1\perp} + \frac{x_{\bar{q}}'}{x_{h_2}}p_{h_2\perp} - p_{2\perp} + x_g u_{\perp}\right) \\
& \times \int d^d p_{1'\perp}  d^d p_{2'\perp} \; \mathbf{F}^*\left(\frac{p_{1'2'\perp}}{2}\right) \delta\left(\frac{x_q'}{x_{h_1}} p_{h_1\perp} - p_{1'\perp} + \frac{x_{\bar{q}}'}{x_{h_2}}p_{h_2\perp} - p_{2'\perp} + x_g u_{\perp}\right) \\
& \times \left \{ \frac{d x_g^2 + 4 x_q'(x_q'+x_g)}{\left(Q^2+ \frac{\left(\frac{x_{\bar{q}}'} {x_{h_2}}\vec{p}_{h_2}-\vec{p}_{2}\right)^2}{(1-x_{\bar{q}}')x_{\bar{q}}'}\right)\left(Q^2+ \frac{\left(\frac{x_{\bar{q}}'}{x_{h_2}}\vec{p}_{h_2}-\vec{p}_{2'}\right)^2}{(1-x_{\bar{q}}')x_{\bar{q}}'}\right) x_q'^2 \left( \vec{u}-\frac{\vec{p}_{h_1}}{x_{h_1}}\right)^2} \right.\\
& + \frac{d x_g^2 + 4x_{\bar{q}}'(x_{\bar{q}}' + x_g)}{\left(Q^2+ \frac{\left(\frac{x_{\bar{q}}'}{x_{h_2}}\vec{p}_{h_2}-\vec{p}_{2}+x_g\vec{u}\right)^2}{(1-x_q')x_q'}\right)\left(Q^2+ \frac{\left(\frac{x_{\bar{q}}'}{x_{h_2}}\vec{p}_{h_2}-\vec{p}_{2'}+x_g\vec{u}\right)^2}{(1-x_q')x_q'}\right)x_{\bar{q}}'^2 \left(\vec{u}-\frac{\vec{p}_{h_2}}{x_{h_2}}\right)^2} \\
& - \frac{[2x_g - d x_g^2 + 4x_q' x_{\bar{q}}'] \left(\vec{u}-\frac{\vec{p}_{h_1}}{x_{h_1}}\right)\cdot \left(\vec{u}-\frac{\vec{p}_{h_2}}{x_{h_2}}\right)}{\left(Q^2+ \frac{\left(\frac{x_{\bar{q}}'}{x_{h_2}}\vec{p}_{h_2}-\vec{p}_{2'}\right)^2}{(1-x_{\bar{q}}')x_{\bar{q}}'}\right)\left(Q^2+ \frac{\left(\frac{x_{\bar{q}}'}{x_{h_2}}\vec{p}_{h_2}-\vec{p}_{2}+x_g\vec{u}\right)^2}{(1-x_q')x_q'}\right) x_q' x_{\bar{q}}' \left(\vec{u}-\frac{\vec{p}_{h_1}}{x_{h_1}}\right)^2 \left(\vec{u}-\frac{\vec{p}_{h_2}}{x_{h_2}}\right)^2} \\
&  \left.  - \frac{[2x_g - d x_g^2 + 4x_q' x_{\bar{q}}'] \left(\vec{u}-\frac{\vec{p}_{h_1}}{x_{h_1}}\right)\cdot \left(\vec{u}-\frac{\vec{p}_{h_2}}{x_{h_2}}\right)}{\left(Q^2+ \frac{\left(\frac{x_{\bar{q}}'}{x_{h_2}}\vec{p}_{h_2}-\vec{p}_{2}\right)^2}{(1-x_{\bar{q}}')x_{\bar{q}}'}\right)\left(Q^2+ \frac{\left(\frac{x_{\bar{q}}'}{x_{h_2}}\vec{p}_{h_2}-\vec{p}_{2'}+x_g\vec{u}\right)^2}{(1-x_q')x_q'}\right) x_q' x_{\bar{q}}' \left(\vec{u}-\frac{\vec{p}_{h_1}}{x_{h_1}}\right)^2 \left(\vec{u}-\frac{\vec{p}_{h_2}}{x_{h_2}}\right)^2} \right\} \\
& + (h_1 \leftrightarrow h_2) \,. 
\end{align*} }
The limit $x_g \rightarrow 0$ in the $F$ function and impact factor can be taken safely in the non- divergent terms of the cross section, as $x_q'$ and $x_{\bar{q}}'$ are limited from below by $x_{h_1}, x_{h_2}$ and so cannot be arbitrary small (ie of order $x_g$). The cross section in the soft limit becomes: 
{\allowdisplaybreaks
\begin{align*}
& \frac{d \sigma_{3LL}^{q \bar{q} \rightarrow h_1 h_2 }}{d x_{h_1}d x_{h_2} d^d p_{h_1\perp} d p_{h_2 \perp}}\Bigg |_{\text{soft}} \\* 
&=  \frac{4  \alpha_{\mathrm{em}} Q^2}{(2\pi)^{4(d-1)} N_c} \sum_{q}   \int_{x_{h_1}}^1 \frac{d x_q'}{x_q'} \int_{x_{h_2}}^1 \frac{d x_{\bar{q}}'}{ x_{\bar{q}}'} Q_q^2 D_q^{h_1}\left(\frac{x_{h_1}}{x_q'},\mu_F\right)D_{\bar{q}}^{h_2}\left(\frac{x_{h_2}}{x_{\bar{q}}'}, \mu_F\right) \\
& \times \left(\frac{x_q'}{x_{h_1}}\right)^d \left(\frac{x_{\bar{q}}'}{x_{h_2}}\right)^d \int_{\alpha}^1 \frac{d x_g}{x_g^{3-d}} \delta(1-x_q'-x_{\bar{q}}'-x_g) \frac{\alpha_s C_F}{\mu^{2\epsilon}} \int \frac{d^d \vec{u}}{(2\pi)^d} \\
& \times \int d^d p_{1\perp} d^d p_{2\perp} \; \mathbf{F}\left(\frac{p_{12\perp}}{2}\right) \delta\left(\frac{x_q'}{x_{h_1}} p_{h_1\perp} - p_{1\perp} + \frac{x_{\bar{q}}'}{x_{h_2}}p_{h_2\perp} - p_{2\perp} \right) \\
& \times \int d^d p_{1'\perp} d^d p_{2'\perp} \; \mathbf{F}^*\left(\frac{p_{1'2'\perp}}{2}\right) \delta\left(\frac{x_q'}{x_{h_1}} p_{h_1\perp} - p_{1'\perp} + \frac{x_{\bar{q}}'}{x_{h_2}}p_{h_2\perp} - p_{2'\perp} \right) \\
& \times \left \{ \frac{ 4 }{\left(Q^2+ \frac{\left(\frac{x_{\bar{q}}'} {x_{h_2}}\vec{p}_{h_2}-\vec{p}_{2}\right)^2}{(1-x_{\bar{q}}')x_{\bar{q}}'}\right)\left(Q^2+ \frac{\left(\frac{x_{\bar{q}}'}{x_{h_2}}\vec{p}_{h_2}-\vec{p}_{2'}\right)^2}{(1-x_{\bar{q}}')x_{\bar{q}}'}\right) \left( \vec{u}-\frac{\vec{p}_{h_1}}{x_{h_1}}\right)^2} \right.\\
& + \frac{ 4 }{\left(Q^2+ \frac{\left(\frac{x_{\bar{q}}'}{x_{h_2}}\vec{p}_{h_2}-\vec{p}_{2}\right)^2}{(1-x_q')x_q'}\right)\left(Q^2+ \frac{\left(\frac{x_{\bar{q}}'}{x_{h_2}}\vec{p}_{h_2}-\vec{p}_{2'}\right)^2}{(1-x_q')x_q'}\right)\left(\vec{u}-\frac{\vec{p}_{h_2}}{x_{h_2}}\right)^2} \\
& - \frac{4 \left(\vec{u}-\frac{\vec{p}_{h_1}}{x_{h_1}}\right)\cdot \left(\vec{u}-\frac{\vec{p}_{h_2}}{x_{h_2}}\right)}{\left(Q^2+ \frac{\left(\frac{x_{\bar{q}}'}{x_{h_2}}\vec{p}_{h_2}-\vec{p}_{2'}\right)^2}{(1-x_{\bar{q}}')x_{\bar{q}}'}\right)\left(Q^2+ \frac{\left(\frac{x_{\bar{q}}'}{x_{h_2}}\vec{p}_{h_2}-\vec{p}_{2}\right)^2}{(1-x_q')x_q'}\right)  \left(\vec{u}-\frac{\vec{p}_{h_1}}{x_{h_1}}\right)^2 \left(\vec{u}-\frac{\vec{p}_{h_2}}{x_{h_2}}\right)^2} \\
&  \left.  - \frac{4 \left(\vec{u}-\frac{\vec{p}_{h_1}}{x_{h_1}}\right)\cdot \left(\vec{u}-\frac{\vec{p}_{h_2}}{x_{h_2}}\right)}{\left(Q^2+ \frac{\left(\frac{x_{\bar{q}}'}{x_{h_2}}\vec{p}_{h_2}-\vec{p}_{2}\right)^2}{(1-x_{\bar{q}}')x_{\bar{q}}'}\right)\left(Q^2+ \frac{\left(\frac{x_{\bar{q}}'}{x_{h_2}}\vec{p}_{h_2}-\vec{p}_{2}'\right)^2}{(1-x_q')x_q'}\right) \left(\vec{u}-\frac{\vec{p}_{h_1}}{x_{h_1}}\right)^2 \left(\vec{u}-\frac{\vec{p}_{h_2}}{x_{h_2}}\right)^2} 
\right\} + (h_1 \leftrightarrow h_2) \\ 
&=  \frac{4  \alpha_{\mathrm{em}} Q^2}{(2\pi)^{4(d-1)} N_c} \sum_{q} \int_{x_{h_1}}^1 \frac{d x_q'}{x_q'} \int_{x_{h_2}}^1 \frac{d x_{\bar{q}}'}{ x_{\bar{q}}'} Q_q^2 D_q^{h_1}\left(\frac{x_{h_1}}{x_q'},\mu_F\right)D_{\bar{q}}^{h_2}\left(\frac{x_{h_2}}{x_{\bar{q}}'}, \mu_F\right) \\
& \times \left(\frac{x_q'}{x_{h_1}}\right)^d \left(\frac{x_{\bar{q}}'}{x_{h_2}}\right)^d \int_{\alpha}^1 \frac{d x_g}{x_g^{3-d}} \delta(1-x_q'-x_{\bar{q}}'-x_g) \frac{\alpha_s C_F}{\mu^{2\epsilon}} \int \frac{d^d \vec{u}}{(2\pi)^d} \\
& \times \int d^d p_{1\perp} d^d p_{2\perp} \; \mathbf{F}\left(\frac{p_{12\perp}}{2}\right) \delta\left(\frac{x_q'}{x_{h_1}} p_{h_1\perp} - p_{1\perp} + \frac{x_{\bar{q}}'}{x_{h_2}}p_{h_2\perp} - p_{2\perp} \right) \\
& \times \int d^d p_{1'\perp}  d^d p_{2'\perp} \; \mathbf{F}^*\left(\frac{p_{1'2'\perp}}{2}\right) \delta\left(\frac{x_q'}{x_{h_1}} p_{h_1\perp} - p_{1'\perp} + \frac{x_{\bar{q}}'}{x_{h_2}}p_{h_2\perp} - p_{2'\perp} \right) \\
& \times \frac{ 4 }{\left(Q^2+ \frac{\left(\frac{x_{\bar{q}}'} {x_{h_2}}\vec{p}_{h_2}-\vec{p}_{2}\right)^2}{(1-x_{\bar{q}}')x_{\bar{q}}'}\right)\left(Q^2+ \frac{\left(\frac{x_{\bar{q}}'}{x_{h_2}}\vec{p}_{h_2}-\vec{p}_{2'}\right)^2}{(1-x_{\bar{q}}')x_{\bar{q}}'}\right) } \left \{ \frac{ 1 }{ \left( \vec{u}-\frac{\vec{p}_{h_1}}{x_h}\right)^2}  + \frac{ 1 }{\left(\vec{u}-\frac{\vec{p}_{h_2}}{x_{h_2}}\right)^2} \right. \\
& \left. - \frac{2 \left(\vec{u}-\frac{\vec{p}_{h_1}}{x_{h_1}}\right)\cdot \left(\vec{u}-\frac{\vec{p}_{h_2}}{x_{h_2}}\right)}{\left(\vec{u}-\frac{\vec{p}_{h_1}}{x_{h_1}}\right)^2 \left(\vec{u}-\frac{\vec{p}_{h_2}}{x_{h_2}}\right)^2} 
\right\} + (h_1 \leftrightarrow h_2) \; . 
\end{align*} }

Then, the cross section is divided into two parts in order to do two different changes of variables, which are the same changes as in eqs.~(\ref{eq:Transbeta}, \ref{eq:transBeta2}), in each part:
\begin{equation}
    x_q' = \beta_1 x_q \,,\hspace{2 cm}  x_g  = (1-\beta_1) x_q \,,
    \label{eq:Transbeta1}
\end{equation}
\begin{equation}
x_{\bar{q}}' = \beta_2 x_{\bar{q}}\,, \hspace{2 cm} x_g = (1-\beta_2) x_{\bar{q}}  \,,
   \label{eq:Transbeta2}
\end{equation}
 where the integration boundaries are calculated following the steps in Eq.~(\ref{eq:change_variable_integral}). This division and changes of variable respect the symmetry between diagrams (1) + (2) on one side, and (3) + (4) on the other side in Fig.~\ref{fig:NLO-b-div}.
The limits $\beta_{1,2} \rightarrow 1 $, corresponding to $x_g \rightarrow 0$ are then taken. The choice of splitting the cross section in this way, comes from 
the will to observe the cancellation of divergences at integrand level.
The first term, after the transformation (\ref{eq:Transbeta1}) and after taking the limit, gives 
{\allowdisplaybreaks
 \begin{align*}
& \frac{d \sigma_{3LL}^{q \bar{q} \rightarrow h_1 h_2 }}{d x_{h_1}d x_{h_2} d^d p_{h_1\perp} d p_{h_2 \perp}}\Bigg |_{\text{soft } \beta_1} \\
&= \frac{2  \alpha_{\mathrm{em}} Q^2}{(2\pi)^{4(d-1)} N_c} \sum_{q} \int_{x_{h_1}}^{1-x_{h_2}} \frac{d x_q}{x_q} \frac{1}{1-x_q}\left(\frac{x_q}{x_{h_1}}\right)^d \left(\frac{1-x_q}{x_{h_2}}\right)^d \\
& \times   Q_q^2 D_q^{h_1}\left(\frac{x_{h_1}}{x_q},\mu_F\right)D_{\bar{q}}^{h_2}\left(\frac{x_{h_2}}{1-x_q}, \mu_F\right) \\
& \times \int  d^d p_{2\perp} \;  \mathbf{F}\left(\frac{x_q}{2x_{h_1}} p_{h_1\perp}  + \frac{1-x_q}{2x_{h_2}}p_{h_2\perp} - p_{2\perp}\right) \\
& \times  \int  d^d p_{2'\perp} \;  \mathbf{F}^*\left(\frac{x_q}{2x_{h_1}} p_{h_1\perp}  + \frac{1-x_q}{2x_{h_2}}p_{h_2\perp} - p_{2'\perp}\right) \\
& \times \int_{\frac{x_{h_1}}{x_q}}^{1-\frac{\alpha}{x_q}} \frac{d\beta_1}{ (1-\beta_1)^{1-2\epsilon} x_q^{1-2\epsilon}} x_q \\
& \times \frac{4 (1-x_q)^2 x_q^2 }{\left((1-x_q)x_q Q^2+ \left(\frac{1-x_q}{x_{h_2}}\vec{p}_{h_2}-\vec{p}_{2}\right)^2\right)\left((1-x_q)x_q Q^2+ \left(\frac{1-x_q}{x_{h_2}}\vec{p}_{h_2}-\vec{p}_{2'}\right)^2\right)} \\
& \times  \frac{\alpha_s C_F}{\mu^{2\epsilon}} \int \frac{d^d \vec{u}}{(2\pi)^d} \left \{ \frac{1}{\left( \vec{u}-\frac{\vec{p}_{h_1}}{x_{h_1}}\right)^2}  + \frac{1}{\left(\vec{u}-\frac{\vec{p}_{h_2}}{x_{h_2}}\right)^2}  - 2 \frac{ \left(\vec{u}-\frac{\vec{p}_{h_1}}{x_{h_1}}\right)\cdot \left(\vec{u}-\frac{\vec{p}_{h_2}}{x_{h_2}}\right)}{\left(\vec{u}-\frac{\vec{p}_{h_1}}{x_{h_1}}\right)^2 \left(\vec{u}-\frac{\vec{p}_{h_2}}{x_{h_2}}\right)^2} \right\} + (h_1 \leftrightarrow h_2) \,.
\numberthis[soft_beta_1]
\end{align*} }
In a similar fashion, the transformation (\ref{eq:Transbeta2}) leads to second contribution, which reads
    \begin{align*}
& \frac{d \sigma_{3LL}^{q \bar{q} \rightarrow h_1 h_2 }}{d x_{h_1}d x_{h_2} d^d p_{h_1\perp} d p_{h_2 \perp}}\Bigg |_{\text{soft } \beta_2} \\
&= \frac{2 \alpha_{\mathrm{em}} Q^2}{(2\pi)^{4(d-1)} N_c} \sum_{q} \int_{x_{h_2}}^{1-x_{h_1}} \frac{d x_{\bar{q}}}{x_{\bar{q}}} \frac{1}{1-x_{\bar{q}}}\left(\frac{x_{\bar{q}}}{x_{h_2}}\right)^d \left(\frac{1-x_{\bar{q}}}{x_{h_1}}\right)^d \\
& \times   Q_q^2 D_q^{h_1}\left(\frac{x_{h_1}}{1-x_{\bar{q}}},\mu_F\right)D_{\bar{q}}^{h_2}\left(\frac{x_{h_2}}{x_{\bar{q}}}, \mu_F\right) \\
& \times  \int  d^d p_{2\perp} \; \mathbf{F}\left(\frac{1-x_{\bar{q}}}{2x_{h_1}} p_{h_1\perp}  + \frac{x_{\bar{q}}}{2x_{h_2}}p_{h_2\perp} - p_{2\perp}\right) \\
& \times  \int  d^d p_{2'\perp} \; \mathbf{F}^*\left(\frac{1-x_{\bar{q}}}{2x_{h_1}} p_{h_1\perp}  + \frac{x_{\bar{q}}}{2x_{h_2}}p_{h_2\perp} - p_{2'\perp}\right) \\
& \times \int_{\frac{x_{h_2}}{x_{\bar{q}}}}^{1-\frac{\alpha}{x_{\bar{q}}}} \frac{d\beta_2}{ (1-\beta_2)^{1-2\epsilon} x_{\bar{q}}^{1-2\epsilon}} x_{\bar{q}} \\
& \times \frac{4 (1-x_{\bar{q}})^2 x_{\bar{q}}^2 }{\left( (1-x_{\bar{q}}) x_{\bar{q}} Q^2+ \left(\frac{x_{\bar{q}}}{x_{h_2}}\vec{p}_{h_2}-\vec{p}_{2}\right)^2\right)\left((1-x_{\bar{q}}) x_{\bar{q}} Q^2+\left(\frac{x_{\bar{q}}}{x_{h_2}}\vec{p}_{h_2}-\vec{p}_{2'}\right)^2\right)} \\
& \times  \frac{\alpha_s C_F}{\mu^{2\epsilon}} \int \frac{d^d \vec{u}}{(2\pi)^d} \left \{ \frac{1}{\left( \vec{u}-\frac{\vec{p}_{h_1}}{x_{h_1}}\right)^2}  + \frac{1}{\left(\vec{u}-\frac{\vec{p}_{h_2}}{x_{h_2}}\right)^2}  - 2 \frac{ \left(\vec{u}-\frac{\vec{p}_{h_1}}{x_{h_1}}\right)\cdot \left(\vec{u}-\frac{\vec{p}_{h_2}}{x_{h_2}}\right)}{\left(\vec{u}-\frac{\vec{p}_{h_1}}{x_{h_1}}\right)^2 \left(\vec{u}-\frac{\vec{p}_{h_2}}{x_{h_2}}\right)^2} \right\} + (h_1 \leftrightarrow h_2) \,.\numberthis[soft_beta_2]
    \end{align*}
Next, we integrate over $\vec{u}$, which gives 
\begin{align*}
    I_u & = \frac{\alpha_s C_F}{\mu^{2\epsilon}} \int \frac{d^d \vec{u}}{(2\pi)^d} \left \{ \frac{1}{\left( \vec{u}-\frac{\vec{p}_{h_1}}{x_h}\right)^2}  + \frac{1}{\left(\vec{u}-\frac{\vec{p}_{h_2}}{x_{h_2}}\right)^2}  - 2 \frac{ \left(\vec{u}-\frac{\vec{p}_{h_1}}{x_{h_1}}\right)\cdot \left(\vec{u}-\frac{\vec{p}_{h_2}}{x_{h_2}}\right)}{\left(\vec{u}-\frac{\vec{p}_{h_1}}{x_{h_1}}\right)^2 \left(\vec{u}-\frac{\vec{p}_{h_2}}{x_{h_2}}\right)^2} \right\} \\
    & = \frac{\alpha_s C_F}{\mu^{2\epsilon}} \left(\frac{\vec{p}_{h_1}}{x_{h_1}} - \frac{\vec{p}_{h_2}}{x_{h_2}}\right)^2 \int \frac{d^d \vec{u}}{(2\pi)^d} \frac{1}{\left( \vec{u}-\frac{\vec{p}_{h_1}}{x_{h_1}}\right)^2 \left(\vec{u}-\frac{\vec{p}_{h_2}}{x_{h_2}}\right)^2 } \\ 
    &= \frac{\alpha_s C_F}{\mu^{2\epsilon}} \left(\frac{\vec{p}_{h_1}}{x_{h_1}} - \frac{\vec{p}_{h_2}}{x_{h_2}}\right)^2 \int \frac{d^d \vec{u}}{(2\pi)^d} \frac{1}{\vec{u}^2 \left(\vec{u}-\left( \frac{\vec{p}_{h_1}}{x_{h_1}} -\frac{\vec{p}_{h_2}}{x_{h_2}} \right)\right)^2 } \\
    &= \frac{\alpha_s C_F}{\mu^{2\epsilon}} \left(\frac{\vec{p}_{h_1}}{x_{h_1}} - \frac{\vec{p}_{h_2}}{x_{h_2}}\right)^2  \frac{1}{(2 \pi)^d} \pi^{1+\epsilon} \Gamma(1-\epsilon) \beta(\epsilon,\epsilon) \left[\left( \frac{\vec{p}_{h_1}}{x_{h_1}} -\frac{\vec{p}_{h_2}}{x_{h_2}} \right)^2\right]^{\epsilon-1} \\
    &= \frac{\alpha_s}{2\pi} C_F \frac{1}{\hat{\epsilon}} \left[ 1 + \epsilon \ln \left(\frac{\left(\frac{\vec{p}_{h_1}}{x_{h_1}} -\frac{\vec{p}_{h_2}}{x_{h_2}} \right)^2}{\mu^2}\right)\right]. \numberthis[integral_u]
    \end{align*}
Finally, the integral over $\beta$ leads to
\begin{align*}
    \int_{\frac{x_h}{x}}^{1-\frac{\alpha}{x}} \frac{d \beta}{(1-\beta)^{3-d}} 
    &=  \int_{\frac{x_h}{x}}^{1-\frac{\alpha}{x}} \frac{d \beta}{1-\beta} \left[1 + 2 \epsilon\ln  (1-\beta)\right] \\
    &= -\ln\left(\frac{\alpha}{x}\right)  + \ln\left(1-\frac{x_h}{x}\right) - \epsilon \ln^2\left(\frac{\alpha}{x}\right) + \epsilon \ln^2\left(1-\frac{x_h}{x}\right) \\
    &=  -\ln \alpha + \ln x + \ln\left(1-\frac{x_h}{x}\right) - \epsilon \left[\ln^2 \alpha -2\ln \alpha \ln x + \ln^2 x\right]  \\
    & + \epsilon \ln^2\left(1-\frac{x_h}{x}\right) \numberthis[integral_beta].
\end{align*}
Combining both integrals over $\beta$ \eqref{eq:integral_beta} and over $\vec{u}$ \eqref{eq:integral_u} and keeping only the divergent terms, eqs.~\eqref{eq:soft_beta_1} and \eqref{eq:soft_beta_2} become respectively
\begin{align*}
& \frac{d \sigma_{3LL}^{q \bar{q} \rightarrow h_1 h_2 }}{d x_{h_1}d x_{h_2} d^d p_{h_1\perp} d p_{h_2 \perp}}\Bigg |_{\text{soft } \beta_1} \\
&= \frac{2  \alpha_{\mathrm{em}} Q^2}{(2\pi)^{4(d-1)} N_c}  \sum_{q}  \int_{x_{h_1}}^{1} d x_q \int_{x_{h_2}}^{1} d x_{\bar{q}} \; x_q x_{\bar{q}} \left(\frac{x_q}{x_{h_1}}\right)^d \left(\frac{x_{\bar{q}}}{x_{h_2}}\right)^d \delta(1-x_q -x_{\bar{q}}) \\
&  \times Q_q^2 D_q^{h_1}\left(\frac{x_{h_1}}{x_q},\mu_F\right)D_{\bar{q}}^{h_2}\left(\frac{x_{h_2}}{x_{\bar{q}}}, \mu_F\right) \mathcal{F}_{LL} \frac{\alpha_S C_F}{2\pi} \frac{4}{\Hat{\epsilon}}  \left[ - \ln \alpha + \ln x_q + \ln\left(1- \frac{x_{h_1}}{x_q}\right) \right. \\
& \left. - \epsilon \ln^2 \alpha - \epsilon \ln \alpha \ln \left(\frac{\left(\frac{\vec{p}_{h_1}}{x_{h_1}} -\frac{\vec{p}_{h_2}}{x_{h_2}} \right)^2}{\mu^2}\right) + \epsilon \ln x_q \Bigg( \ln x_q + 2 \ln \left( 1 - \frac{x_{h_1}}{x_{q}} \right) \right. \\ & \left. \left. + \ln \left(\frac{\left(\frac{\vec{p}_{h_1}}{x_{h_1}} -\frac{\vec{p}_{h_2}}{x_{h_2}} \right)^2}{\mu^2}\right) \right) + \epsilon \ln \left( 1 - \frac{x_{h_1}}{x_q} \right) \left( \ln \left(\frac{\left(\frac{\vec{p}_{h_1}}{x_{h_1}} -\frac{\vec{p}_{h_2}}{x_{h_2}} \right)^2}{\mu^2}\right)  + \ln \left( 1 - \frac{x_{h_1}}{x_q} \right) \right) \right] \\ & + (h_1 \leftrightarrow h_2)\, ,
\end{align*}
and
{\allowdisplaybreaks
\begin{align*}
& \frac{d \sigma_{3LL}^{q \bar{q} \rightarrow h_1 h_2 } }{d x_{h_1}d x_{h_2} d^d p_{h_1\perp} d p_{h_2 \perp}}\Bigg |_{\text{soft } \beta_2} \\
&=  \frac{2  \alpha_{\mathrm{em}} Q^2}{(2\pi)^{4(d-1)} N_c} \sum_{q} \int_{x_{h_1}}^{1} d x_q \int_{x_{h_2}}^{1} d x_{\bar{q}} \;  x_q x_{\bar{q}} \left(\frac{x_q}{x_{h_1}}\right)^d \left(\frac{x_{\bar{q}}}{x_{h_2}}\right)^d   \delta(1-x_q -x_{\bar{q}})    \\
&  \times Q_q^2 D_q^{h_1}\left(\frac{x_{h_1}}{x_q},\mu_F\right)D_{\bar{q}}^{h_2}\left(\frac{x_{h_2}}{x_{\bar{q}}}, \mu_F\right) \mathcal{F}_{LL} \frac{\alpha_S C_F}{2\pi} \frac{4}{\Hat{\epsilon}}  \left[ - \ln \alpha + \ln x_{\bar{q}} + \ln\left(1- \frac{x_{h_2}}{x_{\bar{q}}}\right) \right. \\
& \left. - \epsilon \ln^2 \alpha - \epsilon \ln \alpha \ln \left(\frac{\left(\frac{\vec{p}_{h_1}}{x_{h_1}} -\frac{\vec{p}_{h_2}}{x_{h_2}} \right)^2}{\mu^2}\right) + \epsilon \ln x_{\bar{q}} \Bigg( \ln x_{\bar{q}} + 2 \ln \left( 1 - \frac{x_{h_2}}{x_{\bar{q}}} \right) \right. \\ & \left. \left. + \ln \left(\frac{\left(\frac{\vec{p}_{h_1}}{x_{h_1}} -\frac{\vec{p}_{h_2}}{x_{h_2}} \right)^2}{\mu^2}\right) \right) + \epsilon \ln \left( 1 - \frac{x_{h_2}}{x_{\bar{q}}} \right) \left( \ln \left(\frac{\left(\frac{\vec{p}_{h_1}}{x_{h_1}} -\frac{\vec{p}_{h_2}}{x_{h_2}} \right)^2}{\mu^2}\right) + \ln \left( 1 - \frac{x_{h_2}}{x_{\bar{q}}} \right) \right) \right] \\ & + (h_1 \leftrightarrow h_2) \, . \\
\end{align*} }
As said above, the total soft contribution expression is found by summing the two above equations. As usual, we split the final result into divergent and finite part. In the $LL$ case, we obtain
{\allowdisplaybreaks
\begin{align*}
\frac{d \sigma_{3LL}^{q \bar{q} \rightarrow h_1 h_2 
}}{d x_{h_1}d x_{h_2} d^d p_{h_1\perp} d p_{h_2 \perp}} & \Bigg |_{\text{soft div} } 
=  \frac{4  \alpha_{\mathrm{em}} Q^2}{(2\pi)^{4(d-1)} N_c} \sum_{q} \int_{x_{h_1}}^{1} d x_q \int_{x_{h_2}}^{1} d x_{\bar{q}} x_q x_{\bar{q}} \left(\frac{x_q}{x_{h_1}}\right)^d \left(\frac{x_{\bar{q}}}{x_{h_2}}\right)^d  \\
& \times \delta(1-x_q -x_{\bar{q}})   Q_q^2 D_q^{h_1}\left(\frac{x_{h_1}}{x_q},\mu_F\right)D_{\bar{q}}^{h_2}\left(\frac{x_{h_2}}{x_{\bar{q}}}, \mu_F\right) \mathcal{F}_{LL}\\
& \times \frac{\alpha_s C_F }{2\pi} \frac{1}{\hat{\epsilon}} \left[ - 4 \ln \alpha + 2 \ln x_q + 2 \ln\left(1-\frac{x_{h_1}}{x_q}\right) -4 \epsilon \ln^2 \alpha \right. \\
& \left. - 4 \epsilon \ln \alpha \ln \left(\frac{\left(\frac{\vec{p}_{h_1}}{x_{h_1}} -\frac{\vec{p}_{h_2}}{x_{h_2}} \right)^2}{\mu^2}\right)  + 2 \ln x_{\bar{q}} + 2 \ln \left(1-\frac{x_{h_2}}{x_{\bar{q}}}\right) 
\right] \\
& + (h_1 \leftrightarrow h_2)\, .
\numberthis[total_soft LL]
\end{align*} }
and 
{\allowdisplaybreaks
\begin{align*}
\frac{d \sigma_{3LL}^{q \bar{q} \rightarrow h_1 h_2 
}}{d x_{h_1}d x_{h_2} d^d p_{h_1\perp} d p_{h_2 \perp}} & \Bigg |_{\text{soft fin} } 
=  \frac{4  \alpha_{\mathrm{em}} Q^2}{(2\pi)^{4(d-1)} N_c} \sum_{q} \int_{x_{h_1}}^{1} d x_q \int_{x_{h_2}}^{1} d x_{\bar{q}} x_q x_{\bar{q}} \left(\frac{x_q}{x_{h_1}}\right)^d \left(\frac{x_{\bar{q}}}{x_{h_2}}\right)^d  \\
& \times \delta(1-x_q -x_{\bar{q}})   Q_q^2 D_q^{h_1}\left(\frac{x_{h_1}}{x_q},\mu_F\right)D_{\bar{q}}^{h_2}\left(\frac{x_{h_2}}{x_{\bar{q}}}, \mu_F\right) \mathcal{F}_{LL} \\
& \times \frac{\alpha_s C_F }{\pi} \left[ \ln x_q \left( \ln x_q + 2 \ln \left( 1 - \frac{x_{h_1}}{x_{q}} \right) + \ln \left(\frac{\left(\frac{\vec{p}_{h_1}}{x_{h_1}} -\frac{\vec{p}_{h_2}}{x_{h_2}} \right)^2}{\mu^2} \right) \right) \right. \\ & \left.  + \ln \left( 1 - \frac{x_{h_1}}{x_q} \right) \left( \ln \left(\frac{\left(\frac{\vec{p}_{h_1}}{x_{h_1}} -\frac{\vec{p}_{h_2}}{x_{h_2}} \right)^2}{\mu^2}\right)  + \ln \left( 1 - \frac{x_{h_1}}{x_q} \right) \right) \right. \\ & \left.
+ \ln x_{\bar{q}} \Bigg( \ln x_{\bar{q}} + 2 \ln \left( 1 - \frac{x_{h_2}}{x_{\bar{q}}} \right) \left. + \ln \left(\frac{\left(\frac{\vec{p}_{h_1}}{x_{h_1}} -\frac{\vec{p}_{h_2}}{x_{h_2}} \right)^2}{\mu^2}\right) \right) \right. \\ & \left. + \ln \left( 1 - \frac{x_{h_2}}{x_{\bar{q}}} \right) \left( \ln \left(\frac{\left(\frac{\vec{p}_{h_1}}{x_{h_1}} -\frac{\vec{p}_{h_2}}{x_{h_2}} \right)^2}{\mu^2}\right) + \ln \left( 1 - \frac{x_{h_2}}{x_{\bar{q}}} \right) \right) \right] \\ & + (h_1 \leftrightarrow h_2) \, .
\numberthis[total_soft_fin LL]
\end{align*} }
For the $TL$ and $TT$ case, the calculation leads  respectively to
{\allowdisplaybreaks
\begin{align*}
\frac{d \sigma_{3TL}^{q \bar{q} \rightarrow h_1 h_2 
}}{d x_{h_1}d x_{h_2} d^d p_{h_1\perp} d p_{h_2 \perp}} & \Bigg |_{\text{soft div} } 
=  \frac{2  \alpha_{\mathrm{em}} Q}{(2\pi)^{4(d-1)} N_c} \sum_{q} \int_{x_{h_1}}^{1} \hspace{-0.3 cm} d x_q \int_{x_{h_2}}^{1} \hspace{-0.3 cm} d x_{\bar{q}} (x_{\bar{q}}-x_q)  \left(\frac{x_q}{x_{h_1}}\right)^d \left(\frac{x_{\bar{q}}}{x_{h_2}}\right)^d  \\
& \times \delta(1-x_q -x_{\bar{q}})   Q_q^2 D_q^{h_1}\left(\frac{x_{h_1}}{x_q},\mu_F\right)D_{\bar{q}}^{h_2}\left(\frac{x_{h_2}}{x_{\bar{q}}}, \mu_F\right) \mathcal{F}_{TL}\\
& \times \frac{\alpha_s C_F }{2\pi} \frac{1}{\hat{\epsilon}} \left[ - 4 \ln \alpha + 2 \ln x_q + 2 \ln\left(1-\frac{x_{h_1}}{x_q}\right) -4 \epsilon \ln^2 \alpha \right. \\
& \left. - 4 \epsilon \ln \alpha \ln \left(\frac{\left(\frac{\vec{p}_{h_1}}{x_{h_1}} -\frac{\vec{p}_{h_2}}{x_{h_2}} \right)^2}{\mu^2}\right)  + 2 \ln x_{\bar{q}} + 2 \ln \left(1-\frac{x_{h_2}}{x_{\bar{q}}}\right) 
\right] \\
& + (h_1 \leftrightarrow h_2)\, ,
\numberthis[total_soft TL]
\end{align*} }
{\allowdisplaybreaks
\begin{align*}
\frac{d \sigma_{3TL}^{q \bar{q} \rightarrow h_1 h_2 
}}{d x_{h_1}d x_{h_2} d^d p_{h_1\perp} d p_{h_2 \perp}} & \Bigg |_{\text{soft fin} } 
=  \frac{2 \alpha_{\mathrm{em}} Q}{(2\pi)^{4(d-1)} N_c} \sum_{q} \int_{x_{h_1}}^{1} \hspace{-0.3 cm} d x_q \int_{x_{h_2}}^{1} \hspace{-0.3 cm} d x_{\bar{q}} (x_{\bar{q}} - x_q)  \left(\frac{x_q}{x_{h_1}}\right)^d \left(\frac{x_{\bar{q}}}{x_{h_2}}\right)^d  \\
& \times \delta(1-x_q -x_{\bar{q}})   Q_q^2 D_q^{h_1}\left(\frac{x_{h_1}}{x_q},\mu_F\right)D_{\bar{q}}^{h_2}\left(\frac{x_{h_2}}{x_{\bar{q}}}, \mu_F\right) \mathcal{F}_{TL} \\
& \times \frac{\alpha_s C_F }{\pi} \left[ \ln x_q \left( \ln x_q + 2 \ln \left( 1 - \frac{x_{h_1}}{x_{q}} \right) + \ln \left(\frac{\left(\frac{\vec{p}_{h_1}}{x_{h_1}} -\frac{\vec{p}_{h_2}}{x_{h_2}} \right)^2}{\mu^2} \right) \right) \right. \\ & \left.  + \ln \left( 1 - \frac{x_{h_1}}{x_q} \right) \left( \ln \left(\frac{\left(\frac{\vec{p}_{h_1}}{x_{h_1}} -\frac{\vec{p}_{h_2}}{x_{h_2}} \right)^2}{\mu^2}\right)  + \ln \left( 1 - \frac{x_{h_1}}{x_q} \right) \right) \right. \\ & \left.
+ \ln x_{\bar{q}} \Bigg( \ln x_{\bar{q}} + 2 \ln \left( 1 - \frac{x_{h_2}}{x_{\bar{q}}} \right) \left. + \ln \left(\frac{\left(\frac{\vec{p}_{h_1}}{x_{h_1}} -\frac{\vec{p}_{h_2}}{x_{h_2}} \right)^2}{\mu^2}\right) \right) \right. \\ & \left. + \ln \left( 1 - \frac{x_{h_2}}{x_{\bar{q}}} \right) \left( \ln \left(\frac{\left(\frac{\vec{p}_{h_1}}{x_{h_1}} -\frac{\vec{p}_{h_2}}{x_{h_2}} \right)^2}{\mu^2}\right) + \ln \left( 1 - \frac{x_{h_2}}{x_{\bar{q}}} \right) \right) \right] \\ & + (h_1 \leftrightarrow h_2) \, .
\numberthis[total_soft_fin TL]
\end{align*} }
and
{\allowdisplaybreaks
\begin{align*}
\frac{d \sigma_{3TT}^{q \bar{q} \rightarrow h_1 h_2 
}}{d x_{h_1}d x_{h_2} d^d p_{h_1\perp} d p_{h_2 \perp}} & \Bigg |_{\text{soft div} } 
=  \frac{  \alpha_{\mathrm{em}} }{(2\pi)^{4(d-1)} N_c} \sum_{q} \int_{x_{h_1}}^{1} \frac{d x_q}{x_q} \int_{x_{h_2}}^{1} \frac{d x_{\bar{q}}}{x_{\bar{q}}}  \left(\frac{x_q}{x_{h_1}}\right)^d \left(\frac{x_{\bar{q}}}{x_{h_2}}\right)^d  \\
& \times \delta(1-x_q -x_{\bar{q}})   Q_q^2 D_q^{h_1}\left(\frac{x_{h_1}}{x_q},\mu_F\right)D_{\bar{q}}^{h_2}\left(\frac{x_{h_2}}{x_{\bar{q}}}, \mu_F\right) \mathcal{F}_{TT}\\
& \times \frac{\alpha_s C_F }{2\pi} \frac{1}{\hat{\epsilon}} \left[ - 4 \ln \alpha + 2 \ln x_q + 2 \ln\left(1-\frac{x_{h_1}}{x_q}\right) -4 \epsilon \ln^2 \alpha \right. \\
& \left. - 4 \epsilon \ln \alpha \ln \left(\frac{\left(\frac{\vec{p}_{h_1}}{x_{h_1}} -\frac{\vec{p}_{h_2}}{x_{h_2}} \right)^2}{\mu^2}\right)  + 2 \ln x_{\bar{q}} + 2 \ln \left(1-\frac{x_{h_2}}{x_{\bar{q}}}\right) 
\right] \\
& + (h_1 \leftrightarrow h_2)\, ,
\numberthis[total_soft TT]
\end{align*} }
{\allowdisplaybreaks
\begin{align*}
\frac{d \sigma_{3TT}^{q \bar{q} \rightarrow h_1 h_2 
}}{d x_{h_1}d x_{h_2} d^d p_{h_1\perp} d p_{h_2 \perp}} & \Bigg |_{\text{soft fin} } 
=  \frac{\alpha_{\mathrm{em}} }{(2\pi)^{4(d-1)} N_c} \sum_{q} \int_{x_{h_1}}^{1}  \frac{d x_q}{x_q} \int_{x_{h_2}}^{1}  \frac{d x_{\bar{q}}}{x_{\bar{q}}}  \left(\frac{x_q}{x_{h_1}}\right)^d \left(\frac{x_{\bar{q}}}{x_{h_2}}\right)^d  \\
& \times \delta(1-x_q -x_{\bar{q}})   Q_q^2 D_q^{h_1}\left(\frac{x_{h_1}}{x_q},\mu_F\right)D_{\bar{q}}^{h_2}\left(\frac{x_{h_2}}{x_{\bar{q}}}, \mu_F\right) \mathcal{F}_{TT} \\
& \times \frac{\alpha_s C_F }{\pi} \left[ \ln x_q \left( \ln x_q + 2 \ln \left( 1 - \frac{x_{h_1}}{x_{q}} \right) + \ln \left(\frac{\left(\frac{\vec{p}_{h_1}}{x_{h_1}} -\frac{\vec{p}_{h_2}}{x_{h_2}} \right)^2}{\mu^2} \right) \right) \right. \\ & \left.  + \ln \left( 1 - \frac{x_{h_1}}{x_q} \right) \left( \ln \left(\frac{\left(\frac{\vec{p}_{h_1}}{x_{h_1}} -\frac{\vec{p}_{h_2}}{x_{h_2}} \right)^2}{\mu^2}\right)  + \ln \left( 1 - \frac{x_{h_1}}{x_q} \right) \right) \right. \\ & \left.
+ \ln x_{\bar{q}} \Bigg( \ln x_{\bar{q}} + 2 \ln \left( 1 - \frac{x_{h_2}}{x_{\bar{q}}} \right) \left. + \ln \left(\frac{\left(\frac{\vec{p}_{h_1}}{x_{h_1}} -\frac{\vec{p}_{h_2}}{x_{h_2}} \right)^2}{\mu^2}\right) \right) \right. \\ & \left. + \ln \left( 1 - \frac{x_{h_2}}{x_{\bar{q}}} \right) \left( \ln \left(\frac{\left(\frac{\vec{p}_{h_1}}{x_{h_1}} -\frac{\vec{p}_{h_2}}{x_{h_2}} \right)^2}{\mu^2}\right) + \ln \left( 1 - \frac{x_{h_2}}{x_{\bar{q}}} \right) \right) \right] \\ & + (h_1 \leftrightarrow h_2) \, .
\numberthis[total_soft_fin TT]
\end{align*}  }

At this level, we are already able to observe the full cancellation of soft divergences (and hence the disappearance of $\ln \alpha$-terms). Consider, for instance, the longitudinal cross section.
Combining the divergent soft contribution, coming from the real part, see Eq.~\eqref{eq:total_soft LL}, with the virtual contribution \eqref{eq:virtual div LL} we see the complete cancellation of these $\ln \alpha$-terms and also of $\frac{1}{\epsilon} \ln (x_q x_{\bar{q}})$-term. Moreover, surviving $\frac{1}{\epsilon}$ divergent terms cancel in combination with:
\begin{itemize}
    \item Terms proportional to $\frac{3}{2} \delta(1-\beta_i)$ appearing inside the splitting functions in \eqref{eq:ct_LL}
    \item Term proportional to $ \ln \left(1- \frac{x_{h_1}}{x_q} \right)$ in Eq.~\eqref{eq:coll_div_q}
    \item Term proportional to $ \ln \left(1- \frac{x_{h_2}}{x_{\bar{q}}}\right)$ in \eqref{eq:coll_qbarg_div} 
\end{itemize}
Now, we are only left with collinearly divergent contributions related to the case of fragmentation from quark and gluon or from anti-quark and gluon. These should cancel the only two divergent contributions left in Eq.~\eqref{eq:ct_LL}, i.e., the ones proportional to $P_{gq} (\beta_i)$.

\subsection{Fragmentation from anti-quark and gluon}
\label{sec:qgfrag}
In this section, we deal with extracting the collinear divergences associated with the contribution (d) in Fig. \ref{fig:sigma-NLO}. This contribution corresponds to the situation in which the anti-quark and the gluon fragment, while the quark plays the role of ``spectator" emitted particle. This case is much simpler than before. We do not have to deal with any soft divergence and the only IR divergence that appears is when the fragmenting gluon is emitted by the quark line after the shockwave and the emitted quark and gluon become collinear.
Hence, we can directly compute the contribution due to the first term of Eq.~(\ref{eq:real_div_LL}). \\
We emphasize the difference with the contribution calculated in section \ref{sec:qqbarfragColl-qbarg}. Although at the level of hard computation the term that generates the present divergence is the same as the one that generates the collinear divergence in section \ref{sec:qqbarfragColl-qbarg}, the situation is completely different. In the present case, we integrate out the quark kinematic variables and remain differential in the variable of the emitted gluon, while, in section \ref{sec:qqbarfragColl-qbarg} it was exactly the opposite.

\subsubsection{Collinear contribution: $q$-$g$ splitting}

According to the above discussion, we should focus on the first term of
 Eq.~\eqref{eq: div real impact factor}, which exhibits a collinear pole, namely
{\allowdisplaybreaks
\begin{align*}
    & \left.\frac{d \sigma_{3LL}^{g \bar{q} \rightarrow h_1 h_2} }{d x_{h_1} d x_{h_2} d^d p_{h_1 \perp} d^d p_{h_2\perp}}\right|_{\rm coll\, qg} \\
    &= \frac{4  \alpha_{\mathrm{em}} Q^2}{(2\pi)^{4(d-1)} N_c} \sum_{q} \int_{x_{h_1}}^1 \frac{d x_g}{x_g^2} \int_0^1  d x_q' \int_{x_{h_2}}^1 \frac{d x_{\bar{q}}}{x_{\bar{q}}} \delta(1-x_q'-x_{\bar{q}}-x_g)\\
    & \times  \left(\frac{x_g}{x_{h_1}}\right)^d \left(\frac{x_{\bar{q}}}{x_{h_2}}\right)^d  Q_q^2 D_g^{h_1}\left(\frac{x_{h_1}}{x_g}, \mu_F\right) D_{\bar{q}}^{h_2}\left(\frac{x_{h_2}}{x_{\bar{q}}}, \mu_F\right) \frac{\alpha_s}{\mu^{2\epsilon}} C_F  \int \frac{d^d p_{q\perp}}{(2\pi)^d} \\
    & \times \int   d^d p_{2\perp}     \mathbf{F}  \left(\frac{p_{q\perp}}{2} +\frac{x_{\bar{q}}}{2 x_{h_2}}p_{h_2 \perp} -p_{2\perp} + \frac{x_g}{2 x_{h_1}} p_{h_1\perp}\right) \\
    & \times  \int   d^d p_{2'\perp}  \mathbf{F}^*\left(\frac{p_{q\perp}}{2} + \frac{x_{\bar{q}}}{2 x_{h_2}}p_{h_2 \perp} -p_{2'\perp} + \frac{x_g}{2 x_{h_1}} p_{h_1\perp}\right)  \\
    & \times \frac{1}{\left( x_{\bar{q}} (1-x_{\bar{q}})Q^2 +\left(\frac{x_{\bar{q}}}{x_{h_2}} \vec{p}_{h_2}-\vec{p}_{2}\right)^2\right) \left( x_{\bar{q}} (1-x_{\bar{q}})Q^2 +\left(\frac{x_{\bar{q}}}{x_{h_2}} \vec{p}_{h_2}-\vec{p}_{2'}\right)^2\right)}\\
    & \times \frac{(d x_g^2 + 4 x_q' (x_q' + x_g)) x_{\bar{q}}^2 (1-x_{\bar{q}})^2}{\left(x_q' \frac{x_g}{x_{h_1}} \vec{p}_{h_1} - x_g \vec{p}_q\right)^2} + (h_1 \leftrightarrow h_2)\,.
\end{align*} }
Using a change of variable similar to \eqref{eq:Transbeta}, here
    \begin{align*}
        x_g & = \beta_1 x_q \\ 
        x_q'&= (1-\beta_1) x_q  
    \end{align*}
    with the Jacobian $d x_q' d x_g = d x_q d \beta_1 x_q$ and treating the integration over longitudinal fractions as follows 
\begin{align*}
    & \int_{x_{h_1}}^1 \frac{d x_g}{x_g^2}  \int_{x_{h_2}}^1 \frac{d x_{\bar{q}}}{x_{\bar{q}}} \int_{0}^1  d x_q' \delta(1-x_q'-x_{\bar{q}}-x_g) \\ 
    &= \int_{x_{h_1}}^1 \frac{d x_g}{x_g^2} \int_0^1  d x_q' \int_{- \infty}^{+\infty} \frac{d x_{\bar{q}}}{x_{\bar{q}}} \theta(x_{\bar{q}}-x_{h_2}) \theta(1- x_{\bar{q}}) \delta(1-x_q'-x_{\bar{q}}-x_g) \\
    &= \int_{x_{h_1}}^{1-x_{h_2}} d x_q \frac{1}{x_q(1-x_q)}  \int_{\frac{x_{h_1}}{x_q}}^1 \frac{d \beta_1}{\beta_1^2 } \; , 
\end{align*}
we get 
{\allowdisplaybreaks
\begin{align*}
& \left. \frac{d \sigma_{3LL}^{g \bar{q} \rightarrow h_1 h_2}}{d x_{h_1} d x_{h_2} d p_{h_1 \perp} d^d p_{h_2\perp}} \right|_{\text{coll. qg}}   \\
&= \frac{4  \alpha_{\mathrm{em}} Q^2}{(2\pi)^{4(d-1)} N_c} \sum_{q} \int_{x_{h_1}}^{1-x_{h_2}} d x_q \int_{\frac{x_{h_1}}{x_q}}^1 \frac{d \beta_1}{\beta_1}  x_q (1-x_q)\left(\frac{x_q}{x_{h_1}}\right)^d \left(\frac{1-x_q}{x_{h_2}}\right)^d  \\
& \times   Q_q^2 D_g^{h_1}\left(\frac{x_{h_1}}{\beta_1 x_q}, \mu_F\right) D_{\bar{q}}^{h_2}\left(\frac{x_{h_2}}{1-x_q}, \mu_F\right) \\
& \times \int   d^d p_{2\perp}  \int d^d z_{1\perp} \frac{e^{i z_{1\perp} \cdot  \left(\frac{1-x_q}{2 x_{h_2}}p_{h_2 \perp} -p_{2\perp} + \frac{\beta_1 x_q}{2 x_{h_1}} p_{h_1\perp}\right)}}{x_q (1-x_q)Q^2 +\left(\frac{1-x_q}{x_{h_2}} \vec{p}_{h_2}-\vec{p}_{2}\right)^2} F(z_{1\perp}) \\
& \times \int   d^d p_{2'\perp}  \int d^d z_{2\perp} \frac{e^{-i z_{2\perp} \cdot  \left(\frac{1-x_q}{2 x_{h_2}}p_{h_2 \perp} -p_{2'\perp} + \frac{\beta_1 x_q}{2 x_{h_1}} p_{h_1\perp}\right)}}{x_q (1-x_q)Q^2 +\left(\frac{1-x_q}{x_{h_2}} \vec{p}_{h_2}-\vec{p}_{2'}\right)^2} F^*(z_{2\perp}) \\
&\times  \frac{2 ( (1-\beta_1)^2 +1 ) + 2 \epsilon \beta_1^2}{\beta_1}\beta_1^{d-2}\frac{\alpha_s}{\mu^{2\epsilon}} C_F \int  \frac{d^d p_{q\perp}}{(2\pi)^d} \frac{e^{i \left(\frac{z_{1\perp}-z_{2\perp}}{2}\right) \cdot p_{q\perp}}}{\left(\frac{(1-\beta_1)x_q}{x_{h_1}}\vec{p}_{h_1} - \vec{p}_q\right)^2} \\
& + (h_1 \leftrightarrow h_2) \,.
\numberthis[coll_qg_qbarg_FF_total]
\end{align*} }
Using Eq.~\eqref{eq:expo} in Eq.~\eqref{eq:coll_qg_qbarg_FF_total} to perform the integration over quark transverse momenta, we obtain \newpage
{\allowdisplaybreaks
\begin{align*}
& \left. \frac{d \sigma_{3LL}^{g \bar{q} \rightarrow h_1 h_2}}{d x_{h_1} d x_{h_2} d p_{h_1 \perp} d^d p_{h_2\perp}} \right|_{\text{coll. qg}}  \\  
&= \frac{4  \alpha_{\mathrm{em}} Q^2}{(2\pi)^{4(d-1)} N_c} \sum_{q} \int_{x_{h_1}}^{1-x_{h_2}} d x_q \int_{\frac{x_{h_1}}{x_q}}^1 \frac{d \beta_1}{\beta_1}  x_q (1-x_q)\left(\frac{x_q}{x_{h_1}}\right)^d  \\
& \times \left(\frac{1-x_q}{x_{h_2}}\right)^d   Q_q^2 D_g^{h_1}\left(\frac{x_{h_1}}{\beta_1 x_q}, \mu_F\right) D_{\bar{q}}^{h_2}\left(\frac{x_{h_2}}{1-x_q}, \mu_F\right) \\
& \times \int   d^d p_{2\perp}  \int d^d z_{1\perp} \frac{e^{i z_{1\perp} \cdot  \left(\frac{1-x_q}{2 x_{h_2}}p_{h_2 \perp} -p_{2\perp} + \frac{\beta_1 x_q}{2 x_{h_1}} p_{h_1\perp}\right)}}{x_q (1-x_q)Q^2 +\left(\frac{1-x_q}{x_{h_2}} \vec{p}_{h_2}-\vec{p}_{2}\right)^2} F(z_{1\perp}) \\
& \times \int   d^d p_{2'\perp}  \int d^d z_{2\perp} \frac{e^{-i z_{2\perp} \cdot  \left(\frac{1-x_q}{2 x_{h_2}}p_{h_2 \perp} -p_{2'\perp} + \frac{\beta_1 x_q}{2 x_{h_1}} p_{h_1\perp}\right)}}{x_q (1-x_q)Q^2 +\left(\frac{1-x_q}{x_{h_2}} \vec{p}_{h_2}-\vec{p}_{2'}\right)^2} F^*(z_{2\perp}) \\
&\times  \frac{2 ( (1-\beta_1)^2 +1 ) + 2 \epsilon \beta_1^2 + 4 \epsilon ((1-\beta_1)^2 + 1) \ln \beta_1 }{\beta_1} \\ 
& \times e^{i \left(\frac{z_{1\perp}-z_{2\perp}}{2}\right) \cdot \frac{(1-\beta_1)x_q}{x_{h_1}} p_{h_1\perp}} \frac{\alpha_s C_F}{4\pi} \left( \frac{1}{\hat{\epsilon}} + \ln \left( \frac{c_0^2}{\left(\frac{z_{1\perp} - z_{2\perp}}{2}\right)^2 \mu^2}\right) \right) + (h_1 \leftrightarrow h_2) \\
&= \frac{4  \alpha_{\mathrm{em}} Q^2}{(2\pi)^{4(d-1)} N_c} \sum_{q} \int_{x_{h_1}}^{1-x_{h_2}} d x_q \int_{\frac{x_{h_1}}{x_q}}^1 \frac{d \beta_1}{\beta_1}  x_q (1-x_q)\left(\frac{x_q}{x_{h_1}}\right)^d \\
& \times   \left(\frac{1-x_q}{x_{h_2}}\right)^d  Q_q^2 D_g^{h_1}\left(\frac{x_{h_1}}{\beta_1 x_q}, \mu_F\right) D_{\bar{q}}^{h_2}\left(\frac{x_{h_2}}{1-x_q}, \mu_F\right) \\
& \times \int   d^d p_{2\perp}  \int d^d z_{1\perp} \frac{e^{i z_{1\perp} \cdot  \left(\frac{1-x_q}{2 x_{h_2}}p_{h_2 \perp} -p_{2\perp} + \frac{x_q}{2 x_{h_1}} p_{h_1\perp}\right)}}{x_q (1-x_q)Q^2 +\left(\frac{1-x_q}{x_{h_2}} \vec{p}_{h_2}-\vec{p}_{2}\right)^2} F(z_{1\perp}) \\
& \times \int   d^d p_{2'\perp}  \int d^d z_{2\perp} \frac{e^{-i z_{2\perp} \cdot  \left(\frac{1-x_q}{2 x_{h_2}}p_{h_2 \perp} -p_{2'\perp} + \frac{x_q}{2 x_{h_1}} p_{h_1\perp}\right)}}{x_q (1-x_q)Q^2 +\left(\frac{1-x_q}{x_{h_2}} \vec{p}_{h_2}-\vec{p}_{2'}\right)^2} F^*(z_{2\perp}) \\
& \frac{\alpha_s}{2\pi} C_F \Bigg [\frac{1}{\hat{\epsilon}}\frac{1 + (1-\beta_1)^2}{\beta_1} + \beta_1 + \frac{2  (1 + (1-\beta)^2) \ln \beta_1}{\beta_1} \\
& + \frac{1+ (1-\beta_1)^2}{\beta_1} \ln \left( \frac{c_0^2}{\left(\frac{z_{1\perp} - z_{2\perp}}{2}\right)^2 \mu^2}\right)\Bigg] + (h_1 \leftrightarrow h_2)  \\
&= \frac{d \sigma_{3LL}^{g \bar{q} \rightarrow h_1 h_2}}{ d x_{h_1} d x_{h_2} d p_{h_1 \perp} d^d p_{h_2\perp} } \bigg |_{\text{coll. qg div}}       +  \frac{d \sigma_{3LL}^{g \bar{q} \rightarrow h_1 h_2}}{ d x_{h_1} d x_{h_2} d p_{h_1 \perp} d^d p_{h_2\perp} } \bigg |_{\text{coll.qg fin}} \numberthis[coll_qg_qbar_g_beta]\; ,
\end{align*} }
where the term labeled with ``div" contains the first term of the square bracket. \\
Putting back $x_{\bar{q}}$ using Eq.~\eqref{eq: xq xbarq }, this divergent term takes the form: 
{\allowdisplaybreaks
\begin{align*}
& \left. \frac{d \sigma_{3LL}^{g \bar{q} \rightarrow h_1 h_2}}{d x_{h_1} d x_{h_2} d p_{h_1 \perp} d^d p_{h_2\perp}}\right|_{\text{coll. qg div}} \\
&= \frac{4  \alpha_{\mathrm{em}} Q^2}{(2\pi)^{4(d-1)} N_c}   \sum_{q}  \int_{x_{h_1}}^{1}  d x_q \int_{\frac{x_{h_2}}{x_q}}^1  d x_{\bar{q}} \;  x_q x_{\bar{q}}  \left(\frac{x_q}{x_{h_1}}\right)^d \left(\frac{x_{\bar{q}}}{x_{h_2}}\right)^d \\*
& \times  \delta(1-x_q -x_{\bar{q}}) \mathcal{F}_{LL} \frac{\alpha_s }{2\pi} \frac{1}{\hat{\epsilon}} \int_{\frac{x_{h_1}}{x_q}}^1 \frac{d \beta_1}{\beta_1}  \\
& \times   Q_q^2 D_g^{h_1}\left(\frac{x_{h_1}}{\beta_1 x_q},\mu_F\right) D_{\bar{q}}^{h_2}\left(\frac{x_{h_2}}{x_{\bar{q}}}, \mu_F\right) C_F \frac{1+(1-\beta_1)^2}{\beta_1} + (h_1 \leftrightarrow h_2) \,.\numberthis
\end{align*} }
This is the term needed to cancel the divergent term proportional to $P_{gq} (\beta_1)$ in \eqref{eq:ct_LL}. Instead, the finite part in Eq.~\eqref{eq:coll_qg_qbar_g_beta} reads
{\allowdisplaybreaks
\begin{align*}
& \left. \frac{d \sigma_{3LL}^{g \bar{q} \rightarrow h_1 h_2}}{d x_{h_1} d x_{h_2} d p_{h_1 \perp} d^d p_{h_2\perp}}\right |_{\text{coll. qg fin}}  \\
& = \frac{4  \alpha_{\mathrm{em}} Q^2}{(2\pi)^{4(d-1)} N_c}   \sum_{q}  \int_{x_{h_1}}^{1}  d x_q \int_{\frac{x_{h_2}}{x_q}}^1  d x_{\bar{q}} \;  x_q x_{\bar{q}}  \left(\frac{x_q}{x_{h_1}}\right)^d \left(\frac{x_{\bar{q}}}{x_{h_2}}\right)^d  \\
& \times \delta(1-x_q -x_{\bar{q}})  \int_{\frac{x_{h_1}}{x_q}}^1 \frac{d \beta_1}{\beta_1}  Q_q^2 D_g^{h_1}\left(\frac{x_{h_1}}{\beta_1 x_q}, \mu_F\right) D_{\bar{q}}^{h_2}\left(\frac{x_{h_2}}{1-x_q}, \mu_F\right) \\
& \times \int   d^d p_{2\perp}  \int d^d z_{1\perp} \frac{e^{i z_{1\perp} \cdot  \left(\frac{1-x_q}{2 x_{h_2}}p_{h_2 \perp} -p_{2\perp} + \frac{x_q}{2 x_{h_1}} p_{h_1\perp}\right)}}{x_q (1-x_q)Q^2 +\left(\frac{1-x_q}{x_{h_2}} \vec{p}_{h_2}-\vec{p}_{2}\right)^2} F(z_{1\perp}) \\
& \times \int   d^d p_{2'\perp}  \int d^d z_{2\perp} \frac{e^{-i z_{2\perp} \cdot  \left(\frac{1-x_q}{2 x_{h_2}}p_{h_2 \perp} -p_{2'\perp} + \frac{x_q}{2 x_{h_1}} p_{h_1\perp}\right)}}{x_q (1-x_q)Q^2 +\left(\frac{1-x_q}{x_{h_2}} \vec{p}_{h_2}-\vec{p}_{2'}\right)^2} F^*(z_{2\perp}) \\
& \times \frac{\alpha_s}{2\pi} C_F \Bigg [\beta_1 + \frac{2 (1 + (1-\beta)^2) \ln \beta_1 }{\beta_1}  \\
& + \frac{1+ (1-\beta_1)^2}{\beta_1} \ln \left( \frac{c_0^2}{\left(\frac{z_{1\perp} - z_{2\perp}}{2}\right)^2 \mu^2}\right)\Bigg] + (h_1 \leftrightarrow h_2)\,. \numberthis
\end{align*} }
In a similar way, in the $TL$ case, we get
{\allowdisplaybreaks
\begin{align*}
& \left. \frac{d \sigma_{3TL}^{g \bar{q} \rightarrow h_1 h_2}}{d x_{h_1} d x_{h_2} d p_{h_1 \perp} d^d p_{h_2\perp}}\right |_{\text{coll. qg }}  \\
&= \frac{2  \alpha_{\mathrm{em}} Q}{(2\pi)^{4(d-1)} N_c}  \sum_{q} \int_{x_{h_1}}^{1} d x_q  \int_{x_{h_2}}^1 d x_{\bar{q}}  \left(\frac{x_q}{x_{h_1}}\right)^d \left(\frac{x_{\bar{q}}}{x_{h_2}}\right)^d (x_{\bar{q}}-x_q)  \\
& \times \delta(1-x_q-x_{\bar{q}})  \int_{\frac{x_{h_1}}{x_q}}^1 \frac{d \beta_1}{\beta_1}   Q_q^2 D_g^{h_1}\left(\frac{x_{h_1}}{\beta_1 x_q}, \mu_F\right) D_{\bar{q}}^{h_2}\left(\frac{x_{h_2}}{x_{\bar{q}}}, \mu_F\right)  \\
& \times \int   d^d p_{2\perp}  \int d^d z_{1\perp} \frac{e^{i z_{1\perp} \cdot  \left(\frac{1-x_q}{2 x_{h_2}}p_{h_2 \perp} -p_{2\perp} + \frac{x_q}{2 x_{h_1}} p_{h_1\perp}\right)}}{x_q (1-x_q)Q^2 +\left(\frac{1-x_q}{x_{h_2}} \vec{p}_{h_2}-\vec{p}_{2}\right)^2} F(z_{1\perp}) \\
& \times \int   d^d p_{2'\perp}  \int d^d z_{2\perp} \frac{e^{-i z_{2\perp} \cdot  \left(\frac{1-x_q}{2 x_{h_2}}p_{h_2 \perp} -p_{2'\perp} + \frac{x_q}{2 x_{h_1}} p_{h_1\perp}\right)}}{x_q (1-x_q)Q^2 +\left(\frac{1-x_q}{x_{h_2}} \vec{p}_{h_2}-\vec{p}_{2'}\right)^2} F^*(z_{2\perp})  \left( \frac{x_{\bar{q}}}{x_{h_2}} \vec{p}_{h_2}  - \vec{p}_{2'}\right) \cdot \vec{\varepsilon}_T^{\; *}  \\
& \times \frac{\alpha_s}{2\pi} C_F \Bigg [\frac{1}{\hat{\epsilon}}\frac{1 + (1-\beta_1)^2}{\beta_1} + \beta_1 + \frac{2  (1 + (1-\beta)^2) \ln \beta_1}{\beta_1} \\
& + \frac{1+ (1-\beta_1)^2}{\beta_1} \ln \left( \frac{c_0^2}{\left(\frac{z_{1\perp} - z_{2\perp}}{2}\right)^2 \mu^2}\right)\Bigg] + (h_1 \leftrightarrow h_2)  \\
&= \frac{d \sigma_{3TL}^{g \bar{q} \rightarrow h_1 h_2}}{ d x_{h_1} d x_{h_2} d p_{h_1 \perp} d^d p_{h_2\perp} } \bigg |_{\text{coll. qg div}}       +  \frac{d \sigma_{3TL}^{g \bar{q} \rightarrow h_1 h_2}}{ d x_{h_1} d x_{h_2} d p_{h_1 \perp} d^d p_{h_2\perp} } \bigg |_{\text{coll. qg fin}} .
\numberthis
\end{align*} }
Finally, in the $TT$ case, we have
{\allowdisplaybreaks
\begin{align*}
& \left.  \frac{d \sigma_{3TT}^{g \bar{q} \rightarrow h_1 h_2}}{d x_{h_1} d x_{h_2} d p_{h_1 \perp} d^d p_{h_2\perp}} \right |_{\text{coll. qg }}  \\
&= \frac{  \alpha_{\mathrm{em}} }{(2\pi)^{4(d-1)} N_c} \sum_{q}  \int_{x_{h_1}}^{1} \frac{d x_q }{x_q} \int_{x_{h_2}}^1 \frac{d x_{\bar{q}}}{x_{\bar{q}}} \left(\frac{x_q}{x_{h_1}}\right)^d \left(\frac{x_{\bar{q}}}{x_{h_2}}\right)^d  \delta(1-x_q-x_{\bar{q}}) \\
& \times \int_{\frac{x_{h_1}}{x_q}}^1 \frac{d \beta_1}{\beta_1}   Q_q^2 D_g^{h_1}\left(\frac{x_{h_1}}{\beta_1 x_q}, \mu_F\right) D_{\bar{q}}^{h_2}\left(\frac{x_{h_2}}{x_{\bar{q}}}, \mu_F\right) \left[ (x_{\bar{q}} -x_q)^2 g_{\perp}^{ri}g_{\perp}^{lk} - g_{\perp}^{rk}g_{\perp}^{li} + g_{\perp}^{rl}g_{\perp}^{ik} \right] \\
& \times \int   d^d p_{2\perp}  \int d^d z_{1\perp} \frac{e^{i z_{1\perp} \cdot  \left(\frac{1-x_q}{2 x_{h_2}}p_{h_2 \perp} -p_{2\perp} + \frac{x_q}{2 x_{h_1}} p_{h_1\perp}\right)}}{x_q (1-x_q)Q^2 +\left(\frac{1-x_q}{x_{h_2}} \vec{p}_{h_2}-\vec{p}_{2}\right)^2} F(z_{1\perp}) \left(\frac{x_{\bar{q}}}{x_{h_2}} p_{h_2} - p_{2}\right)_r \varepsilon_{T i}  \\
& \times \int   d^d p_{2'\perp}  \int d^d z_{2\perp} \frac{e^{-i z_{2\perp} \cdot  \left(\frac{1-x_q}{2 x_{h_2}}p_{h_2 \perp} -p_{2'\perp} + \frac{x_q}{2 x_{h_1}} p_{h_1\perp}\right)}}{x_q (1-x_q)Q^2 +\left(\frac{1-x_q}{x_{h_2}} \vec{p}_{h_2}-\vec{p}_{2'}\right)^2} F^*(z_{2\perp}) \left(\frac{x_{\bar{q}}}{x_{h_2}} p_{h_2} - p_{2'}\right)_l \varepsilon_{T k}^*\\
& \times \frac{\alpha_s}{2\pi} C_F \Bigg [\frac{1}{\hat{\epsilon}}\frac{1 + (1-\beta_1)^2}{\beta_1} + \beta_1 + \frac{2 (1 + (1-\beta)^2) \ln \beta_1 }{\beta_1} \\
& + \frac{1+ (1-\beta_1)^2}{\beta_1} \ln \left( \frac{c_0^2}{\left(\frac{z_{1\perp} - z_{2\perp}}{2}\right)^2 \mu^2}\right)\Bigg] + (h_1 \leftrightarrow h_2) \\
&= \frac{d \sigma_{3TT}^{g \bar{q} \rightarrow h_1 h_2}}{ d x_{h_1} d x_{h_2} d p_{h_1 \perp} d^d p_{h_2\perp} } \bigg |_{\text{coll. qg div}}       +  \frac{d \sigma_{3TT}^{g \bar{q} \rightarrow h_1 h_2}}{ d x_{h_1} d x_{h_2} d p_{h_1 \perp} d^d p_{h_2\perp} } \bigg |_{\text{coll. qg fin}} .\numberthis
\end{align*} }
These results conclude the discussion of divergences in the case of fragmentation from antiquark and gluon.

\subsection{Fragmentation from quark and gluon}
In this section, we deal with extracting the collinear divergences associated with the contribution (c) in Fig. \ref{fig:sigma-NLO}. This contribution corresponds to the situation in which the quark and the gluon fragment, while the anti-quark plays the role of the ``spectator" emitted particle.

\subsubsection{Collinear contribution: $\bar{q}$-$g$ splitting}

The term in Eq.~\eqref{eq: div real impact factor} to consider is the third one. The calculation proceeds in the same way as for the anti-quark and gluon fragmentation, but this time the integration is over $p_{2,2' \perp}$ in the $F$ function. 

For the $LL$ case, we get 
\begin{align*}
\allowdisplaybreaks
& \left. \frac{d \sigma_{3LL}^{ q g  \rightarrow h_1 h_2}}{d x_{h_1} d x_{h_2} d p_{h_1 \perp} d^d p_{h_2\perp}} \right|_{\text{coll. } \bar{q}g}  \\
&= \frac{4  \alpha_{\mathrm{em}} Q^2}{(2\pi)^{4(d-1)} N_c} \sum_{q}  \int_{x_{h_1}}^{1} d x_q \int_{x_{h_2}}^1 d x_{\bar{q}} \;  x_q x_{\bar{q}}  \delta(1-x_q -x_{\bar{q}})   \left(\frac{x_q}{x_{h_1}}\right)^d \left(\frac{x_{\bar{q}}}{x_{h_2}}\right)^d  \\*
& \times \int_{\frac{x_{h_2}}{x_{\bar{q}}}}^1 \frac{d \beta_2}{\beta_2}  Q_q^2  D_{q}^{h_2}\left(\frac{x_{h_1}}{x_q}, \mu_F\right)  D_g^{h_2}\left(\frac{x_{h_2}}{\beta_2 x_{\bar{q}}}, \mu_F\right) \\
& \times \int   d^d p_{1\perp}  \int d^d z_{1\perp} \frac{e^{i z_{1\perp} \cdot  \left(-\frac{x_{\bar{q}}}{2 x_{h_2}}p_{h_2 \perp} + p_{1\perp} - \frac{x_q}{2 x_{h_1}} p_{h_1\perp}\right)}}{x_q x_{\bar{q}} Q^2 +\left(\frac{x_q}{x_{h_1}} \vec{p}_{h_1}-\vec{p}_{1}\right)^2} F(z_{1\perp}) \\
& \times \int   d^d p_{1'\perp}  \int d^d z_{2\perp} \frac{e^{-i z_{2\perp} \cdot  \left(-\frac{x_{ \bar{q}}}{2 x_{h_2}}p_{h_2 \perp} -p_{1'\perp} -\frac{x_q}{2 x_{h_1}} p_{h_1\perp}\right)}}{x_q (1-x_q)Q^2 +\left(\frac{x_q}{x_{h_1}} \vec{p}_{h_1}-\vec{p}_{1'}\right)^2} F^*(z_{2\perp}) \\
& \times \frac{\alpha_s}{2\pi} C_F \Bigg [\frac{1}{\hat{\epsilon}}\frac{1 + (1-\beta_2)^2}{\beta_2} + \beta_2 + \frac{2 (1 + (1-\beta_2)^2)  \ln \beta_2 }{\beta_2} \\
& + \frac{1+ (1-\beta_2)^2}{\beta_2} \ln \left( \frac{c_0^2}{\left(\frac{z_{2\perp} - z_{1\perp}}{2}\right)^2 \mu^2}\right)\Bigg] + (h_1 \leftrightarrow h_2) \\
&= \frac{d \sigma_{3LL}^{q g\rightarrow h_1 h_2}}{ d x_{h_1} d x_{h_2} d p_{h_1 \perp} d^d p_{h_2\perp} } \bigg |_{\text{coll. } \bar{q}g \text{ div}}       +  \frac{d \sigma_{3LL}^{q g  \rightarrow h_1 h_2}}{ d x_{h_1} d x_{h_2} d p_{h_1 \perp} d^d p_{h_2\perp} } \bigg |_{\text{coll. } \bar{q}g\text{ fin}} .
\numberthis
\end{align*}  
This term cancels the divergent term proportional to $P_{gq} (\beta_2)$ in \eqref{eq:ct_LL}. This is the last remaining cancellation of divergences, the rest of the cross section is now completely finite.

For the $TL$ case, we get 
{\allowdisplaybreaks
\begin{align*}
& \left. \frac{d \sigma_{3TL}^{ q g  \rightarrow h_1 h_2}}{d x_{h_1} d x_{h_2} d p_{h_1 \perp} d^d p_{h_2\perp}} \right|_{\text{coll. } \bar{q}g}   \\
&= \frac{2 \alpha_{\mathrm{em}} Q}{(2\pi)^{4(d-1)} N_c} \sum_{q} \int_{x_{h_1}}^{1} d x_q \int_{x_{h_2}}^1 d x_{\bar{q}}  \;  (x_{\bar{q}}-x_q) \delta(1-x_q -x_{\bar{q}})   \left(\frac{x_q}{x_{h_1}}\right)^d \left(\frac{x_{\bar{q}}}{x_{h_2}}\right)^d  \\
& \times \int_{\frac{x_{h_2}}{x_{\bar{q}}}}^1 \frac{d \beta_2}{\beta_2}   Q_q^2  D_{q}^{h_2}\left(\frac{x_{h_1}}{x_q}, \mu_F\right)  D_g^{h_2}\left(\frac{x_{h_2}}{\beta_2 x_{\bar{q}}}, \mu_F\right) \\
& \times \int   d^d p_{1\perp}  \int d^d z_{1\perp} \frac{e^{i z_{1\perp} \cdot  \left(-\frac{x_{\bar{q}}}{2 x_{h_2}}p_{h_2 \perp} + p_{1\perp} - \frac{x_q}{2 x_{h_1}} p_{h_1\perp}\right)}}{x_q x_{\bar{q}} Q^2 +\left(\frac{x_q}{x_{h_1}} \vec{p}_{h_1}-\vec{p}_{1}\right)^2} F(z_{1\perp}) \\
& \times \int   d^d p_{1'\perp}  \int d^d z_{2\perp} \frac{e^{-i z_{2\perp} \cdot  \left(-\frac{x_{ \bar{q}}}{2 x_{h_2}}p_{h_2 \perp} -p_{1'\perp} -\frac{x_q}{2 x_{h_1}} p_{h_1\perp}\right)}}{x_q (1-x_q)Q^2 +\left(\frac{x_q}{x_{h_1}} \vec{p}_{h_1}-\vec{p}_{1'}\right)^2} F^*(z_{2\perp})  \left( \frac{x_q}{x_{h_1}} p_{h_1 }  - p_{1'}\right) \cdot \varepsilon^* _{T}  \\
& \times \frac{\alpha_s}{2\pi} C_F \Bigg [\frac{1}{\hat{\epsilon}}\frac{1 + (1-\beta_2)^2}{\beta_2} + \beta_2 + \frac{2 \ln \beta_2 (1 + (1-\beta_2)^2)}{\beta_2}  \\
& + \frac{1+ (1-\beta_2)^2}{\beta_2} \ln \left( \frac{c_0^2}{\left(\frac{z_{2\perp} - z_{1\perp}}{2}\right)^2 \mu^2}\right)\Bigg] + (h_1 \leftrightarrow h_2) \\
&= \frac{d \sigma_{3TL}^{q g\rightarrow h_1 h_2}}{ d x_{h_1} d x_{h_2} d p_{h_1 \perp} d^d p_{h_2\perp} } \bigg |_{\text{coll. } \bar{q}g \text{ div}}       +  \frac{d \sigma_{3TL}^{q g  \rightarrow h_1 h_2}}{ d x_{h_1} d x_{h_2} d p_{h_1 \perp} d^d p_{h_2\perp} }  \bigg |_{\text{coll. } \bar{q}g\text{ fin}}
.
\numberthis
\end{align*}}

Finally, for the $TT$ case, we get 
{ \allowdisplaybreaks
\begin{align*}
 & \left. \frac{d \sigma_{3TT}^{ q g  \rightarrow h_1 h_2}}{d x_{h_1} d x_{h_2} d p_{h_1 \perp} d^d p_{h_2\perp}} \right|_{\text{coll. } \bar{q}g}  \\
&= \frac{ \alpha_{\mathrm{em}} }{(2\pi)^{4(d-1)} N_c} \sum_{q} \int_{x_{h_1}}^{1} \frac{d x_q}{x_q} \int_{x_{h_2}}^1 \frac{d x_{\bar{q}} }{x_{\bar{q}}}  \delta(1-x_q -x_{\bar{q}})   \left(\frac{x_q}{x_{h_1}}\right)^d \left(\frac{x_{\bar{q}}}{x_{h_2}}\right)^d  \\
& \times \int_{\frac{x_{h_2}}{x_{\bar{q}}}}^1 \frac{d \beta_2}{\beta_2}   Q_q^2  D_{q}^{h_2}\left(\frac{x_{h_1}}{x_q}, \mu_F\right)  D_g^{h_2}\left(\frac{x_{h_2}}{\beta_2 x_{\bar{q}}}, \mu_F\right) \left[ (x_{\bar{q}} -x_q)^2 g_{\perp}^{ri}g_{\perp}^{lk} - g_{\perp}^{rk}g_{\perp}^{li} + g_{\perp}^{rl}g_{\perp}^{ik} \right]  \\
& \times \int   d^d p_{1\perp}  \int d^d z_{1\perp} \frac{e^{i z_{1\perp} \cdot  \left(-\frac{x_{\bar{q}}}{2 x_{h_2}}p_{h_2 \perp} + p_{1\perp} - \frac{x_q}{2 x_{h_1}} p_{h_1\perp}\right)}}{x_q x_{\bar{q}} Q^2 +\left(\frac{x_q}{x_{h_1}} \vec{p}_{h_1}-\vec{p}_{1}\right)^2} F(z_{1\perp}) \left(\frac{x_{q}}{x_{h_1}} p_{h_1} - p_{1}\right)_r \varepsilon_{T i}  \\
& \times \int   d^d p_{1'\perp}  \int d^d z_{2\perp} \frac{e^{-i z_{2\perp} \cdot  \left(-\frac{x_{ \bar{q}}}{2 x_{h_2}}p_{h_2 \perp} -p_{1'\perp} -\frac{x_q}{2 x_{h_1}} p_{h_1\perp}\right)}}{x_q (1-x_q)Q^2 +\left(\frac{x_q}{x_{h_1}} \vec{p}_{h_1}-\vec{p}_{1'}\right)^2} F^*(z_{2\perp}) \left(\frac{x_{q}}{x_{h_1}} p_{h_1} - p_{1'}\right)_l \varepsilon_{T k}^* \\
& \times \frac{\alpha_s}{2\pi} C_F \Bigg [\frac{1}{\hat{\epsilon}}\frac{1 + (1-\beta_2)^2}{\beta_2} + \beta_2 + \frac{2 \ln \beta_2 (1 + (1-\beta_2)^2)}{\beta_2} \\
& + \frac{1+ (1-\beta_2)^2}{\beta_2} \ln \left( \frac{c_0^2}{\left(\frac{z_{2\perp} - z_{1\perp}}{2}\right)^2 \mu^2}\right)\Bigg] + (h_1 \leftrightarrow h_2)  \\
&= \frac{d \sigma_{3TT}^{q g\rightarrow h_1 h_2}}{ d x_{h_1} d x_{h_2} d p_{h_1 \perp} d^d p_{h_2\perp} }\bigg |_{\text{coll. } \bar{q}g \text{ div}}      +  \frac{d \sigma_{3TT}^{q g  \rightarrow h_1 h_2}}{ d x_{h_1} d x_{h_2} d p_{h_1 \perp} d^d p_{h_2\perp} } \bigg |_{\text{coll. } \bar{q}g\text{ fin}}.
\numberthis
\end{align*} }
\section{Additional finite terms}
\label{sec:AdditionalFin}

Some of the finite terms of our calculation are presented in previous sections. They come as a result of the extraction of divergences. There are many other terms, completely disconnected from divergences, which however contribute to the final result. We proceed to list them, also emphasizing again what their nature is.
\subsection{Virtual corrections: Dipole $\times$ double-dipole contribution}
The $1$-loop correction to the $\gamma^{*} \rightarrow q \bar{q}$ contains a dipole and double-dipole terms. The first one receives a contribution from all diagrams, while the second one gets contributions only from diagrams where the virtual gluon crosses the shockwave. At the cross section level there will therefore be two contributions: 
\begin{itemize}
    \item[\textbullet] The one due to the interference between the dipole correction and the Born amplitude. This contains divergences and it is the one that we have completely computed in section \ref{sec: VirtualDiv}.
    \item[\textbullet] The one due to the interference between the double-dipole correction and the Born amplitude. Any  rapidity divergence present in this term is completely reabsorbed into the renormalized Wilson operator, at the amplitude level, with the help of the B-JIMWLK evolution. After this operation, this contribution is finite and can be taken in convolution with FFs without any additional manipulation.
\end{itemize}
Starting from Eq.~(5.34) of \cite{Boussarie:2016ogo}, we get
{ \allowdisplaybreaks
\begin{align*}
& \frac{d\sigma_{2LL}^{q \bar{q} \rightarrow  h_1 h_2}}{d x_{h_1 } d^2 p_{h_1 \perp } d x_{h_2 } d^2 p_{h_2 \perp }  } \\
&= \hspace{-0.05 cm} \frac{\alpha_{\mathrm{em}} \alpha_s Q^2}{(2 \pi)^5 N_c x_{h_1}^2 x_{h_2}^2}  \sum_{q} \frac{Q_q^2}{2}  \int_{x_{h_1}}^1 \hspace{-0.3 cm} d x_q \int_{x_{h_2}}^1 \hspace{-0.3 cm} d x_{\bar{q}} \; x_q  x_{\bar{q}} \; \delta(1-x_q -x_{\bar{q}}) D_q^{h_1} \left(\frac{x_{h_1}}{x_q}, \mu_F\right) D_{\bar{q}}^{h_2} \left(\frac{x_{h_2}}{x_{\bar{q}}},\mu_F\right)  \\
& \times  \int d^2 p_{1 \perp} d^2 p_{2 \perp} d^2 p_{1' \perp} d^2 p_{2' \perp} \int \frac{d^2 p_{3 \perp}}{(2\pi)^2} \frac{\tilde{\mathbf{F}}\left(\frac{p_{12\perp}}{2}, p_{3\perp}\right) 
\mathbf{F}^*\left(\frac{p_{1'2'\perp}}{2}\right)}{\left(\frac{x_q}{x_{h_1}} \vec{p}_{h_1}-\vec{p}_{1'} \right)^2 + x_q x_{\bar{q}}Q^2 }\\ 
& \times \delta \left( \frac{x_q}{x_{h_1}} p_{h_1 \perp}- p_{1\perp} + \frac{x_{\bar{q}}}{x_{h_2}} p_{h_2 \perp} - p_{2 \perp} - p_{3\perp} \right)  \delta(p_{11' \perp} + p_{22' \perp}+ p_{3\perp})  \\
& \times \left \{ 4 x_q x_{\bar{q}} \left[  \frac{x_q x_{\bar{q}} \left(\vec{p}_3^2 - \left(\frac{x_{\bar{q}}}{x_{h_2}} \vec{p}_{h_2} - \vec{p}_2\right)^2 - \left(\frac{x_q}{x_{h_1}} \vec{p}_{h_1} - \vec{p}_1 \right)^2 -2 x_q x_{\bar{q}}Q^2  \right)}{\left(\left(\frac{x_{\bar{q}}}{x_{h_2}}\vec{p}_{h_2} - \vec{p}_2 \right)^2 + x_q x_{\bar{q}} Q^2 \right) \left(\left(\frac{x_q}{x_{h_1}} \vec{p}_{h_1} - \vec{p}_1 \right)^2 + x_q x_{\bar{q}} Q^2 \right) - x_q x_{\bar{q}} Q^2 \vec{p}_3^2 } \right. \right. \\
& \times \ln \left(\frac{x_q x_{\bar{q}}}{e^{2 \eta}}\right) \ln \left( \frac{\left(\left(\frac{x_{\bar{q}}}{x_{h_2}}\vec{p}_{h_2} - \vec{p}_2 \right)^2 + x_q x_{\bar{q}} Q^2 \right) \left(\left(\frac{x_q}{x_{h_1}} \vec{p}_{h_1} - \vec{p}_1 \right)^2 + x_q x_{\bar{q}} Q^2 \right)}{x_q x_{\bar{q}} Q^2 \vec{p}_3^2 }\right) \\
& - \left. \left( \frac{2 x_q x_{\bar{q}}}{Q^2 x_q x_{\bar{q}}  + \left(\frac{x_q}{x_{h_1}} \vec{p}_{h_1} - \vec{p}_1\right)^2 }  \ln \left(\frac{x_q}{e^\eta}\right) \ln \left( \frac{\vec{p}_3^2}{\mu^2}\right) + (q \leftrightarrow \bar{q}) \right) \right] \\
& + \left[ Q^2 \int_0^{x_q} d z \left[\left(\phi_5 + \phi_6\right)_{LL}\right]_+ + (q \leftrightarrow \bar{q}) \right] \Bigg \} + h.c. +  (h_{1} \leftrightarrow h_2)  \numberthis \; . 
\end{align*}  }

Concerning other transitions, we have
{\allowdisplaybreaks
\begin{align*}
& \frac{d\sigma_{2TL}^{q \bar{q} \rightarrow h_1 h_2}}{d x_{h_1 } d^2 p_{h_1 \perp } d x_{h_2 } d^2 p_{h_2 \perp }  } \\
& = \hspace{-0.05 cm} \frac{\alpha_{\mathrm{em}} \alpha_s Q}{(2 \pi)^5 N_c x_{h_1}^2 x_{h_2}^2}  \sum_{q} \frac{Q_q^2}{2}  \int_{x_{h_1}}^1 \hspace{-0.3 cm} d x_q \int_{x_{h_2}}^1 \hspace{-0.3 cm} d x_{\bar{q}} \; x_q  x_{\bar{q}} \; \delta(1-x_q -x_{\bar{q}}) \\
& \times D_q^{h_1} \left(\frac{x_{h_1}}{x_q}, \mu_F\right) D_{\bar{q}}^{h_2} \left(\frac{x_{h_2}}{x_{\bar{q}}},\mu_F\right) \int d^2 p_{1 \perp} d^2 p_{2 \perp} d^2 p_{1' \perp} d^2 p_{2' \perp} \frac{d^2 p_{3 \perp} d^2 p_{3' \perp} }{(2\pi)^2} \varepsilon_{T i}^*  \\
& \times \delta \left( \frac{x_q}{x_{h_1}} p_{h_1 \perp }- p_{1\perp} + \frac{x_{\bar{q}}}{x_{h_2}} p_{h_2 \perp} - p_{2 \perp} - p_{3\perp} \right) \delta \left( p_{11'\perp} + p_{22'\perp} + p_{33'\perp} \right) \\
& \times   \left\{ \delta(p_{3' \perp}) \frac{  \tilde{\mathbf{F}}\left(\frac{p_{12\perp}}{2}, p_{3\perp}\right) \mathbf{F}^*\left(\frac{p_{1'2'\perp}}{2}\right)}{\left(\frac{x_q}{x_{h_1}}\vec{p}_{h_1}-\vec{p}_{1'}\right)^2 + x_q x_{\bar{q}} Q^2 }  \Bigg \{  2 (x_{\bar{q}} - x_q )  \left(\frac{x_q}{x_{h_1}} p_{h_1} - p_{1'}\right)^i \right. \\
& \times \left[  \frac{x_q x_{\bar{q}} \left(\vec{p}_3^2 - \left(\frac{x_{\bar{q}}}{x_{h_2}} \vec{p}_{h_2} - \vec{p}_2\right)^2 - \left(\frac{x_q}{x_{h_1}} \vec{p}_{h_1} - \vec{p}_1 \right)^2 -2 x_q x_{\bar{q}}Q^2  \right)}{\left(\left(\frac{x_{\bar{q}}}{x_{h_2}}\vec{p}_{h_2} - \vec{p}_2 \right)^2 + x_q x_{\bar{q}} Q^2 \right) \left(\left(\frac{x_q}{x_{h_1}} \vec{p}_{h_1} - \vec{p}_1 \right)^2 + x_q x_{\bar{q}} Q^2 \right) - x_q x_{\bar{q}} Q^2 \vec{p}_3^2 } \ln\left(\frac{x_q x_{\bar{q}}}{e^{2\eta}}\right) \right. \\
& \times \ln \left( \frac{\left(\left(\frac{x_{\bar{q}}}{x_{h_2}}\vec{p}_{h_2} - \vec{p}_2 \right)^2 + x_q x_{\bar{q}} Q^2 \right) \left(\left(\frac{x_q}{x_{h_1}} \vec{p}_{h_1} - \vec{p}_1 \right)^2 + x_q x_{\bar{q}} Q^2 \right)}{x_q x_{\bar{q}} Q^2 \vec{p}_3^2 }\right) \\
& - \left. \left( \frac{2 x_q x_{\bar{q}}}{Q^2 x_q x_{\bar{q}} + \left(\frac{x_q}{x_{h_1}} \vec{p}_{h_1} - \vec{p}_1\right)^2 }  \ln \left(\frac{x_q}{e^\eta}\right) \ln \left( \frac{\vec{p}_3^2}{\mu^2}\right) + (q \leftrightarrow \bar{q}) \right) \right] \\
&  + \left[ \frac{1}{2 x_q x_{\bar{q}}} \int_0^{x_q} d z \left[\left( \phi^i_5 + \phi^i_6 \right)_{TL}\right]_+  + (q \leftrightarrow \bar{q }) \right] \Bigg \} \\
& + \delta(p_{3 \perp}) \frac{ \mathbf{F}\left(\frac{p_{12\perp}}{2}\right) \tilde{\mathbf{F} }^*\left(\frac{p_{1'2'\perp}}{2}, p_{3'\perp}\right) }{\left(\frac{x_q}{x_{h_1}}\vec{p}_{h_1}-\vec{p}_{1}\right)^2 + x_q x_{\bar{q}} Q^2 }  \Bigg \{  \Bigg [ 2 x_q x_{\bar{q}} (x_{\bar{q}} - x_q )  \left(\frac{x_q}{x_{h_1}} p_{h_1} - p_{1'}\right)^i \\
& \times \left( \frac{-2 }{Q^2  x_q x_{\bar{q}} + \left(\frac{x_q}{x_{h_1}} \vec{p}_{h_1} - \vec{p}_{1'}\right)^2} \ln\left(\frac{x}{e^\eta}\right) \ln\left(\frac{\vec{p}_{3'}^2}{\mu^2}\right) \right. \\
& - \ln \left(\frac{x_q x_{\bar{q}}}{e^{2 \eta }}\right)  \frac{ \left(\frac{x_{\bar{q}}}{x_{h_2}} \vec{p}_{h_2} - \vec{p}_{2'}\right)^2 + x_q x_{\bar{q}} Q^2 }{ \left(\left(\frac{x_{\bar{q}}}{x_{h_2}} \vec{p}_{h_2} - \vec{p}_{2'}\right)^2 + x_q x_{\bar{q}} Q^2\right) \left(\left(\frac{x_q}{x_{h_1}} \vec{p}_{h_1} - \vec{p}_{1'}\right)^2 + x_q x_{\bar{q}} Q^2\right) - x_q x_{\bar{q}} Q^2 \vec{p}_{3'}^2  } \\
& \times \ln \left(\frac{ \left(\left(\frac{x_{\bar{q}}}{x_{h_2}} \vec{p}_{h_2} - \vec{p}_{2'}\right)^2 + x_q x_{\bar{q}} Q^2\right) \left(\left(\frac{x_q}{x_{h_1}} \vec{p}_{h_1} - \vec{p}_{1'}\right)^2 + x_q x_{\bar{q}} Q^2\right)}{x_q x_{\bar{q}} Q^2 \vec{p}_{3'}^2 }\right) \\
& + \left. \left.  \frac{1}{ \left( \frac{x_q}{x_{h_1}} \vec{p}_{h_1} -\vec{p}_{1'}\right)^2 } \ln \left(\frac{x_q x_{\bar{q}}}{e^{2 \eta}}\right) \ln\left(  \frac{ \left( \frac{x_q}{x_{h_1}} \vec{p}_{h_1} -\vec{p}_{1'}\right)^2 + x_q x_{\bar{q}} Q^2 }{x_q x_{\bar{q}} Q^2 } \right) \right) + (q \leftrightarrow \bar{q}) \right] \\
& + \left. \left[ \int_0^{x_q} d z \left[\left(\phi_5^{i*} + \phi_6^{i*} \right)_{LT}\right]_+ + (q \leftrightarrow \bar{q})    \right]  \right. \Bigg \} \Bigg \} \\
& + ( h_1 \leftrightarrow h_2 ) \; , \numberthis
\end{align*} }
and
{\allowdisplaybreaks
\begin{align*}
& \frac{d\sigma_{2TT}^{q \bar{q} \rightarrow h_1 h_2}}{d x_{h_1 } d^2 p_{h_1 \perp } d x_{h_2 } d^2 p_{h_2 \perp }  } \\
&= \hspace{-0.05 cm} \frac{\alpha_{\mathrm{em}} \alpha_s}{(2 \pi)^5 N_c x_{h_1}^2 x_{h_2}^2}  \sum_{q} \frac{Q_q^2}{2}  \int_{x_{h_1}}^1 \hspace{-0.3 cm} d x_q \int_{x_{h_2}}^1 \hspace{-0.3 cm} d x_{\bar{q}} \; x_q  x_{\bar{q}} \; \delta(1-x_q -x_{\bar{q}}) D_q^{h_1} \left(\frac{x_{h_1}}{x_q}, \mu_F\right) D_{\bar{q}}^{h_2} \left(\frac{x_{h_2}}{x_{\bar{q}}},\mu_F\right) \\
& \times \int d^2 p_{1 \perp} d^2 p_{2 \perp} \int \frac{d^2 p_{3 \perp}}{(2\pi)^2} \int d^2 p_{1' \perp} d^2 p_{2' \perp} \frac{\varepsilon_{T i} \varepsilon_{T j}^* }{\left( \frac{x_q}{x_{h_1}} \vec{p}_{h_1}-\vec{p}_{1'}\right)^2 + x_q x_{\bar{q}} Q^2} \\
& \times  \left[ \delta \left( \frac{x_q}{x_{h_1}} p_{h_1 \perp} - p_{1\perp} + \frac{x_{\bar{q}}}{x_{h_2}} p_{h_2 \perp} - p_{2 \perp} - p_{3\perp} \right)  \delta \left( \frac{x_q}{x_{h_1}} p_{h_1 \perp} - p_{1'\perp} + \frac{x_{\bar{q}}}{x_{h_2}} p_{h_2 \perp} - p_{2' \perp}  \right)  \right.  \\
& \times \tilde{\mathbf{F}}\left(\frac{p_{12\perp}}{2}, p_{3\perp}\right) \mathbf{F}^*\left( \frac{p_{1'2'\perp}}{2}\right)  \Bigg \{  \Bigg [ \left( \frac{x_q}{x_{h_1}} p_{h_1} - p_{1'}\right)_l \left( \frac{x_q}{x_{h_1}} p_{h_1} - p_{1}\right)_k   \\
& \times \left( (x_{\bar{q}} -x_q)^2 g_{\perp}^{ki} g_{\perp}^{lj} - g_{\perp}^{kj} g_{\perp}^{li} + g_{\perp}^{kl} g_{\perp}^{ij} \right) \left( \frac{-2}{Q^2  x_q x_{\bar{q}} +  \left( \frac{x_q}{x_{h_1}} \vec{p}_{h_1} - \vec{p}_{1}\right)^2 }  \ln \left( \frac{x_q}{e^\eta} \right)  \ln \left(\frac{\vec{p}_{3}^2 }{\mu^2 }\right) \right. \\
& + \frac{1}{ \left( \frac{x_q}{x_{h_1}} \vec{p}_{h_1} -\vec{p}_{1}\right)^2 } \ln \left(\frac{x_q x_{\bar{q}}}{e^{2 \eta}}\right) \ln\left(  \frac{ \left( \frac{x_q}{x_{h_1}} \vec{p}_{h_1} -\vec{p}_{1}\right)^2 + x_q x_{\bar{q}} Q^2 }{x_q x_{\bar{q}} Q^2 } \right) \\
& - \ln \left(\frac{x_q x_{\bar{q}}}{e^{2 \eta }}\right) \frac{ \left(\frac{x_{\bar{q}}}{x_{h_2}} \vec{p}_{h_2} - \vec{p}_{2}\right)^2 + x_q x_{\bar{q}} Q^2 }{ \left(\left(\frac{x_{\bar{q}}}{x_{h_2}} \vec{p}_{h_2} - \vec{p}_{2}\right)^2 + x_q x_{\bar{q}} Q^2\right) \left(\left(\frac{x_q}{x_{h_1}} \vec{p}_{h_1} - \vec{p}_{1}\right)^2 + x_q x_{\bar{q}} Q^2\right) - x_q x_{\bar{q}} Q^2 \vec{p}_{3}^2  } \\
& \left. \left. \times \ln \left(\frac{ \left(\left(\frac{x_{\bar{q}}}{x_{h_2}} \vec{p}_{h_2} - \vec{p}_{2}\right)^2 + x_q x_{\bar{q}} Q^2\right) \left(\left(\frac{x_q}{x_{h_1}} \vec{p}_{h_1} - \vec{p}_{1}\right)^2 + x_q x_{\bar{q}} Q^2\right)}{x_q x_{\bar{q}} Q^2 \vec{p}_{3}^2 }\right) \right) + (q \leftrightarrow \bar{q}) \right] \\
& \left. +  \left. \frac{1}{x_q x_{\bar{q}}} \left[ \int_{0}^{x_q} d z\left[\left(\phi_{5}^{i j} + \phi_{6}^{i j} \right)_{T T}\right]_{+} d z+ (q \leftrightarrow \bar{q}) \right]  \right \} +h.c \big |_{(p_1, p_3 \leftrightarrow p_{1'}, p_{3'}) ,  (i \leftrightarrow j)} \right]  \\
& + (h_1 \leftrightarrow h_2 )  \; .\numberthis
\end{align*} }

The $\phi$ function are defined in Appendix~\ref{AppendixD1}.

\subsection{Real corrections: Fragmentation from quark and anti-quark}
In this case, we refer to the finite terms related to the contribution (b) of Fig. \ref{fig:sigma-NLO}. In order to better understand what these contributions are, referring to Ref. \cite{Boussarie:2014lxa}, we recall that the impact factor for the transition $\gamma^{*} \rightarrow q  \bar{q} g$ has a double dipole contribution ($\Phi_4^{(+,i)}$) and a single dipole ($\Phi_3^{(+,i)}$) contribution. The finite contributions which we obtain are
\begin{itemize}
    \item Finite terms related to the dipole $\times$ dipole contribution. 
    \item Dipole $\times$ double dipole contribution. 
    \item Double dipole $\times$ double dipole contribution. 
\end{itemize}

\subsubsection{Finite part of dipole $\times$ dipole contribution}
When we square the dipole contribution, using shorthand notation introduced in (\ref{eq:ShortHand}), we obtain the following structure:
{ \allowdisplaybreaks
\begin{align*}
\Phi_3^\alpha(\vec{p}_1,\vec{p}_2) \Phi_3^{\beta*}(\vec{p}_{1'},\vec{p}_{2'}) & = \tilde{\Phi}_3^\alpha(\vec{p}_1,\vec{p}_2) \tilde{\Phi}_3^{\beta*}(\vec{p}_{1'},\vec{p}_{2'}) \\
& + \left( \tilde{\Phi}_3^\alpha(\vec{p}_1, \vec{p}_2) \Phi_4^{\beta *}(\vec{p}_{1'}, \vec{p}_{2'},\vec{0})+\Phi_4^\alpha(\vec{p}_1, \vec{p}_2,\vec{0}) \Phi_3^{\beta *}(\vec{p}_{1'},\vec{p}_{2'})\right)  \\ 
& + \Phi_4^\alpha(\vec{p}_1,\vec{p}_2,\vec{0}) \Phi_4^{\beta*}(\vec{p}_{1'},\vec{p}_{2'},\vec{0}) \,.
\numberthis[phi3_squared]
\end{align*} }
The first term in the RHS is the one containing divergences that we have considered in previous sections. After isolating soft and collinear divergences, finite terms remain. The finite contributions for
$\tilde{\sigma}_{(b)div,1}$ and $\tilde{\sigma}_{(b)div,3}$ have been computed respectively in
sections \ref{sec:qqbarfragColl-qg} and \ref{sec:qqbarfragColl-qbarg}, see eqs.~\eqref{eq:collqg_LL_fin},
\eqref{eq:TL_coll_qg_finite},
\eqref{eq:collqg_TT_fin} and
\eqref{eq:collqbarg_LL_fin}
\eqref{eq:collqbarg_TL_fin}
\eqref{eq:collqbarg_TT_fin}.
Besides,
the terms $(\tilde{\sigma}_{(b)div,2}-\tilde{\sigma}_{(b)div,2}^{soft})$ and $(\tilde{\sigma}_{(b)div,4}-\tilde{\sigma}_{(b)div,4}^{soft})$ in Eq. \eqref{sigmatilde_q-qbar} are finite. Their contribution read
{\allowdisplaybreaks
\begin{align*}
& \frac{d \sigma_{3LL}^{q \bar{q} \rightarrow h_1 h_2}}{d x_{h_1} d x_{h_2} d^d p_{h_1\perp} d^d p_{h_2 \perp}}\Bigg |_{\text{finite, (b) 2,4}} \\*
&= \frac{\alpha_s C_F}{\mu^{2\epsilon}} \frac{4  \alpha_{\mathrm{em}} Q^2}{(2\pi)^{4(d-1)} N_c \; x_{h_1}^d x_{h_2}^d} \sum_{q} \int_{x_{h_1}}^1 \frac{d x_q}{x_q} \int_{x_{h_2}}^1 \frac{d x_{\bar{q}}}{ x_{\bar{q}}} \int_{\alpha}^1 \frac{d x_g}{x_g^{3-d}}  \left(x_qx_{\bar{q}} \right)^{d-1}  \delta(1-x_q-x_{\bar{q}}-x_g)  \\
& \times Q_q^2 D_q^{h_1}\left(\frac{x_{h_1}}{x_q},\mu_F\right)D_{\bar{q}}^{h_2}\left(\frac{x_{h_2}}{x_{\bar{q}}}, \mu_F\right)  \int \frac{d^d \vec{u}}{(2\pi)^d} \int d^d p_{1\perp} d^d p_{2\perp} \mathbf{F} \left(\frac{p_{12\perp}}{2}\right)  \\
& \times \int d^d p_{1'\perp}  d^d p_{2'\perp} \mathbf{F}^*\left( \frac{p_{1'2'\perp}}{2}\right) \delta\left(p_{1 1' \perp} + p_{2 2'\perp} \right) \\
& \times \left \{ \frac{8 x_q x_{\bar{q}} \; \delta \left(\frac{x_q}{x_{h_1}} p_{h_1\perp} - p_{1\perp} + \frac{x_{\bar{q}}}{x_{h_2}} p_{h_2\perp} - p_{2\perp} \right)}{\left(Q^2+ \frac{\left(\frac{x_{\bar{q}}}{x_{h_2}}\vec{p}_{h_2}-\vec{p}_{2'}\right)^2}{x_{\bar{q}} (1-x_{\bar{q}})}\right)\left(Q^2+ \frac{\left(\frac{x_q}{x_{h_1}} \vec{p}_{h_1}-\vec{p}_{1}\right)^2}{x_q (1-x_q)}\right)} \frac{\left(\vec{u}-\frac{\vec{p}_{h_1}}{x_{h_1}}\right)\cdot \left(\vec{u}-\frac{\vec{p}_{h_2}}{x_{h_2}}\right)}{\left(\vec{u}-\frac{\vec{p}_{h_1}}{x_{h_1}}\right)^2 \left(\vec{u}-\frac{\vec{p}_{h_2}}{x_{h_2}}\right)^2} \right. \\ 
& - \frac{(2 x_g - d x_g^2 + 4 x_q x_{\bar{q}}) \delta\left(\frac{x_q}{x_{h_1}} p_{h_1\perp} - p_{1\perp} + \frac{x_{\bar{q}}}{x_{h_2}} p_{h_2\perp} - p_{2\perp} + x_g u_{ \perp} \right)}{\left(Q^2+ \frac{\left(\frac{x_{\bar{q}}}{x_{h_2}}\vec{p}_{h_2}-\vec{p}_{2'}\right)^2}{x_{\bar{q}} (1-x_{\bar{q}})}\right)\left(Q^2+ \frac{\left(\frac{x_q}{x_{h_1}} \vec{p}_{h_1}-\vec{p}_{1}\right)^2}{x_q (1-x_q)}\right)}  \frac{\left(\vec{u}-\frac{\vec{p}_{h_1}}{x_{h_1}}\right)\cdot \left(\vec{u}-\frac{\vec{p}_{h_2}}{x_{h_2}}\right)}{\left(\vec{u}-\frac{\vec{p}_{h_1}}{x_{h_1}}\right)^2 \left(\vec{u}-\frac{\vec{p}_{h_2}}{x_{h_2}}\right)^2} \\
&   - \frac{(2 x_g - d x_g^2 + 4 x_q x_{\bar{q}}) \delta\left(\frac{x_q}{x_{h_1}} p_{h_1\perp} - p_{1\perp} + \frac{x_{\bar{q}}}{x_{h_2}}p_{h_2\perp} - p_{2\perp} + x_g u_{ \perp} \right) }{\left(Q^2+ \frac{\left(\frac{x_{\bar{q}}}{x_{h_2}}\vec{p}_{h_2}-\vec{p}_{2}\right)^2}{x_{\bar{q}} (1-x_{\bar{q}})}\right)\left(Q^2+ \frac{\left(\frac{x_q}{x_{h_1}} \vec{p}_{h_1}-\vec{p}_{1'}\right)^2}{x_q (1-x_q)}\right)} \\
& \times \left. \frac{\left(\vec{u}-\frac{\vec{p}_{h_1}}{x_{h_1}}\right)\cdot \left(\vec{u}-\frac{\vec{p}_{h_2}}{x_{h_2}}\right)}{\left(\vec{u}-\frac{\vec{p}_{h_1}}{x_{h_1}}\right)^2 \left(\vec{u}-\frac{\vec{p}_{h_2}}{x_{h_2}}\right)^2}
\right\} + (h_1 \leftrightarrow h_2) \; , \numberthis \\ 
\end{align*} }
in the $LL$ case. The same contribution in the $TL$ and $TT$ cases is, respectively,
{\allowdisplaybreaks
\begin{align*}
& \frac{d \sigma_{3TL}^{q \bar{q} \rightarrow h_1 h_2}}{d x_{h_1} d x_{h_2} d^d p_{h_1\perp} d^d p_{h_2 \perp}}\Bigg |_{\text{finite, (b) 2,4}} \\*
&= \frac{\alpha_s C_F}{\mu^{2\epsilon}} \frac{2 \alpha_{\mathrm{em}}Q}{(2\pi)^{4(d-1)}N_c x_{h_1}^d x_{h_2}^d} \sum_{q} \int_{x_{h_1}}^1 \frac{d x_q}{x_q} \int_{x_{h_2}}^1 \frac{x_{\bar{q}}}{x_{\bar{q}}} \int_{\alpha}^1 \frac{d x_g}{x_g^{3-d}} (x_q x_{\bar{q}})^{d-1} \delta(1-x_q -x_{\bar{q}}-x_g) \\
&  \times  Q_q^2  D_q^{h_1}\left(\frac{x_{h_1}}{x_q}, \mu_F\right) D_{\bar{q}}^{h_2}\left(\frac{x_{h_2}}{ x_{\bar{q}}},\mu_F \right) \int \frac{d^d \vec{u}}{(2\pi)^d }   \int d^d p_{1 \perp} d^d p_{2 \perp} \mathbf{F}\left(\frac{p_{12 \perp}}{2}\right) \\
& \times  \int d^d p_{1' \perp } d^d p_{2' \perp} \mathbf{F}^*\left(\frac{p_{1'2' \perp}}{2}\right)   \delta \left(p_{11' \perp} + p_{22' \perp} \right) \varepsilon_{T i}^* \\
& \times \left\{   \frac{\delta \left( \frac{x_q}{x_{h_1}} p_{h_1 \perp} - p_{1 \perp} + \frac{x_{\bar{q}}}{x_{h_2}} p_{h_2 \perp} - p_{2 \perp} + x_g u_\perp \right)}{  \left(Q^2 + \frac{\left( \frac{x_{\bar{q}}}{x_{h_2}} \vec{p}_{h_2} - \vec{p}_{2'} \right)^2}{x_{\bar{q}} (1-x_{\bar{q}} )}\right) \left(Q^2 + \frac{\left(\frac{x_q}{x_{h_1}} \vec{p}_{h_1} -\vec{p}_1\right)^2}{x_q (1-x_q) } \right)}  \frac{\left(u_\perp - \frac{p_{h_1 \perp}}{x_{h_1}} \right)_\mu \left(u_\perp - \frac{p_{h_2 \perp}}{x_{h_2}} \right)_\nu}{\left(\vec{u}-\frac{\vec{p}_{h_1}}{x_{h_1}}\right)^2 \left(\vec{u}-\frac{\vec{p}_{h_2}}{x_{h_2}}\right)^2}\right.   \\
& \times \frac{1}{x_{\bar{q}} (x_q + x_g)} \left[ x_g (4 x_{\bar{q}} + d x_g -2) \left( \left(\frac{x_{\bar{q}}}{x_{h_2}} p_{h_2 \perp} - p_{2' \perp} \right)^\mu g_\perp^{i \nu} -  \left(\frac{x_{\bar{q}}}{x_{h_2}} p_{h_2 \perp} - p_{2' \perp} \right)^\nu g_\perp^{i \mu}   \right) \right. \\
& \left. - (2 x_{\bar{q}} -1 ) (4 x_{\bar{q}} x_q + x_g(2-x_g d)) g_\perp^{\mu \nu} \left( \frac{x_{\bar{q}}}{x_{h_2}} p_{h_2 \perp} - p_{2' \perp} \right)^i \right]  \\ 
& +   \frac{\delta \left( \frac{x_q}{x_{h_1}} p_{h_1 \perp} - p_{1 \perp} + \frac{x_{\bar{q}}}{x_{h_2}} p_{h_2 \perp} - p_{2 \perp} + x_g u_\perp \right)}{  \left(Q^2 + \frac{\left( \frac{x_q}{x_{h_1}} \vec{p}_{h_1} - \vec{p}_{1'} \right)^2}{x_q (1-x_q )}\right) \left(Q^2 + \frac{\left(\frac{x_{\bar{q}}}{x_{h_2}} \vec{p}_{h_2} -\vec{p}_2\right)^2}{x_{\bar{q}} (1-x_{\bar{q}}) } \right)}  \frac{\left(u_\perp - \frac{p_{h_2 \perp}}{x_{h_2}} \right)_\mu \left(u_\perp - \frac{p_{h_1 \perp}}{x_{h_1}} \right)_\nu}{\left(\vec{u}-\frac{\vec{p}_{h_1}}{x_{h_1}}\right)^2 \left(\vec{u}-\frac{\vec{p}_{h_2}}{x_{h_2}}\right)^2} \\
& \times \frac{1}{x_q (x_{\bar{q}} + x_g)} \left[ x_g (4 x_q + d x_g -2) \left( \left(\frac{x_q}{x_{h_1}} p_{h_1 \perp} - p_{1' \perp} \right)^\mu g_\perp^{i \nu} -  \left(\frac{x_q}{x_{h_1}} p_{h_1 \perp} - p_{1' \perp} \right)^\nu g_\perp^{i \mu}   \right) \right. \\
& \left. - (2 x_q -1 ) (4 x_{\bar{q}} x_q + x_g(2-x_g d)) g_\perp^{\mu \nu} \left( \frac{x_q}{x_{h_2}} p_{h_1 \perp} - p_{1' \perp} \right)^i \right]  \\ 
& - \frac{ 8 \; \delta \left(\frac{x_q }{x_{h_1}} p_{h_1 \perp} - p_{1 \perp} + \frac{x_{\bar{q}}}{x_{h_2}} p_{h_2 \perp} - p_{2 \perp}  \right) }{\left(Q^2 + \frac{\left( \frac{x_{\bar{q}}}{x_{h_2}} \vec{p}_{h_2} - \vec{p}_{2'} \right)^2}{x_{\bar{q}} (1-x_{\bar{q}} )}\right) \left(Q^2 + \frac{\left(\frac{x_q}{x_{h_1}} \vec{p}_{h_1} -\vec{p}_1\right)^2}{x_q (1-x_q) } \right)} \frac{\left( \vec{u} - \frac{\vec{p}_{h_1}}{x_{h_1}} \right) \cdot \left( \vec{u} - \frac{\vec{p}_{h_2}}{x_{h_2}} \right) }{\left( \vec{u} - \frac{\vec{p}_{h_1}}{x_{h_1}} \right)^2  \left( \vec{u} - \frac{\vec{p}_{h_2}}{x_{h_2}} \right)^2}  \\ 
& \times  \left. \left( \frac{x_{\bar{q}} }{x_{h_2}} p_{h_2 \perp} - p_{2' \perp} \right)^i (x_{\bar{q}} -x_q) \right\} + (h_1 \leftrightarrow h_2) \; ,
\end{align*} }
and
{\allowdisplaybreaks
\begin{align*}
& \frac{d \sigma_{3TT}^{q \bar{q} \rightarrow h_1 h_2}}{d x_{h_1} d x_{h_2} d^d p_{h_1\perp} d^d p_{h_2 \perp}}\Bigg |_{\text{finite, (b) 2,4}} \\*
&= \frac{\alpha_s C_F}{\mu^{2\epsilon}} \frac{\alpha_{\mathrm{em}} }{(2 \pi)^{4(d-1)} N_c x_{h_1}^d x_{h_2}^d } \sum_q \int_{x_{h_1}}^1 \frac{d x_q}{x_q} \int_{x_{h_2}}^1 \frac{d x_{\bar{q}}}{x_{\bar{q}}} \int_\alpha^1 \frac{d x_g}{x_g^{3-d}} \delta(1-x_q -x_{\bar{q}} -x_g) (x_q x_{\bar{q}})^{d-1} \\
& \times Q_q ^2  D_q^{h_1}\left(\frac{x_{h_1}}{x_q}, \mu_F\right) D_{\bar{q}}^{h_2}\left(\frac{x_{h_2} }{ x_{\bar{q}}},\mu_F \right) \int \frac{d^d \vec{u}}{(2\pi)^d} \int d^d p_{1 \perp} d^d p_{2 \perp}  \mathbf{F}\left(\frac{p_{12 \perp}}{2}\right) \\ 
& \times \int d^d p_{1' \perp} d^d p_{2' \perp}  \mathbf{F}^*\left(\frac{p_{1'2' \perp}}{2}\right) \delta (p_{11' \perp} + p_{22' \perp} ) \varepsilon_{T i} \varepsilon_{T k}^* \\ 
&  \times \left\{ \left[ \left( - \frac{ \delta \left( p_{q1\perp} + p_{\bar{q}2 \perp} + x_g u_\perp \right) }{\left(Q^2+\frac{\vec{p}_{\bar{q} 2}^2}{x_{\bar{q}}\left(1-x_{\bar{q}}\right)}\right)\left(Q^2+\frac{\vec{p}_{q 1^{\prime}}^2}{x_q\left(1-x_q\right)}\right)} \frac{\left(u_\perp - \frac{p_{q\perp}}{x_q} \right)_\mu \left(u_\perp - \frac{p_{\bar{q} \perp}}{x_{\bar{q}}} \right)_\nu}{\left(\vec{u} - \frac{\vec{p}_{q}}{x_q} \right)^2 \left(\vec{u} - \frac{p_{\bar{q}}}{x_{\bar{q}} } \right)^2 }   \right. \right.  \right. \\ 
& \times \frac{1}{(x_q + x_g)(x_{\bar{q}} + x_g) x_q x_{\bar{q}} } \left\{ x_g ((d-4)) x_g -2) \left[p_{q 1^{\prime} \perp}^\nu\left(p_{\bar{q} 2 \perp}^\mu g_{\perp}^{i k}+p_{\bar{q} 2 \perp}^k g_{\perp}^{\mu i}\right) \right. \right. \\
& \left. + g_{\perp}^{\mu \nu}\left(\left(\vec{p}_{q 1^{\prime}} \cdot \vec{p}_{\bar{q} 2}\right) g_{\perp}^{i k}+p_{q 1^{\prime} \perp}^i p_{\bar{q} 2 \perp}^k\right) -g_{\perp}^{\nu k} p_{q 1^{\prime} \perp}^i p_{\bar{q} 2 \perp}^\mu -g_{\perp}^{\mu i} g_{\perp}^{\nu k}\left(\vec{p}_{q 1^{\prime}} \cdot \vec{p}_{\bar{q} 2}\right) \right] -g_{\perp}^{\mu \nu} \\
& \times \hspace{-0.15 cm} \left[ \left(2x_q -1 \right) \left(2 x_{\bar{q}} - 1\right) p_{q1'\perp}^k p_{\bar{q}2\perp}^i \left( 4 x_q x_{\bar{q}} + x_g (2 - x_g d)\right)  + 4 x_q x_{\bar{q}} ((\vec{p}_{q1'} \cdot \vec{p}_{\bar{q}2})g_\perp^{ik} + p_{q1'\perp}^i p_{\bar{q}2\perp}^k  )\right] \\
& + \left( p_{q1'\perp}^\mu p_{\bar{q}2\perp}^\nu g_\perp^{ik} - p_{q1'\perp}^\mu p_{\bar{q}2\perp}^k g_\perp^{\nu i } - p_{q1'\perp}^i p_{\bar{q}2\perp}^\nu g_\perp^{\mu k } - g_\perp^{\mu k } g_\perp^{\nu i } (\vec{p}_{q1'} \cdot \vec{p}_{\bar{q}2} ) \right) \\ 
& \times x_g ((d-4)x_g + 2) + x_g (2x_{\bar{q}} - 1 ) (x_g d + 4 x_q -2 ) \left( g_\perp^{\mu k } p_{q1'\perp}^\nu - g_\perp^{\nu k} p_{q1'\perp}^\mu \right) p_{\bar{q}2\perp}^i \\
&   + \left. \left. \left. x_g (2 x_q -1 ) p_{q1'\perp}^k (4 x_{\bar{q}} + x_g d -2) \left( g_\perp^{\nu i } p_{\bar{q}2\perp}^\mu -g_\perp^{\nu k } p_{q1'\perp}^\nu \right) \right\}  \right) + (q \leftrightarrow \bar{q}) \right] \\
& + \frac{8 \; \delta(p_{q1 \perp} + p_{\bar{q}2\perp} ) }{\left(Q^2+\frac{\vec{p}_{\bar{q} 2}^2}{x_{\bar{q}}\left(1-x_{\bar{q}}\right)}\right)\left(Q^2+\frac{\vec{p}_{q 1^{\prime}}^2}{x_q\left(1-x_q\right)}\right)} \frac{\left( \vec{u} - \frac{\vec{p}_{q}}{x_q} \right) \cdot \left( \vec{u} - \frac{\vec{p}_{\bar{q}}}{x_{\bar{q}}} \right) }{\left( \vec{u} - \frac{\vec{p}_{q}}{x_q} \right)^2 \left( \vec{u} - \frac{\vec{p}_{\bar{q}}}{x_{\bar{q}}} \right)^2  } \\
& \times  \left.  \frac{1}{x_q x_{\bar{q}} } \left[ -(x_{\bar{q}} - x_q )^2 g_\perp^{ri} g_\perp^{kl} +  g_\perp^{il} g_\perp^{rk} -  g_\perp^{rl} g_\perp^{ik}\right] p_{\bar{q}2 \perp r} p_{q1' \perp l } 
\right\} + (h_1 \leftrightarrow h_2 ) \; .
\end{align*} }
In the above expression, the following replacement needs to be done: 
\begin{equation}
    p_{q \perp} = \frac{x_q}{x_{h_1}} p_{h_1 \perp} \; , \hspace{0.5cm}  p_{\bar{q} \perp} = \frac{x_{\bar{q}}}{x_{h_2}} p_{h_2 \perp} \; .
\end{equation}
The remaining term in Eq. \eqref{eq:phi3_squared}, for arbitrary polarization, is
{\allowdisplaybreaks
\begin{align*}
   & \frac{d {\sigma}_{3JI}^{q \bar{q} \rightarrow h_1 h_2}}{d x_{h_1} d x_{h_2} d^d p_{h_1} d^d p_{h_2}} \\ & =  \frac{2 \alpha_s \alpha_{\mathrm{em}} C_F }{\mu^{2\epsilon}(2\pi)^{4(d-1)}N_c} \frac{(p_0^-)^2}{s^2 x_{h_1}^d x_{h_2}^d} \sum_q Q_q^2 \int_{x_{h_1}}^1 \frac{d x_q}{x_q} \int_{x_{h_2}}^{1}  \frac{d x_{\bar{q}}}{x_{\bar{q}}} \; (x_q x_{\bar{q}})^{d-1} D_q^{h_1} \left( \frac{x_{h_1}}{x_q} , \mu_F \right)  \\
    & \times D_{\bar{q}}^{h_2} \left( \frac{x_{h_2}}{x_{\bar{q}}}, \mu_F \right) \int_0^1 \frac{d x_g  }{x_g } \int \frac{d^d p_{g\perp}}{(2\pi)^d} \delta (1-x_q-x_{\bar{q}}-x_g) \int  d^d p_{1\perp} d^d p_{2\perp}   d^d p_{1'\perp} d^d p_{2'\perp}  \\
    & \times  \mathbf{F}\left(\frac{p_{12\perp}}{2}\right) \mathbf{F}^*\left(\frac{p_{1'2'\perp}}{2}\right)  \delta \left(\frac{x_q}{x_{h_1} } p_{h_1 \perp}-p_{1\perp} + \frac{x_{\bar{q}}}{x_{h_2}} p_{h_2 \perp} -  p_{2\perp} + p_{g\perp}\right) \delta (p_{11'\perp} + p_{22'\perp} ) \\
   &  \times \varepsilon_{I\alpha} \; \varepsilon_{J\beta}^* \left[ \Phi_3^\alpha (p_{1\perp}, p_{2\perp}) \Phi_3^{\beta*} (p_{1'\perp}, p_{2'\perp}) - \tilde{\Phi}_3^{\alpha} (p_{1\perp}, p_{2\perp}) \tilde{\Phi}_3^{\beta*} (p_{1'\perp}, p_{2'\perp}) \right] \\
   & + (h_1 \leftrightarrow h_2) \; . \numberthis
\end{align*} }

\subsubsection{Dipole $\times$ double-dipole contribution and double-dipole $\times$ double-dipole contribution}

The dipole $\times$ double dipole contribution, for arbitrary polarization, is given by
{\allowdisplaybreaks
\begin{align*}
& \frac{d\sigma_{4JI}^{q \bar{q} \rightarrow h_1 h_2}}{d x_{h_1} d x_{h_2} d^d p_{h_1 \perp} d^d p_{h_2 \perp}} \\  & = \frac{\alpha_{s}\alpha_{\mathrm{em}} }{\mu^{2\epsilon}(2\pi)^{4(d-1)}N_{c}}\frac{(p_{0}%
^{-})^{2}}{s^{2}x_{h_1}^d x_{h_2}^d} \sum_q Q_{q}^{2} \hspace{-0.1 cm} \int_{x_{h_1}}^1  \hspace{-0.25 cm} \frac{d x_q}{x_q} \hspace{-0.15 cm} \int_{x_{h_2}}^1 \hspace{-0.25 cm} \frac{d x_{\bar{q}}}{x_{\bar{q}}} \; (x_q x_{\bar{q}})^{d-1} D_q^{h_1}\left(\frac{x_{h_1}}{x_q},\mu_F\right)D_{\bar{q}}^{h_2}\left(\frac{x_{h_2}}{x_{\bar{q}}}, \mu_F\right) \\ 
& \times \int_0^1 \frac{d x_g }{x_g } \int \frac{d^{d} p_{g\bot}}{(2\pi)^{d}} \delta(1-x_q -x_{\bar{q}}-x_g) \int d^dp_{1\bot}d^dp_{2\bot} d^dp_{1\bot}^\prime d^dp_{2\bot}^\prime \frac{d^dp_{3\bot}d^dp_{3\bot}^\prime}{\left( 2\pi \right)^d} \\ & \times \delta\left(\frac{x_q}{x_{h_1}}p_{h_1\perp} - p_{1\bot}+ \frac{x_{\bar{q}}}{x_{h_2}} p_{h_2 \perp} -p_{2\bot}+p_{g3\bot}\right) \delta(p_{11^\prime \bot}+p_{22^\prime \bot}+p_{33^\prime \bot}) (\varepsilon_{I\alpha} \varepsilon_{J\beta}^\ast) \\
& \times \left[ \Phi_3^\alpha(p_{1\bot},p_{2\bot}) \Phi_4^{\beta\ast}(p_{1\bot}^\prime, p_{2\bot}^\prime, p_{3\bot}^\prime) \mathbf{F}\left(\frac{p_{12\bot}}{2}\right) \mathbf{\tilde{F}}^\ast \left( \frac{p_{1^\prime 2^\prime \bot}}{2}, p_{3\bot}^\prime \right) \delta(p_{3\bot}) \right. \numberthis \\
& + \left. \Phi_4^\alpha (p_{1\bot}, p_{2\bot}, p_{3\bot}) \Phi_3^{\beta\ast} \left(p_{1'\perp},p_{2'\perp} \right) \mathbf{\tilde{F}}\left(\frac{p_{12\bot}}{2}, p_{3\bot}\right) \mathbf{F}^\ast\left(\frac{p_{1^\prime 2^\prime\bot}}{2} \right) \delta(p_{3\bot}^\prime) \right] + (h_1 \leftrightarrow h_2) \; .
\end{align*} }
The double-dipole $\times$ double-dipole contribution, for arbitrary polarization, is given by
\begin{align*}
\allowdisplaybreaks
& \frac{d\sigma_{5JI}^{q \bar{q} \rightarrow h_1 h_2}}{d x_{h_1} d x_{h_2} d^d p_{h_1} d^d p_{h_2}} \\ & = \frac{\alpha_{s} \alpha_{\mathrm{em}} (\varepsilon_{I\alpha} \varepsilon_{J\beta}^*) }{\mu^{2\epsilon}(2\pi)^{4(d-1)} (N_{c}^{2}-1)}\frac{(p_{0}^{-})^{2}}{s^{2}x_{h_1}^d x_{h_2}^d} \sum_q Q_{q}^{2} \int_{x_{h_1}}^1 \hspace{-0.15 cm} \frac{d x_q}{x_q} \hspace{-0.15 cm} \int_{x_{h_2}}^1 \hspace{-0.15 cm} \frac{d x_{\bar{q}}}{x_{\bar{q}}} \; (x_q x_{\bar{q}})^{d-1} D_q^{h_1}\left(\frac{x_{h_1}}{x_q},\mu_F\right) \\ 
& \times D_{\bar{q}}^{h_2}\left(\frac{x_{h_2}}{x_{\bar{q}}}, \mu_F\right) \int_0^1 \frac{d x_g }{x_g } \int \frac{d^{d}p_{g\bot}}{(2\pi)^{d}}\delta(1-x_q -x_{\bar{q}}-x_g) \int d^dp_{1\bot}d^dp_{2\bot} d^dp_{1\bot}^\prime d^dp_{2\bot}^\prime \\
& \times \int \frac{d^dp_{3\bot}d^dp_{3\bot}^\prime}{\left( 2\pi \right)^{2d}} \delta\left(\frac{x_q}{x_{h_1}}p_{h_1\perp} - p_{1\bot}+ \frac{x_{\bar{q}}}{x_{h_2}} p_{h_2 \perp} -p_{2\bot}+p_{g3\bot}\right) \delta(p_{11^\prime \bot}+p_{22^\prime \bot}+p_{33^\prime \bot})  \\
& \times \Phi_4^\alpha(p_{1\bot},p_{2\bot},p_{3\bot}) \Phi_4^{\beta\ast}(p_{1\bot}^\prime,p_{2\bot}^\prime, p_{3\bot}^\prime) \mathbf{\tilde{F}}\left( \frac{p_{12\bot}}{2}, p_{3\bot} \right) \mathbf{\tilde{F}}^\ast \left(\frac{p_{1^\prime 2^\prime \bot}}{2},p_{3\bot}^\prime\right)  \\
& + (h_1 \leftrightarrow h_2)   \; . \numberthis 
\end{align*} 
The expression for the squared impact factors can be found in Appendix \ref{AppendixD2}. They are written in terms of $p_q, p_{\bar{q}}, p_{g}, z$ and the following identification should be done:
\begin{equation}
    p_{q \perp} = \frac{x_q}{x_{h_1}} p_{h_1 \perp} \; , \hspace{0.5cm}  p_{\bar{q} \perp} = \frac{x_{\bar{q}}}{x_{h_2}} p_{h_2 \perp} \; , \hspace{0.5 cm} z=x_g \; .
\end{equation}

\subsection{Real corrections: Fragmentation from anti-quark and gluon}

\subsubsection{Finite part of dipole $\times$ dipole contribution}

When squaring the dipole contribution, we have also finite terms. This time we write separately for each polarization transition. In the $LL$ case, we have
\begin{align*}
\allowdisplaybreaks
   & \frac{d {\sigma}_{3LL}^{g \bar{q} \rightarrow h_1 h_2}}{d x_{h_1} d x_{h_2} d^d p_{h_1} d^d p_{h_2}} \\ & =  \frac{2 \alpha_s \alpha_{\mathrm{em}} C_F }{\mu^{2\epsilon}(2\pi)^{4(d-1)}N_c} \frac{(p_0^-)^2}{s^2 x_{h_1}^d x_{h_2}^d} \sum_q Q_q^2 \int_{x_{h_1}}^{1} \frac{d x_g}{x_g} \int_{x_{h_2}}^{1} \frac{d x_{\bar{q}} }{x_{\bar{q}}} \; (x_g x_{\bar{q}})^{d-1} D_g^{h_1} \left( \frac{x_{h_1}}{x_g} , \mu_F \right) \\
    & \times  D_{\bar{q}}^{h_2} \left( \frac{x_{h_2}}{x_{\bar{q}}}, \mu_F \right) \int_0^1 \frac{d x_q  }{x_q } \int \frac{d^d p_{q\perp}}{(2\pi)^d}\delta (1-x_q-x_{\bar{q}}-x_g) \int d^d p_{1\perp} d^d p_{2\perp}  d^d p_{1'\perp} d^d p_{2'\perp}  \\
    & \times \mathbf{F}\left(\frac{p_{12\perp}}{2}\right)  \mathbf{F}^*\left(\frac{p_{1'2'\perp}}{2}\right)  \delta \left(p_{q1\perp} + \frac{x_{\bar{q}}}{x_{h_2}} p_{h_2 \perp} - p_{2\perp} + \frac{x_g}{x_{h_1}}p_{h_1 \perp}\right)    \delta (p_{11'\perp} + p_{22'\perp} ) \frac{Q^2}{(p_\gamma^+)^2}  \\
    & \times \left[ \Phi_3^{+} (p_{1\perp}, p_{2\perp}) \Phi_3^{+*} (p_{1'\perp}, p_{2'\perp})  - \frac{8 x_q x_{\bar{q}} (p_\gamma^+)^4 \left( d x_g^2 + 4 x_q (x_q + x_g) \right)}{\left(Q^2 + \frac{\left(\frac{x_{\bar{q}}}{x_{h_2}}\vec{p}_{h_2}- \vec{p}_{2}\right)^2}{x_{\bar{q}}(1-x_{\bar{q}})} \right) \left(Q^2 + \frac{\left(\frac{x_{\bar{q}}}{x_{h_2}}\vec{p}_{h_2}- \vec{p}_{2'}\right)^2}{x_{\bar{q}}(1-x_{\bar{q}})} \right)  } \right. \\
   &  \times \left.  \frac{1}{\left(x_q \frac{x_{g}}{x_{h_1}} \vec{p}_{h_1} -x_g \vec{p}_q \right)^2} \right] + (h_1 \leftrightarrow h_2) \; . \numberthis \\ 
\end{align*} 
For the $TL$ case, we have %
\begin{align*}
   \allowdisplaybreaks
   & \frac{d {\sigma}_{3TL}^{g \bar{q} \rightarrow h_1 h_2}}{d x_{h_1} d x_{h_2} d^d p_{h_1} d^d p_{h_2}} \\ & =  \frac{2\alpha_s \alpha_{\mathrm{em}} C_F }{\mu^{2\epsilon}(2\pi)^{4(d-1)}N_c} \frac{(p_0^-)^2}{s^2 x_{h_1}^d x_{h_2}^d} \sum_q Q_q^2 \int_{x_{h_1}}^{1}   \frac{d x_g}{x_g} \int_{x_{h_2}}^{1}  \frac{d x_{\bar{q}}}{x_{\bar{q}}} \; (x_g x_{\bar{q}})^{d-1} D_g^{h_1} \left( \frac{x_{h_1}}{x_g} , \mu_F \right)  \\
    & \times D_{\bar{q}}^{h_2} \left( \frac{x_{h_2}}{x_{\bar{q}}}, \mu_F \right) \int_0^1 \frac{d x_q}{x_q } \int \frac{d^d p_{q\perp}}{(2\pi)^d} \delta (1-x_q-x_{\bar{q}}-x_g) \int d^d p_{1\perp} d^d p_{2\perp}  d^d p_{1'\perp} d^d p_{2'\perp}\\
    & \times  \mathbf{F}\left(\frac{p_{12\perp}}{2}\right)  \mathbf{F}^*\left(\frac{p_{1'2'\perp}}{2}\right) \delta \left(p_{q1\perp} + \frac{x_{\bar{q}}}{x_{h_2}}p_{h_2 \perp} - p_{2\perp} + \frac{x_{g}}{x_{h_1}} p_{h_1\perp}\right) \delta (p_{11'\perp} + p_{22'\perp}) \varepsilon_{T i}^* \frac{Q}{p_\gamma^+} \\
   & \times \left[ \Phi_3^{+} (p_{1\perp}, p_{2\perp}) \Phi_3^{ i *} (p_{1'\perp}, p_{2'\perp})  + \frac{4 x_q \left(p_\gamma^{+}\right)^3\left(2 x_{\bar{q}} -1\right)\left(x_g^2 d+4 x_q \left(x_q +x_g\right)\right) }{\left(Q^2+\frac{\left(\frac{x_{\bar{q}}}{x_{h_2}}\vec{p}_{h_2}- \vec{p}_{2}\right)^2}{x_{\bar{q}}\left(1-x_{\bar{q}}\right)}\right)\left(Q^2+\frac{\left(\frac{x_{\bar{q}}}{x_{h_2}}\vec{p}_{h_2}- \vec{p}_{2'}\right)^2}{x_{\bar{q}}\left(1-x_{\bar{q}}\right)}\right)} \right.\\
   & \left. \times  \frac{\left( \frac{x_{\bar{q}}}{x_{h_2}} p_{h_2 \perp}-p_{ 2' \perp}\right)^i}{\left(x_q+x_g\right) \left(x_q \frac{x_{g}}{x_{h_1}} \vec{p}_{h_1}  - x_g \vec{p}_q\right)^2}  \right]   + (h_1 \leftrightarrow h_2) \; . \numberthis \\
\end{align*} 
Finally, for $TT$ case, we obtain 
\begin{align*}
\allowdisplaybreaks
   & \frac{d {\sigma}_{3TT}^{g \bar{q} \rightarrow h_1 h_2}}{d x_{h_1} d x_{h_2} d^d p_{h_1} d^d p_{h_2}} \\ & =  \frac{2 \alpha_s \alpha_{\mathrm{em}} C_F }{\mu^{2 \epsilon}(2\pi)^{4(d-1)}N_c} \frac{(p_0^-)^2}{s^2 x_{h_1}^d x_{h_2}^d} \sum_q Q_q^2 \int_{x_{h_1}}^{1}  \frac{d x_g}{x_g}\int_{x_{h_2}}^{1}  \frac{d x_{\bar{q}} }{ x_{\bar{q}}} \; (x_g x_{\bar{q}})^{d-1} D_g^{h_1} \left( \frac{x_{h_1}}{x_g} , \mu_F \right)  \\
    & \times D_{\bar{q}}^{h_2} \left( \frac{x_{h_2}}{x_{\bar{q}}}, \mu_F \right) \int_0^1 \frac{d x_q }{x_q} \int \frac{ d^d p_{q\perp}}{ (2\pi)^d} \delta (1-x_q-x_{\bar{q}}-x_g) \int d^d p_{1\perp} d^d p_{2\perp} \int d^d p_{1'\perp} d^d p_{2'\perp}  \\
    & \times \mathbf{F}\left(\frac{p_{12\perp}}{2}\right)   \mathbf{F}^*\left(\frac{p_{1'2'\perp}}{2}\right)  \delta \left(p_{q1\perp} + \frac{x_{\bar{q}}}{x_{h_2}} p_{h_2}-p_{2\perp} +\frac{x_g}{x_{h_1}} p_{h_1 \perp} \right) \delta (p_{11'\perp} + p_{22'\perp} ) \varepsilon_{T i} \; \varepsilon_{T k}^*  \\
   &  \times \left[  \Phi_3^{i} (p_{1\perp}, p_{2\perp}) \Phi_3^{k*} (p_{1'\perp}, p_{2'\perp}) - \frac{2 x_q (p_\gamma^+)^2 (x_g^2 d + 4x_q (x_q + x_g))}{\left(Q^2+\frac{\left(\frac{x_{\bar{q}}}{x_{h_2}}\vec{p}_{h_2}- \vec{p}_{2}\right)^2}{x_{\bar{q}}\left(1-x_{\bar{q}}\right)}\right) \left(Q^2+\frac{\left(\frac{x_{\bar{q}}}{x_{h_2}}\vec{p}_{h_2}- \vec{p}_{2'}\right)^2}{x_{\bar{q}}\left(1-x_{\bar{q}}\right)}\right)  } \right. \\
   & \times \left. \frac{\left( (1-2x_{\bar{q}})^2 g_{\perp}^{ri}  g_{\perp}^{lk} - g_{\perp}^{li} g_{\perp}^{rk} + g_{\perp}^{rl} g_{\perp}^{ik}  \right)\left( \frac{x_{\bar{q}}}{x_{h_2}} p_{h_2 } - p_{2}\right)_r \left( \frac{x_{\bar{q}}}{x_{h_2}}p_{h_2 } - p_{2'} \right)_l}{x_{\bar{q}} (x_q + x_g)^2\left(x_q\frac{x_{g}}{x_{h_1}} \vec{p}_{h_1}  - x_g \vec{p}_q \right)^2}  \right] \\
   & + (h_1 \leftrightarrow h_2) \; . \numberthis
\end{align*} 
\newpage

\subsubsection{Dipole $\times$ double-dipole contribution and double-dipole $\times$ double-dipole contribution}

The dipole $\times$ double dipole contribution, for arbitrary polarization, is given by
\begin{align*}
\allowdisplaybreaks
& \frac{d\sigma_{4JI}^{g \bar{q} \rightarrow h_1 h_2}}{d x_{h_1} d x_{h_2} d^d p_{h_1 \perp} d^d p_{h_2 \perp}} \\  & = \frac{\alpha_{s}\alpha_{\mathrm{em}} }{\mu^{2 \epsilon} (2\pi)^{4(d-1)}N_{c}}\frac{(p_{0}^{-})^{2}}{s^{2}x_{h_1}^d x_{h_2}^d} \sum_q Q_{q}^{2} \int_{x_{h_1}}^1 \frac{d x_g}{x_g} \int_{x_{h_2}}^1 \frac{d x_{\bar{q}} }{x_{\bar{q}} }\; (x_g x_{\bar{q}})^{d-1} D_g^{h_1}\left(\frac{x_{h_1}}{x_g},\mu_F\right)\\ 
& \times D_{\bar{q}}^{h_2}\left(\frac{x_{h_2}}{x_{\bar{q}}}, \mu_F\right)  \int_0^1 \frac{d x_q }{x_q } \int \frac{d^{d} p_{q\bot}}{(2\pi)^{d}} \delta(1-x_q -x_{\bar{q}}-x_g) \int d^dp_{1\bot}d^dp_{2\bot} d^dp_{1\bot}^\prime d^dp_{2\bot}^\prime \\ 
& \times \frac{d^dp_{3\bot}d^dp_{3\bot}^\prime}{\left( 2\pi \right)^d}  \delta\left(p_{q1\bot}+\frac{x_{\bar{q}}}{x_{h_2}} p_{h_2 \perp}-p_{2\bot}+\frac{x_g}{x_{h_1}}p_{h_1 \perp}-p_{3\bot}\right) \delta(p_{11^\prime \bot}+p_{22^\prime \bot}+p_{33^\prime \bot})  \\
& \times (\varepsilon_{I\alpha} \varepsilon_{J\beta}^\ast) \left[ \Phi_3^\alpha(p_{1\bot},p_{2\bot}) \Phi_4^{\beta\ast}(p_{1\bot}^\prime, p_{2\bot}^\prime, p_{3\bot}^\prime) \mathbf{F}\left(\frac{p_{12\bot}}{2}\right) \mathbf{\tilde{F}}^\ast \left( \frac{p_{1^\prime 2^\prime \bot}}{2}, p_{3\bot}^\prime \right) \delta(p_{3\bot}) \right. \numberthis \\
& + \left. \Phi_4^\alpha (p_{1\bot}, p_{2\bot}, p_{3\bot}) \Phi_3^{\beta\ast} \left( p_{1^\prime \perp}, p_{ 2^\prime \bot} \right) \mathbf{\tilde{F}}\left(\frac{p_{12\bot}}{2}, p_{3\bot}\right) \mathbf{F}^\ast\left(\frac{p_{1^\prime 2^\prime\bot}}{2} \right) \delta(p_{3\bot}^\prime) \right]  + (h_1 \leftrightarrow h_2) \; .
\end{align*} 
The double-dipole $\times$ double-dipole contribution, for arbitrary polarization, is given by
\begin{align*}
\allowdisplaybreaks
& \frac{d\sigma_{5JI}^{g \bar{q} \rightarrow h_1 h_2}}{d x_{h_1} d x_{h_2} d^d p_{h_1} d^d p_{h_2}} \\ 
& = \frac{\alpha_{s} \alpha_{\mathrm{em}}}{\mu^{2\epsilon}(2\pi)^{4(d-1)} (N_{c}^{2}-1)}\frac{(p_{0}^{-})^{2}}{s^{2}x_{h_1}^d x_{h_2}^d} \sum_q Q_{q}^{2} \int_{x_{h_1}}^1  \frac{d x_g}{x_g} \int_{x_{h_2}}^{1}  \frac{d x_{\bar{q}} }{x_{\bar{q}}}\; (x_g x_{\bar{q}} )^{d-1} D_g^{h_1}\left(\frac{x_{h_1}}{x_q},\mu_F\right) \\
& \times D_{\bar{q}}^{h_2}\left(\frac{x_{h_2}}{x_{\bar{q}}}, \mu_F\right) \int_0^1 \frac{d x_q}{x_q } \int \frac{ d^{d}p_{q \bot}}{(2\pi)^{d}} \delta(1-x_q -x_{\bar{q}}-x_g) \int d^dp_{1\bot} d^dp_{2\bot} d^dp_{1\bot}^\prime d^dp_{2\bot}^\prime  \\
& \times  \int \frac{d^dp_{3\bot}d^dp_{3\bot}^\prime}{\left( 2\pi \right)^{2d}} \delta\left(p_{q1\bot}+\frac{x_{\bar{q}}}{x_{h_2}}p_{h_2 \perp}-p_{2\bot}+\frac{x_{g}}{x_{h_1}} p_{h_1 \perp}-p_{3\bot}\right) \delta(p_{11^\prime \bot}+p_{22^\prime \bot}+p_{33^\prime \bot}) \\
& \times  (\varepsilon_{I\alpha} \varepsilon_{J\beta}^*)  \Phi_4^\alpha(p_{1\bot},p_{2\bot},p_{3\bot}) \Phi_4^{\beta\ast}(p_{1\bot}^\prime,p_{2\bot}^\prime, p_{3\bot}^\prime) \mathbf{\tilde{F}}\left( \frac{p_{12\bot}}{2}, p_{3\bot} \right) \mathbf{\tilde{F}}^\ast \left(\frac{p_{1^\prime 2^\prime \bot}}{2},p_{3\bot}^\prime\right) \\
& + (h_1 \leftrightarrow h_2) \; .  \numberthis
\end{align*}  
Expression for the squared impact factors can be found in Appendix \ref{AppendixD2}. They are written in terms of $p_q, p_{\bar{q}}, p_{g}, z$ and the following identification should be done:
\begin{equation}
    p_{g \perp} = \frac{x_g}{x_{h_1}} p_{h_1 \perp} \; , \hspace{0.5cm}  p_{\bar{q} \perp} = \frac{x_{\bar{q}}}{x_{h_2}} p_{h_2 \perp} \; , \hspace{0.5 cm} z=x_g \; .
\end{equation}

\subsection{Real corrections: Fragmentation from quark and gluon}

\subsubsection{Finite part of dipole $\times$ dipole contribution}

When squaring the dipole contribution, one also gets finite terms. This time we write separately for each polarization transition. In the $LL$ case, we have
{\allowdisplaybreaks
\begin{align*}
   & \frac{d {\sigma}_{3LL}^{q g \rightarrow h_1 h_2}}{d x_{h_1} d x_{h_2} d^d p_{h_1} d^d p_{h_2}} \\ & =  \frac{2 \alpha_s \alpha_{\mathrm{em}} C_F }{\mu^{2 \epsilon}(2\pi)^{4(d-1)}N_c} \frac{(p_0^-)^2}{s^2 x_{h_1}^d x_{h_2}^d} \sum_q Q_q^2 \int_{x_{h_1}}^{1}   \frac{d x_q}{x_q} \int_{x_{h_2}}^{1}  \frac{d x_{g}}{x_g} \; (x_q x_{g})^{d-1} D_q^{h_1} \left( \frac{x_{h_1}}{x_q} , \mu_F \right) \\
    & \times D_{g}^{h_2} \left( \frac{x_{h_2}}{x_{\bar{q}}}, \mu_F \right)  \int_0^1 \frac{d x_{\bar{q}}  }{x_{\bar{q}}} \int \frac{d^d p_{\bar{q}\perp}}{ (2\pi)^d} \delta (1-x_q-x_{\bar{q}}-x_g) \int d^d p_{1\perp} d^d p_{2\perp} \int d^d p_{1'\perp} d^d p_{2'\perp}    \\
    & \times \mathbf{F}\left(\frac{p_{12\perp}}{2}\right) \mathbf{F}^*\left(\frac{p_{1'2'\perp}}{2}\right) \delta \left( \frac{x_q}{x_{h_1}} p_{h_1 \perp}-p_{1\perp} + p_{\bar{q}2\perp} + \frac{x_g}{x_{h_2}} p_{h_2 \perp}\right) \delta (p_{11'\perp} + p_{22'\perp}) \frac{Q^2}{(p_{\gamma}^+)^2} \\
   & \times \left[ \Phi_3^{+} (p_{1\perp}, p_{2\perp}) \Phi_3^{+*} (p_{1'\perp}, p_{2'\perp}) - \frac{8 x_q x_{\bar{q}} (p_\gamma^+)^4 \left( d x_g^2 + 4 x_{\bar{q}} (x_{\bar{q}} + x_g) \right)}{\left(Q^2 + \frac{\left(\frac{x_q}{x_{h_1}} \vec{p}_{h_1}-\vec{p}_{ 1}\right)^2}{x_q(1-x_{q})} \right) \left(Q^2 + \frac{\left( \frac{x_q}{x_{h_1}}\vec{p}_{h_1}-\vec{p}_{1'}\right)^2}{x_{q}(1-x_{q})} \right) } \right. \\
   & \times \left. \frac{1}{\left(x_{\bar{q}} \frac{x_g}{x_{h_2}} \vec{p}_{h_2}-x_g \vec{p}_{\bar{q}}\right)^2 }  \right]+ (h_1 \leftrightarrow h_2) \; . \numberthis
\end{align*}  }
For the $TL$ case, we have  %
{\allowdisplaybreaks
\begin{align*}
   & \frac{d {\sigma}_{3TL}^{q g \rightarrow h_1 h_2}}{d x_{h_1} d x_{h_2} d^d p_{h_1} d^d p_{h_2}} \\ & =  \frac{2 \alpha_s \alpha_{\mathrm{em}} C_F }{\mu^{2\epsilon}(2\pi)^{4(d-1)}N_c} \frac{(p_0^-)^2}{s^2 x_{h_1}^d x_{h_2}^d} \sum_q Q_q^2 \int_{x_{h_1}}^{1}   \frac{d x_q }{x_q} \int_{x_{h_2}}^{1}  \frac{d x_{g}}{x_g} \; (x_q x_g)^{d-1} D_q^{h_1} \left( \frac{x_{h_1}}{x_q} , \mu_F \right)  \\
    & \times D_{g}^{h_2} \left( \frac{x_{h_2}}{x_{g}}, \mu_F \right) \int_0^1 \frac{d x_{\bar{q}}  }{x_{\bar{q}} } \int \frac{d^d p_{\bar{q}\perp}}{(2\pi)^d}\delta (1-x_q-x_{\bar{q}}-x_g) \int d^d p_{1\perp} d^d p_{2\perp}  d^d p_{1'\perp} d^d p_{2'\perp}   \\
    & \times \mathbf{F}\left(\frac{p_{12\perp}}{2}\right) \mathbf{F}^*\left(\frac{p_{1'2'\perp}}{2}\right) \delta \left( \frac{x_q}{x_{h_1}} p_{h_1 \perp}-p_{1\perp} + p_{\bar{q}2\perp} + \frac{x_g}{x_{h_2}} p_{h_2 \perp}\right) \delta (p_{11'\perp} + p_{22'\perp}) \varepsilon_{T i}^* \frac{Q}{p_{\gamma}^+}  \\
   & \times \left[  \Phi_3^{+} (p_{1\perp}, p_{2\perp}) \Phi_3^{i*} (p_{1'\perp}, p_{2'\perp})  + \frac{4 x_{\bar{q}} \left(p_\gamma^{+}\right)^3\left(2 x_{q} -1\right)\left(x_g^2 d+4 x_{\bar{q}} \left(x_{\bar{q}} +x_g\right)\right) }{\left(Q^2+\frac{\left( \frac{x_q}{x_{h_1}}\vec{p}_{h_1}-\vec{p}_{ 1}\right)^2}{x_{q}\left(1-x_{q}\right)}\right)\left(Q^2+\frac{\left( \frac{x_q}{x_{h_1}}\vec{p}_{h_1}-\vec{p}_{ 1'}\right)^2}{x_{q}\left(1-x_{q}\right)}\right) }  \right.\\
   & \times \left. \frac{\left(\frac{x_q}{x_{h_1}}p_{h_1}- p_{ 1' }\right)^i}{\left(x_{\bar{q}}+x_g\right)\left(x_{\bar{q}} \frac{x_g}{x_{h_2}} \vec{p}_{h_2}- x_g \vec{p}_{\bar{q}}\right)^2} \right] + (h_1 \leftrightarrow h_2) \; . \numberthis
\end{align*}  }
Finally, for the $TT$ case, we obtain
{\allowdisplaybreaks
\begin{align*}
   & \frac{d {\sigma}_{3TT}^{q g \rightarrow h_1 h_2}}{d x_{h_1} d x_{h_2} d^d p_{h_1} d^d p_{h_2}} \\*
   & =  \frac{2 \alpha_s \alpha_{\mathrm{em}} C_F }{\mu^{2 \epsilon} (2\pi)^{4(d-1)}N_c} \frac{(p_0^-)^2}{s^2 x_{h_1}^d x_{h_2}^d} \sum_q Q_q^2 \int_{x_{h_1}}^{1}   \frac{d x_q}{x_q} \int_{x_{h_2}}^{1}  \frac{d x_{g}}{x_g} \; (x_q x_{g})^{d-1} D_q^{h_1} \left( \frac{x_{h_1}}{x_q} , \mu_F \right)  \\
    & \times D_{g}^{h_2} \left( \frac{x_{h_2}}{x_{g}}, \mu_F \right) \int_0^1 \frac{d x_{\bar{q}}  }{x_{\bar{q}} } \int \frac{d^d p_{\bar{q}\perp}}{(2\pi)^d} \delta (1-x_q-x_{\bar{q}}-x_g) \int d^d p_{1\perp} d^d p_{2\perp} \int d^d p_{1'\perp} d^d p_{2'\perp} \\
    & \times   \mathbf{F}\left(\frac{p_{12\perp}}{2}\right)  \mathbf{F}^*\left(\frac{p_{1'2'\perp}}{2}\right) \delta \left( \frac{x_q}{x_{h_1}} p_{h_1 \perp}-p_{1\perp} + p_{\bar{q}2\perp} + \frac{x_g}{x_{h_2}} p_{h_2 \perp}\right)  \delta (p_{11'\perp} + p_{22'\perp} ) \varepsilon_{T i} \; \varepsilon_{T k}^*  \\
   &\times \left[ \Phi_3^{i} (p_{1\perp}, p_{2\perp}) \Phi_3^{k*} (p_{1'\perp}, p_{2'\perp}) - \frac{2 x_{\bar{q}} (p_\gamma^+)^2 (x_g^2 d + 4x_{\bar{q}} (x_{\bar{q}} + x_g))  }{ \left(Q^2 + \frac{\left( \frac{x_q}{x_{h_1}} \vec{p}_{h_1}- \vec{p}_{1}\right)^2 }{x_{q} (1 - x_{q})} \right) \left(Q^2 + \frac{\left( \frac{x_q}{x_{h_1}} \vec{p}_{h_1}- \vec{p}_{1'}\right)^2}{x_{q} (1 - x_{q})} \right)  }  \right. \\
   & \left. \times \frac{\left((1-2 x_q)^2 g_\perp^{ir} g_\perp^{lk} -  g_\perp^{il} g_\perp^{rl} +  g_\perp^{ik} g_\perp^{lr}\right) \left( \frac{x_q}{x_{h_1}} p_{h_1} - p_{1}\right)_r \left( \frac{x_q}{x_{h_1}} p_{h_1} - p_{1'}\right)_l}{x_{q} (x_{\bar{q}} + x_g)^2 \left(x_{\bar{q}} \frac{x_g}{x_{h_2}} \vec{p}_{h_2}- x_g \vec{p}_{\bar{q}} \right)^2}\right] \\
   & + (h_1 \leftrightarrow h_2) \; . \numberthis
\end{align*}  }

\subsubsection{Dipole $\times$ double-dipole contribution and double-dipole $\times$ double-dipole contribution}

The dipole $\times$ double dipole contribution is given, for arbitrary polarization, by
{\allowdisplaybreaks
\begin{align*}
& \frac{d\sigma_{4JI}^{q \bar{q} \rightarrow h_1 h_2}}{d x_{h_1} d x_{h_2} d^d p_{h_1 \perp} d^d p_{h_2 \perp}} \\  & = \frac{\alpha_{s}\alpha_{\mathrm{em}} }{\mu^{2 \epsilon}(2\pi)^{4(d-1)}N_{c}}\frac{(p_{0}^{-})^{2}}{s^{2}x_{h_1}^d x_{h_2}^d} \sum_q Q_{q}^{2} \int_{x_{h_1}}^1  \frac{d x_q}{x_q}  \int_{x_{h_2}}^1  \frac{d x_g }{x_g} \; (x_q x_{g})^{d-1} D_q^{h_1}\left(\frac{x_{h_1}}{x_q},\mu_F\right) \\ 
& \times D_{g}^{h_2}\left(\frac{x_{h_2}}{x_g}, \mu_F\right) \int_0^1 \frac{d x_{\bar{q}} }{x_{\bar{q}}} \int \frac{d^{d} p_{ \bar{q} \bot}}{ (2\pi)^{d}} \delta(1-x_q -x_{\bar{q}}-x_g) \int d^dp_{1\bot}d^dp_{2\bot} d^dp_{1\bot}^\prime d^dp_{2\bot}^\prime  \\ 
& \times \frac{d^dp_{3\bot}d^dp_{3\bot}^\prime}{\left( 2\pi \right)^d} \delta\left(\frac{x_q}{x_{h_1}} p_{h_1 \perp}-p_{1\bot}+p_{\bar{q}2\bot}+ \frac{x_{g}}{x_{h_2}}p_{h_2 \perp}-p_{3\bot}\right) \delta(p_{11^\prime \bot}+p_{22^\prime \bot}+p_{33^\prime \bot})  \\
& \times (\varepsilon_{I\alpha} \varepsilon_{J\beta}^\ast) \left[ \Phi_3^\alpha(p_{1\bot},p_{2\bot}) \Phi_4^{\beta\ast}(p_{1\bot}^\prime, p_{2\bot}^\prime, p_{3\bot}^\prime) \mathbf{F}\left(\frac{p_{12\bot}}{2}\right) \mathbf{\tilde{F}}^\ast \left( \frac{p_{1^\prime 2^\prime \bot}}{2}, p_{3\bot}^\prime \right) \delta(p_{3\bot}) \right. \numberthis \\
& + \left. \Phi_4^\alpha (p_{1\bot}, p_{2\bot}, p_{3\bot}) \Phi_3^{\beta\ast} \left( p_{1'\perp}, p_{2'\perp}\right) \mathbf{\tilde{F}}\left(\frac{p_{12\bot}}{2}, p_{3\bot}\right) \mathbf{F}^\ast\left(\frac{p_{1^\prime 2^\prime\bot}}{2} \right) \delta(p_{3\bot}^\prime) \right]  + (h_1 \leftrightarrow h_2) \; .
\end{align*}  }
The double-dipole $\times$ double-dipole contribution is given, for arbitrary polarization, by
{\allowdisplaybreaks
\begin{align*}
& \frac{d\sigma_{5JI}^{q \bar{q} \rightarrow h_1 h_2}}{d x_{h_1} d x_{h_2} d^d p_{h_1} d^d p_{h_2}} \\* 
& = \frac{\alpha_{s} \alpha_{\mathrm{em}}}{\mu^{2\epsilon}(2\pi)^{4(d-1)} (N_{c}^{2}-1)}\frac{(p_{0}^{-})^{2}}{s^{2}x_{h_1}^d x_{h_2}^d} \sum_q Q_{q}^{2} \int_{x_{h_1}}^1 \frac{d x_q } {x_q} \int_{x_{h_2}}^1  \frac{d x_{g}}{x_g} \; (x_q x_{g})^{d-1} D_q^{h_1}\left(\frac{x_{h_1}}{x_q},\mu_F\right) \\
& \times D_{g}^{h_2}\left(\frac{x_{h_2}}{x_{g}}, \mu_F\right) \int_0^1 \frac{d x_{\bar{q}} }{x_{\bar{q}} } \int \frac{d^{d}p_{ \bar{q} \bot}}{(2\pi)^{d}} \delta(1-x_q -x_{\bar{q}}-x_g) \int d^dp_{1\bot}d^dp_{2\bot} d^dp_{1\bot}^\prime d^dp_{2\bot}^\prime \\
& \times  \int \frac{d^dp_{3\bot}d^dp_{3\bot}^\prime}{\left( 2\pi \right)^{2d}}  \delta \left( \frac{x_q}{x_{h_1}} p_{h_1 \perp}-p_{1\bot}+p_{\bar{q}2\bot}+ \frac{x_g}{x_{h_2}} p_{h_2 \perp}-p_{3\bot}\right) \delta(p_{11^\prime \bot}+p_{22^\prime \bot}+p_{33^\prime \bot})  \\
& \times  (\varepsilon_{I\alpha} \varepsilon_{J\beta}^*)  \Phi_4^\alpha(p_{1\bot},p_{2\bot},p_{3\bot}) \Phi_4^{\beta\ast}(p_{1\bot}^\prime,p_{2\bot}^\prime, p_{3\bot}^\prime) \mathbf{\tilde{F}}\left( \frac{p_{12\bot}}{2}, p_{3\bot} \right) \mathbf{\tilde{F}}^\ast \left(\frac{p_{1^\prime 2^\prime \bot}}{2},p_{3\bot}^\prime\right) \\ 
& + (h_1 \leftrightarrow h_2) \numberthis \; .
\end{align*} }
Expression for the squared impact factors can be found in Appendix \ref{AppendixD2}. They are written in terms of $p_q, p_{\bar{q}}, p_{g}, z$ and the following identification should be done:
\begin{equation}
    p_{q \perp} = \frac{x_q}{x_{h_1}} p_{h_1 \perp} \; , \hspace{0.5cm}  p_{g \perp} = \frac{x_g}{x_{h_2}} p_{h_2 \perp} \; , \hspace{0.5 cm} z=x_g \; .
\end{equation}
\section{Summary and outlook}
We have considered the diffractive production of a pair of hadrons at large $p_T$, in $\gamma^{(*)}$ nucleon/nucleus scattering, at NLO, in the most general kinematics in the eikonal approximation. \\

This new class of processes provides an access to precision physics of gluon saturation dynamics, with very promising future phenomenological studies both at the LHC in UPC (in photoproduction) and at the future EIC (both in photoproduction and leptoproduction). Our main result is the explicit finite result for the cross section at NLO, obtained after showing explicitly the cancellation of rapidity divergences (through the B-JIMWLK equation), soft divergences and collinear divergences between real, virtual contributions, and  DGLAP evolution equation governing fragmentation functions.
Finite contributions and purely divergent contributions have been separated and the sum of the latter has been shown to be zero. Hence, the collection of all terms labeled with ``fin" in sections~\ref{sec:CounterTerms}, \ref{sec: VirtualDiv}, \ref{sec: RealDiv}, plus all the formulas in section~\ref{sec:AdditionalFin} give the final result. The natural continuation of this work is to evaluate the cross sections numerically, employing a model for the description of the target, such as the McLerran Venugopalan (MV)~\cite{McLerran:1993ni,McLerran:1993ka,McLerran:1994vd} or the Golec-Biernat Wusthoff (GBW)~\cite{GolecBiernat:1998js} ones. \\

This NLO result adds a new piece in the list of processes which are very promising to probe gluonic saturation in nucleons and nuclei at NLO, including inclusive DIS~\cite{Beuf:2021srj,Beuf:2022ndu}, photon-dijet production in DIS~\cite{Roy:2019hwr}, dijets in DIS~\cite{Caucal:2021ent,Caucal:2022ulg}, single hadron~\cite{Bergabo:2022zhe}  and dihadrons production in DIS~\cite{Bergabo:2022tcu,Iancu:2022gpw}, diffractive exclusive dijets~\cite{Boussarie:2016ogo,Boussarie:2014lxa,Boussarie:2019ero} and exclusive light meson production~\cite{Boussarie:2016bkq,Mantysaari:2022bsp}, exclusive quarkonium production~\cite{Mantysaari:2021ryb,Mantysaari:2022kdm}, and inclusive DDIS~\cite{Beuf:2022kyp}. \\

\chapter{Final thoughts and outlook}
\begin{flushright}
\emph{\textit{Deep questions answered, deeper questions posed. \\ Sir. Roger Penrose~\cite{Penrose:2005re}}}
\end{flushright}

In the present work, we have presented a series of results aimed at reaching an era of precision in the high-energy factorization (HEF) framework, both in the linear and non-linear evolution regimes. To this aim we provided:
\begin{itemize}
    \item The next-to-leading order correction to the impact factor (vertex) for the production of a forward Higgs boson, in the infinite top-mass limit. We obtained the result both in the momentum representation and as superposition of the eigenfunctions of the leading-order BFKL kernel. As already mentioned earlier, this impact factor allows to describe the inclusive hadroproduction of a forward Higgs in the limit of small Bjorken $x$, as well as the more interesting case of the inclusive forward emissions of a Higgs boson in association with a backward identified object.
    \item Predictions for a number of partially inclusive processes featuring a forward-plus-backward two-particle final-state configuration. They provide stringent tests for BFKL dynamics at the LHC. 
    \item The computation of the Lipatov vertex in QCD within accuracy $\epsilon^2$, which can be used for several purposes. The most important one is the computation of one of the contributions to the BFKL kernel in the NNLLA. Other include the proof of gluon Reggeization and the calculation of discontinuites of multiple production amplitudes in the MRK (which must be taken into account in the NNLLA).
    \item The next-to-leading order cross sections for the diffractive production of a pair of hadrons at large $p_T$, in $\gamma^{(*)}$ nucleon/nucleus scattering, at NLO, in the most general kinematics. This new class of processes gives access to precision physics of gluon saturation dynamics, with very promising phenomenological studies both at the LHC in UPC (in photoproduction) and at the future EIC (both in photoproduction and leptoproduction). \vspace{0.3 cm} 
\end{itemize}

Since the pioneering works of Balitsky, Fadin, Kuraev and Lipatov, the BFKL approach (along with its extensions) and more in general the high-energy factorization framework have proven to be among the most powerful tools for understanding several aspects of QCD and more in general of QFTs, an incomplete list of them includes: 
\begin{itemize}
    \item The proton structure at small Bjorken-$x$ and various class of processes featuring a forward-plus-backward two-particle final-state configuration~\cite{Mueller:1986ey}. 
    \item The asymptotic behaviour of partonic scattering amplitudes in QCD and in $\mathcal{N}=4$ Super Yang-Mills theories~\cite{Kotikov:2000pm,Kotikov:2002ab,Fadin:2007xy,Fadin:2009gh,Balitsky:2009yp}. 
    \item The connection between high-energy scattering and exactly solvable models~\cite{Lipatov:1993yb,Faddeev:1994zg}.
    \item The duality between the Pomeron in the maximally extended $\mathcal{N} = 4$ super symmetry and the Reggeized graviton in the 10-dimensional anti-de-Sitter space~\cite{Kotikov:2013xu}.
\end{itemize}
Regardless of the sheer elegance of its theoretical formulation, as shown in Chapter \ref{Chap:Pheno}, the high-energy factorization is an important tool for achieving accuracy at modern accelerators such as the EIC and the LHC. The impact of high-energy resummation is expected to grow in the future and, as already shown in recent studies~\cite{Bonvini:2018ixe,POWHAD_Moriond_QCD_2023,Celiberto:2023rtu}, important production channels at the new generation of colliders such as the FCC, will receive such large contributions that they will make the physics of those energies the BFKL physics. \\

In order to continue what started in this thesis and more generally by the scientific community of the sector, there are many developments that, in our opinion, it is necessary to consider in the near future.

\subsection*{Combined resummations and central productions}

In Chapter \ref{Chap:Pheno}, and especially when we discussed the $p_T$-distribution of the Higgs, we pointed out how different resummation mechanisms come into play in the description of this observable. It is important to note that while developing calculations at the frontier in high-energy resummation does not guarantee the absence of residual large logarithmic corrections, of a different nature, that can spoil the stability of the results and must therefore be resummed in a closed form.
For example, the jet impact factor at NLO (projected onto the LO eigenfunctions of the BFKL kernel) has contributions of the type
\begin{equation*}
    \int_{\alpha}^1 \frac{d \zeta }{\zeta} f_{q} \left( \frac{\alpha}{\zeta} \right) \left( \frac{\ln(1-\zeta)}{1-\zeta} \right)_{+} \; ,
\end{equation*}
where $f_{q}$ is the quark PDF. These contributions are not divergent due to the $+$-prescription structure, but contain logarithmically large corrections at the threshold $\alpha \rightarrow 1$. Moreover, because of the $+$-prescription structure, these logarithms will multiply the derivative of the PDF (which is not small for $\alpha \rightarrow 1$). Similar contributions can also be found in the NLO hadron impact factor. Another example is represented by the forward Higgs boson cross section, at small transverse momentum. Here, we expect also Sudakov-type logarithmically large contributions. Perhaps, it would be extremely interesting to supplement the calculations with a resummation of such enhanced corrections in a Sudakov-like manner~\cite{Xiao:2018esv}. One of the great challenges for the future is precisely the construction of versatile formalisms that can accommodate different types of resummations. \\

In the thesis, we have extensively used the concept of \textit{hybrid high-energy/collinear factorization}. It is important to stress that the present ansatz of factorization does not provide a complete interpolation between collinear and high-energy resummations. The underlying idea of this factorization is that even if collinear information is inserted for a correct treatment of the initial and/or final state IR-singularities, the dominant dynamic is that of BFKL-type logarithms. In this way, collinear dynamics is considered in the impact factors only. However, it is well known that collinearly enhanced NLO corrections to the BFKL kernel can be sizable. One of the approaches to improve this lack is the BLM procedure for the scale fixing~\cite{Brodsky:1996sg,Brodsky:1997sd,Brodsky:1998kn,Brodsky:2002ka}, that leads to a suppression of the NLO corrections. An alternative is an all order resummation of the collinearly enhanced terms, also called \textit{collinear improvement}~\cite{Salam:1998tj,Ciafaloni:1999au,Ciafaloni:1999yw,Ciafaloni:2003rd,SabioVera:2005tiv}. In the future, it would be interesting to consider the impact of collinear improvement on some of the observables seen in the thesis~\cite{Celiberto:2022rfj,Caporale:2007vs}. \\

Lastly, within the high-energy factorization, remarkable results have been produced in the context of forward and forward/backward productions. However, as highlighted in~\cite{Bonvini:2018ixe}, at the next generation of accelerators, the high energy corrections will be large also in the central productions, making also the calculation of impact factors in the central region of rapidity very important. In this context, to date a single next-to-leading calculation has been performed~\cite{Bartels:2006hg} (the central jet). The calculation of these amplitudes requires evaluating $2 \rightarrow 3$ amplitudes, such as the one used to extract the Lipatov vertex.  

\subsection*{Hypothesis on the remainder function of the BDS ansatz}

In chapter \ref{Chap:Lipatov} we computed the Lipatov vertex in QCD with the required $\epsilon$-accuracy to construct one of the contribution to the BFKL kernel in the next-to-next-to leading logarithmic approximation. The result obtained is suitable for the purpose, since to obtain the part of the kernel containing the product of two one-loop RRG vertices, the integration over the phase space is that over a single particle and therefore completely trivial. This result is furthermore supplemented by the knowledge of the vertex at arbitrary $\epsilon$ in the soft limit of the emited particle. \\

However, we briefly mentioned at the beginning of the chapter that the discontinuities of multiple gluon production amplitudes in the MRK are interesting also from another point of view. They can be used~\cite{Fadin:2014yxa} for a simple demonstration of  violation of the ABDK-BDS (Anastasiou-Bern-Dixon-Kosower --- Bern-Dixon-Smirnov) ansatz~\cite{Anastasiou:2003kj,Bern:2005iz} for amplitudes  with maximal helicity violation (MHV) in Yang-Mills theories with maximal supersymmetry ($\mathcal{N}=4$ SYM) in the planar limit and for the calculations of the remainder functions to this ansatz. There are two hypothesis about the remainder functions: the hypothesis of dual conformal invariance~\cite{Bern:2006ew,Nguyen:2007ya}, which asserts that the MHV amplitudes are given by products of the BDS amplitudes and that the remainder functions depend only on anharmonic ratios of kinematic invariants, and the hypothesis of scattering amplitude/Wilson loop correspondence~\cite{Drummond:2007aua,Drummond:2007cf,Brandhuber:2007yx,Drummond:2008aq}, according to which the remainder functions are given by expectation values of Wilson loops. Both these hypotheses are not proved. They can be tested by comparison of the BFKL discontinuities with the discontinuities calculated with the use of these hypothesis~\cite{Lipatov:2010qg,Fadin:2011we}. \\

To this aim, impact factors for the Reggeon-gluon transition, which are the natural generalization of the impact factors for the particle-particle transitions, entering the discontinuities of elastic amplitudes, must be constructed with higher $\epsilon$-accuracy. Again, the Lipatov vertex (in $\mathcal{N}=4$ SYM) will be part of these impact factors. Also in $\mathcal{N}=4$ SYM, the pentagon in $6+2\epsilon$ contributes to the vertex~\cite{DelDuca:2009ae}, which implies that the great complications that appear in QCD are also present in this theory. For the computation of the remainder functions non-trivial convolutions of impact factors are required, which impliy that a form as simple as possible of the vertex is required. Attacking this calculation, in our opinion, requires finding a more manageable expression of the second contribution in Eq.~(\ref{IterativeRelI5}), which is one of our plans for the future. 

\subsection*{BFKL vs B-JIMWLK and NNLL resummation}
In this thesis, we presented works aimed at testing both BFKL and saturation dynamics. As explained in Chapter 5, the B-JIMWLK hierarchy of equations contains different aspects of small-$x$ physics compared to the BFKL one. One of these is that saturation effects in large nuclei are emphasized, as can be understood from Eq.~(\ref{Int:Eq:SaturationScale}). Here, the atomic number dependence is due to presence of multiple re-scatterings in a large enough nucleus. This leads to an additional resummation parameters $\alpha_s^2 A^{1/3}$. The physical meaning of the parameter $\alpha_s^2 A^{1/3}$ is rather straightforward: at a given impact parameter the dipole interacts with $A^{1/3}$ nucleons, exchanging two gluons with each (two-gluon exchange is parametrically of order $\alpha_s^2$). However, in small nuclei or a proton, the BFKL dynamics is expected to precede the onset of saturation physics. It is the opinion of the author and collaborators that considering observables in which predictions can be built both in the linear regime (à la BFKL) and in the effective Shockwave/CGC approaches, perhaps in a full NLLA, could help to find the point at which saturation effects modify the shape of BFKL predictions.  \\

Last but not least, it is interesting to point out a different connection between the BFKL and the Shockwave approach. To date, after more than twenty-five years since the NLO corrections to BFKL were obtained, as explained in section \ref{BFKL beyond NLLA}, the problem of NNLO corrections remains opened and challenging. From a conceptual point of view, the main reason is that the pole Regge form of QCD amplitudes is violated in the NNLLA, as observed by direct computation at two-and three-loop. There exist two approaches for the explanation of this violation:
\begin{itemize}
    \item The first has been developed in~\cite{Fadin:2016wso,Fadin:2017nka} and employs the contribution of the cut alone to explain this phenomenon, considering triple Reggeon exchange.
    \item The second is an effective approach based on the Shockwave formalism (as formulated in~\cite{Caron-Huot:2013fea}). In this approach, the phenomenon of gluon Reggeization, that we have discussed in Chapter~1 of the thesis essentially comes out from the fundamental ingredients of the Shockwave formalism: rapidity factorization and eikonal Wilson lines. It is revealed by expanding the Wilson lines close to the identity. In particular, the logarithm of a Wilson line is used as a gauge-invariant interpolating operator for a Reggeized gluon. In this context, the above violation is interpreted in terms of triple Reggeon exchange (Regge cut) and, in addition, its mixing with a single Reggeon~\cite{Caron-Huot:2017fxr}.  
\end{itemize}
Even not being a formal proof, the validity of these approaches may be confirmed or questioned in the four loop approximation~\cite{Fadin:2020hwo}, upon comparing their results with the results of four-loop calculations performed by the infrared-factorization method. In any case, the assertion that the QCD amplitudes with gluon quantum numbers in cross-channels and negative signature are given in the NNLLA, to all orders in perturbation theories, by the contributions of the Regge pole and the three-Reggeon cut (with or without mixing) is still only an hypothesis. It is our personal opinion that this problem undoubtedly represents one of the most fascinating aspects in the study of scattering amplitudes in non-Abelian gauge theories.

\begin{appendices}
\chapter{Further details on the Higgs impact factor}
For consistency with individual papers, definitions of some special functions are repeated in different appendices.
\label{AppendixA}
\section{Feynman rules of the gluon-Higgs effective field theory and definitions}
\label{AppendixA1}

  \begin{figure}
  \begin{center}
  \includegraphics[scale=0.65]{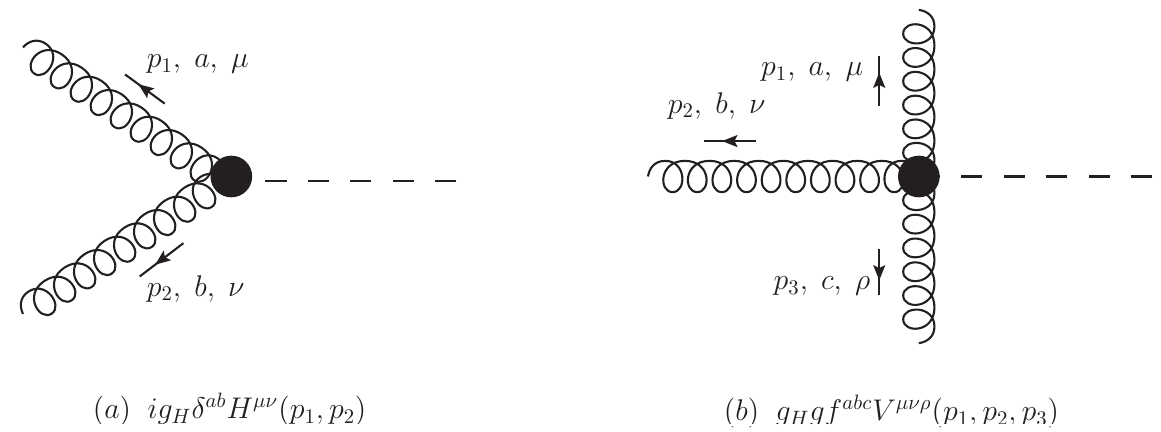}
  \end{center}
  \caption{Feynman rule for the (a) $ggH$ vertex and (b) $gggH$ vertex.}
  \label{HiggsDiagrams}
  \end{figure}

In this Appendix we give more details about the gluon-Higgs effective theory. The Feynman rules associated to the Lagrangian~(\ref{EffLagrangia}) and used in this work are shown in Fig.~\ref{HiggsDiagrams}. The tensor structures appearing in Fig.~\ref{HiggsDiagrams} are
\begin{equation}
    H^{\mu \nu} (p_1,p_2) = g^{\mu \nu} (p_1 \cdot p_2) - p_1^{\nu} p_2^{\mu} \;,
\label{ggH}
\end{equation}
\begin{equation}
    V^{\mu \nu \rho} (p_1, p_2, p_3) = (p_1 - p_2)^{\rho} g^{\mu \nu} + (p_2 - p_3)^{\mu} g^{\nu \rho} + (p_3 - p_1)^{\nu} g^{\rho \mu} \;.
\label{gggH}
\end{equation}
It is important to note that, as in QCD, in using Feynman rules for this theory, symmetry factors must be taken into account correctly. In particular, the first and second diagram in the Fig.~\ref{VirtualDiagrams} require a symmetry factor $S=1/2$. \\
We also define here some useful functions used in the main text:
\begin{equation}
    \psi (z) = \frac{\Gamma'(z)}{\Gamma(z)} \; , \hspace{0.5cm} \psi' (z) = \frac{d}{dz} \psi (z) \; ,\hspace{0.5 cm} H_{n} = \gamma_E + \psi (n+1) \; ,
\end{equation}
\begin{equation}
    {\rm{Li}}_2 (z) = - \int_0^z \frac{\ln (1-t)}{t} dt = - \int_0^1 \frac{\ln (1-zt)}{t} dt \; ,
\end{equation}
\begin{equation}
\, {_2}F_1(a,b,c,z) = \frac{1}{\mathcal{B}(b,c-b)} \int_0^1 dx \, x^{b-1} (1-x)^{c-b-1} (1-zx)^{-a} \;  \hspace{0.2 cm} \rm{for} \; \; \; \Re e{ \{c \} } > \Re e{ \{b \} } > 0 \; ,
\label{hypergeometric}
\end{equation}
where $\mathcal{B}$ is the Euler beta function and $z$ is not a real number such that it is greater than or equal to 1. 
It is useful to show the behaviour for $z \rightarrow 1^{-}$ of the hypergeometric function:
\begin{itemize}
    \item If $\Re e{ \{c \} } > \Re e{ \{ a+b \}}$, then
    \begin{equation}
    {_2}F_1(a,b,c,1) = \frac{\Gamma (c) \Gamma (c-a-b)}{\Gamma (c-a) \Gamma (c-b)} \; .
    \end{equation}
    \item If $\Re e{ \{c \} } = \Re e{ \{ a+b \}}$, then
    \begin{equation}
        \lim_{z \rightarrow 1^{-}} \frac{{_2}F_1(a,b,a+b,z)}{- \ln (1-z)} = \frac{\Gamma (a+b)}{\Gamma(a) \Gamma(b)} \hspace{0.6 cm} {\rm{for}} \; \;  c=a+b \; ,
    \end{equation}
    and
    \begin{equation}
        \lim_{z \rightarrow 1^{-}} (1-z)^{a+b-c} \left( {_2}F_1(a,b,c,z) - \frac{\Gamma (c) \Gamma (c-a-b)}{\Gamma (c-a) \Gamma (c-b)} \right) =  \frac{\Gamma (c) \Gamma (a+b-c)}{\Gamma (a) \Gamma (b)} \hspace{0.6 cm}
    \end{equation}
    for $c \neq a+b$.
    \item If $\Re e{ \{c \} } < \Re e{ \{ a+b \}}$, then
    \begin{equation}
        \lim_{z \rightarrow 1^{-}} \frac{{_2}F_1(a,b,c,z)}{(1-z)^{c-a-b}} = \frac{\Gamma (c) \Gamma (a+b-c)}{\Gamma (a) \Gamma (b)} \; .
        \label{LimitHyper}
    \end{equation}
\end{itemize}

\section{Integrals for the virtual corrections}
\label{AppendixA2}
We give here definition and result of Feynman integrals that appear in the calculation of the virtual corrections. The integrals $B_0$ and $C_0$ can be found also in the Appendix~A of Ref.~\cite{Fadin:2000yp}, the integral $D_0$ in Ref.~\cite{Bern:1993kr} (see also
Ref.~\cite{Zanderighi:2008on}).\footnote{In our notation the subscript $0$ means that all propagators appearing in the integral are massless. The arguments in the definitions of $B_0$ and $C_0$ integrals below represent the squared of the external momenta on which they depend. $D_0$, in addition to the four squared of the external momenta, depends also on the two independent Mandelstam variables typical of a $2 \rightarrow 2$ process.}
Concerning integrals with two denominators, we have
\begin{equation}
    B_0(-\vec{q}^{\; 2}) = \int \frac{d^D k}{i(2 \pi)^D} \frac{1}{k^2 (k+q)^2} = - \frac{1}{(4 \pi)^{2-\epsilon}} \frac{\Gamma
   (1+\epsilon) \Gamma^2 (-\epsilon)}{2 (1-2 \epsilon ) \Gamma (-2 \epsilon )} (\vec{q}^{\; 2})^{-\epsilon } \; ,
\end{equation}
\begin{equation}
    B_0(m_H^2) = \int \frac{d^D k}{i(2 \pi)^D} \frac{1}{k^2 (k+p_H)^2} = - \frac{1}{(4 \pi)^{2-\epsilon}} \frac{\Gamma
   (1+\epsilon) \Gamma^2 (-\epsilon)}{2 (1-2 \epsilon ) \Gamma (-2 \epsilon )} (-m_H^2)^{-\epsilon } \; .
\end{equation}
Independent integrals with three denominators appearing in the computation are
\begin{equation*}
   C_0(m_H^2,0, -\vec{q}^{\; 2}) = C_0(0, -\vec{q}^{\; 2}, m_H^2) = \int \frac{d^D k}{i(2 \pi)^D} \frac{1}{k^2 (k+q)^2 (k+p_H)^2} 
\end{equation*}
\begin{equation}
    = \frac{1}{(4 \pi)^{2-\epsilon}} \frac{ \Gamma (1+\epsilon) \Gamma^2(-\epsilon)}{2 \Gamma (-2 \epsilon )} \frac{1}{\epsilon} \frac{\left((\vec{q}^{\;2})^{-\epsilon }-\left(-m_H^2\right)^{-\epsilon }\right)}{m_H^2 + \vec{q}^{\; 2}} \; , 
\end{equation}
\begin{equation}
   C_0(0,0, -\vec{q}^{\; 2}) = \int \frac{d^D k}{i(2 \pi)^D} \frac{1}{k^2 (k+q)^2 (k+k_2)^2} = \frac{1}{(4 \pi)^{2-\epsilon}} \frac{ \Gamma (1+\epsilon) \Gamma^2(-\epsilon)}{2 \Gamma (-2 \epsilon )} \frac{1}{\epsilon} (\vec{q}^{\;2})^{-\epsilon-1} \; , 
\end{equation}
\begin{equation*}
   C_0(m_H^2,0,s) = \int \frac{d^D k}{i(2 \pi)^D} \frac{1}{k^2 (k+p_H)^2 (k+k_1+k_2)^2}
\end{equation*}
\begin{equation}
    \simeq -\frac{1}{(4 \pi)^{2-\epsilon}} \frac{ \Gamma (1+\epsilon) \Gamma^2(-\epsilon)}{2 \Gamma (-2 \epsilon )} \frac{1}{\epsilon} \frac{\left((-s)^{-\epsilon }-\left(-m_H^2\right)^{-\epsilon }\right)}{s} \; ,
\end{equation}
\begin{equation}
   C_0(0,0,s) = \int \frac{d^D k}{i(2 \pi)^D} \frac{1}{k^2 (k-k_1)^2 (k+k_2)^2}  = \frac{1}{(4 \pi)^{2-\epsilon}} \frac{ \Gamma (1+\epsilon) \Gamma^2(-\epsilon)}{2 \Gamma (-2 \epsilon )} \frac{1}{\epsilon} (-s)^{-\epsilon-1} \; , 
\end{equation}
There is only one relevant four-denominator integral,
\begin{equation*}
   D_0(m_H^2,0,0,0;-\vec{q}^{\; 2},s) = \int \frac{d^D k}{i(2 \pi)^D} \frac{1}{k^2 (k+p_H)^2 (k+q)^2 (k-k_2')^2} 
\end{equation*}
\begin{equation}   
= \frac{\Gamma (1+\epsilon) \Gamma^2(-\epsilon)}{(4 \pi)^{2-\epsilon} \vec{q}^{\; 2} s} 
\frac{1}{\epsilon \Gamma (-2 \epsilon )} \left[(\vec{q}^{\; 2})^{-\epsilon }+(-s)^{-\epsilon}-\left(-m_H^2\right)^{-\epsilon}
\right.\end{equation}
\begin{equation*}
\left. -\epsilon^2 \left(\text{Li}_2\left(1+\frac{m_H^2}{\vec{q}^{\; 2}}\right)+\frac{1}{2} \ln
   ^2\left(-\frac{\vec{q}^{\; 2}}{s}\right)+\frac{\pi ^2}{3}\right) \right]
   + {\cal O}(\epsilon) \; ,
\end{equation*}
where $k_2'=k_2-q$ and $(p_H+k_2')^2=(k_1+k_2)^2=s$.
The definition of the integrals $C_0(m_H^2,0,-s)$, $C_0(0,0,-s)$, $D_0(m_H^2,0,0,0;-\vec{q}^{\; 2},-s)$ can be obtained by exchanging $k_2$ and $-k_2'$; the result for these integrals is obviously obtained through the $s \rightarrow -s$ transformation. We observe that what said above is valid in the Regge limit; it allows us to neglect $m_H^2$ and $\vec{q}^{\; 2}$ when these appear summed to $s$.

\section{Master Integrals for the projection}
\label{AppendixA3}
For the projection onto the eigenfunctions of the BFKL kernel we just need the following integrals:
\[
I_1(\gamma_1, \gamma_2, n, \nu) = \int \frac{d^{2-2 \epsilon} \vec{q}}{\pi \sqrt{2}} (\vec{q}^{\; 2})^{i \nu - \frac{3}{2}} e^{i n \phi} (\vec{q}^{\; 2})^{-\gamma_1} \left[ \left( \vec{q} - \vec{p}_H \right)^2 \right]^{-\gamma_2} \;,
\]    
\[
I_2(\gamma_1, n, \nu) = \int \frac{d^{2-2 \epsilon} \vec{q}}{\pi \sqrt{2}} (\vec{q}^{\; 2})^{i \nu - \frac{3}{2}} e^{i n \phi} (\vec{q}^{\; 2})^{-\gamma_1} \frac{1}{ \left[ (\vec{q}-\vec{p}_H)^2 \right] \left[ (1-z_H) m_H^2 + \left( \vec{p}_H - z_H \vec{q} \right)^2 \right]} \; , 
\]
\[
I_3(\gamma_1, \gamma_2, n, \nu) = \int \frac{d^{2-2 \epsilon} \vec{q}}{\pi \sqrt{2}} (\vec{q}^{\; 2})^{i \nu - \frac{3}{2}} e^{i n \phi} (\vec{q}^{\; 2})^{-\gamma_1} \left[(1-z_H) m_H^2 + \left( \vec{p}_H - z_H \vec{q} \right)^2 \right]^{-\gamma_2}\;.    
\]
We note that
\begin{equation}
    \lim_{z_H \rightarrow 1} I_2 (\gamma_1, n, \nu) = I_1 (\gamma_1, 2, n, \nu) \; ,
\label{First Limit}
\end{equation}
\begin{equation}
    \lim_{z_H \rightarrow 1} I_3 (\gamma_1, \gamma_2, n, \nu) = I_1 (\gamma_1, \gamma_2, n, \nu) \; .
\label{Second Limit}
\end{equation}

We show here the explicit calculation of the most complicated one, $I_2$ and just quote the result for the other two. First, we introduce the vector $l \equiv (1, i)$ and write
\begin{equation}
e^{in\phi}=\left(\cos{\phi}+i\sin{\phi}\right)^n=\left(\frac{q_x+i q_y}{|\vec{q}\;|}\right)^n=\frac{(\vec{q}\cdot\vec{l}\;)^n}{\left(\vec{q}^{\;2}\right)^{\frac{n}{2}}} \; .
\label{ExpTrick}
\end{equation}
Then, after Feynman parameterization and the shift
\begin{equation}
    \vec{q} \rightarrow \vec{q} + \left( x + \frac{y}{z_H} \right) \vec{p}_H
\end{equation}
one finds
\begin{equation*}
    I_2(\gamma_1, n, \nu) = \frac{1}{z_H^2} \frac{\Gamma \left(\frac{7}{2} + \gamma_1 + \frac{n}{2} -i \nu \right)}{\Gamma \left(\frac{3}{2} + \gamma_1 + \frac{n}{2} -i \nu \right)} \int \frac{d^{2-2 \epsilon} \vec{q}}{\pi \sqrt{2}} \int_0^1 dx \int_0^{1-x} dy (1-x-y)^{\frac{1}{2} + \gamma_1 + \frac{n}{2} -i \nu} 
\end{equation*}
\begin{equation}
    \frac{ \left[ \vec{q} \cdot \vec{l} + \left( x + \frac{y}{z_H} \right) \vec{p}_H \cdot \vec{l} \; \right]^n}{\left[ \vec{q}^{\; 2} - \left( x + \frac{y}{z_H} \right)^2 \vec{p}_H^{\; 2} + x \vec{p}_H^{\; 2} + y \left( \Tilde{m}_H^2 + \frac{\vec{p}_H^{\; 2}}{z_H^2} \right) \right]^{\frac{7}{2} + \gamma_1 + \frac{n}{2} -i \nu}} \; ,
\end{equation}
where $\Tilde{m}_H^2 = \frac{(1-z_H)m_H^2}{z_H^2}$. Now, we expand the binomial in the numerator as
\begin{equation}
\left[ \vec{q} \cdot \vec{l} + \left( x + \frac{y}{z_H} \right) \vec{p}_H \cdot \vec{l} \; \right]^n = \sum_{j=0}^{n} \binom{n}{j}  (\vec{q} \cdot \vec{l})^{j} \left( x + \frac{y}{z_H} \right)^{n-j}  (\vec{p}_H \cdot \vec{l})^{n-j} 
\end{equation}
and observe that only the term with $j=0$ survives. At this point, we are able to obtain the simple form,
\[
    I_2(\gamma_1, n, \nu) = \frac{(\vec{p}_H^{\; 2})^{\frac{n}{2}} e^{i n \phi_H}}{z_H^2} \frac{\Gamma \left(\frac{7}{2} + \gamma_1 + \frac{n}{2} -i \nu \right)}{\Gamma \left(\frac{3}{2} + \gamma_1 + \frac{n}{2} -i \nu \right)} 
\]
\begin{equation}
    \times \int_0^1 dx \int_0^{1-x} dy (1-x-y)^{\frac{1}{2} + \gamma_1 + \frac{n}{2} -i \nu} \left( x + \frac{y}{z_H} \right)^n \int \frac{d^{2-2 \epsilon} \vec{q}}{\pi \sqrt{2}} \frac{1}{ \left[ \vec{q}^{\; 2} + L \right]^{\frac{7}{2} + \gamma_1 + \frac{n}{2} -i \nu}} \; ,
\end{equation}
where
\begin{equation}
    L = x \vec{p}_H^{\; 2} + y \left( \Tilde{m}_H^2 + \frac{\vec{p}_H^{\; 2}}{z_H^2} \right) - \left( x + \frac{y}{z_H} \right)^2 \vec{p}_H^{\; 2} \; .
\end{equation}
Finally, we can integrate over $\vec{q}$ and then make the change of variables
\begin{equation}
    x = \lambda \Delta \; , \hspace{0.5 cm} y = \lambda (1-\Delta) \; ,
\end{equation}
to obtain
\begin{equation*}
    I_2(\gamma_1, n, \nu) = \frac{(\vec{p}_H^{\; 2})^{\frac{n}{2}} e^{i n \phi_H}}{z_H^2 \sqrt{2} \pi^{\epsilon}} \frac{\Gamma \left(\frac{5}{2} + \gamma_1 + \frac{n}{2} -i \nu + \epsilon \right)}{\Gamma \left(\frac{3}{2} + \gamma_1 + \frac{n}{2} -i \nu \right)} \int_0^1 d \Delta \left( \Delta + \frac{(1-\Delta)}{z_H} \right)^n \; ,
\end{equation*}
\begin{equation*}
    \left[ \left( \Delta + \frac{(1-\Delta)}{z_H^2} \right) \vec{p}_H^{\; 2} + (1-\Delta) \Tilde{m}_H^{\; 2} \right]^{-\frac{5}{2} - \gamma_1 - \frac{n}{2} +i \nu-\epsilon} \int_0^1 d \lambda \;  \lambda^{-\frac{3}{2} - \gamma_1 + \frac{n}{2} +i \nu-\epsilon}
\end{equation*}
\begin{equation}
   \times (1-\lambda)^{\frac{1}{2} + \gamma_1 + \frac{n}{2} -i \nu} (1-\lambda \zeta)^{-\frac{5}{2} - \gamma_1 -\frac{n}{2} +i \nu -\epsilon} \; ,
\end{equation}
where
\begin{equation}
    \zeta = \frac{\left(\Delta + \frac{(1-\Delta)}{z_H} \right)^2 \vec{p}_H^{\; 2}}{\left[ \left(\Delta + \frac{(1-\Delta)}{z_H^2} \right) \vec{p}_H^{\; 2} + \frac{(1-\Delta)(1-z_H) m_H^2}{z_H^2}  \right]} \; .
\end{equation}
It is easy to see that the integral over $\lambda$ gives a representation of the hypergeometric function. By using Eq.~(\ref{hypergeometric}) we arrive at the final form
\begin{equation*}
    I_2(\gamma_1, n, \nu) = \frac{(\vec{p}_H^{\; 2})^{\frac{n}{2}} e^{i n \phi_H}}{z_H^2 \sqrt{2} \pi^{\epsilon}} \left[ \frac{\Gamma \left( \frac{5}{2} + \gamma_1 + \frac{n}{2} - i \nu + \epsilon \right) \Gamma \left( -\frac{1}{2} - \gamma_1 + \frac{n}{2} + i \nu - \epsilon \right)}{\Gamma \left( 1+n-\epsilon \right)} \right] 
\end{equation*}
\begin{equation*}
    \times \int_0^1 d \Delta \left( \Delta + \frac{(1-\Delta)}{z_H} \right)^n  \left[ \left( \Delta + \frac{(1-\Delta)}{z_H^2} \right) \vec{p}_H^{\; 2} + \frac{(1-\Delta)(1-z_H) m_H^2}{z_H^2} \right]^{- \frac{5}{2} - \gamma_1 + i \nu - \frac{n}{2} - \epsilon } 
\end{equation*}
\begin{equation}
   \times \; _2 F_1 \left( - \frac{1}{2} - \gamma_1 + \frac{n}{2} + i \nu - \epsilon , \frac{5}{2} + \gamma_1 - i \nu + \frac{n}{2}+\epsilon, 1+n-\epsilon, \zeta \right) \; .
\end{equation}
It is very important to note that this integral is indeed divergent for every $\epsilon > -1$ (see the asymptotic behaviors of the hypergeometric function in Appendix~\ref{AppendixA}). It is important to extract the divergent contribution of the integral $I_2$ in the form of a pole in $\epsilon$. For this purpose, we take the limit $\Delta \to 1$ in the integrand of $I_2$. Using Eq.~(\ref{LimitHyper}), we have, up to terms ${\cal O}(\epsilon)$,
\begin{equation*}
  I_{2, \rm{as}}(\gamma_1, n, \nu) =  \frac{(\vec{p}_H^{\; 2})^{-\frac{3}{2}-\gamma_1+i \nu -\epsilon} e^{i n \phi_H} \Gamma(1+\epsilon)}{(1-z_H) \sqrt{2} \pi^{\epsilon}} \frac{1}{\left(m_H^2 + (1-z_H) \vec{p}_H^{\; 2} \right)} \int_{0}^{1} d \Delta (1-\Delta)^{-\epsilon-1}
\end{equation*}
\begin{equation}
     = - \frac{1}{\epsilon} \frac{(\vec{p}_H^{\; 2})^{-\frac{3}{2}-\gamma_1+i \nu -\epsilon} e^{i n \phi_H} \Gamma(1+\epsilon)}{(1-z_H) \sqrt{2} \pi^{\epsilon}} \frac{1}{\left(m_H^2 + (1-z_H) \vec{p}_H^{\; 2} \right)} 
\end{equation}
This simpler integral has the same singular behavior of $I_2$, therefore the combination
    \begin{equation*}
    I_{2, {\rm{reg}}} \equiv I_2 - I_{2,{\rm{as}}} = \frac{(\vec{p}_H^{\; 2})^{\frac{n}{2}} e^{i n \phi_H}}{z_H^2 \sqrt{2}} \left[ \frac{\Gamma \left( \frac{5}{2} + \gamma_1 + \frac{n}{2} - i \nu \right) \Gamma \left( -\frac{1}{2} - \gamma_1 + \frac{n}{2} + i \nu \right)}{\Gamma \left( 1+n \right)} \right] 
\end{equation*}
\begin{equation*}
    \times \int_0^1 d \Delta \left( \Delta + \frac{(1-\Delta)}{z_H} \right)^n  \left[ \left( \Delta + \frac{(1-\Delta)}{z_H^2} \right) \vec{p}_H^{\; 2} + \frac{(1-\Delta)(1-z_H) m_H^2}{z_H^2} \right]^{- \frac{5}{2} - \gamma_1 + i \nu - \frac{n}{2}} 
\end{equation*}
\begin{equation*}
   \times \left \{ _2 F_1 \left( - \frac{1}{2} - \gamma_1 + \frac{n}{2} + i \nu , \frac{5}{2} + \gamma_1 - i \nu + \frac{n}{2}, 1+n, \zeta \right) - \frac{z_H^2 (\vec{p}_H^{\; 2})^{-\frac{3}{2}-\gamma_1 - \frac{n}{2} +i \nu}}{\left(m_H^2 + (1-z_H) \vec{p}_H^{\; 2} \right)} \right. \; 
\end{equation*}
\begin{equation}
    \left. \times \frac{\Gamma(1+n)}{\Gamma(\frac{5}{2} + \gamma_1 + \frac{n}{2} - i \nu) \Gamma (-\frac{1}{2} - \gamma_1 + \frac{n}{2} + i \nu)} \frac{1}{(1-\Delta)(1-z_H)} \right \} \; ,
    \label{TrickAsy}
\end{equation}
is not divergent and has a finite $\epsilon\to 0$ limit.

The other two integrals can be computed by using the same technique:
\begin{equation*}
    I_1(\gamma_1, \gamma_2, n, \nu) = \int \frac{d^{2-2 \epsilon} \vec{q}}{\pi \sqrt{2}} (\vec{q}^{\; 2})^{i \nu - \frac{3}{2}} e^{i n \phi} (\vec{q}^{\; 2})^{-\gamma_1} \left[ \left( \vec{q} - \vec{p}_H \right)^2 \right]^{-\gamma_2} = \frac{(\vec{p}_H^{\; 2})^{-\frac{1}{2} + i \nu - \epsilon - \gamma_1 - \gamma_2} e^{i n \phi_H}}{\sqrt{2} \pi^{\epsilon}} 
\end{equation*}
\begin{equation}
    \times \left[ \frac{\Gamma \left( \frac{1}{2} + \gamma_1 + \gamma_2 + \frac{n}{2} - i \nu + \epsilon \right) \Gamma \left( -\frac{1}{2} - \gamma_1 + \frac{n}{2} + i \nu - \epsilon \right) \Gamma \left( 1- \gamma_2 -\epsilon \right)}{\Gamma \left( \frac{3}{2} + \gamma_1 + \frac{n}{2} - i \nu \right) \Gamma \left( \frac{1}{2} - \gamma_1 - \gamma_2 + \frac{n}{2} + i \nu - 2 \epsilon \right)\Gamma \left( \gamma_2 \right)} \right] \; ,
    \label{FirstMasterIntegral}
\end{equation}
note that for $\gamma_2 \geq 1$ and integer, we cannot put $\epsilon=0$ because the integral is divergent;
\vspace{0.2 cm}
\begin{equation*}
    I_3(\gamma_1, \gamma_2, n, \nu) = \int \frac{d^{2-2 \epsilon} \vec{q}}{\pi \sqrt{2}} (\vec{q}^{\; 2})^{i \nu - \frac{3}{2}} e^{i n \phi} (\vec{q}^{\; 2})^{-\gamma_1} \left[(1-z_H) m_H^2 + \left( \vec{p}_H - z_H \vec{q} \right)^2 \right]^{-\gamma_2}
\end{equation*}
\begin{equation}
    = \frac{(\vec{p}_H^{\; 2})^{\frac{n}{2}} e^{i n \phi_H}}{ (z_H^2)^{\gamma_2 + \frac{n}{2}} \sqrt{2} \pi^{\epsilon}} \left( \frac{\vec{p}_H^{\; 2}}{z_H^2} + \frac{(1-z_H) m_H^2}{z_H^2} \right)^{- \frac{1}{2} - \gamma_1 -\gamma_2 - \frac{n}{2} + i \nu - \epsilon } 
\end{equation}
\begin{equation*}
    \times \left[ \frac{\Gamma \left( \frac{1}{2} + \gamma_1 + \gamma_2 + \frac{n}{2} - i \nu + \epsilon \right) \Gamma (- \frac{1}{2} - \gamma_1 + \frac{n}{2} + i \nu - \epsilon) \Gamma ( \frac{3}{2} + \frac{n}{2} + \gamma_1 - i \nu)}{\Gamma \left( \frac{3}{2} + \gamma_1 + \frac{n}{2} - i \nu \right) \Gamma \left( \gamma_2 \right) \Gamma (1+n - \epsilon)} \right] 
\end{equation*}
\begin{equation*}
    \times \; _2 F_1 \left( - \frac{1}{2} - \gamma_1 + \frac{n}{2} + i \nu - \epsilon , \frac{1}{2} + \gamma_1 + \gamma_2 + \frac{n}{2} - i \nu + \epsilon, 1 + n -\epsilon, \xi \right) \; ,
\end{equation*}
where
\begin{equation}
    \xi = \frac{1}{1 + \frac{(1-z_H) m_H^2}{\vec{p}_H^{\; 2}}} \, .
\end{equation}
\end{appendices}
\begin{appendices}
\chapter{Heavy-quark pair impact factor}
\label{AppendixB}
\section{Impact factor in the transverse momentum space}
In this Section we give the expression of the leading-order impact factor,
together with the functional form of the amplitude for the
$g + R \to q \bar{q}$ subprocess, where $R$ here means ``Reggeized gluon''.
The leading-order impact factor is defined as~\cite{Fadin:1998fv}
\begin{equation}
\begin{split}
  d\Phi_{gg}^{\lbrace{Q\bar{Q}\rbrace}}(z_Q, \vec{p}_Q, \vec{q}) = & \frac{\braket{cc'|\mathcal{\widehat{P}}|0}}{2\left(N^2-1\right)}
 \\ &  \times  \sum_{\lambda_Q\lambda_{\bar{Q}}\lambda_G}\sum_{Q\bar{Q}a} \int\frac{ds_{gR}}{2\pi}d\rho_{\lbrace{Q\bar{Q}\rbrace}}\Gamma_{g \rightarrow \lbrace{Q\bar{Q}\rbrace}}^{ca}\left( \vec{q} \right)\left(\Gamma_{g \rightarrow \lbrace{Q\bar{Q}\rbrace}}^{ac'}\left( \vec{q} \right)\right)^{\ast} \;,
\end{split}
\label{eq:imp.fac}
\end{equation}
where 
\begin{equation}
\braket{cc'|\mathcal{\widehat{P}}|0} = \frac{\delta^{cc'}}{\sqrt{N^2-1}}
\end{equation}
is the projector on the singlet state. We take the sum over helicities,
$\{\lambda_Q,\lambda_{\bar{Q}}\}$, and over color indices, $\{Q,\bar{Q}\}$, of
the two produced particles (quark and antiquark) and average over polarization
and color states of the incoming gluon. In Eq.~(\ref{eq:imp.fac}), $s_{gR}$
denotes the invariant squared mass of the gluon-Reggeon system, while $
d\rho_{\lbrace{Q\bar{Q}\rbrace}}$ is the differential phase space of the outgoing
particles. The amplitude $\Gamma_{g \rightarrow \lbrace{Q\bar{Q}\rbrace}}^{ca}$ describes
the production of quark-antiquark pair in a collision between a gluon and a
Reggeon. 
\begin{figure}
    \begin{picture}(400,160)
\put(30,30){\includegraphics[scale=0.35]{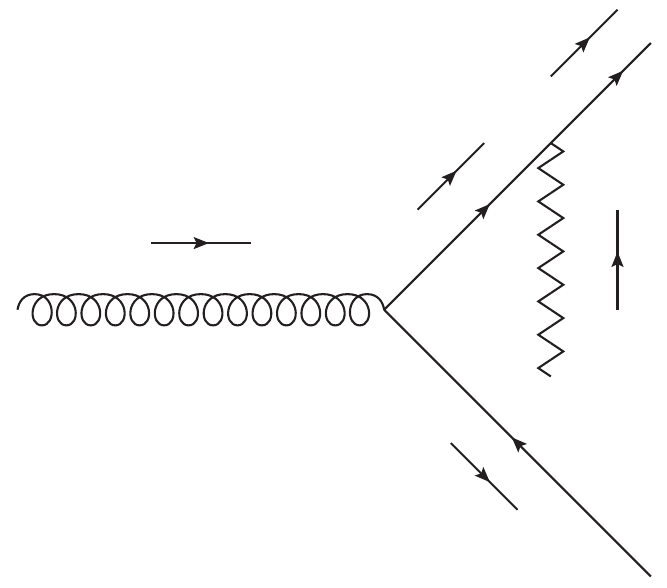}}
\put(90,0){$\mathcal{M}_1$}
\put(115,126){\scalebox{0.8}{$p_Q$}}
\put(140,82){\scalebox{0.8}{$q, c$}}
\put(97,44){\scalebox{0.8}{$p_{\bar{Q}}$}}
\put(55,94){\scalebox{0.8}{$p_1 , a$}}
\put(77,105){\scalebox{0.8}{$p_Q-q$}}
\put(180,30){\includegraphics[scale=0.35]{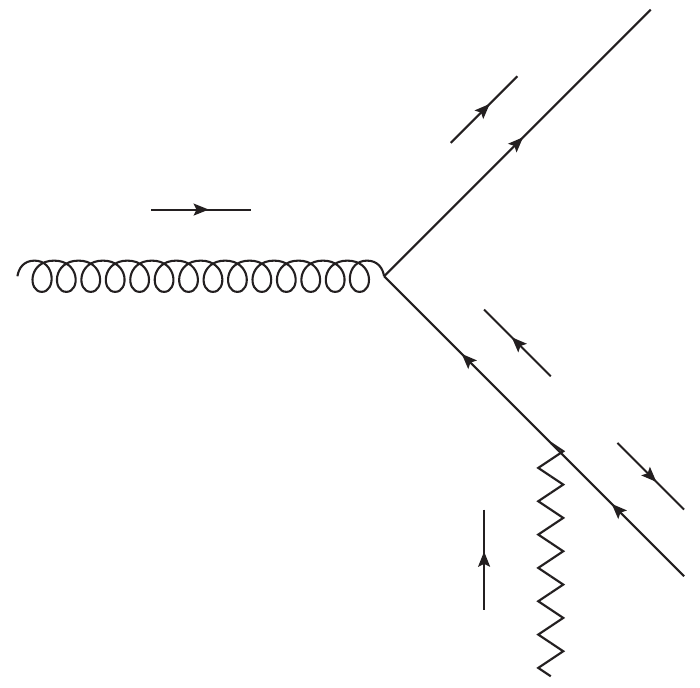}}
\put(240,0){$\mathcal{M}_2$}
\put(247,131){\scalebox{0.8}{$p_Q$}}
\put(245,50){\scalebox{0.8}{$q, c$}}
\put(205,117){\scalebox{0.8}{$p_1 , a$}}
\put(273,89){\scalebox{0.8}{$p_Q-p_1$}}
\put(293,67){\scalebox{0.8}{$p_{\bar{Q}}$}}
\put(330,30){\includegraphics[scale=0.35]{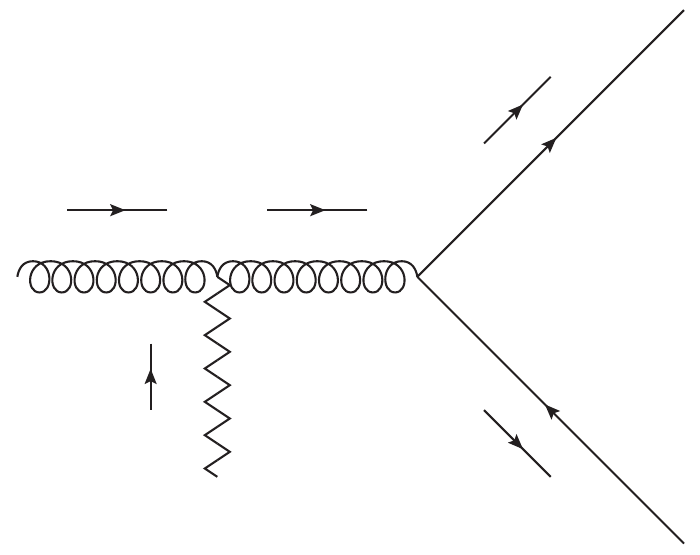}}
\put(400,0){$\mathcal{M}_3$}
\put(342,94){\scalebox{0.8}{$p_1,a$}}
\put(372,94){\scalebox{0.8}{$p_1+q$}}
\put(337,58){\scalebox{0.8}{$q, c$}}
\put(402,108){\scalebox{0.8}{$p_Q$}}
\put(403,44){\scalebox{0.8}{$p_{\bar{Q}}$}}
    \end{picture}
    \caption{Feynman diagrams relevant for the calculation of the impact factor for the
heavy-quark pair hadroproduction.}
    \label{fig:hadroproductionIF}
\end{figure}
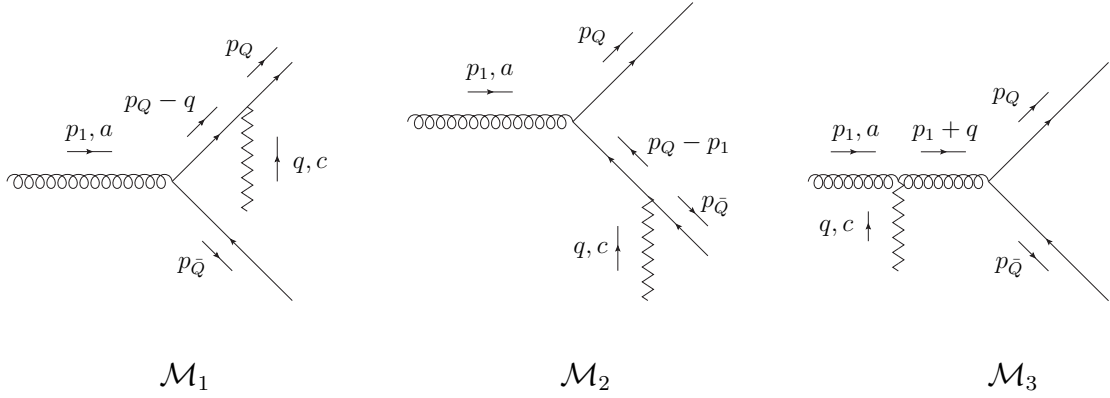
The latter, as before, can be treated as an ordinary gluon in the
``nonsense'' polarization state $\epsilon_R^\mu = - k_2^\mu/s$.
Having two particles produced in the intermediate state, one can write
\begin{equation}
\label{phasespace}
\frac{ds_{gR}}{2 \pi} d\rho_{\lbrace{Q\bar{Q}\rbrace}} = \frac{1}{2\left(2 \pi \right)^3} \delta \left( 1 - z_Q - z_{\bar{Q}} \right) \delta^{(2)} \left( \vec{q}-\vec{p}_Q-\vec{p}_{\bar{Q}} \right) \frac{dz_Q dz_{\bar{Q}}}{z_Q z_{\bar{Q}}} d^2 \vec{p}_Q \; d^2 \vec{p}_{\bar{Q}} \; , 
\end{equation}
with $p_{\bar{Q}}$ the antiquark momentum, and $z_{\bar{Q}}$ its longitudinal momentum fraction (with respect to the incoming gluon).
Summing over the three contributions, $\{{\cal M}_{1,2,3}\}$, of
Fig.~\ref{fig:hadroproductionIF}, one gets
\begin{equation}
\label{amp}
\begin{split}
\Gamma_{g \rightarrow \lbrace{Q\bar{Q}\rbrace}}^{ca} & = -ig^2\left(t^{a} t^{c}\right)\bar{u}\left( p_Q \right)\left(mR\slashed{\epsilon} - 2 z_Q \vec{P} \cdot \vec{\epsilon} - \vec{\slashed{P}} \slashed{\epsilon}\right)\frac{\slashed{k}_2}{s}v\left(p_{\bar{Q}}\right) \\ & - ig^2\left(t^{c} t^{a}\right)\bar{u}\left( p_Q \right)\left(m\bar{R}\slashed{\epsilon} - 2 z_Q \vec{\bar{P}} \cdot \vec{\epsilon}- \vec{\slashed{\bar{P}}} \; \; \slashed{\epsilon} \right)\frac{\slashed{k}_2}{s}v\left(p_{\bar{Q}} \right) \; ,
\end{split}
\end{equation}
where $\epsilon^\mu$ identifies the gluon polarization vector, $\{ t \}$ are
the $SU(3)$ color matrices and $R, \bar{R}, \vec{P}, \vec{\bar{P}}$ are
defined as
\begin{equation}
\label{laR}
R = \frac{1}{m_Q^2 + \vec{p}_Q^{\;2}} - \frac{1}{m_Q^2 + (\vec{p}_Q-z_Q \vec{q})^{2}} \;,  
\end{equation}
\begin{equation}
\label{LaRbar}
\bar{R} = \frac{1}{m_Q^2+(\vec{p}_Q-z_Q \vec{q} )^{2}} - \frac{1}{m_Q^2 + (\vec{p}_Q - \vec{q})^{2}} \;, 
\end{equation}
\begin{equation}
\label{laP}
\vec{P} = \frac{\vec{p}_Q}{m_Q^2+\vec{p}_Q^{\; 2}} - \frac{\vec{p}_Q-z_Q \vec{q}}{m_Q^2+(\vec{p}_Q
  -z_Q \vec{q})^{2}} \;,  
\end{equation}
\begin{equation}
\label{LaPbar}
  \vec{\bar{P}} = \frac{\vec{p}_Q - z_Q \vec{q}}{m_Q^2 + (\vec{p}_Q -z_Q \vec{q} )^{2}}
  -\frac{\vec{p}_Q-\vec{q}}{m_Q^2+(\vec{p}_Q - \vec{q})^{2}} \;.  
\end{equation}
Using Eq.~(\ref{amp}) together with Eq.~(\ref{eq:imp.fac}) and performing
sums and integrations (the latter ones only on the antiquark variables), our
final result reads
\begin{equation}
\begin{split}
d\Phi^{\lbrace{Q\bar{Q}\rbrace}}_{gg}(z_Q, \vec{p}_Q, \vec{q}) & =\frac{\alpha_s^2
\sqrt{N^2-1}}{2\pi N} \left[\left(m_Q^2 \left(R+\bar{R}\right)^2
+\left(z_Q^2 + z_{\bar{Q}}^2 \right) \left(\vec{P}+\vec{\bar{P}}\right)^2\right)
\right. \\ & \left. -\frac{N^2}{N^2-1} \left( 2 m_Q^2 R \bar{R}
+\left(z_Q^2 + z_{\bar{Q}}^2 \right) 2 \vec{P} \cdot \vec{\bar{P}}\right)\right]
\; d^2 \vec{p}_Q \; d z_Q \;.
\end{split}
\end{equation}
\section{Projection onto the eigenfunction of the LO BFKL kernel}
In the following we compute the projection of the impact factors onto the
eigenfunctions of the leading-order BFKL kernel, to get the
$(n, \nu)$-representation. We get 
\begin{equation*}
\frac{d\Phi_{gg}^{\lbrace{Q \bar{Q}\rbrace}}\left(n, \nu, \vec{p}_Q, z_Q \right)}{d^2 \vec{p}_Q
  \; d z_Q} \equiv \int\frac{d^2 \vec{q}}{\pi\sqrt{2}}(\vec{q}^{\;2})^{i\nu-\frac{3}{2}}
e^{in\phi} \frac{d\Phi^{\lbrace{Q\bar{Q}\rbrace}}_{gg}(\vec{q}, \vec{p}_Q,z_Q)}{d^2 \vec{p}_Q
  \; d z_Q} = \frac{\alpha_s^2 \sqrt{N^2-1}}{2\pi N}
\end{equation*}
\begin{equation*}
\times \left\{ m_Q^2 \left( J_3
- 2\frac{J_2(0)}{m_Q^2+\vec{p}_Q^{\;2}} \right) + (z_Q^2 + z_{\bar{Q}}^2)
\left( -m_Q^2 \left(J_3 - 2 \frac{J_2(0)}{m_Q^2+\vec{p}_Q^{\; 2}} \right)
+ \frac{J_2(1)}{m_Q^2 + \vec{p}_Q^{\; 2}} \right) \right.
\end{equation*}
\begin{equation*}
\left. - \frac{N^2}{N^2-1} \Bigg[ 2 m_Q^2 \left[\left( z_Q^2 + z_{\bar{Q}}^2
 - 1 \right) \left( 1 -  \left( z_Q^2 \right)^{\frac{1}{2}-i\nu} \right) \right]
 \frac{J_2(0)}{m_Q^2 + \vec{p}_Q^{\; 2}} + \left[ 2m_Q^2(z_Q^2 + z_{\bar{Q}}^2 -1)
\left( z_Q^2 \right)^{\frac{1}{2}-i\nu} \right] 
\right.
\end{equation*}
\begin{equation}
\left.
\left( J_3 - \frac{J_4(0)}{\left( z_Q^2 \right)^{\frac{1}{2}-i\nu}} \right)
- (z_Q^2 + z_{\bar{Q}}^2) \bigg[ (1-z_Q)^2 J_4(1) - \frac{\left( 1
    - \left(z_Q^2 \right)^{\frac{1}{2}-i\nu} \right)}{m_Q^2 + \vec{p}_Q^{\; 2}} J_2(1) \bigg] 
\Bigg] \right\} \; ,
\label{eq:imp.fac projected1} 
\end{equation}
where $J_2(\lambda)$, $J_3$ and $J_4(\lambda)$ read
\begin{equation}
\label{eq:J2}
\begin{split}
J_2 \left(\lambda\right) & \equiv \int\frac{d^2\vec{q}}{\pi\sqrt{2}}(\vec{q}^{\;2})^{i\nu-\frac{3}{2}}e^{in \phi } \frac{(\vec{q}^{\;2})^{\lambda}}{m_Q^2+(\vec{p}_Q - \vec{q})^2} \\ &  = \frac{\left(\vec{p}_Q^{\; 2}\right)^{\frac{n}{2}}e^{in \phi_{Q}}}
{\sqrt{2}} \frac{1}{\left(m_Q^2 + \vec{p}_Q^{\;2}\right)^{\frac{3}{2}+\frac{n}{2}-i\nu-\lambda}}
\frac{\Gamma\left(\frac{1}{2}+\frac{n}{2}+i\nu+\lambda\right)
  \Gamma\left(\frac{1}{2}+\frac{n}{2}-i\nu-\lambda\right)}{\Gamma\left(1+n\right)}\\ &\times\frac{\left(\frac{1}{2}+\frac{n}{2}-i\nu-\lambda\right)}
     {\left(-\frac{1}{2}+\frac{n}{2}+i\nu+\lambda\right)}\;
     _2F_1\left(-\frac{1}{2}+\frac{n}{2}+i\nu+\lambda,\frac{3}{2}+\frac{n}{2}
     -i\nu-\lambda,1+n,\chi \right) \;,
\end{split} 
\end{equation} 
\vspace{0.5 cm}
\begin{equation}
\label{eq:J3}
\begin{split}
  J_3 & \equiv \int\frac{d^2 \vec{q}}{\pi\sqrt2}(\vec{q}^{\,\,2})^{i\nu-\frac{3}{2}} e^{in \phi}
 \frac{1}{\left(m_Q^2 + (\vec{p}_Q-\vec{q})^2 \right)^2} \\ & = \frac{\left( \vec{p}_Q^{\; 2}\right)^{\frac{n}{2}}e^{in \phi_Q }}{\sqrt{2}} \frac{1}{\left(m_Q^2+\vec{p}_Q^{\;2}\right)^{\frac{5}{2}+\frac{n}{2}-i\nu}}
  \frac{\Gamma\left(\frac{1}{2}+\frac{n}{2}+i\nu\right)
    \Gamma\left(\frac{1}{2}+\frac{n}{2}-i\nu\right)}
       {\Gamma\left(1+n\right)}\frac{\left(\frac{1}{2}+\frac{n}{2}-i\nu\right)}
       {\left(-\frac{1}{2}+\frac{n}{2}+i\nu\right)}\;
       \\ & \times \left(\frac{3}{2}+\frac{n}{2}-i\nu\right)\;
       _2F_1\left(-\frac{1}{2}+\frac{n}{2}+i\nu,\frac{5}{2}+\frac{n}{2}
       -i\nu,1+n,\chi \right) \;,
\end{split} 
\end{equation}
\vspace{0.5 cm}
\begin{equation}
\begin{split}
  J_4\left(\lambda\right) & \equiv \int\frac{d^2\vec{q}}{\pi\sqrt{2}} (\vec{q}^{\;2})^{i\nu-\frac{3}{2}}e^{in\phi}\frac{(\vec{q}^{\;2})^{\lambda}}{(m_Q^2+(\vec{p}_Q-\vec{q})^2)(m_Q^2+(\vec{p}_Q-z_Q \vec{q})^2)} \\ & = \frac{\left(\vec{p}_Q^{\; 2}\right)^{\frac{n}{2}}
    e^{in \phi_Q}}{z_Q^2\sqrt{2}}  \frac{\left(\frac{3}{2}-i\nu-\lambda
+\frac{n}{2}\right)}{\left(m_Q^2+\vec{p}_Q^{\; 2}\right)^{\frac{5}{2}-i\nu-\lambda+\frac{n}{2}}} \frac{\Gamma\left(\frac{1}{2}+\frac{n}{2}+i\nu+\lambda \right)\Gamma\left(\frac{1}{2}+\frac{n}{2}-i\nu-\lambda \right)}{\Gamma\left(1+n\right)} \\ & \times \frac{\left(\frac{1}{2}+\frac{n}{2}-i\nu-\lambda \right)}{\left(-\frac{1}{2}+\frac{n}{2}+i\nu+\lambda \right)} \int_0^1 d\Delta \left(1+\frac{\Delta}{z}-\Delta\right)^n \left(1+\frac{\Delta}{z^2}-\Delta\right)^{-\frac{5}{2}+i\nu+\lambda-\frac{n}{2}} \; \\ & \times \; _2F_1\left(-\frac{1}{2}+i\nu+\lambda+\frac{n}{2},\frac{5}{2}-i\nu-\lambda+\frac{n}{2},1+n,\chi \;  \frac{\left(1+\frac{\Delta}{z}-\Delta\right)^2}{\left(1+\frac{\Delta}{z^2}-\Delta\right)}\right) \;,
\end{split}
\label{J4}
\end{equation}
and $\chi \equiv \frac{\vec{p}_Q^{\; 2}}{m_Q^2+\vec{p}_Q^{\; 2}}$; the azimuthal
angles $\phi$ and $\phi_Q$ are defined as $\cos \phi \equiv q_x/|\vec{q}|$
and $\cos \phi_Q \equiv p_{Q,x}/|\vec{p}_Q|$.

\end{appendices}
\begin{appendices}
\chapter{Further details on the Lipatov vertex}
\section{Polylogarithms, Hypergeometric functions and Nested harmonic sums}
\label{AppendixC1}
In this appendix we give some usual definitions and useful relations. \vspace{0.3 cm} 

\textit{\bf Polylogarithms}~\cite{kolbig:1986ngp}

We define polylogarithms as
\begin{equation}
{\rm{Li}}_{a+1}(z) = \frac{(-1)^a \; z}{a!} \int_0^1 dt \frac{\ln^a (t)}{1-t z} 
\end{equation}
and Nielsen generalized polylogarithms as 
\begin{equation}
    {\rm{S}}_{a,b}(z) = \frac{(-1)^{a+b-1}}{(a-1)!b!} \int_0^1 d t \frac{\ln^{a-1} t \ln^{b} (1 - z t)}{t} \; , \hspace{0.5 cm} {\rm{S}}_{a,1}(z) = {\rm{Li}}_{a+1}(z) \; \, ,
\end{equation}
where $a$ and $b$ are integers. During calculations, the following inversion and reflection formulas for polylogarithms are useful~\cite{kolbig:1986ngp},
\begin{equation*}
    {\rm{Li}}_n (z) + (-1)^n {\rm{Li}}_n \left( \frac{1}{z} \right) = -\frac{1}{n!} \ln^n (-z) - \sum_{j=0}^{n-2} \frac{1}{j!} (1+(-1)^{n-j})(1-2^{1-n+j}) \zeta (n-j) \ln^{j} (-z) \; ,
\end{equation*}
\begin{equation*}
    {\rm{S}}_{n,p} (z) = \sum_{j=0}^{n-1} \frac{\ln^j z}{j!} \left \{ s_{n-j, p} - \sum_{k=0}^{p-1} \frac{(-1)^k \ln^k (1-z)}{k!} {\rm{S}}_{p-k, n-j} (1-z) \right \} + \frac{(-1)^p}{n!p!} \ln^n z \ln^p (1-z) \, ,
\end{equation*}
where 
\begin{equation*}
   s_{n,p} = S_{n,p} (1) \; . 
\end{equation*}
In particular, from these relations, we obtain
\begin{equation*}
    {\rm{Li}}_2 (z) = - {\rm{Li}}_2 \left(\frac{1}{z} \right) - \zeta(2) - \frac{1}{2} \ln^2 (-z) \; , \hspace{0.5 cm} {\rm{Li}}_3 (z) = {\rm{Li}}_3 \left(\frac{1}{z} \right) - \zeta(2) \ln (-z) - \frac{1}{6} \ln^3 (-z) \; ,
\end{equation*}
\begin{equation*}
    {\rm{Li}}_4 (z) = - {\rm{Li}}_4 \left(\frac{1}{z} \right) - \frac{7}{4} \zeta (4) - \frac{1}{2} \zeta(2) \ln^2 (-z) - \frac{1}{4!} \ln^4 (-z) \; , 
\end{equation*}
\begin{equation*}
    {\rm{Li}}_2 (z) = \zeta(2) - {\rm{Li}}_2 (1-z) - \ln z \ln(1-z)  \; , 
\end{equation*}
\begin{equation*}
    {\rm{Li}}_3 (z) = - {\rm{Li}}_3 (1-z) - {\rm{Li}}_3 \left(1-\frac{1}{z} \right) + \zeta (3) + \frac{1}{6} \ln^3 z + \zeta (2) \ln z - \frac{1}{2} \ln^2 z \ln (1-z) \, ,
\end{equation*}
\begin{equation*}
   \text{Li}_4 \left(z \right) = \zeta(4) - \text{S}_{1,3} (1-z) + \ln z ( \zeta (3) - \text{S}_{1,2} (1-z)) + \frac{\ln^2 z}{2} (\zeta(2) - \text{Li}_2 (1-z)) - \frac{1}{6} \ln^3 z \ln (1-z) \; .
\end{equation*}
It is also useful to know that,
\begin{equation}
    \frac{d}{d y} {\rm{Li}}_2 \left( y \right) = - \frac{\ln (1-y)}{y} \; , \hspace{0.5 cm} \frac{d}{d y} {\rm{Li}}_3 \left( y \right) = \frac{{\rm{Li}}_2 \left( y \right)}{y} \; , \hspace{0.5 cm} \frac{d}{d y} {\rm{S}}_{1,2} \left( y \right) = \frac{\ln^2 (1-y)}{2 y} \; ,
\end{equation}
\begin{equation}
    \Im \{ {\rm{Li}}_2 (y + i \varepsilon) \} = \pi \theta (y-1) \ln (y) \; , \hspace{0.5 cm} \Im \{ {\rm{Li}}_3 (y + i \varepsilon) \} = \pi \theta (y-1) \frac{\ln^2 (y)}{2} \;,
\end{equation}
\begin{equation}
    \Im \{ {\rm{S}}_{1,2} (y + i \varepsilon) \} = \pi \theta (y-1) \left[ \zeta (2) - {\rm{Li}}_2 \left( \frac{1}{y} \right) - \frac{\ln^2 (y)}{2} \right]  \;.
\end{equation}

\textit{\bf Hypergeometric function $_2 F_1$}

We can represent the hypergeometric function $_2 F_1$ as
\begin{equation}
    _2 F_1 (a,b,c;z) = \frac{1}{B(b,c-b)} \int_0^1 dx \; x^{b-1} (1-x)^{c-b-1} (1 - z x)^{-a} \; ,
\label{IntReprHyperGau}
\end{equation}
for $\Re \{ c \} > \Re \{ b \} >0 $. The definition is valid in the entire complex $z$-plane with a cut along the real axis from 1 to infinity.
Using the integral representation (\ref{IntReprHyperGau}) and performing the transformation
\begin{equation}
    x = - \frac{y}{z \left[ 1 - \left( 1 + \frac{1}{z} \right) y \right]} \; ,
\end{equation}
one can prove the following identity
\begin{equation}
_2F_1 (a, b, 1 + b; 1+ z ) = (-z)^{-b} \; _2 F_1 \left( 1+b-a, b, 1+b ; 1 + z^{-1} \right) \; .
\label{HyperMoreGenProp}
\end{equation}
Choosing $a=b=\epsilon$, we get
\begin{equation}
_2 F_1 (1, \epsilon, 1+\epsilon ; 1 + z) =\; (-z)^{-\epsilon} \; _2F_1 (\epsilon, \epsilon, 1+\epsilon ; 1+z^{-1}) \; .
\label{HyperProp1}
\end{equation}
One can also prove the following expansions~\cite{Kalmykov:2006hyp}
\begin{equation}
_2 F_1 (1, -\epsilon, 1-\epsilon ; z) = 1 - \sum_{i=1}^{\infty} \epsilon^i {\rm{Li}}_i(z) \; ,
\label{HyperExp2}
\end{equation}
\begin{equation*}
    _2 F_1(2 - 2 \epsilon, 1-\epsilon, 2-\epsilon;z) = \frac{1}{1-z} \left \{ 1 + \left[ 1 + \left( \frac{1+z}{z} \right) \ln(1-z) \right] \epsilon \right.
\end{equation*}
\begin{equation*}
    + \left[ 2 + \left( \frac{1+z}{z} \right)  \ln (1-z) \ln ((1-z)e) - \left( \frac{1-z}{z} \right) {\rm{Li}}_2(z)  \right] \epsilon^2
\end{equation*}
\begin{equation*}
  \left. + \left[ 4 + \ln (1-z) \left( \frac{2(1+z)}{z} + \frac{2(1-z)}{z} \zeta(2) + \frac{\ln (1-z)}{3z} \right.  \right. \right.  
\end{equation*}
\begin{equation*}
   \left. \times \bigg( 3(1+z) + 2(1+z) \ln (1-z) - 3 (1-z) \ln z \right) \bigg) - \frac{1-z}{z} {\rm{Li}}_2 (z) \left( 1+ 2 \ln (1-z) \right)  
\end{equation*}
\begin{equation}
  \left. \left. - \frac{2 (1-z)}{z} \left( {\rm{Li}}_3 (1-z) + \frac{{\rm{Li}}_3 (z)}{2} - \zeta (3) \right) \right] \epsilon^3 \right \} + \mathcal{O}(\epsilon^4) \; .
\label{HyperExp1}
\end{equation}
\vspace{0.2 cm} \ 
\textit{\bf Nested harmonic sums and $\mathcal{M}$ functions}~\cite{DelDuca:2009ac} \vspace{0.2 cm} \\
The nested harmonic sums are defined recursively by
\begin{equation}
    S_i (n) = \sum_{k=1}^n \frac{1}{k^i} \; , \hspace{1 cm} S_{i \vec{j}} (n) = \sum_{k=1}^{n} \frac{S_{\vec{j}} (k)}{k^i} \; ,
\end{equation}
while the $\mathcal{M}$-functions are defined by the double series
\begin{equation}
    \mathcal{M} (\vec{i}, \vec{j}, \vec{k}; x_1, x_2) = \sum_{n_1=0}^{\infty} \sum_{n_2 =0}^{\infty} \binom{n_1+n_2}{n_1}^2 S_{\vec{i}} (n_1) S_{\vec{j}} (n_2) S_{\vec{k}} (n_1+n_2) x_1^{n_1} x_2^{n_2} \; .
\end{equation}

\section{Some useful integrals}
\label{AppendixC2}

\textbf{The integral $I_{a,b,c}$} 
\vspace{0.2 cm}

The integral
\begin{equation}
    I_{a,b,c} = \int_0^1 dx \ \frac{1}{a x + b (1-x) - c x (1-x)} \ln \left( \frac{a x + b(1-x) }{c x(1-x)} \right) 
\tag{\ref{FadinGorba}}
\end{equation}
is invariant with respect to any permutation of its arguments, as it can be seen from the representation
\begin{equation}
    I_{a,b,c} = \int_0^1 d x_1 \int_0^1 d x_2 \int_0^1 d x_3 \ \frac{\delta(1-x_1-x_2-x_3)}{(a x_1 + b x_2 + c x_3)(x_1 x_2 + x_1 x_3 + x_2 x_3)} \;.
    \label{Isimm}
\end{equation}
To prove that (\ref{FadinGorba}) and (\ref{Isimm}) are equivalent, we first integrate over $x_3$ and then perform the change of variables
\begin{equation*}
    x = \frac{x_2}{x_1+x_2} \; , \hspace{0.5 cm} z= \frac{x_1 x_2}{(1-x_1)x_1 + (1-x_2)x_2 -x_1 x_2} \; ,
\end{equation*}
\begin{equation*}
    x_1 = \frac{(1-x)z}{z+x(1-x)(1-z)} \; , \hspace{0.5 cm} x_2 = \frac{x z}{z+x(1-x)(1-z)} \; , \hspace{0.5 cm}  x_3 = \frac{x(1-x)(1- z)}{z+x(1-x)(1-z)} ,
\end{equation*}
with Jacobian
\begin{equation}
    J =  \frac{xz(1-x)}{[z+x(1-x)(1-z)]^{3}} \; .
\end{equation}
We obtain
\begin{equation*}
    I_{a,b,c} = \int_0^1 d x \int_0^{\infty} d z \ \frac{\Theta \left( \frac{x(1-x)(1-z)}{z+x(1-x)(1-z)} \right)}{a z(1-x)  + b x z  + c x(1-x)(1-z) } 
\end{equation*}
\begin{equation}
    = \int_0^1 dx \  \frac{1}{a (1-x) + b x - c x(1-x)} \ln \left( \frac{b  x + a (1-x)}{c x (1-x) }\right) \; ,
\end{equation}
which is equal to (\ref{FadinGorba}) after the trivial change of variables $x \leftrightarrow 1-x$. \\
Another useful representation of immediate proof is 
\begin{equation}
    I_{a,b,c} = \int_0^1 dx \int_1^{\infty} dt \ \frac{1}{t\ [a x(1-x)(t-1) + b (1-x) + c x ]} \vspace{0.2 cm} \; .
\end{equation}

In the case when $a=\vec q_1^{\;2}$, $b=\vec q_2^{\;2}$ and $c=(\vec q_1-\vec q_2)^{2}\equiv \vec p^{\;2}$, the explicit solution of the integral is~\cite{Fadin:2000kx}
\begin{equation*}
    I_{\vec{q}_1^{\;2},\vec{q}_2^{\;2},\vec{p}^{\;2}} = \int_0^1 dx \frac{1}{ \vec{q}_1^{\; 2} x + \vec{q}_2^{\; 2} (1-x) - \vec{p}^{\; 2} x(1-x)} \ln \left( \frac{\vec{q}_1^{\; 2} x + \vec{q}_2^{\; 2} (1-x)}{\vec{p}^{\; 2}x(1-x)} \right) 
\end{equation*}
\begin{equation}
    = - \frac{2}{|\vec{q}_1| |\vec{q}_2| \sin{\phi}} \left[ \ln \rho \arctan \left( \frac{\rho \sin \phi}{1-\rho \cos \phi} \right) + \Im \left( - {\rm{Li}}_2 (\rho e^{i \phi}) \right) \right] \; , 
    \label{Isol}
\end{equation}
where $\phi$ is the angle between $\vec{q}_1$, $\vec{q}_2$ and $\rho = {\rm{min}} \left( \frac{|\vec{q}_1|}{|\vec{q}_2|}, \frac{|\vec{q}_2|}{|\vec{q}_1|}\right)$. \vspace{1.5 cm} \\
\textbf{The box integral with one external mass} 
\vspace{0.3 cm} \\
Here, we derive the result for $I_{4B}$ in Eq.~(\ref{C}), \textit{i.e.} a box integral with massless propagators and one external mass, using direct Feynman technique. The integral is
\begin{equation}
  I_{4B} = \frac{1}{i} \int d^{D}k \frac{1}{(k^{2} + i \varepsilon) [(k+q_1)^{2}+i\varepsilon]
    [(k+q_2)^{2}+i\varepsilon] [(k-p_B)^2 + i \varepsilon]} \; .
  \tag{\ref{I4B}}
\end{equation}
Defining 
\begin{equation}
    d_1 = k^{2} + i \varepsilon \; , \hspace{0.5 cm} d_2 = (k+q_1)^{2}+i\varepsilon \; , \hspace{0.5 cm} d_3 = (k+q_2)^{2}+i\varepsilon \; , \hspace{0.5 cm} d_4 = (k-p_B)^2 + i \varepsilon \; ,
\end{equation}
we can write
\begin{equation}
\frac{1}{d_1 d_4}=\int_0^1 d x \frac{1}{\left((1-x) d_1+x d_4\right)^2} \; , \hspace{0.5 cm} \frac{1}{d_2 d_3}=\int_0^1 d y \frac{1}{\left((1-y) d_2+y d_3\right)^2} \; ,
\end{equation}
and hence
\begin{equation}
    \frac{1}{d_1 d_2 d_3 d_4}=\Gamma(4) \int_0^1 dy \int_0^1 dx \int_0^1 dz \frac{ z (1-z) }{\left[(1-z)\left((1-x) d_1+x d_4\right)+z\left((1-y) d_2+y d_3\right)\right]^4} \; .
\end{equation}
After integration in the $d^D k$, we obtain for $I_{4B}$,
\begin{equation*}
    I_{4 B}=\pi^{D / 2} \Gamma(4-D / 2)  
\end{equation*}
\begin{equation}
 \times \int_0^1  d z \int_0^1 d x \int_0^1 d y \frac{z(1-z)}{ \left[z(1-z)\left(-s_2 x(1-y)+(b y+a(1-y))(1-x) \right)-i 0 \right]^{4-D/2} } \; ,
\end{equation}
where $a=-t_1, \; b= -t_2$. Performing the trivial integrations over $z$ and $x$, we have
\begin{equation*}
I_{4 B}=\pi^{D / 2} \Gamma(1-\epsilon) \frac{\Gamma^2(\epsilon)}{\Gamma(2 \epsilon)} 
\end{equation*}
\begin{equation}
   \times \int_0^1 \frac{d y}{s_2(1-y)+b y+a(1-y)}\left[\left(-s_2(1-y)-i 0\right)^{\epsilon-1}-(b y+a(1-y))^{\epsilon-1}\right] .
\end{equation}
The first term in the square bracket gives
\begin{equation*}
\left(-s_2-i0\right)^{\epsilon-1} \int_0^1 dy \frac{(1-y)^{\epsilon-1}}{\left(s_2+a\right)(1-y)+b y} =\left(-s_2-i0 \right)^{\epsilon-1} \frac{1}{b} \int_0^1 dx \frac{x^{\epsilon-1}}{1-x\left(1- \left(s_2+a\right) / b\right)} 
\end{equation*}
\begin{equation}
    = \left(-s_2-i 0\right)^{\epsilon-1} \frac{1}{b} \frac{1}{\epsilon}\;\;{ }_2 F_1\left(1, \epsilon, 1+\epsilon ; 1- \frac{\left(s_2+a\right)}{b} \right) .
\end{equation}
In the second term, denoting $t = b y+a(1-y) $, one can organize the integral as 
\begin{equation}
\int_a^b d t=\int_0^b d t-\int_0^a d t=b \int_0^1 d x-a \int_0^1 d x \; .
\end{equation}
In this way, we obtain
\begin{equation*}
\int_0^1 \frac{d y}{s_2 (1-y) + b y+a (1-y)}(b y+a(1-y))^{\epsilon-1}=\frac{b^\epsilon}{s_2 b} \int_0^1 d x \frac{x^{\epsilon-1}}{1-x\left(1-\left(b-a\right) / s_2\right)} 
\end{equation*}
\begin{equation*}
-\frac{a^\epsilon}{s_2 b} \int_0^1 d x \frac{x^{\epsilon-1}}{1-x \left(1-(b-a)\left(s_2+a\right) /\left(s_2 b\right)\right)}=\frac{b^\epsilon}{s_2 b \epsilon}{ }_2 F_1\left(1, \epsilon, 1+\epsilon ; 1- \frac{(b-a)}{s_2} \right) 
\end{equation*}
\begin{equation}
-\frac{a^\epsilon}{s_2 b \epsilon}{ }_2 F_1\left(1, \epsilon, 1+\epsilon ; 1- \frac{(b-a)\left(s_2+a\right)}{\left(s_2 b\right)} \right) .
\end{equation}
Finally, restoring $t_1$ and $t_2$, we have
\begin{gather}
I_{4 B}=\frac{\pi^{2+\epsilon}}{s_2 t_2}  \frac{ \Gamma(1-\epsilon) \Gamma^2(1+\epsilon)}{\Gamma(1+2 \epsilon)} \frac{2}{\epsilon^2} \bigg[ \left(-s_2-i 0\right)^\epsilon { }_2 F_1\left(1, \epsilon, 1+\epsilon ; 1- \frac{\left(s_2-t_1 \right)}{(-t_2)} \right)\\  + (-t_2)^\epsilon { }_2 F_1\left(1, \epsilon, 1+\epsilon ; 1- \frac{(t_1-t_2)}{s_2} \right) - (-t_1)^{\epsilon} {}_2 F_1\left(1, \epsilon, 1+\epsilon ; 1- \frac{\left(s_2-t_1 \right)(t_1-t_2)}{s_2 (-t_2)} \right) \bigg] . \nonumber
\end{gather}
This result is exact and coincides with that in~(\ref{C1bis}).
\vspace{0.3 cm} \\
\textbf{Feynman integrals with logarithms} \vspace{0.3 cm} \\
In the calculation it happens to come across momentum integrals that have logarithms in the numerator. We explain below how to evaluate them, considering the example
\begin{equation}
    \int d^{D-2} k \frac{\ln \vec{k}^{\, 2} }{\vec{k}^{\, 2}  (\vec{k}+\vec{q}_2)^2} \;.
\end{equation}
Using $\ln \vec{k}^{\, 2}  = \frac{\partial}{\partial \alpha} (\vec{k}^{\, 2} )^{\alpha} \big |_{\alpha=0}$ and 
exchanging the order of integration and derivative, we get
\begin{equation*}
   \frac{\partial}{\partial \alpha} \left[ \int d^{D-2} k \frac{1}{(\vec{k}^{\, 2} )^{1-\alpha} (\vec{k}+\vec{q}_2)^2} \right]_{\alpha=0}  
\end{equation*}
\begin{equation*}
    = \pi^{1+\epsilon} (\vec{q}_2^{\; 2})^{-1+\epsilon} \frac{\partial}{\partial \alpha} \left[ (\vec{q}_2^{\; 2})^{\alpha} \frac{\Gamma(1-\alpha-\epsilon) \Gamma(\epsilon) \Gamma(\alpha+\epsilon)}{\Gamma(1-\alpha)\Gamma(2 \epsilon+\alpha)} \right]_{\alpha=0}
\end{equation*}
\begin{equation}
    = \pi^{1+\epsilon} (\vec{q}_2^{\; 2})^{-1+\epsilon} \frac{\Gamma(1-\epsilon) \Gamma^2 (\epsilon)}{\Gamma(2 \epsilon)} \bigg [ \ln \vec{q}_2^{\; 2} + \psi (1) + \psi (\epsilon) - \psi (1-\epsilon) - \psi (2 \epsilon) \bigg] \; .
\end{equation}

\section{Relation between Euclidean and Minkowskian integrals}
\label{AppendixC3}
In this appendix, we give an explicit derivation of the following relations:
\begin{equation}
I_5 = \frac{\pi^{2+\epsilon} \Gamma (1-\epsilon)}{s} \left[ \ln \left( \frac{(-s) (\vec{q}_1-\vec{q}_2)^2}{(-s_1)(-s_2)} \right) \mathcal{I}_3 + \mathcal{L}_3 -\mathcal{I}_5 \right] \; ,
\tag{\ref{I5inTransver}}
\end{equation}
\begin{equation}
I_{4B}=- \frac{\pi^{2+\epsilon} \Gamma (1-\epsilon)}{s_2} \left[ \frac{\Gamma^2 (\epsilon)}{\Gamma (2 \epsilon)} (-t_2)^{\epsilon -1} \left( \ln \left( \frac{- s_2}{- t_2} \right) + \psi (1-\epsilon) - 2 \psi (\epsilon) + \psi (2 \epsilon)  \right) + \mathcal{I}_{4B}  \right] \; .
\tag{\ref{I4B-I4Bcal}}
\end{equation}

\vspace{0.2cm}
$\bullet$ Let's start from the definition of $I_5$:
\begin{equation*}
  I_{5} = \frac{1}{i} \int d^{D}k \frac{1}{(k^{2} + i \varepsilon) [(k+q_1)^{2}+i\varepsilon]
    [(k+q_2)^{2}+i\varepsilon] [(k+p_A)^2 + i \varepsilon]
    [(k-p_B)^2 + i \varepsilon]}
  \;.
\end{equation*}
We introduce the standard Sudakov decomposition for momenta,
\begin{equation}
    k = \alpha p_B + \beta p_A + k_{\perp} \; , \hspace{0.5 cm} d^{D} k = \frac{s}{2} d \alpha d \beta d^{D-2} k_{\perp} \; ,
\label{Sudakov1}
\end{equation}
\begin{equation}
    q_1 = p_A-p_{A'} = - \frac{\vec{q}_1^{\; 2}}{s} p_B + \frac{s_2}{s} p_A +q_{1 \perp} \; , \hspace{0.5 cm} q_2 = p_{B'}-p_{B} = - \frac{s_1}{s} p_B + \frac{\vec{q}_2^{\; 2}}{s} p_A + q_{2 \perp} \;.
\label{Sudakov2}
\end{equation}
By expressing denominators in terms of Sudakov variables, we can reduce the integration over $\alpha$ to a simple computation of residues in the complex plane. We have the following five simple poles:
\begin{equation}
    \alpha_1 = \frac{\vec{k}^{\, 2} - i \varepsilon}{\beta s} \; , \hspace{0.5 cm} \alpha_2 = \frac{(k+\vec{q}_1)^{2}-i \varepsilon}{s(\beta + \frac{s_2}{s})} + \frac{\vec{q}_{1}^{\; 2}}{s} \; , \hspace{0.5 cm} \alpha_3 = \frac{(k+\vec{q}_2)^{2}-i \varepsilon}{s(\beta + \frac{\vec{q}_{2}^{\; 2}}{s})} + \frac{s_1}{s} \; ,
\end{equation}
\begin{equation}
    \alpha_4 = \frac{\vec{k}^{2} - i \varepsilon}{(1+\beta) s} \; , \hspace{0.5 cm} \alpha_5 = \frac{\vec{k}^{2} - i \varepsilon}{\beta s} + 1 \; .
\end{equation}
We observe that in the region $\Omega = \{ \beta < -1 \; \;  \vee \; \; \beta > 0 \}$ the integral always vanishes, since all poles are on the same side with respect to the real $\alpha$-axis; 
in the region $\bar{\Omega} = \{ -1 < \beta < 0 \}$ we have contributions from three poles lying in the lower real $\alpha$-axis: $\alpha_4$ in the whole region $-1 < \beta < 0$, $\alpha_2$ in the region $-\frac{s_2}{s} < \beta < 0$ and $\alpha_3$ in the region $-\frac{\vec{q}_2^{\; 2}}{s} < \beta < 0$. The integral can therefore be expressed as
\begin{equation*}
\begin{split}
  I_5 = - \pi \int_{-1}^{0} \frac{d \beta}{1+\beta} & \int d^{D-2} k \frac{1}{\left[\alpha_4 \beta s - \vec{k}^{2} + i \varepsilon \right]} \frac{1}{\left[ (\alpha_4 - \frac{\vec{q}_{1}^{\; 2}}{s})(\beta + \frac{s_2}{s})s - (k+\vec{q}_1)^{2} + i \varepsilon \right]} \\ &
  \times \frac{1}{\left[(\alpha_4-1) \beta s - \vec{k}^{2} + i \varepsilon \right]} \frac{1}{\left[ (\alpha_4 - \frac{s_1}{s})(\beta + \frac{\vec{q}_2^{\; 2}}{s})s - (\vec{k}+\vec{q}_2)^{2} + i \varepsilon \right]}
 \end{split}
\end{equation*}
\begin{equation*}
\begin{split}
     \hspace{1.1 cm} - \pi \int_{-\frac{s_2}{s}}^{0} \frac{d \beta}{\beta + \frac{s_2}{s}} & \int d^{D-2} k \frac{1}{\left[\alpha_2 \beta s - \vec{k}^{2} + i \varepsilon \right]} \frac{1}{\left[ (\alpha_2 - \frac{s_1}{s})(\beta + \frac{\vec{q}_2^{\; 2}}{s})s - (\vec{k}+\vec{q}_2)^{2} + i \varepsilon \right]} \\ &
  \times \frac{1}{\left[\alpha_2 (1+\beta) s - \vec{k}^{2} + i \varepsilon \right]} \frac{1}{\left[(\alpha_2-1) \beta s - \vec{k}^{2} + i \varepsilon \right]} 
 \end{split}
\end{equation*}
\begin{equation}
\begin{split}
     \hspace{1.0 cm} - \pi \int_{-\frac{\vec{q}_2^{\; 2}}{s}}^{0} \frac{d \beta}{\beta + \frac{\vec{q}_2^{\; 2}}{s}} & \int d^{D-2} k \frac{1}{\left[\alpha_3 \beta s - \vec{k}^{2} + i \varepsilon \right]} \frac{1}{\left[ (\alpha_3 - \frac{\vec{q}_{1}^{\; 2}}{s})(\beta + \frac{s_2}{s})s - (\vec{k}+\vec{q}_1)^{2} + i \varepsilon \right]} \\ &
  \times \frac{1}{\left[\alpha_3 (1+\beta) s - \vec{k}^{2} + i \varepsilon \right]}  \frac{1}{\left[(\alpha_3-1) \beta s - \vec{k}^{2} + i \varepsilon \right]} \; .
 \end{split}
\end{equation}
Substituting the explicit values of poles and doing simple algebric manipulations, we end up with
(omitting the $i\varepsilon$'s in the denominators)
\begin{equation*}
    I_5 \simeq - \pi \int_0^1 d \beta \beta^3 \int d^{D-2} k \frac{1}{\vec{
    k}^{\, 2} (\vec{k} + \beta \vec{q}_1)^{2}[(\vec{k} + \beta \vec{q}_2)^{2} + \beta (1-\beta) (-s_1)][\vec{k}^{\, 2}  + \beta (1-\beta) (-s)]} 
\end{equation*}
\begin{equation*}
    \begin{split}
    + \pi \int_0^1 d \beta \beta^3 \int d^{D-2} k \left( \frac{s_2}{s} \right) & \frac{1}{(\vec{
    k}+\vec{q}_1)^{2} (\vec{k} +(1-\beta) \vec{q}_1 + \beta \vec{q}_2)^{2} [(\vec{k} + (1-\beta) \vec{q}_1)^{2}+\beta (1-\beta) \vec{q}_1^{\; 2}]} \\ 
    & \times \frac{1}{[(\vec{k}+(1-\beta) \vec{q}_1)^{2} + \beta (1-\beta) (-s_2)]}
    \end{split}
\end{equation*}
\begin{equation}
    - \pi \int_0^1 d \beta (1-\beta)^3 \int d^{D-2} k \frac{(\vec{q}_{2}^{\; 2})^2}{s_2 s} \frac{1}{[(\vec{k}+\vec{q}_2)^2]^2 (\vec{k} + \beta \vec{q}_2)^{2} [(\vec{k} + \beta \vec{q}_2)^{2}+\beta (1-\beta) \vec{q}_2^{\; 2}]} \; .
\end{equation}
The third integral is suppressed in the high-energy approximation; as for the other two, they can be calculated by first performing the change of variable $\vec k \to \beta \vec k$ in the integration over $\vec k$ and then integrating some terms in $\beta$ by using the following integral:
\begin{equation*}
    \int_0^1 d \beta \frac{(1-\beta)^{D-n}}{\delta + \beta} =  \int_0^1 d \beta \frac{(1-\beta)^{D-n}-1}{\delta + \beta} + \int_0^1 d \beta \frac{1}{\delta + \beta} 
\end{equation*}
\begin{equation}
    \simeq \int_0^1 d\beta \frac{(1-\beta)^{D-n}-1}{\beta} + \int_0^1 d \beta \frac{1}{\delta + \beta} \simeq \psi(1) - \psi(D-n-1) - \ln \delta \; ,
    \label{deltaTrick}
\end{equation}
where $\delta$ is a generic quantity tending to zero, like $\vec{k}^{\, 2} /s$ for instance, and $n=5$ and 6 in the cases we are interested in. We obtain
\begin{equation}
\begin{split}
    I_5 =& \; \frac{\pi}{s} \int d^{D-2} k \frac{1}{\vec{k}^{\, 2}  (\vec{k}+\vec{q}_1)^2(\vec{k}+\vec{q}_2)^2} \left[ \ln \left( \frac{(k+\vec{q}_2)^2 (-s)}{(-s_1)(-s_2)} \right) - ( \psi (1) -\psi (D-5)) \right] \\
    & + \pi \; \frac{\vec{q}_1^{\; 2}}{s} \int_0^1 d \beta \beta^{3} \int d^{D-2} k \frac{1}{\vec{k}^{\, 2}  (\vec{k}+\beta \vec{q}_1)^2 (\vec{k}+\beta \vec{q}_2)^2 [\vec{k}^{\, 2}  + \beta (1-\beta) \vec{q}_1^{\; 2}]} \; .
\end{split}
\end{equation}
To get the desired form, we rewrite the last two terms in the square bracket as an integral over $\beta$,
\[
\psi(1)-\psi(D-5) = \int_0^1 d\beta \, \frac{(1-\beta)^{D-6}-1}{\beta}
= \int_0^1 d\beta \, \frac{\beta^{D-6}}{1-\beta}-\int_0^1 d\beta \,\frac{1}{\beta} \;,
\]
and combine with the last term of the full expression, to get
\begin{equation*}
    I_5 = \; \frac{\pi}{s} \int d^{D-2} k \frac{1}{\vec{k}^{\, 2}  (\vec{k}+\vec{q}_1)^2(\vec{k}+\vec{q}_2)^2} \ln \left( \frac{(k+\vec{q}_2)^2 (-s)}{(-s_1)(-s_2)} \right) 
\end{equation*}
\begin{equation*}
    - \frac{\pi}{s} \hspace{-0.1 cm} \int_0^1 \hspace{-0.2 cm} \frac{d \beta}{1-\beta} \hspace{-0.1 cm} \int \hspace{-0.05 cm} \frac{d^{D-2} k}{\vec{k}^{\, 2} } \hspace{-0.05 cm} \left[ \frac{\beta^2}{(\vec{k}-\beta(\vec{q}_1-\vec{q}_2))^2[(1-\beta) \vec{k}^{\, 2}  + \beta (\vec{k}-\vec{q}_1)^2]} - \frac{1}{(\vec{k}-\vec{q}_1+\vec{q}_2)^2 (\vec{k}-\vec{q}_1)^2} \right] 
\end{equation*}
\begin{equation*}
     = \; \frac{\pi}{s} \int d^{D-2} k \frac{1}{\vec{k}^{\, 2}  (\vec{k}-\vec{q}_1)^2 (\vec{k}-\vec{q}_2)^2} \left[ \ln \left( \frac{(-s)(\vec{q}_1-\vec{q}_2)^2}{(-s_1)(-s_2)} \right) + \ln \left( \frac{(\vec{k}-\vec{q}_1)^2(\vec{k}-\vec{q}_2)^2}{\vec{k}^{\, 2}  (\vec{q}_1-\vec{q}_2)^{\; 2}} \right) \right]
\end{equation*}
\begin{equation*}
    - \frac{\pi}{s} \hspace{-0.1 cm} \int_0^1 \hspace{-0.2 cm} \frac{d \beta}{1-\beta} \hspace{-0.1 cm} \int \hspace{-0.05 cm} \frac{d^{D-2} k}{\vec{k}^{\, 2} } \hspace{-0.05 cm} \left[ \frac{\beta^2}{(\vec{k}-\beta(\vec{q}_1-\vec{q}_2))^2[(1-\beta) \vec{k}^{\, 2}  + \beta (\vec{k}-\vec{q}_1)^2]} - \frac{1}{(\vec{k}-\vec{q}_1+\vec{q}_2)^2 (\vec{k}-\vec{q}_1)^2} \right] 
\end{equation*}
\begin{equation*}
     + \frac{\pi}{s} \int d^{D-2} k \frac{1}{\vec{k}^{\, 2}  (\vec{k}-\vec{q}_1)^2 (\vec{k}-\vec{q}_2)^2} \left[ \ln \left( \frac{\vec{k}^{\, 2} }{(\vec{k}-\vec{q}_1)^2} \right) \right] \; .
\end{equation*}
Expressing the term in the last line as
    \begin{equation*}
     \frac{\pi}{s} \int_0^1 d \beta \frac{1}{1-\beta} \int d^{D-2} k \frac{1}{(\vec{k}-\vec{q}_1)^2 (\vec{k}-\vec{q}_2)^2} \left[ \frac{1}{\beta \vec{k}^{\, 2}  + (1-\beta) (\vec{k}-\vec{q}_1)^{\; 2}} - \frac{1}{\vec{k}^{\, 2} } \right] \, ,
\end{equation*}
and performing simple manipulations, we get
\begin{equation*}
\begin{split}
    I_5 &= \; \frac{\pi}{s} \int d^{D-2} k \frac{1}{\vec{k}^{\, 2}  (\vec{k}-\vec{q}_1)^2 (\vec{k}-\vec{q}_2)^2} \left[ \ln \left( \frac{(-s)(\vec{q}_1-\vec{q}_2)^2}{(-s_1)(-s_2)} \right) \right] \\ & + \frac{\pi}{s} \int d^{D-2} k \frac{1}{\vec{k}^{\, 2}  (\vec{k}-\vec{q}_1)^2 (\vec{k}-\vec{q}_2)^2} \left[\ln \left( \frac{(\vec{k}-\vec{q}_1)^2(\vec{k}-\vec{q}_2)^2}{\vec{k}^{\, 2}  (\vec{q}_1-\vec{q}_2)^{\; 2}} \right) \right]
\end{split}
\end{equation*}
\begin{equation*}
    - \frac{\pi}{s} \; \int_0^1 d \beta \frac{1}{1-\beta} \int d^{D-2} k \frac{1}{\vec{k}^{\, 2} [(1-\beta) \vec{k}^{\, 2}  + \beta (\vec{k}-\vec{q}_1)^2]}\left[ \frac{\beta^2}{(\vec{k}-\beta(\vec{q}_1-\vec{q}_2))^2} - \frac{1}{(\vec{k}-\vec{q}_1+\vec{q}_2)^2} \right] ,
\end{equation*}
which, by using definitions (\ref{I3cal}), (\ref{L3cal}), (\ref{I5cal}), leads exactly to eq.~(\ref{I5inTransver}). \\
\vspace{0.2cm}

$\bullet$ Let's prove the second relation; again we start from the definition of $I_{4B}$:
\begin{equation}
  I_{4B} = \frac{1}{i} \int d^{D}k \frac{1}{(k^{2} + i \varepsilon) [(k+q_1)^{2}+i\varepsilon]
    [(k+q_2)^{2}+i\varepsilon] [(k-p_B)^2 + i \varepsilon]}\;.
\end{equation}
We again introduce the Sudakov decomposition~(\ref{Sudakov1})-(\ref{Sudakov2}) and observe that we have four poles:
\begin{equation*}
   \beta_1 = \frac{\vec{k}^{\, 2} -i \varepsilon}{\alpha s} \; , \hspace{0.5 cm} \beta_2 = \frac{(\vec{k}+\vec{q}_1)^{2}-i \varepsilon}{(\alpha-\frac{\vec{q}_1^{\; 2}}{s}) s} - \frac{s_2}{s} \; , \hspace{0.5 cm} \beta_3 = \frac{(\vec{k}+\vec{q}_2)^{2}-i \varepsilon}{(\alpha-\frac{s_1}{s})s} - \frac{\vec{q}_2^{\; 2}}{s} \; , \hspace{0.5 cm}  \beta_4 = \frac{\vec{k}^{\, 2} -i \varepsilon}{(\alpha-1)s} \; .
\end{equation*}
In the region $\Omega = \{ \alpha < 0 \; \;  \vee \; \; \alpha > 1 \}$ the integral always vanishes, while, in the region $\bar{\Omega} = \{  0 < \alpha < 1\}$, if we close the integration path over $\beta$ in the half-plane $\Im \beta > 0$ and use the residue theorem, we have
\begin{equation*}
I_{4 B} = -\pi \int_0^1 \frac{d \alpha}{1-\alpha} \int d^{D-2} k \ \frac{1}{(\alpha \beta_4 s - \vec{k}^{2} + i \varepsilon )}
\end{equation*}
\begin{equation*}
\times \frac{1}{[\alpha (\beta_4 + \frac{s_2}{s})s - (\vec{k}+\vec{q}_1)^2 + i \varepsilon][ (\alpha-\frac{s_1}{s}) (\beta_4 + \frac{\vec{q}_2^{\; 2}}{s})s - (\vec{k}+\vec{q}_2)^2 + i \varepsilon]}
\end{equation*}
\begin{equation}
    + \pi \int_0^{\frac{s_1}{s}} \frac{d \alpha}{\alpha-\frac{s_1}{s}} \int \frac{d^{D-2} k}{(\alpha \beta_3 s - \vec{k}^{2} + i \varepsilon )[\alpha (\beta_3 + \frac{s_2}{s})s - (\vec{k}+\vec{q}_1)^2 + i \epsilon][ (\alpha-1) \beta_3 s - \vec{k}^2 + i \varepsilon]} \; .
\end{equation}
In the last expression, we neglected the contribution of the pole $\beta_2$, which is present in the region $ \Omega' = \{ 0<\alpha <\vec{q}_1^{\; 2}/s \}$, since it is suppressed in the MRK. For the same reason, in the previous integrals we approximated $\alpha- \vec q_1^{\:2}/s$ with $\alpha$.
Substituting the explicit values of $\beta_3$ and $\beta_4$, one gets (recalling that 
$s_1 s_2/s = (\vec q_1-\vec q_2)^2$ and taking the high-energy limit)
\begin{equation}
    I_{4B} = \pi \int_0^1 d \alpha  \alpha^2 \int d^{D-2} k \frac{1}{\vec{k}^{2} (\vec{k}+\alpha \vec{q}_2)^2 [(\vec{k}+\alpha \vec{q}_1)^2+\alpha(1-\alpha)(-s_2-i \varepsilon)]}
\end{equation}
\begin{equation*}
    - \pi \frac{(\vec{q}_1-\vec{q}_2)^2}{s_2} \int_0^1 \hspace{-0.1 cm} d \alpha \; \alpha^2 \int \frac{ d^{D-2} k}{(\vec{k}+\vec{q}_2)^2[(\vec{k}+(1-\alpha) \vec{q}_2)^2 + \alpha (1-\alpha) \vec{q}_2^{\; 2}](\vec{k}+(1-\alpha) \vec{q}_2 + \alpha \vec{q}_1)^2} \; .
\end{equation*}
Manipulating the first term and using the integral in eq.~(\ref{deltaTrick}), one can obtain the following form:
\begin{equation*}
I_{4 B} = \frac{\pi^{2+\epsilon}}{s_2 t_2} \frac{\Gamma (1-\epsilon) \Gamma^2 (\epsilon)}{\Gamma (2 \epsilon)} (\vec{q}_2^{\; 2})^{\epsilon} \left[ \ln \left( \frac{-s_2}{-t_2} \right) + \psi (1-\epsilon) - \psi (\epsilon) \right] 
\end{equation*}
\begin{equation*}
    + \frac{\pi}{s_2} \int_0^1 \frac{d \alpha}{\alpha} \int d^{D-2} k \frac{1}{(\vec{k}-\vec{q}_1)^2 (\vec{k}-(\vec{q}_1-\vec{q}_2))^2}   
\end{equation*}
\begin{equation*}
    -\frac{\pi}{s_2} \int_0^1 \frac{d \alpha}{1-\alpha} \int d^{D-2} k \frac{1}{(\vec{k}+\vec{q}_2)^2 \left[ \alpha \vec{k}^2 + (1-\alpha) (\vec{k}+\vec{q}_1)^{2} \right]} 
\end{equation*}
\begin{equation}
     - \pi \frac{(\vec{q}_1-\vec{q}_2)^2}{s_2} \int_0^1 d \alpha \; \alpha^2 \int d^{D-2} k \frac{1}{\vec{k}^2[(\vec{k}+\alpha \vec{q}_2)^2 + \alpha (1-\alpha) \vec{q}_2^{\; 2}](\vec{k} + \alpha (\vec{q}_2 - \vec{q}_1))^2} \; .
\end{equation}
In the last integral we perform the transformation 
\begin{equation*}
    \vec{k} \rightarrow \vec{k}' = \vec{k} - \vec{p} \; , \hspace{0.5 cm} \alpha \rightarrow \alpha ' = \frac{\alpha \vec{k'}^{2}}{(1-\alpha) \vec{k}^{2} + \alpha \vec{k'}^{2}} \; \; \; \; {\rm{with}} \; \; \; \; \vec{p} \equiv \vec{q}_1-\vec{q}_2 \;,
\end{equation*}
and get
\begin{equation*}
I_{4B} = \frac{\pi^{2+\epsilon}}{s_2 t_2} \frac{\Gamma (1-\epsilon) \Gamma^2 (\epsilon)}{\Gamma (2 \epsilon)} (\vec{q}_2^{\; 2})^{\epsilon} \left[ \ln \left( \frac{-s_2}{-t_2} \right) + \psi (1-\epsilon) - \psi (\epsilon) \right]  
\end{equation*}
\begin{equation*}
    + \frac{\pi}{s_2} \int_0^1 \frac{d \alpha}{\alpha} \int d^{D-2} k \frac{1}{(\vec{k}-\vec{q}_1)^2 (\vec{k}-(\vec{q}_1-\vec{q}_2))^2}  
\end{equation*}
\begin{equation*}
    -\frac{\pi}{s_2} \int_0^1 \frac{d \alpha}{1-\alpha} \int d^{D-2} k \frac{1}{(\vec{k}+\vec{q}_2)^2 \left[ \alpha \vec{k}^2 + (1-\alpha) (\vec{k}+\vec{q}_1)^{2} \right]} 
\end{equation*}
\begin{equation}
     - \frac{\pi}{s_2} \int_0^1 d \alpha \; \int d^{D-2} k \frac{ \alpha^2 \vec p^{\; 2}} {(\vec{k} + \alpha \vec{p})^2 [(\vec{k}+\alpha \vec{p})^2 + \alpha (1-\alpha) \vec{p}^{\; 2}] [ (1-\alpha) \vec{k}^2 + \alpha (\vec{k} + \vec{q}_1)^2 ]} \; .
\end{equation}
The last two terms can be shown to lead to 
\begin{equation*}
    - \frac{\pi}{s_2} \int_{0}^1 \frac{d \alpha}{\alpha} \int d^{D-2} k \frac{1-\alpha}{(\vec{k}-(1-\alpha) (\vec{q}_1-\vec{q}_2))^2 [\alpha \vec{k}^{\, 2}  + (1-\alpha) (\vec{k}-\vec{q}_1)^2 ]} 
\end{equation*}
\begin{equation}
    + \frac{\pi^{2+\epsilon}}{s_2 t_2} \frac{\Gamma (1-\epsilon) \Gamma^2 (\epsilon)}{\Gamma (2 \epsilon)} (\vec{q}_2^{\; 2})^{\epsilon} (\psi (2 \epsilon) - \psi (\epsilon)) \; ,
\end{equation}
so that we finally find
\begin{equation*}
I_{4B}=- \frac{\pi^{2+\epsilon} \Gamma (1-\epsilon)}{s_2} \left[ \frac{\Gamma^2 (\epsilon)}{\Gamma (2 \epsilon)} (-t_2)^{\epsilon -1} \left( \ln \left( \frac{- s_2}{- t_2} \right) + \psi (1-\epsilon) - 2 \psi (\epsilon) + \psi (2 \epsilon)  \right) \right] 
\end{equation*}
\begin{equation*}
 - \frac{\pi^{2+\epsilon} \Gamma (1-\epsilon)}{s_2} \int_0^1 \frac{d \alpha}{ \alpha} \int \frac{d^{D-2} k}{\pi^{1 + \epsilon} \Gamma (1-\epsilon)}  
\end{equation*}
\begin{equation}
   \times \left[ \frac{1-\alpha}{\left[ \alpha \vec{k}^2 + (1- \alpha) (\vec{k}-\vec{q}_1)^2 \right] (\vec{k}-(1-\alpha)(\vec{q}_1-\vec{q}_2))^2} - \frac{1}{(\vec{k}-\vec{q}_{1})^2 (\vec{k}-(\vec{q}_1-\vec{q}_2))^2} \right] \; ,
\end{equation}
which is exactly eq.~(\ref{I4B-I4Bcal}).

\section{Soft limit}
\label{AppendixC4}
In this appendix we evaluate the soft limit ($\vec{p} \to 0$) of the integrals considered so far. 
We start from $\mathcal{I}_3$, given by eq.~(\ref{I3cal}). In the soft limit, the dominant contribution comes from the region $ \vec{k} \simeq \vec{q}_1 \simeq \vec{q}_2$; hence we can make the replacement
\begin{equation*}
    \frac{1}{\vec{k}^{\, 2} } \to  \frac{1}{\vec{Q}^{\; 2}} \hspace{0.5 cm} \text{with} \hspace{0.5 cm} \vec{Q} \equiv \frac{\vec{q}_1+\vec{q}_2}{2}
\end{equation*}
and obtain 
\begin{equation}
    \mathcal{I}_3 \simeq \frac{1}{\vec{Q}^2} \frac{1}{\pi^{1+\epsilon} \Gamma (1 - \epsilon)} \int d^{2+2 \epsilon} k \frac{1}{\vec{k}^{\, 2}  (\vec{k}+\vec{p})^2} = \frac{1}{\vec{Q}^2} \frac{\Gamma^2 (\epsilon)}{\Gamma (2 \epsilon)} (\vec{p}^{\; 2})^{\epsilon-1} \;.
    \label{I3calSoft}
\end{equation}
To obtain the soft limit of $\mathcal{I}_{4B}$, eq.~(\ref{I4Bcal}), we restart from the representation
\begin{equation*}
    \mathcal{I}_{4B} = \int_0^1 \frac{dx}{x} (J_B(x)-J_B(0)) \; ,
\end{equation*}
where
\begin{equation}
    J_B (x) = \int_0^1 dz \frac{1-x}{\left[ z(1-x)(a x + b (1-x) (1-z)) \right]^{1-\epsilon}} 
    \;, \;\;\;\;\; a= \vec q_1^{\;2}, \; b= \vec q_2^{\;2} \; .
    \label{JB}
\end{equation}
and we observe that, in the soft region
\begin{equation*}
    \vec{q}_1 \simeq \vec{q}_2 \simeq \frac{\vec{q}_1 + \vec{q}_2}{2} = \vec{Q} \; ,
\end{equation*}
$J_B(x)$ becomes
\begin{equation*}
    J_B(x) \simeq \frac{1}{(\vec{Q}^{2})^{1-\epsilon}} \int_0^1 dz \frac{1}{z^{1-\epsilon} (1-x)^{-\epsilon} [1-z(1-x)]^{1-\epsilon}} = \frac{1}{(\vec{Q}^{2})^{1-\epsilon}} \int_0^{1-x} dz \ [z (1-z)]^{\epsilon-1} 
\end{equation*}
and 
\begin{equation*}
    \mathcal{I}_{4B} = - \frac{1}{(\vec{Q}^{2})^{1-\epsilon}} \int_0^1 \frac{dx}{x} \int_{1-x}^1 dz \; [z  (1-z)]^{\epsilon-1} = - \frac{1}{(\vec{Q}^{2})^{1-\epsilon}} \int_0^1 dz \ [z  (1-z)]^{\epsilon-1} \int_{1-z}^1 \frac{dx}{x}
    \end{equation*}
    \begin{equation*}
    = \frac{\Gamma^2(\epsilon)}{\Gamma (2 \epsilon)} \frac{1}{(\vec{Q}^{2})^{1-\epsilon}} (\psi (\epsilon) - \psi (2 \epsilon)) \;.
\end{equation*}
Alternatively one can use the property of the Euler-Beta function to find
\begin{equation*}
    \mathcal{I}_{4B} = - \frac{1}{(\vec{Q}^{2})^{1-\epsilon}} \int_0^1 \frac{dx}{x} \int_{1-x}^1 dz \; [z  (1-z)]^{\epsilon-1} = - \frac{1}{(\vec{Q}^{2})^{1-\epsilon}} \int_0^1 \frac{dx}{x} \left[ B(\epsilon, \epsilon) - B (1-x, \epsilon, \epsilon) \right] \;,
\end{equation*}
where $B(x,a,b)$ is the incomplete Euler beta function and $B(a,b) = B(1,a,b)$. Introducing the regularized Euler beta function,
\begin{equation}
    I_{1-x}(\epsilon, \epsilon) = \frac{B (1-x, \epsilon, \epsilon)}{B (\epsilon, \epsilon)} \;,
\end{equation}
and using the property $1-I_{1-x}(\epsilon, \epsilon)=I_{x}(\epsilon, \epsilon)$, we find
\begin{equation}
     \mathcal{I}_{4B} = - \frac{B (\epsilon, \epsilon)}{(\vec{Q}^{2})^{1-\epsilon}} \int_0^1 \frac{dx}{x} I_{x}(\epsilon, \epsilon) = \frac{\Gamma^2(\epsilon)}{\Gamma (2 \epsilon)} \frac{1}{(\vec{Q}^{2})^{1-\epsilon}} (\psi (\epsilon) - \psi (2 \epsilon)) \;.
     \label{I4BcalSoft}
\end{equation}
The soft limit of the pentagon integral, eq.~(\ref{I5}), can be immediately obtained and reads~\cite{Fadin:2000yp}
\begin{equation}
    I_5 \simeq \frac{\pi^{2+\epsilon} \Gamma (1-\epsilon)}{s} \frac{\Gamma^2 (\epsilon)}{\Gamma (2 \epsilon)} \frac{(\vec{p}^{\; 2})^{\epsilon-1}}{\vec{Q}^{2}} (\psi (\epsilon) - \psi (1-\epsilon) + i \pi) \; .
\end{equation}
From this, by using eq.~(\ref{I5inTransver}), we obtain
\begin{equation}
    \mathcal{I}_5-\mathcal{L}_3 \simeq  \frac{\Gamma^2 (\epsilon)}{\Gamma (2 \epsilon)} \frac{(\vec{p}^{\; 2})^{\epsilon-1}}{\vec{Q}^{2}} ( \psi (1-\epsilon) - \psi (\epsilon)) \;.
    \label{I5cal-L3calSoft}
\end{equation}
\end{appendices}
\begin{appendices}
\chapter{Further details on the di-hadron production}
\section{Finite parts of virtual corrections}
\label{AppendixD1}

\subsection{Building-block integrals}
\label{sec:building_block}

\begin{eqnarray}
I_{1}^{k}(\vec{q}_1,\, \vec{q}_2,\, \Delta_1,\, \Delta_2) & \equiv & \frac{1}{\pi}\int\frac{d^{d}\vec{l}\left(l_{\perp}^{k}\right)}{\left[(\vec{l}-\vec{q}_{1})^{2}+\Delta_{1}\right]\left[(\vec{l}-\vec{q}_{2})^{2}+\Delta_{2}\right]\vec{l}^{^{\, \, 2}}} \label{I1k}, \\
I_2(\vec{q}_1,\, \vec{q}_2,\, \Delta_1,\, \Delta_2) & \equiv & \frac{1}{\pi}\int \frac{d^d \vec{l}}{\left[ (\vec{l}-\vec{q}_1)^2+\Delta_1 \right] \left[ (\vec{l}-\vec{q}_2)^2 +\Delta_2 \right]} \label{I2}, \\
I_3^k(\vec{q}_1,\, \vec{q}_2,\, \Delta_1,\, \Delta_2) & \equiv & \frac{1}{\pi}\int \frac{d^d \vec{l}\left( l_\bot^k \right)}{\left[ (\vec{l}-\vec{q}_1)^2+\Delta_1 \right] \left[ (\vec{l}-\vec{q}_2)^2 +\Delta_2 \right]} \label{I3k}, \\
I^{jk}(\vec{q}_1,\, \vec{q}_2,\, \Delta_1,\, \Delta_2) & \equiv & \frac{1}{\pi}\int\frac{d^{d}\vec{l}\left( l_{\perp}^j l_{\perp}^{k}\right)}{\left[(\vec{l}-\vec{q}_{1})^{2}+\Delta_{1}\right]\left[(\vec{l}-\vec{q}_{2})^{2}+\Delta_{2}\right]\vec{l}^{^{\, \, 2}}} \label{Ijk} \, .
\end{eqnarray}
The arguments of these integrals will be different for each diagram so we will write them explicitly before giving the expression of each diagram, but we will omit them in the equations for the reader's convenience. \\
Explicit results for the first three integrals in (\ref{I1k}-\ref{Ijk}) are obtained by a straightforward Feynman parameter integration. We will express them using the following variables:
\begin{eqnarray}
\rho_{1} & \equiv & \frac{\left(\vec{q}_{12}^{\, \, 2}+\Delta_{12}\right)-\sqrt{\left(\vec{q}_{12}^{\, \, 2}+\Delta_{12}\right)^{2}+4\vec{q}_{12}^{\, \, 2}\Delta_{2}}}{2\vec{q}_{12}^{\, \, 2}} ,\\
\rho_{2} & \equiv & \frac{\left(\vec{q}_{12}^{\, \, 2}+\Delta_{12}\right)+\sqrt{\left(\vec{q}_{12}^{\, \, 2}+\Delta_{12}\right)^{2}+4\vec{q}_{12}^{\, \, 2}\Delta_{2}}}{2\vec{q}_{12}^{\, \, 2}} \, , \label{rhovar}
\end{eqnarray}
where $\Delta_{ij} = \Delta_i - \Delta_j$ . \\
One gets
\begin{eqnarray}
I_{1}^{k} & = & \frac{q_{1\perp}^{k}}{2\left[\vec{q}_{12}^{\, \, 2}\left(\vec{q}_{1}^{\, \, 2}+\Delta_{1}\right)\left(\vec{q}_{2}^{\, \, 2}+\Delta_{2}\right)-\left(\vec{q}_{1}^{\, \, 2}-\vec{q}_{2}^{\, \, 2}+\Delta_{12}\right)\left(\vec{q}_{1}^{\, \, 2}\Delta_{2}-\vec{q}_{2}^{\, \, 2}\Delta_{1}\right)\right]}\\ \nonumber
 & \times & \left\{ \frac{\left(\vec{q}_{2}^{\, \, 2}+\Delta_{2}\right)\vec{q}_{12}^{\, \, 2}+\vec{q}_{2}^{\, \, 2}\left(\Delta_{1}+\Delta_{2}\right)+\Delta_{2}\left(\Delta_{21}-2\vec{q}_{1}^{\, \, 2}\right)}{\left(\rho_{1}-\rho_{2}\right)\vec{q}_{12}^{\, \, 2}}\ln\left[\left(\frac{-\rho_{1}}{1-\rho_{1}}\right)\left(\frac{1-\rho_{2}}{-\rho_{2}}\right)\right]\right.\\ \nonumber
 & \times & \left.\left(\vec{q}_{2}^{\, \, 2}+\Delta_{2}\right)\ln\left[\frac{\Delta_{2}\left(\vec{q}_{1}^{\, \, 2}+\Delta_{1}\right)^{2}}{\Delta_{1}\left(\vec{q}_{2}^{\, \, 2}+\Delta_{2}\right)^{2}}\right]+\left(1\leftrightarrow2\right)\right\} \, ,
\end{eqnarray}

\begin{eqnarray}
I_{2} & = & \frac{1}{\vec{q}_{12}^{\, \, 2}\left(\rho_{1}-\rho_{2}\right)}\ln\left[\left(\frac{-\rho_{1}}{1-\rho_{1}}\right)\left(\frac{1-\rho_{2}}{-\rho_{2}}\right)\right] \, ,
\end{eqnarray}
and

\begin{eqnarray}
I_{3}^{k} & = & \frac{\left(\vec{q}_{12}^{\, \, 2}+\Delta_{12}\right)q_{1}^{k}+\left(\vec{q}_{21}^{\, \, 2}+\Delta_{21}\right)q_{2}^{k}}{2\left(\rho_{1}-\rho_{2}\right)(\vec{q}_{12}^{\, \, 2})^2}\ln\left[\left(\frac{-\rho_{1}}{1-\rho_{1}}\right)\left(\frac{1-\rho_{2}}{-\rho_{2}}\right)\right] \nonumber \\ &-& \frac{q_{12}^{k}}{2\vec{q}_{12}^{\, \, 2}}\ln\left(\frac{\Delta_{1}}{\Delta_{2}}\right) \, .
\end{eqnarray}
Please note that in some cases the real part of $\Delta_1$ or $\Delta_2$ will be negative so the previous results can acquire an imaginary part from the imaginary part $\pm \, i0$ of the arguments. \\ 
The last integral in (\ref{Ijk}) can be expressed in terms of the other ones by writing 
\begin{equation}
I^{jk} = I_{11}\left(q_{1\perp}^{j}q_{1\perp}^{k}\right)+I_{12}\left(q_{1\perp}^{j}q_{2\perp}^{k}+q_{2\perp}^{j}q_{1\perp}^{k}\right)+I_{22}\left(q_{2\perp}^{j}q_{2\perp}^{k}\right) \, ,
\end{equation}
with
\begin{align}
I_{11} & = -\frac{1}{2}\frac{\left[\vec{q}_{2}^{\, \, 2}q_{1\perp k}-\left(\vec{q}_{1}\cdot\vec{q}_{2}\right)q_{2\perp k}\right]}{\left[\vec{q}_{1}^{\, \, 2}\vec{q}_{2}^{\, \, 2}-\left(\vec{q}_{1} \cdot \vec{q}_{2}\right)^{2}\right]^{2}} \\ \nonumber
& \hspace{-0.25 cm} \times  \left[\left(\frac{\vec{q}_{1}\cdot\vec{q}_{2}}{\vec{q}_{1}^{\, \, 2}}\right)\ln\left(\frac{\vec{q}_{1}^{\, \, 2}+\Delta_{1}}{\Delta_{1}}\right)q_{1\perp}^{k}+\left(\vec{q}_{2}\cdot\vec{q}_{21}\right)I_{3}^{k}+\left\{ \vec{q}_{2}^{\, \, 2}\left(\vec{q}_{1}\cdot\vec{q}_{12}\right)+\Delta_{1}\vec{q}_{2}^{\, \, 2}-\Delta_{2}\left(\vec{q}_{1}\cdot\vec{q}_{2}\right)\right\} I_{1}^{k}\right] \; , \\
I_{12} & = \frac{-1}{4\left[\vec{q}_{1}^{\, \, 2} \vec{q}_{2}^{\, \, 2} -\left(\vec{q}_1 \cdot \vec{q}_2\right)^2\right]} \ln \left(\frac{\vec{q}_{1}^{\, \, 2}+\Delta_1}{\Delta_1}\right)  \nonumber \\
&+ \frac{\vec{q}_{2}^{\, \, 2} \left(\vec{q}_1 \cdot \vec{q}_2\right)}{2\left[\vec{q}_{1}^{\, \, 2} \vec{q}_{2}^{\, \, 2} -\left(\vec{q}_1 \cdot \vec{q}_2\right)^2\right]^2}\left[\left(\vec{q}_{1}^{\, \, 2} +\Delta_1\right)\left(q_{1 \perp k} I_1^k\right)+\left(q_{1 \perp k} I_3^k\right)\right] \nonumber\\
& -\frac{\left(\vec{q}_{1}^{\, \, 2} \vec{q}_{2}^{\, \, 2} \right)+\left(\vec{q}_1 \cdot \vec{q}_2\right)^2}{4\left[\vec{q}_{1}^{\, \, 2} \vec{q}_{2}^{\, \, 2} -\left(\vec{q}_1 \cdot \vec{q}_2\right)^2\right]^2}\left[\left(\vec{q}_{2}^{\, \, 2} +\Delta_2\right)\left(q_{1 \perp k} I_1^k\right)+\left(q_{1 \perp k} I_3^k\right)\right]+(1 \leftrightarrow 2),  \\
I_{22}&  = I_{11}|_{1 \leftrightarrow 2} \, .
\end{align}
In what follows, for the $\phi$'s functions we use the variables $x, \bar{x}$. In the main text these should be identified as $x= x_q$ and $\bar{x} = x_{\bar{q}}$.

\subsection{$\phi_4$}

The arguments in the integrals of \ref{sec:building_block} are 
\begin{eqnarray*}
\vec{q}_{1} & = & \vec{p}_{1}-\left(\frac{x-z}{x}\right)\vec{p}_{q}, \quad \, \, \, \, \,
\vec{q}_{2}  =  \left(\frac{x-z}{x}\right)\left(x\vec{p}_{\bar{q}}-\bar{x}\vec{p}_{q}\right) \, ,\\
\Delta_{1} & = & \left(x-z\right)\left(\bar{x}+z\right)Q^{2}, \quad
\Delta_{2}  =  -\frac{x\left(\bar{x}+z\right)}{\bar{x}\left(x-z\right)}\vec{q}^{\; 2}-i0\,.
\end{eqnarray*}
Let us write the impact factors in terms of these variables. \\They read: \vspace{0.2 cm} \\
(longitudinal NLO) $\times$ (longitudinal LO) contribution:
\begin{equation}
\left(\phi_{4}\right)_{LL}=-\frac{4(x-z)(\bar{x}+z)}{z}[-\bar{x}(x-z)(z+1)I_{2}+q_{2\bot k}(2x^{2}-(2x-z)(z+1))I_{1}^{k}] \, ,
\end{equation}
(longitudinal NLO) $\times$ (transverse LO) contribution:
\begin{equation}
\left(\phi_{4}\right)_{LT}^{j}=(1-2x)p_{q1^{\prime}}{}_{\bot}^{j}\left(\phi_{4}\right)_{LL}-4(x-z)(\bar{x}+z)(1-2x+z)[(\vec{q}\cdot\vec{p}_{q1^{\prime}})g_{\bot k}^{j}+q_{2\bot}^{j}p_{q1^{\prime}\bot k}]I_{1}^{k} \, ,
\end{equation}
(transverse NLO) $\times$ (longitudinal LO) contribution:
\begin{align}
\left(\phi_{4}\right)_{TL}^{i} & =2\{[(x-\bar{x}-z)q_{2\bot}^{i}q_{1\bot k}+(-8x\bar{x}-6xz+2z^{2}+3z+1)q_{1\perp}^{j}q_{2\bot k}]I_{1}^{k}\nonumber \\
 & -2[4x^{2}-x(3z+5)+(z+1)^{2}]q_{2\bot k}I^{ik}+(x-\bar{x}-z)\left(\vec{q}_{2}\cdot\vec{q}_{1}\right)I^{i}\nonumber \\
 & +I_{2}[(x-\bar{x}-z)q_{2\bot}^{i}+\bar{x}(2(x-z)^{2}-5x+3z+1)q_{1\perp}^{i}]\nonumber \\
 & -\bar{x}[2(x-z)^{2}-5x+3z+1]I_{3}^{i}\nonumber \\
 & +\frac{x\bar{x}(1-2x)}{z}[2q_{2\bot k}I^{ik}+I_{3}^{i}-q_{1\perp}^{i}(2q_{2\bot k}I_{1}^{k}+I_{2})]\} \, ,
\end{align}
(transverse NLO) $\times$ (transverse LO) contribution:
\begin{eqnarray} 
\allowdisplaybreaks
\nonumber
\left(\phi_{4}\right)_{TT}^{ij} & = & \left[(x-\bar{x}-2z)(x-\bar{x}-z)(\vec{q}_{2}\cdot\vec{p}_{q1^{\prime}})q_{1\perp}^{i}+(z+1)(\left(\vec{q}_{1}\cdot\vec{q}_{2}\right)p_{q1^{\prime}\perp}^{i}-(\vec{q}_{1}\cdot\vec{p}_{q1^{\prime}})q_{2\bot}^{i})\right]I_{1}^{j}\\ \nonumber
 &+& 2\bar{x}[q_{2\bot k}-(x-z)q_{1\perp k}](p_{q1^{\prime}\bot}^{i}I^{jk}-g_{\bot}^{ij}p_{q1^{\prime}\bot l}I^{kl}) \\ \nonumber
 &+& 2(x-z)[(2\bar{x}+z)(\vec{q}_{2}\cdot\vec{p}_{q1^{\prime}})-\bar{x}(\vec{q}_{1}\cdot\vec{p}_{q1^{\prime}})]I^{ij}\\ \nonumber
 &+& [(1-z)((\vec{q}_{1}\cdot\vec{p}_{q1^{\prime}})q_{2\bot}^{j}-(\vec{q}_{2}\cdot\vec{p}_{q1^{\prime}})q_{1\perp}^{j})-(1-2x)(\bar{x}-x+z)\left(\vec{q}_{1}\cdot\vec{q}_{2}\right)p_{q1^{\prime}\perp}^{j}]I_{1}^{i}\\ \nonumber
 &-& 2\left[(x-z)(\bar{x}q_{1\perp}^{j}-(2\bar{x}+z)q_{2\bot}^{j})p_{q1^{\prime}\perp k} \right. \\ \nonumber 
 &+& \left. (1-2x)\left(4x^{2}-(3z+5)x+(z+1)^{2}\right)q_{2\bot k}p_{q1^{\prime}}{}_{\bot}^{j}\right]I^{ik}\\ \nonumber
 &-& \bar{x}\left(\bar{x}-x\right)\left(2(x-z)^{2}-5x+3z+1\right)p_{q1^{\prime}\perp}^{j}I_{3}^{i} \\ \nonumber
 &+& \bar{x}\left(\bar{x}+z\right)(p_{q1^{\prime}\perp}^{i}I_{3}^{j}-g_{\bot}^{ij}p_{q1^{\prime}\perp k}I_{3}^{k})\\ \nonumber
 &+& I_{2}\left[g_{\bot}^{ij}\left((1-z)(\vec{q}_{2}\cdot\vec{p}_{q1^{\prime}})-\bar{x}(1+x-z)(\vec{q}_{1}\cdot\vec{p}_{q1^{\prime}})\right) \right. \\ \nonumber 
 &+& \left.((1-z)q_{2\bot}^{j}-\bar{x}(1+x-z)q_{1\perp}^{j})p_{q1^{\prime}}{}_{\bot}^{i}\right.\\ \nonumber
 &-& \left.(\bar{x}-x)\left((\bar{x}-x+z)q_{2\bot}^{i}-\bar{x}\left(2(x-z)^{2}-5x+3z+1\right)q_{1\perp}^{i}\right)p_{q1^{\prime}}{}_{\bot}^{j}\right]\\ \nonumber
 &+& I_{1}^{k}\left[g_{\bot}^{ij}\left((x-\bar{x}+z)(\vec{q}_{1}\cdot\vec{p}_{q1^{\prime}})q_{2\bot k}+(1-z)(\vec{q}_{2}\cdot\vec{p}_{q1^{\prime}})q_{1\bot k}-(z+1)\left(\vec{q}_{1}\cdot\vec{q}_{2}\right)p_{q1^{\prime}}{}_{\bot k}\right)\right.\\ \nonumber
 &+& q_{1\perp}^{j}((x-\bar{x}+z)q_{2\bot k}p_{q1^{\prime}\perp}^{i}-(z+1)q_{2\bot}^{i}p_{q1^{\prime}}{}_{\bot k})\\ \nonumber
 &+& q_{2\bot}^{j}((x-\bar{x}-2z)(x-\bar{x}-z)q_{1\perp}^{i}p_{q1^{\prime}\perp k}+(1-z)q_{1\perp k}p_{q1^{\prime}}{}_{\bot}^{i})\\ \nonumber
&-&\left.(1-2x)((1-2x+z)q_{2\bot}^{i}q_{1\perp k}-(2z^{2}+3z-x(8\bar{x}+6z)+1)q_{1\perp}^{i}q_{2\bot k})p_{q1^{\prime}}{}_{\bot}^{j}\right]\\ \nonumber
 &+& \frac{x\bar{x}}{z}\left[(x-\bar{x})^{2}p_{q1^{\prime}\perp}^{j}(2q_{2\bot k}I^{ik}+I_{3}^{i}-q_{1\perp}^{i}(I_{2}+2q_{2\bot k}I_{1}^{k}))\right.\\ \nonumber
 &+& p_{q1^{\prime}\perp}^{i}(q_{1\perp}^{j}(I_{2}+2q_{2\bot k}I_{1}^{k})-2q_{2\bot k}I^{jk}-I_{3}^{j})\\
 &+& \left.g_{\bot}^{ij}((\vec{q}_{1}\cdot\vec{p}_{q1^{\prime}})(I_{2}+2q_{2\bot k}I_{1}^{k})+p_{q1^{\prime}\perp k}(2q_{2\bot l}I^{kl}+I_{3}^{k}))\right]\, .
\end{eqnarray} 

\subsection{$\phi_5$}

Here the integrals from \ref{sec:building_block} will have the following arguments :

\begin{equation}
\vec{q}_1 = \left( \frac{x-z}{x} \right) \vec{p}_3 -\frac{z}{x}\vec{p}_1, \quad \vec{q}_2 = \vec{p}_{q1} - \frac{z}{x}\vec{p}_q \, ,\label{var1D5}
\end{equation}
\begin{equation}
 \Delta_1 = \frac{z(x-z)}{x^2\bar{x}} (\vec{p}_{\bar{q}2}^{\, \, 2}+ x\bar{x}Q^2), \quad  \Delta_2 = (x-z)(\bar{x}+z)Q^2 \label{var2D5}\, .
\end{equation}
With such variables, it is easy to see that the argument in the square roots in (\ref{rhovar}) are full squares.
In terms of the variables in (\ref{var1D5}), the impact factors read, \vspace{0.2 cm} \\
(longitudinal NLO) $\times$ (longitudinal LO): 
\begin{gather}
\left(\phi_{5}\right)_{LL}=\frac{4(x-z)(-2x(\bar{x}+z)+z^{2}+z)}{xz}\left[\bar{x}(x-z)I_{2}-\left(zq_{1\perp k}-x\left(\bar{x}+z\right)q_{2\bot k}\right)I_{1}^{k}\right] \, ,
\end{gather}
(longitudinal NLO) $\times$ (transverse LO) : 
\begin{align}
\left(\phi_{5}\right)_{LT}^{j} & =(\bar{x}-x)p_{q1^{\prime}\bot}^{j}\left(\phi_{5}\right)_{LL} \\ \nonumber &+\frac{4(x-z)(x-\bar{x}-z)}{x}\left(zq_{1\perp}^{k}-x(\bar{x}+z)q_{2\perp}^{k}\right)p_{q1^{\prime}\perp l}\left(g_{\perp k}^{j}I_{1}^{l}+I_{1}^{j}\right)\, ,
\end{align}
(transverse NLO) $\times$ (longitudinal LO) : 
\begin{align}
\left(\phi_{5}\right)_{TL}^{i} & =2\left[(x-\bar{x}-z)\left(\vec{q}_{1}\cdot\vec{q}_{2}\right)-\bar{x}(x-z)^{2}Q^{2}+(\frac{z}{x}-x)\vec{q}_{1}^{\,\,2}\right]I_{1}^{i}\nonumber \\
 & +\frac{2}{x}\left[xq_{2\bot k}(-8x\bar{x}-6xz+2z^{2}+3z+1)+2q_{1\bot k}(2xz-2x^{2}+x-z^{2})\right]q_{1\perp}^{i}I_{1}^{k}\nonumber \\
 & +2q_{2\bot}^{i}q_{1\perp k}(x-\bar{x}-z)I_{1}^{k}+2\frac{\bar{x}}{x}(x(8x-3)-6xz+2z^{2}+z)I_{1}^{i}\nonumber \\
 & +\frac{2}{x}\left[xq_{2\bot}^{i}(x-\bar{x}-z)+q_{1\perp}^{i}(8x^{3}-6x^{2}(z+2)+x(z+3)(2z+1)-2z^{2})\right]I_{2}\nonumber \\
 & -\frac{4}{x}\left[(x-z)(\bar{x}+z)q_{1\perp k}+x(4x^{2}-x(3z+5)+(z+1)^{2})q_{2\perp k}\right]I^{ik}\nonumber \\
 & -\frac{4}{z}x\bar{x}(x-\bar{x})\left[q_{2\perp k}I^{ik}+I_{3}^{i}-q_{1\perp}^{i}\left(q_{2\perp k}I_{1}^{k}+I_{2}\right)\right] \, ,
\end{align}
(transverse NLO) $\times$ (transverse LO) : 

\begin{align*}
& \left(\phi_{5}\right)_{TT}^{ij}  =  -2(x-z)\left[\frac{z}{x}(\vec{q}_{1}\cdot\vec{p}_{q1^{\prime}})-(2\bar{x}+z)(\vec{q}_{2}\cdot\vec{p}_{q1^{\prime}})\right]I^{ij}\\ \nonumber
 & +  \left[-\bar{x}(x-z)^{2}Q^{2}p_{q1^{\prime}\perp}^{i}+(\bar{x}-x+2z)(\bar{x}-x+z)(\vec{q}_{2}\cdot\vec{p}_{q1^{\prime}})q_{1\perp}^{i}\right.\\
 & - \left.(\vec{q}_{1}\cdot\vec{p}_{q1^{\prime}})((z+1)q_{2\bot}^{i}-2\frac{z}{x}(2x-z)q_{1\perp}^{i}) \right. \\ \nonumber
 &+ \left. ((z+1)\left(\vec{q}_{1}\cdot\vec{q}_{2}\right)-\left(x+\frac{z}{x}\right)\vec{r}^{\,\,2})p_{q1^{\prime}\bot}^{i}\right]I_{1}^{j}\\ \nonumber
 & -  2\frac{\bar{x}}{x}(xq_{2\perp k}+(x-z)q_{1\perp k})\left(g_{\bot}^{ij}p_{q1^{\prime}\perp l}I^{kl}-p_{q1^{\prime}}{}_{\bot}^{i}I^{jk}\right)\\ \nonumber
 & + \left[\bar{x}\left(x-\bar{x}\right)(x-z)^{2}Q^{2}p_{q1^{\prime}\perp}^{j}-(z-1)(\vec{q}_{1}\cdot\vec{p}_{q1^{\prime}})q_{2\bot}^{j}\right.\\  \nonumber
 & +  \left.(z-1)(\vec{q}_{2}\cdot\vec{p}_{q1^{\prime}})q_{1\perp}^{j}+\frac{x-\bar{x}}{x}\left((x^{2}-z)\vec{q}_{1}^{\,\,2}+x(\bar{x}-x+z)(\vec{q}_{1}\cdot\vec{q}_{2})\right)p_{q1^{\prime}\perp}^{k}\right]I_{1}^{i}\\ \nonumber
 & +  2\left[\frac{x-\bar{x}}{x}\left(x(4x^{2}-(3z+5)x+(z+1)^{2})q_{2\bot k}+(x-z)(\bar{x}+z)q_{1\perp k}\right)p_{q1^{\prime}\perp}^{j}\right.\\
 & -  \left.\frac{x-z}{x}\left(x(2x-z-2)q_{2\bot}^{j}+zq_{1\perp}^{j}\right)p_{q1^{\prime}\perp k}\right]I^{ik} \\ \nonumber 
 & +  \frac{\bar{x}\left(\bar{x}-x\right)}{x}\left(2z^{2}-6xz+z+x(8x-3)\right)p_{q1^{\prime}\perp}^{j}I_{3}^{i}\\ \nonumber
 & + \left[(x-\bar{x})\left((\bar{x}-x+z)q_{2\bot}^{i}+\left(6(z+2)x-8x^{2}-(z+3)(2z+1)+2\frac{z^{2}}{x}\right)q_{1\perp}^{i}r_{\bot}^{i}\right)p_{q1^{\prime}\perp}^{j}\right.\\ \nonumber
 & +  \left.(1-z)(g_{\bot}^{ij}(\vec{q}_{2}\cdot\vec{p}_{q1^{\prime}})+q_{2\bot}^{k}p_{q1^{\prime}\perp}^{i})+(2x+z-3)(g_{\bot}^{ik}(\vec{q}_{1}\cdot\vec{p}_{q1^{\prime}})+q_{1\perp}^{k}p_{q1^{\prime}\perp}^{i})\right]I_{2}\\ \nonumber
 & +  \left(3\bar{x}+z-\frac{z}{x}\right)p_{q1^{\prime}\perp}^{i}I_{3}^{k}-\frac{\bar{x}}{x}(3x-z)g_{\bot}^{ij}p_{q1^{\prime}\perp k}I_{3}^{k}\\
 & +  \left[(x-\bar{x})p_{q1^{\prime}\perp}^{j}\left\{ (\bar{x}-x+z)q_{2\bot}^{i}q_{1\perp k}-(2z^{2}-6xz+3z-8x\bar{x}+1)q_{2\perp k}q_{1\perp}^{i}\right.\right.\\ \nonumber
 & -  \left. 2(\bar{x}-x+2z-\frac{z^{2}}{x})q_{1\perp k}q_{1\perp}^{i}\right\} +\bar{x}(x-z)^{2}Q^{2}g_{\bot}^{ij}p_{q1^{\prime}\perp k} \\ \nonumber 
 &+ (1-z)q_{1\perp k}(g_{\bot}^{ij}(\vec{q}_{2}\cdot\vec{p}_{q1^{\prime}})+q_{2\bot}^{j}p_{q1^{\prime}\perp}^{i})\\ \nonumber
 & +  \left((x-\bar{x}+z)q_{2\bot k}-2q_{1\perp k}\right)(g_{\bot}^{ij}(\vec{q}_{1}\cdot\vec{p}_{q1^{\prime}})+q_{1\perp}^{j}p_{q1^{\prime}\perp}^{i}) \\ \nonumber 
 &+ g_{\bot}^{ij}\left(\left(x+\frac{z}{x}\right)\vec{q}_{1}^{\,\,2}-(z+1)(\vec{q}_{1}\cdot\vec{q}_{2})\right)p_{q1^{\prime}\perp k}\\ \nonumber
 & + \left.\left((x-\bar{x}-2z)(x-\bar{x}-z)q_{1\perp}^{i}q_{2\perp}^{j}-(z+1)q_{2\perp}^{i}q_{1\perp}^{j}+2(2x-z)\frac{z}{x}q_{1\perp}^{i}q_{1\perp}^{j}\right)p_{q1^{\prime}\perp k}\right]I_{1}^{k}\\ \nonumber
 & +  \frac{2x\bar{x}}{z}\left[(x-\bar{x})^{2}p_{q1^{\prime}\perp}^{j}(q_{2\bot k}I^{ik}+I_{3}^{i})-p_{q1^{\prime}\perp}^{i}(q_{2\bot k}I^{jk}+I_{3}^{k})+g_{\bot}^{ij}p_{q1^{\prime}\bot k}(q_{2\bot l}I^{kl}+I_{3}^{k})\right.\\ \nonumber
 & +  \left.(I_{2}+q_{2\perp k}I_{1}^{k})\left(g_{\bot}^{ij}(\vec{q}_{1}\cdot\vec{p}_{q1^{\prime}})+q_{1\perp}^{j}p_{q1^{\prime}\perp}^{i}-(1-2x)^{2}q_{1\perp}^{i}p_{q1^{\prime}\perp}^{j}\right)\right]. \numberthis
\end{align*}

\subsection{$\phi_6$}
We will use the variable
\begin{equation}
\vec{q}=\left( \frac{x-z}{x} \right) \vec{p}_{3}-\frac{z}{x}\vec{p}_1 \, .
\end{equation}
In terms of these variable we have
(longitudinal NLO) $\times $ (longitudinal NLO) :
\begin{equation}
(\phi_6)_{LL}=-4x\bar{x}^2 J_0 \, ,
\end{equation}
(longitudinal NLO) $\times $ (transverse NLO) :
\begin{equation}
(\phi_6)_{LT}^j = (1-2x)p_{q1^\prime\bot}^j(\phi_6)_{LL}\, ,
\end{equation}
(transverse NLO) $\times $ (longitudinal NLO) :
\begin{equation}
(\phi_6)_{TL}^i = 2\bar{x}\left[ (1-2x) p_{\bar{q}2\bot}^{i} J_0 - J_{1\bot}^i \right] \, ,
\end{equation}
(transverse NLO) $\times $ (transverse NLO) :
\begin{align}
(\phi_6)_{TT}^{ij} &  =\bar{x}\left[(x-\bar{x})^{2}p_{\bar{q}2\bot}^{i}%
p_{q1^{\prime}\bot}^{j}-g_{\bot}^{ij}(\vec{p}_{\bar{q}2}
\cdot\vec{p}_{q1^{\prime}})-p_{q1^{\prime}\bot}^{i}p_{\bar{q}2\bot}^{j}\right]J_0 \nonumber\\
&  +\bar{x} \left[(x-\bar{x})p_{q1^{\prime}\bot}^{j}g_{\bot k}^i - p_{q1^\prime\bot k}g_{\bot}^{ij}+p_{q1^{\prime
}\bot}^{i}g_{\bot k}^j\right]J_{1\bot}^k \, .
\end{align}
We introduced
\begin{align}
J_{1\bot}^{k}  &  =\frac{(x-z)^{2}}{x^{2}}\frac{q_{\bot}^{k}}{\vec
{q}^{\,\,2}}\ln\left(  \frac{\vec{p}_{\bar{q}2}^{\,\,2}+x\bar{x}Q^{2}}{\vec{p}_{\bar{q}2}^{\,\,2}+x\bar{x}Q^{2}+	\frac{x^2 \bar{x}}{z(x-z)}\vec{q}^{\,\,2}%
}\right)  ,\\ \nonumber
\mathrm{and} \\
J_0 &  =\frac{z}{x(\vec{p}_{\bar{q}2}^{\,\,2}+x\bar{x}Q^{2})}  -\frac{2x(x-z)+z^{2}}{xz(\vec{p}_{\bar{q}2}^{\,\,2}+x\bar{x}Q^{2})}
\ln\left(  \frac{x^2\bar{x}\mu^{2}}{z(x-z)(\vec{p}_{\bar{q}2}^{\,\,2}+x\bar{x}Q^{2})+x^{2}\bar{x}\vec{q}^{\,\,2}}\right)  .
\end{align}

\section{Finite part of squared impact factors}
\label{AppendixD2}
In this section, we present the finite part of squared impact factors for the real corrections case.
 \subsection{LL transition}
The double-dipole $\times$ double-dipole contribution is
\begin{align}
\label{phi4plus_squared}
\Phi &  _{4}^{+}(p_{1\bot},p_{2\bot},p_{3\bot})\Phi_{4}^{+*}(p_{1\bot}^{\prime
},p_{2\bot}^{\prime},p_{3\bot}^{\prime})=\frac{8p_{\gamma}^{+}{}^{4}%
}{z^{2}\left(  \frac{\vec{p}_{\bar{q}2^\prime}^{\, \, 2}%
}{x_{\bar{q}}\left( 1 - x_{\bar{q}} \right) }+Q^{2}\right)  \left( Q^2+\frac{\vec{p}_{q1^{\prime}}^{\,\,2}}{x_{q}%
} + \frac{\vec{p}_{\bar{q}2^\prime}^{\, \, 2}}{x_{\bar{q}}}+\frac
{\vec{p}_{g3^{\prime}}^{\,\,2}}{z}\right)  }\nonumber\\
\times &  \left[  \frac{x_{\bar{q}}\left(  dz^{2}+4x_{q}\left(  x_{q}%
+z\right)  \right)  \left(  x_{q}\vec{p}_{g3}-z\vec{p}_{q1})(x_{q}\vec
{p}_{g3^{\prime}}-z\vec{p}_{q1^{\prime}}\right)  }{x_{q}\left(  x_{q}%
+z\right)  ^{2}{}\left(  \frac{(\vec{p}_{g3}+\vec{p}_{q1}){}^{2}}{x_{\bar{q}%
}\left(  x_{q}+z\right)  }+Q^{2}\right)  \left(  \frac{(\vec{p}_{g3}+\vec
{p}_{q1}){}^{2}}{x_{\bar{q}}}+\frac{\vec{p}_{g3}^{\,\,2}}{z}+\frac{\vec{p}_{q1}%
^{\,\,2}}{x_{q}}+Q^{2}\right)  }\right. \nonumber\\
-  &  \left.  \frac{(4x_{q}x_{\bar{q}}+2z-dz^{2})(x_{\bar{q}}\vec{p}%
_{g3}-z\vec{p}_{\bar{q}2})(x_{q}\vec{p}_{g3^{\prime}}-z\vec{p}_{q1^{\prime}}%
)}{\left(  x_{\bar{q}}+z\right)  \left(  x_{q}+z\right)  \left(  \frac
{(\vec{p}_{\bar{q}2}+\vec{p}_{g3}){}^{2}}{x_{q}\left(  x_{\bar{q}}+z\right)
}+Q^{2}\right)  \left(  \frac{(\vec{p}_{\bar{q}2}+\vec{p}_{g3}){}^{2}}{x_{q}%
}+\frac{\vec{p}_{g3}^{\,\,2}}{z}+\frac{\vec{p}_{\bar{q}2}^{\,\,2}}{x_{\bar{q}}%
}+Q^{2}\right)  }\right]  +(q\leftrightarrow\bar{q}).
\end{align}
The interference term in the dipole $\times$ dipole contribution reads
{\allowdisplaybreaks
\begin{align*}
& \left( \tilde{\Phi}_3^+(\vec{p}_1, \vec{p}_2) \Phi_4^{+*}(\vec{p}_{1'}, \vec{p}_{2'},\vec{0}) +\Phi_4^+(\vec{p}_1, \vec{p}_2,\vec{0}) \tilde{\Phi}_3^{+*}(\vec{p}_{1'},\vec{p}_{2'})\right) \\
&  =\left[  \frac{8p_{\gamma}^{+}{}^{4}}{z\left(  x_{q}+z\right) \left(  \frac{\vec{p}{}_{\bar{q}2^{\prime}}^{\,\,2}}{x_{\bar{q}}\left(x_{q}+z\right)  }+Q^{2}\right)  \left(  \frac{\vec{p}{}_{q1^{\prime}}^{\,\,2}%
}{x_{q}}+\frac{\vec{p}{}_{\bar{q}2^{\prime}}^{\,\,2}}{x_{\bar{q}}}+\frac
{\vec{p}_{g}{}^{2}}{z}+Q^{2}\right)  }\right. \nonumber\\
&  \times\left\{  \frac{\left(  4x_{q}x_{\bar{q}}+z(2-dz)\right)  (\vec{p}_{g}%
-\frac{z}{x_{\bar{q}}}\vec{p}_{\bar{q}})(x_{q}\vec{p}_{g}-z\vec
{p}_{q1^{\prime}})}{(\vec{p}_{g}-\frac{z\vec{p}_{\bar{q}}}{x_{\bar{q}}}){}%
^{2}\left(  \frac{\vec{p}{}_{q1'}^{\,\,2}}{x_{q}\left(  x_{\bar{q}}+z\right)
}+Q^{2}\right)  }\right. \nonumber\\
&  -\left.  \left.  \frac{x_{\bar{q}}\left(  dz^{2}+4x_{q}\left(
x_{q}+z\right)  \right)  ({}\vec{p}_{g}-\frac{z}{x_{q}}\vec{p}_{q})(\vec{p}_{g}%
-\frac{z}{x_{q}}\vec{p}_{q1^{\prime}})}{(\vec{p}_{g}-\frac{z\vec{p}_{q}%
}{x_{q}}){}^{2}\left(  \frac{\vec{p}{}_{\bar{q}2}^{\,\,2}}{x_{\bar{q}}\left(
x_{q}+z\right)  }+Q^{2}\right)  }\right\}  +(q\leftrightarrow\bar{q})\right]
\nonumber\\
&  +(1\leftrightarrow1^{\prime},2\leftrightarrow2^{\prime}). \numberthis
\end{align*}}
The double-dipole $\times$ dipole contribution has the form 

\begin{equation}
\Phi_{4}^{+}( \vec{p}_1, \vec{p}_2, \vec{p}_3 )\,\Phi_{3}^{+*}(\vec{p}_{1'}, \vec{p}_{2'})=\Phi_{4}^{+} ( \vec{p}_1, \vec{p}_2, \vec{p}_3 ) \Phi_{4}^{+*}(\vec{p}_{1'}, \vec{p}_{2'}, \vec{0})+\Phi_4^+(\vec{p}_1, \vec{p}_2, \vec{p}_3) \tilde{\Phi}_3^{+*}(\vec{p}_{1'}, \vec{p}_{2'}) ,
\end{equation}
where
\begin{align*}
\Phi_4^+(\vec{p}_1, \vec{p}_2, \vec{p}_3) & \tilde{\Phi}_3^{+*}(\vec{p}_{1'}, \vec{p}_{2'})  =\frac{8p_{\gamma}^{+}{}^{4}}{z\left(  x_{q}+z\right)  \left(
\frac{\vec{p}{}_{\bar{q}2}^{\,\,2}}{x_{\bar{q}}\left(  x_{q}+z\right)  }%
+Q^{2}\right)  \left(  \frac{\vec{p}{}_{q1}^{\,\,2}}{x_{q}}+\frac{\vec{p}%
{}_{\bar{q}2}^{\,\,2}}{x_{\bar{q}}}+\frac{\vec{p}_{g3}^{\,\,2}}{z}+Q^{2}\right)
}\nonumber\\
&  \times\left\{  \frac{\left(  4x_{q}x_{\bar{q}}+z(2-dz)\right)  (\vec{p}_{g}%
-\frac{z}{x_{\bar{q}}}\vec{p}_{\bar{q}})(x_{q}\vec{p}_{g3}-z\vec{p}_{q1}%
)}{(\vec{p}_{g}-\frac{z\vec{p}_{\bar{q}}}{x_{\bar{q}}}){}^{2}\left(
\frac{\vec{p}{}_{q1^{\prime}}^{\,\,2}}{x_{q}\left(  x_{\bar{q}}+z\right)
}+Q^{2}\right)  }\right. \nonumber\\
&  -\left.  \frac{x_{\bar{q}}\left(  dz^{2}+4x_{q}\left(  x_{q}+z\right)
\right)  (\vec{p}_{g}-\frac{z}{x_{q}}\vec{p}_{q})(\vec{p}_{g3}-\frac{z}{x_{q}}%
\vec{p}_{q1})}{(\vec{p}_{g}-\frac{z\vec{p}_{q}}{x_{q}}){}^{2}\left(
\frac{\vec{p}{}_{\bar{q}2^{\prime}}^{\,\,2}}{x_{\bar{q}}\left(  x_{q}%
+z\right)  }+Q^{2}\right)  }\right\}  +( q \leftrightarrow \bar{q} ). \numberthis[finite_double_dipole_dipole_LL]
\end{align*}
For the dipole $\times$ double-dipole contribution, one just has to complex conjugate \eqref{eq:finite_double_dipole_dipole_LL} and also invert the name of the momenta i.e. $1',2' \leftrightarrow 1,2$. 

\subsection{LT/TL transition}
The double-dipole $\times$ double-dipole contribution is 
{\allowdisplaybreaks
\begin{align*}
&  \Phi_{4}^{i}(p_{1\bot},p_{2\bot},p_{3\bot})\Phi_{4}^{+*}(p_{1\bot}^{\prime
},p_{2\bot}^{\prime},p_{3\bot}^{\prime}) \\
& =\frac{-4p_{\gamma}^{+}{}^{3}%
}{\left(  Q^{2}+\frac{\vec{p}{}_{g3}^{\,\,2}}{z}+\frac{\vec{p}{}_{q1}^{\,\,2}%
}{x_{q}}+\frac{\vec{p}{}_{{\bar{q}}2}^{\,\,2}}{x_{\bar{q}}}\right)  \left(
Q^{2}+\frac{\vec{p}{}_{g3^{\prime}}^{\,\,2}}{z}+\frac{\vec{p}{}_{q1^{\prime}%
}^{\,\,2}}{x_{q}}+\frac{\vec{p}{}_{{\bar{q}}2^{\prime}}^{\,\,2}}{x_{\bar{q}}%
}\right)  }\nonumber\\
& \hspace{-0.1 cm} \times \hspace{-0.1 cm} \left(  \frac{z\left(  (\vec{P} \hspace{-0.1 cm}  \cdot \hspace{-0.1 cm}  \vec{p}_{q1})G_{\bot}^{i} \hspace{-0.1 cm}  - \hspace{-0.1 cm} (\vec
{G} \hspace{-0.1 cm}  \cdot  \hspace{-0.1 cm} \vec{p}_{q1})P_{\bot}^{i}\right)  \left(  dz+4x_{q}-4\right)  -(\vec{G} \cdot 
\vec{P})p_{q1}^{i}{}_{\bot}\left(  2x_{q}-1\right)  \left(  4\left(
x_{q}-1\right)  x_{\bar{q}}-dz^{2}\right)  }{z^{2}x_{\bar{q}}\left(
z+x_{\bar{q}}\right)  {}^{3}\left(  Q^{2}+\frac{\vec{p}{}_{q1}^{\,\,2}}%
{x_{q}\left(  z+x_{\bar{q}}\right)  }\right)  \left(  Q^{2}+\frac{\vec{p}%
{}_{q1^{\prime}}^{\,\,2}}{x_{q}\left(  z+x_{\bar{q}}\right)  }\right)
}\right. \nonumber\\
 &  + \hspace{-0.05 cm}  \frac{z\left(  (\vec{P} \hspace{-0.1 cm} \cdot \hspace{-0.1 cm}  \vec{p}_{q1})H_{\bot}^{i} \hspace{-0.1 cm} -(\vec{H} \cdot \vec{p}_{q1})P_{\bot}^{i}\right)  \left(  dz+4x_{q}-2\right)  -(\vec{H} \cdot \vec{P}%
)p_{q1}^{i}{}_{\bot}\left(  2x_{q}-1\right)  \left(  z(2-dz)+4x_{q}x_{\bar{q}%
}\right)  }{z^{2}x_{q}\left(  z+x_{q}\right)  \left(  z+x_{\bar{q}}\right)
{}^{2}\left(  Q^{2}+\frac{\vec{p}{}_{\bar{q}2^{\prime}}^{\,\,2}}{\left(
z+x_{q}\right)  x_{\bar{q}}}\right)  \left(  Q^{2}+\frac{\vec{p}{}%
_{q1}^{\,\,2}}{x_{q}\left(  z+x_{\bar{q}}\right)  }\right)  }\nonumber\\
&  +   \left.  \frac{H_{\bot}^{i}\left(  z(zd+d-2)+x_{q}\left(  2-4x_{\bar{q}%
}\right)  \right)  x_{\bar{q}}}{z\left(  z+x_{q}\right)  {}^{2}\left(
z+x_{\bar{q}}\right)  \left(  Q^{2}+\frac{\vec{p}{}_{\bar{q}2^{\prime}%
}^{\,\,2}}{\left(  z+x_{q}\right)  x_{\bar{q}}}\right)  }\right)
+(q\leftrightarrow\bar{q}). \numberthis[phi_4_phi_4_LT]
\end{align*}  }
Here, 
\begin{equation}
G_{\bot}^{i}=x_{\bar{q}}p_{g3^{\prime}\bot}^{i}-zp_{\bar{q}2^{\prime}\bot}%
^{i},\quad H_{\bot}^{i}=x_{q}p_{g3^{\prime}\bot}^{i}-zp_{q1^{\prime}\bot}%
^{i},\quad P_{\bot}^{i}=x_{\bar{q}}p_{g3\bot}^{i}-zp_{\bar{q}2\bot}^{i}.
\end{equation}
The interference term in the dipole $\times$ dipole contribution reads
\begin{align*}
& \left( \Phi_4^{i}(\vec{p}_{1}, \vec{p}_2, \vec{0}) \tilde{\Phi}_3^{+*}(\vec{p}_{1'}, \vec{p}_{2'}) + \tilde{\Phi}_3^{i}(\vec{p}_{1}, \vec{p}_{2}) \Phi_4^{+*} (\vec{p}_{1'}, \vec{p}_{2'}, \vec{0})\right) \\
& =4p_{\gamma}^{+}{}^{3}\left(  \frac{\Delta_{q}{}_{\bot}^{i}%
x_{q}x_{\bar{q}}\left(  dz^{2}+dz-2z+2x_{q}-4x_{q}x_{\bar{q}}\right)  }%
{\vec{\Delta}{}_{q}^{2}\left(  z+x_{q}\right)  {}^{2}\left(  z+x_{\bar{q}%
}\right)  \left(  Q^{2}+\frac{\vec{p}_{g}^{\,\,2}}{z}+\frac{\vec{p}{}%
_{q1}^{\,\,2}}{x_{q}}+\frac{\vec{p}{}_{\bar{q}2}^{\,\,2}}{x_{\bar{q}}}\right)
\left(  Q^{2}+\frac{\vec{p}{}_{\bar{q}2^{\prime}}^{\,\,2}}{\left(
z+x_{q}\right)  x_{\bar{q}}}\right)  }\right. \nonumber\\
& \hspace{-0.5 cm} - \frac{(\vec{J} \cdot \vec{\Delta}_{q})p_{\bar{q}2}^{i}{}_{\bot}\left(
dz^{2}+4x_{q}\left(  z+x_{q}\right)  \right)  \left(  1-2x_{\bar{q}}\right)
+z\left(  (\vec{J} \cdot \vec{p}_{\bar{q}2})\Delta_{q}^{i}{}_{\bot}-(\vec{p}_{\bar
{q}2} \cdot \vec{\Delta}_{q})J_{\bot}^{i}\right)  \left(  dz+4x_{\bar{q}}-4\right)
}{z\left(  z+x_{q}\right)  {}^{3}\vec{\Delta}{}_{q}^{2}\left(  Q^{2}%
+\frac{\vec{p}_{g}^{\,\,2}}{z}+\frac{\vec{p}{}_{q1^{\prime}}^{\,\,2}}{x_{q}%
}+\frac{\vec{p}{}_{\bar{q}2^{\prime}}^{\,\,2}}{x_{\bar{q}}}\right)  \left(
Q^{2}+\frac{\vec{p}{}_{\bar{q}2}^{\,\,2}}{\left(  z+x_{q}\right)  x_{\bar{q}}%
}\right)  \left(  Q^{2}+\frac{\vec{p}{}_{\bar{q}2^{\prime}}^{\,\,2}}{\left(
z+x_{q}\right)  x_{\bar{q}}}\right)  }\nonumber\\
& \hspace{-1 cm} -\frac{x_{q}\left(  z\left(  (\vec{K} \cdot \vec{p}_{\bar{q}2})\Delta_{q}^{i}%
{}_{\bot}-(\vec{p}_{\bar{q}2} \cdot \vec{\Delta}_{q})K_{\bot}^{i}\right)  \left(
dz+4x_{\bar{q}}-2\right)  +(\vec{K} \cdot \vec{\Delta}_{q})p_{\bar{q}2}^{i}{}_{\bot
}\left(  1-2x_{\bar{q}}\right)  \left(  z(dz-2)-4x_{q}x_{\bar{q}}\right)
\right)  }{z\left(  z+x_{q}\right)  {}^{2}x_{\bar{q}}\left(  z+x_{\bar{q}%
}\right)  \vec{\Delta}{}_{q}^{2}\left(  Q^{2}+\frac{\vec{p}_{g}^{\,\,2}}%
{z}+\frac{\vec{p}{}_{q1^{\prime}}^{\,\,2}}{x_{q}}+\frac{\vec{p}{}_{\bar
{q}2^{\prime}}^{\,\,2}}{x_{\bar{q}}}\right)  \left(  Q^{2}+\frac{\vec{p}%
{}_{\bar{q}2}^{\,\,2}}{\left(  z+x_{q}\right)  x_{\bar{q}}}\right)  \left(
Q^{2}+\frac{\vec{p}{}_{q1^{\prime}}^{\,\,2}}{x_{q}\left(  z+x_{\bar{q}%
}\right)  }\right)  }\nonumber\\
& \hspace{-0.6 cm} -\frac{z\left(  (\vec{p}_{q1} \cdot \vec{\Delta}_{q})X_{\bot}^{i}-(\vec{X} \cdot \vec{p}_{q1}) \Delta_{q}^{i}{}_{\bot}\right)  \left(  dz+4x_{q}-2\right)  +(\vec
{X} \cdot \vec{\Delta}_{q})p_{q1}^{i}{}_{\bot}\left(  1-2x_{q}\right)  \left(
z(dz-2)-4x_{q}x_{\bar{q}}\right)  }{z\vec{\Delta}{}_{q}^{2}\left(
z+x_{q}\right)  \left(  z+x_{\bar{q}}\right)  {}^{2}\left(  Q^{2}+\frac
{\vec{p}_{g}^{\,\,2}}{z}+\frac{\vec{p}{}_{q1}^{\,\,2}}{x_{q}}+\frac{\vec{p}%
{}_{\bar{q}2}^{\,\,2}}{x_{\bar{q}}}\right)  \left(  Q^{2}+\frac{\vec{p}%
{}_{\bar{q}2^{\prime}}^{\,\,2}}{\left(  z+x_{q}\right)  x_{\bar{q}}}\right)
\left(  Q^{2}+\frac{\vec{p}{}_{q1}^{\,\,2}}{x_{q}\left(  z+x_{\bar{q}}\right)
}\right)  }\nonumber\\
& \hspace{-0.8 cm} +\left.  \frac{z\left(  (\vec{X} \cdot \vec{p}_{q1})\Delta_{\bar{q}}^{i}{}_{\bot
}-(\vec{p}_{q1} \cdot \vec{\Delta}_{\bar{q}})X_{\bot}^{i}\right)  \left(
dz+4x_{q}-4\right)  -(\vec{X} \cdot \vec{\Delta}_{\bar{q}})p_{q1}^{i}{}_{\bot}\left(
2x_{q}-1\right)  \left(  4\left(  x_{q}-1\right)  x_{\bar{q}}-dz^{2}\right)
}{z\left(  z+x_{\bar{q}}\right)  {}^{3}\vec{\Delta}{}_{\bar{q}}^{2}\left(
Q^{2}+\frac{\vec{p}_{g}^{\,\,2}}{z}+\frac{\vec{p}{}_{q1}^{\,\,2}}{x_{q}}%
+\frac{\vec{p}{}_{\bar{q}2}^{\,\,2}}{x_{\bar{q}}}\right)  \left(  Q^{2}%
+\frac{\vec{p}{}_{q1}^{\,\,2}}{x_{q}\left(  z+x_{\bar{q}}\right)  }\right)
\left(  Q^{2}+\frac{\vec{p}{}_{q1^{\prime}}^{\,\,2}}{x_{q}\left(  z+x_{\bar
{q}}\right)  }\right)  }\right) \nonumber 
\end{align*}
\begin{equation}
    +(q\leftrightarrow\bar{q}) \; , \numberthis[phi_tilde_phi_4_dipole_dipole_LT]
\end{equation}
where 

\begin{equation}
\vec{\Delta}_{q} = \frac{x_q \vec{p}_g - x_g \vec{p}_q}{x_q + x_g} \; , \hspace{0.5 cm}
\vec{\Delta}_{\bar{q}} = \frac{x_{\bar{q}} \vec{p}_g - x_g \vec{p}_{\bar{q}}}{x_q + x_g} \; ,
\end{equation}
\begin{align}
X_{\bot}^{i}   =x_{\bar{q}}p_{g\bot}^{i}-zp_{\bar{q}2\bot}^{i} = & P_{\bot}%
^{i}|_{p_{3}=0},\quad J_{\bot}^{i}=x_{q}p_{g\bot}^{i}-zp_{q1^{\prime}\bot}%
^{i}=H_{\bot}^{i}|_{p_{3}^{\prime}=0},\nonumber\\
K_{\bot}^{i}  &  =x_{\bar{q}}p_{g\bot}^{i}-zp_{\bar{q}2^{\prime}\bot}%
^{i}=G_{\bot}^{i}|_{p_{3}^{\prime}=0}.
\end{align}
The TL transition is obtained from above by complex conjugation and inverting the naming of the different momenta in \eqref{eq:phi_tilde_phi_4_dipole_dipole_LT} and \eqref{eq:phi_4_phi_4_LT}. \\
The double-dipole $\times$ dipole have, respectively, the form 
\begin{equation}
\Phi_4^{i}(\vec{p}_{1}, \vec{p}_{2}, \vec{p}_{3}) \Phi_3^{+*}(\vec{p}_{1'}, \vec{p}_{2'}) = \Phi_4^i(\vec{p}_1, \vec{p}_2, \vec{p}_3) \Phi_4^{+*}(\vec{p}_{1'}, \vec{p}_{2'}, 0) + \Phi_4^i(\vec{p}_1, \vec{p}_2, \vec{p}_3) \tilde{\Phi}_3^{+*}(\vec{p}_{1'}, \vec{p}_{2'})  \; ,
\end{equation}
\begin{equation}
   \Phi_4^{+} (\vec{p}_{1}, \vec{p}_{2}, \vec{p}_{3}) \Phi_3 ^{i*}(\vec{p}_{1'}, \vec{p}_{2'}) = \Phi_4^{+} (\vec{p}_{1}, \vec{p}_{2}, \vec{p}_{3}) \Phi_4^{i*}(\vec{p}_{1'}, \vec{p}_{2'}, \vec{0}) + \Phi_4^{+} (\vec{p}_{1}, \vec{p}_{2}, \vec{p}_{3}) \tilde{\Phi}_3^{i*}(\vec{p}_{1'}, \vec{p}_{2'}) \; ,
\end{equation}
where 
{\allowdisplaybreaks
\begin{align}
& \Phi_4^i(\vec{p}_1, \vec{p}_2, \vec{p}_3) \tilde{\Phi}_3^{+*}(\vec{p}_{1'}, \vec{p}_{2'})  =\frac{4p_{\gamma}^{+}{}^{3}}{\left(  x_{q}+z\right)  \vec{\Delta}_{q}^{2}\left(  \frac{\vec{p}{}_{\bar{q}2^{\prime}}^{\,\,2}}%
{x_{\bar{q}}\left(  x_{q}+z\right)  }+Q^{2}\right)  \left(  \frac{\vec{p}%
{}_{q1}^{\,\,2}}{x_{q}}+\frac{\vec{p}{}_{\bar{q}2}^{\,\,2}}{x_{\bar{q}}}%
+\frac{\vec{p}_{g3}^{\; 2}}{z}+Q^{2}\right)  } \nonumber\\
&  \times\left\{  \frac{x_{q}x_{\bar{q}}\Delta_{q}^{i}\left(
dz(z+1)-2\left(  1-2x_{q}\right)  \left(  x_{q}+z\right)  \right)  }{\left(
x_{q}+z\right)  {}\left(  x_{\bar{q}}+z\right) } + \frac{\left(  dz+4x_{q}-2\right)  \left(  \Delta_{q}^{i}
\vec{P} \cdot \vec{p}_{q1} -P^{i}   \vec{p}_{q1} \cdot
\vec{\Delta}_{q}  \right)  }{\left(  x_{\bar{q}}+z\right)  {}%
^{2}\left(  \frac{\vec{p}_{q1}^{ \; 2}}{x_{q}\left(  x_{\bar{q}}+z\right)
}+Q^{2}\right)  } \right. \nonumber\\
&  +\frac{\left(  2x_{q}-1\right)  p_{q1}^{i} \vec{P} \cdot
\vec{\Delta}_{q} \left(  z(dz-2)-4x_{q}x_{\bar{q}}\right)
}{z\left(  x_{\bar{q}}+z\right)  {}^{2}\left(  \frac{\vec{p}_{q1}^{\; 2}%
}{x_{q}\left(  x_{\bar{q}}+z\right)  }+Q^{2}\right)} - \frac{\left(  (d-4)z-4x_{q}\right)  \left( W^{i}  \vec{p}_{\bar{q}2} \cdot \vec{\Delta}_{q}  -\Delta_{q}^{i} 
\vec{W} \cdot \vec{p}_{\bar{q}2} \right) }{\left(  x_{q}+z\right)
{}^{2}\left(  \frac{\vec{p}_{\bar{q}2}^{ \; 2}}{x_{\bar{q}}\left(
x_{q}+z\right)  }+Q^{2}\right)  } \nonumber\\
&  +\left.  \frac{\left(  2x_{\bar{q}}-1\right)  \left(  dz^{2}+4x_{q}\left(
x_{q}+z\right)  \right)  p_{\bar{q}2}^{i}  \vec{W} \cdot \vec{\Delta}_{q} }{z\left(  x_{q}+z\right)  {}^{2}\left(  \frac
{\vec{p}_{\bar{q}2}^{\; 2}}{x_{\bar{q}}\left(  x_{q}+z\right)  }%
+Q^{2}\right)  }\right\} +(q \leftrightarrow \bar{q}) \; ,
\end{align} }
and
{\allowdisplaybreaks
\begin{align}
\Phi_4^{+} (\vec{p}_{1}, \vec{p}_{2}, \vec{p}_{3}) & \tilde{\Phi}_3^{i*}(\vec{p}_{1'}, \vec{p}_{2'}) =\frac{4p_{\gamma}^{+}{}^{3}}{z\vec{\Delta}{}_{q}^{2}\left(
x_{q}+z\right)  {}^{2}\left(  Q^{2}+\frac{\vec{p}_{g3}^{\,\,2}}{z}+\frac
{\vec{p}{}_{q1}^{\,\,2}}{x_{q}}+\frac{\vec{p}{}_{\bar{q}2}^{\,\,2}}{x_{\bar
{q}}}\right)  \left(  Q^{2}+\frac{\vec{p}{}_{\bar{q}2^{\prime}}^{\,\,2}%
}{\left(  z+x_{q}\right)  x_{\bar{q}}}\right)  }\nonumber\\
&  \times\left[  \frac{x_{q}z\left(  (d-4)z-4x_{q}+2\right)  \left(
P^{i}\left(  \vec{p}_{\bar{q}2^{\prime}} \cdot \vec{\Delta}_{q}\right)
-\Delta_{q}^{i}\left( \vec{P} \cdot \vec{p}_{\bar{q}2^{\prime}}\right)
\right)  }{x_{\bar{q}}\left(  x_{\bar{q}}+z\right)  \left(  \frac
{\vec{p}_{q1}^{\; 2}}{x_{q}\left(  x_{\bar{q}}+z\right)  }+Q^{2}\right)
}\right.  \nonumber\\
&  -\frac{x_{q}\left(  x_{q}-x_{\bar{q}}+z\right)  p_{\bar{q}2^{\prime}}^{i}\left(  \vec{P} \cdot \vec{\Delta}_{q} \right)  \left(  z(dz-2)-4x_{q}%
x_{\bar{q}}\right)}{x_{\bar{q}}\left(  x_{\bar{q}}+z\right)  \left(
\frac{\vec{p}_{q1}^{\; 2}}{x_{q}\left(  x_{\bar{q}}+z\right)  }%
+Q^{2}\right)  }\nonumber\\
&  -\frac{\left(  x_{q}-x_{\bar{q}}+z\right)  \left(  dz^{2}+4x_{q}\left(
x_{q}+z\right)  \right)  p_{\bar{q}2^{\prime}}^{i}\left(  \vec{W} \cdot \vec{\Delta}_{q} \right)  }{\left(  x_{q}+z\right)  {}\left(
\frac{\vec{p}_{\bar{q}2}^{\; 2}}{x_{\bar{q}}\left(  x_{q}+z\right)
}+Q^{2}\right)  }\nonumber\\
&  -\left.  \frac{z\left(  (d-4)z-4x_{q}\right)  \left(  \Delta_{q}^{i} \left( \vec{W} \cdot \vec{p}_{\bar{q}2^{\prime}} \right) - W^{i} \left(
\vec{p}_{\bar{q}2^{\prime}} \cdot \vec{\Delta}_{q} \right)  \right)
}{\left(  x_{q}+z\right)  {}\left(  \frac{\vec{p}_{\bar{q}2}^{\,\, 2}%
}{x_{\bar{q}}\left(  x_{q}+z\right)  }+Q^{2}\right)  }\right] +(q \leftrightarrow \bar{q}) .
\end{align} }
Here, we introduced 
\begin{equation}
W_{\bot}^{i}=x_{q}p_{g3\bot}^{i}-zp_{q1\bot}^{i}.
\end{equation}

\subsection{TT transition}
The double-dipole $\times$ double-dipole contribution is 

\begin{align}
&  \Phi_{4}^{i}(p_{1\bot},p_{2\bot},p_{3\bot})\Phi_{4}^{k}(p_{1\bot}^{\prime
},p_{2\bot}^{\prime},p_{3\bot}^{\prime})^{\ast}=\left(  \frac{p_{\gamma}^{+}%
{}^{2}}{\left(  Q^{2}+\frac{\vec{p}_{g3}^{\;2}}{z}+\frac{\vec{p}_{q1}^{\;2}%
}{x_{q}}+\frac{\vec{p}_{\bar{q}2}^{\,\,2}}{x_{\bar{q}}}\right)  \left(
Q^{2}+\frac{\vec{p}_{g3^{\prime}}^{\;2}}{z}+\frac{\vec{p}_{q1^{\prime}}^{\;2}%
}{x_{q}}+\frac{\vec{p}_{\bar{q}2^{\prime}}^{\,\,2}}{x_{\bar{q}}}\right)
}\right. \nonumber\\
&  \times\left[  -\frac{g_{\bot}^{ik}x_{q}x_{\bar{q}}\left(zd+d-2+2x_{\bar
{q}}\right)  }{\left(  z+x_{q}\right)^{2}\left(z+x_{\bar{q}}\right)
}-\frac{2P_{\bot}^{k} p_{q1\bot}^{i}\left( 1-2x_{q}\right)  }{z\left(
z+x_{\bar{q}}\right)^{2}\left( Q^{2}+\frac{\vec{p}_{q1}^{\;2}}%
{x_{q}\left(  z+x_{\bar{q}}\right)  }\right)  }\left(  \frac{(d-2)z-2x_{\bar
{q}}}{z+x_{\bar{q}}}+\frac{dz+2x_{\bar{q}}}{z+x_{q}}\right)  \right.
\nonumber\\
&  -\frac{2\left(  g_{\bot}^{ik}(\vec{P} \cdot \vec{p}_{q1})+P_{\bot}^{i}p_{q1\bot}%
{}^{k}\right)  }{z\left(  z+x_{\bar{q}}\right)^{2}\left(  Q^{2}+\frac
{\vec{p}_{q1}^{\;2}}{x_{q}\left(  z+x_{\bar{q}}\right)  }\right)  }\left(
\frac{(d-4)z-2x_{\bar{q}}}{z+x_{q}}+\frac{(d-2)z-2x_{\bar{q}}}{z+x_{\bar{q}}%
}\right) \nonumber
\end{align}%
\begin{align}
&  -\frac{1}{z^{2} x_{q}\left(  z+x_{q}\right)^{2} x_{\bar{q}}\left(
z+x_{\bar{q}}\right)^{2}\left(  Q^{2}+\frac{\vec{p}_{\bar{q}2^{\prime}%
}^{\,\,2}}{\left(  z+x_{q}\right)  x_{\bar{q}}}\right)  \left(  Q^{2}%
+\frac{\vec{p}_{q1}^{\;2}}{x_{q}\left(  z+x_{\bar{q}}\right)  }\right)
}\left\{  (\vec{H}  \cdot \vec{P})\left[  p_{q1\bot}^{i}{}p_{\bar{q}2^{\prime}\bot
}^{k}{}\left(  1-2x_{q}\right)  \right.  \right. \nonumber\\
&  \times\left.  \left(  1-2x_{\bar{q}}\right)  \left(  z(2-dz)+4x_{q}%
x_{\bar{q}}\right)  +(g_{\bot}^{ik}(\vec{p}_{q1}  \cdot \vec{p}_{\bar{q}2^{\prime}%
})+p_{q1\bot}^{k} p_{\bar{q}2^{\prime}\bot}^{i})\left(  z(2-(d-4)z)+4x_{q}%
x_{\bar{q}}\right)  \right] \nonumber\\
&  +((d-4)z-2)\left[  z(\vec{H}  \cdot  \vec{p}_{\bar{q}2^{\prime}})(g_{\bot}^{ik}%
(\vec{P}  \cdot  \vec{p}_{q1})+P_{\bot}^{i}p_{q1\bot}^{k})+z H_{\bot}^{k}\left(
(\vec{P}  \cdot \vec{p}_{q1})p_{\bar{q}2^{\prime}\bot}^{i}-(\vec{p}_{q1}  \cdot \vec{p}_{\bar{q}2^{\prime}})P_{\bot}^{i}\right)  \right] \nonumber\\
&  +((d-4)z+2)\left[  zH^{i}\left(  (\vec{P}  \cdot  \vec{p}_{\bar{q}2^{\prime}%
})p_{q1\bot}^{k}-(\vec{p}_{q1}  \cdot \vec{p}_{\bar{q}2^{\prime}}) P_{\bot}%
^{k}\right)  +z(\vec{H}  \cdot  \vec{p}_{q1})(g_{\bot}^{ik}(\vec{P}  \cdot \vec{p}_{\bar{q}2^{\prime}})+P_{\bot}^{k}p_{\bar{q}2^{\prime}\bot}^{i})\right]
\nonumber\\
&  +\left.  2z\left(  (\vec{H}  \cdot  \vec{p}_{\bar{q}2^{\prime}})P_{\bot}^{k}%
-(\vec{P}  \cdot  \vec{p}_{\bar{q}2^{\prime}}) H_{\bot}^{k}\right)  p_{q1\bot}{}%
^{i}\left(  1-2x_{q}\right)  \left(  dz+4x_{\bar{q}}-2\right)  \right\}
\nonumber
\end{align}%
\begin{align}
&  -\frac{1}{z^{2}x_{q}x_{\bar{q}}\left(  z+x_{\bar{q}}\right)  {}^{4}\left(
Q^{2}+\frac{\vec{p}_{q1}^{\;2}}{x_{q}\left(  z+x_{\bar{q}}\right)  }\right)
\left(  Q^{2}+\frac{\vec{p}_{q1^{\prime}}^{\;2}}{x_{q}\left(  z+x_{\bar{q}%
}\right)  }\right)  }\left\{  z\left(  (d-4)z-4x_{\bar{q}}\right)  \frac{{}%
}{{}}\right. \nonumber\\
&  \times\left[  g_{\bot}^{ik}\left(  (\vec{G}  \cdot \vec{p}_{q1^{\prime}}) (\vec{P}  \cdot  \vec{p}_{q1})-(\vec{G}  \cdot  \vec{p}_{q1})(\vec{P}  \cdot  \vec{p}_{q1^{\prime}})\right) +(\vec{p}_{q1}  \cdot  \vec{p}_{q1^{\prime}})\left(  G_{\bot}^{i}P_{\bot}^{k}-G_{\bot}^{k}P_{\bot}^{i}\right)  \right. \nonumber\\
&  +\left.  2(\vec{G}  \cdot  \vec{p}_{q1^{\prime}})\left(  P_{\bot}^{i}p_{q1\bot}^{k}+P_{\bot}^{k} p_{q1\bot}^{i}\left(  1-2x_{q}\right)  \right)
-2(\vec{G}  \cdot  \vec{p}_{q1}) \left(  P_{\bot}^{k} p_{q1^{\prime}\bot}^{i}+P_{\bot}^{i} p_{q1^{\prime}\bot}^{k}\left(  1-2x_{q}\right)  \right)  \right] \nonumber\\
&  +\left.  \left.  (\vec{G}  \cdot  \vec{P})\left[  p_{q1\bot}^{k} p_{q1^{\prime}\bot}^{i}-p_{q1\bot}^{i} p_{q1^{\prime}\bot}^{k}\left(  1-2x_{q}\right)^{2}+g_{\bot}^{ik}\left(  \vec{p}_{q1}  \cdot  \vec{p}_{q1^{\prime}}\right)  \right] \left(  dz^{2}+4x_{\bar{q}}\left(  z+x_{\bar{q}}\right)  \right)  \right\}
\right] \nonumber\\
&  +\left.  \frac{{}}{{}}(1\leftrightarrow1^{\prime},2\leftrightarrow
2^{\prime},3\leftrightarrow3^{\prime},i\leftrightarrow k)\right)
+(q\leftrightarrow\bar{q}).
\end{align}
The interference term in the dipole $\times$ dipole contribution reads
\begin{align*}
& \left( \tilde{\Phi}_3^i(\vec{p}_1,\vec{p}_2)\Phi_4^{k*}(\vec{p}_{1'}, \vec{p}_{2'}, \vec{0}) + \Phi_4^i(\vec{p}_1, \vec{p}_{2}, \vec{0}) \tilde{\Phi}_3^{k*}(\vec{p}_{1'}, \vec{p}_{2'}) \right) \\
&  =\left(  \frac{2p_{\gamma}^{+}{}^{2}}{\vec{\Delta}{}_{q}^{2}\left(
Q^{2}+\frac{\vec{p}_{g}^{\,\,2}}{z}+\frac{\vec{p}{}_{q1}^{\,\,2}}{x_{q}}%
+\frac{\vec{p}{}_{\bar{q}2}^{\,\,2}}{x_{\bar{q}}}\right)  \left(  Q^{2}%
+\frac{\vec{p}{}_{\bar{q}2^{\prime}}^{\,\,2}}{\left(  z+x_{q}\right)
x_{\bar{q}}}\right)  }\right. \\
& \times \left[  \frac{\left(  (d-2)z-2x_{q}\right)  x_{q}}{\left(
z+x_{q}\right)  {}^{3}}\left(  g_{\bot}^{ik}(\vec{p}_{\bar{q}2^{\prime}} \cdot \vec{\Delta}_{q})+p_{\bar{q}2^{\prime}}{}_{\bot}^{i}\Delta_{q\bot}^{k} +p_{\bar{q}2^{\prime}\bot}^{k}\Delta_{q\bot}^{i}\left(  1-2x_{\bar{q}}\right)  \right)  \right. \\
&  +\frac{x_{q}\left(  \left(  (d-4)z-2x_{q}\right)  \left(  g_{\bot}^{ik}(\vec{p}_{\bar{q}2^{\prime}} \cdot \vec{\Delta}_{q})+p_{\bar{q}2^{\prime}\bot}^{i}{}\Delta_{q\bot}^{k}{}\right)  +p_{\bar{q}2^{\prime}\bot}^{k} \Delta_{q\bot}^{i} \left(  dz+2x_{q}\right)  \left(  1-2x_{\bar{q}}\right) \right)  }{\left(  z+x_{q}\right)  {}^{2}\left(  z+x_{\bar{q}}\right)  } \\
& -\frac{1}{z\left(  z+x_{q}\right)  {}^{2}x_{\bar{q}}\left(  z+x_{\bar{q}%
}\right)  {}^{2}\left(  Q^{2}+\frac{\vec{p}_{q1}^{\;2}}{x_{q}\left(
z+x_{\bar{q}}\right)  }\right)  }\left\{  z((d-4)z+2)\frac{{}}{{}}\right. \\
& \hspace{-0.1 cm} \times \hspace{-0.1 cm} \left[  p_{q1}{}_{\bot}^{i} \left(  (\vec{p}_{\bar{q}2^{\prime}}\cdot \vec{\Delta}_{q})X_{\bot}^{k}-(\vec{X} \cdot \vec{p}_{\bar{q}2^{\prime}}) \Delta_{q \bot}^{k}\right)  \left( 2x_{q}-1\right)  -(\vec{X} \cdot \vec{p}_{\bar{q}2^{\prime}})\left(  g_{\bot}^{ik}(\vec{p}_{q1} \cdot \vec{\Delta}_{q})+p_{q1\bot}^{k} \Delta_{q\bot}^{i}\right)  \right. \\
&  \left. - \hspace{-0.05 cm}  X_{\bot}^{k} \hspace{-0.05 cm} \left(  (\vec{p}_{q1} \hspace{-0.05 cm} \cdot \hspace{-0.05 cm} \vec{\Delta}_{q}) p_{\bar
{q}2^{\prime} \bot}^{i} \hspace{-0.1 cm} -(\vec{p}_{q1} \hspace{-0.05 cm} \cdot \vec{p}_{\bar{q}2^{\prime}})\Delta_{q}{}_{\bot}^{i}\right)  \right] \hspace{-0.05 cm} + \hspace{-0.05 cm} 4x_{q}z \left(  1 \hspace{-0.05 cm} - \hspace{-0.05 cm} 2x_{q}\right)
p_{q1\bot}^{i} \hspace{-0.1 cm} \left(  (\vec{p}_{\bar{q}2^{\prime}} \hspace{-0.05 cm} \cdot \hspace{-0.05 cm} \vec{\Delta}%
_{q})X_{\bot}^{k} \hspace{-0.1 cm} - \hspace{-0.1 cm}(\vec{X} \hspace{-0.05 cm} \cdot \vec{p}_{\bar{q}2^{\prime}}) \Delta_{q\bot}%
^{k}\right)  \\
&  +z\left(  1-2x_{\bar{q}}\right)  \left(  dz+4x_{q}-2\right)  p_{\bar
{q}2^{\prime}\bot}^{k}\left(  (\vec{p}_{q1} \cdot \vec{\Delta}_{q}) X_{\bot}%
^{i}-(\vec{X} \cdot \vec{p}_{q1}) \Delta_{q\bot}^{i}\right)  -z((d-4)z-2)\\
&  \times \left[  \left(  g_{\bot}^{ik} (\vec{X} \cdot \vec{p}_{q1})+X_{\bot}^{i} 
p_{q1\bot}^{k}\right)  (\vec{p}_{\bar{q}2^{\prime}} \cdot \vec{\Delta}%
_{q})+ \left(  (\vec{X}\vec{p}_{q1})p_{\bar{q}2^{\prime}}{}_{\bot}^{i}-(\vec
{p}_{q1}\vec{p}_{\bar{q}2^{\prime}})X_{\bot}^{i}\right)  \Delta_{q\bot
}^{k}\right] \\
&  +(\vec{X} \cdot \vec{\Delta}_{q})p_{q1\bot}^{i}p_{\bar{q}2^{\prime} \bot
}^{k}\left( 1-2x_{q}\right)  \left( 1-2x_{\bar{q}}\right)  \left(
z(d z-2)-4x_{q}x_{\bar{q}}\right) \\
&  - \left.  (\vec{X} \cdot \vec{\Delta}_{q})\left(  g_{\bot}^{ik}(\vec{p}_{q1} \cdot \vec{p}_{\bar{q}2^{\prime}})+p_{q1\bot}^{k} p_{\bar{q}2^{\prime}\bot}^{i}\right)  \left(  z(2-(d-4)z)+4x_{q}x_{\bar{q}}\right)  \right\} \\
& -    \frac{1}{z\left(  z+x_{q}\right)  {}^{4}\left(  Q^{2}+\frac{\vec{p}%
{}_{\bar{q}2}^{\,\,2}}{\left(  z+x_{q}\right)  x_{\bar{q}}}\right)  x_{\bar
{q}}}\left\{  z\left(  dz+4x_{\bar{q}}-4\right)  \left[  \left(  1-2x_{\bar
{q}}\right)  \frac{{}}{{}}\right.  \right. \nonumber\\
&  \times  \left(  p_{\bar{q}2^{\prime}\bot}^{k} \left(  (\vec{p}_{\bar
{q}2} \cdot \vec{\Delta}_{q}) V_{\bot}^{i}-(\vec{V} \cdot \vec{p}_{\bar{q}2})\Delta_{q\bot}^{i} \right)  +p_{\bar{q}2\bot}^{i}\left(  (\vec{V}\vec{p}_{\bar{q}2^{\prime}})\Delta_{q\bot}^{k}-(\vec{p}_{\bar{q}2^{\prime}} \cdot \vec{\Delta}_{q}) V_{\bot}^{k}\right)  \right) \nonumber\\
& +    V_{\bot}^{k} \left(  (\vec{p}_{\bar{q}2} \cdot \vec{\Delta}_{q}) p_{\bar
{q}2^{\prime}\bot}^{i}-(\vec{p}_{\bar{q}2} \cdot \vec{p}_{\bar{q}2^{\prime}}) \Delta_{q\bot}^{i}\right)  + \left(  (\vec{p}_{\bar{q}2} \cdot \vec{p}_{\bar
{q}2^{\prime}}) V_{\bot}^{i}-(\vec{V} \cdot \vec{p}_{\bar{q}2}) p_{\bar{q}2^{\prime}\bot}^{i}\right)  \Delta_{q \bot}^{k}\nonumber\\
& +    \left.  g_{\bot}^{ik} \left(  (\vec{V} \cdot \vec{p}_{\bar{q}2^{\prime}}%
)(\vec{p}_{\bar{q}2} \cdot \vec{\Delta}_{q})-(\vec{V} \cdot \vec{p}_{\bar{q}2} )(\vec
{p}_{\bar{q}2^{\prime}} \cdot  \vec{\Delta}_{q}) \right)  + p_{\bar{q}2\bot}^{k}\left(  (\vec{V} \cdot \vec{p}_{\bar{q}2^{\prime}}) \Delta_{q\bot}^{i} 
- (\vec{p}_{\bar{q}2^{\prime}} \cdot \vec{\Delta}_{q}) V_{\bot}^{i}\right)  \right]
\nonumber\\
& +    \left.  \left.  (\vec{V} \cdot \vec{\Delta}_{q})\left(  p_{\bar{q}2\bot
}^{i}p_{\bar{q}2^{\prime}\bot}^{k}\left( 1-2x_{\bar{q}}\right)^{2}-g_{\bot}^{ik}(\vec{p}_{\bar{q}2} \cdot \vec{p}_{\bar{q}2^{\prime}})-p_{\bar{q}2\bot}^{k} p_{\bar{q}2^{\prime} \bot}^{i}\right)  \left(d z^{2}-4x_{q}\left( x_{\bar{q}}-1\right)  \right)  \right\}  \frac{{}}{{}%
}\right] \nonumber \\
&  +   \left.  \frac{{}}{{}}(1\leftrightarrow1^{\prime},2\leftrightarrow
2^{\prime},i\leftrightarrow k)\right)  +(q\leftrightarrow\bar{q}). \numberthis
\end{align*}
Here,
\begin{equation}
V_{\bot}^{i}=x_{q}p_{g\bot}^{i}-zp_{q1\bot}^{i}.
\end{equation}
The double-dipole $\times$ dipole contribution has the form  
\begin{equation}
\Phi_{4}^{i}(\vec{p}_1, \vec{p}_2, \vec{p}_3)\,\Phi_{3}^{k*}(\vec{p}_{1'}, \vec{p}_{2'})=\Phi_{4}^{i}(\vec{p}_1, \vec{p}_2, \vec{p}_3)\Phi_{4}^{k*}(\vec{p}_{1'}, \vec{p}_{2'}, \vec{0})+\Phi_4^i(\vec{p}_1, \vec{p}_2, \vec{p}_3) \tilde{\Phi}_3^{k*}(\vec{p}_{1'}, \vec{p}_{2'}),
\end{equation}
where
\begin{align*}
& \Phi_4^i(\vec{p}_1, \vec{p}_2, \vec{p}_3) \tilde{\Phi}_3^{k*}(\vec{p}_{1'}, \vec{p}_{2'})  \\
&  =\frac{2p_{\gamma}^{+}{}^{2}}{\vec{\Delta}{}_{q}^{2}\left(
Q^{2}+\frac{\vec{p}_{g3}^{\,\,2}}{z}+\frac{\vec{p}{}_{q1}^{\,\,2}}{x_{q}%
}+\frac{\vec{p}{}_{\bar{q}2}^{\,\,2}}{x_{\bar{q}}}\right)  \left(  Q^{2}%
+\frac{\vec{p}{}_{\bar{q}2^{\prime}}^{\,\,2}}{\left(  z+x_{q}\right)
x_{\bar{q}}}\right)  }\nonumber\\
&  \times\left[  \frac{\left(  (d-2)z-2x_{q}\right)  x_{q}}{\left(
z+x_{q}\right)  {}^{3}}\left(  g_{\bot}^{ik}(\vec{p}_{\bar{q}2^{\prime}}%
\vec{\Delta}_{q})+p_{\bar{q}2^{\prime}}{}_{\bot}^{i}\Delta_{q\bot}{}%
^{k}+p_{\bar{q}2^{\prime}\bot}{}^{k}\Delta_{q\bot}{}^{i}\left(  1-2x_{\bar{q}%
}\right)  \right)  \right. \nonumber\\
&  +\frac{x_{q}\left(  \left(  (d-4)z-2x_{q}\right)  \left(  g_{\bot}%
^{ik}(\vec{p}_{\bar{q}2^{\prime}}\vec{\Delta}_{q})+p_{\bar{q}2^{\prime}\bot
}^{i}{}\Delta_{q\bot}^{k}{}\right)  +p_{\bar{q}2^{\prime}\bot}^{k}{}%
\Delta_{q\bot}^{i}{}\left(  dz+2x_{q}\right)  \left(  1-2x_{\bar{q}}\right)
\right)  }{\left(  z+x_{q}\right)  {}^{2}\left(  z+x_{\bar{q}}\right)
}\nonumber\\
&  -\frac{1}{z\left(  z+x_{q}\right)  {}^{2}x_{\bar{q}}\left(  z+x_{\bar{q}%
}\right)  {}^{2}\left(  Q^{2}+\frac{\vec{p}_{q1}^{\;2}}{x_{q}\left(
z+x_{\bar{q}}\right)  }\right)  }\left\{  z((d-4)z+2)\frac{{}}{{}}\right.
\nonumber\\
&  \times\left[  p_{q1}{}_{\bot}^{i}\left(  (\vec{p}_{\bar{q}2^{\prime}}%
\vec{\Delta}_{q})P_{\bot}^{k}-(\vec{P}\vec{p}_{\bar{q}2^{\prime}})\Delta_{q}%
{}_{\bot}^{k}\right)  \left(  2x_{q}-1\right)  -(\vec{P}\vec{p}_{\bar
{q}2^{\prime}})\left(  g_{\bot}^{ik}(\vec{p}_{q1}\vec{\Delta}_{q})+p_{q1}%
{}_{\bot}^{k}\Delta_{q}{}_{\bot}^{i}\right)  \right. \nonumber\\
&  -\!\left.  P_{\bot}^{k}\!\left(  \!(\vec{p}_{q1}\vec{\Delta}_{q})p_{\bar
{q}2^{\prime}}{}_{\bot}^{i}\!-\!(\vec{p}_{q1}\vec{p}_{\bar{q}2^{\prime}%
})\Delta_{q}{}_{\bot}^{i}\!\right)  \!\right]  +4x_{q}z\left(  1-2x_{q}%
\right)  p_{q1}{}_{\bot}^{i}\!\left(  \!(\vec{p}_{\bar{q}2^{\prime}}%
\vec{\Delta}_{q})P_{\bot}^{k}-(\vec{P}\vec{p}_{\bar{q}2^{\prime}})\Delta_{q}%
{}_{\bot}^{k}\!\right) \nonumber\\
&  +z\left(  1-2x_{\bar{q}}\right)  \left(  dz+4x_{q}-2\right)  p_{\bar
{q}2^{\prime}}{}_{\bot}^{k}\left(  (\vec{p}_{q1}\vec{\Delta}_{q})P_{\bot}%
^{i}-(\vec{P}\vec{p}_{q1})\Delta_{q}{}_{\bot}^{i}\right)
-z((d-4)z-2)\nonumber\\
&  \times\left[  \left(  g_{\bot}^{ik}(\vec{P}\vec{p}_{q1})+P_{\bot}^{i}%
p_{q1}{}_{\bot}^{k}\right)  (\vec{p}_{\bar{q}2^{\prime}}\vec{\Delta}%
_{q})+\left(  (\vec{P}\vec{p}_{q1})p_{\bar{q}2^{\prime}}{}_{\bot}^{i}-(\vec
{p}_{q1}\vec{p}_{\bar{q}2^{\prime}})P_{\bot}^{i}\right)  \Delta_{q}{}_{\bot
}^{k}\right] \nonumber\\
&  +(\vec{P}\vec{\Delta}_{q})p_{q1}{}_{\bot}^{i}p_{\bar{q}2^{\prime}}{}_{\bot
}^{k}\left(  1-2x_{q}\right)  \left(  1-2x_{\bar{q}}\right)  \left(
z(dz-2)-4x_{q}x_{\bar{q}}\right) \nonumber\\
&  -\left.  (\vec{P}\vec{\Delta}_{q})\left(  g_{\bot}^{ik}(\vec{p}_{q1}\vec
{p}_{\bar{q}2^{\prime}})+p_{q1}{}_{\bot}^{k}p_{\bar{q}2^{\prime}}{}_{\bot}%
^{i}\right)  \left(  z(2-(d-4)z)+4x_{q}x_{\bar{q}}\right)  \right\} \; . \numberthis
\end{align*}%
As above, the dipole $\times$ double-dipole contribution is obtained by complex conjugation and changing the momenta. 
\end{appendices}

\printbibliography[heading=bibintoc]

\end{document}